\documentclass[12pt,a4paper,oneside]{book} %book ,doublepage,openright

\pdfoutput=1

\usepackage[british]{babel}
\usepackage[utf8x]{inputenc} %ansinew
\usepackage{amsmath,amssymb}
\usepackage{hyperref}
\usepackage{eprint}
\usepackage{subeqn}
\usepackage{graphicx}
\usepackage{verbatim}
\usepackage{subeqn}
\usepackage{mathrsfs}
\usepackage{tikz}
\usepackage{setspace}
\usepackage{floatflt}
\usepackage{array}
\usepackage{geometry}
\let\MakeUppercase\relax % used to write headings lower case, may affect other parts - be careful: could also use \usepackage{fancyhdr} to customize the header

\newtheorem{Def}{Definition}

\DeclareSymbolFontAlphabet{\mathrsfs}{rsfs}
\newcommand{\scri}{\mathrsfs{I}}
\newcommand{\rscri}{r_{\!\!\scri}}
\newcommand{\rtort}{\tilde r_{*}}
\newcommand{\aconf}{\bar\Omega}
\newcommand{\CZ}{Z4c}
\newcommand{\hpt}{h'(\tilde r)}

\newcommand{\mttr}{\bar\gamma}%\,^{(4)}
\newcommand{\Ktr}{\bar K}%\,^{(4)} % used for the conformal trace K, the one mixed with Theta is simply K
\newcommand{\Kc}{K_{CMC}}
\newcommand{\Cc}{C_{CMC}}
\newcommand{\iPhi}{\tilde\Phi}
\newcommand{\iPi}{\tilde\Pi}
\newcommand{\bPhi}{\bar\Phi}
\newcommand{\bPi}{\bar\Pi}
\newcommand{\other}{\check}
\newcommand{\oOmega}{\other\Omega}
\newcommand{\om}{\other g}
\newcommand{\A}{A}
\newcommand{\atscri}[1]{\left.#1\right|_{\scri}}
\newcommand{\atscrip}[1]{\left.#1\right|_{\scri^+}}
\newcommand{\cL}{\xi_{\beta^rBH}}
\newcommand{\pphi}{\xi_{\alpha BH}}

\newcommand{\tildeu}{u}
\newcommand{\tildev}{v}

\newcommand{\K}{\Delta K}
\newcommand{\DPK}{\Delta \tilde{K}}
\newcommand{\odK}{\delta K} % old variable used in -r 629 of paper 1

\newcommand{\puK}{\tilde{\bar K}} % used for the physical trace K
\newcommand{\pK}{\tilde K} % used for the physical trace K mixed with Theta

\newcommand{\cK}{K} % used for the conformal trace K mixed with Theta
\newcommand{\cT}{\Theta}% used for the 'conformal' Theta

\newcommand{\pT}{\tilde \Theta}% used for the 'physical' Theta
\newcommand{\Te}{\tilde \Theta}% used for the 'physical' Theta in the spherically symmetric equations

\newcommand{\rootp}{\ root}

\newcommand{\eref}[1]{(\ref{#1})}
\newcommand{\fref}[1]{figure~\ref{#1}}
\newcommand{\tref}[1]{table~\ref{#1}}
\newcommand{\case}[2]{{\textstyle\frac{#1}{#2}}}\def\pt(#1){({\it #1\/})}

{\cleardoublepage\null \vfill\begin{center}%
\bfseries \abstractname \end{center}}%
{\vfill\null}

\newcommand{\upda}{\textcolor{red}}

\begin{document}

\frontmatter	  % Begin Roman style (i, ii, iii, iv...) page numbering

%\begin{comment}

\title{Free evolution of the hyperboloidal initial value problem in spherical symmetry} % original: Simulation and detection of gravitational waves emitted by black hole binary systems
\author{Alex Vano-Vinuales \\ Director: Sascha Husa}
\date{May 2015}

\begin{titlepage}

\begin{center}

\includegraphics[width=0.4\textwidth]{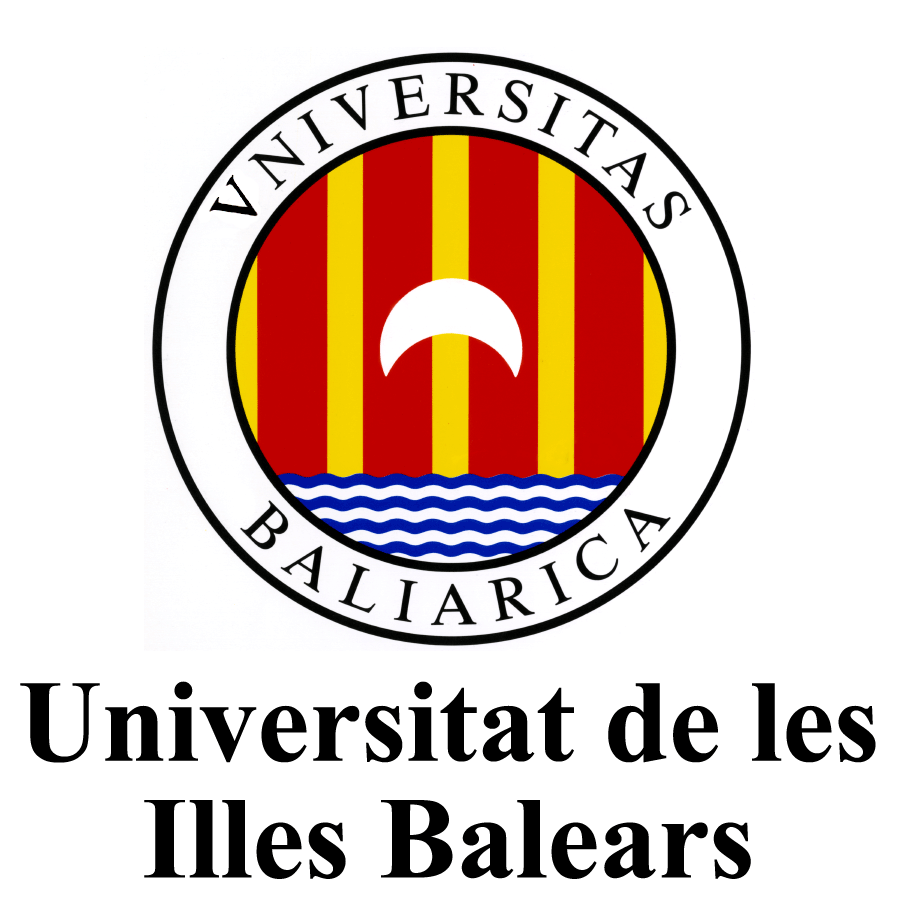}

\vspace{1.cm}

\textsc{\Large DOCTORAL THESIS}\\[0.6cm]

{\LARGE 2015}

\vspace{1.1cm}

% Title
{\begin{spacing}{1.4} 
\Large \bfseries FREE EVOLUTION OF THE HYPERBOLOIDAL INITIAL VALUE PROBLEM IN SPHERICAL SYMMETRY %Free evolution of the hyperboloidal initial value problem in spherical symmetry % original: Simulation and detection of gravitational waves emitted by black hole binary systems
\end{spacing}}

\vspace{9.0cm}

% Author 
{\Large \bfseries Alex Va\~n\'o Vi\~nuales}

\vfill

%\vspace{2.0cm}

\end{center}

\end{titlepage}

\thispagestyle{empty}

\begin{titlepage}

\begin{center}

\includegraphics[width=0.4\textwidth]{figures/uib.png}

\vspace{1.cm}

\textsc{\Large DOCTORAL THESIS}\\[0.6cm]

{\LARGE 2015}

\vspace{1.0cm}

{\Large \bfseries Doctoral Programme of Physics}

\vspace{0.8cm}

% Title
{\begin{spacing}{1.4} 
\Large \bfseries FREE EVOLUTION OF THE HYPERBOLOIDAL INITIAL VALUE PROBLEM IN SPHERICAL SYMMETRY %Free evolution of the hyperboloidal initial value problem in spherical symmetry % original: Simulation and detection of gravitational waves emitted by black hole binary systems
\end{spacing}}

\vspace{1.2cm}

% Author 
{\Large \bfseries Alex Va\~n\'o Vi\~nuales}
\end{center}

\vspace{1.4cm}

{\begin{spacing}{1.1} \Large \bfseries
Thesis Supervisor: Sascha Husa

Thesis Co-supervisor: Alicia M. Sintes Olives

%Thesis tutor: N/A
\end{spacing}}

\vspace{2.0cm}

{\Large \bfseries Doctor by the Universitat de les Illes Balears}

\vfill

\end{titlepage}

\thispagestyle{empty}

%\layout

\clearpage

%\thispagestyle{empty}

%\begin{abstract}abstract...\end{abstract}
\chapter*{Summary} %\large{ }
\addcontentsline{toc}{chapter}{Summary}

The present work deals with the application of conformal compactification methods to the numerical solution of the Einstein equations, the field equations of General Relativity. They form a complex system of non-linear partial differential equations that can only be solved analytically for highly symmetric spacetimes. The most general spacetimes have to be obtained with the help of numerical techniques. In General Relativity, central physical quantities such as the total energy or radiation flux can only be defined unambiguously in the asymptotic region of a spacetime, which calls for the numerical treatment of infinite domains.

The traditional approach in Numerical Relativity codes is based on spacelike slices that are cut at an artificial timelike boundary and whose data are extrapolated to infinity.
The goal of this thesis is to further develop an elegant alternative approach, which aims to efficiently solve the Einstein equations for spacetimes of isolated radiating systems and compute the radiation signal without any approximations. Following a framework by Penrose, we use a finite unphysical spacetime related to the physical one by a conformal rescaling. On this rescaled spacetime, taking limits towards infinity is replaced by local differential geometry and observable physical quantities can be directly evaluated.

In order to compute radiation quantities, it is convenient to foliate spacetime by hyperboloidal slices. These are smooth spacelike slices that reach future null infinity, the ``place'' in spacetime where light rays arrive. Among the advantages of evolving on compactified hyperboloidal slices are that no boundary conditions are required, because future null infinity is an ingoing null surface and it does not allow any information to enter the domain from beyond.
The price to pay is that the conformally rescaled Einstein equations are singular at infinity and need to be regularized. Besides, the nontrivial background geometry of the hyperboloidal slices makes the evolution equations prone to continuum instabilities.

As a first step towards developing numerical algorithms for the hyperboloidal initial value problem for strong field dynamical spacetimes, the numerical work in this thesis is restricted to spherical symmetry. Given that the regularization of the radial direction is common to spherical symmetry and the full three-dimensional case, the results obtained are expected to apply, at least to some degree, to the full system.

This work's approach uses standard unconstrained formulations of General Relativity, specifically the BSSN (Baumgarte-Shapiro-Shibata-Nakamura) equations and the Z4 equations. The derivation of their spherically symmetric component equations will be described, as well as the calculation of appropriate initial data on the hyperboloidal slice given by a constant-mean-curvature foliation. A critical point is the treatment of the gauge conditions: both the specific requirements for the hyperboloidal value problem, such as scri-fixing or the preferred conformal gauge, and the adaptation of currently common gauge choices will be explained.
As expected, the numerical implementation was difficult to stabilize, but by means of a variable transformation on the trace of the extrinsic curvature and the addition of a constraint damping term to the evolution equation of the contracted connection, the implementation finally became well-behaved.
Stable simulations of the Einstein equations coupled to a massless scalar field have been performed with regular and strong field initial data. Small perturbations of regular initial data give stationary data that are stable forever, while larger scalar field perturbations result in the formation of a black hole. Schwarzschild trumpet initial data have been found to slowly drift away from the expected stationary values, but the effect for small perturbations is slow enough to allow the observation of the power-law decay tails of the scalar field.

\chapter*{Acknowledgments}
\addcontentsline{toc}{chapter}{Acknowledgments}

%I am grateful 
%I wish to express my sicere thanks to 
%I place on record, my sincere thank you to 
%I am also grateful to 
%I am extremly thankful and indebted 
%I take this opportunity to express gratitude to 
%I also thank 
%I also place on record, my sense of gratitude to one and all, who directly or indirectly, have lemt their hand in this venture. 
%My sincere thanks also goes to

%\noindent First I would like to thank the Relativity Group of the Universitat de les Illes Balears; especially Alicia Sintes, for giving me useful advice and recommendations during the last years, and my supervisor Sascha Husa, for guiding me through the vast field of Numerical Relativity and helping me find a topic to work on, as well as for all his explanations in both practical and theoretical issues.

%\noindent I am grateful to the Group for Gravitational Physics of the Universit\"at Wien for the interesting discussions and in particular to Mark Hannam for his assistance and comments to my work during my summer research stay. %making me feel at home

%\noindent Finally, I would also like to thank the Relativity Group of the Friedrich-Schiller-Universit\"at Jena for assiting me during my first contact with Numerical Relativity. %first contact with %first introducing me to

%Sascha and Alicia
%rest of the group

%David

%Sergio Dain and group 

%Wien group, Mark - not really here ... 

%An\i l?

%Jena group

%office people

%phds from Córdoba, especially Ivan Gentile de Austria

%friends and family: old friends, trip friends, Chinese classmates, choir, family for support

First I would like to express my sincere thanks to Prof.~Sascha Husa for being my PhD advisor, for his guidance and encouragement and for all his explanations and useful help. Also for his patience and time invested in discussions when I would come up to his office confused and with a crashing code. 
I want to thank Prof.~Alicia M. Sintes Olives for being the co-director of my work and for all of her useful help and comments both in science and in many other academic and research topics. 

I am really grateful to Dr.~David Hilditch, for the interesting work we did together during his stay in Palma, for his help and advice regarding my research and for all the useful lessons I learned from him. 
%his moral and practical support writing letters 
%
I thank Prof.~Sergio Dain for supervising me during my research stay in the Gravitation and Relativity Group of the FaMAF in the Universidad Nacional de Córdoba and for the useful lessons I received. I also want to express my gratitude to the rest of members of the group, for both comments and recommendations on my work and the interesting discussions in the group seminars. 

\vspace{1ex}

I am grateful to the Relativity and Cosmology Group of the Universitat de les Illes Balears, for comments and recommendations to my research work and also for the lectures and help that I received from some of the members while I was still an undergraduate student: it was at that point that my interest in General Relativity started growing. 
It was also thanks to the Relativity Group of the Friedrich-Schiller-Universit\"at Jena that I discovered the field of Numerical Relativity and became interested in this research topic that has accompanied my during my graduate career so far. 

I would like to thank An\i{}l Zengino\u{g}lu for his comments regarding the hyperboloidal initial value approach taken here and for being a source of inspiration. 
I take this opportunity to also thank José M. Martín-García, for the development of the package {\tt xAct}. 

\vspace{1ex}

Sincere thanks to my office and group mates Maria del Mar, Dani, Maria, Lorena, Aquilina, Gemma, Diego, David, Moner, Igor, Juan, Xisco and to those who only stayed here for a shorter time. Thanks for making the long hours spent at the office much lighter and enjoyable, for the coffee breaks (especially the cookies) and the chats and jokes.  

I also want to thank the PhD students that I met at the FaMAF during my stay in Córdoba, for the nice atmosphere during lunch at the office and the things we did together. Special thanks to Iván Gentile, for helping me get settled when I arrived and for making sure I had everything I needed.  

Last but not least I am really thankful to my family, for their continued support, the understanding when I was away for longer than expected, their love and the faith they put in me. 
Thanks a lot to my dear friends: Viki, Maru, Mer, Carmen, ... and many more that I cannot list here for space reasons. I also want to thank the people that I met during trips, hiking, at Chinese class and at the Teatre Principal. Doing one's best at work also requires free time's relax in good company and that is what I got from you. 

{ \ }

I am grateful to Prof.~Sascha Husa for his guidance during the progress of writing this thesis. I also want to thank Dr.~David Hilditch and Dr.~Juan A. Valiente Kroon for their valuable feedback and comments that helped improve this work. 

{ \ }

I acknowledge the support of the FPU-grant AP2010-1697 of the Spanish Ministry of Economy and Competitiveness, the European Union FEDER founds and the MINECO grants of the Spanish Ministry of Economy and Competitiveness FPA2010-16495 and FPA2013-41042-P and the ``Multimessenger Approach for Dark Matter Detection'' CONSO-LIDER-INGENIO 2010 Project CSD2009-00064. % and the Conselleria d'Economia i Competitivitat of the Govern de les Illes Balears. 

%Most of the algebraic derivations were performed using the Mathematica package {\tt xAct} \cite{xAct}. 

%\clearpage

%\end{comment}

\setcounter{tocdepth}{1}
\tableofcontents
%\listoffigures
%\listoftables

\chapter*{Notation and abbreviations}
\addcontentsline{toc}{chapter}{Notation and abbreviations}
\markboth{Notation and abbreviations}{}

%Metric notations: \upda{copied from paper, make sure they're consistent:} the 4-dimensional physical metric is denoted as $\tilde g$, the 4d-conformal metric as $\bar g$, the 3d conformal metric (induced by $\bar g$) as $\bar \gamma$, the 3d twice conformal metric $\gamma$ and the 3d twice conformal background metric $\hat \gamma$.

{\setstretch{1.5}

\begin{center}
{\large{\bf Abbreviations}}

{ \ } 

%\begin{table}[hhh]
%\caption{List of acronyms:}\vspace{-1.5ex}
%\center
\begin{tabular}{ll}
%\hline
GR & General Relativity\\
GW & Gravitational Wave\\
BH & Black Hole\\
%NS & Neutron Star\\
dS & deSitter\\
AdS & Anti-deSitter\\
%CBC & Compact Binary coalescence\\
%PN & post-Newtonian\\
%EOB & Effective-One-Body\\
CMC & Constant Mean Curvature\\
CFEs & Conformal Field Equations\\
BSSN & Baumgarte-Shapiro-Shibata-Nakamura\\
GBSSN & Generalized BSSN \\
%CFT & Conformal Field Theory\\
LHS & left-hand-side (of an equation)\\
RHS & right-hand-side (of an equation)\\
ADM & Arnowitt–Deser–Misner \\
Z4c & Z4 conformal\\
CCZ4 & Conformal and covariant Z4 \\
RN & Reissner-Nordstr\"om\\
MoL & Method of Lines \\
RK & Runge-Kutta \\
PDE & Partial differential equation \\
ODE & Ordinary differential equation \\
CFL & Courant–Friedrichs–Lewy \\
IMEX & Implicit-explicit \\
PIRK & Partially implicit Runge Kutta \\
%\hline
\end{tabular}
%\end{table}

\end{center}
}

\newpage

\begin{center}{ \large{\bf Metric notation}}\end{center}
\begin{table}[hhh]
%\caption{Metric notation:}\vspace{-1.5ex}
\center
{\renewcommand{\arraystretch}{1.3}
\begin{tabular}{@{}cl}
\hline
$\tilde g_{ab}$ & 4-dimensional physical metric\\
$\tilde{\bar\gamma}_{ab}$ & spacelike physical metric (spatial projection of $\tilde g_{ab}$) \\
$\bar g_{ab}$ & 4-dimensional conformal metric ($\bar g_{ab}=\Omega^2\tilde g_{ab}$)\\
$\mttr_{ab}$ & spacelike conformal metric (spatial projection of $\bar g_{ab}$)\\
%$\bar \gamma_{ab}$ & 3-dimensional conformal metric (induced from $\bar g_{ab}$, 3-dim. part of $^{(4)}\bar \gamma_{ab}$)\\
$\gamma_{ab}$ & spacelike twice conformal metric ($\gamma_{ab}=\chi\bar \gamma_{ab}$)\\
$\hat \gamma_{ab}$ & spacelike twice conformal background metric (stationary values of $\gamma_{ab}$) \\
$\hat{\bar g}_{ab}$& 4-dimensional conformal background metric (stationary values of $\bar g_{ab}$)\\
$\hat{\tilde g}_{ab}$& 4-dimensional physical background metric (stationary values of $\tilde g_{ab}$)\\
\hline
\end{tabular}
}
\end{table}

{ \ }

\begin{center}{ \large{\bf Units and index notation}}\end{center}

Geometrized units are used, so that the speed of light is taken to be $c=1$ and the gravitational constant $G=1$. %geometrized units

Abstract indices in tensorial quantities are denoted by the letters $a,b,c,...$ . Component indices in four dimensions (ranging from  0 to 3) are written in Greek letters $\mu,
\nu,\sigma,...$, while the ones that only cover the spatial coordinates (from 1 to 3) use $i,j,k,...$ . %\upda{2d??}
%I will denote indices that can take values from 0 to 3 (that is, including the time coordinate) with letters $a,b,c,...$ and with $i,j,k,...$ the ones that only cover the spatial coordinates (values 1, 2 and 3). have to correct!! }

\mainmatter	  % Begin normal, numeric (1,2,3...) page numbering

\chapter[Introduction]{Introduction}\label{c:introduction}

\section{Basic concepts in General Relativity} 

\subsection{General Relativity}

The theory of General Relativity (GR) relates the distribution of matter and energy to the deformation of spacetime, and describes how the latter reciprocally affects the motion of the existing particles. It is elegantly encoded in the Einstein equations 
\begin{equation}\label{ei:einsteinp} G[\tilde g]_{ab} + \tilde g_{ab}\Lambda = 8 \pi T[\tilde g]_{ab} , \end{equation}
where on the left the Einstein tensor $G[\tilde g]_{ab}$ and the cosmological constant $\Lambda$ characterize the geometry of the spacetime described by the metric $\tilde g_{ab}$, and on the right the stress-energy tensor $T[\tilde g]_{ab}$ completes the picture providing the properties of the objects present in the spacetime. Here the Einstein tensor is given by $G[\tilde g]_{ab}=R[\tilde g]_{ab}-\case{1}{2}\tilde g_{ab}R[\tilde g]$, with the Ricci tensor $R[\tilde g]_{ab}=R[\tilde g]^c{}_{acb}$ and the Ricci scalar $R[\tilde g]=R[\tilde g]^a{}_a=\tilde g^{ab}R[\tilde g]_{ab}$. The Riemann tensor can be computed from the metric $\tilde g_{ab}$ and its Christoffel symbol $\tilde \Gamma^a_{bc} =  \case{1}{2}\tilde g^{ad}(\partial_b \tilde g_{cd} + \partial_c \tilde g_{bd} - \partial_d \tilde g_{bc})$ as 
\begin{equation}\label{ei:Riemannp} 
R[\tilde g]^a{}_{bcd}  =  \partial_c \tilde \Gamma^a_{bd} - \partial_d \tilde \Gamma^a_{bc} + \tilde \Gamma^e_{bd}\tilde \Gamma^a_{ec} 
    - \tilde \Gamma^e_{bc}\tilde \Gamma^a_{ed}  
 . \end{equation}
Writing the Einstein equations in terms of metric components and their partial derivatives manifests their character as non-linear partial differential equations of second differential order in the metric field, the quantity for which they are to be solved in a metric formulation. 

%GR is a relativistic theory, which means that a wave-like behaviour of the dynamical quantities is expected. According to the theory of Special Relativity, the maximum achievable speed is the speed of light, so that all propagation speeds are necessarily finite. The dynamical quantity of GR is the gravitational field (encoded in the metric), so that its perturbations will propagate in the form of radiation. This gravitational radiation, also called gravitational waves, can be imagined as ripples of the spacetime's curvature that travel at the speed of light. 
GR is a relativistic theory, so that the causal structure is defined by the light cone. This implies that the propagation speeds are finite and a wavelike behaviour is expected. The dynamical quantity of GR is the gravitational field (encoded in the metric) and its perturbations propagate in the form of radiation. This gravitational radiation, also called gravitational waves, can be imagined as ripples of the spacetime's curvature that travel at the speed of light. 

\subsection{Gravitational waves} % and the goal of this thesis

The existence of gravitational waves (GWs) was predicted by Einstein \cite{1918SPAW.......154E} by studying wave phenomena in linearized gravity. However, due to the physical and mathematical complexity of the theory, the coordinate dependent quantities involved in the calculations and the background independence of GR, the question whether GWs were actual physical phenomena or just simply coordinate effects was raised and it was not until the 1960s that the physical nature of GWs, interpreted as free gravitational degrees of freedom carrying positive energy \cite{Bondi:1962px,Sachs:1962wk}, was finally understood. For a historical description of the controversy see \cite{Kennefick:1997kb}. 

GWs, which are emitted by non-spherically accelerated massive objects in the universe, have such a small amplitude, that no direct observation has taken place so far. However, they are known to exist thanks to the discovery of the binary pulsar PSR 1913+16 by Hulse and Taylor \cite{Hulse:1974eb}. The pulsar forms a binary system with another neutron star, and the energy loss that causes the shrinking of the distance between both corresponds to the emitted GW radiation predicted by GR. 

Efforts towards a direct detection of GW are currently performed with high sensitivity interferometers, like Advanced LIGO \cite{Abbott:2007kv,TheLIGOScientific:2014jea} and Advanced Virgo \cite{Acernese:2008zzf,TheVirgo:2014hva}. 
The analysis of the experimental data measured in the GW detectors requires waveform models created by numerical methods. 

{ \ }

The goal of this thesis is to contribute to the current waveform modeling efforts by developing useful techniques that will allow to solve the Einstein equations numerically in a more efficient way. More specifically, I have implemented the hyperboloidal initial value problem in spherical symmetry using commonly used formulations and thus showing that such an implementation is feasible. 

\subsection{Equivalence principle and the absence of background}

The equivalence principle, one of the pillars on which General Relativity is based, states that all bodies ``fall'' in the same way in the presence of a gravitational field. 
%In Newtonian gravity it is expressed as the equivalence between the gravitational and the inertial mass of a body. 
This indicates that the gravitational field is a property of the spacetime itself. The paths of freely falling bodies are described by geodesics of the metric which, in presence of massive objects, will no longer correspond to a flat geometry.  

If we were to measure an electromagnetic field, the procedure would be the following: we first set a ``background observer'', which is unaffected by the electromagnetic interaction and follows a geodesic path; then a charged test body is released; finally, by measuring the deviation of the particle's trajectory from the geodesic path (the one followed by the background observer), the electromagnetic field is determined. 

When we try to measure a gravitational field we encounter a problem already in the first step: we cannot set any ``background observer'', as its behaviour will be exactly the same as that of the test particle and no deviation will be detected. Both the observer and the test particle are affected by the gravitational field in the same way. 

In GR the structure of the spacetime is a dynamical quantity itself and, due to its non-linear character, the gravitational field also acts as its own source. The consequence is that a background with respect to which the curvature of spacetime can be measured can only be defined under specific symmetry assumptions, but not in general. 

\subsection{Isolated systems}

Suppose we are interested not in studies of cosmological nature, but of astrophysical processes inside of a gravitating system, such as a single star, black hole (BH) or a binary system. The relevant physics for us is the gravitational interaction of the system. The large-scale structure of the universe will have minimal influence on the properties of our system in study, so that neglecting it is a good idealization for our purposes. %Besides, it is not feasible to study the system in its interaction with the rest of the universe due to its complexity. % Sascha: could embed in FRW

An isolated system will thus allow us to study the physical properties of a system as a whole. 
To isolate it, the system is considered to be embedded in a spacetime with certain asymptotic conditions, which should not depend on the isolated system under consideration. %The physical interpretation is clean: it is as if the system existed alone in the universe. 
The gravitational influence of the isolated system is then expected to fade away as we are infinitely far away from it, and the  metric $\tilde g_{ab}$ should approach the Minkowski metric at infinity, at least in the spacelike and null directions.
Such spacetimes will be referred to as being asymptotically flat.  %If matter or radiation are present, the metric does not become flat in the timelike direction.
%To qualify the asymptotic behaviour of the spacetime, we know that the gravitational field becomes weak away from sources, so that the gravitational influence of the isolated system is expected to fade away as we are infinitely far away from it. This means that the manifold where the isolated system is embedded is classified as asymptotically flat; its metric $\tilde g_{ab}$ will approach the flat metric at infinity, at least in the spacelike and null directions. If matter or radiation is present, the metric does not become flat in the timelike direction.  %asymptotically flat along spatial slices or null slices? see Schwarzschild example [110] in An\il's intro. 

%Nevertheless, a strict definition of asymptotic flatness cannot be formulated. The fall-off of the curvature of the metric $\tilde g_{ab}$ of the asymptotically flat spacetime at large distances cannot be determined by comparison with a background flat metric, because the latter does not exist. There is also no preferred radial coordinate to specify the fall-off rates. Besides, the asymptotic flatness condition can be taken along spatial directions or null directions (light rays). 

\subsection{Gravitational radiation and energy}

The absence of a natural background also renders the unambiguous local determination of quantities such as mass or energy density impossible in general; there is no way of separating the curvature effects from what would be a flat spacetime, where this difference is exactly what qualifies the presence of massive bodes or other energetic perturbations. 

What is indeed possible is to define the total energy of a system by evaluating the gravitational field far away from the sources. %, where their effect has already fallen off. 
Also the energy flux radiated away by an isolated system in the form of GW is well defined asymptotically. For this reason, in GR the energy, mass and radiation flux are global quantities (can only be calculated for the complete spacetime) and are closely related to the asymptotic behaviour of the spacetimes. 
%The same is valid for the flux of gravitational radiation (GW) emitted by the system. 

Work towards a formal characterization of gravitational radiation started in the 1950s - the basic historic development can be found summarized in \cite{lrr-2004-1}. Among the relevant results was the ``peeling property'' obtained by Sachs \cite{Sachs:1961zz}, where the fall-off behaviour of the curvature is described by a decomposition of the Weyl tensor in terms of powers of $1/\tilde r$, with $\tilde r$ an affine parameter along outgoing null geodesics. 
Another important achievement by Bondi, van der Burg and Metzner \cite{Bondi:1962px} was the introduction of inertial coordinates in flat spacetime at infinity along null curves, the so-called Bondi coordinates, that rely on the use of a retarded time function that labels outgoing null hypersurfaces. 
%\cite{Bondi:1962px} Bondi (retarded time function, Bondi coordinates, news functions)

The actual calculation of radiation where it is unambiguously defined (in the asymptotic region) poses considerable difficulties, because it involves using specific coordinate systems and taking limits at infinity. 
An invariant characterization of radiation, where coordinate independent definitions can be performed, would be preferred. 

\section{Conformal compactification}\label{si:confcomp}

A new point of view introduced by Penrose \cite{PhysRevLett.10.66,Penrose:1965am} allows to solve the previously mentioned problems. This new approach takes advantage of the conformal structure of spacetime and uses it to define the notion of asymptotic flatness in a coordinate independent way by adding the ``points at infinity'' as a ``null cone at infinity''. 

%The basic idea comes from realizing that ``infinity'' in a spacetime is infinitely far away with respect to its metric $\tilde g_{ab}$. If distances are measured by means of metre sticks, we need infinitely many metre sticks to reach infinity. However, if the metre sticks are such that they become longer at the appropriate rate as we approach infinity, a finite number of them will be enough to cover the distance. This naive explanation is formally expressed as follows. 
The basic idea is how the distance to infinity is measured. The physical distance is infinite, but the coordinates can be freely chosen in such a way that the coordinate distance is finite. Using this compactification of the coordinates, infinity is set at a finite coordinate location. The coordinate compactification however implies that the metric becomes infinite and this is what is solved by Penrose's idea. 

The physical spacetime is represented by a Lorentzian manifold $\mathcal{\tilde M}$ characterized by a Lorentzian metric $\tilde g_{ab}$, infinite at infinity. %This physical metric becomes infinite at  infinity. 
A new regular metric $\bar g_{ab}$ is introduced with help of a conformal factor $\Omega$: 
\begin{equation}\label{ei:rescmetric}
\bar g_{ab}\equiv\Omega^2\tilde g_{ab} \qquad \textrm{and} \qquad \bar g^{ab}\equiv \frac{\tilde g^{ab}}{\Omega^2} .
\end{equation}
The conformally rescaled metric $\bar g_{ab}$ is defined on a compactified auxiliary manifold $\mathcal{\bar M}$. The physical manifold $\mathcal{\tilde M}$ is given by $\mathcal{\tilde M}=\{p\in\bar{\mathcal{M}}\,|\,\Omega(p)>0\}$, so it is a submanifold of $\mathcal{\bar M}$.  %the extendable 
The conformal factor $\Omega$ is such that it vanishes at the appropriate rate exactly where the physical metric $\tilde g_{ab}$ becomes infinite, thus giving a rescaled metric $\bar g_{ab}$ which is finite everywhere, and so allowing for a conformal extension of $\mathcal{\bar M}$ across the physical infinity. %The physical infinity is given by $\Omega=0$. 
 
This approach has many beneficial properties. The first one is that conformal rescalings leave the angles unaffected, so that the causal structure of $\mathcal{\tilde M}$ and $\mathcal{\bar M}$ is exactly the same. The calculation of limits for the fall-off conditions at $\mathcal{\tilde M}$'s infinity is substituted by simple differential geometry on the extended manifold $\mathcal{\bar M}$, therefore providing a geometric formulation of the fall-off behaviour. The ``peeling properties'' found by Sachs could be deduced by Penrose \cite{PhysRevLett.10.66,Penrose:1965am} from the conformal picture in a coordinate independent way. 

\subsection{Example: compactification of Minkowski spacetime}\label{compMin}

The following textbook example (see e.g. \cite{Wald,lrr-2004-1}) illustrates the conformal compactification procedure in a simple way. 
Let us consider Minkowski spacetime in coordinates adapted to spherical symmetry, with line element
\begin{equation}\label{ei:iniMin}
d \tilde s^2 = -d\tilde t^2+d\tilde r^2+\tilde r^2d\sigma^2 , \quad \textrm{where} \quad d\sigma^2\equiv d\theta^2+\sin^2\theta d\phi^2 . 
\end{equation}
%The ranges allowed for the time and radial coordinates are $\tilde t\in(-\infty,\infty)$ and $\tilde r\in[0,\infty)$. 
We introduce the null coordinates $\tildeu$ and $\tildev$  
\begin{equation}\label{ei:uvtr}
\tildeu=\tilde t-\tilde r , \qquad \tildev=\tilde t+\tilde r . 
\end{equation}
Constant $\tildev$ represents ingoing null rays, while constant $\tildeu$ are outgoing ones. The only restriction on the values that $\tildeu$ and $\tildev$ can take is $\tildev-\tildeu\,(=2\tilde r)\ge0$. 
In terms of the null coordinates the line element takes the form 
\begin{equation}
d \tilde s^2 = -d\tildeu \,d\tildev+\frac{1}{4}(\tildev-\tildeu)^2d\sigma^2 . 
\end{equation}
The infinite range of the null coordinates is compactified by making the substitution 
\begin{equation}\label{ei:UVuv}
U=\arctan \tildeu, \qquad V=\arctan \tildev , 
\end{equation}
with coordinate ranges $U,V\in(-\case{\pi}{2},\case{\pi}{2})$ and  $V-U\ge0$. 
The resulting metric is
\begin{equation}
d \tilde s^2 = \frac{1}{\cos^2U\cos^2V}\left[-dUdV+\frac{1}{4}\sin^2(V-U)d\sigma^2\right] .
\end{equation}
It is not possible to evaluate this line element at the points $U=\pm\case{\pi}{2}$ or $V=\pm\case{\pi}{2}$, which correspond to the infinity along null directions (denoted as $\scri$ (Scri)), due to the vanishing denominator. 
Introducing a conformally rescaled line element $d\bar s^2=\Omega^2 d\tilde s^2$ as indicated in \eref{ei:rescmetric} with conformal factor 
%multiply with conformal rescaling with conformal factor
\begin{equation}
\Omega =2\cos U \cos V , 
\end{equation}
gives 
\begin{equation}
d \bar s^2 = \Omega^2 d\tilde s^2 = -4dUdV+\sin^2(V-U)d\sigma^2 , 
\end{equation}
an expression that can indeed be extended to $U=\pm\case{\pi}{2}$ and $V=\pm\case{\pi}{2}$ and even for $|U|,|V|>\case{\pi}{2}$. 
Defining the new compactified time and space coordinates 
\begin{equation}\label{ei:TRUV}
T=V+U , \qquad R=V-U , 
\end{equation}
we obtain a Lorentz metric on $\mathbb{R}\times S^3$, which is the metric of the Einstein static universe:  
\begin{equation}\label{ei:finEin}
d \bar s^2 = -dT^2+dR^2+\sin^2R\,d\sigma^2 .
\end{equation}
The relations $\tilde t\to \tilde t(T,R)$ and $\tilde r\to \tilde r(R,T)$ substituting \eref{ei:uvtr}, \eref{ei:UVuv} and \eref{ei:TRUV} give the embedding of the initial Minkowski metric \eref{ei:iniMin} into the Einstein universe, so from $\mathcal{\tilde M}=\mathbb{R}^4=\{\tilde t\in(-\infty,\infty),\tilde r\in[0,\infty)\}$ to $\mathcal{\bar M}=\mathbb{R}\times S^3=\{T\in[-\pi,\pi],R\in[0,\pi]\}$, not taking into account the angular coordinates.  % or Einstein static universe

\begin{figure}[htpb!!]
\center
\includegraphics[width=0.25\linewidth]{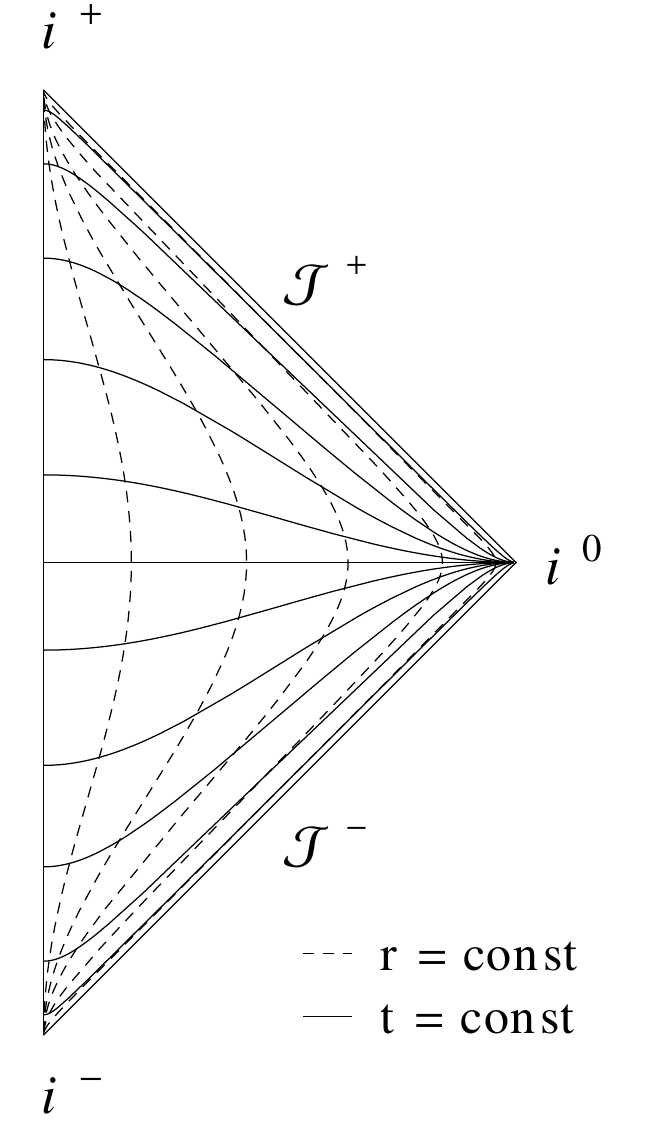}
\vspace{-2ex}
\caption{Penrose diagram showing the compactification of Minkowski spacetime. Null infinity ($\scri$) is denoted in  this and the following diagrams as $\mathcal{J}$.}
\label{fi:Minko}
\end{figure}

A (Carter-)Penrose diagram is used to show the causal structure in a compactified way. The Penrose diagram in \fref{fi:Minko}, where $T$ over $R$ are plotted implicitly, shows the curves of constant Minkowski time and radius in the conformally rescaled picture. Except for the leftmost vertical line connecting $i^-$ and $i^+$, which corresponds to the origin $\tilde r=0$, all other points in the diagram represent a sphere in terms of the angular coordinates that have been suppressed. 
The solid lines in the diagram are spacelike hypersurfaces labeled by a constant value of $\tilde t$. They all extend from the origin $\tilde r=0$ to spacelike infinity, the point denoted by $i^0$ and where $i^0=\{T=0,R=\pi\}$, corresponding to $\tilde r\to\infty$ in the original Minkowski spacetime.  
The dashed lines are timelike hypersurfaces of constant radial coordinate. They all originate at past timelike infinity $i^-=\{T=-\pi,R=0\}$ and end at future timelike infinity $i^+=\{T=\pi,R=0\}$. 
As the causal structure is left unchanged by the conformal transformation, light rays should be depicted as straight lines at $\pm 45º$ in \fref{fi:Minko}. For instance, outgoing light rays are shown with solid lines in diagram b) in \fref{fi:slices}. Ingoing null rays are given by constant $V$ and they all propagate to the left starting from the line labeled with $\scri^-=\{U=-\case{\pi}{2},|V|<\case{\pi}{2}\}$, which is called past null infinity or past lightlike infinity. Equivalently, constant $U$ determines outgoing null geodesics that propagate to the right until they reach $\scri^+=\{V=\case{\pi}{2},|U|<\case{\pi}{2}\}$, future null or lightlike infinity. 
%$\scri^-=\{T=R-\pi,R\in(0,\pi)\}$

The original Minkowski spacetime is mapped to $\{|T+R|<\pi,|T-R|<\pi\}$ in $\mathcal{\bar M}$ and its conformal boundary consists of the pieces $i^0$, $i^\pm$ and $\scri^\pm$. As the conformal metric $\bar g_{ab}$ (used in the line element \eref{ei:finEin}) is regular at the boundary, $\mathcal{\bar M}$ has a conformal extension outside of the boundary. This conformal extension depends on the choice of the conformal factor $\Omega$, while the conformal boundary is uniquely determined by the physical manifold $\mathcal{\tilde M}$, Minkowski in the present example.

\subsection{Asymptotic flatness}

%\begin{comment}
%The notion of asymptotic flatness can be defined in a coordinate independent way \upda{by adding the ``points at infinity'' as an appropriate boundary}
%After introducing the conformal compactification picture by Penrose, asymptotic flatness can be defined in a stricter way. The simple definitions that follow can be found in \cite{lrr-2004-1,Husa:2002zc}: \vspace{-2.5ex}
To generalize from Minkowski spacetime, asymptotic flatness can be defined in the conformal compactification picture as \cite{lrr-2004-1,Husa:2002zc}: %\vspace{-2.5ex}
\begin{Def}[asymptotic simplicity]\label{def:simple}
A smooth spacetime $(\tilde{\cal M}, \tilde g_{ab})$ is
called asymptotically simple, if there exist another smooth manifold
$(\bar{\cal M},\bar g_{ab})$ that satisfies % and a scalar function $\Omega$
\begin{enumerate}
\item 
$\tilde{\cal M}$ is an open submanifold of $\bar{\cal M}$ with smooth boundary $\partial\tilde{\cal M} = \scri$,
\item 
a smooth scalar function $\Omega$ exists on $\bar{\cal M}$, such that $\bar g_{ab} = \Omega^2 \tilde g_{ab}$ on $\tilde{\cal M}$, with $\Omega > 0$ on  $\tilde{\cal M}$, and that both $\Omega=0$ and $\bar\nabla_a \Omega\neq0$ hold on $\scri$,
\item 
every null geodesic in $\tilde {\cal M}$ acquires two end points on $\scri$.
\end{enumerate}
\end{Def}
\begin{Def}[asymptotic flatness]\label{def:flat}
An asymptotically simple spacetime %$(\tilde{\cal M}, \tilde g_{ab})$ 
is called asymptotically flat
if in addition its Ricci tensor $R[\tilde g]_{ab}$ vanishes in a neighborhood of $\scri$.
\end{Def}  
%\end{comment}

Spacetimes which are asymptotically simple but not asymptotically flat are, for instance, deSitter (dS) and Anti-deSitter (AdS) spacetimes, where the cosmological constant $\Lambda$ is positive and negative, respectively. 

\begin{figure}[htpb!!]
\center
\includegraphics[width=0.75\linewidth]{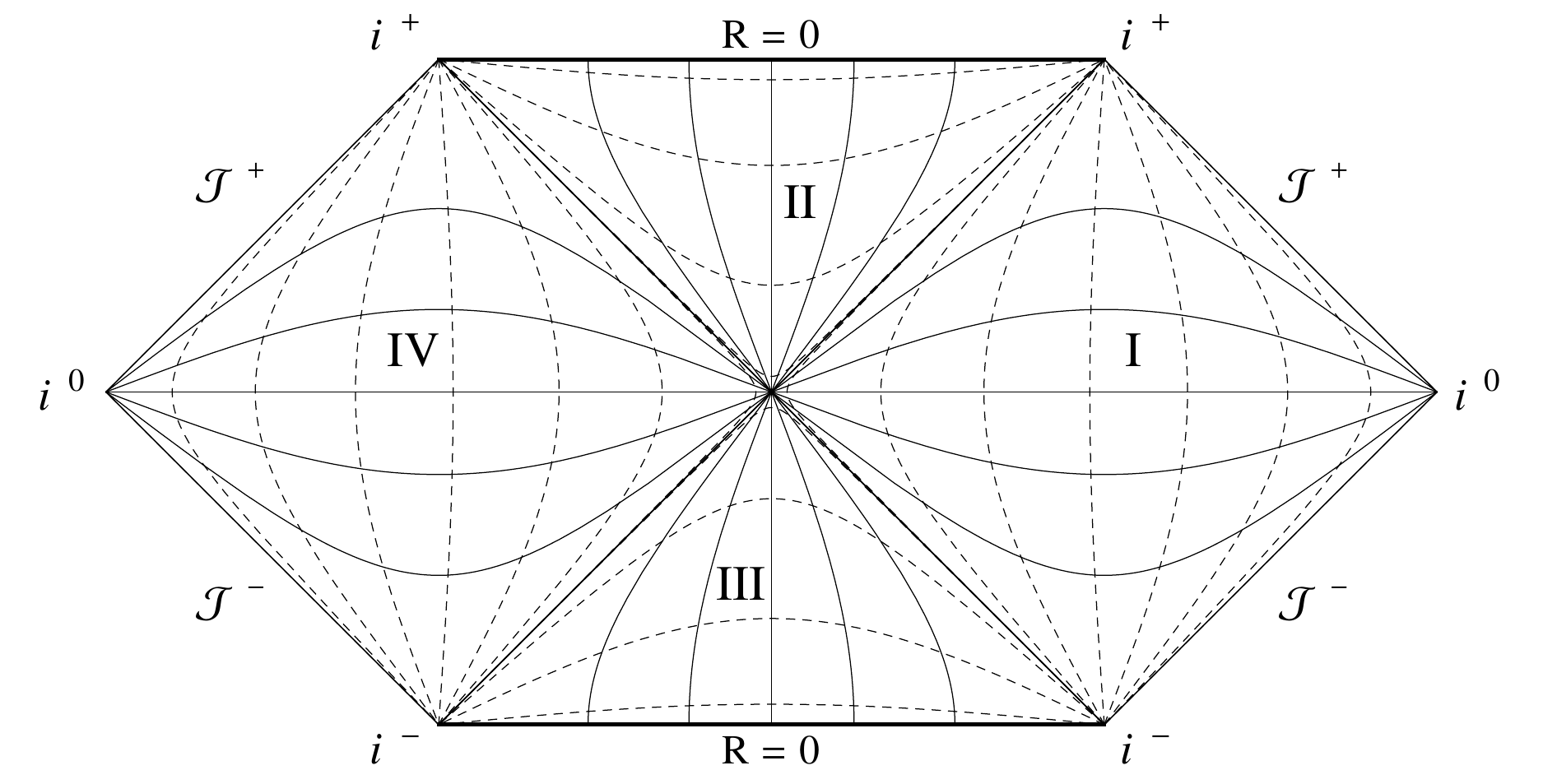}
\vspace{-2ex}
\caption{Penrose diagram showing the compactification of the Schwarzschild spacetime. In the same way as in \fref{fi:Minko}, the dashed lines denote timelike surfaces (of constant radial coordinate) and the solid ones represent constant-time spacelike surfaces.}
\label{fi:Schw}
\end{figure}

A compactification for the Schwarzschild spacetime equivalent to the Minkowski one is shown in \fref{fi:Schw}. The construction of the diagram is described in section \ref{cap:Schw}. 
%The expressions used to construct it are presented in section \ref{cap:Schw}. 
As can be deduced from the diagram, condition 3 in definition \ref{def:simple} excludes BH spacetimes, because the characteristics that enter the BH's horizon will not have an end point on $\scri^+$, but at the singularity (denoted by the upper ``R=0'' in \fref{fi:Schw}). Less restrictive conditions are described in \cite{Penrose:1965am,Wald,stewart1997advanced}. 
This is due to the fact that when matter is present, timelike infinity is not asymptotically flat. 
A BH spacetime is thus asymptotically simple and asymptotically flat in the spatial and future pointing outward null directions.

\subsection{Einstein equations for the conformally rescaled metric}\label{si:confeqs}

%such that the conformally rescaled metric $\bar g_{ab}$ on the compactified auxiliary manifold $\mathcal{\bar M}$ is finite at $\scri$. The conformal factor has to satisfy $\Omega=0$ and $d\Omega\neq0$ \upda{or $\nabla_a \Omega\neq0$?} on $\scri$. Note that the physical spacetime $\mathcal{\tilde M}$ is a submanifold of $\mathcal{\bar M}$; the rescaled metric $\bar g_{ab}$ is defined on $\mathcal{\bar M}$ and $\mathcal{\tilde M}$, but the physical one is only defined on the latter. 

The Einstein equations \eref{ei:einsteinp} expressed in terms of the conformally rescaled metric $\bar g_{ab}$ are given by \cite{Wald}
\begin{equation}\label{ei:einsteinc} 
G[\bar g]_{ab} + {2\over\Omega}(\bar\nabla_a\bar\nabla_b\Omega-\bar g_{ab}\bar\Box \Omega) + {3\over\Omega^2}\bar g_{ab}(\bar\nabla_c\Omega)(\bar\nabla^c\Omega) + {1\over\Omega^2}\bar g_{ab}\Lambda = 8 \pi T[\case{\bar g}{\Omega^2}]_{ab} , 
\end{equation}
where $G[\bar g]_{ab}$ is the Einstein tensor of the conformally rescaled metric. The physical metric appearing in the stress-energy tensor $T_{ab}$ has to be expressed in terms of the conformal metric $\bar g_{ab}$. We now multiply the previous equation by $\Omega^2$ and evaluate it at $\scri$, so setting $\Omega=0$. The stress-energy tensor is supposed to be finite, so that the following relation is obtained: 
\begin{equation}\label{ei:nablaOmega}
\atscri{(\bar\nabla_c\Omega)(\bar\nabla^c\Omega)} = -\Lambda . 
\end{equation}
This indicates what kind of hypersurface null infinity is, depending on the value of the cosmological constant: 
\begin{itemize} 
\item if $\Lambda=0$ (asymptotically flat), then $\bar\nabla^c\Omega$ is a null vector and $\scri$ is a null surface; 
\item if $\Lambda>0$ (asymptotically dS), we have that $\bar\nabla^c\Omega$ points in a timelike direction and $\scri$, being perpendicular to it, is thus spacelike; % deSitter (dS) spacetime
\item if $\Lambda<0$ (asymptotically AdS), $\bar\nabla^c\Omega$ is spacelike and $\scri$ is a timelike surface. % Anti-deSitter (AdS) spacetime
\end{itemize} 
From now on only the asymptotically flat case ($\Lambda=0$) will be considered. 

The second and third terms in \eref{ei:einsteinc} formally diverge at null infinity, as $\atscri{\Omega}=0$ holds there. However, together they attain a regular limit at $\scri$, because the equations are conformally regular \cite{Friedrich:1981wx}. They are also divergence-free and satisfy the Bianchi identities without requiring any additional conditions on $\Omega$ \cite{Zenginoglu:2007it,Zenginoglu:2008pw}. This is important, because it means that we can freely specify the conformal factor $\Omega$. There exists a preferred conformal gauge choice \cite{Tamburino:1966zz,9780511564048,stewart1997advanced}, whose expression \eref{eg:condother} will be later discussed, that ensures that the conformal factor terms have regular limits at null infinity individually. 

\section{Spacetime slices}\label{i:slices}

%In order to solve the Einstein equations in an iterative way, suitable for a numerical implementation, the problem has to be cast into the form of an initial value formulation. 
It is convenient to be able to solve the Einstein equations in an iterative way, preferably as an evolution in time. For this, the problem is cast into the form of an initial value formulation, which requires breaking the coordinate invariance of the Einstein equations and slicing the spacetime to obtain an appropriate hypersurface where initial data can be specified. This is also a common approach taken in numerical implementations. 

\begin{figure}[htpb!!]
\center
\begin{tabular}{@{}c@{}@{}c@{}@{}c@{}@{}c@{}}
a)&b)&c)&d)\vspace{-3ex}\\
\includegraphics[width=0.25\linewidth]{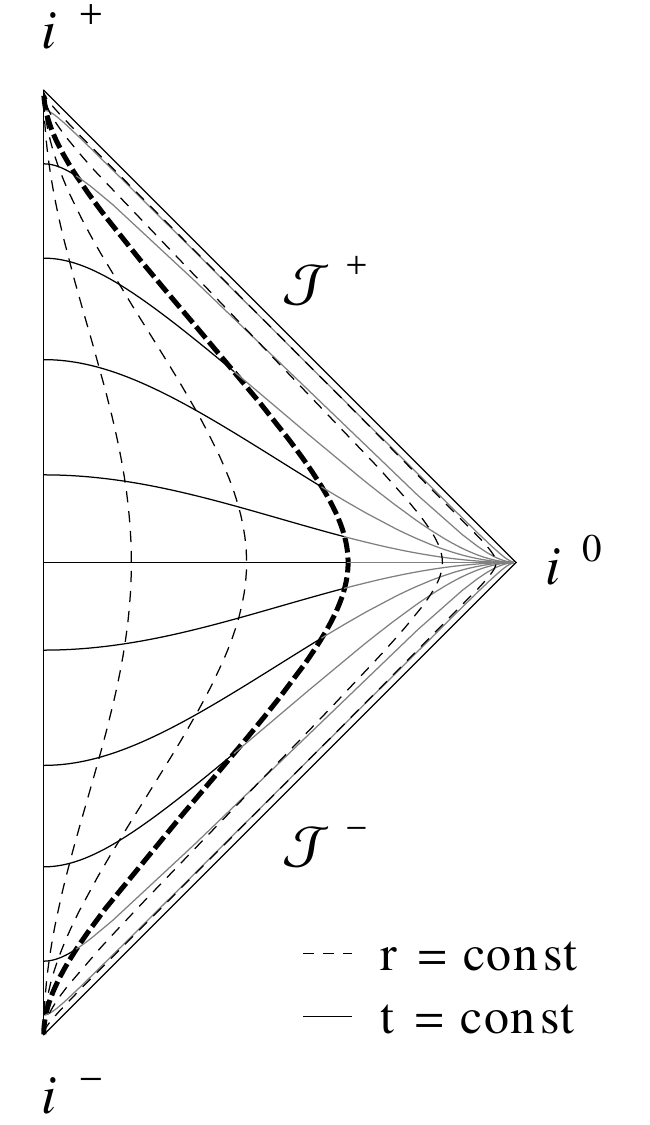}&
\includegraphics[width=0.25\linewidth]{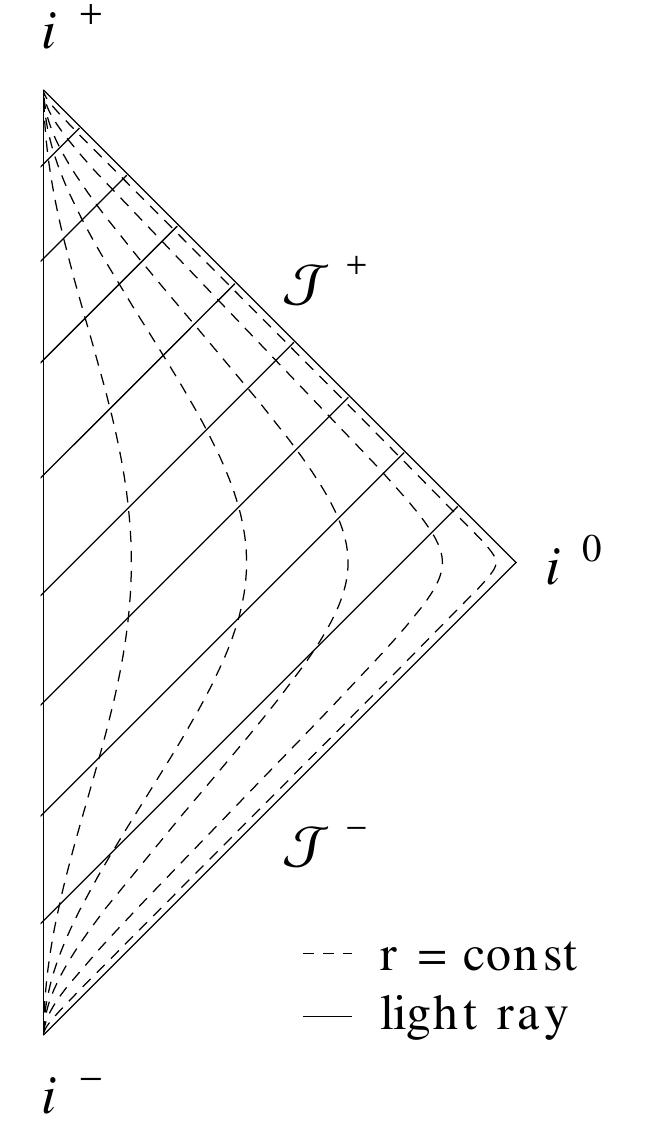}&
\includegraphics[width=0.25\linewidth]{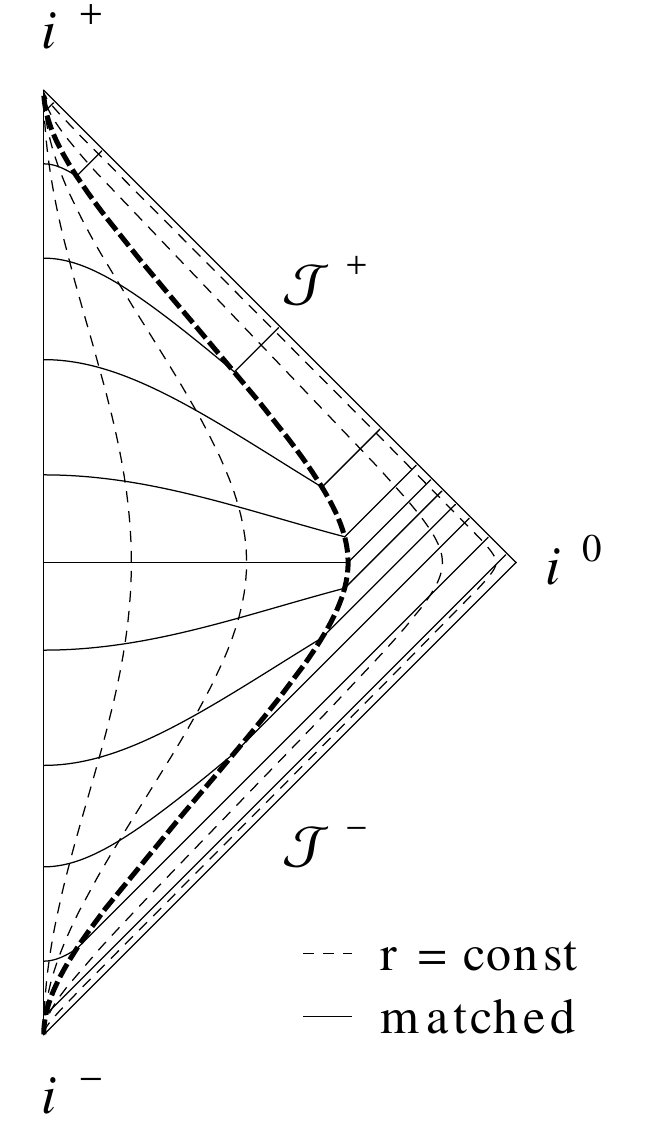}&
\includegraphics[width=0.25\linewidth]{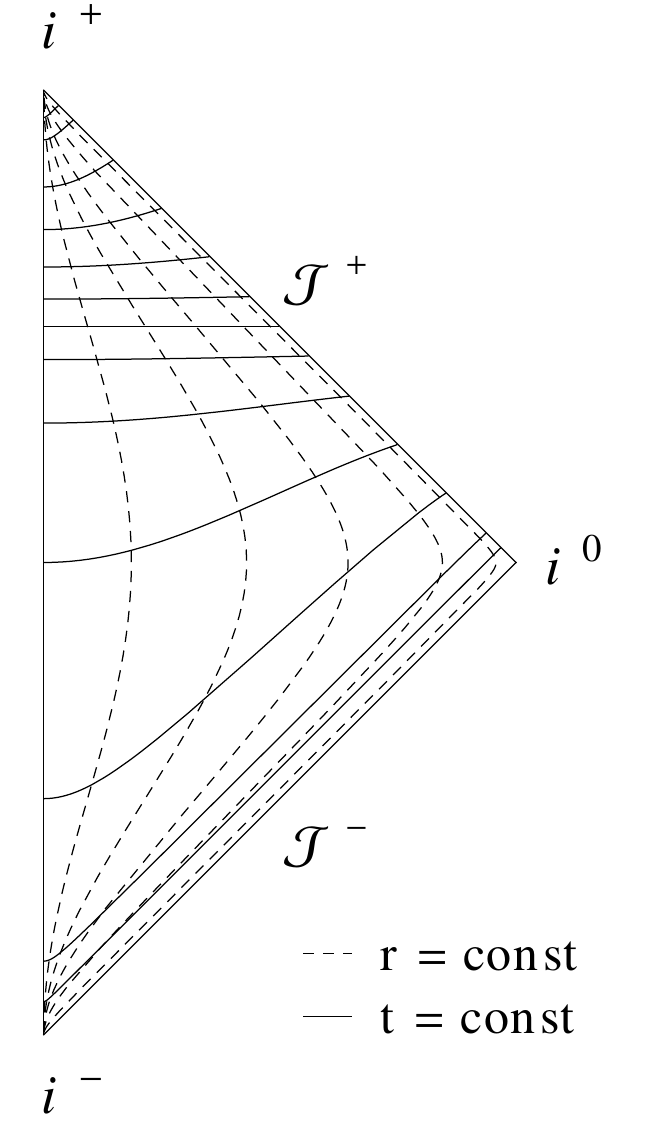}
\end{tabular}
\caption{Penrose diagrams showing Minkowski spacetime foliated along different types of hypersurfaces used in initial value formulations: a) Cauchy slices, b) characteristic slices, c) Cauchy-Characteristic matching and d) hyperboloidal slices.} % Spatial infinity $i^0$, past and future timelike infinity $i^\pm$ and past and future null infinity $\scri^\pm$ are also indicated.
\label{fi:slices}
\end{figure}

Different possible types of foliations of Minkowski spacetime are displayed in \fref{fi:slices} in the form of Penrose diagrams: %, where the time and radial coordinates are compactified so that the whole spacetime can be included in a finite size. Note that each point in the diagram corresponds to a 2-sphere. 
\begin{description}\itemsep-0.2ex

\item[a) Standard Cauchy slices:] 
the solid lines represent constant-time spacelike slices (also called Cauchy slices). They extend from the origin (the vertical line on the left) to spatial infinity ($i^0$), so that they are asymptotically Euclidean. This type of slices are commonly used in numerical simulations, but they have two considerable drawbacks regarding their use in numerical simulations. The first one is that the slices actually extend to infinity, whereas an infinite number of gridpoints cannot be simulated computationally. Usually one would cut the slices at a certain value of the radial coordinate and only evolve the interior part. In the diagram, a possible cut is denoted by the thick dashed line. The solid gray lines denote the part of the slices that is cut off in this procedure. However, setting a timelike boundary introduces two extra problems: one of them is that the extraction of the radiation signals can no longer be performed at infinity, where the global quantities are defined, so that the calculation of the waveform at an finite distance from the source will necessarily incorporate some error; the other problem is the treatment of the boundary. In a numerical implementation, extra points outside of the integration domain have to be filled in according to given boundary conditions to evaluate the derivatives, see section \ref{sn:bcs}. This transforms the problem into an initial boundary value problem, whose boundary conditions are very difficult to specify properly: they have to allow radiation to leave the domain, preserve the constraints and not ruin the numerical stability of the system, among other requirements. 
Regarding the extraction at a finite radius, the accuracy of the extracted signals has been widely studied \cite{Reisswig:2009us,Pollney:2009ut,Pollney:2009yz,Taylor:2013zia}: even if some of the methods come quite close to the expected values, there is a systematic error that cannot be estimated by convergence tests. % \cite{Pazos:2006kz} [106] in An\i l's intro; \cite{Berti:2007fi,Marronetti:2007ya} [16,97] in An\i l's introduction; \cite{Taylor:2013zia}; \cite{Kocsis:2007zz}

A possible way around the timelike boundary would be the compactification of the spacelike slices. Nevertheless, this takes us to the second problem: as radiation approaches spacelike infinity $i^0$, its speed tends to zero and the waves start to pile up. In a numerical simulation with limited resolution, at some point the waves will not be resolved anymore, causing a loss of accuracy and even instabilities in the simulation. Radiation should be naturally measured at future null infinity ($\scri^+$) \cite{Frauendiener:1998yi}, as it also corresponds to the appropriate idealization of gravitational observers. 

A possibility to overcome this problem is to extrapolate to $\scri^+$, e.g. as in \cite{Boyle:2009vi}. This is done by extracting the wave signal at various radii, either along a series of concentric spheres or along an outgoing null ray, and using it to calculate the radiation signal at future null infinity. 

%cannot simulate infinity bcs problem in extraction
%initital boundary value problem, constraint pereserving bcs (define bc)
%The drawback they have regarding GW extraction is that the wave signal has to be extracted at a finite value of the radius. 
%if compactification of spatial infinity: piling up of the radiation waves as they approach spacelike infinity, waves not resolved after some time, loss of accuracy and instabilities. 
%the natural place to measure radiation is $\scri^+$ \cite{Frauendiener:1998yi}

\item[b) Characteristic slices:] 
here the solid lines show outgoing null or characteristic slices, the paths followed by light rays that leave the spacetime. Ingoing null slices would be represented by straight lines perpendicular to the solid ones. Expressed in null coordinates the equations simplify significantly and compactified outgoing characteristic slices are well suited for GW extraction, as they reach $\scri^+$. However, the coordinates are not flexible and if the gravitational field in the interior part of the spacetime becomes strong, they are likely to create caustics, where the generators of the null geodesics become tangent. A treatment of caustics has been proposed at the theoretical level \cite{stefried} with views towards a numerical implementation \cite{1983RSPSA.385..345F}, although to my knowledge it has not been attempted yet. The idea is to include the singularities that arise at the caustics as part of the evolution. For more details, see e.g. \cite{Winicour:2005ge}. 

\item[c) Cauchy-Characteristic matching / extraction:] 
in this case the solid lines that represent the slices are a match between spacelike and null slices \cite{CBO9780511524639A010,Bishop:1998ukk}, performed at a timelike interface that is indicated by a thick dashed line. The interior of the spacetime (where the source dynamics takes place) is described by Cauchy slices, but the GW radiation emitted by the central system is tracked and extracted using characteristic slices. The Cauchy data serve as an interior boundary to the characteristic part and the characteristic data are used in the boundary conditions for the interior Cauchy slices. %The complete slice is continuous, but not differentiable, at the matching point, where the causal nature of the slice changes.
The causal nature of the slice changes at the matching point. 
Stability problems may arise in a numerical evolution due to the interpolation at the interface between the interior Cartesian code (Cauchy part) and the exterior spherical code (characteristic part). 
%For a review in the Characteristic initial value problem and in Cauchy-Characteristic matching see for instance \cite{Winicour:2005ge}.

Another approach that takes advantage of Cauchy and null slices is the Cauchy-characteristic extraction \cite{Bishop:1996gt,Bishop:1997ik,Zlochower:2003yh}. Here the Cauchy evolution supplies data for the inner boundary of the characteristic evolution, but the characteristic evolution does not provide data for the boundary conditions of the Cauchy slices. This avoids the instability problems present in the Cauchy-characteristic matching procedure and allows a radiation extraction free of finite radius approximations. However, the Cauchy integration domain becomes smaller in time, because the part affected by the boundary grows with time. This implies that a long evolution requires a large initial Cauchy slice, which can become quite expensive computationally. 

%the causal nature of the slice changes at the interface
%stability problems due to the interpolation between the interior Cartesian code (Cauchy part) and the exterior spherical code (characteristic part)

\item[d) Hyperboloidal slices:] 
hyperboloidal slices are spacelike slices that tend asymptotically to null slices and reach $\scri$ (while being spacelike). The solid line shows a family of them. A property of hyperboloidal slices is that asymptotically the metric that describes the system has constant non-vanishing curvature. More specifically, the slices shown here have a constant trace of the extrinsic curvature and are called constant mean curvature (CMC) slices. Hyperboloidal slices are spacelike slices and thus do not have the coordinate problems that characteristic ones have, while also allowing for an appropriate GW extraction. The boundary of the integration domain can be put either right on $\scri^+$ or in the unphysical extended region. As $\scri^+$ is an ingoing null surface, no radiation can enter the integration domain from the unphysical part. Numerically this is not completely true, as numerical modes with superluminal speeds may indeed enter through $\scri^+$, but their effect is expected to converge away with resolution. 
Unlike the Cauchy problem, which is global in the sense that a single slice determines the entire spacetime, the hyperboloidal one is semi-global, because a slice intersecting $\scri^+$ can only predict the future, not its past. 

%boundary either right on $\scri^+$ or in the unphysical region, so influence from the boundary conditions should converge away. 
\end{description}

Hyperboloidal foliations are especially interesting and convenient compared to the other possible foliation options. They have the ideal asymptotic behaviour, reaching future null infinity and allowing for an unambiguous radiation extraction; the propagation speed of the radiation is always finite; the causal character of the foliation does not change, the slice is always smooth and spacelike and thus as flexible as a Cauchy slice. 
%list advantages: reach $\scri^+$, thus allowing for an unambiguous radiation extraction; finite speeds; as flexible as Cauchy slices
The numerical implementation of the problem poses however several difficulties, such as the treatment of the divergent terms, thus making it a challenging problem that motivates the work contained in this thesis. 

\section{Ingredients for the hyperboloidal approach}

\subsubsection{Behaviour of $\scri^+$}

In the numerical setup the location of $\scri^+$ is given by $\Omega=0$. 
For this reason, an important point in the implementation is the behaviour of the conformal factor $\Omega$ in time, which we can choose. It can be evolved in time with the rest of evolution variables, so that the position of $\scri^+$ will move in the integration domain, or its prescribed value can be kept fixed in time. 

The asymptotically flat case of \eref{ei:nablaOmega} implies that future null infinity is an ingoing null surface. When evolved in time, the physical domain to the left of $\scri^+$ will become smaller in time, as if it was moving outwards through $\scri^+$ at the speed of light. In a numerical simulation this translates to a loss of resolution, because the number of grid-points in the integration domain decreases with the iterations. 

A simple solution is to fix the position of $\scri^+$ in the numerical grid, where an especially interesting choice is to make it coincide with the outer boundary of the integration domain. This can be achieved using a scri-fixing gauge \cite{Frauendiener:1997ze}, which is obtained by imposing certain conditions onto the gauge variables.

%Fixing the position of $\scri^+$ in the grid is a simple implementation that can be achieved using a scri-fixing gauge \cite{Frauendiener:1997ze}, which is obtained by imposing certain conditions onto the gauge variables. %It fixes the location of $\scri^+$ to a certain position in the grid (which can be made to coincide with the outer boundary of the integration domain), so that it does not change with time and the resolution in maintained during the whole simulation. To obtain the scri-fixing gauge certain conditions have to be imposed onto the gauge variables. This will be discussed in detail in section \ref{cg:scrifix}. 

%scri-fixing Frauendiener [49] An\i l's thesis \cite{Frauendiener:1997ze}

\subsubsection{Bondi coordinates}

Null infinity is a shear-free null hypersurface. It can also be made expansion free by satisfying the preferred conformal gauge condition \eref{eg:condother}. 
The most convenient gauge at null infinity is to choose Bondi coordinates, because they correspond to inertial observers. %expressed in them the spacetime will be equivalent to Minkowski spacetime. 

If the time vector flows along $\scri^+$ and the preferred conformal gauge is satisfied, the null generators of $\scri^+$ are geodesic and their affine parameter can be identified with the Bondi time, the only type of time coordinate which will not deform the signal. 

%Bondi coordinates, affinely parametrized Bondi time 
%preferred conformal gauge
%scri is a shear-free null surface

\section[Brief history of the numerical hyperboloidal initial value problem]{Brief history of the hyperboloidal initial value problem from a numerical perspective} % in Numerical Relativity

%The motivation for what would afterwards develop into the hyperboloidal initial value problem and other practical applications was idea by Penrose \cite{PhysRevLett.10.66,Penrose:1965am}, originated as a result to the ongoing studies of isolated systems and fall-off behaviour of the gravitational fields. This idea suggested that the asymptotic behaviour of gravitational fields could be analyzed in terms of extensions to null infinity of the conformal spacetime described by the rescaled metric $\bar g_{ab}$ \eref{ei:rescmetric}. 

Taking Penrose's idea \cite{PhysRevLett.10.66,Penrose:1965am} as a starting point, Friedrich pioneered the work on the conformal extended spacetime \cite{Fredrich:aaa}. Starting from \eref{ei:einsteinc} he derived the Conformal Field Equations (CFEs) and showed that their solution for the conformally rescaled metric $\bar g_{ab}$ on the conformal extension $\mathcal{\bar M}$ transforms to a solution of the physical metric $\tilde g_{ab}$ on the physical spacetime $\mathcal{\tilde M}$. Imposing general gauge conditions to obtain the reduced system of the CFEs, the regularity of the equations could be manifestly shown; it was found that $\scri^+$ is smooth during evolution provided that the initial data on the hyperboloidal slice are smooth \cite{friedrich1983,Friedrich:2003fq}. The final system of equations includes about 60 component equations in total, more than usually solved by conventional codes. %rendering a numerical implementation quite complicated, as conventional codes solve approximately 10 component equations. 

The CFEs have been first tested numerically in a metric-based formulation by H\"ubner:
in spherical symmetry \cite{Hubner:1994pd} and in a more general three-dimensional framework \cite{Hubner:1998hn,Hubner:1999th,Hubner:2000zn,Hubner:2000pb}. 
% \cite{Hubner:1999th} and Husa \cite{Husa:2002kk}; %, where the evolved variables are the metric and extrinsic curvature; %,Husa:2002zc
%\upda{papers by H\"ubner \cite{Hubner:1998hn,Hubner:1999th,Hubner:2000zn,Hubner:2000pb}, others \cite{Hubner:1997xa}}
The simulations by H\"ubner had a small initial perturbation amplitude, so that Husa \cite{Husa:2002kk} studied the system for larger amplitudes. It was found that the equations were prone to continuum instabilities \cite{Husa:2005ns,Husa:2002zc}. 
Another numerical approach by Frauendiener \cite{Frauendiener:1997zc,Frauendiener:1997ze,Frauendiener:1998yi} took the form of a tetrad formalism, where a tetrad and its connection coefficients were used as variables. These simulations were performed in axisymmetry. 

%For a complete introduction to the CFEs and the geometry they describe see \cite{Friedrich:2002xz,lrr-2004-1}. 

An alternative approach was taken by Zengino\u{g}lu \cite{Zenginoglu:2007jw,Zenginoglu:2008wc,Zenginoglu:2008pw,Zenginoglu:2008uc}, who implemented a free evolution in spherical symmetry using the generalized harmonic formulation. The conformal factor $\Omega$ was chosen to be time independent and the coordinate gauge satisfied the preferred conformal gauge condition \eref{eg:condother}, that imposes Bondi coordinates at $\scri^+$. %, see chapter \ref{c:gauge} for more information. 
Preliminary results for Schwarzschild in spherical symmetry, as well as the detailed implementation that lead to them, are found in chapter 2 of \cite{Zenginoglu:2007it}. 
This approach has also become the standard method for solving the Teukolsky equation \cite{Zenginoglu:2011zz,Harms:2014dqa}. %\cite{Bernuzzi:2011aj} hyperboloidal layer method
%fixed conformal factor and generalized harmonic formulation. Free evolution with generalized harmonic

Regarding a constrained evolution, appropriate conditions for regularity at $\scri^+$ and indications on how to solve the conformal constraint equations are given in \cite{Andersson:2002gn}. %\cite{Andersson:springer}
A numerical implementation of the hyperboloidal initial value problem, using a partially constrained evolution scheme (solving a constraint for the conformal factor at every time-step), was performed by Rinne and Moncrief \cite{Rinne:2009qx,Rinne:2013qc}. The resulting axisymmetric code allowed the calculation of the outgoing radiation field at $\scri^+$. 
%Constrained evolution by Moncrief and Rinne : boundary conditions can be imposed, but have to solve elliptic equations at every timestep. Working axisymmetric code. 

Another approach based on the tetrad formalism was suggested by Bardeen, Sarbach and Buchman \cite{Bardeen:2011ip}. It consists of a first order constrained hyperbolic system with the connection coefficients as evolution variables. %Here the freedom in the tetrad definition is fixed by imposing the 3D Nester gauge. The regularity conditions of the resulting equations are discussed, but no numerical results have been presented yet. 

Efforts on hyperboloidal initial data  have been performed in spherically and axially symmetric spacetimes in \cite{Schneemann}, and as hyperboloidal Bowen-York initial data in \cite{Buchman:2009ew}. %, but this will be briefly described in the initial data calculation in chapter \ref{c:initial}.} 

\section{The approach of this work}

%We can derive several approach considerations from the results of previous work on the hyperboloidal value problem. Due to the divergent terms at null infinity that appear in the equations \eref{ei:einsteinc}, obtaining a well-behaved system of equations that allows for a non-crashing numerical evolution is not simple, especially in the case of an unconstrained evolution, where boundary conditions cannot be imposed. 

Here we will mainly follow Zengino\u{g}lu using a time-independent conformal factor $\Omega$ and a free (unconstrained) evolution. A considerable difference to the Conformal Field Equation's approach is that here we will first set our conformal and coordinate gauges (scri-fixing and gauge evolution equations with appropriate source functions) and then make sure that the resulting set of equations is regular. 3+1 decomposed formulations of the Einstein equations, namely the Baumgarte-Shapiro-Shibata-Nakamura (BSSN) \cite{PhysRevD.52.5428,Baumgarte:1998te} and Z4 \cite{bona-2003-67,Bona:2003qn} formalisms, are chosen with the aim of testing the hyperboloidal initial value problem with a setup commonly used in current numerical codes. A description of the approach used and the results obtained, that will be extensively explained in this thesis, are presented in \cite{Vano-Vinuales:2014koa,Vano-Vinuales:2014ada}. 

%in the simple example described in \ref{compMin} both space and time were compactified, but in the compactifications used in this work time will not be compactified

A sensible starting point is to test the system in spherical symmetry. Expressing the equations in spherical coordinates simplifies them drastically, makes their numerical implementation quite simple (a one-dimensional spatial grid evolved in time) and the resulting code provides results much faster than a three-dimensional one. In spite of the simplifications, the critical part embodied in the regularization of the radial coordinate is maintained even when imposing spherical symmetry, so that the methods and results from the spherically symmetric case are expected to also apply to the full three-dimensional case, at least to some degree. 
Besides being a first step towards solving the three-dimensional problem and so allowing to extract the GWs signal at null infinity, the spherically symmetric approach can also provide results for other interesting aspects of gravity, such as the collapse of scalar fields into BHs, perturbations of single BHs, which can be either Schwarzschild or Reissner-Nordstr\"om BHs, etc. 

%\upda{ From spherical symmetry derivation:
%Solving the Einstein equations in spherical symmetry (1+1 dimensions) enables us to deal with interesting aspects of the theory and to reach conclusions that in many cases can be easily extended to the full 3+1 dimensional problem. The advantage of restricting to spherical symmetry is that the computational resources required for the simulations are much lower than in the general case and results can be obtained in a much shorter time. 
%}

\section{Outline of the thesis}

The thesis is structured as follows: %in chapter \ref{c:introduction} I will introduce the concepts on which the hyperboloidal initial value formulation is based and the ingredients used to construct it, as well as presenting a brief review of the previous efforts and work on this topic and some of its useful applications. 
in chapter \ref{c:eqs} I will review the derivation of the 3+1 decomposed formulation of the Einstein equations including the divergent terms at null infinity, then transform them into their Generalized BSSN (GBSSN) and Z4 conformal (\CZ) formulations, which are commonly used in current numerical simulations, and finally present the spherically symmetric reduction that will be implemented. Chapter \ref{c:initial} is devoted to describing the calculation of appropriate initial data, as well as the study of hyperboloidal foliations in spacetimes including a spherically symmetric BH. A very important ingredient are the gauge conditions, which play a critical role in the simulations and require special adjustment and tuning to obtain well-behaved numerical evolutions; they are discussed in chapter \ref{c:gauge}. In chapter \ref{c:reg} I present the conditions that the equations have to satisfy at the continuum level to result in a well-behaved evolution, as well as the regularity conditions that have to hold at null infinity. Chapter \ref{c:num} describes the numerical implementation in the code. The main experiments performed are explained in chapter \ref{c:exper} and the results obtained are presented in chapter \ref{c:results}. A discussion of the achieved goals and future prospects of this work follow in chapter \ref{c:discussion}. The expressions used to construct the Penrose diagrams are included in appendix \ref{c:diagr}. 

\chapter{Initial value formulation}\label{c:eqs} %General Relativity % of the conformally rescaled equations

%\section{Introduction}

%One of the features of these equations is that time and space are mixed together. In order to find a systematic way of solving them, typically evolving some appropriate initial data in time (initial value problem), the time coordinate has to be separated from the spatial ones and treated in a different way. A brief description of this procedure is presented, following \cite{Wald,Alcubierre,Bona,Baumgarte:2002jm,Gourgoulhon:2007ue}. 

We will adopt the abstract index notation for the derivations in this chapter. Abstract tensor indices will be denoted by $a, b, c, ...$, four-dimensional tensor components by $\mu, \nu, \sigma, ...$ and three-dimensional tensor components by $i, j, k, ...$ . Most of the algebraic derivations were performed using the {\tt Mathematica} package {\tt xAct} \cite{xAct}.

\section{Conformally rescaled equations}

The Einstein equations written in terms of the rescaled metric $\bar g_{ab}=\Omega^2\tilde g_{ab}$ \eref{ei:rescmetric} have already been presented in section \ref{si:confeqs} as \eref{ei:einsteinc}. In this work we will restrict to the case of a vanishing cosmological constant $\Lambda=0$. We will derive the equations for our initial value problem within the Z4 formalism \cite{bona-2003-67,Bona:2003qn}. More specifically we will derive the conformally rescaled equations starting from the Einstein equations for the physical metric $\tilde g_{ab}$: 
\begin{equation}\label{c3:einsteinp}
G[\tilde g]_{ab}+2\tilde\nabla_{(a}\bar Z_{b)} -\tilde g_{ab}\tilde\nabla^c\bar Z_c - \kappa_1\left(2\,\tilde n_{(a}\bar Z_{b)}+\kappa_2\,\tilde g_{ab}\,\tilde n^c\bar Z_c\right) =8\pi T[\tilde g]_{ab}. 
\end{equation}
Here again $G[\tilde g]_{ab}= R[\tilde g]_{ab}-\case{1}{2}\tilde g_{ab} R[\tilde g]$ is the Einstein tensor constructed from the physical metric and $T[\tilde g]_{ab}$ is the stress-energy-momentum tensor. The extra dynamical quantity $\bar Z_a$ introduced in the Z4 formalism appears in the constraint propagation terms of the Z4 formulation (second and third terms in \eref{c3:einsteinp}'s left-hand-side (LHS)) and its damping terms \cite{Gundlach:2005eh} proportional to the timelike normal vector $\tilde n^a$ (with the parameters $\kappa_1$ and $\kappa_2$ chosen empirically). The Einstein equations are satisfied when the field $\bar Z_a$ vanishes.

The vector $\tilde n^a$ is defined as the future-directed normal to a three-dimensional spacelike hypersurface $\tilde\Sigma_t$ labeled with a constant value of the parameter $t$ (that will be interpreted as the time). The normal vector $\tilde n^a$ is such that $\tilde n^a\tilde n_a=-1$ is satisfied (it is a timelike unit vector). Under the conformal rescaling for the metric \eref{ei:rescmetric} the unit normal vector transforms as: 
\begin{equation}\label{e3:barn}
\bar n^a = \frac{\tilde n^a}{\Omega} \qquad \textrm{and} \qquad \bar n_a = \Omega \, \tilde n_a .
\end{equation}
A conformal transformation leaves the orientation of the objects (and thus the causal structure of the spacetime) invariant, so that $\bar n^a$ continues to be perpendicular to the now transformed hypersurface $\bar\Sigma_t$ and the transformations in \eref{e3:barn} are set in such a way that $\bar n^a\bar n_a=-1$ is satisfied. 

The transformation of the Ricci tensor due to the conformal rescaling of the metric is (see a standard textbook like \cite{Wald})
\begin{equation}\label{c3:einsteinR}
R[\tilde g]_{ab} = R[\bar g]_{ab} + {1\over\Omega}(2\bar\nabla_a\bar\nabla_b\Omega+\bar g_{ab}\bar\Box \Omega) - {3\over\Omega^2}\bar g_{ab}(\bar\nabla_c\Omega)(\bar\nabla^c\Omega) . 
\end{equation}
It is calculated by means of the transformation of the connection, which is also shown here for completeness: 
\begin{equation}\label{c3:Chrischange}
\tilde\Gamma_{ab}^c = \bar\Gamma_{ab}^c -\frac{1}{\Omega}\left(\delta_a^c\bar\nabla_b\Omega+\delta_b^c\bar\nabla_a\Omega-\bar g_{ab}\bar g^{cd}\bar\nabla_d\Omega\right) .  
\end{equation}

Under the conformal rescalings of the metric and the normal unit vector the Einstein equations become
\begin{equation}\label{c3:einstein}
 G[\tilde g]_{ab}+2\bar\nabla_{(a} \bar Z_{b)} -\bar g_{ab}\bar\nabla^c\bar Z_c + \frac{4}{\Omega} \bar Z_{(a}\bar\nabla_{b)}\Omega - \frac{\kappa_1}{\Omega}\left(2\,\bar n_{(a}\bar Z_{b)}+\kappa_2\,\bar g_{ab}\,\bar n^c\bar Z_c\right)=8\pi T[\case{\bar g}{\Omega^2}]_{ab} ,  
\end{equation}
where the physical metric appearing in the stress-energy tensor is expressed in terms of the rescaled one, $\bar g_{ab}$, and all indices are raised and lowered with the conformal metric $\bar g_{ab}$, whose covariant derivative is denoted by $\bar\nabla$ and $\bar\Box\equiv\bar g^{ab}\bar\nabla_a\bar\nabla_b$. The Einstein tensor of the physical metric, $G[\tilde g]_{ab}$, is related to that of the conformal metric, $G[\bar g]_{ab}$, as
\begin{equation}\label{c3:einsteinG}
G[\tilde g]_{ab} = G[\bar g]_{ab} + {2\over\Omega}(\bar\nabla_a\bar\nabla_b\Omega-\bar g_{ab}\bar\Box \Omega) + {3\over\Omega^2}\bar g_{ab}(\bar\nabla_c\Omega)(\bar\nabla^c\Omega) .
\end{equation}
By setting $\bar Z_a$ to zero the two previous equations reduce to \eref{ei:einsteinc} with vanishing cosmological constant. 

The Z4 quantities were introduced in the physical Einstein equations \eref{c3:einsteinp}, but adding them at the level of the conformal metric equations is in principle also feasible. In this case, \eref{c3:einstein} would look like
\begin{equation}\label{c3:einsteinZ4conf}
 G[\tilde g]_{ab}+2\bar\nabla_{(a} \bar Z_{b)} -\bar g_{ab}\bar\nabla^c\bar Z_c - \kappa_1\left(2\,\bar n_{(a}\bar Z_{b)}+\kappa_2\,\bar g_{ab}\,\bar n^c\bar Z_c\right)=8\pi T[\case{\bar g}{\Omega^2}]_{ab} . 
\end{equation}
There are no divergent conformal factor terms multiplying the $Z4$ variable. Although this last expression could a priori be expected to present better stability properties than \eref{c3:einstein}, the divergent damping terms appearing in \eref{c3:einstein} actually play a decisive role in controlling the continuum instabilities that arise in the equations. This will be explained in subsection \ref{se:gbssnstabi}. 

\section{3+1 decomposition}\label{s3:decomp}

\subsection{3+1 foliations}

We will now slice the conformally compactified spacetime into three-dimensional spacelike hypersurfaces following the common procedure. 
The normal to the spacelike hypersurfaces $\bar\Sigma_t$, defined by a constant value of the parameter $t$, was introduced in \eref{e3:barn}. It is expressed in terms of the parameter $t$ as a future pointing vector: 
\begin{equation}\label{c3:nalpha}
\bar n_a=-\alpha\bar\nabla_at \quad \textrm{or equivalently}\quad \bar n^a=-\alpha \bar g^{ab}\bar\nabla_bt  .
\end{equation}

The scalar quantity $\alpha$ is called the lapse function and satisfies $\alpha=(-\bar g^{ab}\bar\nabla_at\bar\nabla_bt)^{-1/2}$. It can be interpreted as the proper time elapsed between the hypersurfaces as seen by an observer moving along the normal direction ($d\tau=\alpha dt$).

\vspace{1ex}

The change of coordinates between two hypersurfaces $\bar\Sigma_t$ and $\bar\Sigma_{t+dt}$ can be expressed as $x^i_{t+dt}=x^i_t-\beta^idt$. The three components of a vector $\beta^i$ control the change in coordinates in the three spatial dimensions from one spacelike hypersurface to the next and belong to the vector $\beta^\mu=(0,\beta^i)^T$, called the shift vector. The shift vector is spacelike, so that it is orthogonal to the timelike normal: $\bar n_a\beta^a=0$. 

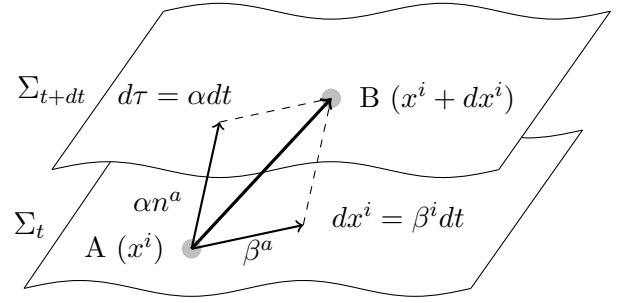
\begin{floatingfigure}[p]{0.5\linewidth}                         
    \center
    \begin{tikzpicture}[scale=1.05]
\draw[x=0.7cm,y=0.1cm] (0.1,8) node {$\Sigma_{t}$};
\draw[x=0.7cm,y=0.1cm] (0,0) sin (1,1) cos (2,0) sin (3,-1) cos (4,0) sin (5,1) cos (6,0) sin (7,-1) cos (8,0);
\draw[x=0.7cm,y=0.1cm] (8.9,19.1) sin (9,19) cos (10,20);%(3,20) sin (3,21) cos (4,20) sin (5,19) cos (6,20) sin (7,21) cos (8,20) sin (9,19)
\draw[x=0.7cm,y=0.1cm] (0,0) -- (1.55,16); \draw[x=0.7cm,y=0.1cm] (8,0) -- (10,20); 
\draw[x=0.7cm,y=0.1cm] (0.5,25) node {$\Sigma_{t+dt}$};
\draw[x=0.7cm,y=0.1cm] (0.5,15) sin (1.5,16) cos (2.5,15) sin (3.5,14) cos (4.5,15) sin (5.5,16) cos (6.5,15) sin (7.5,14) cos (8.5,15);
\draw[x=0.7cm,y=0.1cm] (2.5,35) sin (3.5,36) cos (4.5,35) sin (5.5,34) cos (6.5,35) sin (7.5,36) cos (8.5,35) sin (9.5,34) cos (10.5,35);
\draw[x=0.7cm,y=0.1cm] (0.5,15) -- (2.5,35); \draw[x=0.7cm,y=0.1cm] (8.5,15) -- (10.5,35); 
\fill[x=0.7cm,y=0.1cm, color=lightgray] (3, 5) circle (3.5pt); \fill[x=0.7cm,y=0.1cm, color=lightgray] (5.5, 24) circle (3.5pt);
\draw[x=0.7cm,y=0.1cm] (1.8,5) node {\small A ($x^i$)}; \draw[x=0.7cm,y=0.1cm] (7.4,24) node {\small B ($x^i+dx^i$)};
\draw[x=0.7cm,y=0.1cm, very thick, ->] (3, 5) -- (5.5, 24);
\draw[x=0.7cm,y=0.1cm, dashed](3.5, 21) -- (5.5, 24); \draw[x=0.7cm,y=0.1cm, dashed](5, 8) -- (5.5, 24);
\draw[x=0.7cm,y=0.1cm, thick, ->](3, 5) -- (3.5, 21); \draw[x=0.7cm,y=0.1cm, thick,  ->](3, 5) -- (5, 8);
%\draw[x=0.7cm,y=0.1cm, thin] (3.0625, 7) -- (3.2625,7); \draw[x=0.7cm,y=0.1cm] (3.2,6.2) -- (3.2625,8); 
\draw[x=0.7cm,y=0.1cm] (2.7,24.5) node {\small $d\tau=\alpha dt$}; \draw[x=0.7cm,y=0.1cm] (6.7,9) node {\small $dx^i=\beta^i dt$};
\draw[x=0.7cm,y=0.1cm] (2.4,11) node {\small $\alpha n^a$}; \draw[x=0.7cm,y=0.1cm] (4.2,4.5) node {\small $\beta^a$};
    \end{tikzpicture}
    \caption{Two spacelike slices $\Sigma_{t}$ and $\Sigma_{t+dt}$ and the change in the coordinates between the points $A$ and $B$. Neither tildes nor overbars are added to the symbols, as this decomposition is valid both in the physical (denoted by tildes) and in the conformal (denoted by overbars) pictures.}
    \label{n:3+1foliations}
\vspace{3ex}
\end{floatingfigure}

\vspace{2ex}

The time vector $t^a$ is defined in terms of the previous quantities as: 
\begin{equation} \label{c3:tvect}
t^a=\alpha \bar n^a+\beta^a  .
\end{equation}
It is tangential to the time lines, the lines with constant spatial coordinates. In general $t_a\neq\bar\nabla_at$, because the previous relation and \eref{c3:nalpha} yield $t_a=-\alpha^2\bar \nabla_at+\beta_a$ (with $\beta_a=\bar g_{ab}\beta^b$). Using the definitions of $t^a$ and $\bar n_a$ we obtain that $t^a\bar\nabla_at=1$, which means that $t^a$ is a basis vector and $\bar\nabla_at$ is a basis covector. 

The interpretation of (\ref{c3:tvect}) is that the evolution from one slice $\bar\Sigma_t$ to the next one is determined by the lapse and the shift: The first determines the proper time along the vector $\bar n^a$ and the second regulates how spatial coordinates are shifted with respect to the normal vector. 

\subsection{3+1 decomposition of the variables}

The tensors that appear in the equations are projected perpendicular to the spacelike surfaces or tangential to them. To project them perpendicular to $\bar\Sigma_t$ (that is, parallel to $\bar n^a$), the tensor has to be contracted with $\bar n^a$. The projection tangential to $\bar \Sigma_t$ (normal to $\bar n^a$) is performed using the projection operator \begin{equation}\bar\perp^a_b\equiv\delta^a_b+\bar n^a\bar n_b\end{equation} to contract the quantities. 
To illustrate the 3+1 projection procedure, the decomposition of a tensor $T_a\,^b$ is given by 
\begin{eqnarray}
T_a\,^b= \delta_a^c\delta_d^bT_c\,^d&=&(\bar\perp_a^c-\bar n_a\bar n^c)(\bar\perp_d^b-\bar n_d\bar n^b)T_c\,^d \nonumber\\&=& \bar\perp_a^c\bar\perp_d^bT_c\,^d - \bar\perp_a^c\bar n_d\bar n^bT_c\,^d - \bar\perp_d^b\bar n_a\bar n^cT_c\,^d + \bar n_a\bar n^c\bar n_d\bar n^bT_c\,^d .\label{e3:tdecomp}
\end{eqnarray}
The first term in \eref{e3:tdecomp}'s right-hand-side (RHS) is the tangential term and the last one is the normal term, while the other two terms are mixed. Each index of the tensor has to be projected. 

\subsubsection{Metric}

The result of applying the projection operator to the metric $\bar\perp^c_a\bar\perp^d_b \bar g_{cd}\equiv\mttr_{ab}$ gives a spacelike projection of the metric $\mttr_{ab}$, induced on $\bar \Sigma_t$ from the four-dimensional metric $\bar g_{ab}$. The relation between the original metric and the spacelike projected one is %Note that despite being spacelike, the projected metric is a four-dimensional object.
\begin{equation}\label{e3:induc}
\mttr_{ab}\equiv \bar g_{ab}+\bar n_a\bar n_b , 
\end{equation}
where we used the relation $\bar n^a\, \mttr_{ab}=0$, which holds because $\mttr_{ab}$ is spacelike. 

The three dimensional space on the spacelike hypersurface $\bar \Sigma_t$ can be described using the three dimensional spatial part of $\mttr_{ab}$, that will be denoted by $\bar\gamma_{ab}$: 
\begin{equation}
dl^2=\bar\gamma_{ij}dx^idx^j  .
\end{equation}
Note that $\mttr^{ab}$ is not the inverse of the projected metric $\mttr_{ab}$. Their relation is
\begin{equation}
\mttr^{ac}\mttr_{cb}=\bar g^{ac}\mttr_{cb}=\mttr^a_b\equiv\bar \perp^a_b=\delta^a_b+\bar n^a\bar n_b  . 
\end{equation}
The spacelike four-dimensional metric $\mttr^{ab}$ and $\mttr_{ab}$ can be considered as the projection operator with indices raised or lowered with the four-dimensional metric $\bar g_{ab}$. 

%\vspace{2ex}

%\upda{Have not changed the indices i's to a's from the following part yet.}

The components of $\bar g_{ab}$ are expressed in terms of $\alpha$, $\beta^i$ and $\bar\gamma_{ij}$ for its decomposition. For this, each of its components has to be determined. They are given by $\bar g_{\mu\nu}=\bar g_{ab}e^a_\mu e^b_\nu$ as follows, where the relations  $\bar n^a\bar n_a=-1$, $\beta^a\bar n_a=0$, $\beta^\mu=(0,\beta^i)^T$, $\bar n_a\bar \perp^a_i=0$ and $\beta_a\bar \perp^a_i=\beta_i$ have been used:
\begin{subequations}\begin{eqnarray}
\bar g_{00} &=& \bar g_{ab}e^a_0e^b_0 = \bar g_{ab}t^at^b = t^at_a = (\alpha \bar n^a+\beta^a)(\alpha \bar n_a+\beta_a) = -\alpha^2+\beta^k\beta_k  , \\
\bar g_{0j} &=& \bar g_{ab}e^a_0e^b_j = \bar g_{ab}t^ae^b_j  = t_be^b_j = (\alpha \bar n_b+\beta_b)\bar \perp^b_j = \beta_b\bar \perp^b_j = \beta_j  , \\
\bar g_{i0} &=& \bar g_{ab}e^a_ie^b_0 = ... = \beta_i  , \\
\bar g_{ij} &=& \bar g_{ab}e^a_ie^b_j = \bar g_{ab}\bar \perp^a_i\bar \perp^b_j = \bar \gamma_{ij}  . 
\end{eqnarray}\end{subequations}

The four dimensional metric $\bar g_{ab}$ and its inverse can be written as
\begin{equation}\label{e3:fourmetric}
\bar g_{\mu\nu}=\left( \begin{array}{cc} -\alpha^2+\beta_k\beta^k&\beta_j\\ \beta_i&\bar \gamma_{ij}  \end{array} \right) \qquad \textrm{and} \qquad
\bar g^{\mu\nu}=\frac{1}{\alpha^2}\left( \begin{array}{cc} -1&\beta^j\\ \beta^i&\alpha^2\bar \gamma^{ij}-\beta^i\beta^j  \end{array} \right)  . 
\end{equation}
Using \eref{e3:induc} and the fact that now the normal timelike vector in components takes the form $\bar n_\mu=(-\alpha,0)$ and $\bar n^\mu=\case{1}{\alpha}(1,-\beta^i)^T$, the spacelike projection $\mttr_{ab}$ and inverse are given by
\begin{equation}
\mttr_{\mu\nu}=\left( \begin{array}{cc} \beta_k\beta^k&\beta_j\\ \beta_i&\bar \gamma_{ij}  \end{array} \right) \qquad \textrm{and} \qquad
\mttr^{\mu\nu}=\left( \begin{array}{cc} 0&0\\ 0&\bar \gamma^{ij}  \end{array} \right)  .
\end{equation}
The line element now takes the form
\begin{equation}
d\bar s^2=\left(-\alpha^2+\beta_i\beta^i\right)dt^2+2\beta_idtdx^i+\bar\gamma_{ij}dx^idx^j . %\upda{\textrm{ use a's or i's??}} .
\end{equation}
The index of the three-dimensional shift vector has to be lowered with the three-dimensional spatial metric $\beta_i\equiv\bar \gamma_{ij}\beta^j$. % \upda{\textrm{ use a's or i's??}}

\subsubsection{Extrinsic curvature}

The Einstein equations contain second derivatives of the metric, so that second order derivatives in time will appear in the decomposition. To express them as a first order in time system a new variable has to be introduced: the extrinsic curvature $\Ktr_{ab}$ is defined as 
\begin{equation}
\Ktr_{ab}\equiv-\bar \perp^c_a\bar\nabla_cn_b = -(\bar \nabla_a\bar n_b+\bar n_a\bar n^c\bar \nabla_c\bar n_b)  .
\end{equation}
It is a purely spacelike and symmetric tensor that describes the curvature of the spacelike hypersurface $\bar\Sigma_t$ with respect to its embedding in the four-dimensional spacetime. The sign convention is the common one in Numerical Relativity, opposite to the one used in \cite{Wald}. 

The extrinsic curvature $\Ktr_{ab}$ can also be expressed using the Lie derivative along the normal direction ${\cal L}_{\bar n}$:  
\begin{equation}
{\cal L}_{\bar n}\mttr_{ab} = \bar n^c\bar \nabla_c\mttr_{ab}+\mttr_{ac}\bar\nabla_b\bar n^c+\mttr_{cb}\bar\nabla_a\bar n^c = ... =\, \mttr^c_a\bar\nabla_c\bar n_b + \mttr^c_b\bar\nabla_c\bar n_a = - 2\Ktr_{ab}  .
\end{equation}
Here the relations $\bar n^a\bar\nabla_b\bar n_a=0$, $\bar \nabla_c \bar g_{ab}=0$ and the symmetry property of $\Ktr_{ab}$ have been used. The extrinsic curvature can now be expressed in terms of the spacelike projection of the metric:
\begin{equation}\label{e3:K}
\Ktr_{ab}=-\frac{1}{2}{\cal L}_{\bar n}\mttr_{ab}  .
\end{equation}

\subsection{3+1 decomposition of the equations}

%\subsection{Derivatives}

The spacelike equivalent to the covariant derivative $\bar\nabla_a$ that acts on spatial tensors is the ``projected'' covariant derivative $\bar D_a\equiv\bar\perp_a^b\bar\nabla_b$.  %It can only be applied to a spatial tensor. 
The projection operator $\bar\perp_a^b$ has to be applied to all the indices in the expression, not just $\bar\nabla_a$. %As an example
The new three-dimensional spatial covariant derivative applied to the three-dimensional spacial metric vanishes:  
\begin{equation} \bar D_k\bar \gamma_{ij}=0 . \end{equation}

\subsubsection{Decomposition of the Riemann tensor}

An intermediate step to decomposing the Einstein equations in \eref{c3:einstein} is the decomposition of the four-dimensional Riemann tensor $R[\bar g]^a{}_{bcd}$. We perform three independent projections: 
\begin{itemize}

\item Full projection onto the spacelike hypersurface $\bar\Sigma_t$ - Gauss-Codazzi equation: 
\begin{equation}
\bar \perp_a^e\bar \perp_b^f\bar \perp_c^g\bar \perp_d^hR[\bar g]_{efgh}= R[\mttr]_{abcd}+\, \Ktr_{ac}\, \Ktr_{bd}-\, \Ktr_{ad}\, \Ktr_{bc}  .\label{e3:4perp}
\end{equation}

\item Projection onto $\bar \Sigma_t$ of the Riemann tensor contracted once with the normal vector - Codazzi-Mainardi equations: 
\begin{equation}
\bar \perp_a^e\bar \perp_b^f\bar \perp_c^g\bar n^dR[\bar g]_{efgd}=\bar D_b\, \Ktr_{ac}-\bar D_a\, \Ktr_{bc}  . \label{e3:3perp}
\end{equation}

\item Projection onto $\bar \Sigma_t$ of the Riemann tensor contracted twice with the normal vector - Ricci equations: 
\begin{equation}
\bar \perp_a^e\bar \perp_c^f\bar n^b\bar n^dR[\bar g]_{ebfd}={\cal L}_{\bar n}\, \Ktr_{ac}+\frac{1}{\alpha}\bar D_a\bar D_c\alpha+\, \Ktr_{ad}\, \Ktr_c^d  . \label{e3:2perp}
\end{equation}
\end{itemize}

The Riemann tensor $R[\bar g]_{abc}{}^d$ is a function of the spacetime metric $\bar g_{ab}$, whereas the ``induced'' Riemann tensor $R[\mttr]_{abc}{}^d$ that describes $\bar\Sigma_t$'s curvature is expressed in terms of the projected metric $\mttr_{ij}$. %\upda{?? $R[\mttr]_{abc}^d$ in indeed 4D, but later reinterpreted as 3D, right?}

%\vspace{2ex}

\subsubsection{Derivation of the 3+1 equations}\label{s3:deriv31}

The following relations will be used in the decomposition of the Einstein equations: 
\begin{subequations}
\begin{eqnarray}
%\begin{equation}
\bar \perp^{ac}\bar \perp^{bd}R[\bar g]_{abcd} &=& R[\bar g]+2\bar n^a\bar n^bR[\bar g]_{ab} = 2\bar n^a\bar n^bG[\bar g]_{ab} , \label{e3:2rel} \\
\bar \perp^{ab}\bar n^cR[\bar g]_{bc} &=&\bar \perp^{ab}\bar n^cG[\bar g]_{bc}  , \label{e3:1rel}
%\end{equation}
\end{eqnarray}
\end{subequations}
where $R[\bar g]_{ab}$ and $R[\bar g]$ are the four dimensional Ricci tensor and scalar and $G[\bar g]_{ab}$ is the four dimensional Einstein tensor $G[\bar g]_{ab}=R[\bar g]_{ab}-\frac{1}{2}\bar g_{ab}R[\bar g]$, all of them expressed in terms of the conformal metric $\bar g_{ab}$. 

\vspace{1ex}

Contracting the Gauss-Codazzi equation (\ref{e3:4perp}) twice (using $\bar \perp^{ab}\equiv\bar \gamma^{ab}$) we find that 
\begin{equation}
\bar \perp^{ac}\bar \perp^{bd}R[\bar g]_{abcd}=\ R[\mttr]+\bar K^2-\Ktr_{ab}\Ktr^{ab}  ,
\end{equation}
where $\bar K\equiv \Ktr_a^a$ the trace of the extrinsic curvature tensor. Substituting (\ref{e3:2rel}) gives
\begin{equation}
2\bar n^a\bar n^bG[\bar g]_{ab}=R[\mttr]+\bar K^2-\Ktr_{ab}\Ktr^{ab} .
\end{equation}
Finally we use the Einstein equations for the conformal metric (\eref{c3:einstein} and \eref{c3:einsteinG})
\begin{eqnarray}
R[\mttr]+\bar K^2-\Ktr_{ab}\Ktr^{ab} && \nonumber \\
 +\frac{4\left(\mttr^{ab}\bar D_a\bar D_b\Omega+\bar K \mathcal{L}_{\bar n}\Omega\right)}{\Omega}+\frac{6\left[(\mathcal{L}_{\bar n}\Omega)^2-\mttr^{ab}(\bar D_a\Omega)(\bar D_b\Omega)\right]}{\Omega^2}&& \nonumber \\
-2\bar K\Theta+2\bar D_a Z^a-\frac{2Z^a\bar D_a\alpha}{\alpha} -\frac{2\kappa_1\left(2+\kappa_2\right)\Theta}{\Omega}-2\mathcal{L}_{\bar n}\Theta -\frac{8\Theta\mathcal{L}_{\bar n}\Omega}{\Omega} &=& 16\pi\rho. \label{e3:3eqTh}
\end{eqnarray}
The scalar quantity $\rho$ denotes the local energy density as measured by observers following the normal trajectories to the spacelike hypersurfaces and is defined as $\rho\equiv \bar n^a\bar n^bT[\bar g]_{ab}$. 
The variables $\Theta$ and $Z_a$ are introduced as the projections of $\bar Z_a$ along the normal and tangential directions to $\bar\Sigma_t$ respectively. The decomposition of the Z4 quantity is thus $\bar Z_a=Z_a+\bar n_a\Theta$, where $\Theta=-\bar n^a \bar Z_a$ and $Z_a = \bar\perp_a^b\bar Z_b$. 
%Given that no derivatives along the normal $\bar n^a$ are present, which are related to the time derivatives, this is not an evolution equation. It is a constraint equation, a relation between different variables that has to be satisfied at all times during the evolution. This equation is called the Hamiltonian or energy constraint. 

\vspace{1ex}

From the Codazzi-Mainardi equations (\ref{e3:3perp}) and using (\ref{e3:1rel}) one obtains
\begin{equation}
\bar \perp^{ab}\bar n^cG[\bar g]_{bc}=\bar \perp^{ab}\bar n^cR[\bar g]_{bc}=\bar D^a\bar K-\bar D_b\Ktr^{ab} .
\end{equation}
Substituting the field equations and writing $ J^a\equiv-\bar \perp^{ab}\bar n^cT[\bar g]_{bc}$ for the momentum density measured along the normal direction gives
\begin{eqnarray}\label{e3:3eqZ}
\bar D_b\Ktr^{ab}-\bar D^a\bar K&& \nonumber \\
-\frac{2\left[\Ktr^{ab}\bar D_b\Omega+\mttr^{ab}\bar D_b(\mathcal{L}_{\bar n}\Omega)\right]}{\Omega} 
-2\Ktr^{ab}Z_b -\frac{\kappa_1Z^a}{\Omega} \nonumber \\ + \mttr^{ab}\bar D_b\Theta -\frac{\mttr^{ab}\Theta\bar D_b\alpha}{\alpha}+\frac{2\mttr^{ab}\Theta\bar D_b\Omega}{\Omega}-\mttr^{ab}\mathcal{L}_{\bar n}Z_b -\frac{2Z^a\mathcal{L}_{\bar n}\Omega}{\Omega}
 &=&8\pi  J^a .
\end{eqnarray}
%As the case $a=0$ is trivial, three constraint equations are obtained, known as the momentum constraints. 

\vspace{1ex}

Contracting only once the Gauss-Codazzi equation yields the following relation:
\begin{equation}
\bar \perp^d_a\bar \perp^f_b(R[\bar g]_{df}+\bar n^c\bar n^eR[\bar g]_{cdef})= R[\mttr]_{ab}+\bar K\Ktr_{ab}-\Ktr_{ac}\Ktr^c_b . 
\end{equation}
The second term in its LHS can be substituted by the Ricci equations (\ref{e3:2perp}), while the first one appears in the projection of the Einstein equations written in terms of the Ricci tensor, which is
{\small
\begin{eqnarray}
\bar \perp^c_a\bar \perp^d_bR[\bar g]_{cd}&=&\bar \perp^c_a\bar \perp^d_b\left[8\pi\left(T[\bar g]_{cd}-\case{1}{2}\bar g_{cd}\bar g^{ef}T[\bar g]_{ef}\right)
%\right.\nonumber\\&&\left.
-\left(  \frac{2\bar\nabla_a\bar\nabla_b\Omega+\bar g_{ab}\bar\Box \Omega}{\Omega} - \frac{3\bar g_{ab}(\bar\nabla_c\Omega)(\bar\nabla^c\Omega)}{\Omega^2}\right) 
\right.\nonumber\\&&\left.
-\left( 2\bar\nabla_{(a} \bar Z_{b)} + \frac{4 \bar Z_{(a}\bar\nabla_{b)}\Omega}{\Omega} -\frac{2\bar g_{ab}\bar Z^c\bar\nabla_c\Omega}{\Omega} - \frac{\kappa_1\left(2\,\bar n_{(a}\bar Z_{b)}-(1+\kappa_2)\bar g_{ab}\,\bar n^c\bar Z_c\right)}{\Omega}\right) 
\right] . \qquad \ \ 
\end{eqnarray}
}%

Performing these substitutions and introducing a new variable $S_{ab}\equiv\bar \perp^c_a\bar \perp^d_bT[\bar g]_{cd}$ as the spatial stress tensor ($S\equiv S_a\,^a$), an evolution equation for the extrinsic curvature can be derived: 
\begin{eqnarray}\label{e3:3eqK}
{\cal L}_{\bar n}\Ktr_{ab}&=&-\frac{1}{\alpha}\bar D_a\bar D_b\alpha+R[\mttr]-2\Ktr_{ac}\Ktr_b^c + \Ktr_{ab} (\bar K-2\Theta)
+ 2 \bar D_{(a}Z_{b)} -  \frac{\kappa_1 (1+\kappa_2) \bar\gamma_{ab} \Theta}{\Omega} 
\nonumber \\ &&
 + \frac{3 \bar\gamma_{ab} \left[(\partial_\perp\Omega)^2-\alpha^2\bar D^{c}\Omega \bar D_{c}\Omega\right]}{\alpha^2 \Omega^2}   + \frac{4 Z_{(a} \bar D_{b)}\Omega}{\Omega} + \frac{2  \bar D_{b}\bar D_{a}\Omega}{\Omega} -  \frac{2  \bar\gamma_{ab} Z^{c} \bar D_{c}\Omega}{\Omega}\nonumber \\  && 
 + \frac{\bar\gamma_{ab} \bar D^{c}\alpha \bar D_{c}\Omega}{\alpha\Omega} + \frac{\bar\gamma_{ab} \bar\bigtriangleup\Omega}{\Omega} 
+ \frac{2 \bar K_{ab} {\cal L}_{\bar n}\Omega}{\Omega} + \frac{\bar\gamma_{ab} (\bar K-2\Theta) {\cal L}_{\bar n}\Omega}{\Omega}
 -  \frac{\bar\gamma_{ab} {\cal L}_{\bar n}{\cal L}_{\bar n}\Omega}{\Omega}
\nonumber \\ &&
 + 4\pi[\mttr_{ab}(S-\rho)-2S_{ab}] .
\end{eqnarray}

%all indices up down with $\mttr_{ab}$

\subsubsection{Decomposed evolution and constraint equations}

Even if the induced quantities on the spacelike hypersurface $\bar \Sigma_t$ are four-dimensional, the information is encoded exclusively in their spatial components. This means that in the adapted coordinate system that has been chosen, we do not need to consider the timelike components. From now on the indices in the equations will thus only cover the three spatial coordinates. The Lie derivative along the normal direction ${\cal L}_n$ can also be expressed in adapted coordinates as ${\cal L}_{\bar n}\equiv\frac{1}{\alpha}\partial_\perp$, with $\partial_\perp= \partial_t-{\cal L}_\beta$. The three-dimensional spatial metric $\bar\gamma_{ab}$ is used to raise and lower indices, and we use the notation $\bar\bigtriangleup\equiv\bar\gamma^{ab}\bar D_a\bar D_b$. 

Using \eref{e3:K} as evolution equation for the physical metric and solving \eref{e3:3eqTh}, \eref{e3:3eqZ} and \eref{e3:3eqK} for $\partial_\perp\Theta$, $\partial_\perp Z_a$ and $\partial_\perp\Ktr_{ab}$ respectively, the equations of motion are finally 
{\small
\begin{subequations}\label{e3:adm}
\begin{eqnarray}
 \partial_\perp \bar\gamma_{ab} &= & -2\alpha \bar K_{ab} \label{e3:admg} , \\ 
 \partial_\perp \bar K_{ab} &= & \alpha\left[R[\bar\gamma]_{ab}-2 \bar K_{a}^{c} \bar K_{bc} + \bar K_{ab} (\bar K-2C_{Z4c}\Theta) + 2 \bar D_{(a}Z_{b)} -  \frac{\kappa_1 (1+\kappa_2) \bar\gamma_{ab} \Theta}{\Omega}  \right] -  \bar D_{b}\bar D_{a}\alpha     \nonumber \\  && 
 + \frac{3 \bar\gamma_{ab} \left[(\partial_\perp\Omega)^2-\alpha^2\bar D^{c}\Omega \bar D_{c}\Omega\right]}{\alpha \Omega^2}   + \frac{4 \alpha Z_{(a} \bar D_{b)}\Omega}{\Omega} + \frac{2 \alpha \bar D_{b}\bar D_{a}\Omega}{\Omega} -  \frac{2 \alpha \bar\gamma_{ab} Z^{c} \bar D_{c}\Omega}{\Omega}\nonumber \\  && 
 + \frac{\bar\gamma_{ab} \bar D^{c}\alpha \bar D_{c}\Omega}{\Omega} + \frac{\alpha \bar\gamma_{ab} \bar\bigtriangleup\Omega}{\Omega} + \frac{2 \bar K_{ab} \partial_\perp\Omega}{\Omega} + \frac{\bar\gamma_{ab} (\bar K-2C_{Z4c}\Theta) \partial_\perp\Omega}{\Omega} \nonumber \\   && 
 + \frac{\bar\gamma_{ab} \partial_\perp\alpha \partial_\perp\Omega}{\alpha^2 \Omega}  -  \frac{\bar\gamma_{ab} \partial_\perp\partial_\perp\Omega}{\alpha \Omega} +4\pi\alpha\left[\bar\gamma_{ab}(S-\rho) - 2\bar  S_{ab}\right], \label{e3:admK} \\
 \partial_\perp \Theta &= & \frac{\alpha}{2}\left[R[\bar\gamma] - \bar K_{ab} \bar K^{ab} +\bar K(\bar K-2C_{Z4c}\Theta) + 2 \bar D_{a}Z^{a} - \frac{2\kappa_1 (2+\kappa_2) \Theta}{\Omega}\right]  -  C_{Z4c} Z^{a} \bar D_{a}\alpha   \nonumber \\   && 
+ \frac{3 \left[(\partial_\perp\Omega)^2-\alpha^2 \bar D^{a}\Omega \bar D_{a}\Omega\right]}{\alpha \Omega^2} + \frac{2 \alpha \bar\bigtriangleup\Omega}{\Omega}  + \frac{2 (\bar K-2C_{Z4c}\Theta) \partial_\perp\Omega}{\Omega} - 8 \pi \alpha \rho  , \\
 \partial_\perp Z_{a} &= & \alpha\left[\bar D_{b}\bar K_{a}^{b} - \bar D_{a}\bar K + \bar D_{a}\Theta - 2 \bar K_{ab} Z^{b} - \frac{\kappa_1 Z_{a}}{\Omega} \right]   -  C_{Z4c} \Theta \bar D_{a}\alpha + \frac{2 \alpha \Theta \bar D_{a}\Omega}{\Omega}   \nonumber \\   && 
 -  \frac{2 \bar D_{a}\partial_\perp\Omega}{\Omega}  -  \frac{2 \alpha \bar K_{a}{}^{b} \bar D_{b}\Omega}{\Omega}  -  \frac{2 Z_{a} \partial_\perp\Omega}{\Omega} + \frac{2 \bar D_{a}\alpha \partial_\perp\Omega}{\alpha \Omega} -8 \pi \alpha  J_{a} ,
\end{eqnarray}
\end{subequations}
}%
while the Hamiltonian and momentum constraints are given by
{\small
\begin{subequations}\label{e3:admc}
\begin{eqnarray}
 \mathcal{H} = R[\bar\gamma] - \bar K_{ab} \bar K^{ab} + \bar K^2  + \frac{6 \left[(\partial_\perp\Omega)^2-\alpha^2\bar D^{a}\Omega \bar D_{a}\Omega\right]}{\alpha^2 \Omega^2} 
 + \frac{4 \bar\bigtriangleup\Omega}{\Omega} + \frac{4 \bar K \partial_\perp\Omega}{\alpha \Omega}  - 16 \pi \rho  , \quad  \label{e3:admcH} \\
 \mathcal{M}^a =  \bar D_{b}\bar K^{ab} -  \bar\gamma^{ab} \bar D_{b}\bar K -  \frac{2 \bar K^{ab} \bar D_{b}\Omega}{\Omega} -  \frac{2 \bar\gamma^{ab} \bar D_{b}\partial_\perp\Omega}{\alpha \Omega} + \frac{2 \bar\gamma^{ab} \bar D_{b}\alpha \partial_\perp\Omega}{\alpha^2 \Omega} -8 \pi  J^{a}  . \quad \ \label{e3:admcM}
\end{eqnarray}
\end{subequations}
}%

The evolution equations of the Z4 variables can also be expressed as
{\small
\begin{subequations}\label{c3:admZ4c}
\begin{eqnarray}
 \partial_\perp \Theta &= & \frac{\alpha}{2}\left[\mathcal{H} -2C_{Z4c}\bar K\Theta + 2 \bar D_{a}Z^{a} - \frac{2\kappa_1 (2+\kappa_2) \Theta}{\Omega}\right]  -  C_{Z4c} Z^{a} \bar D_{a}\alpha  
 - \frac{4C_{Z4c}\Theta \partial_\perp\Omega}{\Omega}   , \qquad \quad \\
 \partial_\perp Z_{a} &= &  \alpha\left[\mathcal{M}_a + \bar D_{a}\Theta - 2 \bar K_{ab} Z^{b} - \frac{\kappa_1 Z_{a}}{\Omega} \right]   -  C_{Z4c} \Theta \bar D_{a}\alpha + \frac{2 \alpha \Theta \bar D_{a}\Omega}{\Omega}  -  \frac{2 Z_{a} \partial_\perp\Omega}{\Omega}  , \qquad
\end{eqnarray}
\end{subequations}
}%
where the dependence on the constraints has been explicitly written. 

Dropping the conformal factor terms and the Z4 terms, the equations \eref{e3:admg}, \eref{e3:admK}, \eref{e3:admcH} and \eref{e3:admcM} are the Arnowitt–Deser–Misner (ADM) equations \cite{Arnowitt:1962hi}, specifically in the form obtained by York \cite{york}. In the original ADM formulation a multiple of the Hamiltonian constraint  appears in \eref{e3:admK}'s RHS. %$-\case{\alpha\bar\gamma_{ab}}{2}\mathcal{H}$

The coefficient $C_{Z4c}$ is introduced to label the Z4 non-damping non-principal-part that are treated differently in variations of the Z4 formulation. For $C_{Z4c} = 0$ those terms are dropped in an equivalent way as done in the Z4 conformal (Z4c) formulation \cite{Bernuzzi:2009ex,Weyhausen:2011cg}, while for $C_{Z4c} = 1$ all Z4 terms are kept as in the conformal and covariant Z4 (CCZ4) formulation \cite{Alic:2011gg}. %More details \upda{in the following section}.

\subsection{Relation between physical and conformal quantities}\label{se:rel3+1}

The 3+1 decomposition just performed is formally the same in the conformal and in the physical picture. The physical extrinsic curvature $\tilde {\bar K}_{ab}$ can also be expressed in terms of the physical projected metric metric $\tilde{\bar\gamma}_{ab}$ as  
\begin{equation}\label{e3:physK}
\tilde{\bar K}_{ab} = -\case{1}{2}\mathcal{L}_{\tilde n}\tilde{\bar\gamma}_{ab} =-\case{1}{2\tilde\alpha}\partial_\perp\tilde{\bar\gamma}_{ab} \qquad \textrm{or equivalently} \qquad 
\tilde{\bar K}_{ab} = -\tilde{\bar{\perp_a^c}}\tilde\nabla_c \tilde n_b . 
\end{equation}
Using the transformation from the physical to the conformal metric \eref{ei:rescmetric} and \eref{e3:K}, the relation between the physical and conformal extrinsic curvatures and their traces $\puK=\tilde {\bar K}_a^a$ can be found. 

A list of the transformations between the conformal and physical (denoted by a tilde) versions of the most relevant quantities presented so far is % four- and three-dimensional 
\begin{subequations}
\begin{eqnarray}
\bar g_{ab} = \Omega^2\tilde g_{ab} &\leftrightarrow& \tilde g_{ab} =\frac{\bar g_{ab}}{\Omega^2} , \\
\bar \gamma_{ab} = \Omega^2\tilde{\bar \gamma}_{ab} &\leftrightarrow& \tilde {\bar \gamma}_{ab} =\frac{\bar \gamma_{ab}}{\Omega^2} , \\
\alpha = \Omega\tilde\alpha &\leftrightarrow& \tilde\alpha=\frac{\alpha}{\Omega}, \\
\bar K_{ab} =\Omega\tilde{\bar K}_{ab} -\frac{\tilde {\bar \gamma}_{ab}\partial_\perp\Omega}{\tilde\alpha} &\leftrightarrow& \tilde {\bar K}_{ab} = \frac{1}{\Omega}\bar K_{ab} +\frac{\bar \gamma_{ab}}{\alpha}\frac{\partial_\perp\Omega}{\Omega^2} , \\
\bar K = \frac{\puK}{\Omega}-\frac{3\partial_\perp\Omega}{\tilde\alpha\Omega^2}&\leftrightarrow& \puK=\Omega\bar K+\frac{3\partial_\perp\Omega}{\alpha} , \label{e3:physconfK} \\
\cT = \frac{\pT}{\Omega} &\leftrightarrow& \pT=\Omega\cT  . \label{e3:physconfTheta}
\end{eqnarray}
\end{subequations}
Note that the shift vector $\beta^a$ does not change under the conformal rescaling. 

%\upda{copy contents from gbssneqs.tex here to avoid starting it in a new page and the bibliography problems.}

%\renewcommand\bibname{{References}}
%\bibliographystyle{../../master/thesis/tocunsrt}
%\bibliography{../articles/hypcomp} 

%\chapter{Conformally rescaled equations in GBSSN and Z4c form}\label{c:gbssnz4}

%\upda{generalize {\huge everything} to Z4c}

\section{Generalized BSSN and conformal Z4}

The 3+1 decomposition just performed yields the ADM-York equations (+Z4 terms). In general, the numerical implementation of the ADM equations in unconstrained form behaves badly due to the fact that their initial value problem is not well-posed \cite{Kidder:2001tz} in absence of spherical symmetry. They can however be used as a starting point to derive some well-posed formulations of the Einstein equations. The aim of the work presented here is to test a numerical implementation of the hyperboloidal initial value problem in spherical symmetry using common techniques. For this purpose we will choose two formulations with well-posed initial value problem widely used in numerical codes. The reason for implementing two different formulations is that, when the numerical simulations are unstable, they can provide information about these instabilities from two different sides, helping to understand how they are originated and what can be done to cure them. 

The mentioned formulations are the Baumgarte-Shapiro-Shibata-Nakamura (BSSN) equations \cite{PhysRevD.52.5428,Baumgarte:1998te}, also called BSSNOK for the first formulation \cite{NOK} by Nakamura, Oohara and Kojima, and the Z4 formalism \cite{bona-2003-67,Bona:2003qn}. A useful feature of the BSSN formulation is that the spatial metric is split off into a conformal factor and the remaining metric part. This has also been introduced in some versions of Z4, like the Z4c formulation \cite{Bernuzzi:2009ex,Weyhausen:2011cg} or the CCZ4 system \cite{Alic:2011gg}. 

In the traditional BSSN formulation, the determinant $\gamma$ of the conformal metric $\gamma_{ab}$, defined in \eref{ce:chimetr}, is assumed to be unity. This is appropriate in the case of Cartesian coordinates, but it is incompatible with coordinates adapted to spherical symmetry. The Generalized BSSN (GBSSN) \cite{Brown:2007nt,Brown:2009dd} formalism does not impose this restriction on the determinant of the conformal metric and is thus very useful for spherically symmetric simulations. 

The formulations of the Einstein equations that will be used for the numerical work presented here are the GBSSN system and the Z4c system, expressed in such a way that on can easily switch between one an the other when performing the simulations. We will now proceed to the derivation of these formulations, starting from \eref{e3:adm} and \eref{e3:admc}. 

\subsection{Formulation and variables}

The GBSSN and conformal Z4 systems evolve a conformally rescaled version of the spatial quantities. This spatial rescaling introduces a conformal factor $\chi$, a scalar quantity that can be expressed in terms of the determinants of the spatial metric $\bar\gamma$ and of the conformally rescaled spatial metric $\gamma$ as 
\begin{equation}\label{et:chidef}
\chi = \left(\frac{\gamma}{\bar\gamma}\right)^{\frac{1}{3}} .
\end{equation}
Instead of $\chi$, the spatial conformal factor can also be expressed using the variables $\varphi$, related as $\varphi=-\case{1}{4}\ln\chi$ and $\chi= e^{-4\varphi}$, or $\psi$, with $\psi=\chi^{-1/4}$ and $\chi=\psi^{-4}$. Given that the determinants of the metric vary during evolution, the quantity $\chi$ is evolved with the rest of evolution variables. 

In this way a conformally rescaled spatial metric $\gamma_{ab}$ is defined by the rescaling of the spatial metric $\bar\gamma_{ab}$ (the induced spatial metric from the conformally compactified $\bar g_{ab}$) and the spatial extrinsic curvature $\bar K_{ab}$ defined in \eref{e3:K} is also conformally rescaled into a new $K_{ab}$ in a similar way: 
\begin{eqnarray}
\gamma_{ab} = \chi\, \bar\gamma_{ab} \qquad &\textrm{and}& \qquad \gamma^{ab}=\frac{\bar\gamma^{ab}}{\chi}, \label{ce:chimetr} \\
K_{ab} = \chi\, \bar K_{ab} \qquad &\textrm{and}& \qquad K^{ab}=\frac{\bar K^{ab}}{\chi} . \label{ce:chiK}
\end{eqnarray}

The rescaled extrinsic curvature tensor is divided into two parts, given by its trace $\bar K$ and a trace-free tensor $A_{ab}$: $K_{ab}=A_{ab} + \frac{1}{3} \gamma_{ab}\bar K$. The trace-free part is defined as
\begin{equation}\label{et:defA}
A_{ab}=K_{ab}-\frac{1}{3}\gamma_{ab}\bar K .
\end{equation}
The transformation rules of the introduced quantities are derived from the previous relations: the trace of the extrinsic curvature is left unchanged by the spatial rescaling, $\bar K=\bar K_{ab}\bar\gamma^{ab} = K_{ab}\gamma^{ab}$, while the trace-free part transforms in the same way as the extrinsic curvature itself: $A_{ab}=\chi\,\bar A_{ab}$, having defined $\bar A_{ab}=\bar K_{ab}-\frac{1}{3}\bar \gamma_{ab}\bar K$. 

Introducing a new variable is required for hyperbolicity (see section \ref{sn:wellposed}). Doing so will allow us to write the Ricci tensor in such a way that it resembles the ``spatial part'' of the wave equation (the first term in the RHS of \eref{et:ricciexpr}). This variable is used in \cite{Brown:2009dd,Alcubierre:2010is} and is defined as 
\begin{equation}
\Delta\Gamma^a = \Gamma^a-\hat\Gamma^a . \label{g:Lambdaa}
\end{equation}

The quantity $\Gamma^a$ was included in the original BSSN formulation \cite{Baumgarte:1998te} and is called the conformal connection. It is a contraction of the Christoffel symbols $\Gamma_{bc}^a$ constructed from the conformal metric $\gamma_{ab}$: 
\begin{equation}
\Gamma^a = \gamma^{bc}\Gamma^a_{bc} = -\frac{1}{\sqrt{\gamma}}\partial_b\left(\sqrt{\gamma} \, \gamma^{ab} \right), \label{et:Gammaa}
\end{equation}
Similarly, we define $\hat\Gamma^a = \gamma^{bc}\hat \Gamma^a_{bc}$, where $\hat \Gamma_{bc}^a$ is connection constructed from a time-independent background metric $\hat \gamma_{ab}$. It is useful to make the value of $\hat\gamma_{ab}$'s components coincide with those of a stationary solution of the spacetime. For instance, choosing the flat metric in Cartesian coordinates gives $\hat\Gamma^a_{bc}=0$ and $\Delta\Gamma^a=\Gamma^a$, like in the original BSSN formulation. 

An equivalent way of defining $\Delta\Gamma^a$ is with help of the quantity $\Delta\Gamma_{bc}^a$: 
\begin{equation}\label{et:DefineGamma}
\Delta\Gamma^a  = \gamma^{bc}\Delta\Gamma^a_{bc}, \qquad \textrm{where}\qquad \Delta\Gamma^a_{bc} = \Gamma^a_{bc} - \hat\Gamma_{bc}^a .
\end{equation} 

The actual variable that will be evolved by the equations is not $\Delta\Gamma^a$ but $\Lambda^a$, which will substitute the Z4 quantity $Z_a$ as evolution variable by 
\begin{equation}\label{et:LDZ}
\Lambda^a=\Delta\Gamma^a+2\gamma^{ab}Z_b .
\end{equation}

Optionally the trace of the extrinsic curvature $\bar K$ can be mixed with the scalar Z4 variable $\Theta$ providing a new variable $K$, which is the approach we will take here for similarity with the original BSSN formulation:  
\begin{equation}\label{et:KTmix}
K=\bar K -2\Theta .
\end{equation}

%The evolution variables are ...
\begin{comment}
\upda{
The resulting evolution variables $\gamma_{ab}$, $A_{ab}$, $K$, $\chi$, $\Theta$, $\Lambda^a$ (and $\bar Z_a$ - not evolved). They are respectively tensors, scalars, a contravariant vector and a covariant covector with no density weights. 
%
A tensor density is a ``tensor'' that transforms from one coordinate system to another like a normal tensor except for the fact that it is also multiplied by a power of the Jacobian determinant of the coordinate transformation. Given that the physical metric $\bar g_{ab}$ and extrinsic curvature $\bar K_{ab}$ are normal tensors, imposing in the original BSSN formulation that $g=1$ for the conformal metric turns $g_{ab}$ and $K_{ab}$ into tensors with a weight $\bar g^{-1/3}$ according to (\ref{g:physical}). 
%
In GBSSN, as $g$ is not set to any value, the weights of $\bar g$ and $g$ cancel and the metric and the extrinsic curvature become normal tensors. 
}
\end{comment}

\subsection{Derivation of the equations}

The starting point are the 3+1 decomposed equations \eref{e3:adm} and \eref{e3:admc}. 
%Note that $\partial_\perp \equiv \partial_t-\mathcal{L}_{\beta}$, where $\mathcal{L}_{\beta}$ is the Lie derivative along the shift. 
%In order to write the equations in a covariant form, instead of writing partial derivatives in the expression for $\mathcal{L}_{\beta}$, it is more convenient to use $\hat D_a$, the covariant derivative built from the background metric $\hat g_{ab}$. 
%The procedure to derive each of the equations in GBSSN/\CZ{} form, presented in \eref{et:tensoreqs} and \eref{et:tensorceqs}, is the following. 
Common instructions for the derivation of all equations are the substitution of $\bar \gamma_{ab} = \case{\bar\gamma_{ab}}{\chi}$ and its inverse, as well as $\bar K_{ab}=\case{1}{\chi}\left[A_{ab} + \frac{1}{3} \gamma_{ab}(K+2\Theta\right]$. The covariant derivatives $\bar D_a$ associated to $\bar\gamma_{ab}$ are to be transformed to the derivatives $D_a$ associated with $\gamma_{ab}$. Their transformation when applied to a scalar quantity, e.g. $\alpha$, is trivial: $\bar D_a\alpha\equiv D_a\alpha$. When applied to tensors,
%the transformation has to be performed with help of \eref{et:relchris}, that relates the transformation between the Christoffel symbols associated to $\bar\gamma_{ab}$ and to $\gamma_{ab}$: 
the difference between the connections associated to $\bar\gamma_{ab}$ and to $\gamma_{ab}$, given by the following relation, has to be taken into account: 
\begin{equation}
	\bar \Gamma^a_{bc}  =  \Gamma^a_{bc} -\frac{2\delta^a_{(b}\partial_{c)}\chi-\gamma^{ad}\gamma_{bc}\partial_d\chi}{2\chi}  . \label{et:relchris}
\end{equation}

Some useful relations that relate the ADM quantities to the GBSSN ones are:
{\small
\begin{eqnarray}
	R[\bar\gamma]_{ab} & = & R[\gamma]_{ab} + \frac{D_aD_b\chi+\gamma_{ab}\bigtriangleup\chi}{2\chi}-\frac{D_a\chi D_b\chi+3\gamma_{ab}D^c\chi D_c\chi}{4\chi^2}, \label{et:relrab}\\
	\bar R[\bar\gamma] & = & \chi R[\gamma] +2\bigtriangleup\chi-\frac{5D^a\chi D_a\chi}{2\chi}  , \label{et:relr} \\
	\bar D_a\bar D_b\alpha & = & D_a D_b \alpha +\frac{2D_{(a}\alpha D_{b)}\chi-\gamma_{ab}D^c\alpha D_c\chi}{2\chi} , \label{et:relaab} \\
	\bar \bigtriangleup\alpha & = & \chi\bigtriangleup\alpha-\case{1}{2}D^a\alpha D_a\chi , \qquad \bar \bigtriangleup\Omega  =  \chi\bigtriangleup\Omega-\case{1}{2}D^a\Omega D_a\chi , \label{et:rela} %\\
%  \left[ {\alpha} R[\bar\gamma]_{ab} - \bar D_a \bar D_b {\alpha} \right]^{\rm TF} & = &
%  \left[ {\alpha} R[\gamma]_{ab} - D_a D_b{\alpha} +\frac{\alpha D_a D_b\chi-2D_{(a}\alpha D_{b)}\chi}{2\chi}-\frac{\alpha D_a\chi D_b\chi}{4\chi^2}   \right]^{\rm TF}  , \label{et:reltf} \\
\end{eqnarray}
}%
Similarly as before, $\bigtriangleup\equiv\gamma^{ab} D_a D_b$. 
The spatial conformal metric $\gamma_{ab}$ is used to raise and lower indices for all quantities, except those of $Z_a$, $J^a$ and $\mathcal{M}^a$, which are moved with $\bar \gamma_{ab}=\case{\gamma_{ab}}{\chi}$ and its inverse. 

\subsubsection{Ricci tensor}

The quantity $\Delta\Gamma^a_{bc}$ can be introduced in the equations by expressing the spatial Ricci tensor related to $\gamma_{ab}$ in the following way \cite{Alcubierre:2010is,Brown:2009dd}
\begin{eqnarray}\label{et:ricciexpr}
	  R[\gamma]_{ab} & = & -\frac{1}{2} \gamma^{cd} \hat D_c \hat D_d \gamma_{ab} 
	+ \gamma_{c(a}D_{b)}\Delta\Gamma^c - \gamma^{cd} \gamma_{e(a} R[\hat \gamma]_{b)cd}{}^e \nonumber\\
%+ \gamma_{c(a}\hat D_{b)}\Delta\Gamma^c+ \Delta\Gamma^c \Delta\Gamma_{(ab)c}
       && + \gamma^{cd} \left( 2 \Delta\Gamma^e_{c(a} \Delta\Gamma_{b)ed} 
	+ \Delta\Gamma^e_{ac} \Delta\Gamma_{ebd} \right)  ,
\end{eqnarray}
where $\Delta\Gamma^a_{bc}$ is defined in \eref{et:DefineGamma}. The covariant derivative $\hat D_a$ is associated to the time-independent background metric $\hat\gamma_{ab}$ and thus relies on the connection functions $\hat\Gamma^a_{bc}$. The quantity $R[\hat \gamma]^a{}_{bcd}$ is the Riemann tensor built from the background metric $\hat \gamma_{ab}$. The contraction of this Ricci tensor will be substituted in the Hamiltonian constraint \eref{et:H} (and \eref{et:pKH}) and is given by
\begin{eqnarray}\label{et:ricciscaexpr}
	  R[\gamma]  =  \gamma^{ab}R[\hat \gamma]_{ab} = -\frac{1}{2}\gamma^{ab}  \gamma^{cd} \hat D_c \hat D_d \gamma_{ab} 
	+ D_{a}\Delta\Gamma^a  
	+ \gamma^{ab} \gamma^{cd} \Delta\Gamma^e_{ac}\left( 2  \Delta\Gamma_{bed} 
	+  \Delta\Gamma_{ebd} \right)  .
\end{eqnarray}

For the Ricci tensor and scalar in the evolution equations \eref{et:A} and \eref{et:T} (and \eref{et:pKA} and \eref{et:pKT}) respectively, we will use the following expression and its contraction: 
\begin{eqnarray}\label{et:riccizexpr}
	  R[\gamma]_{ab} + 2D_{(a}Z_{b)} & = & -\frac{1}{2} \gamma^{cd} \hat D_c \hat D_d \gamma_{ab} 
	+ \gamma_{c(a}D_{b)}\Lambda^c - \gamma^{cd} \gamma_{e(a} R[\hat \gamma]_{b)cd}{}^e \nonumber\\
%+ \gamma_{c(a}\hat D_{b)}\Delta\Gamma^c+ \Delta\Gamma^c \Delta\Gamma_{(ab)c}
       && + \gamma^{cd} \left( 2 \Delta\Gamma^e_{c(a} \Delta\Gamma_{b)ed} 
	+ \Delta\Gamma^e_{ac} \Delta\Gamma_{ebd} \right) . 
\end{eqnarray}
The reason for this difference is that in this way the $Z_a$ dependence is included in the evolution variable $\Lambda^a$ according to \eref{et:LDZ}. 
%The Z4 variable $Z_a$ is eliminated in favour of $\Lambda^a$ according to \eref{tensorZ}. 

\subsubsection{Evolution equations}

\begin{itemize}
\item
In the calculation of the evolution equation for $\chi$ \eref{et:chi}, the operator $\partial_\perp$ is applied to its definition in terms of \eref{et:chidef}. The equation of motion of the determinant of the spatial metric $\bar \gamma_{ab}$ is to be substituted with
\begin{equation} \label{g:lngp}
\partial_\perp\ln\bar \gamma = -2\alpha \bar K . 
\end{equation}
This expression is derived by contracting \eref{e3:admg} with $\bar\gamma^{ab}$ and using the relation $\bar\gamma^{ab}\partial_\perp\bar\gamma_{ab}=\partial_\perp\ln\bar\gamma$, valid for any metric and any derivative operator. 

\item
For the evolution equation of the rescaled spatial metric $\gamma_{ab}$ \eref{et:g}, the substitution $\bar \gamma_{ab}  =  \case{\gamma_{ab}}{\chi}$ is performed in \eref{e3:admg}. The just obtained evolution equation for $\chi$ has to be inserted to obtain the equation for $\gamma_{ab}$. 

\item 
We continue with the derivation of $\Theta$'s equation of motion \eref{et:T}. The transformations to the conformal new quantities and covariant derivatives (using \eref{et:relchris}) are performed and relations \eref{et:relr} and \eref{et:rela} are used to obtain the final expression. %The Ricci scalar appearing in the final equation is to be substituted by the contraction of \eref{et:riccizexp}. 

\item
In the case of $K$ with final equation \eref{et:K}, first the operator $\partial_\perp$ has to be applied to $\bar K=\bar \gamma^{ab}\bar K_{ab}$. Then we use $\partial_\perp\bar\gamma^{ab}=-\bar\gamma^{ac}\bar\gamma^{bd}\partial_\perp\bar\gamma_{cd}$ and substitute the evolution equations for $\bar \gamma_{ab}$ (\ref{e3:admg}) and for $\bar K_{ab}$ (\ref{e3:admK}). This yields the evolution equation of the trace of the extrinsic curvature $\bar K$. To obtain the final equation of motion of $K$, defined in \eref{et:KTmix}, the evolution equation of $\Theta$ has to be substituted. This is equivalent to adding $-\alpha\mathcal{H}$ and substituting the Hamiltonian constraint \eref{e3:admcH}. Only expressing all derivatives in terms of $D_a$ using \eref{et:relchris} and \eref{et:rela} is left. %\upda{Write eq for $\partial_\perp\bar K$?}

\item
For the $A_{ab}$ equation \eref{et:A}, $\partial_\perp$ is applied to its definition \eref{et:defA} and $\partial_\perp \chi$, $\partial_\perp \bar K_{ab}$ and $\partial_\perp \bar K$ (the last one has been computed as an intermediate step towards $\partial_\perp K$) are substituted.  Relation \eref{et:relr} is used for the Ricci tensor and \eref{et:relaab} is used on $\bar D_a\bar D_a\alpha$ and equivalently on $\bar D_a\bar D_a\Omega$ as part of the transformation from $\bar D_a$ to $D_a$. The notation with a superindex ${}^{\rm TF}$ is introduced to present the equation in a more compact way and means ``trace-free'', so that for a given tensor $T_{ab}$ we would have  $[T_{ab}]^{\rm TF}=T_{ab}-\case{1}{3}\gamma_{ab}\gamma^{cd}T_{cd}$. 
%For the trace-free part that will appear, the relation (\ref{et:reltf}) is to be used. 

\item 
Obtaining the evolution equation for $\Lambda^a$ is more complicated. From the definition of $\Delta\Gamma^a$ \eref{g:Lambdaa} we can write
\begin{equation}
\partial_\perp(\Delta\Gamma^a)=\partial_\perp\Gamma^a-\partial_\perp\hat\Gamma^a  . \label{g:eqdgam}
\end{equation}

To obtain the equation of motion for $\Gamma^a$ we apply $\partial_\perp$ to the second equality of its definition \eref{et:Gammaa}. The resulting $\partial_\perp\gamma_{ab}$ terms are substituted with the derived evolution equation for the conformal metric \eref{et:g}.
After this calculation the evolution equation for $\Gamma^a$ is obtained:  %(the first term comes from the fact that this variable has a weight): 
\begin{eqnarray}
  \partial_\perp { \Gamma}^a &=& g^{bc}\partial_b\partial_c\beta^a-\frac{1}{3} { \Gamma}^a \partial_\perp {\ln g}  - \frac{1}{6} { g}^{ab}\partial_b \partial_\perp{\ln g}  - 2{ A}^{ab}\partial_b \alpha \nonumber \\ & & 
-2\alpha\left(\partial_bA^{ab}+ \Gamma^c_{bc}A^{ab}\right)
 .
%\upda{\textrm{ updated from equation leq (not form lceq as previously thought)}}
\end{eqnarray}
%Dropping the Z4 and the conformal factor terms, it can be compared to the one shown in \cite{Brown:2009dd,Alcubierre:2010is}. 
The first term appears because the variable $\Gamma^a$ is a vector density with a weight.

Using the fact that the background metric $\hat\gamma_{ab}$ is time-independent, we can write the right term in (\ref{g:eqdgam})'s RHS as 
\begin{equation}
\partial_\perp\hat\Gamma^a = \partial_t(\gamma^{bc}\hat\Gamma^a_{bc})-{\cal L}_\beta(\gamma^{bc}\hat\Gamma^a_{bc})= \hat\Gamma^a_{bc}\partial_\perp \gamma^{bc} - \gamma^{bc}{\cal L}_\beta\hat\Gamma^a_{bc} , \label{et:hG}
\end{equation}
Here we have used that $\partial_\perp \equiv \partial_t-\mathcal{L}_{\beta}$, where $\mathcal{L}_{\beta}$ is the Lie derivative along the shift. The part $\partial_\perp \gamma^{bc}$ is substituted first with $-\gamma^{bd}\gamma^{ce}\partial_\perp \gamma_{de}$ and then using the corresponding equation of motion. 

The Lie derivative of $\hat\Gamma^a_{bc}$ in \eref{et:hG} is substituted using the following relation \cite{Alcubierre:2010is}
\begin{equation}\label{g:derivsbeta}
\gamma^{bc}\hat D_b\hat D_c\beta^a=\gamma^{bc}\partial_b\partial_c\beta^a+\gamma^{bc}{\cal L}_\beta\hat\Gamma^a_{bc}+\beta^d\gamma^{bc} R[\hat\gamma]^a{}_{bcd}  . 
\end{equation}

The expression of the evolution equation for $\Delta\Gamma^a$ is
\begin{eqnarray}
	\partial_\perp (\Delta\Gamma^a)  & = & g^{bc} {\hat D}_b{\hat D}_c \beta^a 
	- g^{bc} {R[\hat\gamma]}^a{}_{bcd}\beta^d 
		-\frac{1}{3} \Delta\Gamma^a \,\partial_\perp\ln g 
        - \frac{1}{6} g^{ab} \partial_b \partial_\perp\ln g -2A^{ab}\partial_b\alpha \nonumber\\ & & 
-2\alpha\left(\partial_bA^{ab} + \Delta\Gamma^c_{bc}A^{ab} + \hat\Gamma^c_{bc}A^{ab} + \hat\Gamma^a_{bc}A^{bc}\right)
.
%\upda{\textrm{ update equation (DGammaceq)}} 
\end{eqnarray}

The equation of motion of $Z_a$ in terms of the new variables is obtained by simple variable and derivative transformations. 
From $\Lambda^a$'s definition \eref{et:LDZ} we obtain: 
\begin{equation}
\partial_\perp\Lambda^a=\partial_\perp (\Delta\Gamma^a)+2\left(\gamma^{ab}\partial_\perp Z_b-Z_b\gamma^{ac}\gamma^{bd}\partial_\perp\gamma_{cd}\right) ,
\end{equation}
and only substituting the corresponding evolution equations is left. %\upda{$Z_a$ also includes the appropriate amount of momentum constraint. Look again at the derivation!}

\end{itemize}

\subsubsection{Constraint equations}

To obtain the Hamiltonian constraint \eref{et:H} one has to write $\bar K_{ab}$ in terms of $A_{ab}$, $K$ and $\Theta$, use (\ref{et:relr}) to substitute $R[\bar\gamma]$ and \eref{et:rela} for the Laplacian of $\Omega$. 
The momentum constraint \eref{et:M} is also obtained by simple variable substitution and derivative transformation. 
The third constraint \eref{et:Z} is the Z4 quantity $Z_a$, expressed in terms of $\Lambda^a$ and $\Delta\Gamma^a$ according to \eref{et:LDZ}. Note that $Z_a$ will appear in the final equations, but it is not to be regarded as an evolution equation (it has been eliminated in terms of $\Lambda^a$) but as a shortcut to be substituted by \eref{et:Z}.

\subsection{GBSSN and \CZ{} formulation}

%\upda{Old:} After making the calculations described in the previous section, the evolution equations of the GBSSN system finally give \cite{Brown:2009dd,Brown:2009ki}

\subsubsection{Tensorial equations using the conformal $\cK$ and conformal $\cT$}

The evolution equations are
{\small
\begin{subequations}\label{et:tensoreqs}
\begin{eqnarray}
\partial_\perp\chi& = & \case{2}{3} \alpha \chi (K+2\Theta) + \case{1}{3} \chi \partial_\perp\ln\gamma
 \label{et:chi} , \\ 
\partial_\perp\gamma_{ab}& = & -2 A_{ab} \alpha + \case{1}{3} \gamma_{ab} \partial_\perp\ln\gamma
  \label{et:g}, \\ 
\partial_\perp A_{ab}& = & \left[   \alpha \chi \left(R[\gamma]_{ab} + 2D_{(a}Z_{b)} \right) -  \chi D_{a}D_{b}\alpha  - D_{(a}\alpha D_{b)}\chi  -  \frac{\alpha D_{a}\chi D_{b}\chi}{4 \chi} + \case{1}{2} \alpha D_{a}D_{b}\chi \right. \nonumber \\  
&& \left. + 2 Z_{(a} \alpha D_{b)}\chi  + \frac{2\alpha D_{(a}\chi D_{b)}\Omega}{\Omega} + \frac{2 \alpha \chi D_{a}D_{b}\Omega}{\Omega}   + \frac{4\alpha \chi  Z_{(a} D_{b)}\Omega}{\Omega}  - 8 \pi \alpha \chi S_{ab} \right]^{TF} \nonumber \\ 
&& -2 \alpha A_{a}^{c} A_{bc} + \alpha A_{ab} [K+2(1-C_{Z4c})\Theta] + \frac{2 A_{ab} \partial_\perp\Omega}{\Omega} + \case{1}{3} A_{ab} \partial_\perp\ln\gamma
  , \label{et:A} \\
\partial_\perp K& = & \alpha\left[A_{ab} A^{ab} + \case{1}{3} (K+2\Theta)^2  + \frac{\kappa_1(1-\kappa_2) \Theta}{\Omega}\right] - \chi \bigtriangleup\alpha  + \case{1}{2} D^{a}\alpha D_{a}\chi + 2 C_{Z4c} Z^{a} D_{a}\alpha  \nonumber \\ 
&& + \frac{3 [(\partial_\perp\Omega)^2-\alpha^2\chi D^{a}\Omega D_{a}\Omega]}{\Omega^2 \alpha} -  \frac{2 C_{Z4c} \alpha Z^{a} D_{a}\Omega}{\Omega} + \frac{3 \chi D^{a}\alpha D_{a}\Omega}{\Omega} -  \frac{\alpha D^{a}\chi D_{a}\Omega}{2 \Omega}  \nonumber \\ 
&& + \frac{\alpha \chi \bigtriangleup\Omega}{\Omega} + \frac{[K+4\Theta] \partial_\perp\Omega}{\Omega} + \frac{3 \partial_\perp\alpha \partial_\perp\Omega}{\Omega \alpha^2} -  \frac{3 \partial_\perp\partial_\perp\Omega}{\Omega \alpha} + 4 \pi \alpha (\rho+S)
 \label{et:K} , \\ 
\partial_\perp \Lambda^a& = &  \frac{2Z^b \tilde D_{b}\beta^{a}}{\chi} +\alpha\left[ 2 A^{bc} \Delta \Gamma^{a}_{bc} -  \case{2}{3} D^{a}(2K+\Theta)  -  \frac{3 A^{ab} D_{b}\chi}{\chi} -  \frac{4 Z^{a}(K+2\Theta)}{3 \chi} -  \frac{2 \kappa_1 Z^{a} }{\Omega \chi} \right] \nonumber \\ 
&&  + \gamma^{bc} \hat D_{b}\hat D_{c}\beta^{a} - \gamma^{bc} R[\hat \gamma]^{a}{}_{bcd} \beta^{d}  - 2 A^{ab} D_{b}\alpha - 2 C_{Z4c} \Theta D^{a}\alpha  -  \frac{4 \alpha A^{ab} D_{b}\Omega}{\Omega} \nonumber \\ 
&& -  \frac{2 \alpha (2K+\Theta) D^{a}\Omega}{3 \Omega} +  \frac{2 C_{Z4c} \alpha \Theta D^{a}\Omega}{\Omega}  -  \frac{4 D^{a}\partial_\perp\Omega}{\Omega}  + \frac{4 D^a\alpha \partial_\perp\Omega}{\Omega \alpha}  -  \frac{4 Z^{a} \partial_\perp\Omega}{\Omega \chi} \nonumber \\ 
&& -  \case{1}{6} D^a\partial_\perp\ln\gamma -  \case{1}{3} \Delta \Gamma^{a} \partial_\perp\ln\gamma -  \frac{2 Z^{a} \partial_\perp\ln\gamma}{3 \chi} -  \frac{16 \pi J^{a} \alpha}{\chi}
  , \label{et:L}\\ 
\partial_\perp \Theta& = & \case{\alpha}{2}\left[ \chi (R[\gamma]+ 2 D^{a}Z_{a}) -  A_{ab} A^{ab} + \case{2}{3} (K+2\Theta)^2  - 2 C_{Z4c} \Theta (K+2\Theta)  -  \frac{2\kappa_1(2+\kappa_2) \Theta}{\Omega} \right]  \nonumber \\ 
&& + \alpha\bigtriangleup\chi -  \frac{5 \alpha D^{a}\chi D_{a}\chi}{4 \chi}  -  C_{Z4c} Z^{a} D_{a}\alpha -  \frac{C_{Z4c}\alpha Z^{a} D_{a}\chi}{2 \chi} + \frac{2 \alpha \chi \bigtriangleup\Omega}{\Omega}  -  \frac{\alpha D^{a}\chi D_{a}\Omega}{\Omega} \nonumber \\ 
&& + \frac{3 [(\partial_\perp\Omega)^2-\alpha^2\chi D^a\Omega D_a\Omega]}{\Omega^2 \alpha}   + \frac{2 K\partial_\perp\Omega}{\Omega} - 8 \pi \alpha \rho
  . \label{et:T} 
\end{eqnarray}
\end{subequations}
}%

The Ricci tensor that appears in \eref{et:A} is to be substituted by \eref{et:riccizexpr} and the Ricci scalar in \eref{et:T} by its contraction. 
The first term in \eref{et:L}'s RHS is added to cancel a potentially bad-behaved term. 
Note that the equations are not complete without specifying an evolution equations for the gauge conditions $\alpha$ and $\beta^a$. The question of which properties these gauge conditions must have will be addressed in chapter \ref{c:gauge}. 
%
%\upda{From paper 1:}
%The coefficient $C_{Z4c}$ denotes the Z4 non-damping non-principal-part terms that are dropped in the Z4c formulation \cite{Bernuzzi:2009ex,Weyhausen:2011cg}. 
%-They are kept in the CCZ4 one \cite{Alic:2011gg}, but here we are not able to compare with this last formulation, as there the Ricci scalar in the equation of motion of the trace of the extrinsic curvature is not substituted, while we eliminate it using the evolution equation of $\Theta$. Note that this difference does not play a role if the algebraic constraints are satisfied, because the properties of the continuum partial differential equations are unaffected by a change of variables. By choosing $C_{Z4c}=0$ our \CZ{} equations are the close as possible to the Z4c formulation, whereas setting $C_{Z4c}=1$ maintains all the non-principal-part terms. 
%
The constraint equations are given by
{\small
\begin{subequations}\label{et:tensorceqs}
\begin{eqnarray}
\mathcal{H}& = & \chi R[\gamma] - A_{ab} A^{ab} + \case{2}{3} (K+2\Theta)^2 + 2 \bigtriangleup\chi -  \frac{5 D^{a}\chi D_{a}\chi}{2 \chi} + \frac{6 [(\partial_\perp\Omega)^2-\alpha^2\chi D^{a}\Omega D_{a}\Omega]}{\Omega^2 \alpha^2}  \nonumber \\ 
&&  -  \frac{2  D^{a}\chi D_{a}\Omega}{\Omega} + \frac{4 \chi \bigtriangleup\Omega}{\Omega}  + \frac{4 (K+2\Theta) \partial_\perp\Omega}{\Omega \alpha} - 16 \pi \rho , \label{et:H} \\ 
\mathcal{M}_{a}& = & D_{b}A_{a}^{b}  -  \case{2}{3} D_{a}(K+2\Theta)  -  \frac{3 A_{a}^{b} D_{b}\chi}{2 \chi}  -  \frac{2 A_{a}^{b} D_{b}\Omega}{\Omega} -  \frac{2 (K+2\Theta) D_{a}\Omega}{3 \Omega} -  \frac{2 D_{a}\partial_\perp\Omega}{\Omega \alpha} \nonumber \\ 
&& + \frac{2 D_{a}\alpha \partial_\perp\Omega}{\Omega \alpha^2} 
-8 \pi J_{a} \label{et:M} , \\
%
%\mathcal{C}^{a}& = &  \Lambda^{a} - \Delta \Gamma^{a} -  \frac{2 Z^{a}}{\chi}
Z_{a}& = & \frac{\gamma_{ab}}{2}\left(\Lambda^{b} - \Delta \Gamma^{b}\right) \label{et:Z}. 
\end{eqnarray}
\end{subequations}
}%
The Ricci scalar in the Hamiltonian constraint has to be substituted with \eref{et:ricciscaexpr}.

\subsubsection{Tensorial equations for the physical $\pK$ and physical $\pT$}

The evolution equations equivalent to \eref{et:tensoreqs} in terms of $\pK$ and $\pT$ are
{\small
\begin{subequations}\label{et:pKtensoreqs}
\begin{eqnarray}
\partial_\perp\chi& = & \frac{2 \alpha \chi (\pK+2\pT)}{3\Omega} + \case{1}{3} \chi \partial_\perp\ln\gamma -\frac{2\chi\partial_\perp\Omega}{\Omega}
 \label{et:pKchi} , \\ 
\partial_\perp\gamma_{ab}& = & -2 A_{ab} \alpha + \case{1}{3} \gamma_{ab} \partial_\perp\ln\gamma
  \label{et:pKg}, \\ 
\partial_\perp A_{ab}& = & \left[ \alpha \chi \left(R[\gamma]_{ab} + 2D_{(a}Z_{b)} \right) -  \chi D_{a}D_{b}\alpha  - D_{(a}\alpha D_{b)}\chi  -  \frac{\alpha D_{a}\chi D_{b}\chi}{4 \chi} + \case{1}{2} \alpha D_{a}D_{b}\chi \right. \nonumber \\  
&& \left. + 2 Z_{(a} \alpha D_{b)}\chi  + \frac{2\alpha D_{(a}\chi D_{b)}\Omega}{\Omega} + \frac{2 \alpha \chi D_{a}D_{b}\Omega}{\Omega}   + \frac{4\alpha \chi  Z_{(a} D_{b)}\Omega}{\Omega}  - 8 \pi \alpha \chi S_{ab} \right]^{TF} \nonumber \\ 
&& -2 \alpha A_{a}^{c} A_{bc} + \frac{\alpha A_{ab} [\pK+2(1-C_{Z4c})\pT]}{\Omega} - \frac{A_{ab} \partial_\perp\Omega}{\Omega} + \case{1}{3} A_{ab} \partial_\perp\ln\gamma
  , \label{et:pKA} \\
\partial_\perp \pK& = & \alpha\left[\Omega A_{ab} A^{ab} + \frac{(\pK+2\pT)^2}{3\Omega}  + \frac{\kappa_1(1-\kappa_2) \pT}{\Omega}\right] - \Omega \chi \bigtriangleup\alpha  + \case{1}{2}\Omega D^{a}\alpha D_{a}\chi  \nonumber \\ 
&& + 2 C_{Z4c}\Omega Z^{a} D_{a}\alpha - \frac{3 \alpha\chi D^{a}\Omega D_{a}\Omega}{\Omega} -  2 C_{Z4c} \alpha Z^{a} D_{a}\Omega + 3 \chi D^{a}\alpha D_{a}\Omega - \case{1}{2}\alpha D^{a}\chi D_{a}\Omega \nonumber\\&& 
 + \alpha \chi \bigtriangleup\Omega + 4 \pi \alpha \Omega(\rho+S)
 \label{et:pKK}, \\ 
\partial_\perp \Lambda^a& = &  \frac{2Z^b \tilde D_{b}\beta^{a}}{\chi} +\alpha\left[ 2 A^{bc} \Delta \Gamma^{a}_{bc} -  \frac{2 D^{a}(2\pK+\pT)}{3\Omega}  -  \frac{3 A^{ab} D_{b}\chi}{\chi} -  \frac{4 Z^{a}(\pK+2\pT)}{3 \chi \Omega} -  \frac{2 \kappa_1 Z^{a} }{\Omega \chi} \right] \nonumber \\ 
&&  + \gamma^{bc} \hat D_{b}\hat D_{c}\beta^{a} - \gamma^{bc} R[\hat \gamma]^{a}{}_{bcd} \beta^{d}  - 2 A^{ab} D_{b}\alpha - \frac{2 C_{Z4c} \pT D^{a}\alpha}{\Omega}  -  \frac{4 \alpha A^{ab} D_{b}\Omega}{\Omega} \nonumber \\ 
&& +  \frac{2 C_{Z4c} \alpha \pT D^{a}\Omega}{\Omega^2}   -  \case{1}{6} D^a\partial_\perp\ln\gamma -  \case{1}{3} \Delta \Gamma^{a} \partial_\perp\ln\gamma -  \frac{2 Z^{a} \partial_\perp\ln\gamma}{3 \chi} -  \frac{16 \pi J^{a} \alpha}{\chi}
  , \label{et:pKL}\\ 
\partial_\perp \pT& = & \case{\alpha}{2}\left[ \Omega\chi (R[\gamma]+ 2 D^{a}Z_{a}) -  \Omega A_{ab} A^{ab} + \frac{2(\pK+2\pT)^2}{3\Omega}  - \frac{2 C_{Z4c} \pT (\pK+2\pT)}{\Omega}  -  \frac{2\kappa_1(2+\kappa_2) \pT}{\Omega} \right]  \nonumber \\ 
&& + \Omega\alpha\bigtriangleup\chi -  \frac{5\Omega \alpha D^{a}\chi D_{a}\chi}{4 \chi}  -  C_{Z4c}\Omega Z^{a} D_{a}\alpha -  \frac{C_{Z4c}\Omega\alpha Z^{a} D_{a}\chi}{2 \chi} + 2 \alpha \chi \bigtriangleup\Omega  -  \alpha D^{a}\chi D_{a}\Omega \nonumber \\ 
&& - \frac{3 \alpha\chi D^a\Omega D_a\Omega}{\Omega}  - 8 \pi \alpha \Omega \rho
  . \label{et:pKT} 
\end{eqnarray}
\end{subequations}
}%

The previous equations are the same as \eref{et:tensoreqs}, but using $\pK$ instead of $\cK$ as evolution variable, and  $\pT$ instead of $\cT$. %The inclusion of this set of equations is included for 
The numerical results that will be described in chapter \ref{c:exper} motivate the use of the trace of the physical extrinsic curvature $\tilde{\bar K}$ instead of the conformal one $\bar K$ as an evolution variable. The relation between both is given by \eref{e3:physconfK}. A ``physical'' Z4 quantity $\pT$ is also used; it is related to $\cT$ via \eref{e3:physconfTheta}. Mixing $\tilde{\bar K}$ and $\pT$ in the same way as was done in \eref{et:KTmix} gives rise to the actual evolution variable $\pK$:
\begin{equation}\label{et:KTmixphys}
\pK=\puK -2\pT .
\end{equation}

Note that some terms in the $\dot\pK$ and $\dot\pT$ evolution equations are degenerate at $\scri^+$ (they are multiplied by $\Omega$). Nevertheless, no problems at the numerical level have been detected in the simulations performed for this work. % actually principal part terms

The constraint equations expressed in $\pK$ and $\pT$ take the form
{\small
\begin{subequations}\label{et:pKtensorcqs}
\begin{eqnarray}
\mathcal{H}& = & \chi R[\gamma] - A_{ab} A^{ab} + \frac{2}{3} \left(\frac{\pK+2\pT}{\Omega}\right)^2 + 2 \bigtriangleup\chi -  \frac{5 D^{a}\chi D_{a}\chi}{2 \chi} - \frac{6\chi D^{a}\Omega D_{a}\Omega}{\Omega^2 } -  \frac{2  D^{a}\chi D_{a}\Omega}{\Omega} \nonumber \\ 
&&   + \frac{4 \chi \bigtriangleup\Omega}{\Omega}  - 16 \pi \rho , \label{et:pKH} \\ 
\mathcal{M}_{a}& = & D_{b}A_{a}^{b}  -  \frac{2 D_{a}(\pK+2\pT)}{3\Omega} -  \frac{3 A_{a}^{b} D_{b}\chi}{2 \chi}  -  \frac{2 A_{a}^{b} D_{b}\Omega}{\Omega}  
-8 \pi J_{a} \label{et:pKM} , \\
%
%\mathcal{C}^{a}& = &  \Lambda^{a} - \Delta \Gamma^{a} -  \frac{2 Z^{a}}{\chi}
Z_{a}& = & \frac{\gamma_{ab}}{2}\left(\Lambda^{b} - \Delta \Gamma^{b}\right) \label{et:pKZ}. 
\end{eqnarray}
\end{subequations}
}%

\subsubsection{Lagrangian and Eulerian conditions}

The evolution in time of the determinant of the spatial metric $\bar\gamma$ is determined by the equations, but the evolution of the determinant of the rescaled spatial metric $\gamma$ is not prescribed, so we can choose freely how it will behave in time. 
A natural choice would be to keep it constant in time. Nevertheless, in presence of a shift vector there are two possible choices of ``constant in time''. They are given by the Lagrangian condition $\partial_t\ln \gamma=0$ and the Eulerian condition $\partial_\perp\ln \gamma=0$ \cite{Brown:2005aq,Brown:2007nt,Brown:2009dd}.

Substituting $\partial_\perp\ln \gamma=-2vD_a\beta^a$ provides a convenient way of switching between both: $v=1$ is the Lagrangian case ($\partial_t\ln \gamma=0 \Rightarrow \partial_\perp\ln \gamma=-{\cal L}_\beta \ln \gamma=-2D_a\beta^a$) and $v=0$ the Eulerian one. For the calculation of the equations in spherical symmetry and the rest of the work presented here we will adopt the Lagrangian condition, because it keeps the appropriate initial data stationary. 

\subsection{Relation between physical and conformal quantities}\label{se:reltens}

For comparison we now define the spatial conformal metric $\tilde\gamma_{ab}$ rescaled from the physical spatial metric $\tilde{\bar\gamma}_{ab}$ by means of a conformal factor $\tilde\chi$ as $\tilde\gamma_{ab}=\tilde\chi\,\tilde{\bar\gamma}_{ab}$. In an equivalent way we also define $\tilde A_{ab}=\tilde\chi\tilde{\bar K}_{ab}-\case{1}{3}\tilde\gamma_{ab}\tilde{\bar K}$. The relations between the physical quantities and those derived from the conformal rescaling by $\Omega$ are 
\begin{subequations}
\begin{eqnarray}
\gamma_{ab} = \tilde\gamma_{ab} &\leftrightarrow& \tilde \gamma_{ab} =\gamma_{ab} , \\ \chi = \frac{\tilde\chi}{\Omega^2} &\leftrightarrow& \tilde\chi=\Omega^2\chi, \\A_{ab} = \frac{\tilde A_{ab}}{\Omega}  &\leftrightarrow& \tilde A_{ab} = \Omega\bar A_{ab}  .
%\gamma_{ab} = \tilde\gamma_{ab} \leftrightarrow \tilde \gamma_{ab} =\gamma_{ab} , \quad \chi = \frac{\tilde\chi}{\Omega^2} \leftrightarrow \tilde\chi=\Omega^2\chi, \quad A_{ab} = \frac{\tilde A_{ab}}{\Omega}  \leftrightarrow \tilde A_{ab} = \Omega\bar A_{ab}  .
\end{eqnarray}
\end{subequations}
%\upda{Add the shift, which does not change. Also include alpha and trace K again? Maybe put everything together in the end ...}

%\begin{absolutelynopagebreak}

\section{GBSSN and Z4c in spherical symmetry} \label{g:secsphersym}

Standard spherical spatial coordinates ($r$, $\theta$ and $\phi$) will be employed. They relate to the Cartesian ones ($x$, $y$ and $z$) as: $x = r\sin\theta\cos\phi$, $y = r\sin\theta\sin\phi $ and $z = r\cos\theta $ .
%\begin{subequations}\label{g:carspher}\begin{eqnarray}x &=& r\sin\theta\cos\phi  , \\y &=& r\sin\theta\sin\phi  , \\z &=& r\cos\theta  .\end{eqnarray}\end{subequations}

\subsection{Variables}\label{ss:ansaetze}

In spherical symmetry all variables will only depend on the radial coordinate $r$ and the time coordinate $t$. To simplify the notation, this dependence will not be explicitly written. 
The components chosen here for each of the quantities are the same as in \cite{Brown:2007nt}.
The following ansatz is made for the metric:
\begin{equation}\label{g:g_ss}
{ \gamma}_{ij} = \left( \begin{array}{ccc} { \gamma}_{rr} & 0 & 0 \\
                                         0 & r^2{ \gamma}_{\theta\theta} & 0 \\
                                         0 & 0 & r^2{ \gamma}_{\theta\theta}\sin^2{\theta} 
                            \end{array}  \right) .
\end{equation}
It is expressed in terms of two different components $\gamma_{rr}$ and $\gamma_{\theta\theta}$. Given that the conformal factor $\chi$ was introduced, the determinant freedom will only be fixed if one of the metric components is eliminated in terms of the other. The most natural substitution is
\begin{equation}\label{es:delgtt}
\gamma_{\theta\theta} = \gamma_{rr}^{-1/2} , 
\end{equation}
but in the equations presented in the next subsections the component $\gamma_{\theta\theta}$ will be kept for generality. 
In a similar way we use for the background metric  
\begin{equation}\label{es:gb}
  {\hat\gamma}_{ij} = \left( \begin{array}{ccc} {\hat\gamma}_{rr} & 0 & 0 \\
                                         0 & r^2\hat{ \gamma}_{\theta\theta} & 0 \\
                                         0 & 0 & r^2\hat{ \gamma}_{\theta\theta}\sin^2{\theta} 
                            \end{array}  \right)  =  \left( \begin{array}{ccc} 1 & 0 & 0 \\
                                         0 & r^2 & 0 \\
                                         0 & 0 & r^2\sin^2{\theta} 
                            \end{array}  \right) , 
\end{equation}
where the natural choice (consistent with our conformally flat initial data, see subsection \ref{sin:compact}) ${\hat\gamma}_{rr}=\hat{ \gamma}_{\theta\theta}=1$ is made. A more general form may be preferred in other cases. %\upda{set to 1 or change in the equations!! }

Imposing that the trace-free part of the extrinsic curvature expressed in components actually has vanishing trace, we obtain a spherically symmetric ansatz with a single independent component
\begin{equation}\label{g:A_ss}
{ A}_{ij} = { A}_{rr} \left( \begin{array}{ccc} 1 & 0 & 0 \\
                      0 & -\case{r^2{ \gamma}_{\theta\theta}}{{2{ \gamma}_{rr}}} & 0 \\
		      0 & 0 & -\case{r^2{ \gamma}_{\theta\theta}\sin^2{\theta}}{{2{ \gamma}_{rr}}}
		      \end{array} \right) . 
\end{equation}
The vector quantities $\beta^a$, $\Lambda^a$ and $\Delta\Gamma^a$ and the covector $Z_a$ only have a radial non-vanishing component. It will be denoted by $\beta^r$, $\Lambda^r$ and $Z_r$, whereas for $\Delta\Gamma^a$ the substitution is %in the equations will be
\begin{equation}
\Delta\Gamma^i=\left( \case{2}{\gamma_{\theta\theta} r}- \case{2}{\gamma_{rr} r} + \case{\gamma_{rr}'}{2 \gamma_{rr}^2} -  \case{\gamma_{\theta\theta}'}{\gamma_{rr} \gamma_{\theta\theta}}, 0, 0 \right)^T , 
\end{equation}
where the components of the background metric $\hat\gamma_{ij}$ have already been set to unity.  
The full 4-dimensional line element (where $d\sigma^2\equiv d\theta^2 + \sin^2\theta d\phi^2$) is given by
\begin{equation}\label{es:liel}
d\bar s^2 = - \left(\alpha^2-\chi^{-1}\gamma_{rr}{\beta^r}^2\right) dt^2 + \chi^{-1}\left(2\, \gamma_{rr}\beta^r dt\,dr +  \gamma_{rr}\, dr^2 +  \gamma_{\theta\theta}\, r^2\, d\sigma^2\right) .
\end{equation}
The substitution of these ans\"atze into the tensorial equations \eref{et:tensoreqs} and \eref{et:tensorceqs} will give us the spherically symmetric reduction of the GBSSN and \CZ{} equations.  

\subsection{Spherically symmetric equations}\label{appspher}

\subsubsection{Component equations with conformal $\cK$ and conformal $\cT$}

The component evolution equations derived from \eref{et:tensoreqs} are

%\end{absolutelynopagebreak}

{\small
\begingroup
\allowdisplaybreaks
\begin{subequations}\label{es:eeqs}
\begin{eqnarray}
 \dot{\chi } &=& {\beta^r} \chi ' + \frac{2}{3} \alpha  \chi (\cK+2\cT)-\frac{{\beta^r} \gamma_{rr}' \chi }{3 \gamma_{rr}}-\frac{2 {\beta^r} \gamma_{\theta\theta}' \chi }{3
   \gamma_{\theta\theta}}   -\frac{4 {\beta^r} \chi }{3 r}-\frac{2 {\beta^r}' \chi }{3} , \\ 
 \dot{\gamma_{rr}} &=& \frac{2 {\beta^r} \gamma_{rr}'}{3} -2 A_{rr} \alpha -\frac{2 \gamma_{rr} {\beta^r} \gamma_{\theta\theta}'}{3 \gamma_{\theta\theta}} +  \frac{4
   \gamma_{rr} {\beta^r}'}{3}-\frac{4 \gamma_{rr} {\beta^r}}{3 r}   , \\ 
 \dot{\gamma_{\theta\theta}} &= & \frac{{\beta^r} \gamma_{\theta\theta}'}{3} + \frac{A_{rr} \gamma_{\theta\theta} \alpha }{\gamma_{rr}}-\frac{\gamma_{\theta\theta} {\beta^r} \gamma_{rr}'}{3 \gamma_{rr}} +  \frac{2 \gamma_{\theta\theta} {\beta^r}}{3 r}-\frac{2 \gamma_{\theta\theta} {\beta^r}'}{3}  , \\ 
 \dot{A_{rr}} &= &  {\beta^r} A_{rr}'+\frac{2}{3} \gamma_{rr} \alpha  \chi  {\Lambda^r}'-\frac{\alpha  \chi  \gamma_{rr}''}{3 \gamma_{rr}}+\frac{\alpha  \chi  \gamma_{\theta\theta}''}{3 \gamma_{\theta\theta}}-\frac{2 \chi  \alpha ''}{3}+\frac{\alpha  \chi ''}{3} + \alpha A_{rr}\left[\cK +2(1-C_{Z4c}) \cT\right] \nonumber \\ 
  &&  -\frac{2 \alpha  A_{rr}^2}{\gamma_{rr}}+\frac{4 {\beta^r}' A_{rr}}{3}-\frac{4 {\beta^r} A_{rr}}{3 r}-\frac{{\beta^r} \gamma_{rr}' A_{rr}}{3 \gamma_{rr}}-\frac{2 {\beta^r} \gamma_{\theta\theta}' A_{rr}}{3 \gamma_{\theta\theta}}+\frac{\alpha  \chi  \left(\gamma_{rr}'\right)^2}{2 \gamma_{rr}^2}-\frac{2 \alpha  \chi  \left(\gamma_{\theta\theta}'\right)^2}{3 \gamma_{\theta\theta}^2} \nonumber \\ 
  &&-\frac{\alpha \left(\chi '\right)^2}{6 \chi }+\frac{2 \gamma_{rr} \alpha  \chi }{\gamma_{\theta\theta} r^2}-\frac{2 \alpha  \chi }{r^2}-\frac{2 \alpha  {\Lambda^r} \chi \gamma_{rr} }{3 r}+\frac{\alpha  {\Lambda^r} \chi  \gamma_{rr}'}{3}-\frac{\alpha  {\Lambda^r} \chi \gamma_{rr}  \gamma_{\theta\theta}'}{3 \gamma_{\theta\theta}}+\frac{2 \alpha  \chi \gamma_{rr}  \gamma_{\theta\theta}'}{\gamma_{\theta\theta}^2 r}\nonumber \\ 
  &&-\frac{2 \alpha  \chi  \gamma_{rr}'}{3 \gamma_{\theta\theta} r} -\frac{4 \alpha  \chi  \gamma_{\theta\theta}'}{3 \gamma_{\theta\theta} r}+\frac{2 \chi  \alpha '}{3 r}+\frac{\chi  \gamma_{rr}' \alpha '}{3 \gamma_{rr}}+\frac{\chi  \gamma_{\theta\theta}' \alpha '}{3 \gamma_{\theta\theta}}-\frac{\alpha  \chi '}{3r}-\frac{\alpha  \gamma_{rr}' \chi '}{6 \gamma_{rr}}-\frac{\alpha  \gamma_{\theta\theta}' \chi '}{6 \gamma_{\theta\theta}}-\frac{2 \alpha ' \chi '}{3} \nonumber \\ 
  &&  +\frac{4}{3} Z_r \alpha  \chi '  -\frac{2 \alpha  \chi  \gamma_{rr}' \Omega '}{3 \gamma_{rr} \Omega }-\frac{2 \alpha  \chi  \gamma_{\theta\theta}' \Omega'}{3 \gamma_{\theta\theta} \Omega }+\frac{4 \alpha  \chi ' \Omega '}{3 \Omega }-\frac{4 \alpha  \chi  \Omega '}{3 r \Omega }-\frac{2A_{rr} {\beta^r} \Omega '}{\Omega } +\frac{4 \alpha  \chi  \Omega ''}{3 \Omega }  \nonumber \\   &&
 +\frac{8 Z_r \alpha  \chi  \Omega '}{3 \Omega }   , \\ 
 \dot{\cK} &= &  {\beta^r} \cK'-\frac{\chi  \alpha ''}{\gamma_{rr}} +\frac{\alpha}{3} (\cK+2\cT)^2 +  \frac{3 \alpha  A_{rr}^2}{2 \gamma_{rr}^2}-\frac{2 \chi  \alpha '}{\gamma_{rr} r}+\frac{\chi \gamma_{rr}' \alpha '}{2 \gamma_{rr}^2}-\frac{\chi  \gamma_{\theta\theta}' \alpha '}{\gamma_{rr} \gamma_{\theta\theta}}+\frac{\alpha ' \chi '}{2 \gamma_{rr}} - \frac{3\beta^r\dot\alpha\Omega'}{\alpha^2\Omega} \nonumber \\ 
  && + \frac{3\dot\beta^r\Omega'}{\alpha\Omega}  +\frac{\kappa_1(1-\kappa_2) \alpha  \cT }{\Omega } +\frac{2 C_{Z4c} Z_r \chi  \alpha '}{\gamma_{rr}} 
  -\frac{\alpha  \chi  \gamma_{rr}' \Omega '}{2 \gamma_{rr}^2 \Omega}+\frac{\alpha  \chi  \gamma_{\theta\theta}' \Omega '}{\gamma_{rr} \gamma_{\theta\theta} \Omega}+\frac{3 \chi  \alpha ' \Omega '}{\gamma_{rr} \Omega }   -\frac{\alpha  \chi ' \Omega '}{2 \gamma_{rr} \Omega} \nonumber \\ 
  &&  +\frac{2 \alpha  \chi  \Omega '}{\gamma_{rr} r \Omega}+\frac{\alpha  \chi  \Omega ''}{\gamma_{rr} \Omega} -\frac{2C_{Z4c} Z_r \alpha  \chi  \Omega '}{\gamma_{rr} \Omega}  +\frac{3\alpha\left(\Omega '\right)^2}{ \Omega ^2}\left(\frac{{\beta^r}^2}{\alpha^2}-\frac{\chi  }{\gamma_{rr}}\right)  + \frac{3{\beta^r}^2\alpha'\Omega'}{\alpha^2\Omega} -\frac{3\beta^r{\beta^r}'\Omega'}{\alpha\Omega}
 \nonumber \\  &&
 -\frac{3{\beta^r}^2\Omega''}{\alpha\Omega} -\frac{\left(\cK+4\cT\right)\beta^r\Omega'}{\Omega} , \\ 
 \dot{{\Lambda^r}} &= & {\beta^r} {\Lambda^r}' -\frac{2 \alpha (2 \cK'+\cT')}{3 \gamma_{rr}}+\frac{{\beta^r} \gamma_{rr}''}{6 \gamma_{rr}^2}+\frac{{\beta^r} \gamma_{\theta\theta}''}{3 \gamma_{rr} \gamma_{\theta\theta}} +\frac{4 {\beta^r}''}{3 \gamma_{rr}}   +\frac{2 A_{rr} \alpha }{\gamma_{rr}^2 r}  -\frac{2 A_{rr} \alpha }{\gamma_{rr} \gamma_{\theta\theta} r}  -\frac{2 A_{rr} \alpha '}{\gamma_{rr}^2}  \nonumber \\ 
  && -\frac{3 A_{rr} \alpha  \chi '}{\gamma_{rr}^2 \chi } +\frac{A_{rr} \alpha  \gamma_{rr}'}{\gamma_{rr}^3}+\frac{A_{rr} \alpha  \gamma_{\theta\theta}'}{\gamma_{rr}^2 \gamma_{\theta\theta}}  -\frac{{\beta^r} \left(\gamma_{\theta\theta}'\right)^2}{\gamma_{rr} \gamma_{\theta\theta}^2} +\frac{2{\beta^r} \gamma_{rr}'}{3 \gamma_{rr} \gamma_{\theta\theta} r}  -\frac{8 {\beta^r} \gamma_{\theta\theta}'}{3 \gamma_{rr} \gamma_{\theta\theta} r}+\frac{4 {\beta^r} \gamma_{\theta\theta}'}{3 \gamma_{\theta\theta}^2 r} \nonumber \\ 
  && +\frac{2 \gamma_{\theta\theta}' {\beta^r}'}{3 \gamma_{rr} \gamma_{\theta\theta}}  -\frac{10 {\beta^r}}{3 \gamma_{rr} r^2} +\frac{2 {\beta^r}}{3 \gamma_{\theta\theta} r^2}+\frac{4 {\beta^r}'}{3 \gamma_{rr} r} +\frac{4 {\beta^r}'}{3 \gamma_{\theta\theta} r} -\frac{4 (\cK+2\cT) Z_r \alpha }{3 \gamma_{rr}}+\frac{2Z_r {\beta^r} \gamma_{rr}'}{3 \gamma_{rr}^2} \nonumber \\ 
  && +\frac{4 Z_r {\beta^r} \gamma_{\theta\theta}'}{3 \gamma_{rr} \gamma_{\theta\theta}} +\frac{8 Z_r {\beta^r}}{3 \gamma_{rr} r}+\frac{4 Z_r {\beta^r}'}{3 \gamma_{rr}}-\frac{2 C_{Z4c} \cT  \alpha '}{\gamma_{rr}} -\frac{2\alpha(2\cK+\cT)\Omega'}{3\gamma_{rr}\Omega} +\frac{2 C_{Z4c} \alpha \cT \Omega '}{\gamma_{rr} \Omega } -\frac{4 A_{rr} \Omega ' \alpha }{\gamma_{rr}^2 \Omega } \nonumber \\ 
  && - \frac{4\beta^r\alpha'\Omega'}{\alpha\gamma_{rr}\Omega} +\frac{4{\beta^r}'\Omega'}{\gamma_{rr}\Omega} +\frac{4\beta^r\Omega''}{\gamma_{rr}\Omega}
+\frac{4Z_r\beta^r\Omega'}{\gamma_{rr}\Omega}
-\frac{2 \kappa_1 \, \alpha Z_r }{\gamma_{rr} \Omega }  , \label{es:Lambdardot}  \\  %-\frac{2 \kappa_1 \, \alpha\left(\frac{1}{2}\gamma_{rr}\Lambda^r+\frac{1}{r}-\frac{\gamma_{rr}}{\gamma_{\theta\theta}r}-\frac{\gamma_{rr}'}{4\gamma_{rr}}+\frac{\gamma_{\theta\theta}'}{2\gamma_{\theta\theta}}\right) }{\gamma_{rr} \Omega }
 \dot{\cT } &= &  {\beta^r} \cT '+\frac{1}{2} \alpha  \chi  {\Lambda^r}'-\frac{\alpha  \chi  \gamma_{rr}''}{4 \gamma_{rr}^2}-\frac{\alpha  \chi  \gamma_{\theta\theta}''}{2\gamma_{rr} \gamma_{\theta\theta}}+\frac{\alpha  \chi ''}{\gamma_{rr}}  +\frac{\alpha}{3} (\cK+2\cT)^2  -C_{Z4c} \alpha  \cT (\cK+2\cT)  \nonumber \\* 
  && -\frac{C_{Z4c} Z_r \chi  \alpha '}{\gamma_{rr}}-\frac{C_{Z4c} Z_r \alpha  \chi '}{2 \gamma_{rr}} -\frac{\kappa_1(2+\kappa_2) \alpha  \cT }{\Omega } -\frac{3 \alpha  A_{rr}^2}{4 \gamma_{rr}^2} +\frac{\alpha  {\Lambda^r} \chi }{r}+\frac{3 \alpha  \chi  \left(\gamma_{rr}'\right)^2}{8 \gamma_{rr}^3}\nonumber \\* 
  &&+\frac{\alpha  \chi  \left(\gamma_{\theta\theta}'\right)^2}{4 \gamma_{rr} \gamma_{\theta\theta}^2}-\frac{5 \alpha  \left(\chi '\right)^2}{4 \gamma_{rr} \chi }-\frac{\alpha  \chi  \gamma_{rr}'}{2 \gamma_{rr} \gamma_{\theta\theta} r}+\frac{\alpha  {\Lambda^r} \chi  \gamma_{rr}'}{4\gamma_{rr}}-\frac{\alpha  \chi  \gamma_{\theta\theta}'}{\gamma_{rr} \gamma_{\theta\theta} r}+\frac{\alpha  {\Lambda^r} \chi  \gamma_{\theta\theta}'}{2 \gamma_{\theta\theta}}+\frac{2 \alpha  \chi '}{\gamma_{rr} r}  \nonumber \\* 
  && -\frac{\alpha  \gamma_{rr}' \chi '}{2 \gamma_{rr}^2}+\frac{\alpha  \gamma_{\theta\theta}' \chi '}{\gamma_{rr} \gamma_{\theta\theta}} -\frac{\alpha \chi  \gamma_{rr}' \Omega '}{\gamma_{rr}^2 \Omega }+\frac{2 \alpha  \chi  \gamma_{\theta\theta}' \Omega '}{\gamma_{rr} \gamma_{\theta\theta} \Omega}-\frac{\alpha  \chi ' \Omega '}{\gamma_{rr} \Omega}+\frac{4 \alpha  \chi  \Omega '}{\gamma_{rr} r \Omega} +\frac{2 \alpha  \chi  \Omega ''}{\gamma_{rr} \Omega}  \nonumber \\  &&  %+\frac{(C_{Z4c}-1) {\beta^r} \cT  \Omega '}{\Omega }
 +\frac{3\alpha\left(\Omega '\right)^2}{ \Omega ^2}\left(\frac{{\beta^r}^2}{\alpha^2}-\frac{\chi  }{\gamma_{rr}}\right) -\frac{2\cK\beta^r\Omega'}{\Omega}  . 
\end{eqnarray}
\end{subequations}
\endgroup
}%
The corresponding spherically symmetric constraint equations are the following
{\small
\begin{subequations}\label{es:ceqs}
\begin{eqnarray}
 \mathcal{H} &= &   -\frac{3 A_{rr}^2}{2 \gamma_{rr}^2}+\frac{2}{3}(\cK+2\cT)^2+\frac{\chi  \left(\gamma_{\theta\theta}'\right)^2}{2 \gamma_{rr} \gamma_{\theta\theta}^2}-\frac{5 \left(\chi '\right)^2}{2
   \gamma_{rr} \chi }-\frac{2 \chi }{\gamma_{rr} r^2}+\frac{2 \chi }{\gamma_{\theta\theta} r^2}+\frac{2 \chi  \gamma_{rr}'}{\gamma_{rr}^2 r}-\frac{6 \chi  \gamma_{\theta\theta}'}{\gamma_{rr} \gamma_{\theta\theta} r} \nonumber \\ 
  &&+\frac{\chi  \gamma_{rr}' \gamma_{\theta\theta}'}{\gamma_{rr}^2 \gamma_{\theta\theta}}-\frac{\gamma_{rr}' \chi '}{\gamma_{rr}^2}+\frac{2 \gamma_{\theta\theta}' \chi '}{\gamma_{rr} \gamma_{\theta\theta}}+\frac{4 \chi '}{\gamma_{rr} r}-\frac{2 \chi  \gamma_{\theta\theta}''}{\gamma_{rr} \gamma_{\theta\theta}}+\frac{2 \chi ''}{\gamma_{rr}} -\frac{2 \chi  \gamma_{rr}' \Omega '}{\gamma_{rr}^2 \Omega }+\frac{4 \chi 
   \gamma_{\theta\theta}' \Omega '}{\gamma_{rr} \gamma_{\theta\theta} \Omega } -\frac{2 \chi ' \Omega '}{\gamma_{rr} \Omega } \nonumber \\ 
  &&  +\frac{6\left(\Omega '\right)^2}{ \Omega ^2}\left(\frac{{\beta^r}^2}{\alpha^2}-\frac{\chi  }{\gamma_{rr}}\right)+\frac{8 \chi  \Omega '}{\gamma_{rr} r \Omega }+\frac{4 \chi 
   \Omega ''}{\gamma_{rr} \Omega }  -\frac{4\left(\cK+2\cT\right)\beta^r\Omega'}{\alpha\Omega}  , \\
 \mathcal{M}_r &= &   -\frac{\gamma_{rr}' A_{rr}}{\gamma_{rr}^2}+\frac{3 \gamma_{\theta\theta}' A_{rr}}{2 \gamma_{rr} \gamma_{\theta\theta}}-\frac{3 \chi ' A_{rr}}{2 \gamma_{rr} \chi
   }+\frac{3 A_{rr}}{\gamma_{rr} r}+\frac{A_{rr}'}{\gamma_{rr}}-\frac{2}{3} (\cK'+2\cT') -\frac{2 A_{rr} \Omega '}{\gamma_{rr} \Omega } \nonumber \\ &&
-\frac{2\left(\cK+2\cT\right)\Omega'}{3\Omega} - \frac{2\beta^r\alpha'\Omega'}{\alpha^2\Omega} +\frac{2{\beta^r}'\Omega'}{\alpha\Omega} +\frac{2\beta^r\Omega''}{\alpha\Omega}
, \\
Z_r &= & \frac{1}{2}\gamma_{rr}\Lambda^r+\frac{1}{r}-\frac{\gamma_{rr}}{\gamma_{\theta\theta}r}-\frac{\gamma_{rr}'}{4\gamma_{rr}}+\frac{\gamma_{\theta\theta}'}{2\gamma_{\theta\theta}}.
 \label{es:Zrdef}
\end{eqnarray}
\end{subequations}
}%

The GBSSN system is obtained by setting the Z4 quantities $\cT$ and $Z_r$ to zero and not evolving $\cT$. For the \CZ{} one, $\cT$ is evolved in time and $Z_r$ is substituted in the RHSs using \eref{es:Zrdef}.
The matter terms present in the tensorial equations have not been included here, because their specific form for a massless scalar field will be given in the next section. % \ref{ss:scalar}. 

The path followed here to obtain the component equations for the conformally rescaled quantities started with the 3+1 decomposition of the conformally rescaled equations and performed the tensorial calculations in the conformal picture. Nevertheless, knowing the relations between the physical and conformal quantities (listed in subsections \ref{se:rel3+1} and \ref{se:reltens}, from which the component relations can be easily obtained) it is also possible to start with the component equations of the physical quantities and transform those into the conformally rescaled quantities via the corresponding variable transformations. % involving the appropriate powers of $\Omega$. 
%\upda{comment that equations for the conformally rescaled variables can also be obtained by transforming the physical equations}

\subsubsection{Component equations with physical $\pK$ and physical $\pT$}

Here I present the component equations of \eref{et:pKtensoreqs} and \eref{et:pKtensorcqs}, where the physical quantities $\pK$ and $\pT$ as defined in \eref{et:KTmixphys} are used. 
The component $Z_r$ in the last term in \eref{es:Lambdardot} is explicitly substituted by \eref{es:pKZrdef} in \eref{es:pKLambdardot}, to avoid dropping it when the GBSSN equations are used. This term is important for stability, as will be described in subsection \ref{se:gbssnstabi}.   
The component evolution equations in terms of the physical $\pK$ and $\pT$ are 

{\small
\begin{subequations}\label{es:pKeeqs}
\begin{eqnarray}
 \dot{\chi } &=& {\beta^r} \chi ' + \frac{2 \alpha  \chi (\pK+2\Te)}{3\Omega}-\frac{{\beta^r} \gamma_{rr}' \chi }{3 \gamma_{rr}}-\frac{2 {\beta^r} \gamma_{\theta\theta}' \chi }{3
   \gamma_{\theta\theta}}   -\frac{4 {\beta^r} \chi }{3 r}-\frac{2 {\beta^r}' \chi }{3} +  \frac{2 {\beta^r} \chi  \Omega '}{\Omega } , \\ 
 \dot{\gamma_{rr}} &=& \frac{2 {\beta^r} \gamma_{rr}'}{3} -2 A_{rr} \alpha -\frac{2 \gamma_{rr} {\beta^r} \gamma_{\theta\theta}'}{3 \gamma_{\theta\theta}} +  \frac{4
   \gamma_{rr} {\beta^r}'}{3}-\frac{4 \gamma_{rr} {\beta^r}}{3 r}   , \\ 
 \dot{\gamma_{\theta\theta}} &= & \frac{{\beta^r} \gamma_{\theta\theta}'}{3} + \frac{A_{rr} \gamma_{\theta\theta} \alpha }{\gamma_{rr}}-\frac{\gamma_{\theta\theta} {\beta^r} \gamma_{rr}'}{3 \gamma_{rr}} +  \frac{2 \gamma_{\theta\theta} {\beta^r}}{3 r}-\frac{2 \gamma_{\theta\theta} {\beta^r}'}{3}  , \\ 
 \dot{A_{rr}} &= &  {\beta^r} A_{rr}'+\frac{2}{3} \gamma_{rr} \alpha  \chi  {\Lambda^r}'-\frac{\alpha  \chi  \gamma_{rr}''}{3 \gamma_{rr}}+\frac{\alpha  \chi  \gamma_{\theta\theta}''}{3 \gamma_{\theta\theta}}-\frac{2 \chi  \alpha ''}{3}+\frac{\alpha  \chi ''}{3} + \frac{\alpha A_{rr}\left[\pK +2(1-C_{Z4c}) \Te\right]}{\Omega} \nonumber \\ 
  &&  -\frac{2 \alpha  A_{rr}^2}{\gamma_{rr}}+\frac{4 {\beta^r}' A_{rr}}{3}-\frac{4 {\beta^r} A_{rr}}{3 r}-\frac{{\beta^r} \gamma_{rr}' A_{rr}}{3 \gamma_{rr}}-\frac{2 {\beta^r} \gamma_{\theta\theta}' A_{rr}}{3 \gamma_{\theta\theta}}+\frac{\alpha  \chi  \left(\gamma_{rr}'\right)^2}{2 \gamma_{rr}^2}-\frac{2 \alpha  \chi  \left(\gamma_{\theta\theta}'\right)^2}{3 \gamma_{\theta\theta}^2} \nonumber \\ 
  &&-\frac{\alpha \left(\chi '\right)^2}{6 \chi }+\frac{2 \gamma_{rr} \alpha  \chi }{\gamma_{\theta\theta} r^2}-\frac{2 \alpha  \chi }{r^2}-\frac{2 \alpha  {\Lambda^r} \chi \gamma_{rr} }{3 r}+\frac{\alpha  {\Lambda^r} \chi  \gamma_{rr}'}{3}-\frac{\alpha  {\Lambda^r} \chi \gamma_{rr}  \gamma_{\theta\theta}'}{3 \gamma_{\theta\theta}}+\frac{2 \alpha  \chi \gamma_{rr}  \gamma_{\theta\theta}'}{\gamma_{\theta\theta}^2 r}\nonumber \\ 
  &&-\frac{2 \alpha  \chi  \gamma_{rr}'}{3 \gamma_{\theta\theta} r} -\frac{4 \alpha  \chi  \gamma_{\theta\theta}'}{3 \gamma_{\theta\theta} r}+\frac{2 \chi  \alpha '}{3 r}+\frac{\chi  \gamma_{rr}' \alpha '}{3 \gamma_{rr}}+\frac{\chi  \gamma_{\theta\theta}' \alpha '}{3 \gamma_{\theta\theta}}-\frac{\alpha  \chi '}{3r}-\frac{\alpha  \gamma_{rr}' \chi '}{6 \gamma_{rr}}-\frac{\alpha  \gamma_{\theta\theta}' \chi '}{6 \gamma_{\theta\theta}}-\frac{2 \alpha ' \chi '}{3} \nonumber \\ 
  &&  +\frac{4}{3} Z_r \alpha  \chi '  -\frac{2 \alpha  \chi  \gamma_{rr}' \Omega '}{3 \gamma_{rr} \Omega }-\frac{2 \alpha  \chi  \gamma_{\theta\theta}' \Omega'}{3 \gamma_{\theta\theta} \Omega }+\frac{4 \alpha  \chi ' \Omega '}{3 \Omega }-\frac{4 \alpha  \chi  \Omega '}{3 r \Omega }+\frac{A_{rr} {\beta^r} \Omega '}{\Omega } +\frac{4 \alpha  \chi  \Omega ''}{3 \Omega }  \nonumber \\   &&
 +\frac{8 Z_r \alpha  \chi  \Omega '}{3 \Omega }   , \\ 
 \dot{\pK} &= &  {\beta^r} \pK'-\frac{\chi  \alpha ''\Omega}{\gamma_{rr}} +\frac{\alpha (\pK+2\Te)^2}{3\Omega} +  \frac{3 \alpha  A_{rr}^2\Omega}{2 \gamma_{rr}^2}-\frac{2 \chi  \alpha '\Omega}{\gamma_{rr} r}+\frac{\chi \gamma_{rr}' \alpha '\Omega}{2 \gamma_{rr}^2}-\frac{\chi  \gamma_{\theta\theta}' \alpha '\Omega}{\gamma_{rr} \gamma_{\theta\theta}}+\frac{\alpha ' \chi '\Omega}{2 \gamma_{rr}} \nonumber \\ 
  &&  +\frac{\kappa_1(1-\kappa_2) \alpha  \Te }{\Omega } +\frac{2 C_{Z4c} Z_r \chi  \alpha '\Omega}{\gamma_{rr}} 
  -\frac{\alpha  \chi  \gamma_{rr}' \Omega '}{2 \gamma_{rr}^2 }+\frac{\alpha  \chi  \gamma_{\theta\theta}' \Omega '}{\gamma_{rr} \gamma_{\theta\theta} }+\frac{3 \chi  \alpha ' \Omega '}{\gamma_{rr}  }   -\frac{\alpha  \chi ' \Omega '}{2 \gamma_{rr} } \nonumber \\ 
  &&  +\frac{2 \alpha  \chi  \Omega '}{\gamma_{rr} r }+\frac{\alpha  \chi  \Omega ''}{\gamma_{rr} } -\frac{2C_{Z4c} Z_r \alpha  \chi  \Omega '}{\gamma_{rr} }  -\frac{3 \alpha  \chi \left(\Omega '\right)^2}{\gamma_{rr} \Omega}   , \\ 
 \dot{{\Lambda^r}} &= & {\beta^r} {\Lambda^r}' -\frac{2 \alpha (2 \pK'+\Te')}{3 \gamma_{rr}\Omega}+\frac{{\beta^r} \gamma_{rr}''}{6 \gamma_{rr}^2}+\frac{{\beta^r} \gamma_{\theta\theta}''}{3 \gamma_{rr} \gamma_{\theta\theta}} +\frac{4 {\beta^r}''}{3 \gamma_{rr}}   +\frac{2 A_{rr} \alpha }{\gamma_{rr}^2 r}  -\frac{2 A_{rr} \alpha }{\gamma_{rr} \gamma_{\theta\theta} r}  -\frac{2 A_{rr} \alpha '}{\gamma_{rr}^2}  \nonumber \\ 
  && -\frac{3 A_{rr} \alpha  \chi '}{\gamma_{rr}^2 \chi } +\frac{A_{rr} \alpha  \gamma_{rr}'}{\gamma_{rr}^3}+\frac{A_{rr} \alpha  \gamma_{\theta\theta}'}{\gamma_{rr}^2 \gamma_{\theta\theta}}  -\frac{{\beta^r} \left(\gamma_{\theta\theta}'\right)^2}{\gamma_{rr} \gamma_{\theta\theta}^2} +\frac{2{\beta^r} \gamma_{rr}'}{3 \gamma_{rr} \gamma_{\theta\theta} r}  -\frac{8 {\beta^r} \gamma_{\theta\theta}'}{3 \gamma_{rr} \gamma_{\theta\theta} r}+\frac{4 {\beta^r} \gamma_{\theta\theta}'}{3 \gamma_{\theta\theta}^2 r} \nonumber \\ 
  && +\frac{2 \gamma_{\theta\theta}' {\beta^r}'}{3 \gamma_{rr} \gamma_{\theta\theta}}  -\frac{10 {\beta^r}}{3 \gamma_{rr} r^2} +\frac{2 {\beta^r}}{3 \gamma_{\theta\theta} r^2}+\frac{4 {\beta^r}'}{3 \gamma_{rr} r} +\frac{4 {\beta^r}'}{3 \gamma_{\theta\theta} r} -\frac{4 (\pK+2\Te) Z_r \alpha }{3 \gamma_{rr}\Omega}+\frac{2Z_r {\beta^r} \gamma_{rr}'}{3 \gamma_{rr}^2} \nonumber \\ 
  && +\frac{4 Z_r {\beta^r} \gamma_{\theta\theta}'}{3 \gamma_{rr} \gamma_{\theta\theta}} +\frac{8 Z_r {\beta^r}}{3 \gamma_{rr} r}+\frac{4 Z_r {\beta^r}'}{3 \gamma_{rr}}-\frac{2 C_{Z4c} \Te  \alpha '}{\gamma_{rr}\Omega}  +\frac{2 C_{Z4c} \alpha \Te \Omega '}{\gamma_{rr} \Omega^2 } -\frac{4 A_{rr} \Omega ' \alpha }{\gamma_{rr}^2 \Omega } \nonumber \\ 
  && -\frac{2 \kappa_1 \, \alpha\left(\frac{1}{2}\gamma_{rr}\Lambda^r+\frac{1}{r}-\frac{\gamma_{rr}}{\gamma_{\theta\theta}r}-\frac{\gamma_{rr}'}{4\gamma_{rr}}+\frac{\gamma_{\theta\theta}'}{2\gamma_{\theta\theta}}\right) }{\gamma_{rr} \Omega }  , \label{es:pKLambdardot} \\ 
 \dot{\Te } &= &  {\beta^r} \Te '+\frac{1}{2} \alpha  \chi  {\Lambda^r}'\Omega-\frac{\alpha  \chi  \gamma_{rr}''\Omega}{4 \gamma_{rr}^2}-\frac{\alpha  \chi  \gamma_{\theta\theta}''\Omega}{2\gamma_{rr} \gamma_{\theta\theta}}+\frac{\alpha  \chi ''\Omega}{\gamma_{rr}}  +\frac{ \alpha (\pK+2\Te)^2}{3\Omega}  -\frac{C_{Z4c} \alpha  \Te (\pK+2\Te)}{\Omega}  \nonumber \\ 
  && -\frac{C_{Z4c} Z_r \chi  \alpha '\Omega}{\gamma_{rr}}-\frac{C_{Z4c} Z_r \alpha  \chi '\Omega}{2 \gamma_{rr}} -\frac{\kappa_1(2+\kappa_2) \alpha  \Te }{\Omega } -\frac{3 \alpha  A_{rr}^2\Omega}{4 \gamma_{rr}^2} +\frac{\alpha  {\Lambda^r} \chi \Omega}{r}+\frac{3 \alpha  \chi  \left(\gamma_{rr}'\right)^2\Omega}{8 \gamma_{rr}^3}\nonumber \\ 
  &&+\frac{\alpha  \chi  \left(\gamma_{\theta\theta}'\right)^2\Omega}{4 \gamma_{rr} \gamma_{\theta\theta}^2}-\frac{5 \alpha  \left(\chi '\right)^2\Omega}{4 \gamma_{rr} \chi }-\frac{\alpha  \chi  \gamma_{rr}'\Omega}{2 \gamma_{rr} \gamma_{\theta\theta} r}+\frac{\alpha  {\Lambda^r} \chi  \gamma_{rr}'\Omega}{4\gamma_{rr}}-\frac{\alpha  \chi  \gamma_{\theta\theta}'\Omega}{\gamma_{rr} \gamma_{\theta\theta} r}+\frac{\alpha  {\Lambda^r} \chi  \gamma_{\theta\theta}'\Omega}{2 \gamma_{\theta\theta}}+\frac{2 \alpha  \chi '\Omega}{\gamma_{rr} r}  \nonumber \\ 
  && -\frac{\alpha  \gamma_{rr}' \chi '\Omega}{2 \gamma_{rr}^2}+\frac{\alpha  \gamma_{\theta\theta}' \chi '\Omega}{\gamma_{rr} \gamma_{\theta\theta}} -\frac{\alpha \chi  \gamma_{rr}' \Omega '}{\gamma_{rr}^2  }+\frac{2 \alpha  \chi  \gamma_{\theta\theta}' \Omega '}{\gamma_{rr} \gamma_{\theta\theta} }-\frac{\alpha  \chi ' \Omega '}{\gamma_{rr} }+\frac{4 \alpha  \chi  \Omega '}{\gamma_{rr} r } +\frac{2 \alpha  \chi  \Omega ''}{\gamma_{rr} }  \nonumber \\  &&  %+\frac{(C_{Z4c}-1) {\beta^r} \Te  \Omega '}{\Omega }
 -\frac{3 \alpha  \chi  \left(\Omega '\right)^2}{\gamma_{rr} \Omega } , 
\end{eqnarray}
\end{subequations}
}%
and the corresponding constraint equations are given by
{\small
\begin{subequations}\label{es:pKceqs}
\begin{eqnarray}
 \mathcal{H} &= &   -\frac{3 A_{rr}^2}{2 \gamma_{rr}^2}+\frac{2(\pK+2\Te)^2}{3\Omega^2}+\frac{\chi  \left(\gamma_{\theta\theta}'\right)^2}{2 \gamma_{rr} \gamma_{\theta\theta}^2}-\frac{5 \left(\chi '\right)^2}{2
   \gamma_{rr} \chi }-\frac{2 \chi }{\gamma_{rr} r^2}+\frac{2 \chi }{\gamma_{\theta\theta} r^2}+\frac{2 \chi  \gamma_{rr}'}{\gamma_{rr}^2 r}-\frac{6 \chi  \gamma_{\theta\theta}'}{\gamma_{rr} \gamma_{\theta\theta} r} \nonumber \\ 
  &&+\frac{\chi  \gamma_{rr}' \gamma_{\theta\theta}'}{\gamma_{rr}^2 \gamma_{\theta\theta}}-\frac{\gamma_{rr}' \chi '}{\gamma_{rr}^2}+\frac{2 \gamma_{\theta\theta}' \chi '}{\gamma_{rr} \gamma_{\theta\theta}}+\frac{4 \chi '}{\gamma_{rr} r}-\frac{2 \chi  \gamma_{\theta\theta}''}{\gamma_{rr} \gamma_{\theta\theta}}+\frac{2 \chi ''}{\gamma_{rr}} -\frac{2 \chi  \gamma_{rr}' \Omega '}{\gamma_{rr}^2 \Omega }+\frac{4 \chi 
   \gamma_{\theta\theta}' \Omega '}{\gamma_{rr} \gamma_{\theta\theta} \Omega } -\frac{2 \chi ' \Omega '}{\gamma_{rr} \Omega } \nonumber \\ 
  &&  -\frac{6 \chi  \left(\Omega '\right)^2}{\gamma_{rr} \Omega ^2}+\frac{8 \chi  \Omega '}{\gamma_{rr} r \Omega }+\frac{4 \chi 
   \Omega ''}{\gamma_{rr} \Omega }    , \\
 \mathcal{M}_r &= &   -\frac{\gamma_{rr}' A_{rr}}{\gamma_{rr}^2}+\frac{3 \gamma_{\theta\theta}' A_{rr}}{2 \gamma_{rr} \gamma_{\theta\theta}}-\frac{3 \chi ' A_{rr}}{2 \gamma_{rr} \chi
   }+\frac{3 A_{rr}}{\gamma_{rr} r}+\frac{A_{rr}'}{\gamma_{rr}}-\frac{2 (\pK'+2\Te')}{3\Omega} -\frac{2 A_{rr} \Omega '}{\gamma_{rr} \Omega } , \\
Z_r &= & \frac{1}{2}\gamma_{rr}\Lambda^r+\frac{1}{r}-\frac{\gamma_{rr}}{\gamma_{\theta\theta}r}-\frac{\gamma_{rr}'}{4\gamma_{rr}}+\frac{\gamma_{\theta\theta}'}{2\gamma_{\theta\theta}} 
. \label{es:pKZrdef}
\end{eqnarray}
\end{subequations}
}%

\subsubsection{Component equations in terms of the variation $\DPK$ of the physical $\pK$}

In the numerical tests that will be described in chapter \ref{c:exper}, the evolution variable chosen is not the trace of the physical extrinsic curvature $\pK$, but its variation $\DPK$ with respect to the stationary value $\Kc$: 
\begin{equation} \label{ee:DPKdef}
\DPK=\pK-\Kc . 
\end{equation}
Here I include the complete set of equations written in terms of $\DPK$ because they are exactly those used in the simulations, even if they only differ from \eref{es:pKeeqs} and \eref{es:pKceqs} by a few terms. 

The constraint equations expressed in terms of $\DPK$ are
{\small
\begin{subequations}\label{es:DPKceqs}
\begin{eqnarray}
 \mathcal{H} &= &   -\frac{3 A_{rr}^2}{2 \gamma_{rr}^2}+\frac{2(\DPK+2\Te)^2}{3\Omega^2}+\frac{\chi  \left(\gamma_{\theta\theta}'\right)^2}{2 \gamma_{rr} \gamma_{\theta\theta}^2}-\frac{5 \left(\chi '\right)^2}{2
   \gamma_{rr} \chi }-\frac{2 \chi }{\gamma_{rr} r^2}+\frac{2 \chi }{\gamma_{\theta\theta} r^2}+\frac{2 \chi  \gamma_{rr}'}{\gamma_{rr}^2 r}-\frac{6 \chi  \gamma_{\theta\theta}'}{\gamma_{rr} \gamma_{\theta\theta} r} \nonumber \\ 
  &&+\frac{\chi  \gamma_{rr}' \gamma_{\theta\theta}'}{\gamma_{rr}^2 \gamma_{\theta\theta}}-\frac{\gamma_{rr}' \chi '}{\gamma_{rr}^2}+\frac{2 \gamma_{\theta\theta}' \chi '}{\gamma_{rr} \gamma_{\theta\theta}}+\frac{4 \chi '}{\gamma_{rr} r}-\frac{2 \chi  \gamma_{\theta\theta}''}{\gamma_{rr} \gamma_{\theta\theta}}+\frac{2 \chi ''}{\gamma_{rr}} -\frac{2 \chi  \gamma_{rr}' \Omega '}{\gamma_{rr}^2 \Omega }+\frac{4 \chi 
   \gamma_{\theta\theta}' \Omega '}{\gamma_{rr} \gamma_{\theta\theta} \Omega } -\frac{2 \chi ' \Omega '}{\gamma_{rr} \Omega } \nonumber \\ 
  &&  -\frac{6 \chi  \left(\Omega '\right)^2}{\gamma_{rr} \Omega ^2}+\frac{8 \chi  \Omega '}{\gamma_{rr} r \Omega }+\frac{4 \chi 
   \Omega ''}{\gamma_{rr} \Omega }+\frac{4 \Kc (\DPK+2\Te)}{ 3 \Omega^2 }+\frac{2\Kc{}^2}{ 3 \Omega ^2}  , \\
 \mathcal{M}_r &= &   -\frac{\gamma_{rr}' A_{rr}}{\gamma_{rr}^2}+\frac{3 \gamma_{\theta\theta}' A_{rr}}{2 \gamma_{rr} \gamma_{\theta\theta}}-\frac{3 \chi ' A_{rr}}{2 \gamma_{rr} \chi
   }+\frac{3 A_{rr}}{\gamma_{rr} r}+\frac{A_{rr}'}{\gamma_{rr}}-\frac{2 (\DPK'+2\Te')}{3\Omega} -\frac{2 A_{rr} \Omega '}{\gamma_{rr} \Omega } 
, \\
%
% \mathcal{C}^r &= &   {\Lambda^r}-\frac{\gamma_{rr}'}{2 \gamma_{rr}^2}+\frac{\gamma_{\theta\theta}'}{\gamma_{rr} \gamma_{\theta\theta}}+\frac{2}{\gamma_{rr} r}-\frac{2}{\gamma_{\theta\theta} r}  -\frac{2 Z_r}{\gamma_{rr}}   .
Z_r &= & \frac{1}{2}\gamma_{rr}\Lambda^r+\frac{1}{r}-\frac{\gamma_{rr}}{\gamma_{\theta\theta}r}-\frac{\gamma_{rr}'}{4\gamma_{rr}}+\frac{\gamma_{\theta\theta}'}{2\gamma_{\theta\theta}}. \label{Zrdef}
\end{eqnarray}
\end{subequations}
}%

The evolution equations take the form

{\small
\begin{subequations}\label{es:DPKeeqs}
\begin{eqnarray}
 \dot{\chi } &=& {\beta^r} \chi ' + \frac{2 \alpha  \chi (\DPK +\Kc+2\Te)}{3\Omega}-\frac{{\beta^r} \gamma_{rr}' \chi }{3 \gamma_{rr}}-\frac{2 {\beta^r} \gamma_{\theta\theta}' \chi }{3
   \gamma_{\theta\theta}}   -\frac{4 {\beta^r} \chi }{3 r}-\frac{2 {\beta^r}' \chi }{3} +  \frac{2 {\beta^r} \chi  \Omega '}{\Omega } %\nonumber \\&&
%+\frac{2 \Kc \alpha  \chi }{ 3  \Omega }
, \qquad\quad \\ 
 \dot{\gamma_{rr}} &=& \frac{2 {\beta^r} \gamma_{rr}'}{3} -2 A_{rr} \alpha -\frac{2 \gamma_{rr} {\beta^r} \gamma_{\theta\theta}'}{3 \gamma_{\theta\theta}} +  \frac{4
   \gamma_{rr} {\beta^r}'}{3}-\frac{4 \gamma_{rr} {\beta^r}}{3 r}   , \\ 
 \dot{\gamma_{\theta\theta}} &= & \frac{{\beta^r} \gamma_{\theta\theta}'}{3} + \frac{A_{rr} \gamma_{\theta\theta} \alpha }{\gamma_{rr}}-\frac{\gamma_{\theta\theta} {\beta^r} \gamma_{rr}'}{3 \gamma_{rr}} +  \frac{2 \gamma_{\theta\theta} {\beta^r}}{3 r}-\frac{2 \gamma_{\theta\theta} {\beta^r}'}{3}  , \\ 
 \dot{A_{rr}} &= &  {\beta^r} A_{rr}'+\frac{2}{3} \gamma_{rr} \alpha  \chi  {\Lambda^r}'-\frac{\alpha  \chi  \gamma_{rr}''}{3 \gamma_{rr}}+\frac{\alpha  \chi  \gamma_{\theta\theta}''}{3 \gamma_{\theta\theta}}-\frac{2 \chi  \alpha ''}{3}+\frac{\alpha  \chi ''}{3} + \frac{\alpha A_{rr}\left[\DPK +2(1-C_{Z4c}) \Te\right]}{\Omega} \nonumber \\ 
  &&  -\frac{2 \alpha  A_{rr}^2}{\gamma_{rr}}+\frac{4 {\beta^r}' A_{rr}}{3}-\frac{4 {\beta^r} A_{rr}}{3 r}-\frac{{\beta^r} \gamma_{rr}' A_{rr}}{3 \gamma_{rr}}-\frac{2 {\beta^r} \gamma_{\theta\theta}' A_{rr}}{3 \gamma_{\theta\theta}}+\frac{\alpha  \chi  \left(\gamma_{rr}'\right)^2}{2 \gamma_{rr}^2}-\frac{2 \alpha  \chi  \left(\gamma_{\theta\theta}'\right)^2}{3 \gamma_{\theta\theta}^2} \nonumber \\ 
  &&-\frac{\alpha \left(\chi '\right)^2}{6 \chi }+\frac{2 \gamma_{rr} \alpha  \chi }{\gamma_{\theta\theta} r^2}-\frac{2 \alpha  \chi }{r^2}-\frac{2 \alpha  {\Lambda^r} \chi \gamma_{rr} }{3 r}+\frac{\alpha  {\Lambda^r} \chi  \gamma_{rr}'}{3}-\frac{\alpha  {\Lambda^r} \chi \gamma_{rr}  \gamma_{\theta\theta}'}{3 \gamma_{\theta\theta}}+\frac{2 \alpha  \chi \gamma_{rr}  \gamma_{\theta\theta}'}{\gamma_{\theta\theta}^2 r}\nonumber \\ 
  &&-\frac{2 \alpha  \chi  \gamma_{rr}'}{3 \gamma_{\theta\theta} r} -\frac{4 \alpha  \chi  \gamma_{\theta\theta}'}{3 \gamma_{\theta\theta} r}+\frac{2 \chi  \alpha '}{3 r}+\frac{\chi  \gamma_{rr}' \alpha '}{3 \gamma_{rr}}+\frac{\chi  \gamma_{\theta\theta}' \alpha '}{3 \gamma_{\theta\theta}}-\frac{\alpha  \chi '}{3r}-\frac{\alpha  \gamma_{rr}' \chi '}{6 \gamma_{rr}}-\frac{\alpha  \gamma_{\theta\theta}' \chi '}{6 \gamma_{\theta\theta}}-\frac{2 \alpha ' \chi '}{3} \nonumber \\ 
  &&  +\frac{4}{3} Z_r \alpha  \chi '  -\frac{2 \alpha  \chi  \gamma_{rr}' \Omega '}{3 \gamma_{rr} \Omega }-\frac{2 \alpha  \chi  \gamma_{\theta\theta}' \Omega'}{3 \gamma_{\theta\theta} \Omega }+\frac{4 \alpha  \chi ' \Omega '}{3 \Omega }-\frac{4 \alpha  \chi  \Omega '}{3 r \Omega }+\frac{A_{rr} {\beta^r} \Omega '}{\Omega } +\frac{\Kc \alpha  A_{rr}}{ \Omega } \nonumber \\ 
  &&+\frac{4 \alpha  \chi  \Omega ''}{3 \Omega } +\frac{8 Z_r \alpha  \chi  \Omega '}{3 \Omega }  , \\ 
 \dot{\DPK} &= &  {\beta^r} \DPK'-\frac{\chi  \alpha ''\Omega}{\gamma_{rr}} +\frac{\alpha (\DPK+2\Te)^2}{3\Omega} +  \frac{3 \alpha  A_{rr}^2\Omega}{2 \gamma_{rr}^2}-\frac{2 \chi  \alpha '\Omega}{\gamma_{rr} r}+\frac{\chi \gamma_{rr}' \alpha '\Omega}{2 \gamma_{rr}^2}-\frac{\chi  \gamma_{\theta\theta}' \alpha '\Omega}{\gamma_{rr} \gamma_{\theta\theta}}+\frac{\alpha ' \chi '\Omega}{2 \gamma_{rr}} \nonumber \\ 
  &&  +\frac{\kappa_1(1-\kappa_2) \alpha  \Te }{\Omega } +\frac{2 C_{Z4c} Z_r \chi  \alpha '\Omega}{\gamma_{rr}} 
  -\frac{\alpha  \chi  \gamma_{rr}' \Omega '}{2 \gamma_{rr}^2 }+\frac{\alpha  \chi  \gamma_{\theta\theta}' \Omega '}{\gamma_{rr} \gamma_{\theta\theta} }+\frac{3 \chi  \alpha ' \Omega '}{\gamma_{rr}  }   -\frac{\alpha  \chi ' \Omega '}{2 \gamma_{rr} } -\frac{3 \alpha  \chi \left(\Omega '\right)^2}{\gamma_{rr} \Omega}  \nonumber \\ 
  &&  +\frac{2 \Kc \alpha (\DPK+2\Te)}{3\Omega }+\frac{2 \alpha  \chi  \Omega '}{\gamma_{rr} r }+\frac{\alpha  \chi  \Omega ''}{\gamma_{rr} } -\frac{2C_{Z4c} Z_r \alpha  \chi  \Omega '}{\gamma_{rr} } +\frac{\Kc{}^2 \alpha }{ 3 \Omega} 
 , \\ 
 \dot{{\Lambda^r}} &= & {\beta^r} {\Lambda^r}' -\frac{2 \alpha (2 \DPK'+\Te')}{3 \gamma_{rr}\Omega}+\frac{{\beta^r} \gamma_{rr}''}{6 \gamma_{rr}^2}+\frac{{\beta^r} \gamma_{\theta\theta}''}{3 \gamma_{rr} \gamma_{\theta\theta}} +\frac{2 A_{rr} \alpha }{\gamma_{rr}^2 r}  -\frac{2 A_{rr} \alpha }{\gamma_{rr} \gamma_{\theta\theta} r}  -\frac{2 A_{rr} \alpha '}{\gamma_{rr}^2} -\frac{3 A_{rr} \alpha  \chi '}{\gamma_{rr}^2 \chi } \nonumber \\ 
  && +\frac{A_{rr} \alpha  \gamma_{rr}'}{\gamma_{rr}^3}+\frac{A_{rr} \alpha  \gamma_{\theta\theta}'}{\gamma_{rr}^2 \gamma_{\theta\theta}}  -\frac{{\beta^r} \left(\gamma_{\theta\theta}'\right)^2}{\gamma_{rr} \gamma_{\theta\theta}^2} +\frac{2{\beta^r} \gamma_{rr}'}{3 \gamma_{rr} \gamma_{\theta\theta} r}  -\frac{8 {\beta^r} \gamma_{\theta\theta}'}{3 \gamma_{rr} \gamma_{\theta\theta} r}+\frac{4 {\beta^r} \gamma_{\theta\theta}'}{3 \gamma_{\theta\theta}^2 r}+\frac{2 \gamma_{\theta\theta}' {\beta^r}'}{3 \gamma_{rr} \gamma_{\theta\theta}}  \nonumber \\ 
  &&-\frac{10 {\beta^r}}{3 \gamma_{rr} r^2} +\frac{2 {\beta^r}}{3 \gamma_{\theta\theta} r^2}+\frac{4 {\beta^r}'}{3 \gamma_{rr} r} +\frac{4 {\beta^r}'}{3 \gamma_{\theta\theta} r}+\frac{4 {\beta^r}''}{3 \gamma_{rr}}  -\frac{4 (\DPK+2\Te) Z_r \alpha }{3 \gamma_{rr}\Omega}+\frac{2Z_r {\beta^r} \gamma_{rr}'}{3 \gamma_{rr}^2} \nonumber \\ 
  && +\frac{4 Z_r {\beta^r} \gamma_{\theta\theta}'}{3 \gamma_{rr} \gamma_{\theta\theta}} +\frac{8 Z_r {\beta^r}}{3 \gamma_{rr} r}+\frac{4 Z_r {\beta^r}'}{3 \gamma_{rr}}-\frac{2 C_{Z4c} \Te  \alpha '}{\gamma_{rr}\Omega}  +\frac{2 C_{Z4c} \alpha \Te \Omega '}{\gamma_{rr} \Omega^2 } -\frac{4 A_{rr} \Omega ' \alpha }{\gamma_{rr}^2 \Omega }-\frac{4 \Kc Z_r \alpha }{ 3  \gamma_{rr} \Omega } \nonumber \\ 
  && -\frac{2 \kappa_1 \left(\frac{1}{2}\gamma_{rr}\Lambda^r+\frac{1}{r}-\frac{\gamma_{rr}}{\gamma_{\theta\theta}r}-\frac{\gamma_{rr}'}{4\gamma_{rr}}+\frac{\gamma_{\theta\theta}'}{2\gamma_{\theta\theta}}\right) \alpha }{\gamma_{rr} \Omega }   , \label{es:DPKLambdardot} \\ 
 \dot{\Te } &= &  {\beta^r} \Te '+\frac{1}{2} \alpha  \chi  {\Lambda^r}'\Omega-\frac{\alpha  \chi  \gamma_{rr}''\Omega}{4 \gamma_{rr}^2}-\frac{\alpha  \chi  \gamma_{\theta\theta}''\Omega}{2\gamma_{rr} \gamma_{\theta\theta}}+\frac{\alpha  \chi ''\Omega}{\gamma_{rr}}  +\frac{ \alpha (\DPK+2\Te)^2}{3\Omega}  -\frac{C_{Z4c} \alpha  \Te (\DPK+2\Te)}{\Omega}  \nonumber \\ 
  && -\frac{C_{Z4c} Z_r \chi  \alpha '\Omega}{\gamma_{rr}}-\frac{C_{Z4c} Z_r \alpha  \chi '\Omega}{2 \gamma_{rr}} -\frac{\kappa_1(2+\kappa_2) \alpha  \Te }{\Omega } -\frac{3 \alpha  A_{rr}^2\Omega}{4 \gamma_{rr}^2} +\frac{\alpha  {\Lambda^r} \chi \Omega}{r}+\frac{3 \alpha  \chi  \left(\gamma_{rr}'\right)^2\Omega}{8 \gamma_{rr}^3}\nonumber \\ 
  &&+\frac{\alpha  \chi  \left(\gamma_{\theta\theta}'\right)^2\Omega}{4 \gamma_{rr} \gamma_{\theta\theta}^2}-\frac{5 \alpha  \left(\chi '\right)^2\Omega}{4 \gamma_{rr} \chi }-\frac{\alpha  \chi  \gamma_{rr}'\Omega}{2 \gamma_{rr} \gamma_{\theta\theta} r}+\frac{\alpha  {\Lambda^r} \chi  \gamma_{rr}'\Omega}{4\gamma_{rr}}-\frac{\alpha  \chi  \gamma_{\theta\theta}'\Omega}{\gamma_{rr} \gamma_{\theta\theta} r}+\frac{\alpha  {\Lambda^r} \chi  \gamma_{\theta\theta}'\Omega}{2 \gamma_{\theta\theta}}+\frac{2 \alpha  \chi '\Omega}{\gamma_{rr} r}  \nonumber \\ 
  && -\frac{\alpha  \gamma_{rr}' \chi '\Omega}{2 \gamma_{rr}^2}+\frac{\alpha  \gamma_{\theta\theta}' \chi '\Omega}{\gamma_{rr} \gamma_{\theta\theta}} -\frac{\alpha \chi  \gamma_{rr}' \Omega '}{\gamma_{rr}^2  }+\frac{2 \alpha  \chi  \gamma_{\theta\theta}' \Omega '}{\gamma_{rr} \gamma_{\theta\theta} }-\frac{\alpha  \chi ' \Omega '}{\gamma_{rr} }+\frac{4 \alpha  \chi  \Omega '}{\gamma_{rr} r } +\frac{2 \alpha  \chi  \Omega ''}{\gamma_{rr} }\nonumber \\ %+\frac{(C_{Z4c}-1) {\beta^r} \Te  \Omega '}{\Omega }
  &&+\frac{2 \Kc \alpha (\DPK+2\Te)}{ 3 \Omega }-\frac{\Kc C_{Z4c} \alpha  \Te }{\Omega }+\frac{\Kc{}^2 \alpha }{ 3 \Omega }-\frac{3 \alpha  \chi  \left(\Omega '\right)^2}{\gamma_{rr} \Omega }  . 
\end{eqnarray}
\end{subequations}
}%

%\subsubsection{Relation between physical and conformal quantities}\upda{ ??? They are the same as the tensor quantities in the previous section ...}

\section{Scalar field}\label{ss:scalar}

The simplest non-vacuum content that can be added to the Einstein equations is a massless scalar field $\iPhi$. Its corresponding energy-momentum tensor $T[\tilde g]_{ab}$ is
\begin{equation}
T[\tilde g]_{ab} = \tilde \nabla_a\iPhi\tilde\nabla_b\iPhi-\case{1}{2}\tilde g_{ab}\,(\tilde\nabla_c\iPhi)(\tilde\nabla^c\iPhi) ,   
\end{equation}
which expressed in terms of the conformal metric $\bar g_{ab}$ and its corresponding covariant derivatives takes the form
\begin{equation}
T[\case{\bar g}{\Omega^2}]_{ab}  = \bar \nabla_a\iPhi\bar\nabla_b\iPhi-\case{1}{2}\bar g_{ab}\,(\bar\nabla_c\iPhi)(\bar\nabla^c\iPhi)  .  
\end{equation}

In spherical symmetry, the matter terms $\rho$, $J^a$, $S_{ab}$ and $S$ that appear in the 3+1 decomposed equations \eref{e3:adm} and \eref{e3:admc} have the following expressions in terms of the spherically symmetric ansatz components, where $\iPi\equiv\dot\iPhi$ has been substituted: 
{\small
\begin{subequations}\label{mmtermspher}
\begin{eqnarray}
\rho &=& \frac{S^+}{2}  ,\quad \label{esca:rho} %\\  \frac{1}{2}\left( \frac{(\iPi-\beta^r\iPhi')^2}{\alpha^2}+\frac{\chi(\iPhi')^2}{\gamma_{rr}} \right)
J_i = (-\frac{\iPhi'(\iPi-\beta^r\iPhi')}{\alpha},0,0) \quad \textrm{and}\quad 
J^i = (-\frac{\chi\iPhi'(\iPi-\beta^r\iPhi')}{\gamma_{rr}\alpha},0,0)^T  , \\
S_{ij} &=& \frac{1}{2\chi}\left(\begin{array}{ccc}\gamma_{rr}S^+&0&0\\0&\gamma_{\theta\theta}r^2S^-&0\\0&0&\gamma_{\theta\theta}r^2\sin^2\theta S^-\end{array}\right)  , \ S = \frac{3}{2}\left( \frac{(\iPi-\beta^r\iPhi')^2}{\alpha^2}-\frac{\chi(\iPhi')^2}{3\gamma_{rr}} \right) , \qquad \quad \\
\textrm{with}&& \quad S^+ = \frac{(\iPi-\beta^r\iPhi')^2}{\alpha^2}+\frac{\chi(\iPhi')^2}{\gamma_{rr}} \quad \textrm{and} \quad S^- = \frac{(\iPi-\beta^r\iPhi')^2}{\alpha^2}-\frac{\chi(\iPhi')^2}{\gamma_{rr}}  . \nonumber %\\
\end{eqnarray}
\end{subequations}
}%

The terms to add to the RHSs of the GBSSN and \CZ{} equations \eref{es:eeqs} and \eref{es:ceqs}, \eref{es:pKeeqs} and \eref{es:pKceqs}, \eref{es:DPKeeqs} and \eref{es:DPKceqs} are the following: 
{\small
\begin{subequations}\label{esca:terms}
\begin{eqnarray}
\dot \chi &\to& 0  , \qquad
\dot \gamma_{rr} \to 0  , \qquad
\dot \gamma_{\theta\theta} \to 0  , \\
\dot A_{rr} &\to& -\frac{16\pi\chi\alpha(\iPhi')^2}{3}  , \\
\dot \cK &\to& 4\pi\alpha\left(\frac{2(\iPi-\beta^r\iPhi')^2}{\alpha^2}\right)  , \quad 
\dot \pK \textrm{ or } \dot \DPK \to 4\pi\alpha\Omega\left(\frac{2(\iPi-\beta^r\iPhi')^2}{\alpha^2}\right) , \\
\dot \Lambda^r &\to& \frac{16\pi(\iPi-\beta^r\iPhi')\iPhi'}{g_{rr}}  , \\
\dot \cT &\to& -4\pi\alpha\left(\frac{(\iPi-\beta^r\iPhi')^2}{\alpha^2}+\frac{\chi(\iPhi')^2}{\gamma_{rr}}\right) , \ 
\dot \pT \to -4\pi\alpha\Omega\left(\frac{(\iPi-\beta^r\iPhi')^2}{\alpha^2}+\frac{\chi(\iPhi')^2}{\gamma_{rr}}\right) , \quad \\
{\cal H} &\to& -8\pi\left(\frac{(\iPi-\beta^r\iPhi')^2}{\alpha^2}+\frac{\chi(\iPhi')^2}{\gamma_{rr}}\right)  , \quad
{\cal M}_r \to \frac{8\pi(\iPi-\beta^r\iPhi')\iPhi'}{\alpha}  , \quad
Z_r \to 0  .
\end{eqnarray}
\end{subequations}
}%

The equation of motion for the scalar field is given by
\begin{equation}\label{esca:tildephi}
\tilde\Box\iPhi=0  ,
\end{equation} 
where $\tilde \Box = \tilde g_{ab}\tilde\nabla_a\tilde\nabla_b$. 
Expressed in terms of the conformally rescaled quantities it transforms to the following equation, where equivalently  $\bar \Box = \bar g_{ab}\bar\nabla_a\bar\nabla_b$, 
\begin{equation}\label{esca:barphi}
\bar\Box\iPhi-\frac{2}{\Omega}(\bar\nabla_a\iPhi)(\bar\nabla^a\Omega)=0 . 
\end{equation}
An alternative choice for the conformally invariant scalar field equation would have been $\left(\tilde\Box-\case{R[\tilde g]}{6}\right)\iPhi=0$. It is such that under the transformations \eref{ei:rescmetric} and  
\begin{equation}\label{esca:tildebarphi}
\bPhi= \frac{\iPhi}{\Omega} , 
\end{equation} 
it becomes $\Omega^3\left(\bar\Box-\case{R[\bar g]}{6}\right)\bPhi=0$ and the equation remains formally the same. Nevertheless, \eref{esca:tildephi} was finally chosen here, with the idea that the divergent terms at $\scri^+$ pose extra difficulties on the stability of the simulations to test the robustness of our implementation. 

Using \eref{e3:fourmetric} and the spherically symmetric ans\"atze in subsection \ref{ss:ansaetze}, as well as introducing $\iPi=\dot\iPhi$ to write the system as first order in time, the spherically symmetric reduction of \eref{esca:barphi} is 
{\small
\begin{subequations}\label{esca:eqs}
\begin{eqnarray}
\dot{\iPhi } &= &  \iPi    , \\ 
\dot{\iPi } &= & -\frac{{\beta^r} \dot{\alpha } \iPhi '}{\alpha }+\frac{\iPi  \dot{\alpha }}{\alpha }+\frac{{\beta^r} \iPhi ' \dot{\gamma _{\theta \theta }}}{\gamma _{\theta \theta }}-\frac{3 {\beta^r} \dot{\chi } \iPhi '}{2 \chi }+\dot{{\beta^r}} \iPhi '-\frac{\iPi  \dot{\gamma _{\theta \theta }}}{\gamma _{\theta \theta }}+\frac{3 \iPi  \dot{\chi }}{2 \chi }+\frac{{\beta^r} \iPhi ' \dot{\gamma _{rr}}}{2 \gamma _{rr}}-\frac{\iPi  \dot{\gamma _{rr}}}{2 \gamma _{rr}}
\nonumber \\ && 
+\frac{{\beta^r}^2 \alpha ' \iPhi '}{\alpha }-\frac{{\beta^r} \iPi  \alpha '}{\alpha }-\frac{{\beta^r}^2 \iPhi ' \gamma _{\theta \theta }'}{\gamma _{\theta \theta }}-{\beta^r}^2 \iPhi ''+\frac{3 {\beta^r}^2 \iPhi ' \chi '}{2 \chi }+\frac{2 {\beta^r}^2 \iPhi ' \Omega '}{\Omega }+\iPi  {\beta^r}'-2 {\beta^r} {\beta^r}' \iPhi '
\nonumber \\ && 
+\frac{{\beta^r} \iPi  \gamma _{\theta \theta }'}{\gamma _{\theta \theta }}+2 {\beta^r} \iPi '-\frac{3 {\beta^r} \iPi  \chi '}{2 \chi }-\frac{2 {\beta^r} \iPi  \Omega '}{\Omega }-\frac{2 {\beta^r}^2 \iPhi '}{r}+\frac{2 {\beta^r} \iPi }{r}+\frac{2 \alpha ^2 \chi  \iPhi '}{r \gamma _{rr}}+\frac{\alpha ^2 \chi  \iPhi ' \gamma _{\theta \theta }'}{\gamma _{\theta \theta } \gamma _{rr}}
\nonumber \\ && 
+\frac{\alpha ^2 \chi  \iPhi ''}{\gamma _{rr}}-\frac{\alpha ^2 \iPhi ' \chi '}{2 \gamma _{rr}}-\frac{2 \alpha ^2 \chi  \iPhi ' \Omega '}{\Omega  \gamma _{rr}}-\frac{\alpha ^2 \chi  \iPhi ' \gamma _{rr}'}{2 \gamma _{rr}^2}+\frac{\alpha  \chi  \alpha ' \iPhi '}{\gamma _{rr}}-\frac{{\beta^r}^2 \iPhi ' \gamma _{rr}'}{2 \gamma _{rr}}+\frac{{\beta^r} \iPi  \gamma _{rr}'}{2 \gamma _{rr}}
 . \qquad \quad 
\end{eqnarray}
\end{subequations}
}%
The quantities $\dot\gamma_{rr}$, $\dot\gamma_{\theta\theta}$, $\dot\chi$, $\dot\alpha$ and $\dot\beta^r$ have to be substituted with the corresponding evolution equations. They have been left unsubstituted to avoid restricting the scalar field equations to a specific choice of variables or gauge conditions. 

Instead of the physical quantities, whose amplitude vanishes at $\scri^+$, it is more convenient to evolve $\bPhi$, as defined in \eref{esca:tildebarphi}, and equivalently $\bPi={\iPi}/{\Omega}$, because non-zero fields at $\scri^+$ are clearer to visualize. The corresponding equations of motion are
{\small
\begin{subequations}\label{esca:beqs}
\begin{eqnarray}
\dot{\bPhi } &= &  \bPi  
  , \\ 
\dot{\bPi } &= &  -\frac{{\beta^r} \dot{\alpha } \bPhi '}{\alpha }-\frac{{\beta^r} \bPhi  \dot{\alpha } \Omega '}{\alpha  \Omega }+\frac{\bPi  \dot{\alpha }}{\alpha }+\frac{{\beta^r} \bPhi ' \dot{\gamma _{\theta \theta }}}{\gamma _{\theta \theta }}+\frac{{\beta^r} \bPhi  \Omega ' \dot{\gamma _{\theta \theta }}}{\Omega  \gamma _{\theta \theta }}-\frac{3 {\beta^r} \dot{\chi } \bPhi '}{2 \chi }+\dot{{\beta^r}} \bPhi '-\frac{3 {\beta^r} \bPhi  \dot{\chi } \Omega '}{2 \chi  \Omega }
\nonumber \\ && 
+\frac{\bPhi  \dot{{\beta^r}} \Omega '}{\Omega }-\frac{\bPi  \dot{\gamma _{\theta \theta }}}{\gamma _{\theta \theta }}+\frac{3 \bPi  \dot{\chi }}{2 \chi }+\frac{{\beta^r} \bPhi ' \dot{\gamma _{rr}}}{2 \gamma _{rr}}+\frac{{\beta^r} \bPhi  \Omega ' \dot{\gamma _{rr}}}{2 \Omega  \gamma _{rr}}-\frac{\bPi  \dot{\gamma _{rr}}}{2 \gamma _{rr}}+\frac{{\beta^r}^2 \alpha ' \bPhi '}{\alpha }+\frac{{\beta^r}^2 \bPhi  \alpha ' \Omega '}{\alpha  \Omega }
\nonumber \\ && 
-\frac{{\beta^r} \bPi  \alpha '}{\alpha }-\frac{{\beta^r}^2 \bPhi ' \gamma _{\theta \theta }'}{\gamma _{\theta \theta }}-\frac{{\beta^r}^2 \bPhi  \Omega ' \gamma _{\theta \theta }'}{\Omega  \gamma _{\theta \theta }}-{\beta^r}^2 \bPhi ''+\frac{3 {\beta^r}^2 \bPhi ' \chi '}{2 \chi }+\frac{3 {\beta^r}^2 \bPhi  \chi ' \Omega '}{2 \chi  \Omega }-\frac{{\beta^r}^2 \bPhi  \Omega ''}{\Omega }
\nonumber \\ && 
+\frac{2 {\beta^r}^2 \bPhi  \left(\Omega '\right)^2}{\Omega ^2}+\bPi  {\beta^r}'-2 {\beta^r} {\beta^r}' \bPhi '-\frac{2 {\beta^r} \bPhi  {\beta^r}' \Omega '}{\Omega }+\frac{{\beta^r} \bPi  \gamma _{\theta \theta }'}{\gamma _{\theta \theta }}+2 {\beta^r} \bPi '-\frac{3 {\beta^r} \bPi  \chi '}{2 \chi }
\nonumber \\ && 
-\frac{2 {\beta^r}^2 \bPhi '}{r}-\frac{2 {\beta^r}^2 \bPhi  \Omega '}{r \Omega }+\frac{2 {\beta^r} \bPi }{r}+\frac{2 \alpha ^2 \chi  \bPhi '}{r \gamma _{rr}}+\frac{2 \alpha ^2 \bPhi  \chi  \Omega '}{r \Omega  \gamma _{rr}}+\frac{\alpha ^2 \chi  \bPhi ' \gamma _{\theta \theta }'}{\gamma _{\theta \theta } \gamma _{rr}}+\frac{\alpha ^2 \bPhi  \chi  \Omega ' \gamma _{\theta \theta }'}{\Omega  \gamma _{\theta \theta } \gamma _{rr}}
\nonumber \\ && 
+\frac{\alpha ^2 \chi  \bPhi ''}{\gamma _{rr}}-\frac{\alpha ^2 \bPhi ' \chi '}{2 \gamma _{rr}}-\frac{\alpha ^2 \chi  \bPhi ' \gamma _{rr}'}{2 \gamma _{rr}^2}-\frac{\alpha ^2 \bPhi  \chi ' \Omega '}{2 \Omega  \gamma _{rr}}+\frac{\alpha ^2 \bPhi  \chi  \Omega ''}{\Omega  \gamma _{rr}}-\frac{\alpha ^2 \bPhi  \chi  \Omega ' \gamma _{rr}'}{2 \Omega  \gamma _{rr}^2}-\frac{2 \alpha ^2 \bPhi  \chi  \left(\Omega '\right)^2}{\Omega ^2 \gamma _{rr}}
\nonumber \\ && 
+\frac{\alpha  \chi  \alpha ' \bPhi '}{\gamma _{rr}}+\frac{\alpha  \bPhi  \chi  \alpha ' \Omega '}{\Omega  \gamma _{rr}}-\frac{{\beta^r}^2 \bPhi ' \gamma _{rr}'}{2 \gamma _{rr}}-\frac{{\beta^r}^2 \bPhi  \Omega ' \gamma _{rr}'}{2 \Omega  \gamma _{rr}}+\frac{{\beta^r} \bPi  \gamma _{rr}'}{2 \gamma _{rr}}
. 
\end{eqnarray}
\end{subequations}
}%
The physical scalar field quantities $\iPhi$ and $\iPi$ in the matter terms \eref{esca:terms} to be added to the component Einstein equations also have to be expressed in terms of $\bPhi$ and $\bPi$. 

Another possible definition for the auxiliary variable of the rescaled scalar field, $\bPi=\dot{\bPhi}$, is $\bPi_{adv}=\case{1}{\alpha}\left(\dot{\bPhi}-\beta^r\bPhi'\right)$. The evolution equations for the scalar field in terms of this auxiliary variable are given by 
{\small
\begin{subequations}\label{esca:badveqs}
\begin{eqnarray}
\dot{\bPhi } &= & \alpha \bPi_{adv} + \beta^r\bPhi'  
  , \\ 
\dot{\bPi }_{adv} &= &  -\frac{{\beta^r} \bPhi  \dot{\alpha } \Omega '}{\alpha ^2 \Omega }+\frac{{\beta^r} \bPhi  \Omega ' \dot{\gamma _{\theta \theta }}}{\alpha  \Omega  \gamma _{\theta \theta }}-\frac{3 {\beta^r} \bPhi  \dot{\chi } \Omega '}{2 \alpha  \chi  \Omega }+\frac{\bPhi  \dot{{\beta^r}} \Omega '}{\alpha  \Omega }-\frac{\bPi_{adv}  \dot{\gamma _{\theta \theta }}}{\gamma _{\theta \theta }}+\frac{3 \bPi_{adv}  \dot{\chi }}{2 \chi }+\frac{{\beta^r} \bPhi  \Omega ' \dot{\gamma _{rr}}}{2 \alpha  \Omega  \gamma _{rr}}
\nonumber \\ && 
-\frac{\bPi_{adv}  \dot{\gamma _{rr}}}{2 \gamma _{rr}}+\frac{{\beta^r}^2 \bPhi  \alpha ' \Omega '}{\alpha ^2 \Omega }-\frac{{\beta^r}^2 \bPhi  \Omega ' \gamma _{\theta \theta }'}{\alpha  \Omega  \gamma _{\theta \theta }}+\frac{3 {\beta^r}^2 \bPhi  \chi ' \Omega '}{2 \alpha  \chi  \Omega }-\frac{{\beta^r}^2 \bPhi  \Omega ''}{\alpha  \Omega }+\frac{2 {\beta^r}^2 \bPhi  \left(\Omega '\right)^2}{\alpha  \Omega ^2}
\nonumber \\ && 
-\frac{2 {\beta^r} \bPhi  {\beta^r}' \Omega '}{\alpha  \Omega }+\bPi_{adv}  {\beta^r}'+\frac{{\beta^r} \bPi_{adv}  \gamma _{\theta \theta }'}{\gamma _{\theta \theta }}+{\beta^r} \bPi_{adv} '-\frac{3 {\beta^r} \bPi_{adv}  \chi '}{2 \chi }-\frac{2 {\beta^r}^2 \bPhi  \Omega '}{\alpha  r \Omega }+\frac{2 {\beta^r} \bPi_{adv} }{r}
\nonumber \\ && 
+\frac{2 \alpha  \chi  \bPhi '}{r \gamma _{rr}}+\frac{2 \alpha  \bPhi  \chi  \Omega '}{r \Omega  \gamma _{rr}}+\frac{\chi  \alpha ' \bPhi '}{\gamma _{rr}}+\frac{\bPhi  \chi  \alpha ' \Omega '}{\Omega  \gamma _{rr}}-\frac{{\beta^r}^2 \bPhi  \Omega ' \gamma _{rr}'}{2 \alpha  \Omega  \gamma _{rr}}+\frac{\alpha  \chi  \bPhi ' \gamma _{\theta \theta }'}{\gamma _{\theta \theta } \gamma _{rr}}+\frac{\alpha  \bPhi  \chi  \Omega ' \gamma _{\theta \theta }'}{\Omega  \gamma _{\theta \theta } \gamma _{rr}}+\frac{\alpha  \chi  \bPhi ''}{\gamma _{rr}}
\nonumber \\ && 
-\frac{\alpha  \bPhi ' \chi '}{2 \gamma _{rr}}-\frac{\alpha  \chi  \bPhi ' \gamma _{rr}'}{2 \gamma _{rr}^2}-\frac{\alpha  \bPhi  \chi ' \Omega '}{2 \Omega  \gamma _{rr}}+\frac{\alpha  \bPhi  \chi  \Omega ''}{\Omega  \gamma _{rr}}-\frac{\alpha  \bPhi  \chi  \Omega ' \gamma _{rr}'}{2 \Omega  \gamma _{rr}^2}-\frac{2 \alpha  \bPhi  \chi  \left(\Omega '\right)^2}{\Omega ^2 \gamma _{rr}}+\frac{{\beta^r} \bPi_{adv}  \gamma _{rr}'}{2 \gamma _{rr}}
,  \qquad \quad 
\end{eqnarray}
\end{subequations}
}%
and in \eref{esca:terms} the substitution $\bPi=\alpha \bPi_{adv} + \beta^r\bPhi' $ has to be performed. 

%\begin{comment}

\section{Properties of the spacetime} %Energy, mass of the system, apparent horizon finder, ...

%\upda{These quantities diverge at $\scri^+$ at some point during the numerical evolutions, eventhough for the initial data they look perfectly ok. Check again in code!} Ok, bad behaviour at scri converges away. 

\begin{comment}
\subsubsection{Energy of the scalar field}

The energy density associated to the massless scalar field is given by \eref{esca:rho}, that is 
\begin{equation}\label{een:endens}
\rho = \frac{1}{2}\left( \frac{(\iPi-\beta^r\iPhi')^2}{\alpha^2}+\frac{\chi(\iPhi')^2}{\gamma_{rr}} \right) . 
\end{equation}
The integrated energy contained inside of a sphere of radius $r$ is given by
\begin{equation}
\int\sqrt{-\bar g}\,\rho\, dx^i = 4\pi\int_0^r\frac{\alpha\gamma_{\theta\theta}\sqrt{\gamma_{rr}}}{\chi^{3/2}}\rho \,{r'}^2 dr'. 
\end{equation}
In case a BH is present, its mass has to be added in order to obtain the total integrated energy of the spacetime. Note that both the energy and its integral have been given in terms of the rescaled quantities, as the physical ones diverge at $\scri^+$. 
\end{comment}

\subsubsection{Schwarzschild or areal radial coordinate}

The physical Schwarzschild radial coordinate expressed in our rescaled quantities is given by 
\begin{equation}
R_{\rm{Schw}}= \frac{r}{\Omega}\sqrt{\frac{\gamma_{\theta\theta}}{\chi}} . 
\end{equation}
The previous expression is obtained by comparing the angular part of the physical Schwarz-schild line element,
\begin{equation}
d\tilde s^2 = R_{\rm{Schw}}^2 d\sigma^2 , 
\end{equation}
with the same quantity expressed in the spherically symmetric metric components of our ansatz \eref{es:liel}
\begin{equation}
d\tilde s^2 = \frac{d\bar s^2}{\Omega^2} = \frac{\gamma_{\theta\theta}}{\chi} \frac{r^2}{\Omega^2} d\sigma^2. 
\end{equation}

\subsubsection{Misner-Sharp mass}

In general, the concept of mass cannot be defined quasi-locally due to the equivalence principle. 
However, in cases with high symmetry, where all degrees of freedom are fixed - like in spherical symmetry, where no gravitational radiation exists -, it is possible to define a quasi-local concept of mass. The Misner-Sharp mass function \cite{Misner:1964je}, a special case of the Hawking quasi-local mass \cite{Hawking:1968qt}, is constant on each round sphere and represents the gravitational mass contained by the sphere of areal radius $R_{\rm{Schw}}$. Its definition is 
\begin{equation}
\tilde g^{ab}(\tilde\nabla_aR_{\rm{Schw}})(\tilde\nabla_bR_{\rm{Schw}})= 1-\frac{2 M_{MS}}{R_{\rm{Schw}}} , 
\end{equation}
so that its actual expression is given by
\begin{equation}\label{em:msmass}
M_{MS}=\frac{R_{\rm{Schw}}}{2}\left(1-\tilde g^{ab}(\tilde\nabla_aR_{\rm{Schw}})(\tilde\nabla_bR_{\rm{Schw}})\right) = \frac{R_{\rm{Schw}}}{2}\left(1-\Omega^2\bar g^{ab}(\bar\nabla_aR_{\rm{Schw}})(\bar\nabla_bR_{\rm{Schw}})\right) .
\end{equation}
The Misner-Sharp mass coincides with the ADM mass at spacelike infinity and with the Bondi mass at null infinity. % Have to fix relation!: Its spatial derivative compares to the energy density \eref{een:endens}, $\case{M_{MS}'}{4\pi}\equiv\rho$. %\upda{- extra factors?}. 

The explicit expression of \eref{em:msmass} in our evolution quantities is
{\small
\begin{eqnarray}\label{em:msmassexpl}
M_{MS}&=& \frac{r}{2\Omega}\sqrt{\frac{\gamma_{\theta\theta}}{\chi}}\left(1 + \frac{{\beta^r}^2 \gamma _{\theta \theta }}{\alpha ^2 \chi }+\frac{{\beta^r}^2 r^2 \gamma _ {\theta \theta } \left(\chi '\right)^2}{4 \alpha ^2 \chi ^3}+\frac{{\beta^r}^2 r^2 \gamma _ {\theta \theta } \chi ' \Omega '}{\alpha ^2 \chi ^2 \Omega }-\frac{{\beta^r}^2 r^2 \chi ' \gamma _ {\theta \theta }'}{2 \alpha ^2 \chi ^2}-\frac{{\beta^r}^2 r^2 \Omega ' \gamma _ {\theta \theta }'}{\alpha ^2 \chi  \Omega }
\right. \nonumber \\ && 
+\frac{{\beta^r}^2 r^2 \gamma _ {\theta \theta } \left(\Omega '\right)^2}{\alpha ^2 \chi  \Omega ^2}+\frac{{\beta^r}^2 r^2 \left(\gamma _ {\theta \theta }'\right){}^2}{4 \alpha ^2 \chi  \gamma _{\theta \theta }}-\frac{{\beta^r} r^2 \dot{\chi } \gamma _ {\theta \theta } \Omega '}{\alpha ^2 \chi ^2 \Omega }+\frac{{\beta^r} r^2 \dot{\chi } \gamma _ {\theta \theta }'}{2 \alpha ^2 \chi ^2}-\frac{{\beta^r} r^2 \dot{\chi } \gamma _ {\theta \theta } \chi '}{2 \alpha ^2 \chi ^3}+\frac{{\beta^r} r^2 \dot{\gamma _{\theta \theta }} \chi '}{2 \alpha ^2 \chi ^2}
\nonumber \\ && 
+\frac{{\beta^r} r^2 \dot{\gamma _{\theta \theta }} \Omega '}{\alpha ^2 \chi  \Omega }-\frac{{\beta^r} r^2 \dot{\gamma _{\theta \theta }} \gamma _ {\theta \theta }'}{2 \alpha ^2 \chi  \gamma _{\theta \theta }}+\frac{r^2 \dot{\chi }^2 \gamma _{\theta \theta }}{4 \alpha ^2 \chi ^3}-\frac{r^2 \dot{\chi } \dot{\gamma _{\theta \theta }}}{2 \alpha ^2 \chi ^2}+\frac{r^2 \dot{\gamma _{\theta \theta }}{}^2}{4 \alpha ^2 \chi  \gamma _{\theta \theta }}-\frac{r^2 \gamma _ {\theta \theta } \chi ' \Omega '}{\chi  \Omega  \gamma _{rr}}+\frac{r^2 \chi ' \gamma _ {\theta \theta }'}{2 \chi  \gamma _{rr}}
\nonumber \\ && 
-\frac{r^2 \gamma _ {\theta \theta } \left(\chi '\right)^2}{4 \chi ^2 \gamma _{rr}}+\frac{r^2 \Omega ' \gamma _ {\theta \theta }'}{\Omega  \gamma _{rr}}-\frac{r^2 \gamma _ {\theta \theta } \left(\Omega '\right)^2}{\Omega ^2 \gamma _{rr}}-\frac{r^2 \left(\gamma _ {\theta \theta }'\right){}^2}{4 \gamma _ {\theta \theta } \gamma _{rr}}-\frac{{\beta^r}^2 r \gamma _ {\theta \theta } \chi '}{\alpha ^2 \chi ^2}-\frac{2 {\beta^r}^2 r \gamma _ {\theta \theta } \Omega '}{\alpha ^2 \chi  \Omega }
\nonumber \\ && \left.
+\frac{{\beta^r}^2 r \gamma _ {\theta \theta }'}{\alpha ^2 \chi }+\frac{{\beta^r} r \dot{\chi } \gamma _{\theta \theta }}{\alpha ^2 \chi ^2}-\frac{{\beta^r} r \dot{\gamma _{\theta \theta }}}{\alpha ^2 \chi }+\frac{r \gamma _ {\theta \theta } \chi '}{\chi  \gamma _{rr}}+\frac{2 r \gamma _ {\theta \theta } \Omega '}{\Omega  \gamma _{rr}}-\frac{r \gamma _ {\theta \theta }'}{\gamma _{rr}}-\frac{\gamma _{\theta \theta }}{\gamma _{rr}}  \right).
\end{eqnarray}
}%

\subsubsection{Apparent horizon finder}

In simulations involving a BH or a scalar field strong enough to collapse into a BH it is very useful to have a quantity that detects the creation of a horizon and locates it in the domain. The actual event horizon can only be determined a posteriori, because it is a global quantity of the spacetime, but the apparent horizon, located inside of the event horizon if weak cosmic censorship holds, can be calculated along with the simulation. 

The apparent horizon is observer dependent and is defined as the outermost marginally trapped surface, a smooth closed 2-surface whose outgoing null geodesics have zero expansion. 
%\upda{add derivation of expression.}
In spherical symmetry it can be calculated explicitly in terms of the Misner-Sharp mass and the areal radius.
The horizon will be located at the outermost point along the radial coordinate where the following expression equals zero: 
\begin{equation}
1-\frac{M_{MS}}{\frac{R_{\rm{Schw}}}{2}}=\Omega^2\bar g^{ab}(\bar\nabla_aR_{\rm{Schw}})(\bar\nabla_bR_{\rm{Schw}}) .
\end{equation}
The final explicit expression of this quantity is straightforwardly calculated from \eref{em:msmassexpl}. 

%\end{comment}

%\renewcommand\bibname{{References}}
%\bibliographystyle{../../master/thesis/tocunsrt}
%\bibliography{../articles/hypcomp} 

\chapter{Initial data}\label{c:initial}

\section{Solving the constraints}
%\subsubsection{General procedures}

In a Cauchy formulation, the Einstein equations are decomposed into evolution equations and constraint equations, as derived in section \ref{s3:decomp}. The constraint equations are such that if satisfied by the initial data, they will be satisfied at all times at the continuum level due to the Bianchi identities (see e.g. \cite{Alcubierre}). They are commonly expressed as elliptic equations, so that they have to be solved globally and require boundary conditions.

The constraint equations are given by the scalar Hamiltonian constraint and the vector momentum constraint, that is, four equations in total. The metric and extrinsic curvature have a total of twelve components to be determined, which means that there are eight degrees of freedom to be fixed and the solution of the constraint equations will provide the four remaining quantities.

In order to separate the freely specifiable data and provide a convenient set of elliptic equations to solve, variable transformations are performed on some of the quantities.
%This involves a conformal decomposition of the spatial metric and splitting the extrinsic curvature into its trace and trace-free part. \upda{Include basic equations?}
Among the widely used constraint decompositions are the York-Lichnerowicz conformal decompositions.
Their first ingredient is the conformal decomposition of the Hamiltonian constraint proposed by Lichnerowicz \cite{Lichnerowicz:d}. This is performed via a decomposition of the spatial metric $\bar\gamma_{ab}$ into a conformal factor $\psi$ and an auxiliary conformal background spatial metric $\gamma_{ab}$ \cite{Lichnerowicz:d,York:1971hw,York:1972sj}:
\begin{equation} \label{ein:lich}
\bar\gamma_{ab} = \psi^4 \gamma_{ab} .
\end{equation}
The Hamiltonian constraint is expressed in terms of the new variables as
\begin{equation}
\nabla^a\nabla_a\psi - \case{1}{8}\psi R[\gamma] + \case{1}{8}\psi^5\left(\bar K_{ab}\bar K^{ab}-\bar K^2\right) - 2\pi\psi^5\rho = 0 .
\end{equation}
The second ingredient is the decomposition of the extrinsic curvature suggested by York \cite{York:1973ia,York:1974psa}, which takes the form
\begin{equation}
\bar K_{ab} = \bar A_{ab}+\case{1}{3}\bar \gamma_{ab} \bar K .
\end{equation}
The trace-free part $A^{ab}$ is split into a transverse-traceless tensor and a longitudinal part.
%\begin{equation}(\mathbb{L}X)^{ab} = \ \end{equation}
The longitudinal part can be defined in terms of the original metric $\bar\gamma_{ab}$ or the conformal one $\gamma_{ab}$, thus giving the physical \cite{O'Murchadha:1974nc,O'Murchadha:1974nd} and conformal \cite{1979sgrr.work...83Y,1982sag..conf..147Y} transverse-traceless variants, respectively. Making the same initial choices for the freely specifiable data in the two formulations will in principle provide results with different physical properties.

The conformal thin-sandwich decomposition by York \cite{York:1998hy} takes into account the change in the metric between two neighboring hypersurfaces in the form of the quantity
\begin{equation}
u_{ab} = \partial_t \gamma_{ab} ,
\end{equation}
which is used in the decomposition of the trace-free part of the extrinsic curvature.
This formulation provides a more direct relation between the choices for the freely specifiable data and their effects on the initial physics of the system.

For more details about these formulation see e.g. \cite{Cook:2000vr} or \cite{Alcubierre}.

\subsubsection{Bowen-York initial data}

The Bowen-York initial data solution \cite{Bowen:1980yu} is obtained under the assumptions of conformal and asymptotic flatness and maximal slicing. This conditions allow to find an analytical solution to the momentum constraint for a BH with given spin and linear momentum. Moreover, as the momentum constraint is linear under these assumptions, it allows to obtain initial data for more than one BH by superposition. %, obtained with the so-called conformal-imaging method

Another important ingredient is the puncture approach by Brandt and Br\"ugmann \cite{Brandt:1997tf}, a generalization of the Brill-Lindquist data \cite{1963PhRv..131..471B}. It assumes a certain form for the conformal factor and solves the Hamiltonian equation for a smoother quantity.

%\subsubsection{High spin initial data}

The assumptions used to obtain the Bowen-York puncture initial data are too restrictive for some initial data configurations. For instance, these methods are not suited to obtain initial data for a very rapidly spinning BH, because the Kerr-Newman solution accepts no conformally flat slices \cite{ValienteKroon:2004gj}. The Bowen-York solution of a spinning BH is a superposition a Kerr BH and some gravitational radiation, commonly called ``junk'' radiation. A consequence is that there is a maximum limit in the BH's spin achievable with this initial data calculation, namely 0.93 out of 1 \cite{Cook:1989fb,Dain:2002ee,Hannam:2009ib}. Going beyond this limit requires dropping the conformal flatness assumption and solving the constraints with a more general approach \cite{Lovelace:2010ne,Scheel:2014ina}.

%high-spin initial data (Novikov and Thorne's estimate about realistic value of high spins, "Black Holes", 1973, 1974, a=0.9982)
%Bowen-York limit of a=0.92, junk radiation
%non-corformally flat initial data (1412.1803), efforts in this direction, difficulties with high spins

%\upda{
%\vspace{1ex}
%Although not included in this thesis, I also performed some work on initial data calculation. More specifically, it involved getting familiar with an initial data code written in Mathematica with the constraint equations in conformal thin-sandwich decomposition form and generalizing the numerical calculations to an independent number of grid-points in any of the three spatial dimensions. % the author
%}

\subsubsection{Black hole excision and moving punctures}

The singularities present in the BHs require a special numerical treatment. The most common choices are excision and puncture evolution.

BH excision was first used in spherical symmetry in \cite{Seidel:1992vd}. It consists basically of two ingredients: first the central singularity is excised by setting a boundary inside of the BH's (apparent) horizon and then the shift vector is given non-zero values such that it fixes the BH's horizon to a given coordinate location. The excision boundary is spacelike (it is located inside of the BH) so no boundary conditions are required, provided that no gauge modes with superluminal speeds have outgoing characteristics at the excision surface.
The excision approach has been successful in spherical symmetry \cite{Anninos:1994dj,Marsa:1996fa,Scheel:1997kb}, but its implementation into 3+1 three-dimensional codes becomes more complicated \cite{Anninos:1995am} due to the difficulty in expressing the spheroidal excision surface in the Cartesian coordinates used in the code. The excision method is used by the Spectral Einstein Code \cite{spec,Ossokine:2013zga}.

%The puncture evolution method treats the BH's singularity by factoring out the singular behaviour of the metric and evolving the regular part separately. Factoring out a factor with singular terms was first done in \cite{1964AnPhy..29..304H,Smarr:1976qy}, and later the puncture idea was applied to a single Schwarzschild BH in \cite{Anninos:1995am}. The use of the puncture method to evolve Bowen-York initial data with a conformally flat metric was proposed by \cite{Bruegmann:1997uc}.
%The puncture (the factored out singularity) cannot be located directly on a grid point, so that its location on the grid must be staggered.
%If the singular conformal factor is fixed in time, it is called a static puncture. Its main drawback is that they require coordinate systems that follow the motion of the individual BHs. A large breakthrough was the development of moving punctures by \cite{Campanelli:2006gf,Campanelli:2005dd} and \cite{Baker:2006yw,Baker:2005vv}: in this case, the singularity is not factored out, but evolved as a dynamical quantity and its position moves on the grid during evolution in the same way as the actual BH would do.
%singularity avoiding slicings, as those given by the maximal slicing and the 1+log gauge conditions.

Unlike the static puncture method, where the singular behaviour of the metric at the BH's singularity is factored out and a regular part is evolved separately \cite{1964AnPhy..29..304H,Smarr:1976qy}, the moving puncture approach puts the singular part into a dynamical conformal factor.
The use of the puncture method to evolve Bowen-York initial data with a conformally flat metric was proposed by \cite{Bruegmann:1997uc}.
In the BSSN formulation, the singularity can be evolved in the conformal factor $\chi$ \eref{et:chidef} as done in \cite{Campanelli:2006gf,Campanelli:2005dd} or using the variable $\varphi$ as in \cite{Baker:2006yw,Baker:2005vv}.
It is important that the puncture is staggered, so that it never coincides with an actual gridpoint.
Singularity avoiding slicings, as those given by maximal slicing and the 1+log gauge condition (see subsection \ref{gs:1plog}), are needed in order to prevent the slices from reaching the singularity and ruining the simulation. Also a non-vanishing shift, like e.g. the Gamma-driver condition (see subsection \ref{sg:gammadriver}), is required to allow the punctures to move across the numerical grid. In unconstrained evolutions, the 1+log slicing condition and the Gamma-driver shift are common choices, as in \cite{Bruegmann:2006at}.

\subsubsection{Hyperboloidal initial data}

%Yamabe equation, Schneeman's thesis, put in initial data chapter? (Yamabe equation degenerates on the boundary)
%solving the Yamabe equation for hyperboloidal, degenerate at scri initial data \cite{Frauendiener:2002iw}

%black hole initial data on hyperboloidal slices \cite{Buchman:2009ew}
%Schneemann's Diplomarbeit \cite{Schneemann}
%Bondi-Sachs Energy-Momentum for the CMC Initial Value Problem \cite{Bardeen:2011pd}

%\subsubsection{Initial data calculation: an example} ! moved here from intro

%\upda{does it make sense to put this here? the 3+1 decomp. has not been performed yet ... }

% mainly copied from Sascha's 0204057
To illustrate, following \cite{Frauendiener:2002iw,Husa:2002zc}, the kind of problems that may arise when solving the constraint equations to obtain initial data from the conformal equations, let us consider the subclass of hyperboloidal slices where the initial value of the extrinsic curvature is pure trace, $\tilde K_{ab}=\case{1}{3}\tilde \gamma_{ab}\tilde K$. The vacuum momentum constraint is
\begin{equation}
\tilde \nabla^b\tilde K_{ab}-\tilde\nabla_a\tilde K=0 ,
\end{equation}
and it requires that the trace of the extrinsic curvature $\tilde K$ is a non-vanishing constant.
The Hamiltonian constraint
\begin{equation}
R[\tilde \gamma]-\tilde K_{ab}\tilde K^{ab}+{\tilde K}^2=0 ,
\end{equation}
is reduced to a single second order elliptic equation by the modified Lichnerowicz ansatz
\begin{equation}
\tilde \gamma_{ab} = \Omega^{-2} \bar\gamma_{ab} = \omega^{-2}\phi^4 \bar\gamma_{ab}.
\end{equation}
Of the two conformal factors introduced, an appropriate value that vanishes at $\scri^+$ will be set for $\omega$ and the Hamiltonian constraint, now transformed into the form of the Yamabe equation \cite{yamabe1960}, will be solved for $\phi$:
\begin{equation}\label{ei:Yamabe}
4\,\omega^2\bar\bigtriangleup\phi-4\,\omega(\bar\nabla^a\omega)(\bar\nabla_a\phi)-\left[\frac{1}{2}R[\bar\gamma]\omega^2+2\omega\bar\bigtriangleup\omega-3(\bar\nabla^a\omega)(\bar\nabla_a\omega)\right]\phi=\frac{1}{3}{\tilde K}^2\phi^5
\end{equation}
The Yamabe equation is degenerate at $\scri^+$, as $\atscri{\omega}=0$ holds. It allows to determine the boundary value as
\begin{equation}
\phi^4=\frac{9}{{\tilde K}^2}(\bar\nabla^a\omega)(\bar\nabla_a\omega) .
\end{equation}
A positive and unique solution for the Yamabe equation \eref{ei:Yamabe} has been proven to exist in \cite{Andersson:1992yk}. In spite of the equation's degeneracy, its numerical resolution does not pose any problems and the Yamabe equation can be solved using standard procedures.

%Yamabe equation, Schneeman's thesis, put in initial data chapter? (Yamabe equation degenerates on the boundary)
%solving the Yamabe equation for hyperboloidal, degenerate at scri initial data \cite{Frauendiener:2002iw}
%existence of solution to the Yamabe equation in \cite{Andersson:1992yk}

%Some of the problem encountered when solving the constraints on a hyperboloidal slice are described in \cite{Frauendiener:2002iw}: the Yamabe equation is degenerate on $\scri^+$.
%
Among the efforts to obtain initial data on hyperboloidal slices are the first implementations by Frauendiener \cite{Frauendiener1999475,Frauendiener:2002iw} and  H\"ubner \cite{Hubner:1998hn,Hubner:2000zn}; the work in Schneemann's Diploma thesis \cite{Schneemann}, where solutions in spherical and in axial symmetry are presented; the generalization of Bowen-York initial data to hyperboloidal slices for binaries of boosted and spinning BHs in \cite{Buchman:2009ew}; and perturbed Kerr initial data in \cite{Schinkel:2013zm}.

\section{Compactified hyperboloidal vacuum initial data}\label{cin:indata}

The procedure to obtain vacuum initial data for the spherically symmetric equations \eref{es:eeqs} (or \eref{es:pKeeqs} or \eref{es:DPKeeqs}), which will of course satisfy \eref{es:ceqs} (or \eref{es:pKceqs} or \eref{es:DPKceqs}), is described in detail in this section. It follows very similar steps as in \cite{Zenginoglu:2007jw}.

\subsection{General procedure}

\subsubsection{Transformations of the initial line element}

We will calculate spherically symmetric vacuum initial data on a hyperboloidal slice starting with the line element on a Cauchy slice:
\begin{equation}\label{ein:lielphys}
d\tilde s^2 = -A(\tilde r)d\tilde t^2+\frac{1}{A(\tilde r)}d\tilde r^2+\tilde r^2 d\sigma^2  , \quad \textrm{where} \quad d\sigma^2\equiv d\theta^2+\sin^2\theta d\phi^2  .
\end{equation}
Both the line element and the coordinates include a tilde to indicate that they measure distances on the physical domain (and not on the conformal one). The form of the initial metric is general enough to consider flat spacetime, the Schwarzschild spacetime and the Reissner-Nordstr\"om (RN) spacetime, among others.

The time coordinate $\tilde t$ is transformed into a new time coordinate $t$ using a height function $h(\tilde r)$:
\begin{equation}
t = \tilde t-h(\tilde r) ,
\end{equation}
and the line element now reads (substituting $d\tilde t = dt + \hpt d\tilde r$)
\begin{equation}\label{ein:lielt}
d\tilde s^2 = -A(\tilde r)dt^2-2A(\tilde r)\hpt dt\, d\tilde r+\frac{\left[1-\left(A(\tilde r)\hpt\right)^2\right]}{A(\tilde r)}d\tilde r^2+\tilde r^2 d\sigma^2   .
\end{equation}
The hypersurfaces of constant time $t$ are now hyperboloidal slices that reach $\scri^+$. However, $\scri^+$ is still infinitely far away, so that the next step is to compactify the hyperboloidal slices. This is done by rescaling the radial component $\tilde r$ by a compactifying factor $\aconf$. Do not confuse the compactifying factor $\aconf$ with the conformal factor $\Omega$ that rescales the metric in \eref{ei:rescmetric}, as they are not necessarily the same. The new radial coordinate $r$ is given by
\begin{equation}\label{ein:compact}
\tilde r=\frac{r}{\aconf(r)}  ,
\end{equation}
where $\aconf$ is such that it vanishes at the value of $r$ where the infinity of $\tilde r$ ($\scri^+$ in this slice) is mapped to, that is $\tilde r\to \infty \Leftrightarrow r\to\rscri$ so that $\aconf(\rscri)=0$.
After the radial coordinate transformation $d\tilde r= \case{\aconf-r\,\aconf'}{\aconf^2}d r$ the initial line element reads
\begin{equation}
%d\tilde s^2 = -A(\case{r}{\aconf})dt^2-2A(\case{r}{\aconf})\hp  \frac{\aconf-r\,\aconf'}{\aconf^2}dt\,dr+\frac{\left[1-\left(A(\case{r}{\aconf})\hp\right)^2\right]}{A(\case{r}{\aconf})}\frac{(\aconf-r\,\aconf')^2}{\aconf^4}dr^2+\frac{r^2}{\aconf^2}  d\sigma^2   .
d\tilde s^2 = -Adt^2-2A\,h'\frac{\aconf-r\,\aconf'}{\aconf^2}dt\,dr+\frac{\left[1-\left(A\,h'\right)^2\right]}{A}\frac{(\aconf-r\,\aconf')^2}{\aconf^4}dr^2+\frac{r^2}{\aconf^2}  d\sigma^2   .
\end{equation}
Both $A$ and $h'$ are functions of $\case{r}{\aconf}$ and $\aconf$ depends on $r$, but for reasons of space and clarity this dependence is not explicitly written.
Finally the complete line element is conformally rescaled by $\Omega^2$, in an equivalent way as done with the metric in \eref{ei:rescmetric}, as $d\bar s^2 = \Omega^2d\tilde s^2$:
\noindent{\small\begin{equation}\label{fsthyp}
d\bar s^2= -A\,\Omega^2dt^2+\frac{\Omega^2}{\aconf^2}\left[-2A\,h'\,(\aconf-r\,\aconf')dt\,dr+\frac{\left[1-\left(A\,h'\right)^2\right]}{A}\frac{(\aconf-r\,\aconf')^2}{\aconf^2}d r^2 + r^2 d\sigma^2\right] .
\end{equation}}%
Here the overbar indicates that this line element measures distances in the conformally rescaled spacetime.

\subsubsection{Initial data for the metric components}

Comparing with the line element written in terms of the component variables \eref{es:liel}, shown here again for convenience,
\begin{equation}
d\bar s^2 = - \left(\alpha^2-\chi^{-1}\gamma_{rr}{\beta^r}^2\right) dt^2 + \chi^{-1}\left[2\, \gamma_{rr}\beta^r dt\,dr +  \gamma_{rr}\, dr^2 +  \gamma_{\theta\theta}\, r^2\, d\sigma^2\right] .
\end{equation}
the initial values of each of the metric components can be directly read off. A convenient choice is %, listing the variables in the order of assignment,
\begin{subequations}\label{ein:metrinih}
\begin{eqnarray}
{\gamma_{\theta\theta}}_0 &=& 1 , \\
\chi{}_0 &=& \frac{\aconf^2}{\Omega^2} , \\
{\gamma_{rr}}_0 &=& \frac{\left[1-\left(A\,h'\right)^2\right]}{A}\frac{(\aconf-r\,\aconf')^2}{\aconf^2} , \\
{\beta^r}_0&=& - \frac{A^2\aconf^2\,h'}{\left[1-\left(A\,h'\right)^2\right](\aconf-r\,\aconf')}, \\
\alpha_0&=& \Omega\sqrt{\frac{A}{1-\left(A\,h'\right)^2}},
\end{eqnarray}
\end{subequations}
where the subscript $_0$ indicates that these are the expressions for the initial values.

\subsubsection{Initial data for the derived quantities}

%The Z4 quantites $\Theta$ and $Z_r$ are zero initially, because that corresponds to the real solution. \upda{rewrite}
The solution of the Z4 equations only coincides with a solution of the Einstein equations when the constraint fields $\Theta$ and $Z_r$ are zero. For this reason, their stationary value is expected to vanish and their initial values will also be zero.

The initial values of the component $A_{rr}$ of the trace-free part of the extrinsic curvature and its trace $\bar K$ are expressed in terms of the metric components as
\begin{subequations} \label{ein:curv0}
\begin{eqnarray}
A_{rr}{}_0 &=& \frac{{\beta^r}_0 {\gamma_{rr}}_0'}{3 \alpha_0 }+\frac{2 {\gamma_{rr}}_0 {\beta^r}_0'}{3 \alpha_0 }-\frac{{\beta^r}_0 {\gamma_{rr}}_0 {\gamma_{\theta\theta}}_0'}{3 \alpha_0  {\gamma_{\theta\theta}}_0}-\frac{2 {\beta^r}_0 {\gamma_{rr}}_0}{3 \alpha_0  r} , \label{ein:Acurv0} \\
\bar K_0 &=& \frac{{\beta^r}_0'}{\alpha_0 }-\frac{3 {\beta^r}_0 \chi '}{2 \alpha_0  \chi }+\frac{{\beta^r}_0 {\gamma_{\theta\theta}}_0'}{\alpha_0  {\gamma_{\theta\theta}}_0}+\frac{{\beta^r}_0 {\gamma_{rr}}_0'}{2 \alpha_0  {\gamma_{rr}}_0}+\frac{2 {\beta^r}_0}{\alpha_0  r} .
\end{eqnarray}
\end{subequations}
They were calculated from the decomposition of \eref{e3:K} supposing that the initial values are time-independent.
Substituting the initial values for the metric components will provide the explicit expressions for the initial values of the extrinsic curvature. They are not shown here because they are lengthy expressions, but they will be presented in subsection \ref{sin:compact} after performing some simplifications. Note that the initial value of $K$ is the same as $\bar K$, as $\Theta_0=0$.

If the background metric is set to the initial values of the evolved metric,
\begin{equation}
\hat\gamma_{rr} = {\gamma_{rr}}_0 =\frac{\left[1-\left(A\,h'\right)^2\right]}{A}\frac{(\aconf-r\,\aconf')^2}{\aconf^2} \qquad \textrm{and} \qquad \hat\gamma_{\theta\theta}={\gamma_{\theta\theta}}_0=1 ,
\end{equation}
then by definition \eref{g:Lambdaa} ${\Delta\Gamma^r}_0 = 0$, which together with ${Z_r}_0=0$ sets ${\Lambda^r}_0=0$.

\subsection{Height function approach}

The derivation of a height function that provides CMC slices will follow \cite{Gentle:2000aq,Malec:2003dq}, also consider \cite{Iriondo:1995ar}.
First we compare \eref{ein:lielt} to the line element
\begin{equation}
d\tilde s^2 =\tilde g_{\mu\nu}dx^\mu dx^\nu = - \left(\tilde\alpha^2-\tilde{\bar{\gamma}}_{rr}\tilde{\beta^r}^2\right) dt^2 + 2\, \tilde{\bar\gamma}_{rr}\tilde\beta^r dt\,d\tilde r +  \tilde{\bar\gamma}_{rr}\, d\tilde r^2 +  \tilde{\bar\gamma}_{\theta\theta}\, \tilde r^2\, d\sigma^2 ,
\end{equation}
and thus see that \eref{ein:lielt} corresponds to the metric
\begin{equation}\label{eil:mat4t}
{ \tilde g}_{\mu\nu} = \left( \begin{array}{cccc}-A(\tilde r) & -A(\tilde r)\hpt & 0 & 0 \\
                                       -A(\tilde r)\hpt & \frac{1-\left(A(\tilde r)\hpt\right)^2}{A(\tilde r)} & \tilde r^2 & 0 \\
                                       0 & 0 & 0 & \tilde r^2 \sin^2{\theta}
                            \end{array}  \right) ,
\end{equation}
with determinant  $\tilde g=-\tilde r^4\sin^2\theta$.
The normal vector in adapted coordinates can be written as $\tilde n^\mu =  \frac{1}{\tilde\alpha}\left(1,-\tilde\beta^r,0,0\right)^T$. Its expression according to \eref{ein:lielt} is
\begin{equation}\label{ein:nexpr}
\tilde n^\mu =  \left(\sqrt{\frac{1-\left(A(\tilde r)\hpt\right)^2}{A(\tilde r)}} , \frac{A^{3/2}(\tilde r)\,\hpt}{\sqrt{1-\left(A(\tilde r)\hpt\right)^2}},0,0\right)^T .
\end{equation}
Contracting the right equation in \eref{e3:physK}, an expression for the trace of the physical extrinsic curvature $\tilde{\bar K}$ is obtained:
\begin{equation}
\tilde g^{ab}\tilde{\bar K}_{ab} = -\tilde g^{ab}\tilde{\bar{\perp_a^c}}\tilde\nabla_c \tilde n_b \quad\to\quad\tilde{\bar K} = - \tilde{\bar\perp}_a^b\tilde{\nabla}_b \tilde n^a = -\frac{1}{\sqrt{-\tilde g}}\partial_a\left(\sqrt{-\tilde g}\,\tilde n^a\right) .
\end{equation}
Substituting the determinant of \eref{eil:mat4t} and the expression of $\tilde n^\mu$ \eref{ein:nexpr}, the previous relation now reads
\begin{equation}
\tilde{\bar K} = -\frac{1}{r^2}\partial_r\left[ r^2 \frac{A^{3/2}(\tilde r)\,\hpt}{\sqrt{1-\left(A(\tilde r)\hpt\right)^2}} \right] .
\end{equation}
This expression can be integrated by setting the value of the trace of the extrinsic curvature to a constant value $\tilde{\bar K} = \Kc$:
\begin{equation}\label{ein:ninteg}
\frac{A^{3/2}(\tilde r)\,\hpt}{\sqrt{1-\left(A(\tilde r)\hpt\right)^2}} =-\frac{1}{r^2}\int \Kc\, r^2 dr - \frac{\Cc}{r^2} = -\frac{\Kc\, r}{3} - \frac{\Cc}{r^2} ,
\end{equation}
where the parameter $\Kc$ and the integration constant $\Cc$ are set in such a way that $\Kc<0$ (according to the convention chosen for the extrinsic curvature) and $\Cc>0$. Solving for $\hpt$ and choosing the convenient sign will give us its value to calculate the initial data:
\begin{equation}\label{ein:heightp}
\hpt = - \frac{\frac{\Kc\,\tilde  r}{3} + \frac{\Cc}{\tilde r^2} }{A(\tilde r)\sqrt{A(\tilde r)+\left(\frac{\Kc\,\tilde  r}{3} + \frac{\Cc}{\tilde r^2}\right) ^2}}.
\end{equation}
This expression with $\Kc=0$ set (maximal slicing) has been widely used in relation with trumpet \cite{Hannam:2006vv} initial and stationary data. More details are given in subsection \ref{sin:strong}.

The components of the rescaled spatial conformal metric obtained before, \eref{ein:metrinih}, turn into the following after setting the previous expression for $\hpt$ expressed in terms of $\case{r}{\aconf}$:
\begin{subequations}\label{ein:metrinia}
\begin{eqnarray}
{\gamma_{\theta\theta}}_0 &=& 1 , \\
\chi{}_0 &=& \frac{\aconf^2}{\Omega^2} , \\
{\gamma_{rr}}_0 &=& \frac{(\aconf-r\,\aconf')^2}{\left[A(\case{r}{\aconf})+\left(\frac{\Kc\,r}{3\aconf}+\frac{\Cc\aconf^2}{r^2}\right)^2\right] \aconf^2} , \\
{\beta^r}_0&=&\frac{\left(\frac{\Kc\,r}{3\aconf}+\frac{\Cc\aconf^2}{r^2}\right)\sqrt{A(\case{r}{\aconf})+\left(\frac{\Kc\,r}{3\aconf}+\frac{\Cc\aconf^2}{r^2}\right)^2} \, \aconf^2}{(\aconf-r\,\aconf')}, \\
\alpha_0&=& \Omega\sqrt{A(\case{r}{\aconf})+\left(\frac{\Kc\,r}{3\aconf}+\frac{\Cc\aconf^2}{r^2}\right)^2} .
\end{eqnarray}
\end{subequations}

%\subsubsection{Calculation of $C_{CMC}$ value}

\subsection{Compactification}\label{sin:compact}

The compactification of the hyperboloidal slices has been performed by rescaling the radial coordinate as in \eref{ein:compact}. In principle, the only conditions that the function $\aconf(r)$ has to satisfy is being smooth, positive, going to zero as $r$ goes to the radial location assigned to $\scri^+$, which is equivalent to $\tilde r\to \infty$, and with $\aconf'\neq0$ to ensure that the transformation is invertible.

A very convenient choice is choosing $\aconf$ such that the initial spatial metric is conformally flat, equivalent to imposing initial isotropic coordinates. This translates to setting
\begin{equation}\label{ein:conflat}
{\gamma_{rr}}_0 = \frac{(\aconf-r\,\aconf')^2}{\left[A(\case{r}{\aconf})+\left(\frac{\Kc\,r}{3\aconf}+\frac{\Cc\aconf^2}{r^2}\right)^2\right] \aconf^2} = 1 .
\end{equation}
This condition can indeed be satisfied and the corresponding $\aconf$ can be calculated analytically for flat spacetime and numerically for the Schwarzschild and RN cases (more details in the next section). The derivative $\aconf'$ can be isolated from condition \eref{ein:conflat} and expressed in terms of $r$ and $\aconf$. Substituting it into the initial values for the metric components \eref{ein:metrinia} yields
\begin{subequations}\label{ein:metrinio}
\begin{eqnarray}
{\gamma_{\theta\theta}}_0 &=& 1 , \\
\chi{}_0 &=& \frac{\aconf^2}{\Omega^2} , \\
{\gamma_{rr}}_0 &=& 1 , \\
{\beta^r}_0&=& \frac{\Kc\,r}{3}+\frac{\Cc\aconf^3}{r^2} , \label{ein:betarinio} \\ %\left(\frac{\Kc\,r}{3\aconf}+\frac{\Cc\aconf^2}{r^2}\right)\aconf, \\
\alpha_0&=& \Omega\sqrt{A(\case{r}{\aconf})+\left(\frac{\Kc\,r}{3\aconf}+\frac{\Cc\aconf^2}{r^2}\right)^2},
\end{eqnarray}
\end{subequations}

Under the assumption of this condition the initial values for the extrinsic curvature \eref{ein:curv0} simplify considerably and are written as
\begin{subequations}\label{ein:curvinio}
\begin{eqnarray}
A_{rr}{}_0 &=& -\frac{2\Cc \aconf^3}{r^3\Omega} , \\
\bar K_0 &=& \frac{\Kc}{\Omega} +\frac{\frac{\Kc\,r}{3}+\frac{\Cc\aconf^3}{r^2}}{\sqrt{A(\case{r}{\aconf})+\left(\frac{\Kc\,r}{3\aconf}+\frac{\Cc\aconf^2}{r^2}\right)^2}}\frac{3\Omega'}{\Omega^2} . \label{ein:Kini}
\end{eqnarray}
\end{subequations}

This choice for the compactification factor $\aconf$ suggests setting the flat spatial metric in spherical coordinates for the background metric $\hat\gamma_{ij}=diag(1,r^2,r^2\sin^2\theta)$, as was done in \eref{es:gb}. The Z4 quantities $\Theta$ and $Z_r$ have to vanish initially and the conformal flatness of the initial data also implies that $\Lambda^r{}_0=0$.
The initial value of the trace of the physical extrinsic curvature is $\tilde{\bar K}_0=\Kc$ (as was set to integrate \eref{ein:ninteg}), for any choice of $A(\case{r}{\aconf})$.

\section{Spacetimes considered}

The vacuum initial data procedure will be applied to the spherically symmetric cases of flat spacetime and also to Schwarzschild and RN BHs.

\subsection{Flat spacetime}

The metric of Minkowski spacetime is recovered in \eref{ein:lielphys} by setting $A(\tilde r)=1$. This simplifies the expressions, even allowing to integrate $\hpt$ and obtain an analytic expression for the height function. Choosing a value for the integration constant $\Cc$ is actually a critical part, but this will be explained in subsection \ref{sin:Ccmccalcul}, as it is much more relevant in the presence of singularities. For regular initial data it is simply set to zero, $\Cc=0$. %\upda{It flat spacetime it will be simply set to zero $\Cc=0$.}

\begin{figure}[htpb!!]
\center
\begin{tabular}{@{}c@{}@{}c@{}@{}c@{}@{}c@{}}
\includegraphics[width=0.25\linewidth]{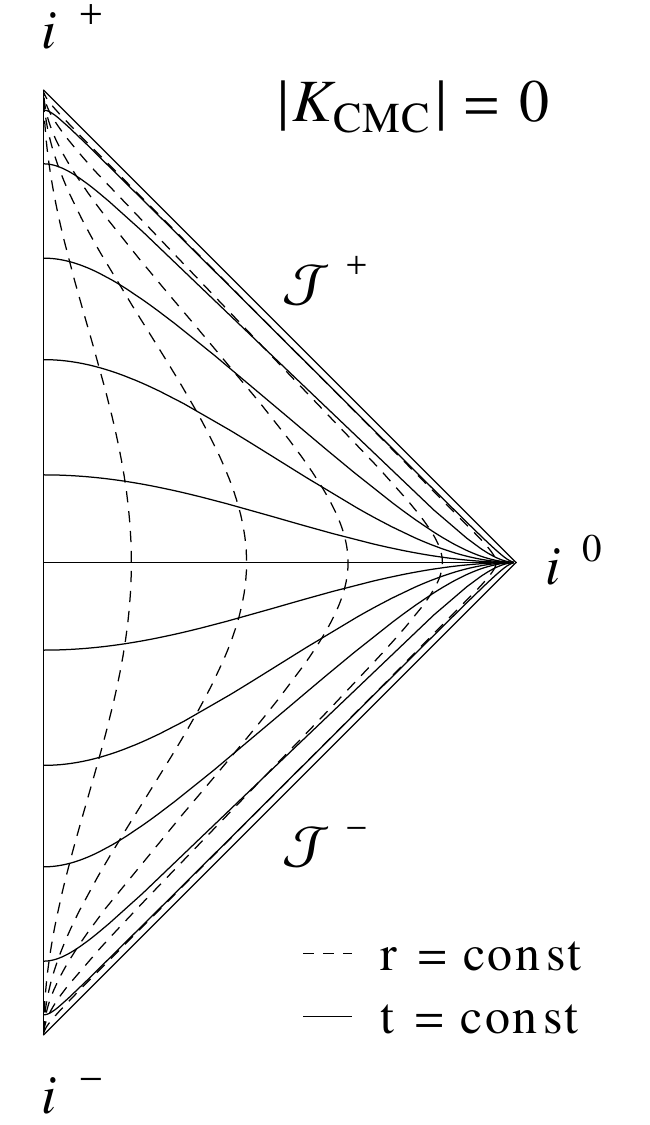}&
\includegraphics[width=0.25\linewidth]{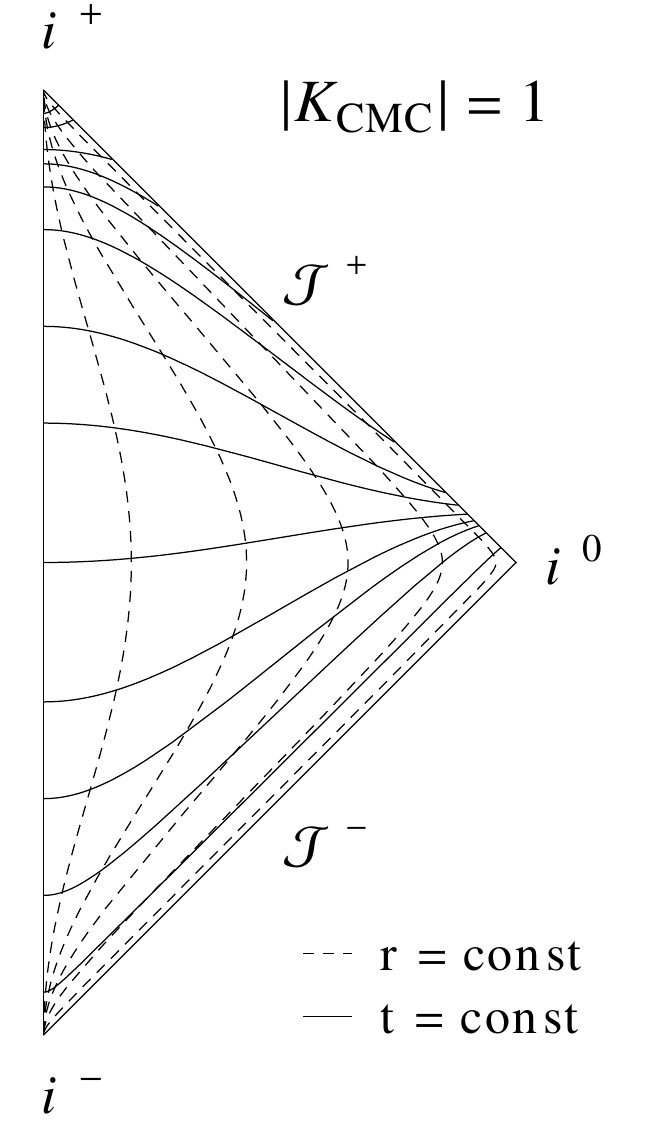}&
\includegraphics[width=0.25\linewidth]{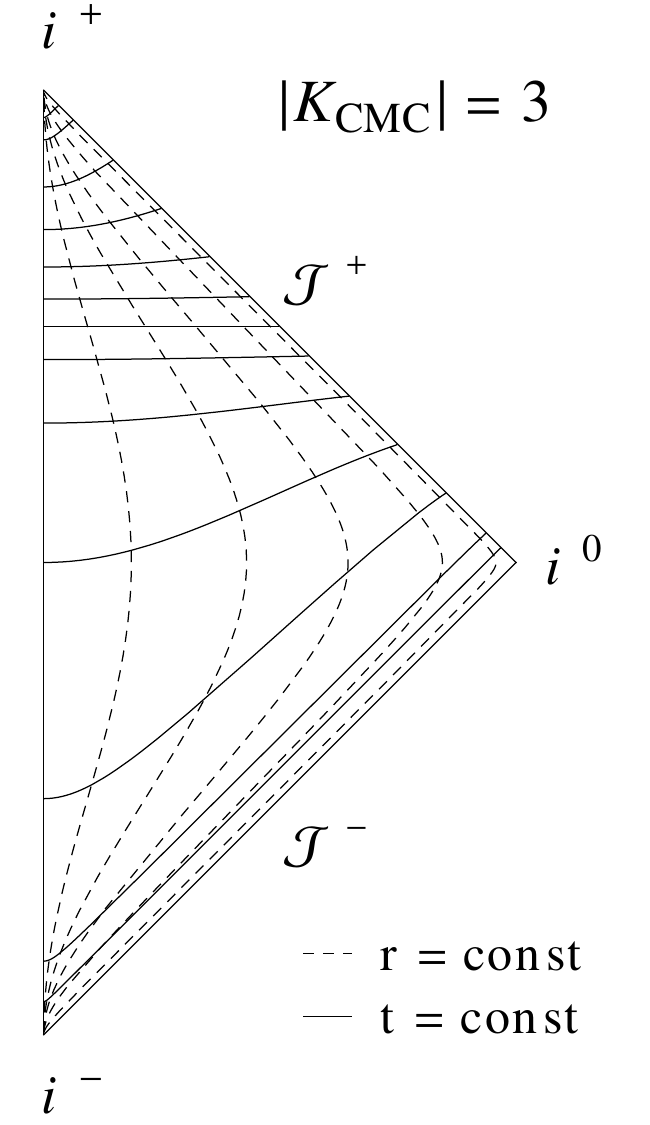}&
\includegraphics[width=0.25\linewidth]{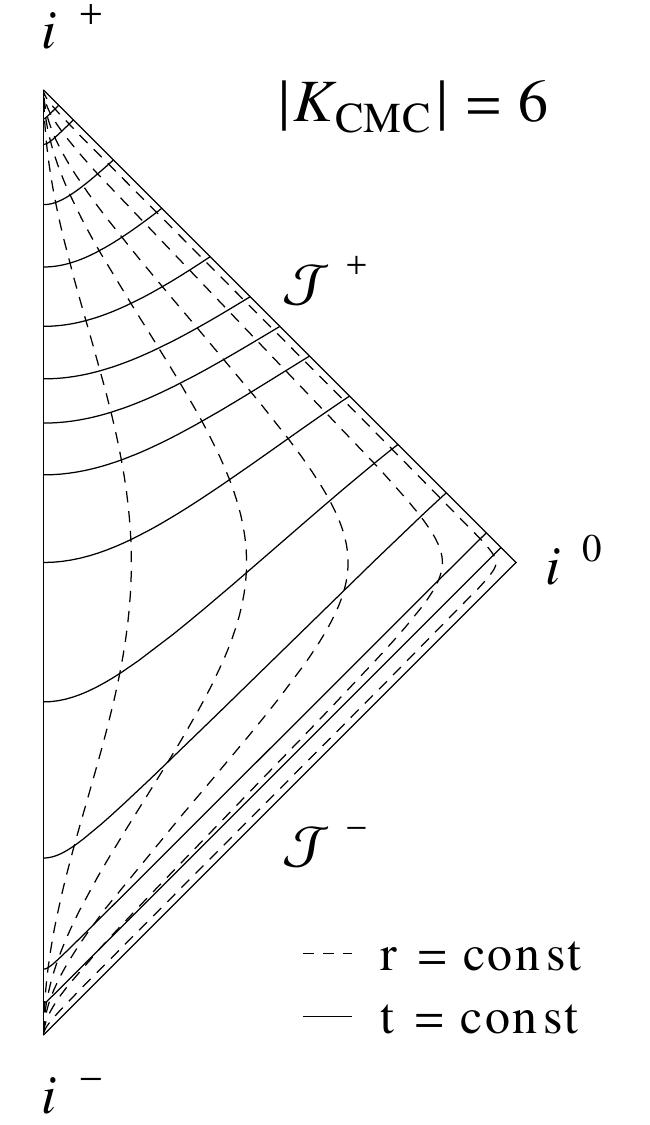}
\end{tabular}
\caption{Foliations of flat spacetime for different values of the constant trace of the extrinsic curvature $K_{CMC}$: the leftmost diagram shows maximal slices (Cauchy slices), while the rest show hyperboloidal ones. Compare to figure 4 in \cite{Zenginoglu:2007jw}.} %Compare to \cite{Zenginoglu:2007jw}.
\label{fin:flat}
\end{figure}

After the simplifications, the height function for flat spacetime is given by
\begin{equation}
\hpt = - \frac{\frac{\Kc\,\tilde  r}{3}  }{\sqrt{1+\left(\frac{\Kc\,\tilde  r}{3} \right) ^2}} ,
\end{equation}
and by straightforward integration we obtain
\begin{equation}\label{ein:flath}
h(\tilde r) = \sqrt{\tilde r^2+\left(\frac{3}{\Kc}\right)^2}.
\end{equation}
In \fref{fin:flat} we can see Penrose diagrams showing foliations with different values of the parameter $\Kc$. The larger its absolute value, the closer to light rays the slices become. If positive values of $\Kc$ were chosen, the height function would have the opposite sign and the foliations would intersect $\scri^-$ (past null infinity) instead of $\scri^+$. %If positive values of $\Kc$ were chosen, the foliations would intersect $\scri^-$ (past null infinity) instead of $\scri^+$ - not completely true, the integrated height function depends on $\Kc^2$, but changing the sign in front of $h(\tilde r)$ does make it happen.

The calculation of the compactifying factor $\aconf$ from \eref{ein:conflat}, with the condition that $\aconf(\rscri)=0$, is also simplified and yields the analytic expression
\begin{equation}\label{ein:aconfflat}
\frac{(\aconf-r\,\aconf')^2}{\left[1+\left(\frac{\Kc\,r}{3\aconf}\right)^2\right] \aconf^2} = 1 \quad \Rightarrow \quad \aconf = (-\Kc)\frac{\rscri^2-r^2}{6\, r_{\!\!\scri}} ,
\end{equation}
where $\rscri$ is the radial location of $\scri^+$ and the sign of the compactification factor has been chosen so that $\aconf\geq0$ for $r\in[0,\rscri]$. This expression was already used in \cite{Husa:2002zc,Schneemann}. % also supposed to be used in http://online.kitp.ucsb.edu/online/numrel00/moncrief/ (Moncrief, conformally regular adm evolution equations, talk at KITP Santa Barbara, 2000)

The properties of the expression obtained for $\aconf$ also satisfy the requirements for the conformal factor $\Omega$, so that the simplest choice in flat spacetime is $\Omega=\aconf$. The explicit expression for the conformal factor is then
\begin{equation}\label{ein:omega}
\Omega = (-\Kc)\frac{\rscri^2-r^2}{6\, r_{\!\!\scri}} .
\end{equation}

The flat spacetime initial data on the hyperboloidal slice for the metric components \eref{ein:metrinia} are
\begin{subequations}\label{ein:metrinioflat}
\begin{equation}
{\gamma_{\theta\theta}}_0 = 1 , \quad
\chi{}_0 = 1 , \quad
{\gamma_{rr}}_0 = 1 , \quad
{\beta^r}_0= \frac{\Kc\,r}{3} ,  \quad
\alpha_0= \sqrt{\Omega^2+\left(\frac{\Kc\,r}{3}\right)^2} ,
\end{equation}
and for the extrinsic curvature quantities
\begin{equation}\label{ein:curvinioflat}
A_{rr}{}_0 = 0 , \quad
\bar K_0 = \frac{\Kc}{\Omega} +\frac{\Kc\,r}{\sqrt{\Omega^2+\left(\frac{\Kc\,r}{3}\right)^2}}\frac{\Omega'}{\Omega} .
\end{equation}
\end{subequations}
The rest of the evolution variables ($\Lambda^r$ and $\Theta$) have vanishing initial values.

\subsection{Strong field initial data}\label{sin:strong}

The value of $\Cc$ set in the initial data has a great influence on the properties of the initial data in the case of a spacetime where a singularity is present.

To illustrate this, let us consider puncture initial data of a Schwarzschild BH. A transformation to isotropic coordinates introduces a new radial coordinate that, instead of entering the BH's horizon, covers two copies of the exterior spacetime. In these coordinates the initial spatial metric is conformally flat and the singularity information is encoded in a conformal factor. Nevertheless, the derived initial value for the lapse is not appropriate because it is negative in a part of the integration domain, so that $\alpha_0=1$ is chosen as the new initial value. These initial data are not a time-independent solution for the Einstein equations and when evolved numerically with common gauge conditions (1+log slicing \cite{Bona:1994dr} and $\tilde\Gamma$-freezing shift \cite{Alcubierre:2002kk} conditions) as done in \cite{Hannam:2006vv,Hannam:2006xw}, the equations find a stationary solution where the slice has detached from the asymptotically flat end of the spacetime's second copy and now ends at the value $\case{3}{2}M$ of the Schwarzschild radius (inside of the horizon). This new stationary solution is called a trumpet: it has an asymptotically flat end at one side and an infinitely long cylinder of radius $\case{3}{2}M$ at the other side. In \cite{Baumgarte:2007ht} it was pointed out that both the puncture initial data and the trumpet stationary data are respectively described by the cases $\Cc=0$ or $\Cc=\case{3\sqrt{3}}{4}M^2$ for the time independent maximal slices given in \cite{PhysRevD.7.2814}.

The aim of this work is to perform simulations on a hyperboloidal slice and the numerical evolution is likely to be unstable at first. Stationary initial data will make it easier to spot unexpected behaviours and changes in the variables. For this reason having an equivalent to the trumpet geometry on the hyperboloidal slice (that will hopefully be a stationary solution) seems like a good option when evolving strong field initial data. Besides, in this way the use of excision, even if it is probably an easier approach, will not be necessary. Note that the implementation of the gauge conditions also has to be compatible to keep the initial data stationary (that is, a time-independent solution of the implemented Einstein equations), see sections \ref{sg:lapse}, \ref{sg:shift} and \ref{sg:source}. The derivation of the trumpet initial data, for both Schwarzschild and RN, will follow parts of \cite{Baumgarte:2007ht,Buchman:2009ew,Tuite:2013hza} and starts by determining the value of $\Cc$ that provides trumpet initial data.

%\upda{interested in checking stability, stationary initial data are most useful. also prefer not to use excision.  - trumpets, these data are stationary due to the source functions in the gauge conditions. }

\subsubsection{Calculation of $C_{CMC}$ value}\label{sin:Ccmccalcul}

%The existence of a ``critical'' value of the integration constant $\Cc$

We take the spatial part of the line element \eref{ein:lielt}, having substituted $\hpt$ with \eref{ein:heightp} and want to express it in isotropic coordinates. Then, the following relation has to be satisfied:
\begin{equation}
d\tilde l^2 = \frac{1}{A(\tilde r)+\left(\frac{\Kc\,\tilde r}{3}+\frac{\Cc}{\tilde r^2}\right)^2}d\tilde r^2+\tilde r^2 d\sigma^2= \frac{1}{{\aconf}^2}\left(dr^2 + r^2d\sigma^2\right) .
\end{equation}
The compactification factor $\aconf$ has been introduced in such a way that \eref{ein:compact} is re-obtained when comparing the coefficient of $d\sigma^2$:
\begin{equation}\label{ein:dsigmacoeff}
\tilde r^2=\frac{r^2}{\aconf^2} .
\end{equation}
The other part gives
\begin{equation}
\frac{1}{A(\tilde r)+\left(\frac{\Kc\,\tilde r}{3}+\frac{\Cc}{\tilde r^2}\right)^2}d\tilde r^2= \frac{1}{{\aconf}^2}dr^2 .
\end{equation}
Substituting $\aconf=\case{r}{\tilde r}$ (taken from \eref{ein:compact} or \eref{ein:dsigmacoeff}) the two radial coordinates can be separated,
\begin{equation}
\frac{1}{\tilde r^2\left[A(\tilde r)+\left(\frac{\Kc\,\tilde r}{3}+\frac{\Cc}{\tilde r^2}\right)^2\right]}d\tilde r^2= \frac{1}{{r}^2}dr^2 ,
\end{equation}
and the expression can be written as
\begin{equation}\label{ein:integralgamma}
\pm \int\frac{dr}{{r}} =  \int\frac{d\tilde r}{\tilde r \sqrt{A(\tilde r)+\left(\frac{\Kc\,\tilde r}{3}+\frac{\Cc}{\tilde r^2}\right)^2}} .
\end{equation}
The integral in the RHS will simplify if the expression in the denominator has a root. This will happen if $\tilde r^2\left[A(\tilde r)+\left(\frac{\Kc\,\tilde r}{3}+\frac{\Cc}{\tilde r^2}\right)^2\right]$ has a double root. To calculate this we will now substitute the RN expression for $A(\tilde r) = 1 -\case{2M}{\tilde r}+\case{Q^2}{\tilde r^2}$, where $M$ is the mass of the BH and $Q$ its electric charge. The Schwarzschild version is simply obtained by setting $Q=0$. The expression is
\begin{equation}\label{ein:exproots}
%\tilde r^2\left[\left(1 -\frac{2M}{\tilde r}+\frac{Q^2}{\tilde r^2}\right)+\left(\frac{\Kc\,\tilde r}{3}+\frac{\Cc}{\tilde r^2}\right)^2\right] =
\frac{1}{\tilde r^2}\left[\frac{\Kc^2\tilde r^6}{9} + \tilde r^4  +2\left(\frac{\Kc\Cc}{3}- M\right) \tilde r^3 +Q^2\tilde r^2 +\Cc^2\right] .
\end{equation}
The double root of $\tilde r$ will appear if the discriminant of the polynomial between brackets vanishes.
This condition on the discriminant is the following,
{\small
\begin{eqnarray} \label{ein:discrim}
0 &=&\Cc^2 \Kc^2 \left(729 \Cc^6 \Kc^4+1944 \Cc^5 \Kc^5 M^3-1458 \Cc^5 \Kc^5 M Q^2 \right.  \nonumber \\ && \left. -2916 \Cc^5
   \Kc^3 M+81 \Cc^4 \Kc^6 Q^6-8748 \Cc^4 \Kc^4 M^4\right.  \nonumber \\ && \left.+7047 \Cc^4 \Kc^4 M^2 Q^2-243 \Cc^4
   \Kc^4 Q^4+3402 \Cc^4 \Kc^2 M^2\right.  \nonumber \\ && \left.-972 \Cc^4 \Kc^2 Q^2+1296 \Cc^4-324 \Cc^3 \Kc^5 M Q^6+13122 \Cc^3
   \Kc^3 M^5\right.  \nonumber \\ && \left.-14580 \Cc^3 \Kc^3 M^3 Q^2+3078 \Cc^3 \Kc^3 M Q^4+2916 \Cc^3 \Kc M^3\right.  \nonumber \\ && \left.-1944
   \Cc^3 \Kc M Q^2+378 \Cc^2 \Kc^4 M^2 Q^6-108 \Cc^2 \Kc^4 Q^8\right.  \nonumber \\ && \left.-6561 \Cc^2 \Kc^2 M^6+10935
   \Cc^2 \Kc^2 M^4 Q^2-4617 \Cc^2 \Kc^2 M^2 Q^4\right.  \nonumber \\ && \left.+603 \Cc^2 \Kc^2 Q^6-2187 \Cc^2 M^4+2916 \Cc^2
   M^2 Q^2-648 \Cc^2 Q^4\right.  \nonumber \\ && \left.+324 \Cc \Kc^3 M^3 Q^6-216 \Cc \Kc^3 M Q^8+54 \Cc \Kc M Q^6+16
   \Kc^4 Q^{12}\right.  \nonumber \\ && \left.-243 \Kc^2 M^4 Q^6+324 \Kc^2 M^2 Q^8-72 \Kc^2 Q^{10}-81 M^2 Q^6+81 Q^8\right)  ,
\end{eqnarray}
}%
and solving it for $\Cc$ will give the value of this parameter that simplifies the integral in \eref{ein:integralgamma}. An analytical expression for this critical value of $\Cc$ is obtained for $Q=0,M$ (Schwarzschild and extreme RN), but for a different value of the charge I have only found numerical solutions.

\begin{figure}[h!!]
\center
\includegraphics[width=0.9\linewidth]{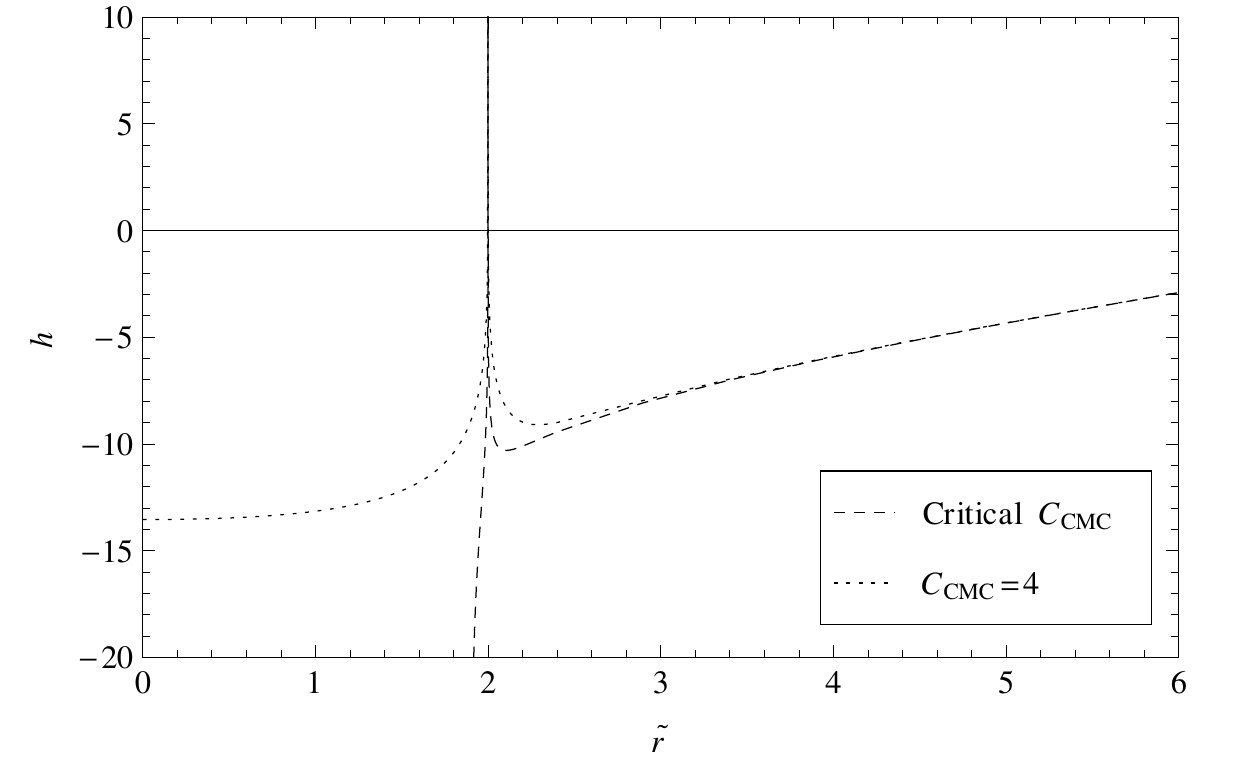}\vspace{-3ex}
\caption{Example of two numerically integrated height functions for the Schwarzschild spacetime with $M=1$ and $\Kc=-1$. The integration constant has been set in such a way that as $\tilde r\to\infty$ the Schwarzschild height function approaches the flat spacetime one \eref{ein:flath}. In the case with non-critical $\Cc=4$, the height function attains a finite value at the singularity $\tilde r=0$. This height function corresponds to diagram d) in \fref{fin:varC}. In the critical $\Cc=3.11$ case (corresponding to the outer slices in \fref{fin:varC} or to the $|\Kc|=1$ spacetime in \fref{fin:schwK}), the height function goes to $-\infty$ at the root $\tilde r = R_0=1.905$.} %Example of a height function for $M=1, \Cc=2, \Kc=-0.3$ plotted in terms of $\tilde r$.
\label{fin:ploth}
\end{figure}

In \cite{Baumgarte:2007ht}, under the assumptions of maximal slicing $\Kc=0$ and Schwarzschild $Q=0$ spacetime and using the value $\Cc=\case{3\sqrt{3}}{4}M^2$ obtained from \eref{ein:discrim}, \eref{ein:integralgamma} can be integrated and so an analytical expression for the relation between the Schwarzschild radial coordinate and the isotropic one is found. For the more general case of $\Kc\neq0$ considered here I have not been able to integrate it analytically. Examples of numerically integrated height functions for the Schwarzschild case are shown in \fref{fin:ploth}. The main difference between the maximal case and a finite value of $\Kc$ is that in the first case the height function remains constant for values of $\tilde r$ larger than the location of the peak (the horizon), whereas for a hyperboloidal slice $h(\tilde r)$ continues to grow linearly with the radius towards $\scri^+$.

Another important part in this calculation is related to the roots of \eref{ein:exproots}. For the critical value of $\Cc$ a double root $R_0$ is obtained. For a smaller value of $\Cc$ two different roots $R_1$ and $R_2$ appear, while for a larger value of $\Cc$ there are no real solutions to \eref{ein:exproots}. This effect can be seen in figure 1 in \cite{Tuite:2013hza}, where the expression in the square root in \eref{ein:integralgamma} is plotted against $\tilde r$ for these three different options for $\Cc$. A property of the root values $R_0$, $R_1$ and $R_2$ is that they are always located between the horizons $\tilde r_\pm=M\pm\sqrt{M^2-Q^2}$ of the BH.

There is also a minimal value of $\Cc$ that makes the slides intersect the BH horizon instead of the white hole one. This value can be easily determined by considering radial null rays on the horizon. In order to obtain an expression valid for both Schwarzschild and RN cases, the calculation will be performed on the outer horizon of RN. The condition is that at $\tilde r = M+\sqrt{M^2-Q^2}$ the radial component of the normal vector $\tilde n^a$ has to be negative, i.e. $\left.\tilde n^r\right|_{\tilde r=M+\sqrt{M^2-Q^2}}<0$. As can be deduced from \eref{ein:nexpr} and \eref{ein:ninteg}, $\tilde n^r=-\frac{\Kc\, r}{3} - \frac{\Cc}{r^2}$, so that the condition for intersecting the BH horizon translates to
\begin{equation} \label{ein:BHintersect}
\Cc>-\case{1}{3}\Kc \left(M + \sqrt{M^2-Q^2}\right)^3 .
\end{equation}
In the Schwarzschild case this reduces to the commonly used $\Cc>-\case{8}{3}\Kc M^3$.

\begin{figure}[htbp!!]
\center
\begin{tabular}{@{}r@{}@{}r@{}}
\vspace{1ex}&\\
a) $\qquad$&b) $\qquad$\vspace{-5ex}\\
\mbox{\includegraphics[width=0.5\linewidth]{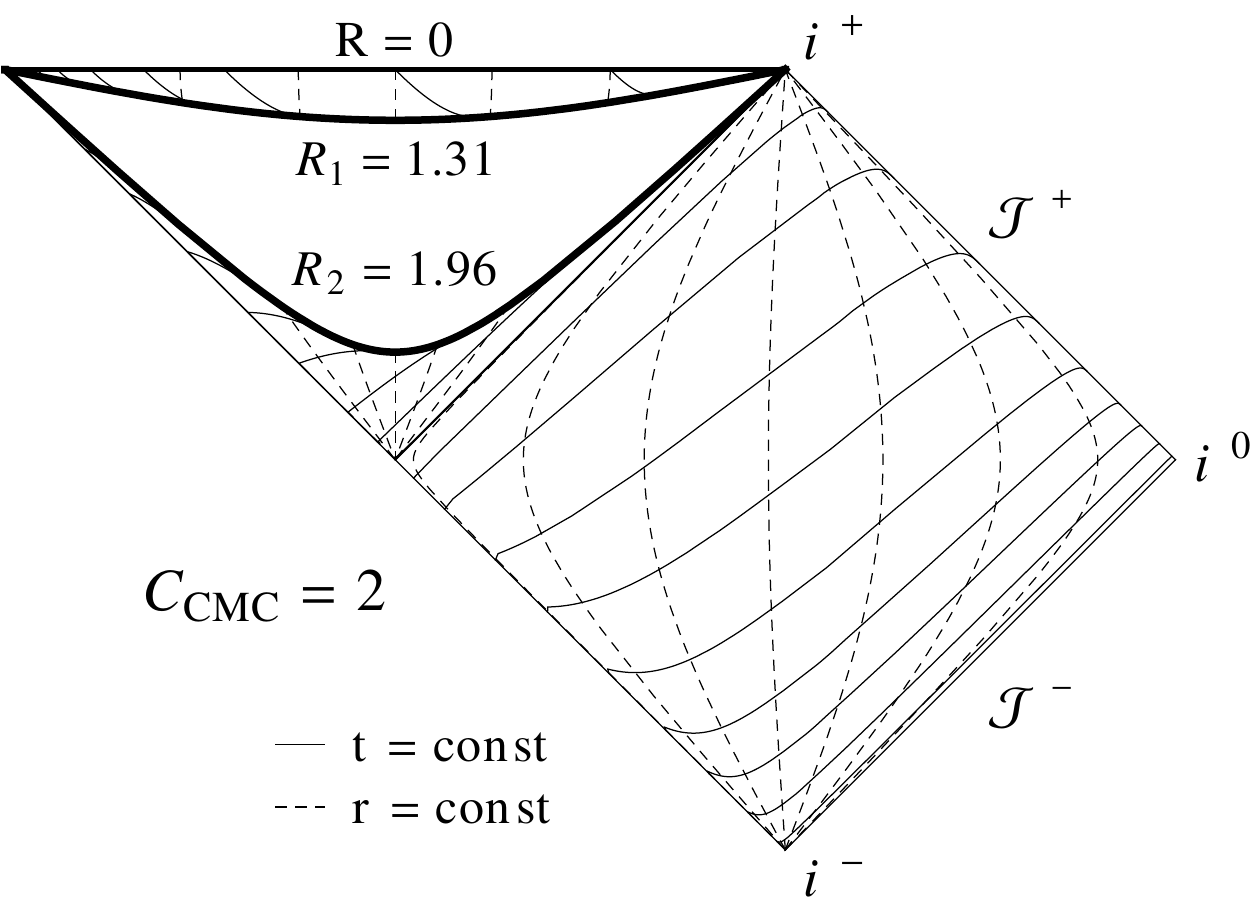}}&
\mbox{\includegraphics[width=0.5\linewidth]{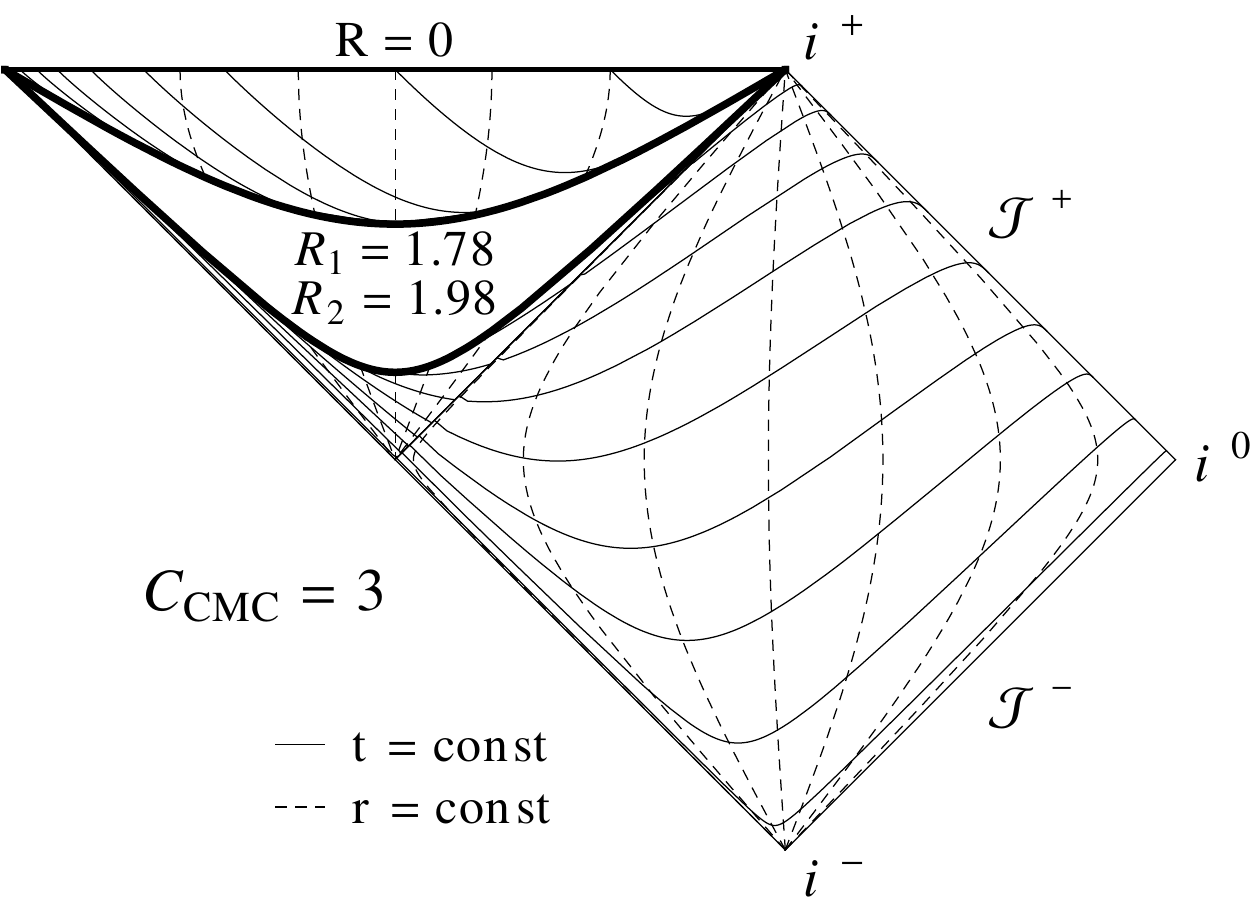}} \\
\vspace{1ex}&\\
c) $\qquad$&d) $\qquad$\vspace{-5ex}\\
\mbox{\includegraphics[width=0.5\linewidth]{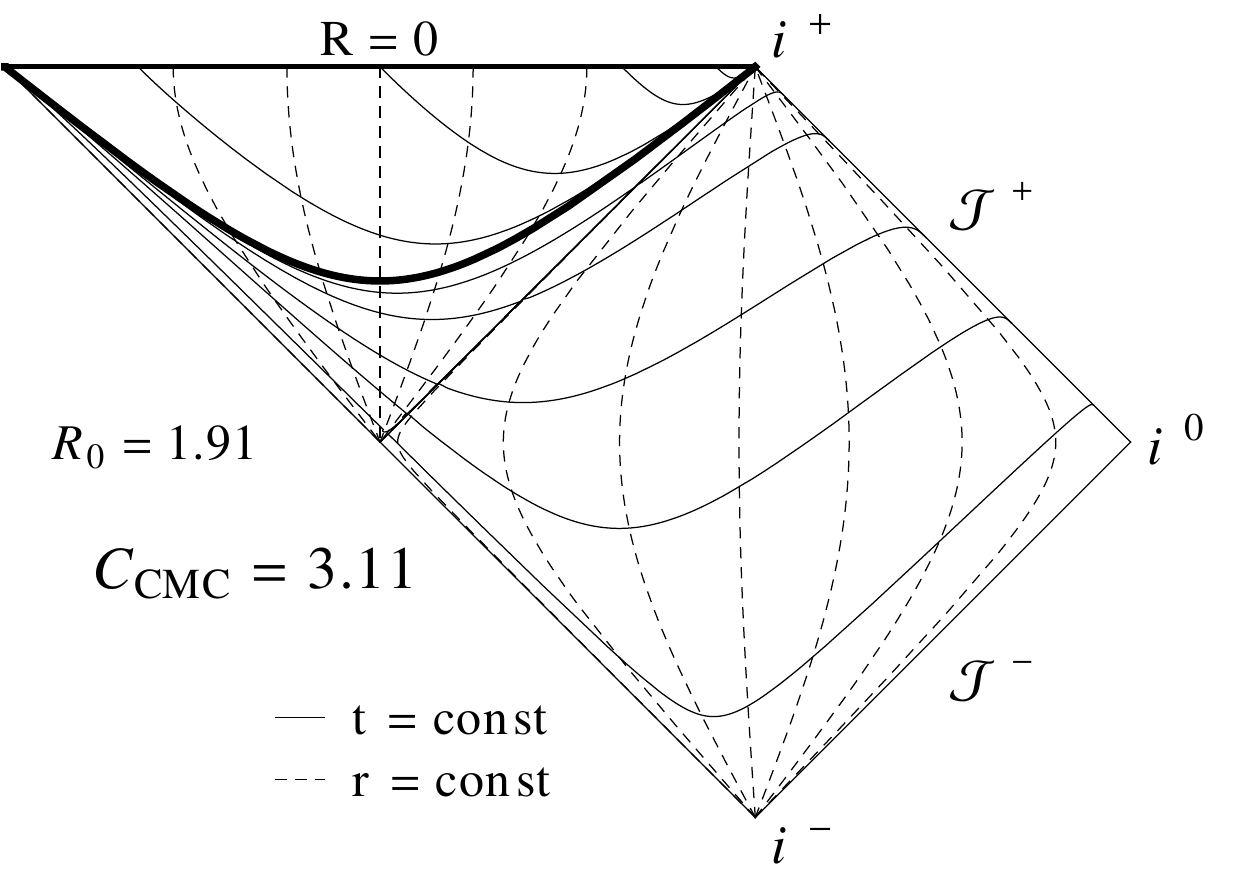}}&
\mbox{\includegraphics[width=0.5\linewidth]{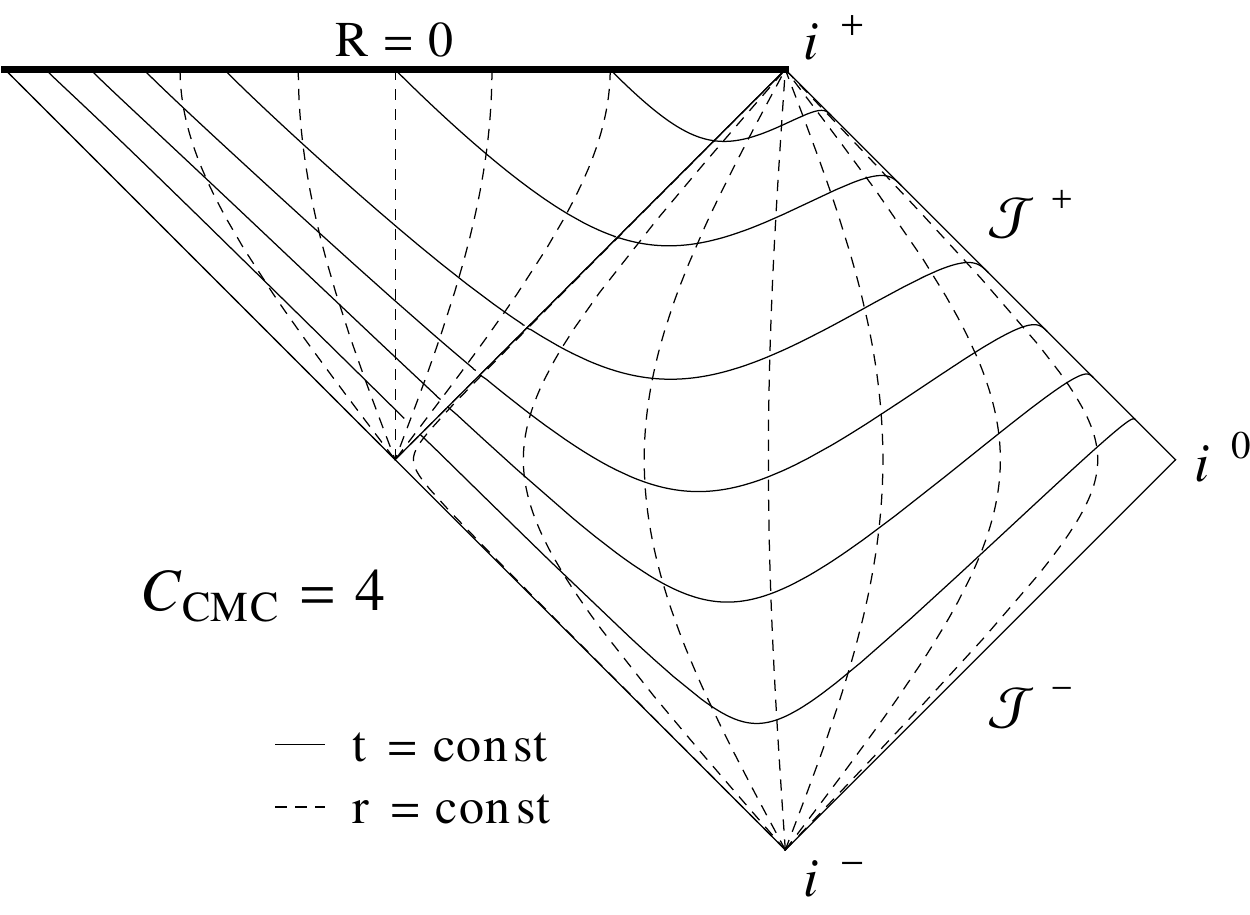}}
\end{tabular}
\caption{Carter-Penrose diagrams showing foliations with $M=1$, $Q=0$ (Schwarzschild), $\Kc=-1$ and different values of $\Cc$. Diagram c) corresponds to the trumpet geometry: here all outer slices reach the symmetric point to $i^+$ on the left, whereas the inner slices all reach $i^+$, and unlike the other cases, the inner and outer lines correspond to different slices. Compare to Penrose diagrams in \cite{Zenginoglu:2007jw,Ohme:2009gn}.}
\label{fin:varC}
\end{figure}

\begin{figure}[htbp!!]
\center
	\includegraphics[width=0.75\linewidth]{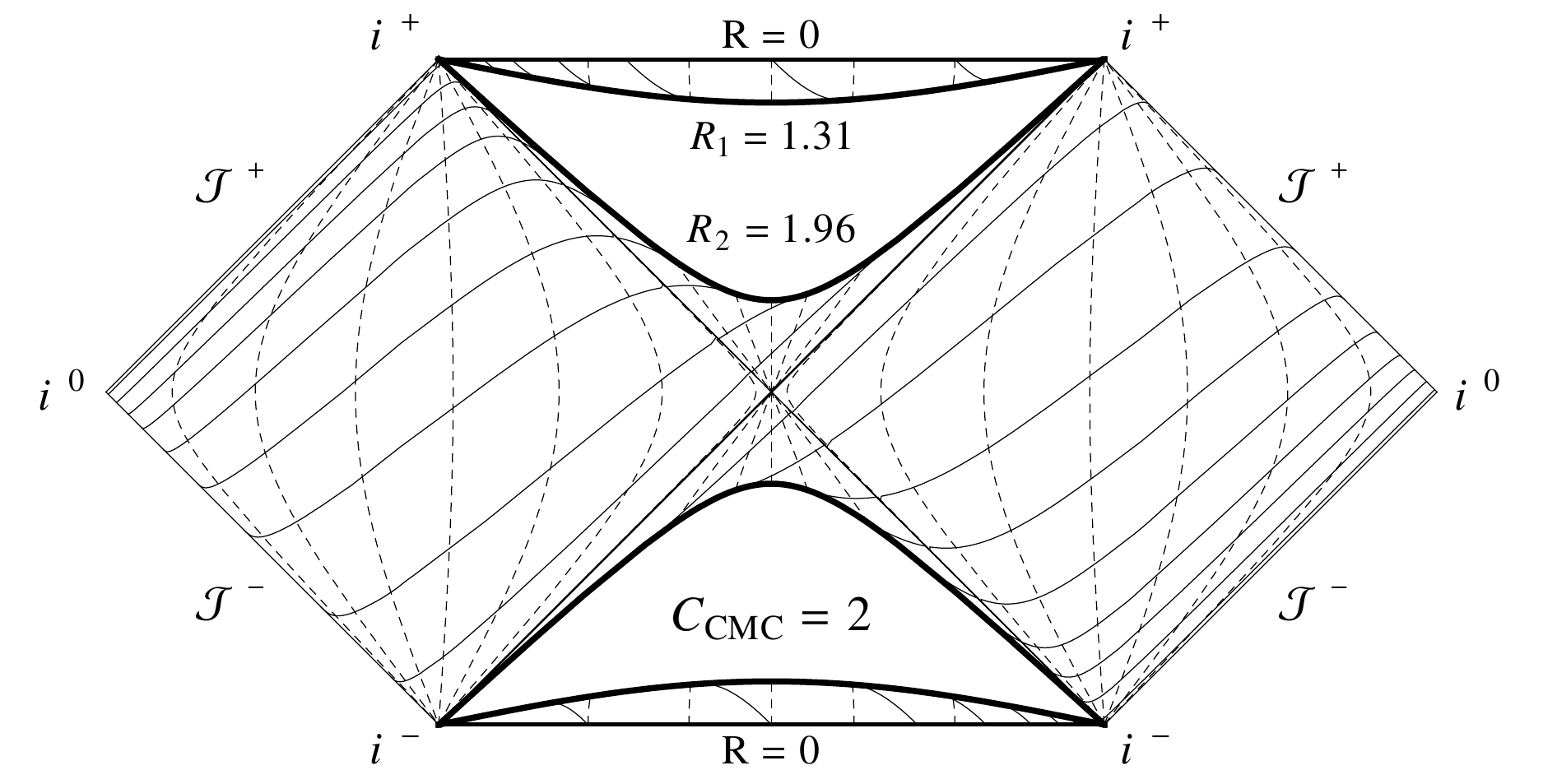}
	\vspace{-2ex}
\caption{Complete spacetime diagram for the case $\Cc=2$, where the slices intersect the white hole horizon.}
\label{fin:C2compl}
\end{figure}

The effect of the value of $\Cc$ on the geometry of Schwarzschild slices is shown in \fref{fin:varC}. To obtain these diagrams, the height function \eref{ein:heightp} has been integrated numerically choosing $M=1$, $Q=0$, $\Kc=-1$ and the indicated value of $\Cc$. The diagrams a) and b) correspond to the case where $\Cc$ is smaller than the critical value and two different roots $R_1$ and $R_2$ exist. In a) the value of $\Cc<8/3\approx2.67$ and the foliations go across the white hole horizon (see \fref{fin:C2compl} for a diagram including the complete spacetime), whereas in b) they do intersect the BH horizon. In both cases the region between the Schwarzschild radii $R_1$ and $R_2$ is not covered by any foliation (this is where the square root in \eref{ein:integralgamma} takes complex values), but note that the slices shown for $\tilde r<R_1$ are the continuation of those plotted for $\tilde r>R_2$. Case c) corresponds to the critical value of $\Cc$ and the slices connect $\scri^+$ with the trumpet and $i^+$ with the singularity at $R=0$. The value of the double root is $R_0$ and it is the innermost Schwarzschild radius that the outer slices can reach. However, it is infinitely far away from the apparent horizon measured in proper distance. Inversely, $R_0$ is also the largest radius achievable by the inner slices that reach the singularity. Note that the curves at $\tilde r<R_0$ and $\tilde r>R_0$ are actually different slices, unlike those in the a) and b) cases. In diagram d) the value of $\Cc$ is larger than the critical one and no real roots appear, so that all slices run from $\scri^+$ uninterrupted into the singularity. The effect of the value of $\Cc$ on non-extremal RN foliations is shown in an equivalent way in figures \ref{fin:RNvarC} and \ref{fin:RNCcompl}.


The reason why $\Cc=0$ was chosen in the flat spacetime case described in the previous subsection is that it is the only real solution to the zero discriminant condition \eref{ein:discrim} with $M=Q=0$.

The chosen value for $\Cc$ also affects the properties of the conformal factor $\aconf$, to be determined from \eref{ein:conflat}. If $\Cc$ is smaller than the critical value given by the discriminant condition, $\aconf$ is not defined near the origin of the isotropic radius, whereas if it is larger than the critical value, $\aconf$ will diverge close to $r=0$.

The trumpet solutions given by the critical values of $\Cc$ are especially interesting for the purpose of this work, because they provide regular initial values for the variables, the slices avoid the singularity so that the use of excision is not necessary and, as will be illustrated in the following spacetime examples, a compactification factor that maps the trumpet and $\scri^+$ to points in the integration domain can be found.

\subsubsection{Schwarzschild}

\begin{figure}[htbp!!]
\center
\begin{tabular}{@{}c@{}@{}c@{}}
\mbox{\includegraphics[width=0.5\linewidth]{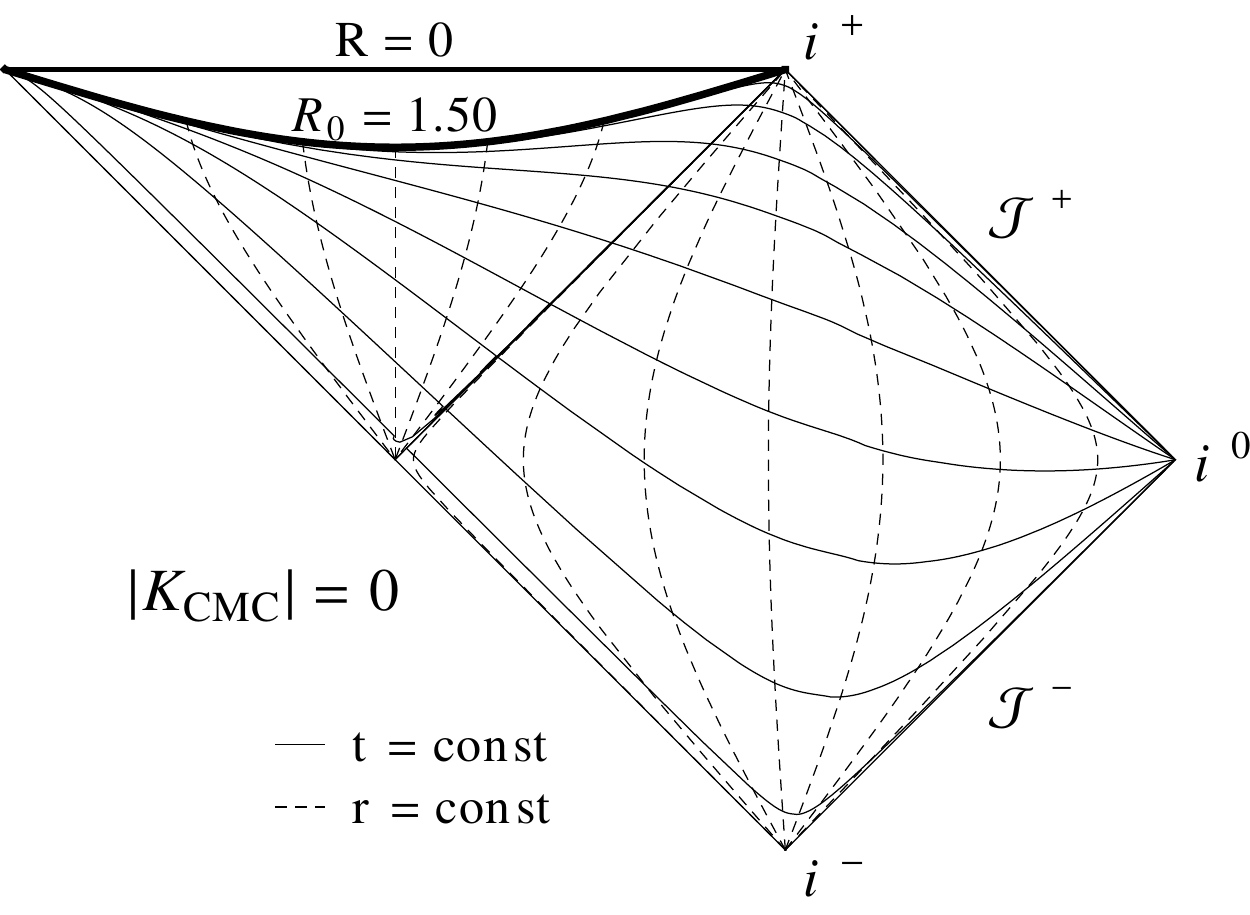}}&
\mbox{\includegraphics[width=0.5\linewidth]{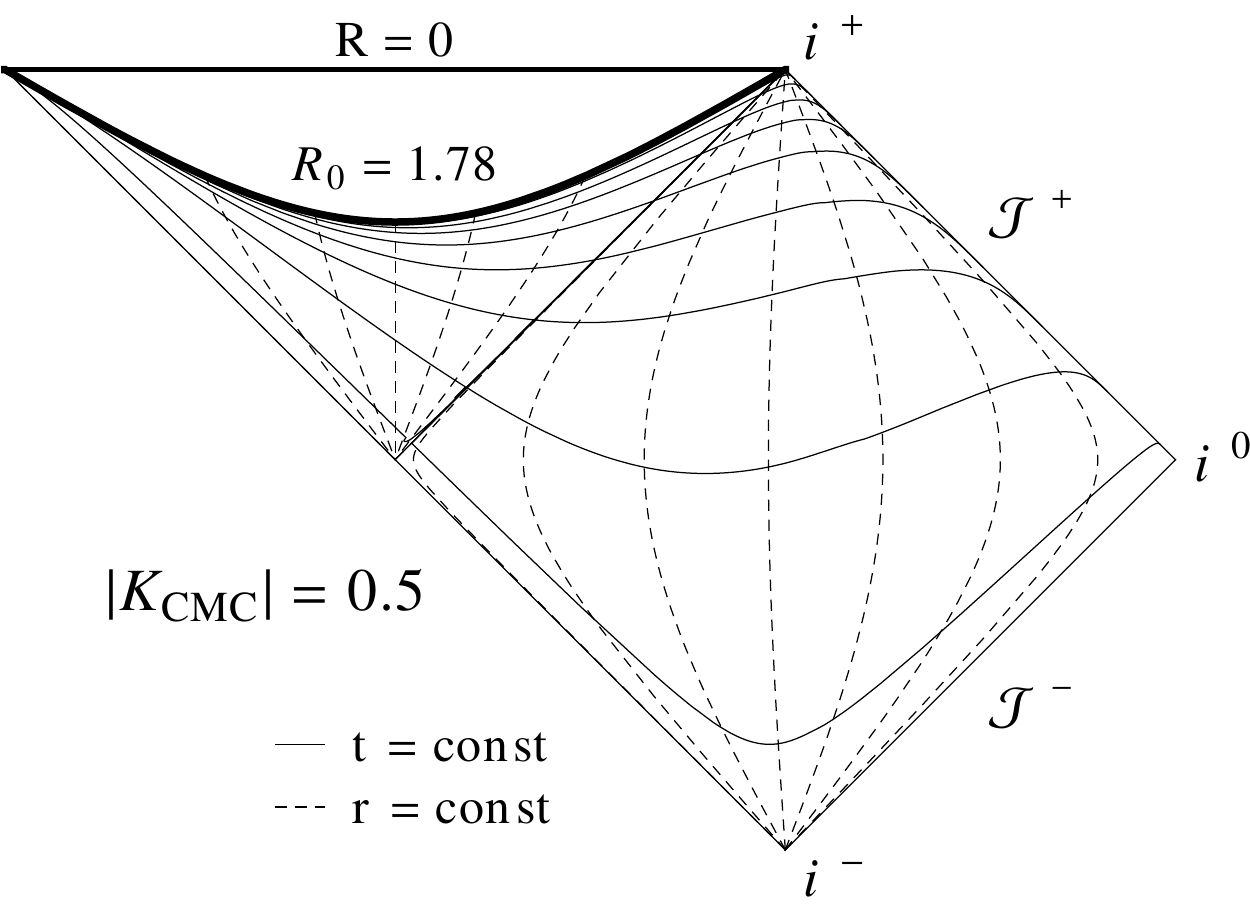}} \\
\mbox{\includegraphics[width=0.5\linewidth]{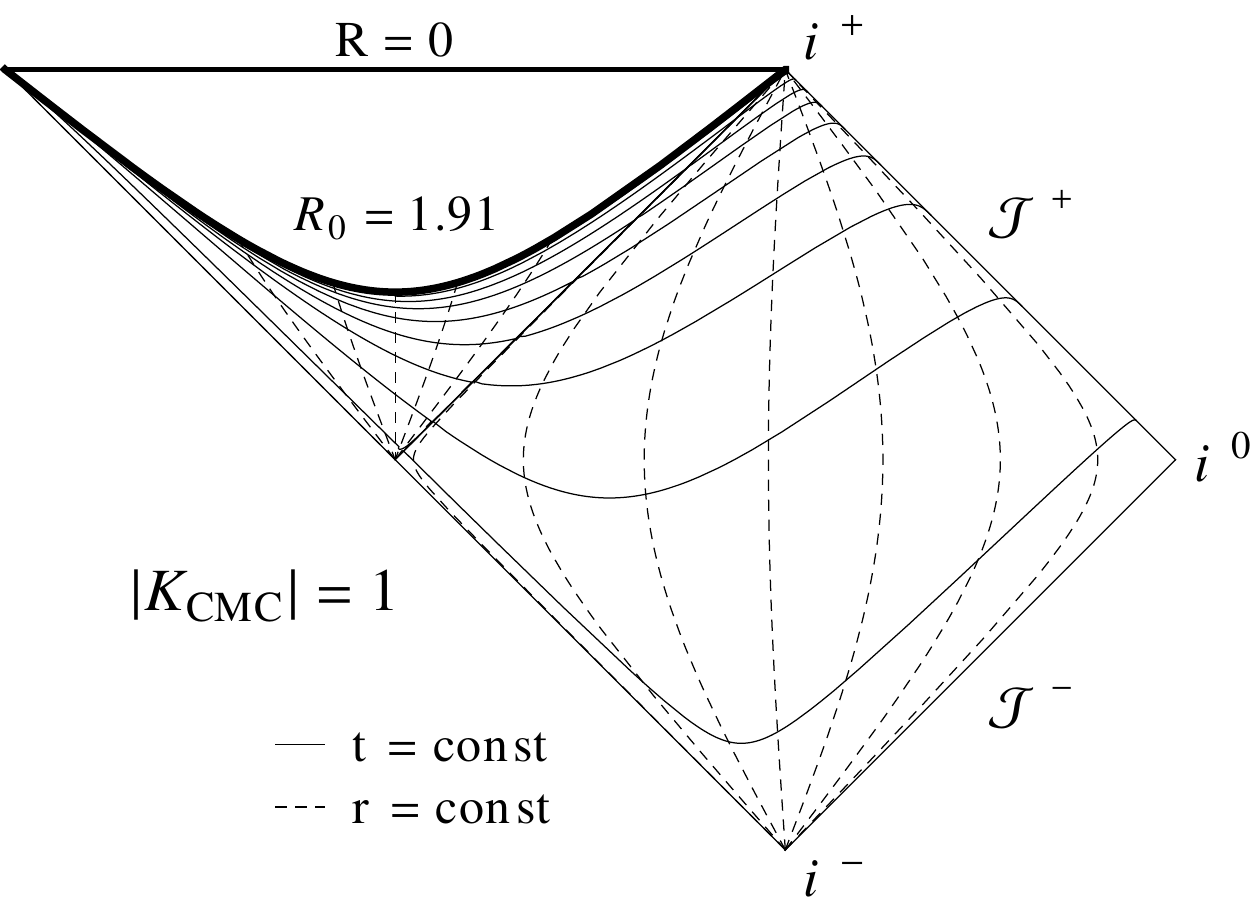}}&
\mbox{\includegraphics[width=0.5\linewidth]{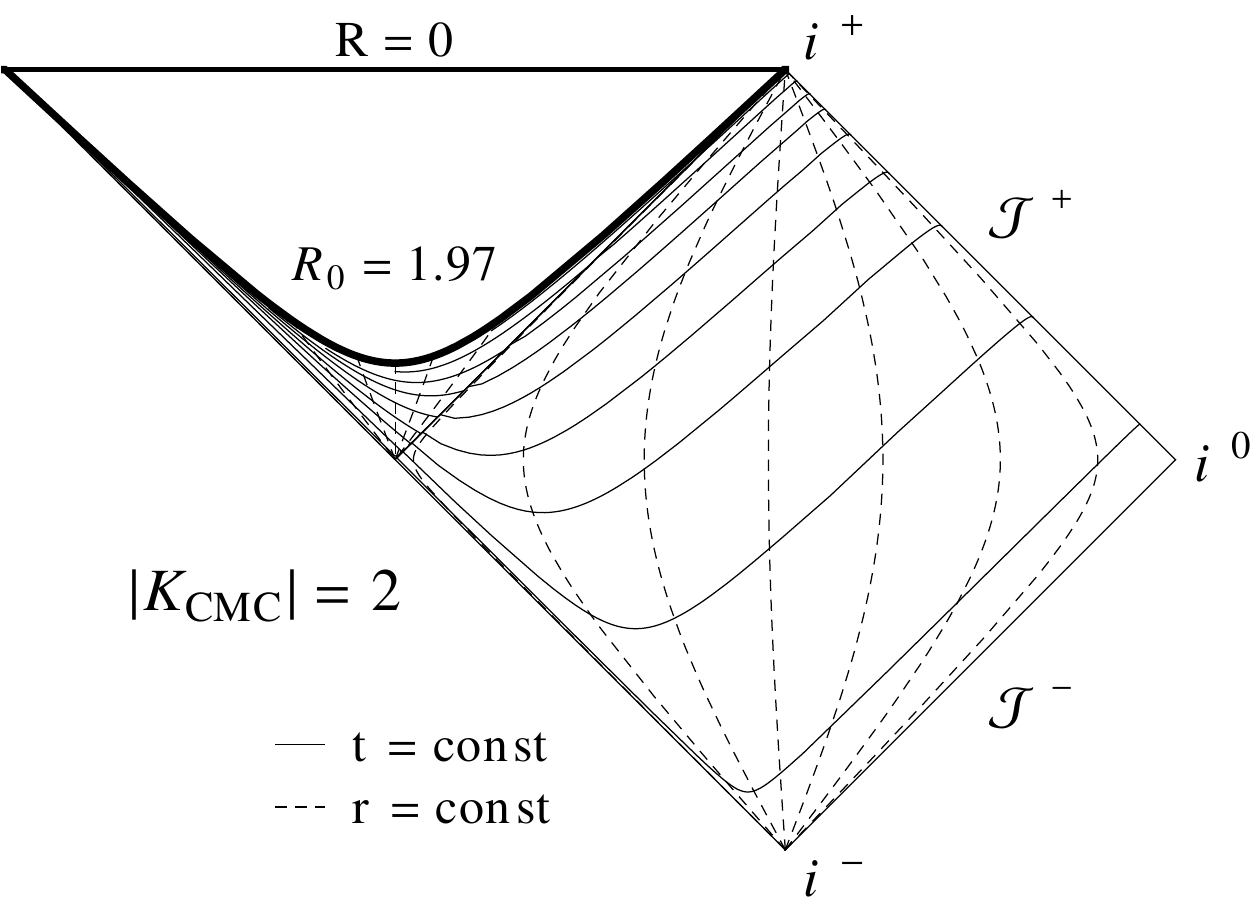}}
\end{tabular}
\caption{Penrose diagrams showing foliations with $M=1$ and critical $\Cc$ for different values of $K_{CMC}$. The thick line corresponds to the innermost reachable value of the Schwarzschild radial coordinate, where the trumpet is located. Only the outer slices of those shown in diagram c) in \fref{fin:varC} are displayed, because they are the ones physically relevant for our studies.}
\label{fin:schwK}
\end{figure}

The previous calculations apply to the Schwarzschild case for $Q=0$ and $A(\tilde r)=1-\case{2M}{\tilde r}$. In \fref{fin:schwK} we can see the trumpet solution for different values of $\Kc$, given by the corresponding critical value of $\Cc$ in each case. The smallest value of the Schwarzschild radial coordinate reachable is labeled by $R_0$ in each diagram.

\begin{figure}[htbp!!]
\center
	\includegraphics[width=0.6\linewidth]{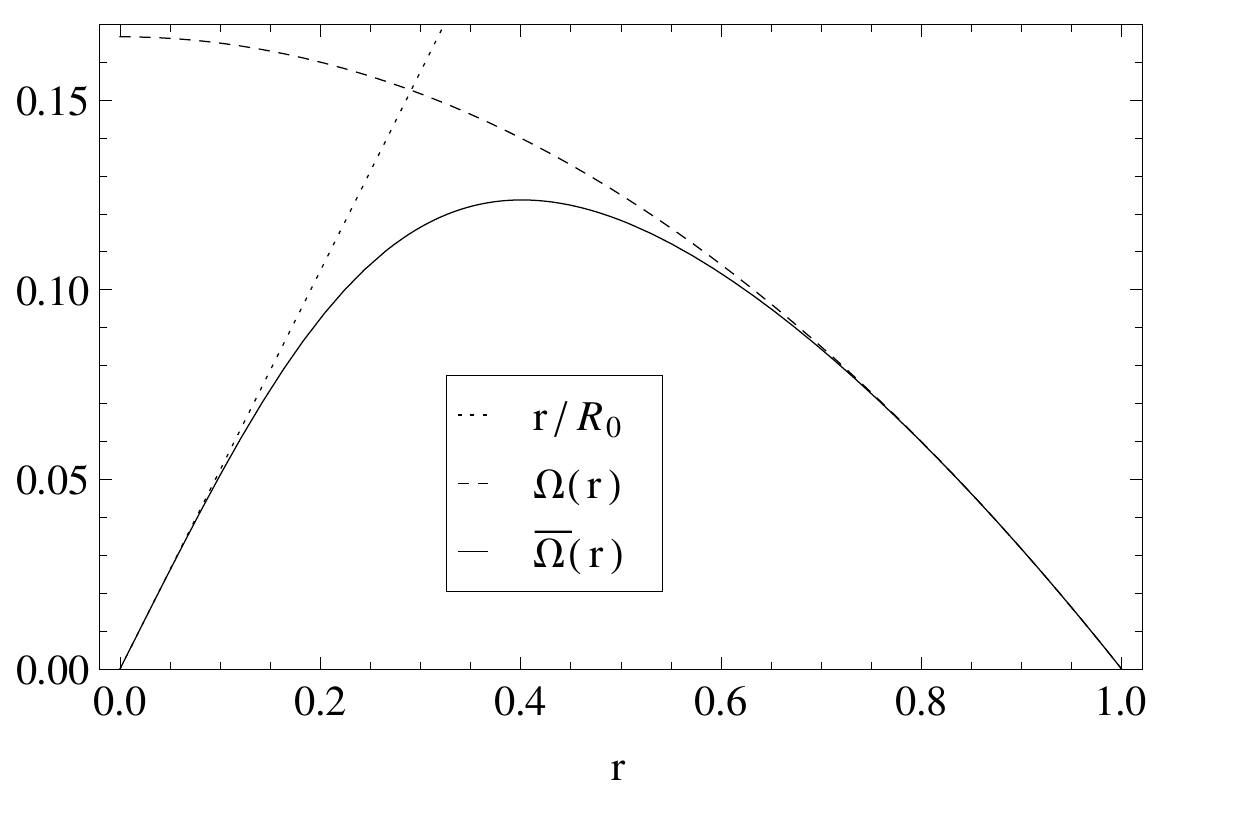}
	\vspace{-2ex}
\caption{Compactification factor $\aconf$ for $K_{CMC}=-1$, $M=1$ and critical $\Cc$. Near the origin it is linear in $r$ and near $\rscri=1$ it behaves like $\Omega$.}\label{fin:omegas}
\end{figure}

The compactification factor $\aconf$ in terms of $r$ is obtained by solving \eref{ein:conflat} by means of numerical techniques as described in subsection \ref{sn:aconf}. A useful property of $\aconf$ is that for small values of the isotropic radius $r$ it behaves linearly with the inverse of $R_0$ as proportionality factor: $\aconf\to\case{r}{R_0}$ for $r\to 0$, as can be deduced taking the limit in \eref{ein:dsigmacoeff}. %\upda{True, but something missing in the explanation: how do we know that $r=0\Leftrightarrow \tilde r=R_0$?}
Close to $r=\rscri$ is is expected to behave in the same way as flat spacetime's $\aconf$, that is \eref{ein:aconfflat}. The metric conformal factor $\Omega$ only has to vanish at $\scri^+$, so that the choice made for flat spacetime \eref{ein:omega} is also made here. In \fref{fin:omegas} the numerically determined $\aconf$ for the trumpet geometry with $K_{CMC}=-1$ and $M=1$ is displayed together with its approximation at the origin and the chosen expression for the conformal factor $\Omega$.

\begin{figure}[htbp!!]
\center
	\includegraphics[width=0.6\linewidth]{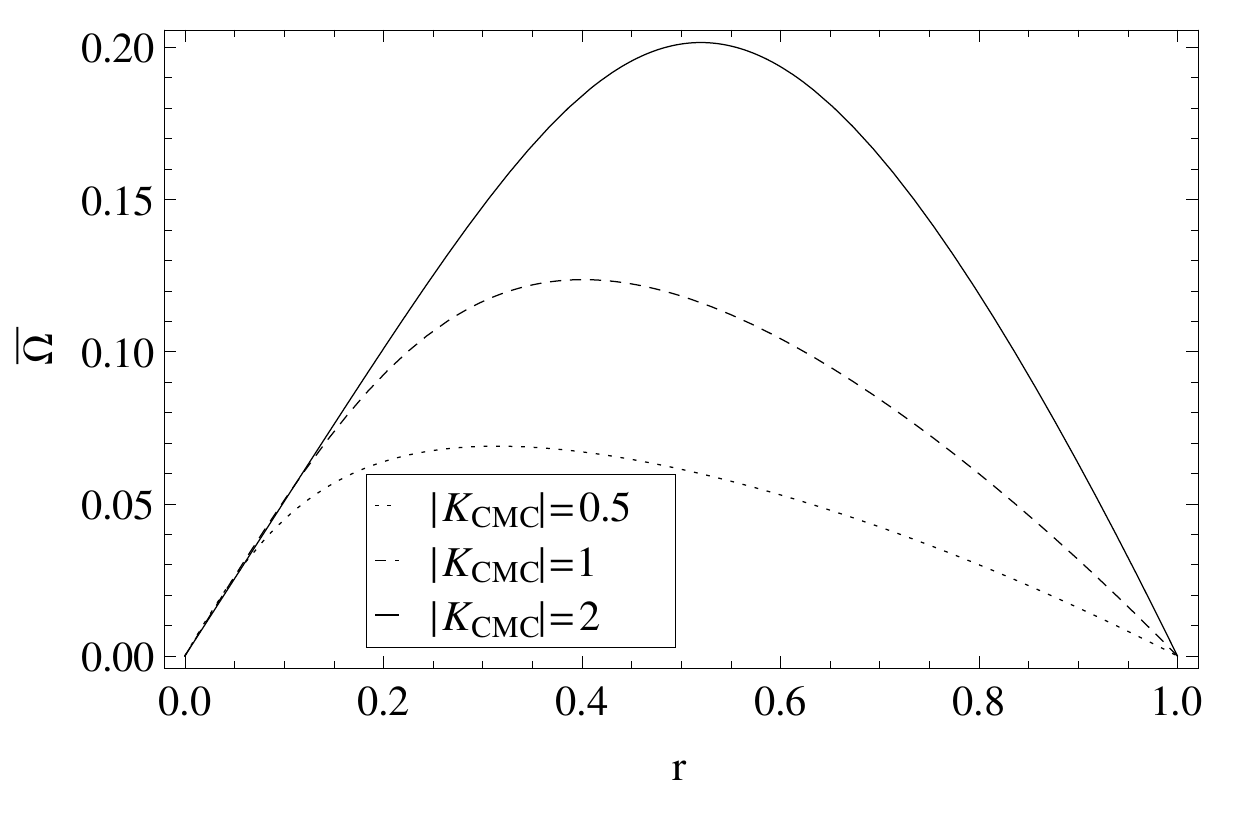}
	\vspace{-2ex}
\caption{Compactification factors $\aconf$ for $M=1$ and different values of $K_{CMC}$ (and their corresponding critical $\Cc$s).}\label{fin:Komegas}
\end{figure}

Figure \ref{fin:Komegas} shows the change in $\aconf$ due to the variation in the value of $\Kc$ (and accordingly of $\Cc$). The larger $|\Kc|$, the steeper the slope close to $\rscri$, the larger the maximum value achieved by $\aconf$ and the larger the isotropic radius where it is located. %and the larger the value of the isotropic radius where this maximum is located. %\upda{show plot with change in mass?}

\subsubsection{Reissner-Nordström}

%K C considerations in \cite{Buchman:2009ew} -> minimal surface for trumpet  (critical minimal surface $f^2=0$ and $\partial_rf^2=0$ equivalent to having a double root for the trumpet location), also similar convention for $\aconf$

Using $A(\tilde r)=1-\case{2M}{\tilde r}+\case{Q^2}{\tilde r^2}$ and choosing a value for the charge $Q\in(0,M]$ will give us RN initial data. The treatment is very similar to Schwarzschild, but the extreme RN case (Q=M) requires special attention. In the extreme case the discriminant equation for $\Cc$ simplifies substantially, giving as solution
\begin{equation}\label{ein:ernCcmc}
\Cc=-\case{1}{3}\Kc M^3 \qquad \textrm{for} \qquad Q=M .
\end{equation}
This value also coincides with the limit in $\Cc$ such that the slices intersect the BH, meaning that at the horizon $\tilde n^r=0$. The expression \eref{ein:exproots} will only provide a real root choosing \eref{ein:ernCcmc} (so no $R_1$ and $R_2$ will exist) and it is the double root $R_0=M$. This is to be expected from the fact that the roots are always located between the horizons of the BH, and in the extreme RN case both horizons coincide at $\tilde r_\pm=M$, so $R_0=M$ necessarily.

\begin{figure}[htbp!!]
\center
\includegraphics[width=0.8\linewidth]{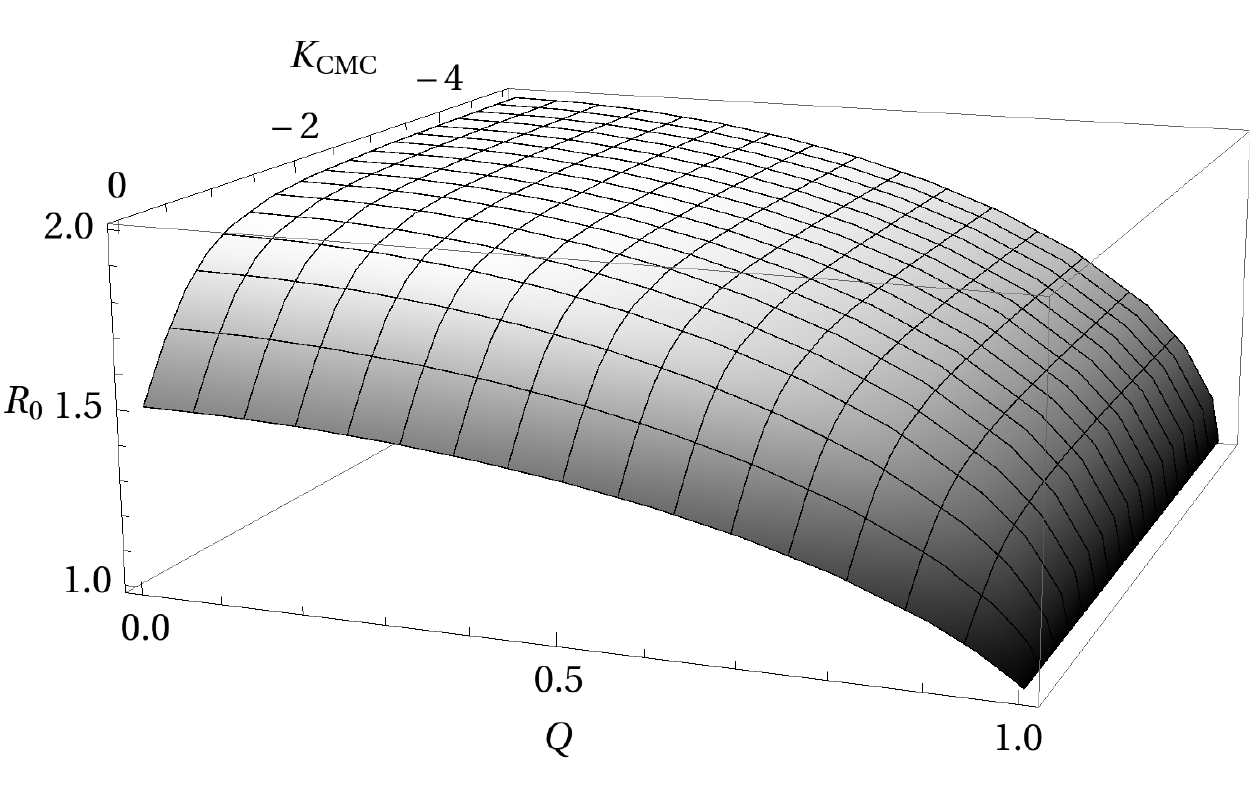}
\caption{Innermost achievable value of the Schwarzschild radial coordinate (the double root) $R_0$ for $M=1$ as a function of $\Kc$ and $Q$.}
\label{fin:R0vsKQ}
\end{figure}

In \fref{fin:R0vsKQ} the dependence of the critical root $R_0$ in terms of the hypersurface parameter $\Kc$ and the charge $Q$ can be seen, for a fixed $M=1$. In the extreme RN case, $R_0$ is equal to $M$ independently of the value of $\Kc$. For maximal slicing ($\Kc=0$) this value increases as $Q$ decreases and in the Schwarzschild case the value $R_0=\case{3}{2}M$ \cite{PhysRevD.7.2814,Malec:2003dq,Baumgarte:2007ht} is obtained. Finally, as $K_{CMC}\to-\infty$ the value of $R_0$ comes closer to $\tilde r_+=M+\sqrt{M^2-Q^2}$ ($2M$ for Schwarzschild).

\begin{figure}[htbp!!]\label{omegas}
\center
	\includegraphics[width=0.7\linewidth]{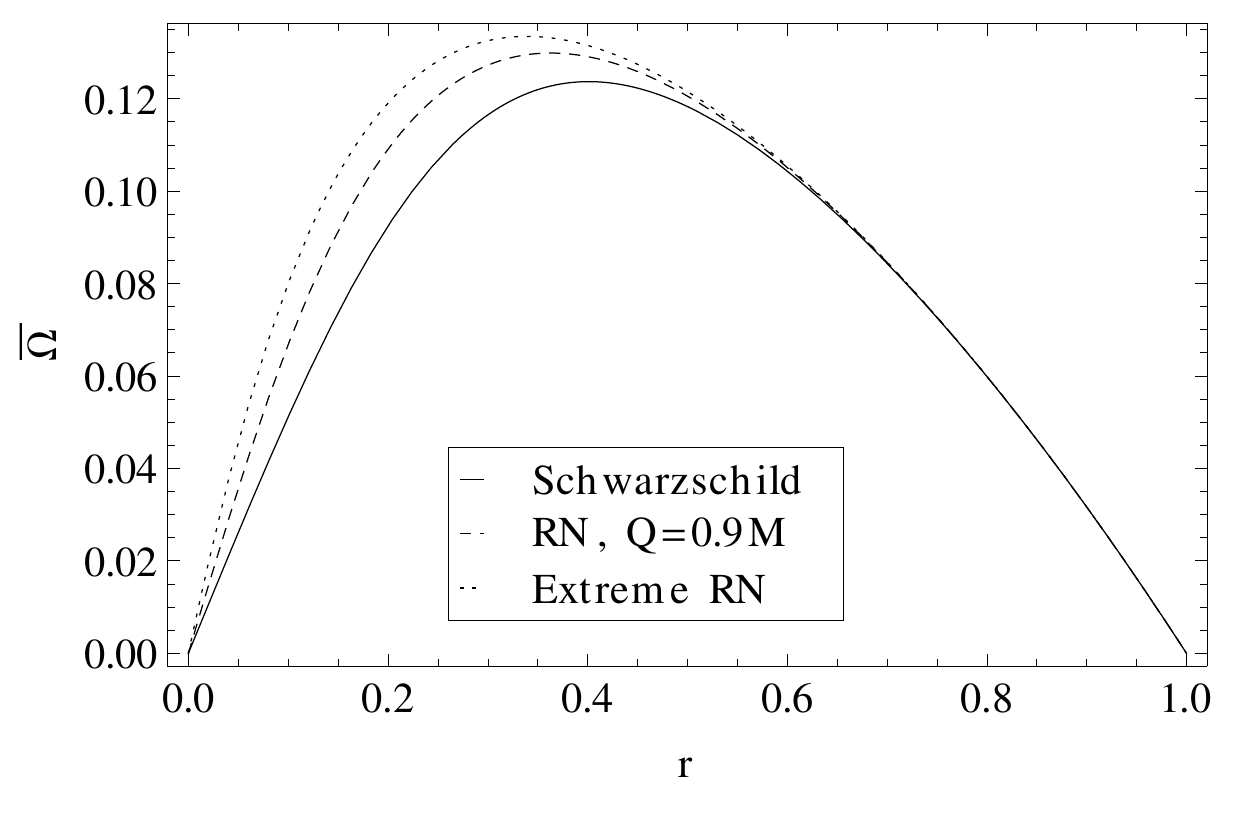}\vspace{-2ex}
\caption{Compactification factor $\aconf$ with $M=1$ and $K_{CMC}=-1$ for Schwarzschild, RN with $Q=0.9$ and extreme RN CMC trumpet geometries.}
\end{figure}

The numerically calculated compactification factors $\aconf$ corresponding to RN are qualitatively the same as in the Schwarzschild case. A comparison among Schwarzschild, RN with $Q=0.9$ and extreme RN compactification factors is shown in \fref{omegas}. The parameters $M=1$ and $K_{CMC}=-1$ are fixed and the effect between the curves is given by the variation in the charge $Q$ (and accordingly the critical value of $\Cc$). The slope of the curves near the origin becomes steeper as a larger $Q$ is chosen. This is due to the decreasing value of $R_0$ (inverse of the slope) with increasing $Q$.

\begin{figure}[h!!!!]
\center
\begin{tabular}{@{}c@{}@{}c@{}}
\includegraphics[width=0.5\linewidth]{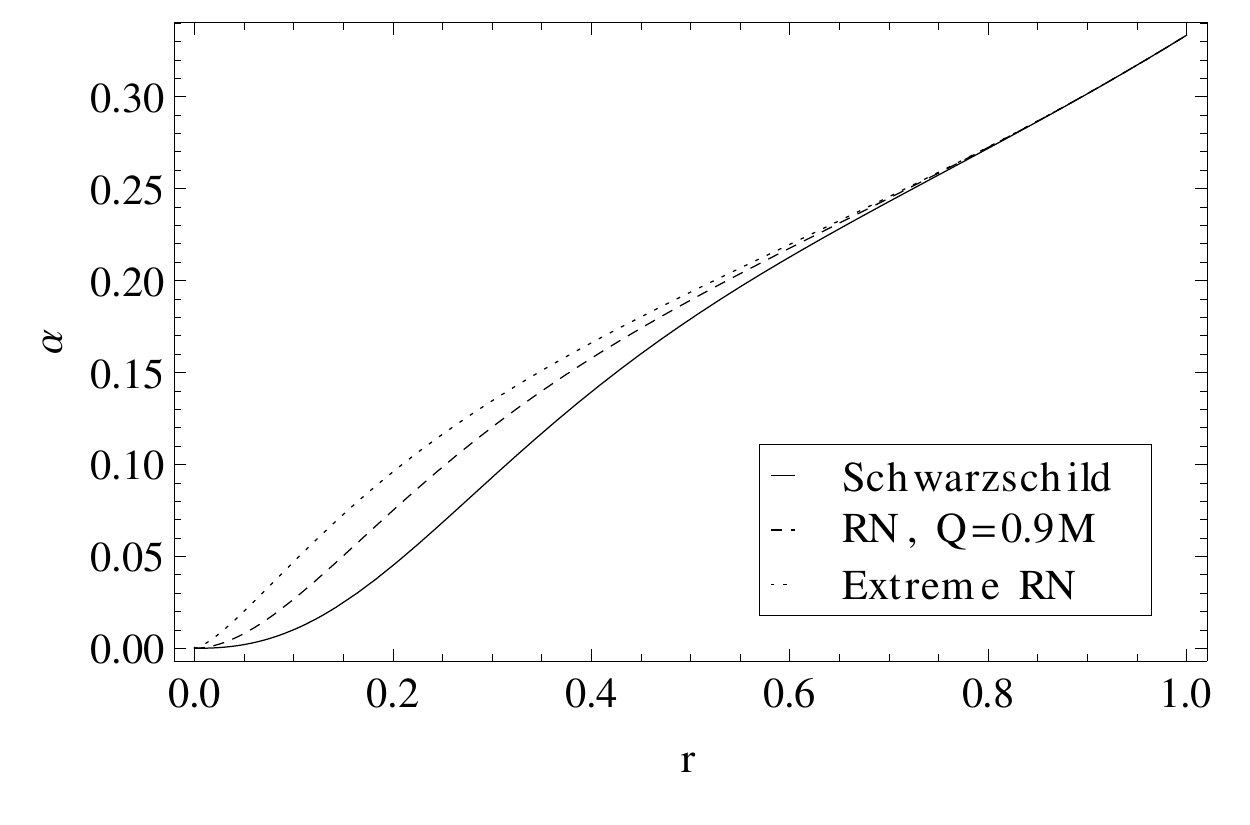}&
\includegraphics[width=0.5\linewidth]{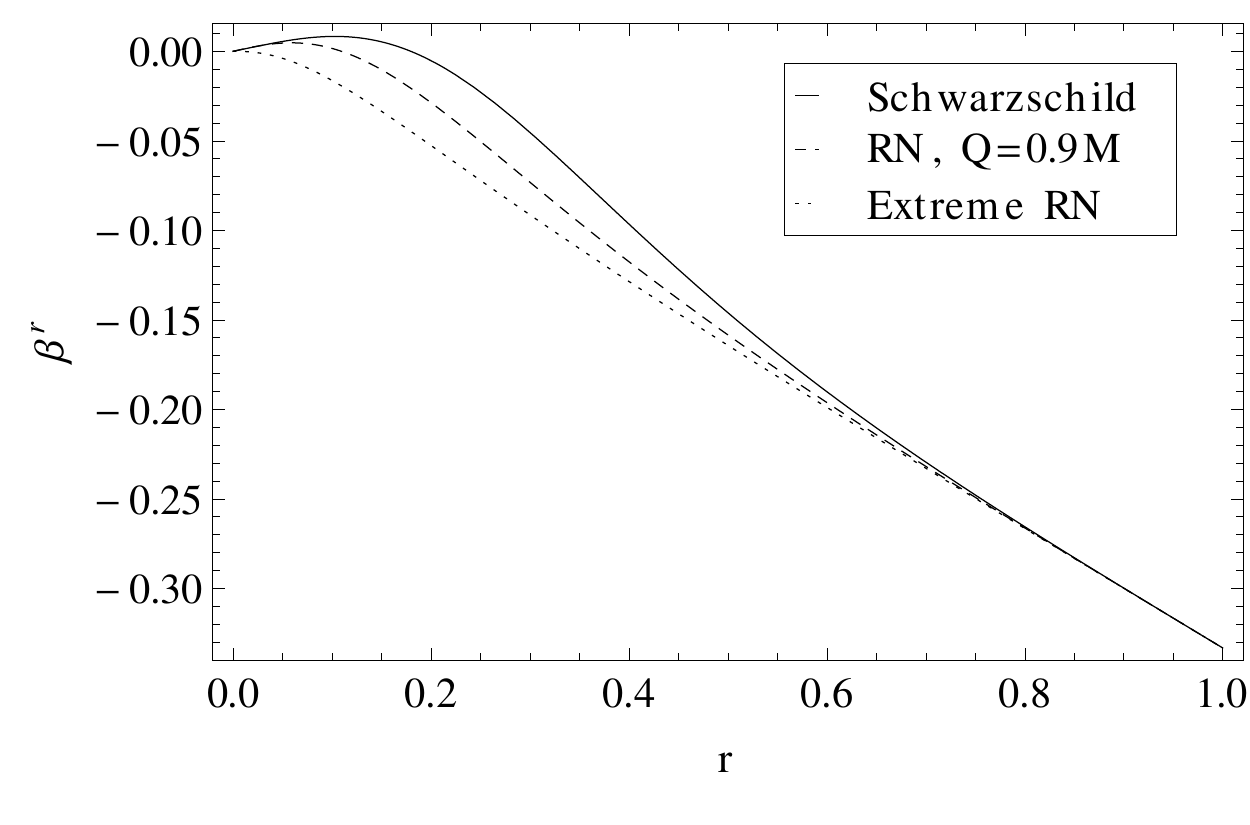} \\
\vspace{-3.8cm}\\&\includegraphics[width=0.20\linewidth]{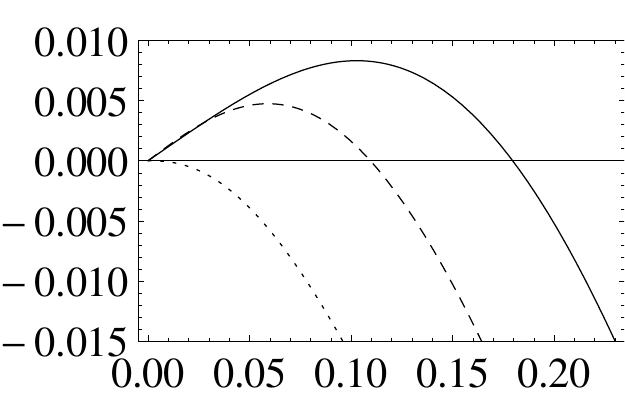}$\quad\qquad\qquad$\\ \vspace{1.5ex}
\end{tabular}
\caption{Trumpet values for the gauge variables $\alpha$ and $\beta^r$ for Schwarzschild, RN with $Q=0.9$ and extreme RN geometries. The detail in $\beta^r$'s plot shows how the shift in the extreme RN case is never positive, unlike the other cases.}
\label{fin:alphabetar}
\end{figure}

Another interesting effect of the extremality of the $Q=M$ case is that the cylindrical infinity of the trumpet and the BH horizon are mapped to the same point $r=0$ of the isotropic radius. This can be easily seen by looking at the profile of the shift $\beta^r$, displayed on the right in \fref{fin:alphabetar}. In the Schwarzschild and non-extreme RN cases the shift is positive at the horizon (mapped to $r_{Schw}\approx0.13$ and $r_{RN+}\approx0.071$ for $Q=0.9$ respectively), and this is actually something necessary if excision has to be used. However, in the extreme case the horizon is mapped to the same point as the origin and the shift never becomes positive.

The curves corresponding to the Schwarzschild case of the $\alpha$ and $\beta^r$ quantities in \fref{fin:alphabetar} compare to figure 5 in \cite{Hannam:2006xw} and figure 2 in \cite{Baumgarte:2007ht}, with the difference that in our case the data are compactified on a hyperboloidal slice.

%\upda{add equivalent diagrams to \fref{fin:schwK} for eRN - working on it!!}

\begin{figure}[htbp!!]
\center
\begin{tabular}{@{}r@{}@{}r@{}}
\vspace{1ex}&\\
a) $\qquad$&b) $\qquad$\vspace{-5ex}\\
\mbox{\includegraphics[width=0.5\linewidth]{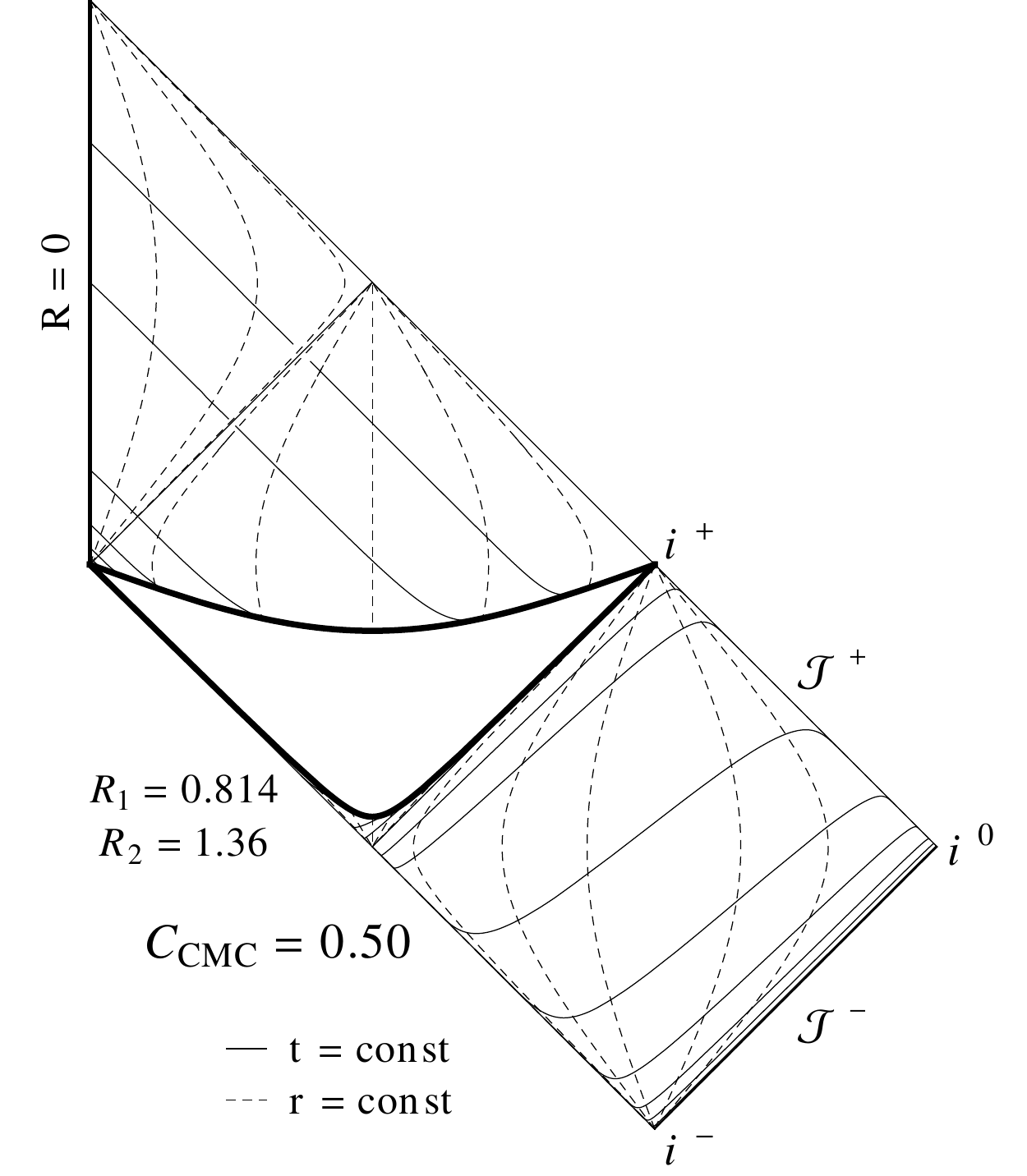}}&
\mbox{\includegraphics[width=0.5\linewidth]{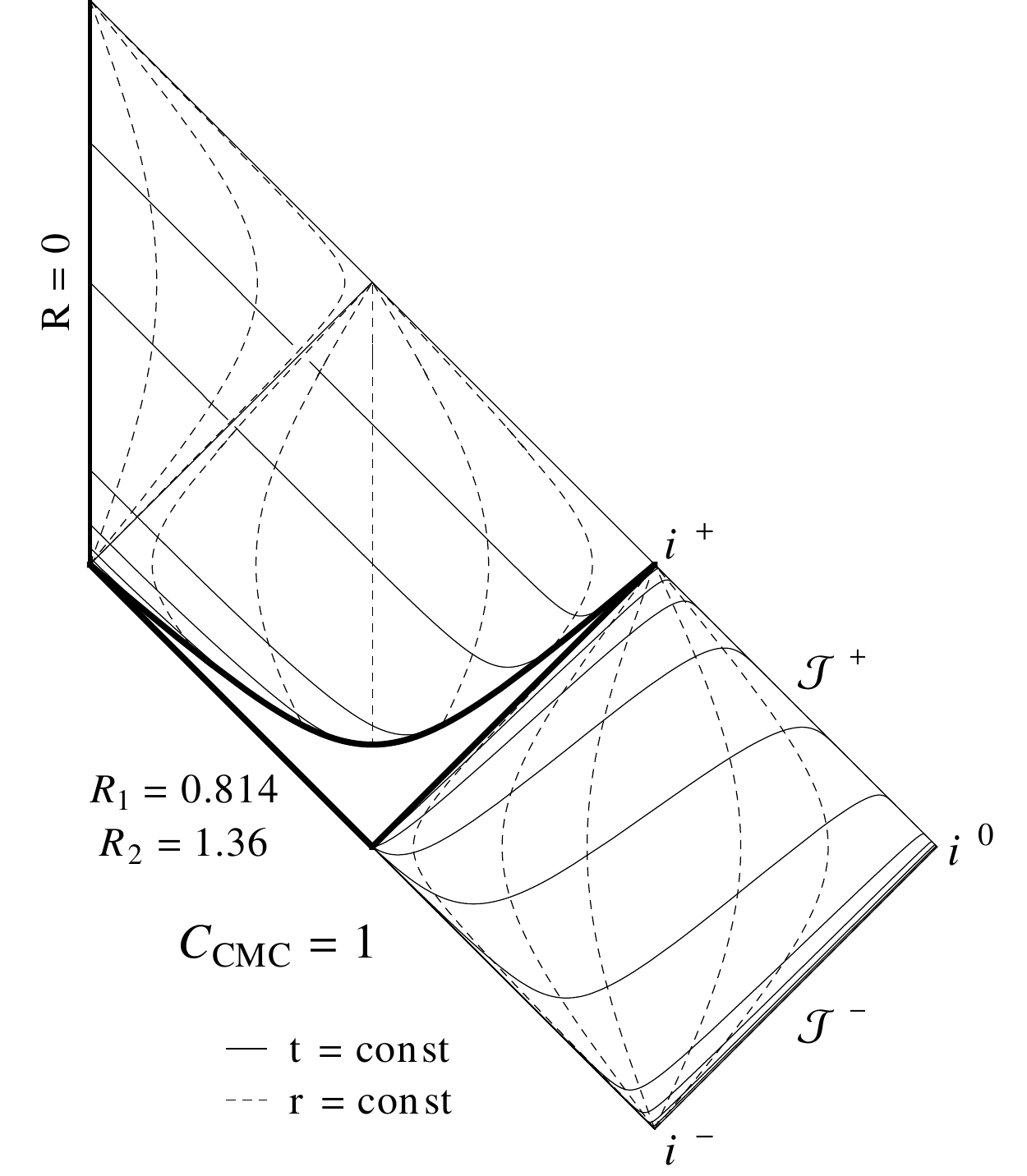}} \\
\vspace{1ex}&\\
c) $\qquad$&d) $\qquad$\vspace{-5ex}\\
\mbox{\includegraphics[width=0.5\linewidth]{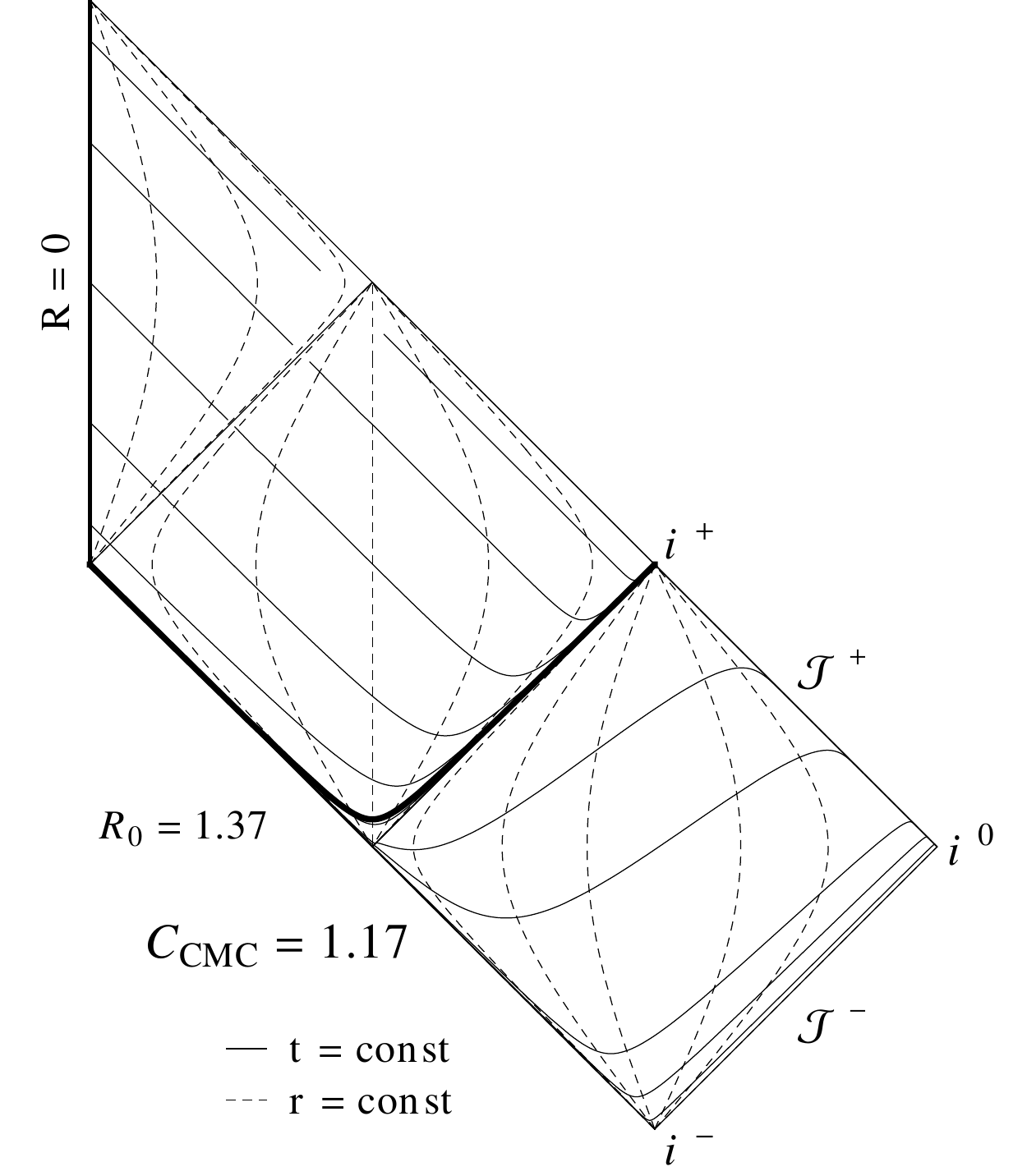}}&
\mbox{\includegraphics[width=0.5\linewidth]{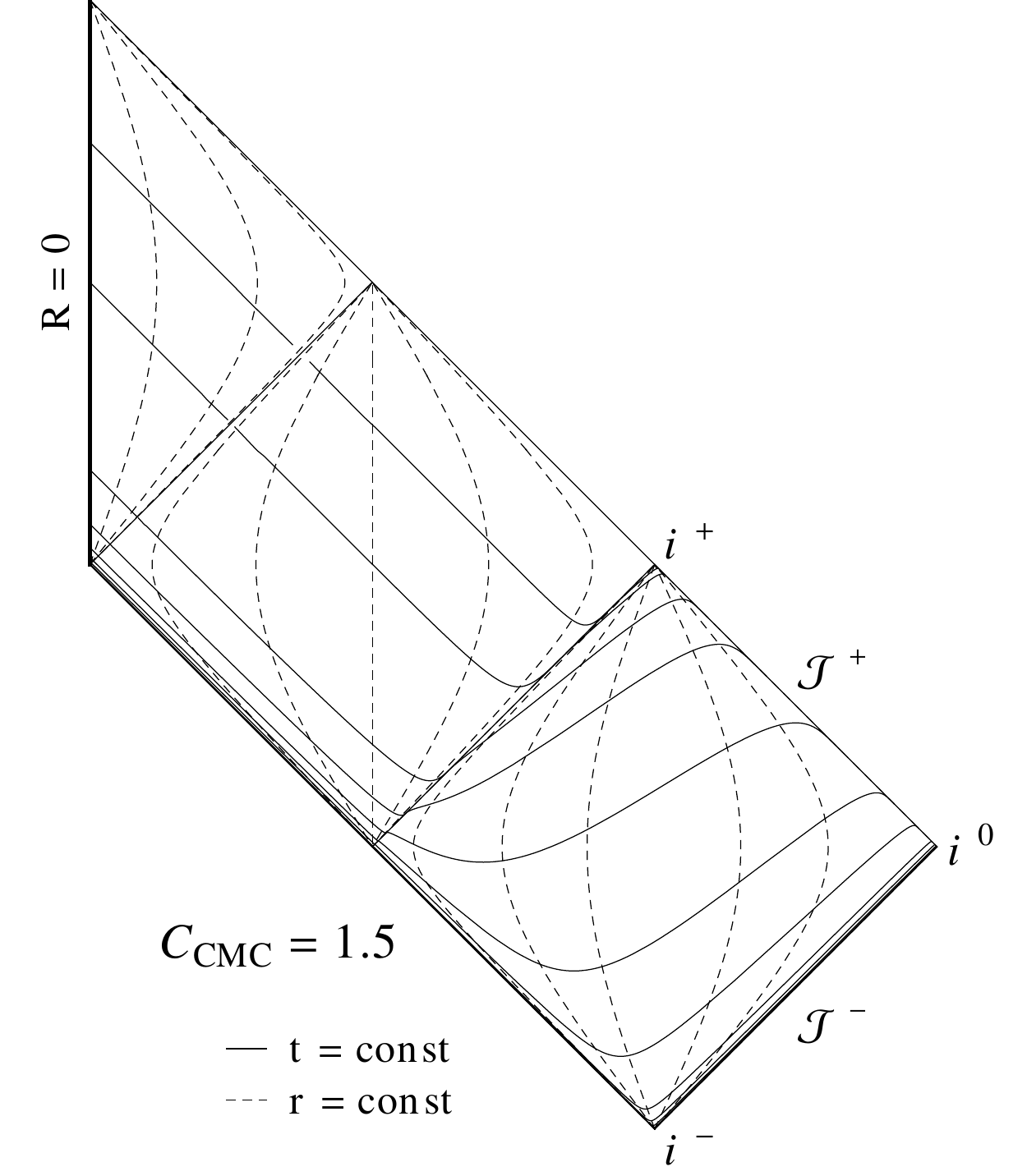}}
\end{tabular}
\caption{Penrose diagrams showing foliations with $M=1$, $Q=0.9$, $\Kc=-1$ and different values of $\Cc$. Diagram c) corresponds to the trumpet geometry and in the same way as with the Schwarzschild case in \fref{fin:varC}, the inner and outer foliation lines correspond to different slices. Compare to Penrose diagrams in \cite{Tuite:2013hza}.}
\label{fin:RNvarC}
\end{figure}

\begin{figure}[htbp!!]
\center
	\includegraphics[width=0.65\linewidth]{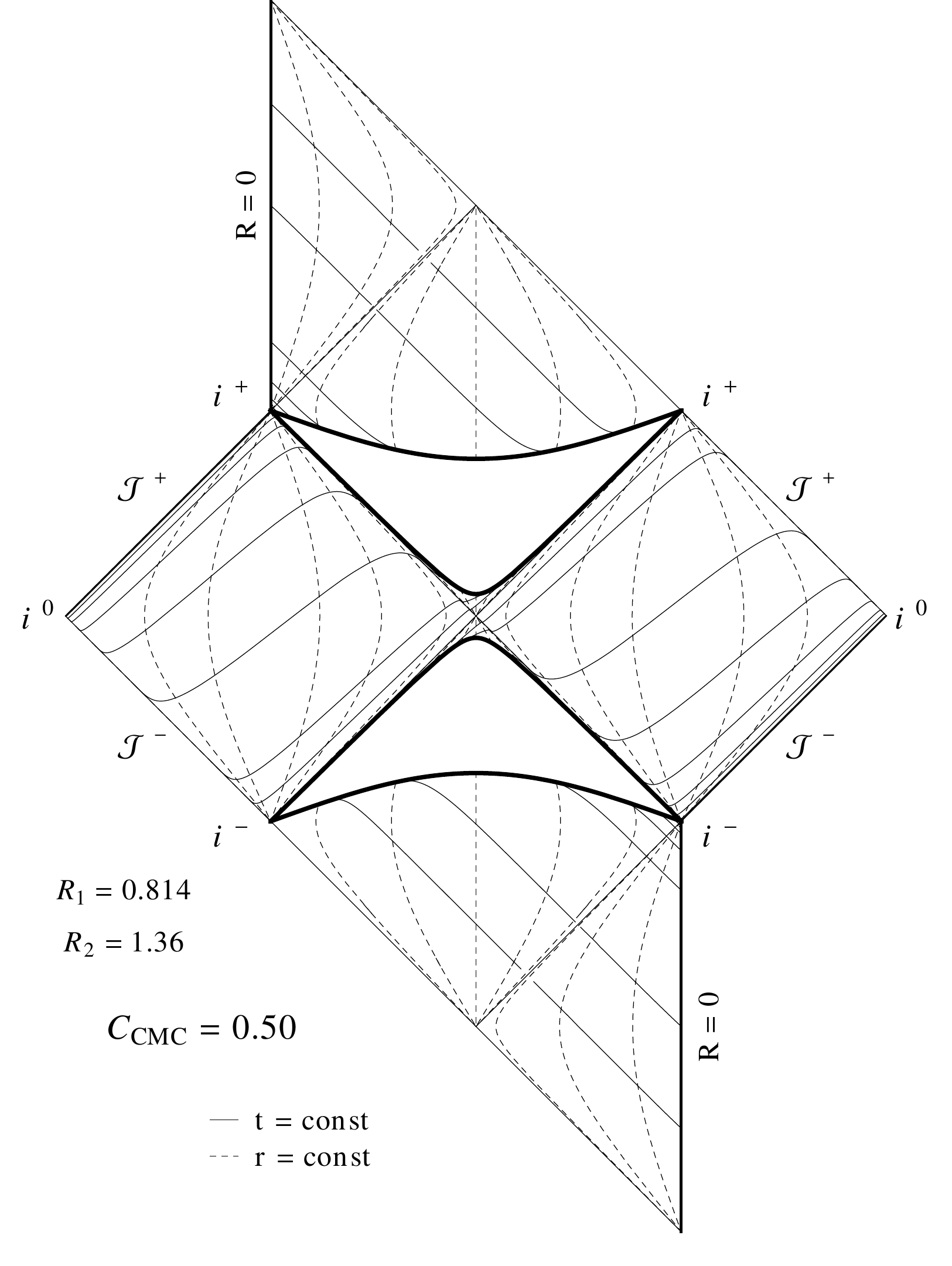}
	\vspace{-2ex}
\caption{Complete spacetime diagram for the case $\Cc=0.5$, where the slices intersect the white hole horizon.} %\upda{The constant radius slices (dashed lines) between the horizond do not reach the inner horizon.}
\label{fin:RNCcompl}
\end{figure}

The equivalent to \fref{fin:varC} in the non-extreme RN spacetime is shown in \fref{fin:RNvarC}. The construction of the diagrams is explained in subsection \ref{cap:RN}. The same considerations of the Schwarzschild case also apply here. The inner hyperboloidal slices look discontinuous around $r_-$ (represented by the diagonal line at 45º from the vertical singularity line labeled by $R=0$), but this is purely an effect of the numerical integration at the divergence at $\tilde r = r_-$ of the height function.
The complete diagram for the a) case (where the hyperboloidal slices intersect the white hole) is shown in \fref{fin:RNCcompl}.

%\upda{I am not showing the effect of varying $\Cc$ on the extreme RN spacetime.}

\begin{figure}[htbp!!]
\center
\begin{tabular}{@{}c@{}@{}c@{}}
\mbox{\includegraphics[width=0.5\linewidth]{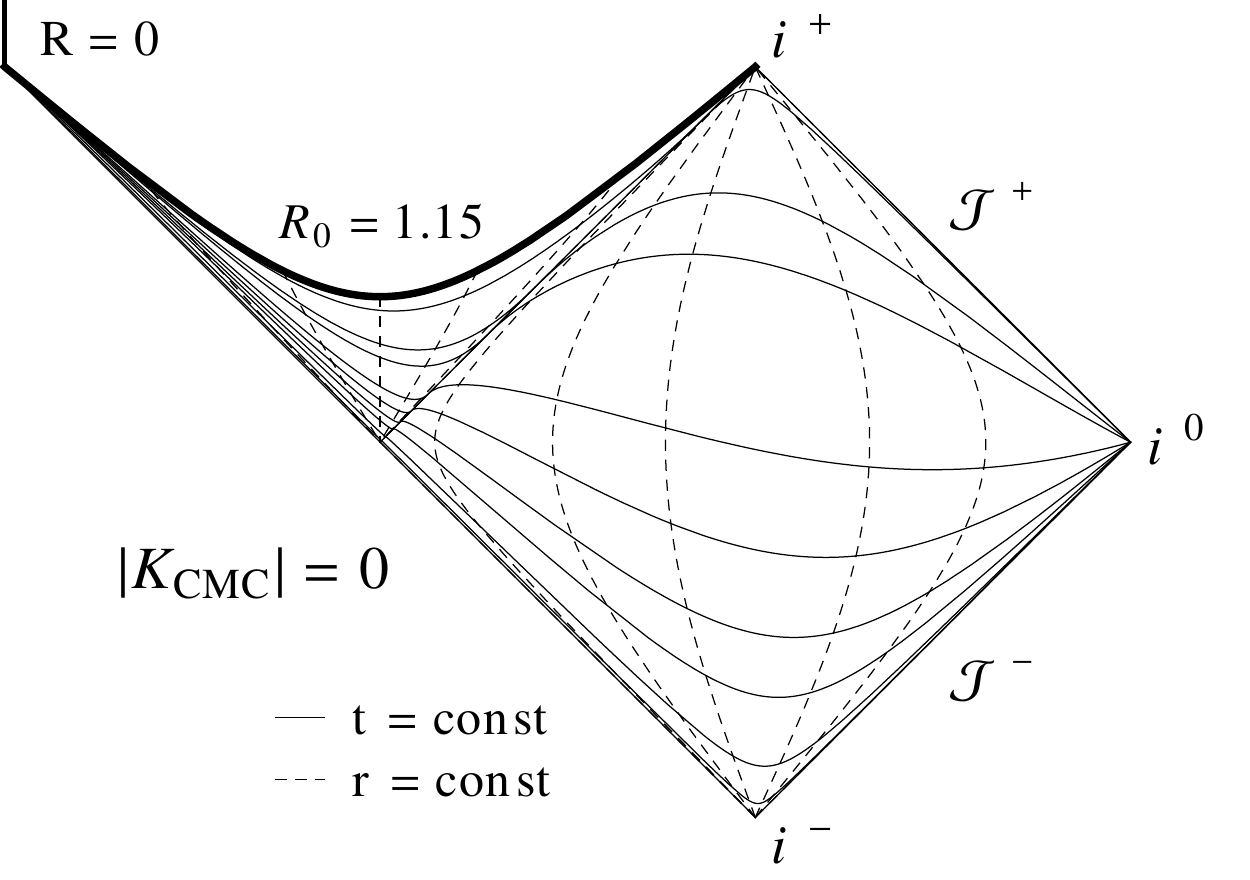}}&
\mbox{\includegraphics[width=0.5\linewidth]{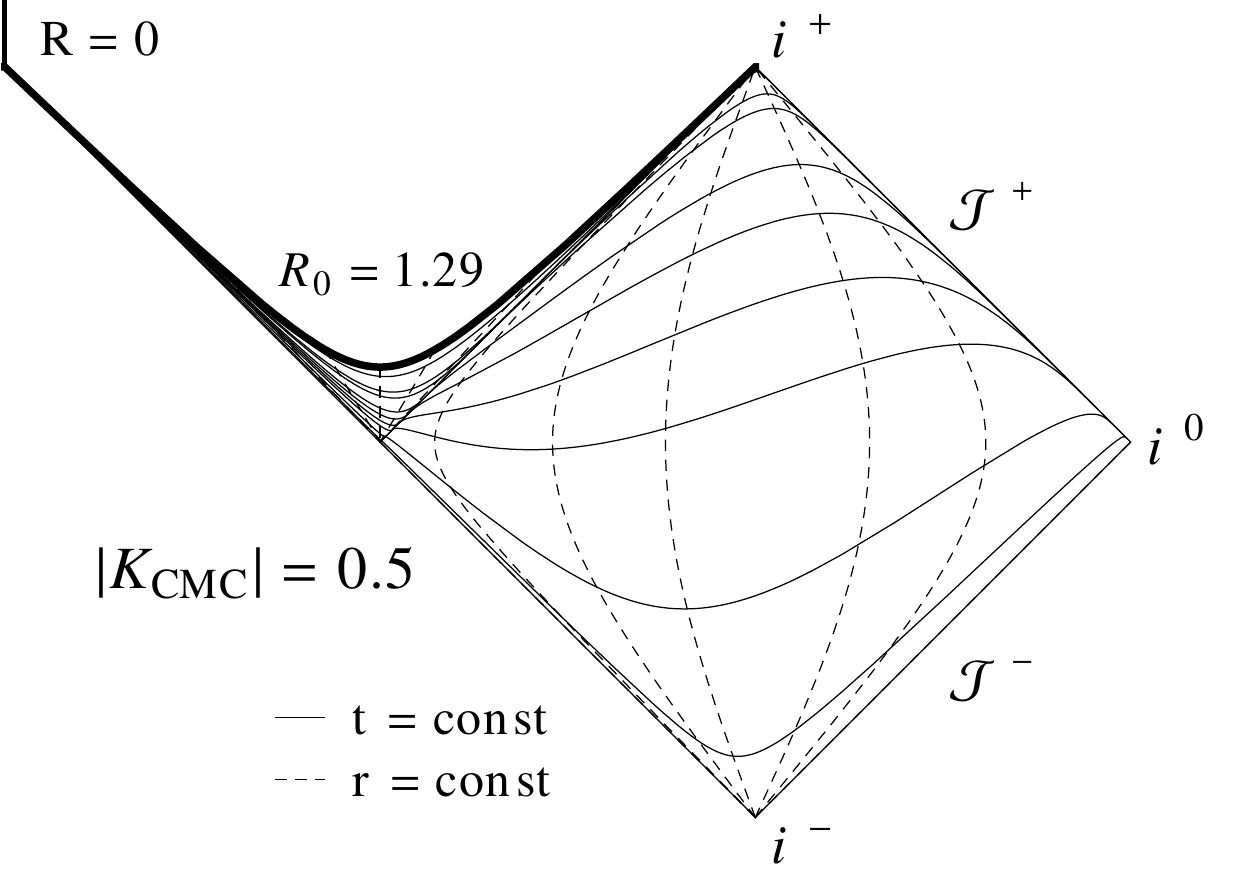}} \\
\mbox{\includegraphics[width=0.5\linewidth]{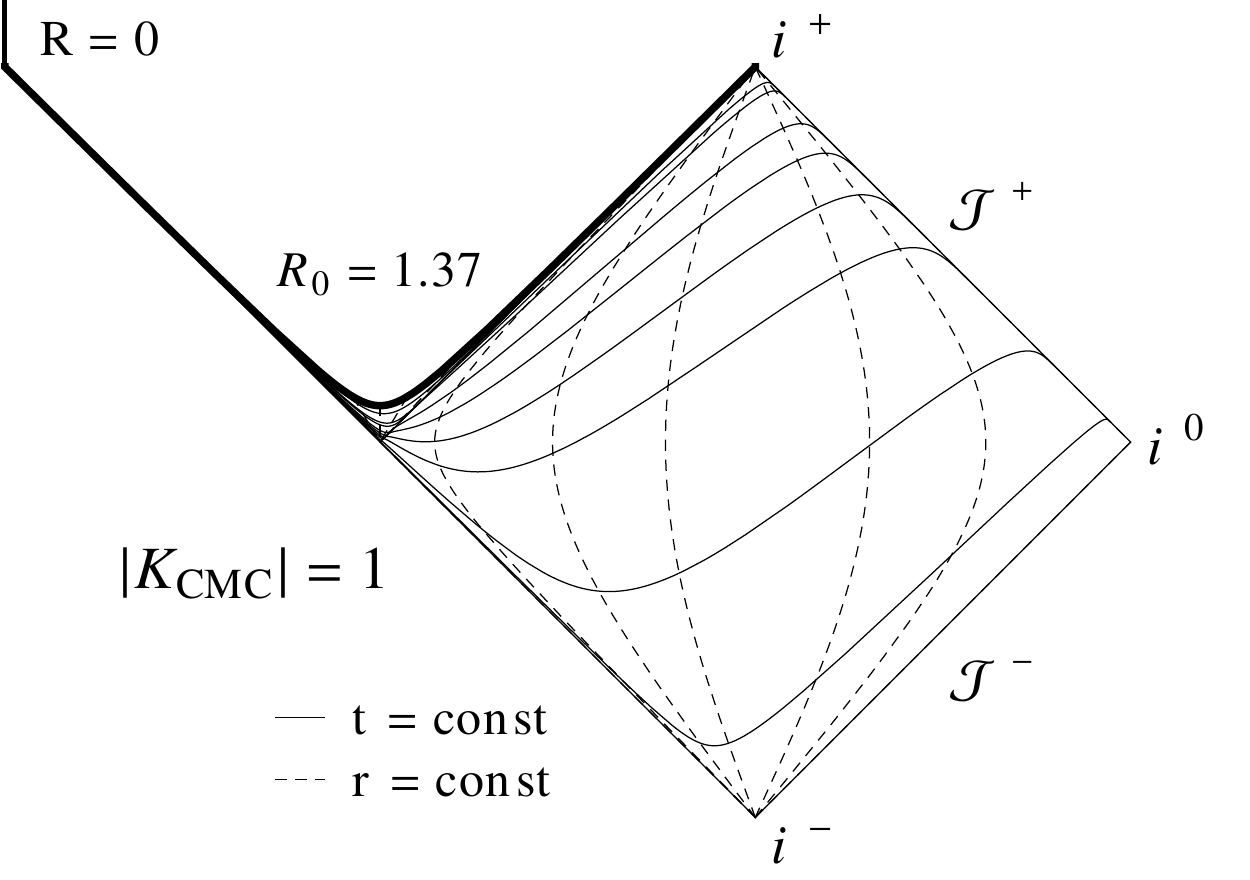}}&
\mbox{\includegraphics[width=0.5\linewidth]{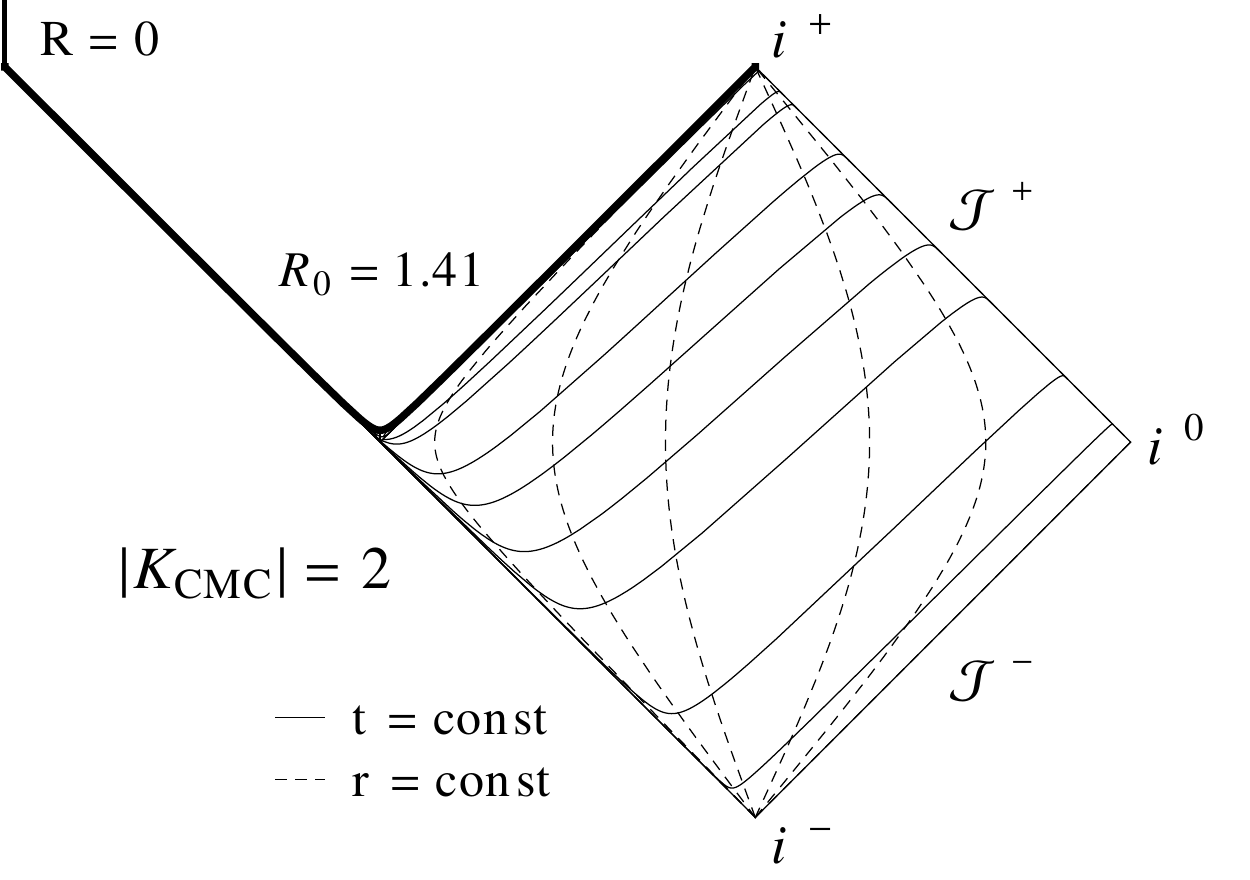}}
\end{tabular}
\caption{Penrose diagrams showing RN foliations with different values of $K_{CMC}$. Only the slices that reach $\scri^+$ (the outer slices of diagram c) in \fref{fin:RNvarC}) are displayed. The thick line corresponds to the innermost possible value of the Schwarzschild radial coordinate, where the trumpet is located.}
\label{fin:RNK}
\end{figure}

The effect of the variation of $\Kc$ on the non-extreme RN spacetime is presented in \fref{fin:RNK}. In the same way as done in \fref{fin:schwK} for the Schwarzschild case, only the outer slices (the ones covering the outer spacetime) are displayed. The behaviour of the foliations is equivalent to the Schwarzschild case.
\newpage

\begin{figure}[htbp!!]
\center
\begin{tabular}{@{}c@{}@{}c@{}}
\mbox{\includegraphics[width=0.5\linewidth]{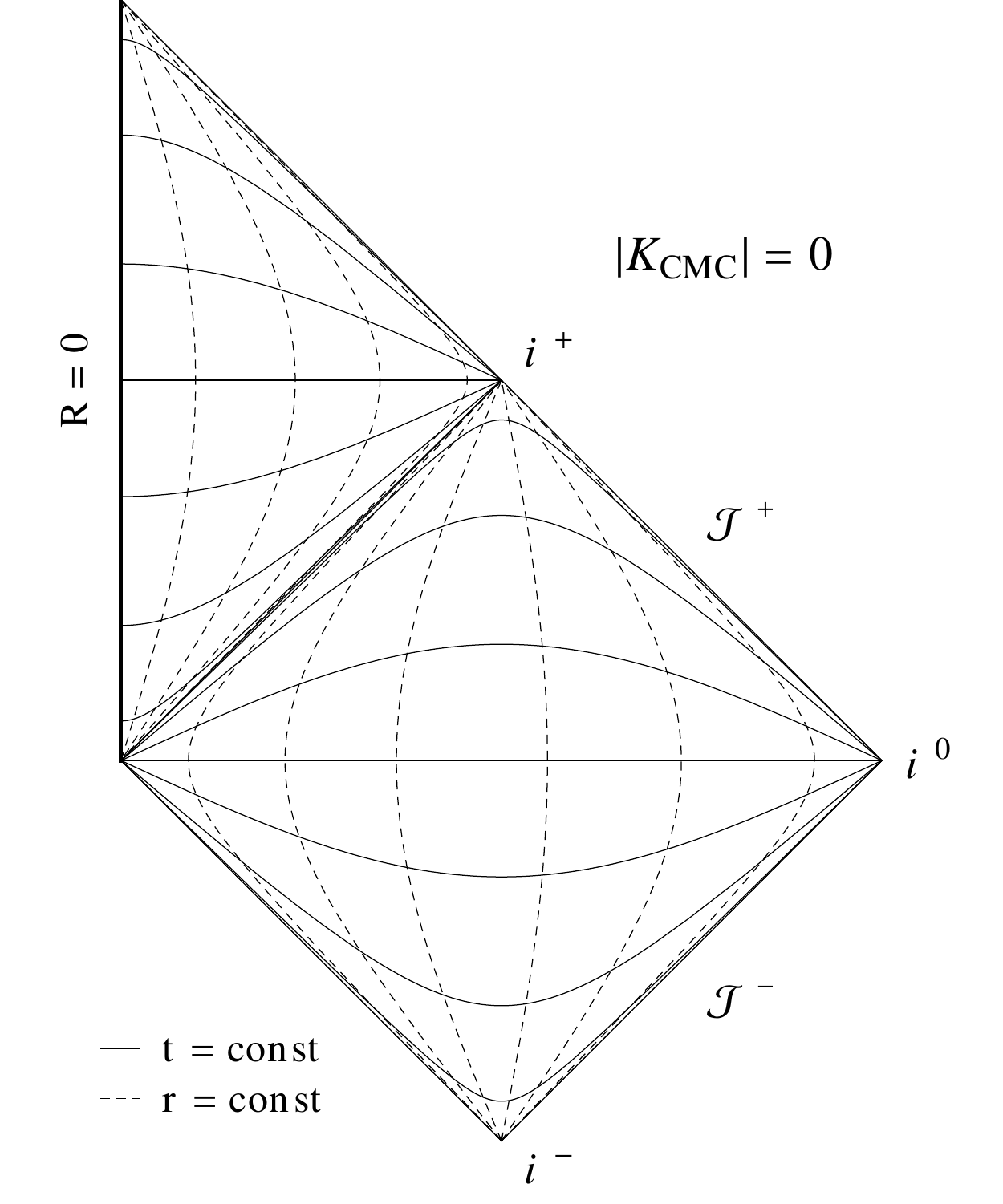}}&
\mbox{\includegraphics[width=0.5\linewidth]{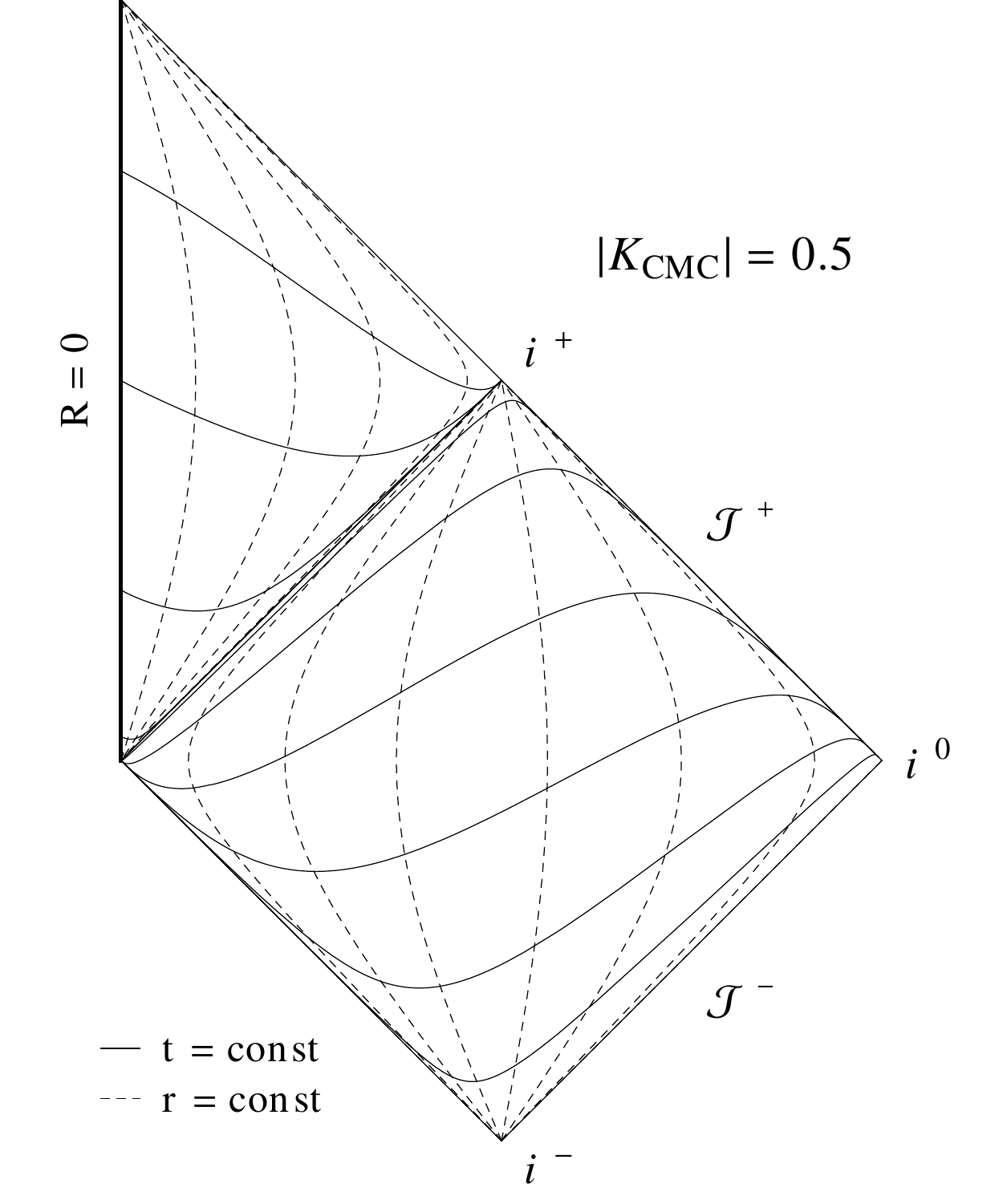}}\\
\mbox{\includegraphics[width=0.5\linewidth]{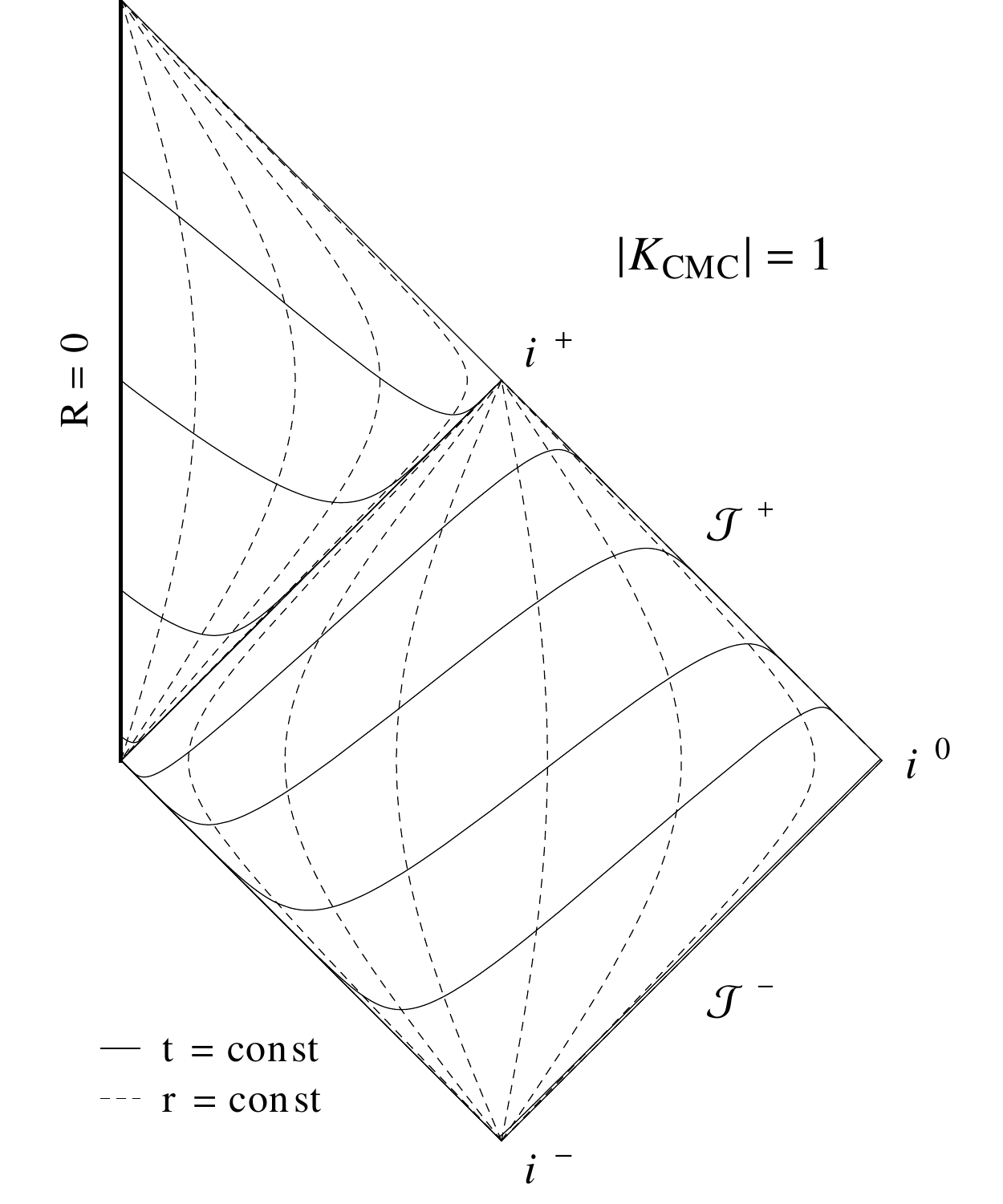}}&
\mbox{\includegraphics[width=0.5\linewidth]{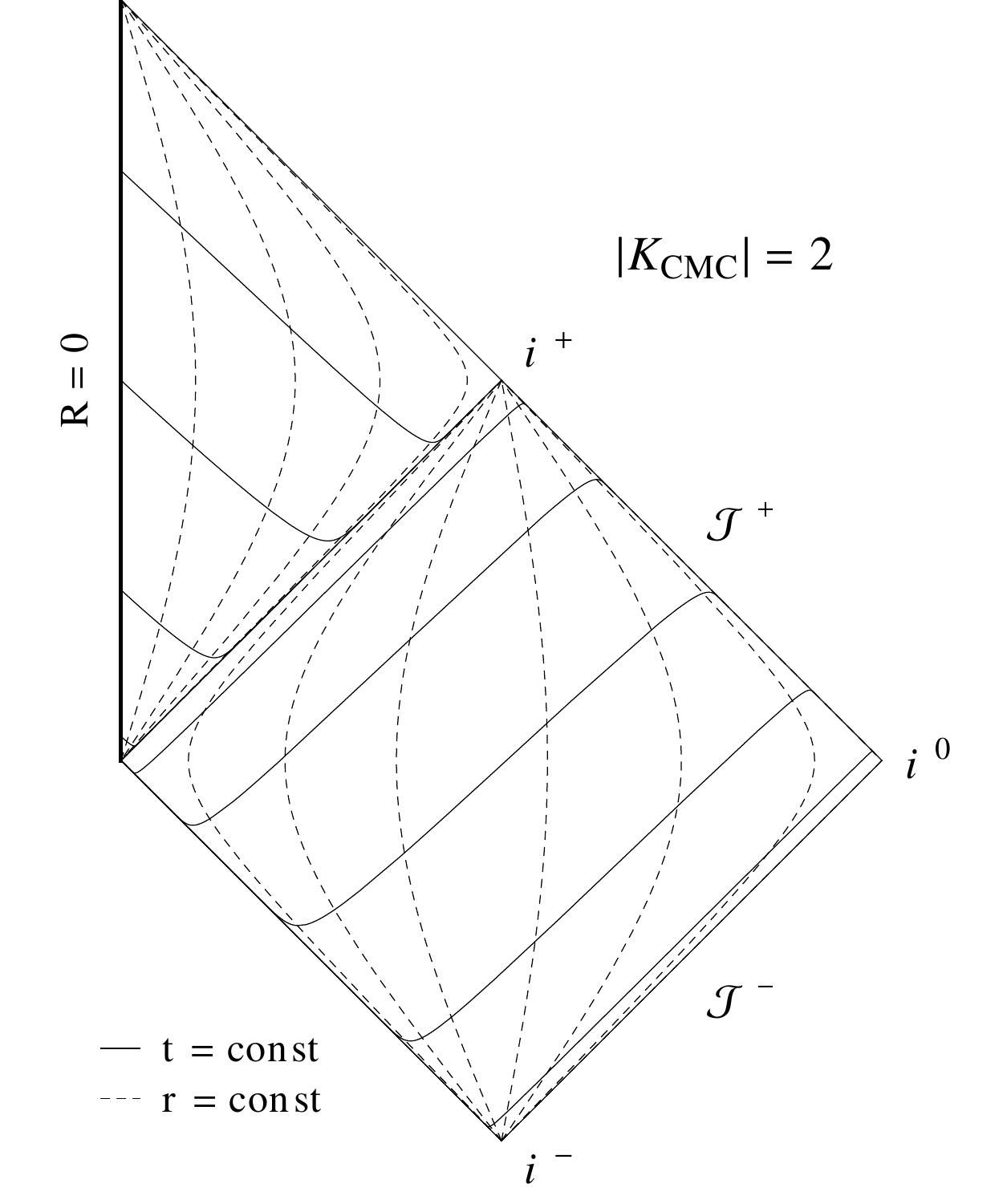}}
\end{tabular}
\caption{Penrose diagrams showing extreme RN foliations with $M=1$ and different values of $K_{CMC}$. In the $|\Kc|=0$ case, the critical value of $\Cc=0$, so that the height function vanishes everywhere. The value of $R_0$ is always $R_0 = M=1$.}
\label{fin:eRNK}
\end{figure}

The behaviour of extreme RN foliations for different values of $\Kc$ and the critical choice of $\Cc$ is presented in \fref{fin:eRNK}. The diagrams include the slices inside of the horizon (as in c) in \fref{fin:varC} and \fref{fin:RNvarC}). The main feature of the extreme case is that $R_0=M$ always and all hyperboloidal foliations corresponding to the critical $\Cc$ cross the horizon $r_+=r_-=M$ at the point symmetric to $i^0$.

\section[Solving the constraint equations for scalar field initial data]{Solving the spherically symmetric constraint equations for scalar field initial data}\label{sin:constrsolv}

Non-trivial scalar field initial data require solving the constraints. For a certain combination of the initial values of $\iPhi$ and $\iPi$, namely $\iPi_0=\beta^r{}_0\iPhi'_0$, the momentum constraint is automatically satisfied and only the Hamiltonian constraint needs to be solved. Those initial values correspond to a scalar field perturbation that will split into an ingoing pulse traveling towards the origin and an outgoing one moving towards $\scri^+$. Given that more general initial data configurations such as mainly ingoing or outgoing pulses may be desirable, a simple procedure to solve both Hamiltonian and constraint equations will be described.
%\hfill

%rescale $\iPi$
For the calculation of the initial data it is especially convenient to use the following definition for the scalar field auxiliary variable:
\begin{equation}\label{ein:piconstr}
%\dot\iPhi \equiv \frac{\alpha\chi^{3/2}}{\gamma_{\theta\theta}\sqrt{\gamma_{rr}}}\iPi+\beta^r\iPhi'  . \label{ein:piconstr}
\breve\Pi = \frac{\gamma_{\theta\theta}\sqrt{\gamma_{rr}}}{\alpha\chi^{3/2}}\left(\dot\iPhi-\beta^r\iPhi'\right)
\end{equation}
Then it is useful to perform the following variable transformations for the conformal factor (following common prescriptions as in \eref{ein:lich}) and the trace-free part of the extrinsic curvature
\begin{equation}\label{ein:constrvars}
\chi\to\chi_0\psi^{-4} \qquad \textrm{and} \qquad A_{rr}\to (A_{rr0}+\psi_A)\psi^{-6}  .
\end{equation}
As initial value of $\breve\Pi$ we choose %propsign*valuedphiomtbh(x,a,center,sigma)
\begin{equation}
\breve\Pi_0 = sign\frac{\gamma_{\theta\theta}{}_0\sqrt{\gamma_{rr}{}_0}}{\alpha_0\chi^{3/2}_0}\beta^r{}_0\iPhi'_0 ,
\end{equation}
where $sign$ is -1 gives a mainly ingoing pulse and +1 a mainly outgoing one. %, although by tuning its value experimentally the ratio between the ingoing and outgoing energies can be varied.

Setting the corresponding initial values for all the remaining variables (and substituting the value of $\aconf'$ determined from \eref{ein:conflat} and $\aconf''$ obtained by deriving $\aconf'$), the resulting Hamiltonian and momentum equations become respectively
{\small
\begin{subequations}\label{ein:constr}
\begin{eqnarray}
\psi'' &=& -\pi\left[\psi({\iPhi_0}')^2-\frac{1}{\psi^7}\left(\frac{\aconf^2}{\Omega^2}\breve\Pi_0\right)^2\right] -\frac{\psi'}{r}\left( 1+\sqrt{\left(1-\frac{2M\aconf}{r}\right)+\left(\frac{K_{CMC}\,r}{3\aconf}+\frac{C_{CMC}\aconf^2}{r^2}\right)^2} \right)  \nonumber \\
&& +\frac{K_{CMC}^2\psi(\psi^4-1)}{12\aconf^2} -\frac{3}{16\psi^7}\left(\frac{\Omega}{\aconf}\psi_A-\frac{2C_{CMC}\aconf^2}{r^3}\right)^2 + 3\psi\left(\frac{C_{CMC}\aconf^2}{2r^3}\right)^2   , \label{ein:hamil}\\
\psi_A' &=& -\frac{8\pi \aconf^3\breve\Pi_0\iPhi_0'}{\Omega^3} -\psi_A\left( \frac{3}{r}\sqrt{\left(1-\frac{2M\aconf}{r}\right)+\left(\frac{K_{CMC}\,r}{3\aconf}+\frac{C_{CMC}\aconf^2}{r^2}\right)^2} +\frac{\Omega'}{\Omega}\right)   . \label{ein:momen}
\end{eqnarray}
\end{subequations}
}%
The momentum constraint is independent of $\psi$ thanks to the specific choice of variables. This simplifies the procedure a lot, because it allows to first determine $\psi_A$ from \eref{ein:momen} and then set the obtained function into the Hamiltonian constraint \eref{ein:hamil} to calculate $\psi$.
The initial value of $\iPi$ ($\dot\iPhi=\iPi$) has to be calculated undoing the transformation \eref{ein:piconstr}:
%-(-a*exp(-(x**2-center**2)**2/(4._wp*sigma**4))*x*(x**2-center**2)/sigma**4) * ( Ccmc*psic**3/x**2 + Kcmc*x/3._iwp )*(-propsign/psi**6-1)
\begin{equation}
\iPi_0=\beta^r{}_0\iPhi'_0(1+\frac{sign}{\psi^6}) .
\end{equation}
A completely outgoing or ingoing pulse will not be obtained with the choices $sign=\pm1$, because $\chi$ is involved in the change of variables \eref{ein:piconstr}, but the value of $sign$ can be tuned experimentally to vary the ratio between the ingoing and outgoing amplitudes and so the undesired pulses can be minimized.

\section{Initial data for the simulations}

\subsection{Initial data perturbations}\label{sin:iniperturbs}

\subsubsection{Gauge wave initial data}\label{initialgauge}

Choosing the trace of the physical extrinsic curvature $\pK$ as variable makes the vacuum constraint equations \eref{es:pKceqs} independent of the gauge quantities. This allows us to introduce a perturbation in one of the gauge variables, e.g. the lapse $\alpha$, without having to solve the constraints for the initial data. The resulting evolution will of course only correspond to gauge dynamics - no actual physical processes take place -, but even these suppose a strong test for our equations.

The initial data for the variables is set as in \eref{ein:metrinio} and \eref{ein:curvinio}, or by the simpler \eref{ein:metrinioflat} in the flat spacetime case. To the initial value of the lapse we add a Gaussian-like compact-support perturbation of the form
\begin{equation}\label{initialgaussian}
\delta_{\alpha_0}=A_{\alpha}e^{-\frac{(r^2-c^2)^2}{4\sigma^4}}  .
\end{equation}
The reason for choosing this particular form for the initial perturbation is its even parity with respect to the origin.

\subsubsection{Scalar field initial data}\label{initialwave}

Including non-trivial scalar field initial data requires solving the constraints. For this, first a Gaussian in $r^2$, the same function as for the lapse perturbation, is set as initial value for the scalar field,
\begin{equation}\label{initialgaussianphi}
\Phi_0=A_{\Phi}e^{-\frac{(r^2-c^2)^2}{4\sigma^4}}  ,
\end{equation}
and then the constraint equations are solved as described in section \ref{sin:constrsolv}.

\subsection{Initial data plots}\label{sin:inidataplots}

Examples of the initial data used for the simulations that gave the results presented in this work are displayed in the following figures.
%As described in the previous section, in presence of a scalar field the constraints have to be solved and the changes in the variables $\chi$ and if corresponding $A_{rr}$ can be seen as deviations from their flat value.
%The amplitude of the scalar field perturbations shown in the Schwarzschild case is exaggerated to emphasize the effect on the other quantities.
%
The initial data corresponding to some of the variables are not shown in the plots due to their simplicity: $\gamma_{rr}$ and $\gamma_{\theta\theta}$ (if evolved by some reason) are always initially 1, which implies that $\Lambda^r$ vanishes. The Z4 quantity $\cT$ (or used in its ``physical'' form $\pT$) is also initially zero.

In the flat spacetime case, the generic choice for the CMC parameter is $\Kc=-3$, while Schwarzschild trumpet initial data use $\Kc=-1$, $M=1$ and the critical value of $\Cc$. The reason for this smaller $|\Kc|$ in the BH case is the numerical precision limitation in the calculation of $\aconf$ that will be explained in subsection \ref{sn:aconf}.
The initial perturbations of the scalar field (if present) in the following plots use $A_{\Phi}=0.035$, $\sigma=0.1$ and $c=0.25$, although these values may differ from those used in the actual simulations. The ``mostly ingoing'' data use $sign=-1$ and the ``mostly outgoing'' ones, $sign=+1$.

\subsubsection{Flat and regular spacetime}

%\upda{put in general for all plots} $\gamma_{rr}$ and $\gamma_{\theta\theta}$ (if evolved by some reason) are always initially 1. Consequently, $\Lambda^r$ vanishes. The Z4 quantity $\cT$ or $\pT$ are also initially zero. These quantities will not be shown in the plots.

\begin{figure}[htbp!!]
\center
	\includegraphics[width=1.0\linewidth]{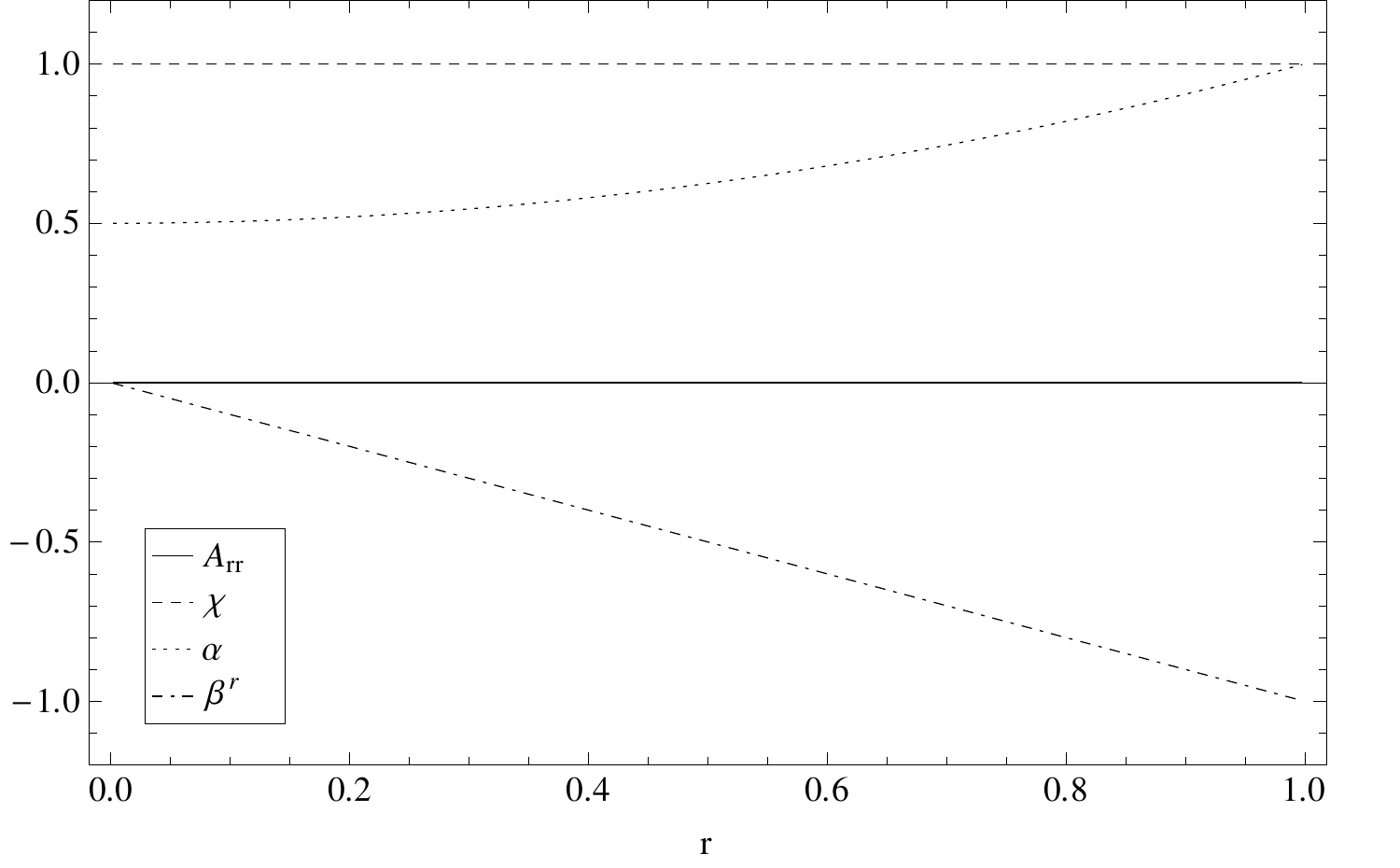}
\caption{Stationary flat initial data (with $\Kc=-3$).}
\label{fin:Fini}
\end{figure}

Figure \ref{fin:Fini} shows the stationary data corresponding to flat spacetime on the hyperboloidal slice. The value of lapse and shift at $\scri^+$ depends on the choice of parameter $\Kc$ (compare to \fref{fin:Bini}).

\begin{figure}[htbp!!]
\center
	\includegraphics[width=1.0\linewidth]{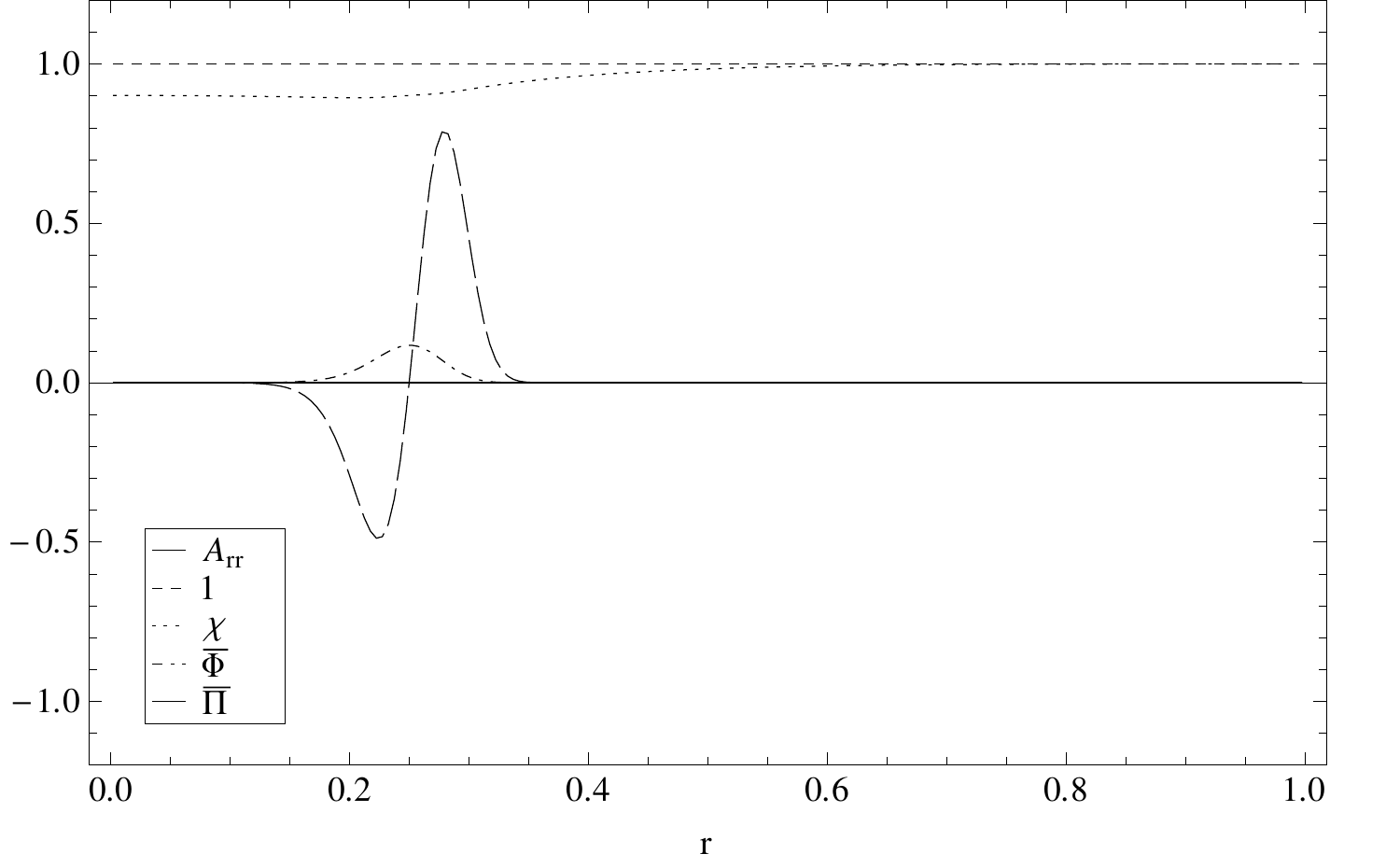}
\caption{Regular initial data perturbed by a massless scalar field. The effect of the perturbation appears in $\chi$.}
\label{fin:Finic}
\end{figure}

The perturbation of flat spacetime by an initially symmetric scalar field is plotted in the initial data in \fref{fin:Finic}. The scalar field's presence affects the conformal factor $\chi$, so that it differs from unity in the region close to the origin. The horizontal line at 1 is plotted to mark this difference.

\begin{figure}[htbp!!]
\center
	\includegraphics[width=1.0\linewidth]{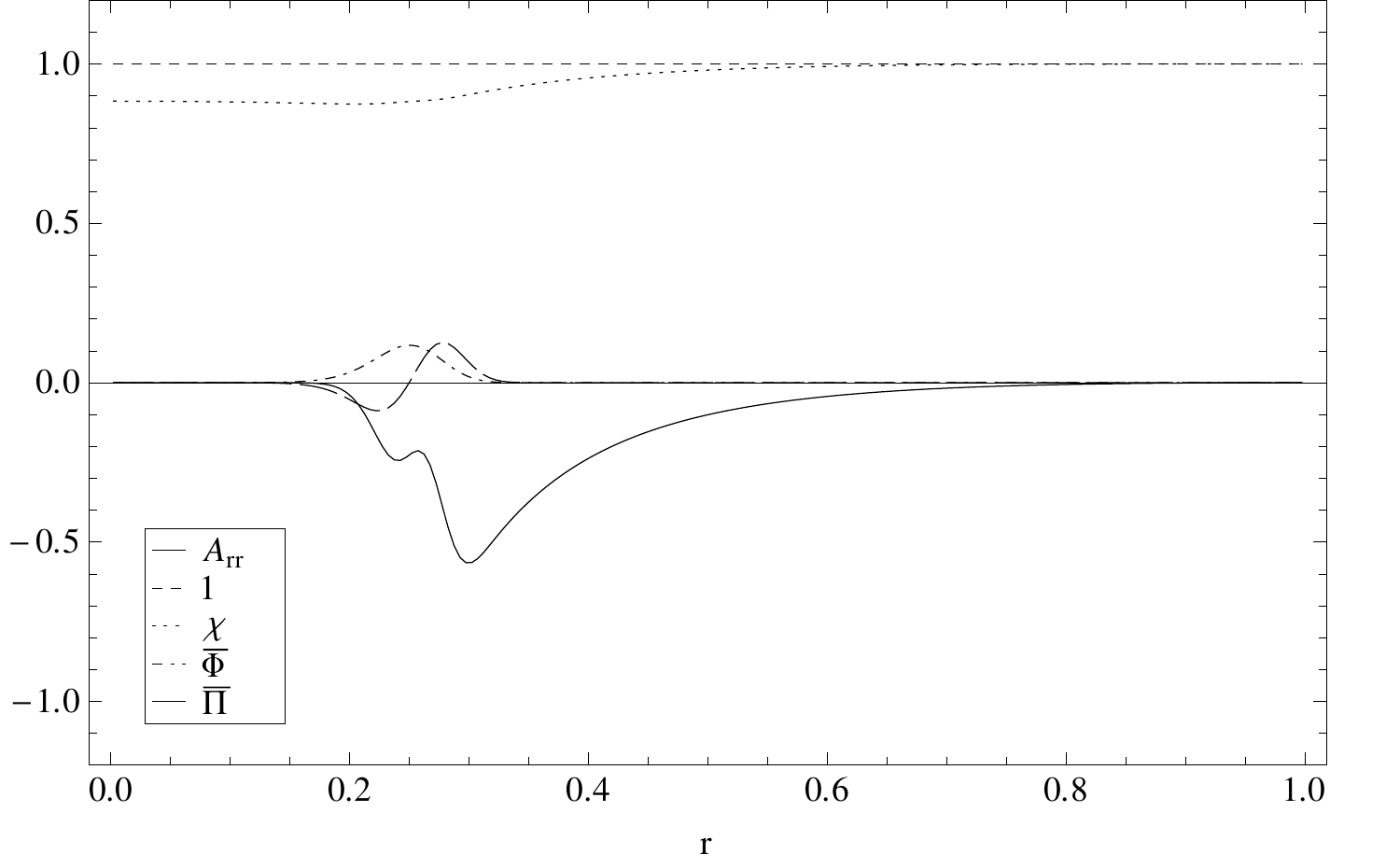}
\caption{Regular initial data perturbed by a mostly ingoing massless scalar field. The effect of the perturbation appears in $\chi$ and $A_{rr}$.}
\label{fin:FinicAm}
\end{figure}

If time asymmetric initial data are chosen for the scalar field, like an ingoing or an outgoing pulse, then the time derivatives of the metric (embodied in the extrinsic curvature) are also affected and the momentum constraint has to be solved for the perturbation on $A_{rr}$, as described in section \ref{sin:constrsolv}. An example of initial data corresponding to a mostly ingoing scalar field is plotted in \fref{fin:FinicAm}. If the initial scalar field was an outgoing pulse, the value of $A_{rr}$ would positive and $\chi$'s profile would also change.

\subsubsection{Schwarzschild spacetime}

%\upda{add $A_{rr}$ complete curve! comment on betar positive at horizon and negative at scri}

\begin{figure}[htbp!!]
\center
	\includegraphics[width=1.0\linewidth]{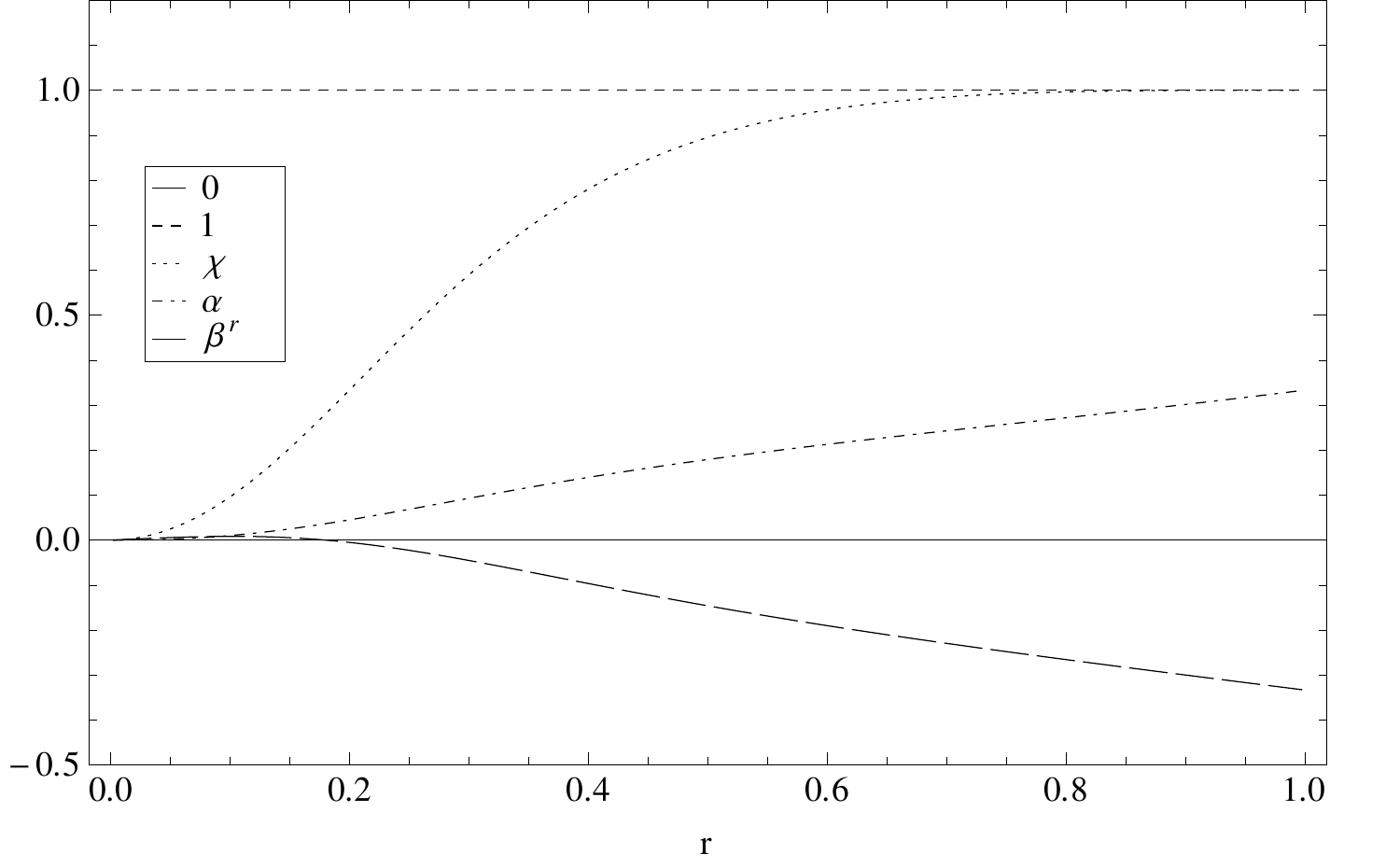}
\caption{Schwarzschild trumpet BH initial data (with $\Kc=-1$).}
\label{fin:Bini}
\end{figure}

\begin{figure}[htbp!!]
\center
	\includegraphics[width=1.0\linewidth]{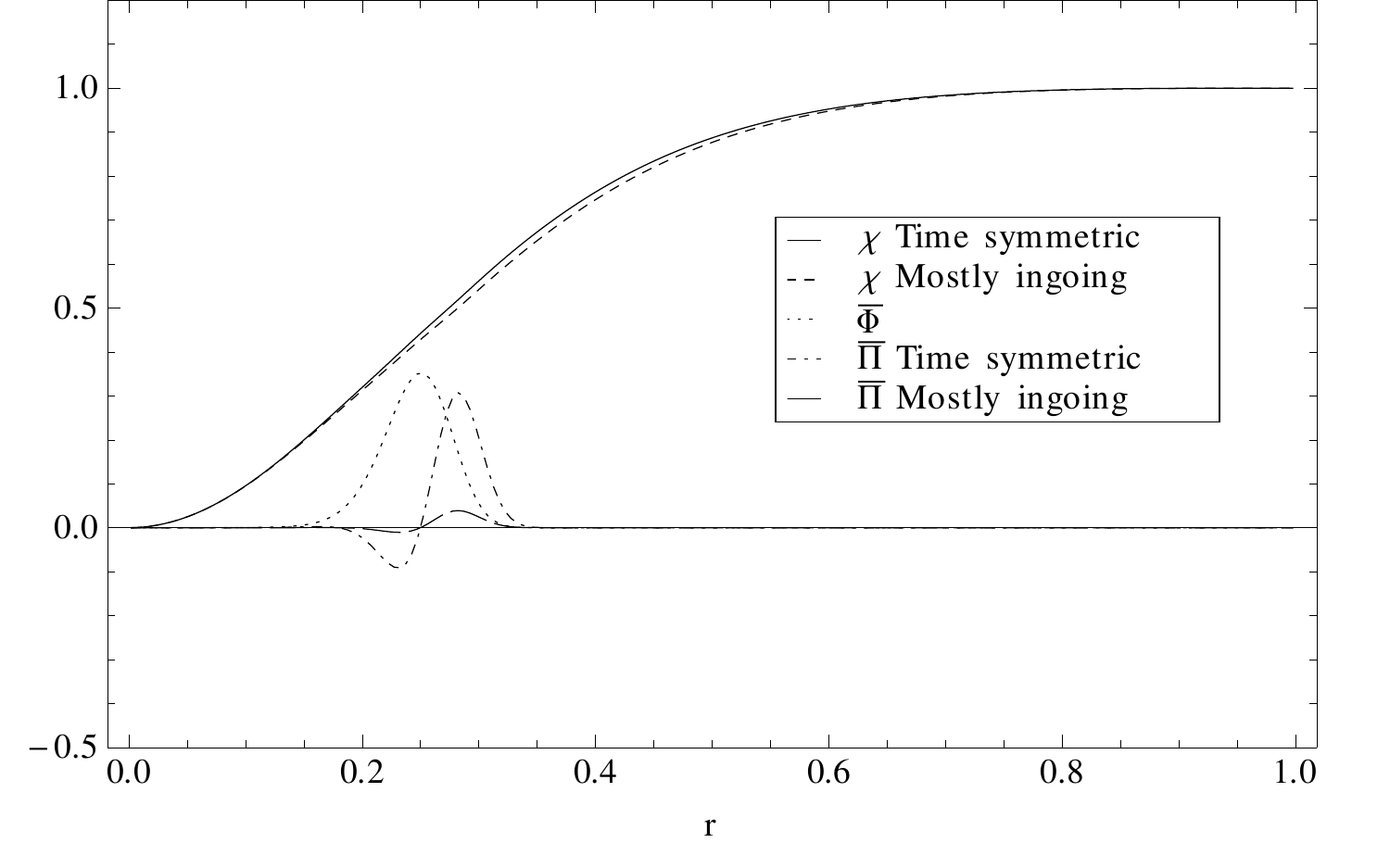}
\caption{Schwarzschild trumpet BH initial data perturbed by a massless scalar field. }
\label{fin:Binichiphipi}
\end{figure}

\begin{figure}[htbp!!]
\center
	\includegraphics[width=1.0\linewidth]{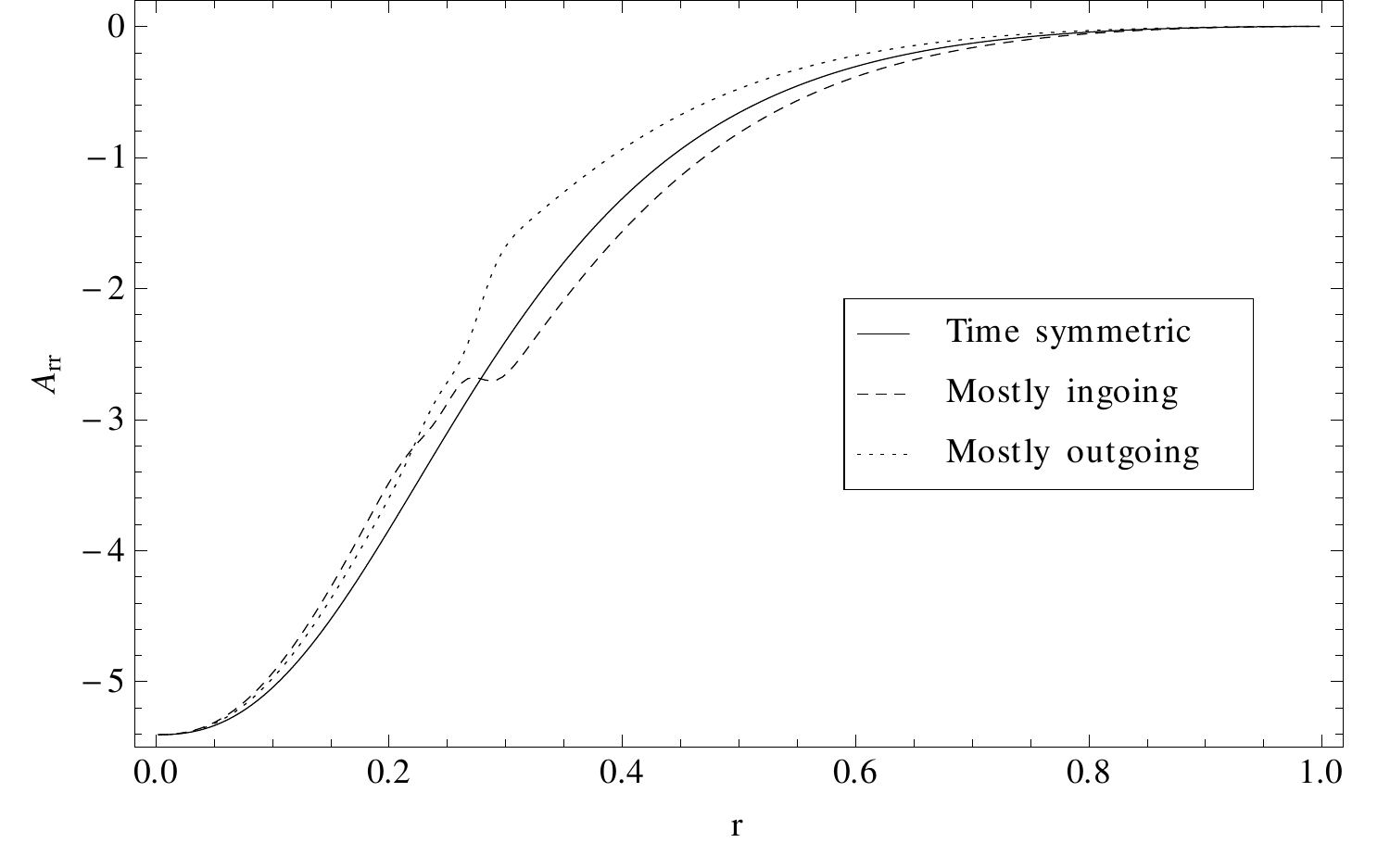}
\caption{Schwarzschild trumpet BH initial data for the variable $A_{rr}$.}
\label{fin:BiniArr}
\end{figure}

The trumpet values of a BH spacetime with $M=1$ are displayed in \fref{fin:Bini}. The conformal factor $\chi$, the lapse $\alpha$ and the shift $\beta^r$ all become zero at the origin, which is where the trumpet is mapped to. The shift is positive inside of some radius around the origin (including the BH horizon, which in this case is located at $r_{Schw}\approx0.13$) and then becomes negative.
The initial value of the variable $A_{rr}$ is plotted in \fref{fin:BiniArr}.
%\begin{figure}[htbp!!]\label{}
%\center
%	\includegraphics[width=1.0\linewidth]{figures/Binic.pdf}
%\caption{Schwarzschild trumpet black hole initial data perturbed by a massless scalar field. The effect appears in $\chi$.}
%\end{figure}
%\begin{figure}[htbp!!]\label{}
%\center\includegraphics[width=1.0\linewidth]{figures/inischwkgfin.png}
%\caption{Schwarzschild trumpet black hole initial data perturbed by a mostly ingoing massless scalar field. The effect appears in $\chi$ and $A_{rr}$.}
%\end{figure}
Figure \ref{fin:Binichiphipi} shows the influence of a time symmetric or a mostly ingoing scalar field perturbation on the conformal factor $\chi$.
The Schwarzschild and perturbed trumpet values of $A_{rr}$ are presented in \fref{fin:BiniArr}.

\subsubsection{Variable choice for the trace of the extrinsic curvature}

\begin{figure}[htbp!!]
\center
	\includegraphics[width=1.0\linewidth]{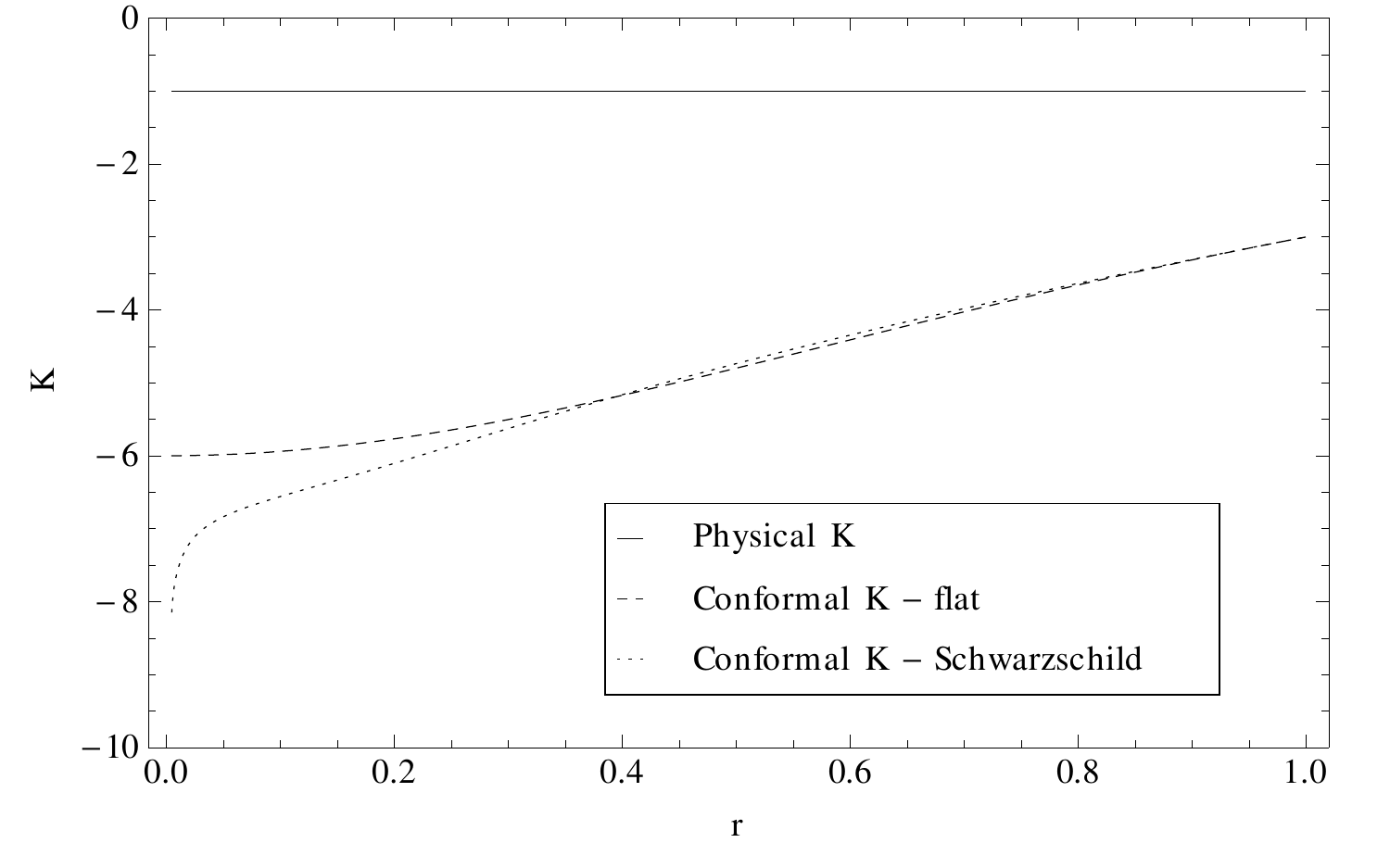}
\caption{Initial data for the trace of the physical extrinsic curvature ($\pK$) and of the conformal one ($\cK$), both with $\Kc=-1$. The first does not change in presence of a BH, but the latter does.}
\label{fin:iniK}
\end{figure}

The initial values of $\cK$, with expression \eref{ein:Kini}, and $\pK$, which is simply $\pK_0=\Kc$, are shown in \fref{fin:iniK}. The latter has a much simpler stationary value, more appropriate for numerical and visualization purposes. The quantity $\DPK$ introduced as the variation of $\pK$ in \eref{ee:DPKdef} has a vanishing initial value.

%\renewcommand\bibname{{References}}
%\bibliographystyle{../../master/thesis/tocunsrt}
%\bibliography{../articles/hypcomp}

\chapter{Gauge conditions}\label{c:gauge}

%\section{Importance of the gauge conditions}

The gauge variables determine how the spacetime is sliced for the decomposition of the equations. The lapse $\alpha$ measures the separation of the slices along the proper time of an observer and the shift $\beta^a$ controls the change in spatial coordinates from one slice to the next. How these gauge quantities behave during the evolution is not prescribed by the Einstein equations. For a practical implementation this gauge freedom has to be fixed by imposing some conditions on the lapse and shift. These conditions will depend on the physical problem that has to be solved and are usually motivated by either a substantial simplification in the equations and problem setup or by a good numerical behaviour, or both at the same time. In any case, the effect of the gauge conditions on the evolution is huge and the choice has to be made carefully. 

%Some common choices are the geodesic slicing, the maximal slicing, the harmonic slicing and the 1+log slicing, which mainly set conditions for the lapse, whereas the conditions for the shift are contained in the normal coordinates, minimal distortion, Gamma freezing and Gamma driver. 

In the rest of this chapter I will describe which specific conditions are convenient for the hyperboloidal initial value problem and how some commonly used choices (see \cite{Alcubierre} for more details) can be adapted to it. 

\section{Scri-fixing condition}\label{cg:scrifix}

In the conformal picture future null infinity is an ingoing null surface. From the point of view of a set of hyperboloidal slices it means that the position of $\scri^+$ on the slice moves inwards with the speed of light as time goes by. In a numerical implementation this effect translates to a loss of resolution on the integration domain. A way of avoiding this and setting the location of $\scri^+$ to a fixed position of the spatial domain is by means of the ``scri-fixing'' condition \cite{Frauendiener:1997ze}. 

The basic idea of the scri-fixing condition is to choose the coordinate time vector in such a way that it flows along $\scri^+$. The time vector is given by $t^a=\alpha n^a+\beta^a$ \eref{c3:tvect}, and it has to be chosen in such a way that it is an ingoing null vector at $\scri^+$. For simplicity and to avoid the confusion with complex conjugates we will drop the overbars on the vectors and covectors in this section, but note that all calculations are performed in the conformal spacetime. 

\subsubsection{Newman-Penrose null tetrad}

Applying the Newman-Penrose formalism \cite{newman1962approach} at $\scri^+$ we introduce two real null vectors $k^a$ and $l^a$. The vector $l^a$ is defined as the ingoing null vector and $k^a$ as the outgoing one, as displayed in \fref{fg:scridiagram}. The tetrad is completed by two complex tangential null vectors, $m^a$ and its complex conjugate $\bar m^a$. The only nonzero inner products of the tetrad vectors are $l^ak_a=-1$ and $=m^a\bar m_a=+1$ (for the metric convention used, $(-,+,+,+)$). 

The tetrad vectors can be expressed in terms of the timelike unit vector $n^a$ and the spacelike unit vectors $s^a$, $e^a_{\theta}$ and $e^a_{\phi}$, for which the only non-vanishing inner products give
\begin{equation} n^a n_a = -1 , \quad s^a s_a = +1 , \quad e_{\theta}^a e_{\theta\, a} = +1 , \quad e_{\phi}^a e_{\phi\, a} = +1 .
\end{equation}
The expressions of the tetrad null vectors are 
\begin{equation} k^a = \frac{1}{\sqrt{2}} (n^a + s^a) , \quad l^a = \frac{1}{\sqrt{2}} (n^a - s^a) , \quad m^a = \frac{1}{\sqrt{2}} (e_{\theta}^a + i e_{\phi}^a), \quad %\textrm{and}\quad \bar m^a = \frac{1}{\sqrt{2}} (e_{\theta}^a - i e_{\phi}^a) .
\end{equation}

\begin{figure}[htbp]
\center
	\begin{tikzpicture}[scale=1.2]
		\draw (0cm, 0cm) -- (8cm, 0cm); \draw [thick,->] (5cm, 0cm) -- (1.5cm, 3.5cm);
		\draw (1.3cm, 3.9cm) node {$\Omega=0$}; \draw (3.3cm, 2.4cm) node {$\scri^+$};
		\draw [very thick,->] (1cm, 0cm) -- (2.41421cm, 0cm); \draw [very thick,->] (1cm, 0cm) -- (1cm, 1.41421cm);
		\draw (0.7cm, 0.7cm) node {$ n^a$}; \draw (1.7cm, 0.3cm) node {$ s^a$};
		\draw [very thick,->] (5cm, 0cm) -- (4cm, 1cm); \draw [very thick,->] (5cm, 0cm) -- (6cm, 1cm);
		\draw (4.3cm, 1.3cm) node {$l^a$}; \draw (6.3cm, 1.3cm) node {$k^a$}; 
		\draw [very thick,->] (6.5cm, 2.8cm) -- (5.5cm, 3.8cm); \draw [very thick,->] (6.5cm, 2.8cm) -- (6.5cm, 3.8cm); \draw [very thick,->] (6.5cm, 2.8cm) -- (5.5cm, 2.8cm);
		\draw (5.2cm, 3.8cm) node {$t^a$}; \draw (7cm, 3.3cm) node {$\alpha n^a$}; \draw (6cm, 2.4cm) node {$\beta^a$};
	\end{tikzpicture}
	\caption{Timelike, spacelike and null vectors near $\scri^+$.}\label{fg:scridiagram}
\end{figure}

\subsubsection{Scri-fixing conditions}

Null infinity is given by $\Omega=0$ and $\bar \nabla_a\Omega\neq0$. At $\scri^+$ we have that $\atscrip{l^a}=\atscrip{\A\bar\nabla^a\Omega}$ (where $A$ is a constant positive proportionality factor) and the condition that the time vector flows along $\scri^+$ is given by $\atscrip{t^a}=B \atscrip{l^a}$ (with $B$ another constant factor). It can be written as 
\begin{equation}
\atscrip{t^at_a}=
\atscrip{\left(\nabla^a\Omega\right)\left(\nabla_a\Omega\right)}=0. 
\end{equation}
Using \eref{c3:tvect} it translates to 
\begin{equation}
\atscrip{t^at_a}=\atscrip{\left(\alpha n^a+\beta^a\right)\left(\alpha n_a+\beta_a\right)}=\atscrip{-\alpha^2+\beta^a\beta_a}=0 . 
\end{equation}
In our spherically symmetric ansatz (see section \ref{ss:ansaetze}) it is expressed as
\begin{equation} \label{eg:scrifixspher}
\atscrip{-\alpha^2+\chi^{-1}\gamma_{rr}{\beta^r}^2}=0  .
\end{equation}

% In a more general geometric setup, we could also have chosen a shift vector which has non-radial components, including at null infinity.

%Since $\scri^+$ is a null surface, $k^a$ is parallel to $\nabla^a\Omega$, and we will choose our shift vector such that they are also parallel to $\left(\frac{\partial}{\partial t}\right)_a$ at $\scri^+$. Consequently, we obtain $\atscrip{\partial_t\Omega}=0$, consistent with the value of $\Omega$ not being evolved in time. 

The scri-fixing condition in our approach can be divided into two parts: 
\begin{itemize}
\item The behaviour in time of $\Omega$: if the value of $\Omega$ changes with time, it may affect the coordinate position of $\Omega=0$. In the approach taken here following Zengino\u{g}lu \cite{Zenginoglu:2007it,Zenginoglu:2008pw}, the conformal factor $\Omega$ is a function of the radial coordinate $r$ fixed in time, so the coordinate location of $\scri^+$ does not change during the evolution.
\item Condition \eref{eg:scrifixspher}: the evolution quantities $\chi$, $\gamma_{rr}$, $\alpha$ and $\beta^r$ have to satisfy at all times the condition that the time vector is null at the position to which $\scri^+$ has been fixed. 
\end{itemize}
In a more general setup other than spherical symmetry, corotation is also possible, in which case the scri-fixing condition is not as strict as $t^a$ being a null geodesic generator. 

\section{Preferred conformal gauge}\label{sg:preferred}

Here we will follow chapter 11 of \cite{Wald} and \cite{Zenginoglu:2008pw}. We start by studying the gauge freedom in the choice of the conformal factor $\Omega$. It allows to rescale $\Omega$ by some strictly positive function $\omega$, such that $\oOmega=\omega\Omega$. Using this new conformal factor $\oOmega$, another conformally rescaled metric $\om_{ab}$ can be defined:
\begin{equation}\label{eg:oO}
\om_{ab}=\oOmega^2\tilde g_{ab}, \qquad\textrm{which implies}\qquad  \om_{ab} = \omega^2\bar g_{ab} . 
\end{equation}

\begin{comment}
We will now consider the transformations of the Einstein equations due to the conformal rescaling of the physical metric, both expressed in terms of \upda{the Ricci tensor\eref{???}} or the Einstein tensor \eref{c3:einsteinG} in \eref{c3:einstein}. The Z4 quantities are assumed to be exactly zero and suppose that the matter fields encoded in the energy-momentum tensor either have compact support or their fall-off behaviour is strong enough that $\atscrip{T[\case{\bar g}{\Omega^2}]_{ab}}=\mathcal{O}(\Omega^{0})$ at least. The expressions are included here again for convenience: 
\begin{eqnarray}
G[\bar g]_{ab} + { 2\over\Omega}(\bar\nabla_a\bar\nabla_b\Omega-\bar g_{ab}\bar\Box \Omega) + {3\over\Omega^2}\bar g_{ab}(\bar\nabla_c\Omega)(\bar\nabla^c\Omega)&=&8\pi T[\case{\bar g}{\Omega^2}]_{ab} ,  \\
R[\bar g]_{ab} + {1\over\Omega}(2\bar\nabla_a\bar\nabla_b\Omega+\bar g_{ab}\bar\Box \Omega) - {3\over\Omega^2}\bar g_{ab}(\bar\nabla_c\Omega)(\bar\nabla^c\Omega)&=&8\pi\left(T[\case{\bar g}{\Omega^2}]_{ab}-\right) . 
\end{eqnarray}
\end{comment}
We will now consider the transformations of the Einstein tensor \eref{c3:einsteinG} (or equivalently of the Ricci tensor \eref{c3:einsteinR}, as they differ by pure trace terms) due to the conformal rescaling of the metric \eref{ei:rescmetric}. For simplicity we assume a vacuum solution without Z4 terms, so that $G[\tilde g]_{ab}=0$ or $R[\tilde g]_{ab}=0$. However, the following considerations will also apply if the Z4 quantities are such that $\bar Z_a=\mathcal{O}(\Omega^{1})$ or higher at $\scri^+$ and if the matter fields encoded in the energy-momentum tensor either have compact support or their fall-off behaviour is strong enough that $\atscrip{T[\case{\bar g}{\Omega^2}]_{ab}}=\mathcal{O}(\Omega^{0})$ at least. 
As already mentioned in subsection \ref{si:confeqs}, if \eref{c3:einsteinG} is multiplied by $\Omega^2$ and evaluated at $\scri^+$, the result $\atscrip{(\bar\nabla_c\Omega)(\bar\nabla^c\Omega)}=0$ also states that the vector $\bar\nabla^c\Omega$ is null at $\scri^+$. 
Now we multiply \eref{c3:einsteinG} (or \eref{c3:einsteinR}) by $\Omega$, take its trace-free part and evaluate it at $\scri^+$. The result is 
\begin{equation}\label{eg:invariant}
\atscrip{\left(\bar\nabla_a\bar\nabla_b\Omega-\frac{1}{4}\bar g_{ab}\bar\Box\Omega\right)}=0 . 
\end{equation}
%\upda{The procedure is equivalent to doing it with the Ricci tensor, as the difference between both approaches are only pure trace terms.}
The relation obtained is independent of the conformal gauge, because it was derived purely from the transformation behaviour of the Einstein or Ricci tensors. Using the transformation \eref{eg:oO} and the definition of $\oOmega$ the following transformations are valid at $\scri^+$
\begin{eqnarray}
\atscrip{\other\nabla_a\other\nabla_b\oOmega} &=& \atscrip{\left(\omega\bar\nabla_a\bar\nabla_b\Omega+\bar g_{ab}\bar\nabla^c\omega\bar\nabla_c\Omega\right)} , \\
\atscrip{\other\Box\oOmega}&=&\atscrip{\frac{1}{\omega^2}\left(\omega\bar\Box\Omega+4\bar\nabla^a\omega\bar\nabla_a\Omega\right)} , \label{eg:boxo}
\end{eqnarray} 
and using them we obtain 
\begin{equation}\label{eg:invariantother}
\atscrip{\left(\other\nabla_a\other\nabla_b\oOmega-\frac{1}{4}\other g_{ab}\other\Box\oOmega\right)}=\omega\atscrip{\left(\bar\nabla_a\bar\nabla_b\Omega-\frac{1}{4}\bar g_{ab}\bar\Box\Omega\right)}=0 . 
\end{equation}
The invariance of \eref{eg:invariant} under a conformal rescaling is thus shown. It also implies shear-freeness of $\scri^+$. To see this we use the previously used identification of the null vector $l^a$ with the gradient of $\Omega$ at $\scri^+$, such that $\atscrip{l^a}=\atscrip{\A\bar\nabla^a\Omega}$. If $l^a$ is tangent to an affinely parametrized null geodesic, the shear of the null geodesic congruence given by $l^a$ is expressed by one of the twelve complex spin coefficients introduced by Newman and Penrose \cite{newman1962approach}, namely by $\sigma=-m^am^b\bar\nabla_bl_a$. Evaluating this quantity at $\scri^+$ we have
\begin{equation}\label{eg:sigmazero}
\atscrip{\sigma}=\atscrip{-m^am^b\bar\nabla_bl_a }=\atscrip{ -A m^am^b\bar\nabla_b\bar\nabla_a\Omega} =\atscrip{-\frac{\A}{4}m^am^b\bar g_{ab}\bar\Box\Omega} = 0 , 
\end{equation}
where relation \eref{eg:invariant} and $m^a m_a=0$ have been used. Instead of $l^a$, this calculation can also be performed using a rescaled null vector $\other l^a=C l^a$. This new vector $\other l^a$ is chosen to be proportional to $\other\nabla^a\oOmega$ at $\scri^+$, such that 
\begin{equation}
\atscrip{\other l^a}=\atscrip{\A\other\nabla^a\oOmega} =\atscrip{ \A\bar\nabla^a(\omega\Omega)} = \atscrip{ \A\omega\bar\nabla^a\Omega } =\atscrip{\omega l^a}, 
\end{equation}
so that at future null infinity $C=\omega$ and the relation between ingoing null vectors is $\atscrip{\other l^a}=\atscrip{\omega l^a}$. Calculating $\atscrip{\sigma}$ in terms of $\other l^a$ and using \eref{eg:invariantother} gives obviously the same result
\begin{equation}\label{eg:sigmazero}
\atscrip{\sigma}=\atscrip{-m^am^b\other\nabla_b\other l_a }
%=\atscrip{ -A m^a \bar m^b\other\nabla_b\other\nabla_a\oOmega} 
=\atscrip{-\frac{\A}{4}m^am^b\other g_{ab}\other\Box\oOmega}=-\omega^2\atscrip{\frac{\A}{4}m^am^b\bar g_{ab}\other\Box\oOmega} = 0 . 
\end{equation}

We will now turn our attention to another of the Newman-Penrose spin coefficients, $\rho=-m^a\bar m^b\bar\nabla_al_b$ (not to be confused with the energy density $\rho$ defined in section \ref{s3:deriv31}). The spin coefficient $\rho$ characterizes the expansion of the null geodesic congruence given by $l^a$. Evaluating it at $\scri^+$ and using \eref{eg:invariant} and $m^a\bar m^b\bar g_{ab}=1$ yields
\begin{equation}\label{eg:rho}
\atscrip{\rho}=\atscrip{-m^a\bar m^b\bar\nabla_bl_a }=\atscrip{ -A m^am^b\bar\nabla_b\bar\nabla_a\Omega} =\atscrip{-\frac{\A}{4}m^a\bar m^b\bar g_{ab}\bar\Box\Omega} = \atscrip{-\frac{\A}{4}\bar\Box\Omega} . 
\end{equation}
As $\bar\Box\Omega$ does not necessarily vanish at $\scri^+$, the expansion can take nonzero values there. Following the same procedure, the expression of $\rho$ calculated using $\other l^a$ is 
\begin{equation}\label{eg:rhoother}
\atscrip{\rho}=\atscrip{-m^a\bar m^b\other\nabla_b\other l_a }= ... = -\atscrip{\omega^2\frac{\A}{4}\bar\Box\Omega} . 
\end{equation}
The conformal gauge freedom represented by $\omega$ can now be used to impose 
\begin{equation}\label{eg:condother}
\atscrip{\other\Box\oOmega}=0 . 
\end{equation}
This is the preferred conformal gauge condition \cite{Tamburino:1966zz,9780511564048,stewart1997advanced}. Setting it in \eref{eg:boxo} translates it to a condition on $\omega$: 
\begin{equation}\label{eg:condbar}
\atscri{\bar\nabla^a\Omega\bar\nabla_a\ln\omega}=\atscrip{-\frac{1}{4}\bar\Box\Omega} . 
\end{equation}
From \eref{eg:invariantother} the preferred conformal gauge condition \eref{eg:condother} implies 
\begin{equation}\label{eg:nabnabother}
\atscrip{\other\nabla_a\other\nabla_b\oOmega}=0 , 
\end{equation}
and these two together give 
\begin{equation}\label{eg:vvnullother}
\lim_{\oOmega\to0}\frac{1}{\oOmega}\other g^{ab}\other\nabla_a\oOmega\other\nabla_b\oOmega=0 , 
\end{equation}
calculated from the equivalent vacuum equation to \eref{c3:einsteinG} where the conformal rescalings have been performed with $\oOmega$ instead of with $\Omega$. The three conditions \eref{eg:condother}, \eref{eg:nabnabother} and \eref{eg:vvnullother} imply that in the preferred conformal gauge all conformal factor terms resulting from the transformation of the Einstein (or Ricci) tensor attain a regular limit at $\scri^+$ individually. 

\subsection{Bondi time}

The preferred conformal gauge \eref{eg:condother} also implies that the null tangent $\other l^a$ to the null geodesic generators of $\scri^+$ is geodesic, so that it satisfies 
\begin{equation}\label{eg:geodesic_generators}
\atscrip{\other l^a \other\nabla_a \other l^b} = 0 \; .
\end{equation}
The null geodesic generators are parametrized by an affine parameter $t_B$, scaled such that 
\begin{equation}\label{eg:scatB}
\left(\partial_{t_B}\right)^a \other\nabla_a t_B = 1 . 
\end{equation}
This parameter is generally known as Bondi parameter or Bondi time and it gives the proper time measured by a free-falling observer moving along $\scri^+$.

\subsection{Implementation of the preferred conformal gauge}

The quantity $\bar\Box\Omega$, expressed in our spherically symmetric variables and under the assumption that the conformal factor is time-independent, takes the form 
{\small
\begin{eqnarray}\label{eg:boxomegacomps}
\bar\Box\Omega&=&\Omega ' \left(\frac{\dot{\beta ^r}}{\alpha ^2}-\frac{\dot{\alpha } \beta ^r}{\alpha ^3}+\frac{\beta ^r \dot{\gamma _{\theta \theta }}}{\alpha ^2 \gamma _{\theta \theta }}+\frac{\beta ^r \dot{\gamma _{rr}}}{2 \alpha ^2 \gamma _{rr}}-\frac{3 \dot{\chi } \beta ^r}{2 \alpha ^2 \chi }-\frac{{\beta^r}^2 \gamma _{\theta \theta }'}{\alpha ^2 \gamma _{\theta \theta }}+\frac{3 {\beta^r}^2 \chi '}{2 \alpha ^2 \chi }-\frac{2 {\beta^r}^2}{\alpha ^2 r}-\frac{2 \beta ^r \left(\beta ^r\right)'}{\alpha ^2} \right. \nonumber \\ &&\left.+\frac{\alpha ' {\beta^r}^2}{\alpha ^3}-\frac{{\beta^r}^2 \gamma _{rr}'}{2 \alpha ^2 \gamma _{rr}}+\frac{2 \chi }{r \gamma _{rr}}+\frac{\chi  \alpha '}{\alpha  \gamma _{rr}}+\frac{\chi  \gamma   _{\theta \theta }'}{\gamma _{\theta \theta } \gamma _{rr}}-\frac{\chi '}{2 \gamma _{rr}}-\frac{\chi  \gamma _{rr}'}{2 \gamma _{rr}^2}\right)+\Omega '' \left(\frac{\chi }{\gamma _{rr}}-\frac{{\beta^r}^2}{\alpha ^2}\right) . \qquad  \ 
\end{eqnarray}
}%
As a reminder about the notation, the primes denote derivatives with respect to $r$ and the dots indicate time derivatives. The quantities $\dot\gamma_{rr}$, $\dot \gamma_{\theta\theta}$ and $\dot \chi$ are given by the evolution equations. However, the gauge freedom allows us to choose the evolution of the gauge quantities $\dot\alpha$ and $\dot\beta^r$ such that \eref{eg:boxomegacomps}'s RHS vanishes at $\scri^+$ and so $\Omega$ satisfies the preferred conformal gauge. I will explain in more detail how this can be done in section \ref{sg:source}.

\section{Slicing conditions}\label{sg:lapse}

An initial value formulation can be treated as hyperbolic-elliptic problem, where the constraints are solved at each time-step thus giving what is called a constrained evolution, or as a purely hyperbolic problem, where the evolution equations alone determine the evolution of the system and the constraints are used to monitor the quality of the solution. 
The development of evolved slicing conditions was especially motivated by the work on hyperbolic reformulations of the 3+1 evolution equations \cite{PhysRevD.40.1022,Bona:1992zz,bonmas}.

The Bona-Mass\'o family of slicing conditions \cite{Bona:1994dr} is widely used in current numerical simulations. 
To adapt it to the hyperboloidal initial value problem, two extra source functions $K_0$ and $L_0$ that depend on the radial coordinate $r$ have been added to it:
\begin{equation}\label{eg:bonamasso}
\dot \alpha=\beta^r\alpha'-\alpha^2f(\alpha)\left(K-K_0\right)+L_0 . 
\end{equation}
The presence of $K_0$ is necessary because the initial and stationary value of $K$ on the hyperboloidal slice is negative, as can be seen in the initial data plots in section \ref{sin:inidataplots}. As $f(\alpha)=\alpha^n$, a negative value of $K$ is very likely to cause an exponential growth of the $\alpha$ variable. For a more detailed description of this phenomenon see subsection \ref{se:single}. The time independent function $K_0$ is thus introduced to counteract the effect of $K$'s stationary value. 

The Bona-Mass\'o slicing conditions used in traditional Cauchy slices do not need any source functions. As the radial coordinate goes to infinity, $\alpha\to1$ and the trace of the extrinsic curvature vanishes, so that the lapse condition already provides the appropriate stationary solution. However, the behaviour on a hyperboloidal slice is different: even if the rescaled lapse attains a finite value at $\scri^+$ (the physical one becomes infinite there), its flat initial values are not a stationary solution of the lapse evolution equation without source terms. What is more, such an evolution equation tends to make $\alpha$ vanish at $\scri^+$, thus going back to a Cauchy slice. 
Therefore, the presence of a source function $L_0$ that makes the desired value of $\alpha$ stationary seems necessary and it has been used in all the successful tests presented here. 

Note that the slicing condition has been expressed directly in terms of the conformal quantities ($\alpha$, $\cK$) instead of the physical ones ($\tilde \alpha$, $\pK$). The second option is also feasible a priori and has been considered for instance in \cite{Ohme:2009gn}. Nevertheless, for consistency with the relations at $\scri^+$ presented in the previous section, which are expressed in terms of the conformal quantities, the gauge conditions will be imposed on the conformal variables. 

\subsection{Generalized 1+log condition}\label{gs:1plog}

The 1+log slicing condition is obtained by substituting $f(\alpha)=\case{n}{\alpha}$ (here $n$ denotes a real positive number) in \eref{eg:bonamasso}, 
\begin{equation}\label{eg:1ploglapse}
\dot \alpha=\beta^r\alpha'-n\,\alpha\left(K-K_0\right)+L_0 .
\end{equation}
The commonly used choice of $n=2$ has proven to be well-behaved in numerical evolutions of spacetimes with strong gravitational fields \cite{Arbona:1999ym,Alcubierre:2001vm,Alcubierre:2002kk}. It has not been directly set in the previous expression because it does not provide the appropriate eigenspeeds at future null infinity for any value of $\Kc$ - a property of the 1+log slicing condition is that gauge speeds can easily become superluminal. Some eigenspeeds at $\scri^+$ may become larger than the lightspeed there and this causes the appearance of negative characteristic speeds (incoming modes) at $\scri^+$. To see the behaviour of the lightspeeds see \fref{fr:lightspeeds}. As already explained in detail, $\scri^+$ is an ingoing null hypersurface, which implies that no information from the outside can cross it. So, the slicing condition given by $n=2$ is not appropriate. The choice of $n$ that makes the 1+log expression compatible with the appropriate eigenspeeds at $\scri^+$ is
\begin{equation}\label{eg:nokval}
n= -\frac{\Kc\ \rscri}{3} .
\end{equation}

The question of stationary hyperboloidal slices was already raised in \cite{Ohme:2009gn}. The results for the 1+log case were that no stationary solution with nonzero offset $K_0$ could be found, so this slicing condition did not seem appropriate for a hyperboloidal evolution. However, the calculations in \cite{Ohme:2009gn} were performed with the physical quantities on a non-compactified hyperboloidal slice and no source function $L_0$ was considered, so not all of their conclusions necessarily carry over to the case considered here. 

\subsection{Generalized harmonic lapse}

The harmonic slicing condition ($f(\alpha)=1$) was also considered in \cite{Ohme:2009gn} and in this case stationary hyperboloidal slices were found. The approach to the hyperboloidal initial value problem taken in \cite{Zenginoglu:2007it} used the generalized harmonic formulation of the Einstein equations, so that implementing the harmonic lapse condition in our case is likely to provide successful results. In our formulation, the generalized harmonic slicing condition takes the form
\begin{equation}\label{eg:harmlapse}
\dot \alpha=\beta^r\alpha'-\alpha^2\left(K-K_0\right)+L_0  ,
\end{equation}
and the eigenspeeds provided by this equation of motion are in perfect agreement with the requirements at $\scri^+$, as the gauge speed coincides with the speed of light (the effect that the tilting of the causal cone along the hyperboloidal slice has on the speed of light is shown in \fref{fr:lightspeeds}). Actually, the choice in \eref{eg:nokval} is nothing but the value of $\atscrip{\alpha}$ in its stationary initial value. Nevertheless, harmonic slicing is only marginally singularity avoiding, which means that the singularity is reached 
and this is not useful in presence of a BH if we are not using excision.  %\upda{describe more about how $K_0$ and $L_0$ are calculated here or later?}

\subsection{Matching}

\begin{figure}[htbp!!]\label{fg:match}
\center
	\includegraphics[width=0.75\linewidth]{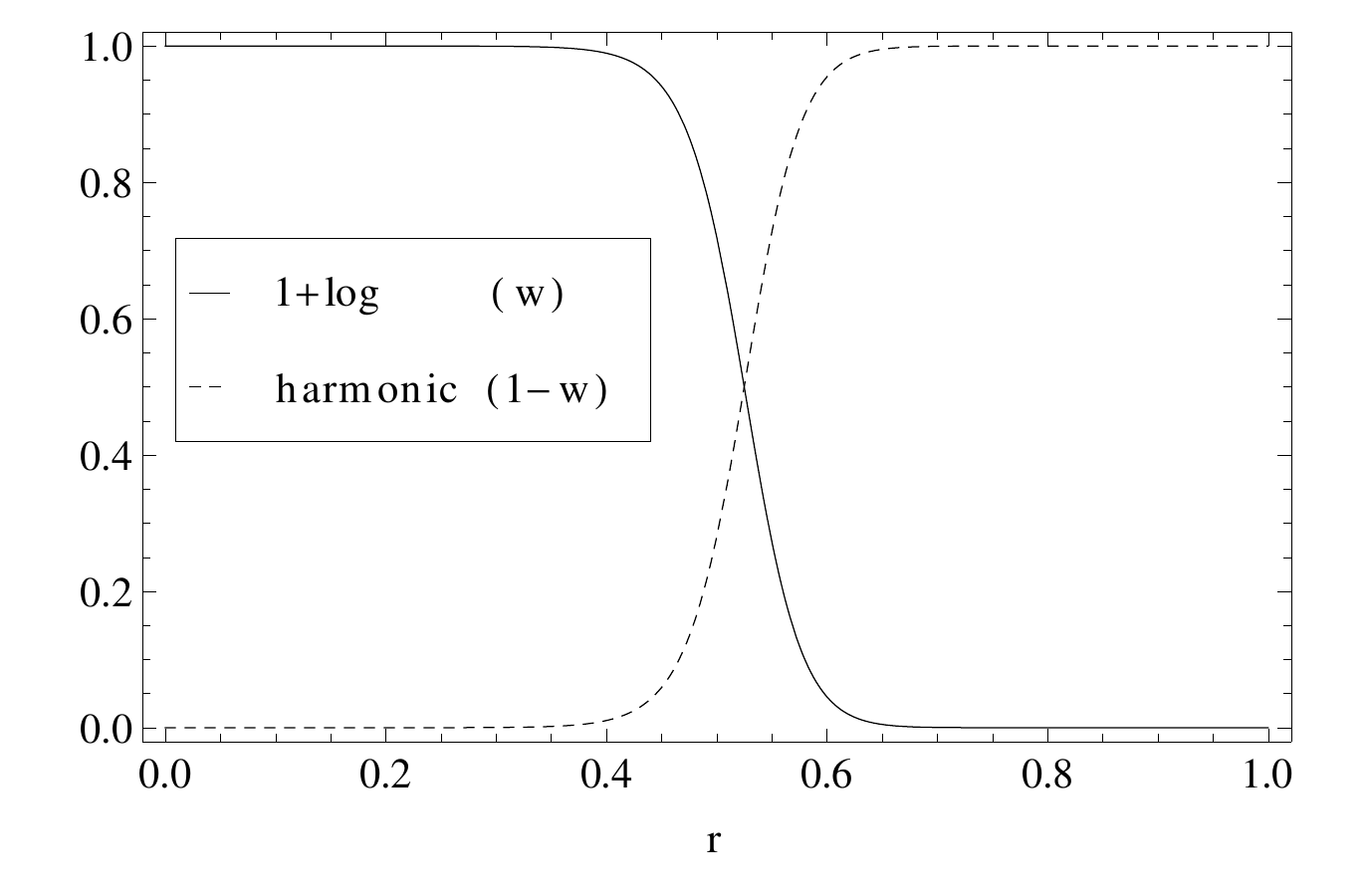}\vspace{-3ex}
\caption{Example of matching weight function \eref{eg:matchw} with $o=2$ and $p=8$.}
\end{figure}

A way to profit from the singularity-avoidance property of the 1+log slicing and the more natural behaviour at null infinity of the harmonic slicing condition is to use the first in the interior part of the domain and the latter in the region close to $\scri^+$. This would allow to use any convenient value of $n$. There are several things to take into account when performing the matching: where in the domain it is done (e.g. inside or outside of the horizon), how wide the matching region is, make sure the transition is smooth, check if the change in eigenspeeds is reasonable, how the source functions are calculated, etc. An example of a match could be using the weight
\begin{equation}\label{eg:matchw}
w=\frac{1}{1+\left(\frac{r}{(\rscri^2-r^2)^o}\right)^p}  ,
\end{equation}
with parameters $o$ and $p$ that control the location and the amplitude of the match. The final slicing condition is obtained by multiplying the 1+log and harmonic parts by 
\begin{equation}
\dot \alpha=\beta^r\alpha' +w\left[-n\alpha\left(K-K_{0\,log}\right)+L_{0\,log}\right] +(1-w)\left[-\alpha^2\left(K-K_{0\,har}\right)+L_{0\,har}\right] . 
\end{equation}
An example of the corresponding weight functions with parameters $o=2$ and $p=8$ is plotted in \fref{fg:match}. Close to the origin the contribution of the harmonic condition is set to zero, while near $\scri^+$ (located at $r=1$) the 1+log condition will have no effect. 

\section{Shift conditions}\label{sg:shift}

The shift condition controls the change in the spatial coordinates during the evolution. The shift plays a very important role in the scri-fixing condition, because its profile ensures that the time vector flows along $\scri^+$. In the evolved shift conditions that will be considered in this section, a source term will be added to the RHSs for the same reason as done previously with the lapse conditions. In some cases, a damping term for the shift will also be included. 

\subsection{Fixed shift}

A time-independent non-vanishing shift compatible with the scri-fixing condition \eref{eg:scrifixspher} is the simplest possible choice. 
A numerical evolution with a fixed shift is appropriate for flat initial data or including a scalar field perturbation that does not collapse into a BH, but strong field initial data do require a more sophisticated shift condition. However, the fixed shift option allowed to simplify the system in the first tests in flat space time. 

The initial value of the shift in flat spacetime is ${\beta^r}_0= \frac{\Kc\,r}{3}$, calculated from \eref{ein:betarinio}. It is negative at $\scri^+$, a requirement for the shift that satisfies the scri-fixing condition as indicated by the diagram in \fref{fg:scridiagram}. 

\subsection{Generalized Gamma-driver}\label{sg:gammadriver}

The ``driver'' conditions have their origin in relaxation techniques to solve elliptic shift equations \cite{Smarr:1977uf,Smarr:1978dia}, like the minimal strain, minimal distortion and Gamma-freezing conditions. 
Among the hyperbolic driver conditions there is the Gamma-driver shift condition \cite{Alcubierre:2002kk}, which seeks to mimic the effect of the Gamma-freezing condition. It has been successful in BH simulations with moving punctures, where the inclusion of advection terms to fix the structure of light cones seems beneficial \cite{Baker:2006yw,vanMeter:2006vi,Gundlach:2006tw}. % simulations with moving punctures: \cite{Baker:2006yw}

The generalized Gamma-driver shift condition adapted to our setup is expressed in terms of $\Lambda^r$ instead of $\Gamma^r$. 
%This is actually an interesting point, because using $\Lambda^a$ (a tensor) allows to write the shift conditions in a covariant form. 
According to common use, a contravariant vector $B^a$ is introduced as an auxiliary variable. The equations of motion are given by: 
\begin{subequations}\label{eg:Gammadriver}
\begin{eqnarray}
\dot \beta^r & = & \beta^r{\beta^r}'+\frac{3}{4} \mu B^r + L_0 -\frac{\xi_{\beta^r}}{\Omega}\beta^r , \\ %(\mu+4(1-r^2)) -\frac{K_{CMC}^2\xi_{\beta^r}}{9\Omega}\beta^r
\dot B^r & = & \beta^r {B^r}'-\eta B^r +\lambda\left(\dot \Lambda^r-\beta^r{\Lambda^r}'\right)  . 
\end{eqnarray}	
\end{subequations}
In the evolution equation for $B^a$, $\partial_\perp \Lambda^a$ has to be substituted by the RHS of the equation of motion for $\Lambda^r$, either \eref{es:Lambdardot}, \eref{es:pKLambdardot} or \eref{es:DPKLambdardot}. The parameter $\eta$ is especially relevant when evolving BHs, see \cite{Husa:2007zz}, and the parameters $\lambda$ and $\mu$ have to be chosen carefully, because they determine the eigenspeeds. The source function $L_0$ and the damping term ($-\case{\xi_{\beta^r}}{\Omega}\beta^r$) ensure  that the value of $\beta^r$ stays fixed at $\scri^+$, a similar behaviour as in the fixed shift case. The initial and stationary value of the auxiliary variable is $B^r{}_0=0$. %Appropriate values for our simulations are: $\xi_{\beta^r}=5$, $\lambda=3/4$, $\mu=0.15+4(1-r^2)$ and $\eta=0.1$. 

The Gamma-driver condition can also be implemented in its integrated form: 
\begin{equation}\label{eg:integGammadriver}
\dot \beta^r = \beta^r{\beta^r}' + \lambda \, \Lambda^r - \eta \,\beta^r  + L_0 -\frac{\xi_{\beta^r}}{\Omega}\beta^r . 
\end{equation}
Written in this way it has similar structure to that of the harmonic shift condition, which will be described next. 

%\upda{mention something about the rescaled shift: $\sigma^i=\beta^i/\alpha$??}

\subsection{Generalized harmonic shift}\label{sg:harmshi}

The harmonic shift condition is a generalization of the condition for harmonic spatial coordinates \cite{Alcubierre:2005gh}. 
% equation (4.3.37) in page 149 of Alcubierre "has been known for a long time" \cite{Garfinkle:2001ni,york}. include eq? 
In our case, the generalized harmonic lapse and shift conditions are derived from the harmonic coordinate condition on the scalar coordinate functions $\varphi^a$, $\bar\Box \varphi^c = F^c$, where a generalization to a non-vanishing source function $F^c$ has been included. After identifying $\varphi^\mu=x^\mu$, the generalized harmonic condition simply becomes
\begin{equation}\label{eg:gharm}
\bar g^{ab}\bar\Gamma^c_{ab}=F^c . 
\end{equation}
The generalized harmonic lapse condition is given by the time component and the shift condition in spherical symmetry by the radial component: 
{\small
\begin{subequations}\label{eg:harmtwo}
\begin{eqnarray}
\dot\alpha &=& \alpha ' \beta ^r + \frac{\alpha  \dot{\gamma _{\theta \theta }}}{\gamma _{\theta \theta }}-\frac{3 \alpha  \dot{\chi }}{2 \chi }+\frac{\alpha  \dot{\gamma _{rr}}}{2 \gamma _{rr}}-\frac{\alpha  \beta ^r \gamma _{\theta \theta }'}{\gamma _{\theta \theta }}+\frac{3 \alpha  \beta ^r \chi '}{2 \chi }-\alpha {\beta  ^r}'-\frac{2 \alpha  \beta ^r}{r}-\frac{\alpha  \beta ^r \gamma _{rr}'}{2 \gamma _{rr}} -\alpha ^3 F^t ,\qquad \quad \label{eg:harmderlapse} \\
\dot \beta^r &=& 2 \beta ^r {\beta ^r}'+\frac{\dot{\alpha } \beta ^r}{\alpha }-\frac{\beta ^r \dot{\gamma _{\theta \theta }}}{\gamma _{\theta \theta }}+\frac{3 \dot{\chi } \beta ^r}{2 \chi }-\frac{\beta ^r \dot{\gamma _{rr}}}{2 \gamma _{rr}}-\frac{\alpha ' {\beta^r}^2}{\alpha }+\frac{{\beta^r}^2 \gamma _{\theta \theta }'}{\gamma _{\theta \theta }}-\frac{3 {\beta^r}^2 \chi '}{2 \chi }+\frac{2 {\beta^r}^2}{r} -\frac{2 \alpha ^2 \chi }{r \gamma _{rr}} \nonumber \\ && +\frac{{\beta^r}^2 \gamma _{rr}'}{2 \gamma _{rr}}-\frac{\alpha ^2 \chi  \gamma _{\theta \theta }'}{\gamma _{\theta \theta } \gamma _{rr}}+\frac{\alpha ^2 \chi '}{2 \gamma _{rr}}+\frac{\alpha ^2 \chi  \gamma _{rr}'}{2 \gamma _{rr}^2}-\frac{\alpha  \chi  \alpha '}{\gamma _{rr}} -\alpha ^2 F^r.  \label{eg:harmdershift}
\end{eqnarray}	
\end{subequations}
}%
Substituting the corresponding evolution equations from \eref{es:eeqs}, the previous \eref{eg:harmderlapse} will become \eref{eg:harmlapse}, without the $K_0$ function and a differently introduced source term. In \eref{eg:harmlapse} the term with $\cT$ was dropped, for similarity with the implementation in \cite{Weyhausen:2011cg}. %because in earlier tests it was causing the simulation to crash. 

The harmonic shift condition is obtained from \eref{eg:harmdershift}. First \eref{eg:harmderlapse} is set and then $\gamma_{rr}'$ and $\gamma_{\theta\theta}'$ are substituted in terms of $\Lambda^r$ using the RHS of \eref{Zrdef} set to zero. This is necessary for hyperbolicity reasons. Finally the following expression is obtained, with the only difference that the expected $-\alpha ^2 \beta ^rF^t-\alpha ^2 F^r$ have been substituted by the last two terms in \eref{eg:integGammadriver}
\begin{equation}
\dot \beta^r = \beta^r{\beta^r}'+\alpha^2\chi\Lambda ^r+\frac{\alpha ^2 \chi'}{2 \gamma_{rr}}-\frac{\alpha\chi\alpha'}{\gamma_{rr}}-\frac{2 \alpha^2\chi}{r\gamma_{\theta\theta}} + L_0 -\frac{\xi_{\beta^r}}{\Omega}\beta^r . 
\end{equation}
%Note that this equation includes a term with $r$ in its denominator that will diverge at the origin. This term does not belong to the principal part (the terms in the equations that determine hyperbolicity and the eigenspeeds and eigenfields of the system, see section \ref{sn:wellposed}), so it should be dropped to avoid problems at the origin. 
This equation includes a term with $r$ in its denominator that will diverge at the origin and that is therefore likely to raise an instability there. As it does not belong to the principal part (the terms in the equations that determine hyperbolicity and the characteristic behaviour of the system, see section \ref{sn:wellposed}), it should be dropped to avoid problems at the origin. 

\section[Generalized harmonic gauge]{Generalized harmonic gauge conditions compatible with the preferred conformal gauge}\label{sg:source}

Aiming to systematize the calculation of the source functions in the equations, a new approach will be taken. Motivated by the definition of the spatial quantity $\Delta\Gamma^a$ \eref{et:DefineGamma}, we will consider the difference of four-dimensional Christoffel symbols, so that the source functions will be expressed in terms of the components of a time-independent background metric. 
The spatial part has also been used in a discussion about harmonic conformal spatial coordinates in \cite{talkmoncrief}. % leading to an elliptic equation

The expression can be defined in terms of the conformal or the physical metric (so far all the gauge conditions have been set using the conformal quantities). 

\subsection{In the conformal picture}

The source functions added to the generalized harmonic condition \eref{eg:gharm} are expressed in terms of the background connection $\hat{\bar\Gamma}^c_{ab}$ calculated from a time independent 4-dimensional metric $\hat{\bar g}_{ab}$. 
This new generalized harmonic gauge can be written as  
\begin{equation}\label{eg:gharmb}
%\bar g^{ab}\left( \bar\Gamma^c_{ab}-\hat{\bar\Gamma}^c_{ab} \right)=\bar F^c . %, \qquad \textrm{with}\quad \bar\Lambda^c \bar\Lambda^c = \bar\Gamma^c_{ab}-\hat{\bar\Gamma}^c_{ab}
\bar\Lambda^c=\bar F^c , \qquad \textrm{with}\quad \bar\Lambda^c=\bar g^{ab}\bar\Lambda^c_{ab}  \quad \textrm{and} \quad \bar\Lambda^c_{ab} = \bar\Gamma^c_{ab}-\hat{\bar\Gamma}^c_{ab} . 
\end{equation}
Following the notation used in \eref{et:DefineGamma}, it may seem more appropriate to denote $\bar\Lambda^c$ by $\Delta\bar{\Gamma}^c$, but using $\bar\Lambda^c$ will explicitly remind us to perform the substitution of $\gamma_{rr}'$ and $\gamma_{\theta\theta}'$ in terms of $\Lambda^r$ mentioned in subsection \ref{sg:harmshi}. 

We will now show how these gauge conditions are a priori compatible with the preferred conformal gauge, whose condition can be expanded as
\begin{equation}\label{eg:prefdecomp}
\bar\Box\Omega = \bar g^{ab}\bar\nabla_a\bar\nabla_b\Omega= \bar g^{ab}\partial_a\partial_b\Omega-\bar g^{ab}\bar\Gamma^c_{ab}\partial_c\Omega
\end{equation}
Isolating $\bar g^{ab}\bar\Gamma^c_{ab}$ from \eref{eg:gharmb} and substituting it into \eref{eg:prefdecomp} gives the following relation, where in the second equality we have used the fact that in our approach the conformal factor $\Omega$ only depends on the radius, 
\begin{equation}\label{eg:prefdecompr}
\bar\Box\Omega = \bar g^{ab}\partial_a\partial_b\Omega-\bar g^{ab}\hat{\bar\Gamma}^c_{ab}\partial_c\Omega -\bar F^c\partial_c\Omega =  \bar g^{rr}\Omega''-\bar g^{ab}\hat{\bar\Gamma}^r_{ab}\Omega' -\bar F^r\Omega'. 
\end{equation}
The quantity $\atscri{\Omega'}\neq0$ by definition and in principle also $\atscri{\Omega''}\neq0$. The metric component $\bar g^{rr}$ can be read off from \eref{e3:fourmetric} giving $\bar g^{rr}=\frac{\chi}{\gamma_{rr}}-\frac{{\beta^r}^2}{\alpha^2}$, which vanishes at $\scri^+$ by means of the scri-fixing condition \eref{eg:scrifixspher} - the stationary values satisfy $\frac{\chi}{\gamma_{rr}}-\frac{{\beta^r}^2}{\alpha^2}=\frac{\chi\Omega^2}{\gamma_{rr}\alpha^2}$, so $\atscrip{\bar g^{rr}}=\mathcal{O}(\Omega^2)$. The background Christoffel symbol $\hat{\bar\Gamma}^r_{ab}$ is calculated from the initial data and it is such that at $\scri^+$ the only nonzero component is $\atscrip{\hat{\bar\Gamma}^r_{rr}}=\atscrip{\case{1}{r}}$. Then $\bar g^{rr}\hat{\bar\Gamma}^r_{rr}$ will be zero at future null infinity by virtue of the scri-fixing condition. The only remaining term is $-\bar F^r\Omega'$: if $\atscrip{\bar F^r}\propto\Omega^q$ with $q>0$, then \eref{eg:prefdecompr}'s RHS will vanish at $\scri^+$ and the preferred conformal gauge condition $\atscri{\bar\Box\Omega}=0$ will hold. 

The explicit gauge evolution equations are given by 
{\small
\begingroup
\allowdisplaybreaks
\begin{subequations}\label{eg:harmtwob}
\begin{eqnarray}
\dot\alpha &=& \frac{\alpha  \dot{\gamma _{\theta \theta }}}{\gamma _{\theta \theta }}-\frac{3 \alpha  \dot{\chi }}{2 \chi }+\frac{\alpha  \dot{\gamma _{rr}}}{2 \gamma _{rr}}-\alpha ^3 \bar F^t-\frac{\alpha ^3 \chi  \hat{\gamma _{\theta \theta }} \hat{\beta ^r} \hat{\chi }'}{\hat{\alpha }^2 \hat{\chi }^2 \gamma _{\theta \theta }}+\frac{\alpha ^3 \chi  \hat{\beta ^r} \hat{\gamma _{\theta \theta }}'}{\hat{\alpha }^2 \hat{\chi } \gamma _{\theta \theta }}+\frac{2 \alpha ^3 \chi  \hat{\gamma _{\theta \theta }} \hat{\beta ^r}}{\hat{\alpha }^2 r \hat{\chi } \gamma _{\theta \theta }}+\alpha ' \beta ^r \nonumber \\*
 && -\frac{\alpha  \beta ^r \gamma _{\theta \theta }'}{\gamma _{\theta \theta }}+\frac{3 \alpha  \beta ^r \chi '}{2 \chi }-\frac{2 \alpha  \hat{\alpha }' \beta ^r}{\hat{\alpha }}-\alpha  \left(\beta ^r\right)'+\frac{\alpha  \hat{\alpha }' \hat{\beta ^r}}{\hat{\alpha }}-\frac{2 \alpha  \beta ^r}{r}+\frac{\alpha ^3 \chi  \hat{\gamma _{rr}} \hat{\beta ^r}'}{\hat{\alpha }^2 \hat{\chi } \gamma _{rr}}-\frac{\alpha ^3 \chi  \hat{\beta ^r} \hat{\gamma _{rr}} \hat{\chi }'}{2 \hat{\alpha }^2 \hat{\chi }^2 \gamma _{rr}}\nonumber \\*
 &&+\frac{\alpha ^3 \chi  \hat{\beta ^r} \hat{\gamma _{rr}}'}{2 \hat{\alpha }^2 \hat{\chi } \gamma _{rr}}+\frac{2 \alpha  \hat{\beta ^r} \beta ^r \hat{\gamma _{rr}} \hat{\beta ^r}'}{\hat{\alpha }^2 \hat{\chi }}-\frac{\alpha  \hat{\beta ^r}^2 \beta ^r \hat{\gamma _{rr}} \hat{\chi }'}{\hat{\alpha }^2 \hat{\chi }^2}+\frac{\alpha  \hat{\beta ^r}^2 \beta ^r \hat{\gamma _{rr}}'}{\hat{\alpha }^2 \hat{\chi }}-\frac{\alpha  \hat{\beta ^r}^2 \hat{\gamma _{rr}} \hat{\beta ^r}'}{\hat{\alpha }^2 \hat{\chi }}+\frac{\alpha  \hat{\beta ^r}^3 \hat{\gamma _{rr}} \hat{\chi }'}{2 \hat{\alpha }^2 \hat{\chi }^2}\nonumber \\*
 &&-\frac{\alpha  \hat{\beta ^r}^3 \hat{\gamma _{rr}}'}{2 \hat{\alpha }^2 \hat{\chi }}-\frac{\alpha  \beta ^r \gamma _{rr}'}{2 \gamma _{rr}}-\frac{\alpha  \beta ^{2 r} \hat{\gamma _{rr}} \hat{\beta ^r}'}{\hat{\alpha }^2 \hat{\chi }}+\frac{\alpha  \hat{\beta ^r} \beta ^{2 r} \hat{\gamma _{rr}} \hat{\chi }'}{2 \hat{\alpha }^2 \hat{\chi }^2}-\frac{\alpha  \hat{\beta ^r} \beta ^{2 r} \hat{\gamma _{rr}}'}{2 \hat{\alpha }^2 \hat{\chi }} , \label{eg:harmderlapseb} \\
\dot \beta^r &=&  \frac{\dot{\alpha } \beta ^r}{\alpha }+\frac{3 \dot{\chi } \beta ^r}{2 \chi }+2 \left(\beta ^r\right)' \beta ^r+\frac{2 \hat{\beta ^r} \hat{\alpha }' \beta ^r}{\hat{\alpha }}-\frac{2 \hat{\beta ^r}^2 \hat{\gamma _{rr}} \hat{\beta ^r}' \beta ^r}{\hat{\alpha }^2 \hat{\chi }}+\frac{\hat{\beta ^r}^3 \hat{\gamma _{rr}} \hat{\chi }' \beta ^r}{\hat{\alpha }^2 \hat{\chi }^2}-\frac{\hat{\beta ^r}^3 \hat{\gamma _{rr}}' \beta ^r}{\hat{\alpha }^2 \hat{\chi }}-\frac{\dot{\gamma _{rr}} \beta ^r}{2 \gamma _{rr}}\nonumber \\*
 &&-\frac{\dot{\gamma _{\theta \theta }} \beta ^r}{\gamma _{\theta \theta }}-\frac{\alpha ' \beta ^{2 r}}{\alpha }-\frac{3 \chi ' \beta ^{2 r}}{2 \chi }+\frac{\hat{\beta ^r} \hat{\gamma _{rr}} \hat{\beta ^r}' \beta ^{2 r}}{\hat{\alpha }^2 \hat{\chi }}-\frac{\hat{\beta ^r}^2 \hat{\gamma _{rr}} \hat{\chi }' \beta ^{2 r}}{2 \hat{\alpha }^2 \hat{\chi }^2}-\frac{\hat{\chi }' \beta ^{2 r}}{2 \hat{\chi }}+\frac{\hat{\beta ^r}^2 \hat{\gamma _{rr}}' \beta ^{2 r}}{2 \hat{\alpha }^2 \hat{\chi }}\nonumber \\*
 &&+\frac{\hat{\gamma _{rr}}' \beta ^{2 r}}{2 \hat{\gamma _{rr}}}+\frac{\gamma _{rr}' \beta ^{2 r}}{2 \gamma _{rr}}+\frac{\gamma _{\theta \theta }' \beta ^{2 r}}{\gamma _{\theta \theta }}+\frac{2 \beta ^{2 r}}{r}-\bar F^r \alpha ^2-\frac{\alpha  \chi  \alpha '}{\gamma _{rr}}+\frac{\alpha ^2 \chi '}{2 \gamma _{rr}}-\frac{\hat{\beta ^r}^2 \hat{\alpha }'}{\hat{\alpha }}+\frac{\hat{\alpha } \hat{\chi } \hat{\alpha }'}{\hat{\gamma _{rr}}}\nonumber \\*
 &&-\hat{\beta ^r} \hat{\beta ^r}'+\frac{\hat{\beta ^r}^3 \hat{\gamma _{rr}} \hat{\beta ^r}'}{\hat{\alpha }^2 \hat{\chi }}-\frac{\alpha ^2 \chi  \hat{\beta ^r} \hat{\gamma _{rr}} \hat{\beta ^r}'}{\hat{\alpha }^2 \hat{\chi } \gamma _{rr}}-\frac{\hat{\beta ^r}^4 \hat{\gamma _{rr}} \hat{\chi }'}{2 \hat{\alpha }^2 \hat{\chi }^2}+\frac{\hat{\beta ^r}^2 \hat{\chi }'}{2 \hat{\chi }}+\frac{\alpha ^2 \chi  \hat{\beta ^r}^2 \hat{\gamma _{rr}} \hat{\chi }'}{2 \hat{\alpha }^2 \hat{\chi }^2 \gamma _{rr}}+\frac{\alpha ^2 \chi  \hat{\chi }'}{2 \hat{\chi } \gamma _{rr}}\nonumber \\*
 &&-\frac{\alpha ^2 \chi  \hat{\gamma _{\theta \theta }} \hat{\chi }'}{\hat{\chi } \hat{\gamma _{rr}} \gamma _{\theta \theta }}+\frac{\alpha ^2 \chi  \hat{\beta ^r}^2 \hat{\gamma _{\theta \theta }} \hat{\chi }'}{\hat{\alpha }^2 \hat{\chi }^2 \gamma _{\theta \theta }}+\frac{\hat{\beta ^r}^4 \hat{\gamma _{rr}}'}{2 \hat{\alpha }^2 \hat{\chi }}-\frac{\hat{\beta ^r}^2 \hat{\gamma _{rr}}'}{2 \hat{\gamma _{rr}}}-\frac{\alpha ^2 \chi  \hat{\beta ^r}^2 \hat{\gamma _{rr}}'}{2 \hat{\alpha }^2 \hat{\chi } \gamma _{rr}}-\frac{\alpha ^2 \chi  \hat{\gamma _{rr}}'}{2 \hat{\gamma _{rr}} \gamma _{rr}}\nonumber \\*
 &&-\frac{\alpha ^2 \chi  \hat{\beta ^r}^2 \hat{\gamma _{\theta \theta }}'}{\hat{\alpha }^2 \hat{\chi } \gamma _{\theta \theta }}+\frac{\alpha ^2 \chi  \hat{\gamma _{\theta \theta }}'}{\hat{\gamma _{rr}} \gamma _{\theta \theta }}+\frac{\alpha ^2 \chi  \gamma _{rr}'}{2 \gamma _{rr}^2}-\frac{\alpha ^2 \chi  \gamma _{\theta \theta }'}{\gamma _{rr} \gamma _{\theta \theta }}-\frac{2 \alpha ^2 \chi }{r \gamma _{rr}}-\frac{2 \alpha ^2 \chi  \hat{\beta ^r}^2 \hat{\gamma _{\theta \theta }}}{r \hat{\alpha }^2 \hat{\chi } \gamma _{\theta \theta }}+\frac{2 \alpha ^2 \chi  \hat{\gamma _{\theta \theta }}}{r \hat{\gamma _{rr}} \gamma _{\theta \theta }} .  \label{eg:harmdershiftb}
\end{eqnarray}	
\end{subequations}
\endgroup
}%
The background metric components are denoted by $\hat\alpha$, $\hat{\beta^r}$, $\hat\chi$, $\hat{\gamma_{rr}}$ and $\hat{\gamma_{\theta\theta}}$. The two last ones coincide with those of the background spatial metric introduced in \eref{g:Lambdaa}. In the following simplified version of the previous equations, $\hat{\gamma_{rr}}$ and $\hat{\gamma_{\theta\theta}}$ are set to unity, as done in \eref{es:gb}, the evolution equation $\dot\alpha$ is substituted in $\dot\beta^r$ and $\Lambda^r$ is also introduced: % there for hyperbolicity reasons (see subsection \ref{sg:harmshi}): 
{\small
\begin{subequations}\label{eg:harmtwobsimpl}
\begin{eqnarray}
\dot\alpha &=& 
{\beta^r} \alpha '-\alpha  {\beta^r}'-\frac{\alpha  {\beta^r} \gamma _{\theta \theta }'}{\gamma _{\theta \theta }}+\frac{3 \alpha  {\beta^r} \chi '}{2 \chi }-\frac{2 \alpha  {\beta^r} \hat{\alpha }'}{\hat{\alpha }}+\frac{\alpha  \dot{\gamma _{\theta \theta }}}{\gamma _{\theta \theta }}-\frac{3 \alpha  \dot{\chi }}{2 \chi }-\alpha ^3 \bar F^t-\frac{\alpha ^3 \chi  \hat{\beta ^r} \hat{\chi }'}{\hat{\alpha }^2 \hat{\chi }^2 \gamma _{\theta \theta }}+\frac{2 \alpha ^3 \chi  \hat{\beta ^r}}{\hat{\alpha }^2 r \hat{\chi } \gamma _{\theta \theta }}
\nonumber \\ && 
-\frac{\alpha  {\beta^r}^2 \hat{\beta ^r}'}{\hat{\alpha }^2 \hat{\chi }}+\frac{\alpha  {\beta^r}^2 \hat{\beta ^r} \hat{\chi }'}{2 \hat{\alpha }^2 \hat{\chi }^2}+\frac{2 \alpha  {\beta^r} \hat{\beta ^r} \hat{\beta ^r}'}{\hat{\alpha }^2 \hat{\chi }}-\frac{\alpha  {\beta^r} \hat{\beta ^r}^2 \hat{\chi }'}{\hat{\alpha }^2 \hat{\chi }^2}-\frac{\alpha  \hat{\beta ^r}^2 \hat{\beta ^r}'}{\hat{\alpha }^2 \hat{\chi }}+\frac{\alpha  \hat{\beta ^r}^3 \hat{\chi }'}{2 \hat{\alpha }^2 \hat{\chi }^2}+\frac{\alpha  \hat{\alpha }' \hat{\beta ^r}}{\hat{\alpha }}-\frac{2 \alpha  {\beta^r}}{r}
\nonumber \\ && 
+\frac{\alpha ^3 \chi  \hat{\beta ^r}'}{\hat{\alpha }^2 \hat{\chi } \gamma _{rr}}-\frac{\alpha ^3 \chi  \hat{\beta ^r} \hat{\chi }'}{2 \hat{\alpha }^2 \hat{\chi }^2 \gamma _{rr}}-\frac{\alpha  {\beta^r} \gamma _{rr}'}{2 \gamma _{rr}}+\frac{\alpha  \dot{\gamma _{rr}}}{2 \gamma _{rr}}
 , \label{eg:harmderlapsebsimpl} \\
\dot \beta^r &=&  
-\frac{\alpha ^2 \chi  \hat{\chi }'}{\hat{\chi } \gamma _{\theta \theta }}-\frac{2 {\beta^r}^2 \hat{\alpha }'}{\hat{\alpha }}+\hat{\alpha } \hat{\chi } \hat{\alpha }'-\frac{{\beta^r}^2 \hat{\chi }'}{2 \hat{\chi }}+{\beta^r} {\beta^r}'-\alpha ^2 {\beta^r} \bar F^t-\alpha ^2 \bar F^r-\frac{\alpha ^2 {\beta^r} \chi  \hat{\beta ^r} \hat{\chi }'}{\hat{\alpha }^2 \hat{\chi }^2 \gamma _{\theta \theta }}+\frac{2 \alpha ^2 {\beta^r} \chi  \hat{\beta ^r}}{\hat{\alpha }^2 r \hat{\chi } \gamma _{\theta \theta }}
\nonumber \\ && 
+\frac{\alpha ^2 \chi  \hat{\beta ^r}^2 \hat{\chi }'}{\hat{\alpha }^2 \hat{\chi }^2 \gamma _{\theta \theta }}-\frac{2 \alpha ^2 \chi  \hat{\beta ^r}^2}{\hat{\alpha }^2 r \hat{\chi } \gamma _{\theta \theta }}+\alpha ^2 \chi  \Lambda ^r-\frac{{\beta^r}^3 \hat{\beta ^r}'}{\hat{\alpha }^2 \hat{\chi }}+\frac{{\beta^r}^3 \hat{\beta ^r} \hat{\chi }'}{2 \hat{\alpha }^2 \hat{\chi }^2}+\frac{3 {\beta^r}^2 \hat{\beta ^r} \hat{\beta ^r}'}{\hat{\alpha }^2 \hat{\chi }}-\frac{3 {\beta^r}^2 \hat{\beta ^r}^2 \hat{\chi }'}{2 \hat{\alpha }^2 \hat{\chi }^2}
\nonumber \\ && 
-\frac{3 {\beta^r} \hat{\beta ^r}^2 \hat{\beta ^r}'}{\hat{\alpha }^2 \hat{\chi }}+\frac{3 {\beta^r} \hat{\beta ^r}^3 \hat{\chi }'}{2 \hat{\alpha }^2 \hat{\chi }^2}+\frac{3 {\beta^r} \hat{\alpha }' \hat{\beta ^r}}{\hat{\alpha }}+\frac{\hat{\beta ^r}^3 \hat{\beta ^r}'}{\hat{\alpha }^2 \hat{\chi }}-\frac{\hat{\beta ^r}^4 \hat{\chi }'}{2 \hat{\alpha }^2 \hat{\chi }^2}-\frac{\hat{\alpha }' \hat{\beta ^r}^2}{\hat{\alpha }}+\frac{\hat{\beta ^r}^2 \hat{\chi }'}{2 \hat{\chi }}-\hat{\beta ^r} \hat{\beta ^r}'
\nonumber \\ && 
+\frac{\alpha ^2 {\beta^r} \chi  \hat{\beta ^r}'}{\hat{\alpha }^2 \hat{\chi } \gamma _{rr}}-\frac{\alpha ^2 {\beta^r} \chi  \hat{\beta ^r} \hat{\chi }'}{2 \hat{\alpha }^2 \hat{\chi }^2 \gamma _{rr}}-\frac{\alpha ^2 \chi  \hat{\beta ^r} \hat{\beta ^r}'}{\hat{\alpha }^2 \hat{\chi } \gamma _{rr}}+\frac{\alpha ^2 \chi  \hat{\beta ^r}^2 \hat{\chi }'}{2 \hat{\alpha }^2 \hat{\chi }^2 \gamma _{rr}}+\frac{\alpha ^2 \chi '}{2 \gamma _{rr}}+\frac{\alpha ^2 \chi  \hat{\chi }'}{2 \hat{\chi } \gamma _{rr}}-\frac{\alpha  \chi  \alpha '}{\gamma _{rr}}
 .  \label{eg:harmdershiftbsimpl}
\end{eqnarray}	
\end{subequations}
}%
The quantities $\dot\chi$, $\dot\gamma_{rr}$ and $\dot\gamma_{\theta\theta}$ have not been substituted to maintain the freedom to choose any of the evolution systems \eref{es:eeqs}, \eref{es:pKeeqs} or \eref{es:DPKeeqs}. 

\subsection{In the physical picture}

If instead of using conformal quantities in the definition we use the physical ones, the gauge condition is formally the same,
\begin{equation}\label{eg:gharmpb}
\tilde\Lambda^c=\tilde F^c , \qquad \textrm{with}\quad \tilde\Lambda^c=\tilde g^{ab}\tilde\Lambda^c_{ab}  \quad \textrm{and} \quad \tilde\Lambda^c_{ab} = \tilde\Gamma^c_{ab}-\hat{\tilde\Gamma}^c_{ab} . 
\end{equation}
The physical background connection $\hat{\tilde\Gamma}^c_{ab}$ is constructed from a time-independent physical metric $\hat{\tilde g}_{ab}$. 
Transforming the above condition to the conformal quantities (using \eref{ei:rescmetric}, \eref{c3:Chrischange} for the evolved and background metric and Christoffel symbols and the quantities defined in \eref{eg:gharmb}) gives
 \begin{equation}\label{eg:gharmpbtranf}
\bar\Lambda^c=-\frac{\partial_d\Omega}{\Omega}\left(4\bar g^{cd}-\bar g^{ab} \hat{\bar g}_{ab} \hat{\bar g}^{cd}\right) + \frac{\tilde F^c}{\Omega^2} . 
\end{equation}
The principal part of the equation is exactly the same as is \eref{eg:gharmb}, but the source terms involve factors of $\Omega$. Using \eref{eg:gharmpb} on the preferred conformal gauge now yields 
\begin{eqnarray}\label{eg:prefdecomprp}
\bar\Box\Omega &=& \bar g^{ab}\partial_a\partial_b\Omega-\bar g^{ab}\hat{\bar\Gamma}^c_{ab}\partial_c\Omega +\frac{\partial_c\Omega\partial_d\Omega}{\Omega}\left(4\bar g^{cd}-\bar g^{ab} \hat{\bar g}_{ab} \hat{\bar g}^{cd}\right) -\frac{\tilde F^c\partial_c\Omega}{\Omega^2} \nonumber \\
&=&  \bar g^{rr}\Omega''-\bar g^{ab}\hat{\bar\Gamma}^r_{ab}\Omega'+\frac{(\Omega')^2}{\Omega}\left(4\bar g^{rr}-\bar g^{ab} \hat{\bar g}_{ab} \hat{\bar g}^{rr}\right) -\frac{\tilde F^r\Omega'}{\Omega^2}. 
\end{eqnarray}
The vanishing of the first and second terms in \eref{eg:prefdecomprp} at $\scri^+$ has been shown in the previous subsection. The third term also becomes zero at null infinity, because $\atscrip{\bar g^{rr}}=\atscrip{\hat{\bar g}^{rr}}=\mathcal{O}(\Omega^2)$ and $\bar g^{ab} \hat{\bar g}_{ab}$, a combination of conformally rescaled metrics, is expected to attain a finite value. The condition on the function $\tilde F^r$ now is $\atscrip{\tilde F^r}\propto\Omega^q$ with $q>2$. 

The decomposition of \eref{eg:gharmpb} into lapse and shift evolution equations is 
{\small
\begingroup
\allowdisplaybreaks
\begin{subequations}\label{eg:harmtwobp}
\begin{eqnarray}
\dot\alpha &=& {\beta^r} \alpha '-\alpha  {\beta^r}'-\frac{\alpha  {\beta^r} \gamma _{\theta \theta }'}{\gamma _{\theta \theta }}+\frac{3 \alpha  {\beta^r} \chi '}{2 \chi }+\frac{4 \alpha  {\beta^r} \Omega '}{\Omega }-\frac{2 \alpha  {\beta^r} \hat{\alpha }'}{\hat{\alpha }}+\frac{\alpha  \dot{\gamma _{\theta \theta }}}{\gamma _{\theta \theta }}-\frac{3 \alpha  \dot{\chi }}{2 \chi }-\frac{\alpha ^3 \tilde F^t}{\Omega ^2}
\nonumber \\* && 
-\frac{2 \alpha ^3 \chi  \hat{\gamma _{\theta \theta }} \hat{\beta ^r} \Omega '}{\hat{\alpha }^2 \hat{\chi } \Omega  \gamma _{\theta \theta }}-\frac{\alpha ^3 \chi  \hat{\gamma _{\theta \theta }} \hat{\beta ^r} \hat{\chi }'}{\hat{\alpha }^2 \hat{\chi }^2 \gamma _{\theta \theta }}+\frac{\alpha ^3 \chi  \hat{\beta ^r} \hat{\gamma _{\theta \theta }}'}{\hat{\alpha }^2 \hat{\chi } \gamma _{\theta \theta }}+\frac{2 \alpha ^3 \chi  \hat{\gamma _{\theta \theta }} \hat{\beta ^r}}{\hat{\alpha }^2 r \hat{\chi } \gamma _{\theta \theta }}-\frac{\alpha  \hat{\beta ^r} \Omega '}{\Omega }+\frac{\alpha  \hat{\alpha }' \hat{\beta ^r}}{\hat{\alpha }}
\nonumber \\* && 
-\frac{2 \alpha  {\beta^r}}{r}-\frac{\alpha ^3 \chi  \hat{\beta ^r} \hat{\gamma _{rr}} \Omega '}{\hat{\alpha }^2 \hat{\chi } \Omega  \gamma _{rr}}+\frac{\alpha ^3 \chi  \hat{\gamma _{rr}} \hat{\beta ^r}'}{\hat{\alpha }^2 \hat{\chi } \gamma _{rr}}-\frac{\alpha ^3 \chi  \hat{\beta ^r} \hat{\gamma _{rr}} \hat{\chi }'}{2 \hat{\alpha }^2 \hat{\chi }^2 \gamma _{rr}}+\frac{\alpha ^3 \chi  \hat{\beta ^r} \hat{\gamma _{rr}}'}{2 \hat{\alpha }^2 \hat{\chi } \gamma _{rr}}+\frac{\alpha  {\beta^r}^2 \hat{\beta ^r} \hat{\gamma _{rr}} \Omega '}{\hat{\alpha }^2 \hat{\chi } \Omega }
\nonumber \\* && 
-\frac{\alpha  {\beta^r}^2 \hat{\gamma _{rr}} \hat{\beta ^r}'}{\hat{\alpha }^2 \hat{\chi }}+\frac{\alpha  {\beta^r}^2 \hat{\beta ^r} \hat{\gamma _{rr}} \hat{\chi }'}{2 \hat{\alpha }^2 \hat{\chi }^2}-\frac{\alpha  {\beta^r}^2 \hat{\beta ^r} \hat{\gamma _{rr}}'}{2 \hat{\alpha }^2 \hat{\chi }}-\frac{2 \alpha  {\beta^r} \hat{\beta ^r}^2 \hat{\gamma _{rr}} \Omega '}{\hat{\alpha }^2 \hat{\chi } \Omega }+\frac{2 \alpha  {\beta^r} \hat{\beta ^r} \hat{\gamma _{rr}} \hat{\beta ^r}'}{\hat{\alpha }^2 \hat{\chi }}
\nonumber \\* && 
-\frac{\alpha  {\beta^r} \hat{\beta ^r}^2 \hat{\gamma _{rr}} \hat{\chi }'}{\hat{\alpha }^2 \hat{\chi }^2}+\frac{\alpha  {\beta^r} \hat{\beta ^r}^2 \hat{\gamma _{rr}}'}{\hat{\alpha }^2 \hat{\chi }}+\frac{\alpha  \hat{\beta ^r}^3 \hat{\gamma _{rr}} \Omega '}{\hat{\alpha }^2 \hat{\chi } \Omega }-\frac{\alpha  \hat{\beta ^r}^2 \hat{\gamma _{rr}} \hat{\beta ^r}'}{\hat{\alpha }^2 \hat{\chi }}+\frac{\alpha  \hat{\beta ^r}^3 \hat{\gamma _{rr}} \hat{\chi }'}{2 \hat{\alpha }^2 \hat{\chi }^2}-\frac{\alpha  \hat{\beta ^r}^3 \hat{\gamma _{rr}}'}{2 \hat{\alpha }^2 \hat{\chi }}
\nonumber \\ && 
-\frac{\alpha  {\beta^r} \gamma _{rr}'}{2 \gamma _{rr}}+\frac{\alpha  \dot{\gamma _{rr}}}{2 \gamma _{rr}}
 , \label{eg:harmderlapsebp} \\
\dot \beta^r &=& -\frac{\alpha ^2 \tilde F^r}{\Omega ^2}+\frac{2 {\beta^r}^2}{r}+\frac{{\beta^r} \dot{\alpha }}{\alpha }+\frac{3 {\beta^r} \dot{\chi }}{2 \chi }-\frac{{\beta^r}^2 \alpha '}{\alpha }-\frac{\alpha  \chi  \alpha '}{\gamma _{rr}}+2 {\beta^r} {\beta^r}'-\frac{3 {\beta^r}^2 \chi '}{2 \chi }+\frac{\alpha ^2 \chi '}{2 \gamma _{rr}}
\nonumber \\* && 
+\frac{2 \hat{\beta ^r}^2 \Omega '}{\Omega }-\frac{2 {\beta^r} \hat{\beta ^r} \Omega '}{\Omega }-\frac{\hat{\beta ^r}^4 \hat{\gamma _{rr}} \Omega '}{\Omega  \hat{\alpha }^2 \hat{\chi }}+\frac{2 {\beta^r} \hat{\beta ^r}^3 \hat{\gamma _{rr}} \Omega '}{\Omega  \hat{\alpha }^2 \hat{\chi }}-\frac{{\beta^r}^2 \hat{\beta ^r}^2 \hat{\gamma _{rr}} \Omega '}{\Omega  \hat{\alpha }^2 \hat{\chi }}-\frac{3 {\beta^r}^2 \Omega '}{\Omega }-\frac{\hat{\alpha }^2 \hat{\chi } \Omega '}{\Omega  \hat{\gamma _{rr}}}
\nonumber \\* && 
+\frac{\alpha ^2 \chi  \hat{\beta ^r}^2 \hat{\gamma _{rr}} \Omega '}{\Omega  \hat{\alpha }^2 \hat{\chi } \gamma _{rr}}+\frac{3 \alpha ^2 \chi  \Omega '}{\Omega  \gamma _{rr}}+\frac{2 \alpha ^2 \chi  \hat{\beta ^r}^2 \hat{\gamma _{\theta \theta }} \Omega '}{\Omega  \hat{\alpha }^2 \hat{\chi } \gamma _{\theta \theta }}-\frac{2 \alpha ^2 \chi  \hat{\gamma _{\theta \theta }} \Omega '}{\Omega  \hat{\gamma _{rr}} \gamma _{\theta \theta }}-\frac{\hat{\beta ^r}^2 \hat{\alpha }'}{\hat{\alpha }}+\frac{2 {\beta^r} \hat{\beta ^r} \hat{\alpha }'}{\hat{\alpha }}+\frac{\hat{\alpha } \hat{\chi } \hat{\alpha }'}{\hat{\gamma _{rr}}}
\nonumber \\* && 
-\hat{\beta ^r} \hat{\beta ^r}'+\frac{\hat{\beta ^r}^3 \hat{\gamma _{rr}} \hat{\beta ^r}'}{\hat{\alpha }^2 \hat{\chi }}-\frac{2 {\beta^r} \hat{\beta ^r}^2 \hat{\gamma _{rr}} \hat{\beta ^r}'}{\hat{\alpha }^2 \hat{\chi }}+\frac{{\beta^r}^2 \hat{\beta ^r} \hat{\gamma _{rr}} \hat{\beta ^r}'}{\hat{\alpha }^2 \hat{\chi }}-\frac{\alpha ^2 \chi  \hat{\beta ^r} \hat{\gamma _{rr}} \hat{\beta ^r}'}{\hat{\alpha }^2 \hat{\chi } \gamma _{rr}}-\frac{\hat{\beta ^r}^4 \hat{\gamma _{rr}} \hat{\chi }'}{2 \hat{\alpha }^2 \hat{\chi }^2}
\nonumber \\* && 
+\frac{{\beta^r} \hat{\beta ^r}^3 \hat{\gamma _{rr}} \hat{\chi }'}{\hat{\alpha }^2 \hat{\chi }^2}-\frac{{\beta^r}^2 \hat{\beta ^r}^2 \hat{\gamma _{rr}} \hat{\chi }'}{2 \hat{\alpha }^2 \hat{\chi }^2}-\frac{{\beta^r}^2 \hat{\chi }'}{2 \hat{\chi }}+\frac{\hat{\beta ^r}^2 \hat{\chi }'}{2 \hat{\chi }}+\frac{\alpha ^2 \chi  \hat{\beta ^r}^2 \hat{\gamma _{rr}} \hat{\chi }'}{2 \hat{\alpha }^2 \hat{\chi }^2 \gamma _{rr}}+\frac{\alpha ^2 \chi  \hat{\chi }'}{2 \hat{\chi } \gamma _{rr}}-\frac{\alpha ^2 \chi  \hat{\gamma _{\theta \theta }} \hat{\chi }'}{\hat{\chi } \hat{\gamma _{rr}} \gamma _{\theta \theta }}
\nonumber \\* && 
+\frac{\alpha ^2 \chi  \hat{\beta ^r}^2 \hat{\gamma _{\theta \theta }} \hat{\chi }'}{\hat{\alpha }^2 \hat{\chi }^2 \gamma _{\theta \theta }}+\frac{\hat{\beta ^r}^4 \hat{\gamma _{rr}}'}{2 \hat{\alpha }^2 \hat{\chi }}-\frac{{\beta^r} \hat{\beta ^r}^3 \hat{\gamma _{rr}}'}{\hat{\alpha }^2 \hat{\chi }}+\frac{{\beta^r}^2 \hat{\beta ^r}^2 \hat{\gamma _{rr}}'}{2 \hat{\alpha }^2 \hat{\chi }}+\frac{{\beta^r}^2 \hat{\gamma _{rr}}'}{2 \hat{\gamma _{rr}}}-\frac{\hat{\beta ^r}^2 \hat{\gamma _{rr}}'}{2 \hat{\gamma _{rr}}}-\frac{\alpha ^2 \chi  \hat{\beta ^r}^2 \hat{\gamma _{rr}}'}{2 \hat{\alpha }^2 \hat{\chi } \gamma _{rr}}
\nonumber \\* && 
-\frac{\alpha ^2 \chi  \hat{\gamma _{rr}}'}{2 \hat{\gamma _{rr}} \gamma _{rr}}-\frac{\alpha ^2 \chi  \hat{\beta ^r}^2 \hat{\gamma _{\theta \theta }}'}{\hat{\alpha }^2 \hat{\chi } \gamma _{\theta \theta }}+\frac{\alpha ^2 \chi  \hat{\gamma _{\theta \theta }}'}{\hat{\gamma _{rr}} \gamma _{\theta \theta }}+\frac{{\beta^r}^2 \gamma _{rr}'}{2 \gamma _{rr}}+\frac{\alpha ^2 \chi  \gamma _{rr}'}{2 \gamma _{rr}^2}+\frac{{\beta^r}^2 \gamma _{\theta \theta }'}{\gamma _{\theta \theta }}-\frac{\alpha ^2 \chi  \gamma _{\theta \theta }'}{\gamma _{rr} \gamma _{\theta \theta }}-\frac{2 \alpha ^2 \chi }{r \gamma _{rr}}
\nonumber \\* && 
-\frac{{\beta^r} \dot{\gamma _{rr}}}{2 \gamma _{rr}}-\frac{{\beta^r} \dot{\gamma _{\theta \theta }}}{\gamma _{\theta \theta }}-\frac{2 \alpha ^2 \chi  \hat{\beta ^r}^2 \hat{\gamma _{\theta \theta }}}{r \hat{\alpha }^2 \hat{\chi } \gamma _{\theta \theta }}+\frac{2 \alpha ^2 \chi  \hat{\gamma _{\theta \theta }}}{r \hat{\gamma _{rr}} \gamma _{\theta \theta }}
 .  \label{eg:harmdershiftbp}
\end{eqnarray}	
\end{subequations}
\endgroup
}

The simplified version of the equations equivalent to \eref{eg:harmtwobsimpl} is given by
{\small
\begin{subequations}\label{eg:harmtwobpsimpl}
\begin{eqnarray}
\dot\alpha &=& {\beta^r} \alpha '-\alpha  {\beta^r}'-\frac{\alpha  {\beta^r} \gamma _{\theta \theta }'}{\gamma _{\theta \theta }}+\frac{3 \alpha  {\beta^r} \chi '}{2 \chi }+\frac{4 \alpha  {\beta^r} \Omega '}{\Omega }-\frac{2 \alpha  {\beta^r} \hat{\alpha }'}{\hat{\alpha }}+\frac{\alpha  \dot{\gamma _{\theta \theta }}}{\gamma _{\theta \theta }}-\frac{3 \alpha  \dot{\chi }}{2 \chi }-\frac{\alpha ^3 \tilde F^t}{\Omega ^2}
\nonumber \\ && 
-\frac{2 \alpha ^3 \chi  \hat{\beta ^r} \Omega '}{\hat{\alpha }^2 \hat{\chi } \Omega  \gamma _{\theta \theta }}-\frac{\alpha ^3 \chi  \hat{\beta ^r} \hat{\chi }'}{\hat{\alpha }^2 \hat{\chi }^2 \gamma _{\theta \theta }}+\frac{2 \alpha ^3 \chi  \hat{\beta ^r}}{\hat{\alpha }^2 r \hat{\chi } \gamma _{\theta \theta }}+\frac{\alpha  {\beta^r}^2 \hat{\beta ^r} \Omega '}{\hat{\alpha }^2 \hat{\chi } \Omega }-\frac{\alpha  {\beta^r}^2 \hat{\beta ^r}'}{\hat{\alpha }^2 \hat{\chi }}+\frac{\alpha  {\beta^r}^2 \hat{\beta ^r} \hat{\chi }'}{2 \hat{\alpha }^2 \hat{\chi }^2}-\frac{2 \alpha  {\beta^r} \hat{\beta ^r}^2 \Omega '}{\hat{\alpha }^2 \hat{\chi } \Omega }
\nonumber \\ && 
+\frac{2 \alpha  {\beta^r} \hat{\beta ^r} \hat{\beta ^r}'}{\hat{\alpha }^2 \hat{\chi }}-\frac{\alpha  {\beta^r} \hat{\beta ^r}^2 \hat{\chi }'}{\hat{\alpha }^2 \hat{\chi }^2}+\frac{\alpha  \hat{\beta ^r}^3 \Omega '}{\hat{\alpha }^2 \hat{\chi } \Omega }-\frac{\alpha  \hat{\beta ^r}^2 \hat{\beta ^r}'}{\hat{\alpha }^2 \hat{\chi }}+\frac{\alpha  \hat{\beta ^r}^3 \hat{\chi }'}{2 \hat{\alpha }^2 \hat{\chi }^2}-\frac{\alpha  \hat{\beta ^r} \Omega '}{\Omega }+\frac{\alpha  \hat{\alpha }' \hat{\beta ^r}}{\hat{\alpha }}-\frac{2 \alpha  {\beta^r}}{r}
\nonumber \\ && 
-\frac{\alpha ^3 \chi  \hat{\beta ^r} \Omega '}{\hat{\alpha }^2 \hat{\chi } \Omega  \gamma _{rr}}+\frac{\alpha ^3 \chi  \hat{\beta ^r}'}{\hat{\alpha }^2 \hat{\chi } \gamma _{rr}}-\frac{\alpha ^3 \chi  \hat{\beta ^r} \hat{\chi }'}{2 \hat{\alpha }^2 \hat{\chi }^2 \gamma _{rr}}-\frac{\alpha  {\beta^r} \gamma _{rr}'}{2 \gamma _{rr}}+\frac{\alpha  \dot{\gamma _{rr}}}{2 \gamma _{rr}}
 , \label{eg:harmderlapsebpsimpl} \\ 
\dot \beta^r &=&  -\frac{\alpha ^2 \tilde F^r}{\Omega ^2}-\frac{\alpha ^2 {\beta^r} \tilde F^t}{\Omega ^2}+\alpha ^2 \Lambda ^r \chi -\frac{\alpha  \chi  \alpha '}{\gamma _{rr}}+{\beta^r} {\beta^r}'+\frac{\alpha ^2 \chi '}{2 \gamma _{rr}}+\frac{2 \hat{\beta ^r}^2 \Omega '}{\Omega }-\frac{3 {\beta^r} \hat{\beta ^r} \Omega '}{\Omega }-\frac{\hat{\alpha }^2 \hat{\chi } \Omega '}{\Omega }
\nonumber \\ && 
+\frac{{\beta^r}^2 \Omega '}{\Omega }-\frac{\hat{\beta ^r}^4 \Omega '}{\Omega  \hat{\alpha }^2 \hat{\chi }}+\frac{3 {\beta^r} \hat{\beta ^r}^3 \Omega '}{\Omega  \hat{\alpha }^2 \hat{\chi }}-\frac{3 {\beta^r}^2 \hat{\beta ^r}^2 \Omega '}{\Omega  \hat{\alpha }^2 \hat{\chi }}+\frac{{\beta^r}^3 \hat{\beta ^r} \Omega '}{\Omega  \hat{\alpha }^2 \hat{\chi }}+\frac{3 \alpha ^2 \chi  \Omega '}{\Omega  \gamma _{rr}}+\frac{\alpha ^2 \chi  \hat{\beta ^r}^2 \Omega '}{\Omega  \hat{\alpha }^2 \hat{\chi } \gamma _{rr}}
\nonumber \\ && 
-\frac{\alpha ^2 {\beta^r} \chi  \hat{\beta ^r} \Omega '}{\Omega  \hat{\alpha }^2 \hat{\chi } \gamma _{rr}}-\frac{2 \alpha ^2 \chi  \Omega '}{\Omega  \gamma _{\theta \theta }}+\frac{2 \alpha ^2 \chi  \hat{\beta ^r}^2 \Omega '}{\Omega  \hat{\alpha }^2 \hat{\chi } \gamma _{\theta \theta }}-\frac{2 \alpha ^2 {\beta^r} \chi  \hat{\beta ^r} \Omega '}{\Omega  \hat{\alpha }^2 \hat{\chi } \gamma _{\theta \theta }}-\frac{\hat{\beta ^r}^2 \hat{\alpha }'}{\hat{\alpha }}+\frac{3 {\beta^r} \hat{\beta ^r} \hat{\alpha }'}{\hat{\alpha }}+\hat{\alpha } \hat{\chi } \hat{\alpha }'
\nonumber \\ && 
-\frac{2 {\beta^r}^2 \hat{\alpha }'}{\hat{\alpha }}-\hat{\beta ^r} \hat{\beta ^r}'+\frac{\hat{\beta ^r}^3 \hat{\beta ^r}'}{\hat{\alpha }^2 \hat{\chi }}-\frac{3 {\beta^r} \hat{\beta ^r}^2 \hat{\beta ^r}'}{\hat{\alpha }^2 \hat{\chi }}+\frac{3 {\beta^r}^2 \hat{\beta ^r} \hat{\beta ^r}'}{\hat{\alpha }^2 \hat{\chi }}-\frac{{\beta^r}^3 \hat{\beta ^r}'}{\hat{\alpha }^2 \hat{\chi }}-\frac{\alpha ^2 \chi  \hat{\beta ^r} \hat{\beta ^r}'}{\hat{\alpha }^2 \hat{\chi } \gamma _{rr}}+\frac{\alpha ^2 {\beta^r} \chi  \hat{\beta ^r}'}{\hat{\alpha }^2 \hat{\chi } \gamma _{rr}}
\nonumber \\ && 
-\frac{{\beta^r}^2 \hat{\chi }'}{2 \hat{\chi }}+\frac{\hat{\beta ^r}^2 \hat{\chi }'}{2 \hat{\chi }}+\frac{\alpha ^2 \chi  \hat{\chi }'}{2 \hat{\chi } \gamma _{rr}}+\frac{\alpha ^2 \chi  \hat{\beta ^r}^2 \hat{\chi }'}{2 \hat{\alpha }^2 \hat{\chi }^2 \gamma _{rr}}-\frac{\alpha ^2 {\beta^r} \chi  \hat{\beta ^r} \hat{\chi }'}{2 \hat{\alpha }^2 \hat{\chi }^2 \gamma _{rr}}-\frac{\alpha ^2 \chi  \hat{\chi }'}{\hat{\chi } \gamma _{\theta \theta }}+\frac{\alpha ^2 \chi  \hat{\beta ^r}^2 \hat{\chi }'}{\hat{\alpha }^2 \hat{\chi }^2 \gamma _{\theta \theta }}-\frac{\alpha ^2 {\beta^r} \chi  \hat{\beta ^r} \hat{\chi }'}{\hat{\alpha }^2 \hat{\chi }^2 \gamma _{\theta \theta }}
\nonumber \\ && 
-\frac{\hat{\beta ^r}^4 \hat{\chi }'}{2 \hat{\alpha }^2 \hat{\chi }^2}+\frac{3 {\beta^r} \hat{\beta ^r}^3 \hat{\chi }'}{2 \hat{\alpha }^2 \hat{\chi }^2}-\frac{3 {\beta^r}^2 \hat{\beta ^r}^2 \hat{\chi }'}{2 \hat{\alpha }^2 \hat{\chi }^2}+\frac{{\beta^r}^3 \hat{\beta ^r} \hat{\chi }'}{2 \hat{\alpha }^2 \hat{\chi }^2}-\frac{2 \alpha ^2 \chi  \hat{\beta ^r}^2}{r \hat{\alpha }^2 \hat{\chi } \gamma _{\theta \theta }}+\frac{2 \alpha ^2 {\beta^r} \chi  \hat{\beta ^r}}{r \hat{\alpha }^2 \hat{\chi } \gamma _{\theta \theta }}
 .  \label{eg:harmdershiftbpsimpl}
\end{eqnarray}
\end{subequations}
}%

\section{Implementation}

The gauge conditions described above have been presented from a theoretical point of view. Their implementation in the numerical code will require further tuning and experimenting, changing the source functions and damping coefficients in order to obtain a stable numerical simulation. 
The tuning and changes performed in practice to the gauge conditions have to respect certain requirements:
\begin{itemize}
\item the gauge equations of motion have to be well-behaved as single equations: no exponential growths should be present; 
\item they have to provide an appropriate stationary solution, i.e. that is compatible with the rest of the system and ensures that the slices in the evolution continue to be hyperboloidal; 
\item they form a hyperbolic system (see section \ref{sn:wellposed}) with the Einstein equations; 
\item they have appropriate eigenspeeds - this has to be checked especially at $\scri^+$, where no incoming speeds should appear, and at the horizon (if present and if excision is being used). 
\end{itemize}

I have not presented here the exact form of the source functions and damping coefficients because they are tuned according to the rest of equations involved in the simulations. The actual gauge evolution equations used in the numerical tests are presented in section \ref{se:tests}.

%\upda{gauge conditions on the physical or on the compactified picture? what is the difference? explain and put in the correct place!}  

%\renewcommand\bibname{{References}}
%\bibliographystyle{../../master/thesis/tocunsrt}
%\bibliography{../articles/hypcomp} 

\chapter{Well-posedness and regularity}\label{c:reg} % of the equations

In this chapter we will treat two important properties of the continuum equations: the well-posedness of their initial value formulation, which ensures the existence of a unique solution, and their regularity, especially important at future null infinity, where the conformal factor terms diverge. If the system is not well-posed or the regularity conditions are not satisfied, even with a perfect numerical implementation the simulations will be unstable and crash at some point.

\section{Treatment of hyperbolic equations} \label{sn:wellposed} 

In chapter \ref{c:eqs} the Einstein equations were decomposed into a system of elliptic (constraints) and hyperbolic (equations of motion) differential equations. 

Hyperbolic equations usually describe dynamical processes that involve the propagation of information in the form of waves from one place to another following a time evolution. The difficulty is that not any choice of variables or equations is appropriate; they need to be formulated as a well-posed problem to avoid the instabilities that will appear otherwise. 

\subsection{Well-posedness} % well-posed problem, not equations: hyperbolic equations

Well-posedness is a necessary requirement for successfully solving a system of equations iteratively in time. 
It implies that an unique solution to the equations exists and that this solution depends continuously on the initial data provided. This last condition can be expressed as (see e.g. \cite{gustafsson1995time})
\begin{equation}||u(t)||\le Ke^{at}||u(0)||  . \end{equation}
The growth of the solution $u(t)$ with respect to the initial data $u(0)$, calculated using an appropriate norm $||.||$, is bounded by certain finite values of $K$ (not to be confused with the trace of the extrinsic curvature) and $a$, so that it cannot be infinite. 

Quasilinear systems of equations are systems which are linear in the highest derivatives. This is the case of the Einstein equations considered here. The study of well-posedness can be performed on this type of systems using a simple standard method. 
In this method it is sufficient to consider the principal part of the equations. The spherically symmetric reduced equations that will be used here are expressed as a mixed first order in time and second order in space system. The exact terms that belong to the principal part of such a system are specified in the following subsection, but the basic rough idea is that it involves only the highest derivatives of each type of variables. General theorems like theorem 4.3.2 in \cite{gustafsson1995time} state that lower order terms do not contribute to the well-posedness of the system. %4.6.1
The algebraic analysis performed on the principal part will determine the hyperbolicity of the system of equations, which allows to make a claim on the well-posedness of the corresponding initial value problem. 

%If we have a set of non-linear equations and want to see whether they are well-posed, a simple way of analyzing it is to first linearize them around a given background and then select the principal part terms in each of the equations. The procedure to finish the analysis is described in more detail in the following subsections. 

\subsection{First order in time and second order in space systems}\label{sr:1st2nd}

The equations obtained from the 3+1 decomposition contain first order in time derivatives and second order in space derivatives. 
Their hyperbolicity analysis is more complicated than in the first order in space reduction, whose features are described by plenty of textbooks, but an implementation of the equations as a second order in space system has a similar numerical behaviour and avoids introducing extra variables and constraints. The analysis summarized here follows \cite{Calabrese:2005ft,Husa:2007zz}. 

A common first order in time and second order in space system of hyperbolic differential equations has the following structure:
\begin{equation}
\partial_t\boldsymbol u=P\boldsymbol{u} \qquad \textrm{with} \qquad \boldsymbol u=\left(\begin{array}{c}u\\v\end{array}\right)  ,
\end{equation}
with the matrix $P$ given by
\begin{equation}\label{def:principal_part}
P=\left(\begin{array}{cc}A^i\partial_i+B&C\\D^{ij}\partial_i\partial_j+E^i\partial_i+F&G^i\partial_i+J\end{array}\right)  . 
\end{equation}
There are two different types of variables: $u$ represents those which are derived twice in space, whereas $v$ stands for the ones which are derived at most once.  

% introducing n_i means choosing one coordinate
A Fourier transformation is now performed on the system, using $\omega_i\equiv\omega\,n_i$ and $M^n\equiv M^in_i$. This gives the equivalent matrix to $P$ in the Fourier space, $\hat P$. Its principal part is given by the higher order derivatives in each of the four blocks and is denoted as $\hat P'$, the second order principal symbol: 
\begin{equation}
\hat P=\left(\begin{array}{cc}i\omega A^n+B&C\\-\omega^2D^{nn}+i\omega E^n+F&i\omega G^n+J\end{array}\right) , \qquad \hat P'=\left(\begin{array}{cc}i\omega A^n&C\\-\omega^2D^{nn}&i\omega G^n\end{array}\right)  . 
\end{equation}

The system will now be reduced to first order in the Fourier space using a pseudo-differential reduction. To do so, the new variable $\hat w\equiv i\omega\hat u$ is introduced, giving rise to the following system (and the constraint ${\cal C}=\hat w- i\omega\hat u$): 
\begin{equation}
\partial_t\hat {\boldsymbol u}_R=\hat P_R\hat {\boldsymbol u}_R \qquad \textrm{with} \qquad \hat {\boldsymbol u}_R=\left(\begin{array}{c}\hat u\\\hat v\\\hat w\end{array}\right)  
\end{equation}
and a new matrix 
\begin{equation}
\hat P_R=\left(\begin{array}{ccc}B&A^n&C\\0&i\omega A^n+B&i\omega C\\F&i\omega D^{nn}+ E^n&i\omega G^n+J\end{array}\right)  . 
\end{equation}
The principal symbol of the first order is given by the elements in $\hat P_R$ that are multiplied by $i\omega$: 
\begin{equation}\label{er:principalm}
\hat P'_R=\left(\begin{array}{ccc}0&0&0\\0&i\omega A^n&i\omega C\\0&i\omega D^{nn}&i\omega G^n\end{array}\right)  . 
\end{equation}
Now writing $E=\hat P'_R/(i\omega)$ to simplify the notation, the principal part of the system is formally solved as: 
\begin{equation}
\partial_t\hat {\boldsymbol u}_R=\hat P'_R\hat {\boldsymbol u}_R=i\omega E\hat {\boldsymbol u}_R\quad\Rightarrow\quad \hat {\boldsymbol u}_R(t)=e^{i\omega E t}\hat {\boldsymbol u}_R(0)  .
\end{equation}
If $E$ is diagonalizable and has real eigenvalues, the solution will be purely oscillatory and bounded by the initial data. In this case, the initial value problem is well-posed. 

\subsection{Hyperbolicity}

The analysis of the principal part matrix $E$ is performed using a Jordan decomposition. It can be considered a generalization of the Eigen decomposition: apart from the elements in the diagonal, the Jordan canonical form can also have non-zero values in the superdiagonal. These appear due to the existence of Jordan-blocks, which are placed in the diagonal of the Jordan canonical form. %An example of Jordan block is \begin{equation}\left(\begin{array}{cc}1&1\\0&1\end{array}\right)  . \end{equation}
If a matrix is diagonalizable, its diagonal and Jordan canonical forms coincide. 

Both complex eigenvalues or Jordan-blocks in the Jordan canonical form of the matrix $E$ can cause growing modes that will give an ill-posed problem. The properties of the matrix $E$ determine the hyperbolicity of the equations, and thus the well-posedness (provided that the similarity matrix is regular). The different possibilities are: 
\begin{itemize}
\item If all eigenvalues (the characteristic speeds) of the principal part matrix are real, the system is called (weakly) hyperbolic, and it is well-posed in absence of lower order terms.
\item If a complete set of (left) eigenvectors exists and no Jordan-blocks appear, the system is called strongly hyperbolic, and admits a well-posed initial value problem.
\item If the system is strongly hyperbolic and also admits a (direction-independent) conserved energy it is called symmetric hyperbolic. In one dimension strongly hyperbolic systems are always also symmetric hyperbolic, because they are not direction-dependent and a conserved energy is to be found without exception.
\end{itemize}
The BSSN-type formulations of the Einstein equations are known to be strongly hyperbolic \cite{Sarbach:2002bt,Nagy:2004td,Gundlach:2004jp,Beyer:2004sv}. % \cite{Gundlach:2004ri} considers simpler systems like the Maxwell equations
Indeed the systems of evolution equations \eref{es:eeqs}, \eref{es:pKeeqs} and \eref{es:DPKeeqs} together with the gauge conditions considered in chapter \ref{c:gauge} have a diagonalizable principal part matrix with real eigenspeeds and a full set of eigenfields. Adding spherical symmetry to it means that the system is symmetric hyperbolic and thus the problem is well-posed. 

\subsection{Characteristic decomposition of the equations}\label{sr:eigen}

The system of spherically symmetric GBSSN and Z4c equations presented in \eref{es:eeqs} can be analyzed with the tools presented in the previous subsection. As a first order in time and second order in space system, its variables are divided into 
\begin{description}
\item[$\boldsymbol u$ variables,] those which appear spatially derived twice: $\chi$, $\gamma_{rr}$, $\gamma_{\theta\theta}$, $\alpha$, $\beta^r$ and $\iPhi$. 
\item[$\boldsymbol v$ variables,] with only first spatial derivatives: $A_{rr}$, $K$, $\Lambda^r$, $\Theta$, $B^r$ (if present) and $\iPi$. 
\end{description}
The eigenfields of the system will include the spatial derivatives of the $u$ variables, denoted by $\chi'$, $\gamma_{rr}'$, $\gamma_{\theta\theta}'$, $\alpha'$, ${\beta^r}'$ and $\iPhi'$.

\begin{figure}[htbp!!]
\center
\begin{tabular}{@{}c@{}@{}c@{}}
\includegraphics[width=0.5\linewidth]{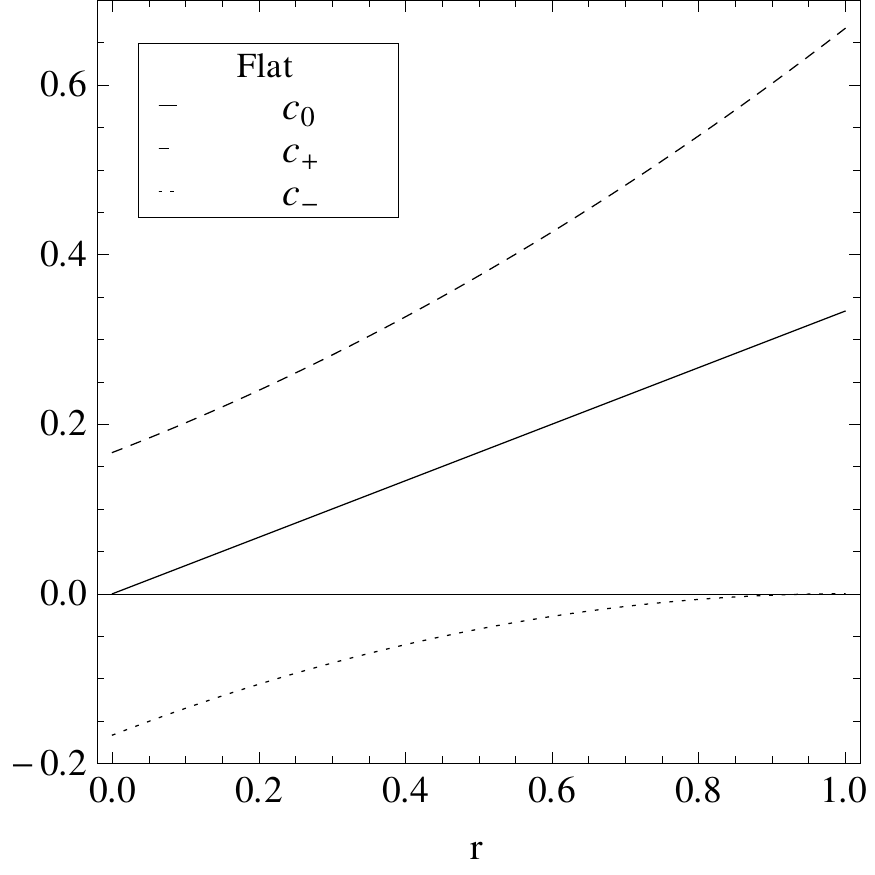} &
\includegraphics[width=0.5\linewidth]{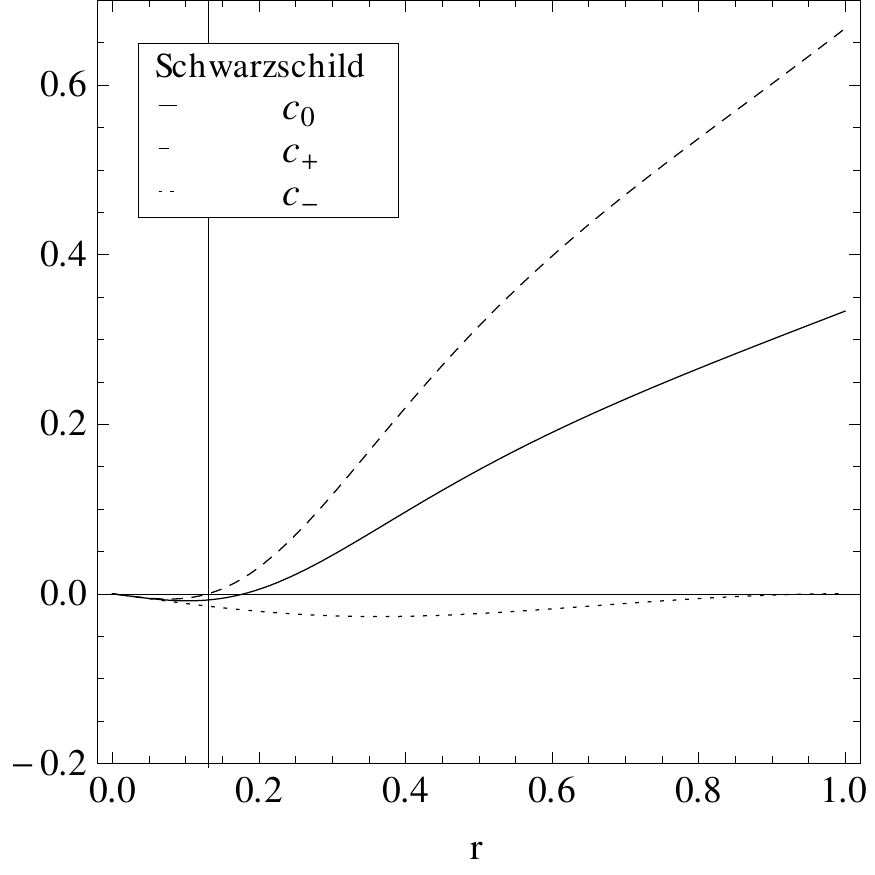}
\end{tabular}
\caption{Zero speed $c_0=-\beta^r$ and lightspeeds $c_\pm=-\beta^r\pm\alpha\sqrt{\case{\chi}{\gamma_{rr}}}$ plotted for flat spacetime (left) and a Schwarzschild BH with $M=1$ (right), using $\Kc=-1$ for both. Again, $\rscri=1$. The vertical line in the right plot indicates the radial position of the horizon, where $c_0,c_-<0$ and $c_+=0$. At $\scri^+$ we have $c_0,c_+>0$ and $c_-=0$.}
%\mbox{\includegraphics[width=0.7\linewidth]{../articles/hypcomp1/figures/light_speeds.pdf}}
%\caption{The zero speed $c_0$ is $-\beta^r$ and the lightspeeds are $c_\pm=-\beta^r\pm\alpha\sqrt{\case{\chi}{\gamma_{rr}}}$. In the plot we have used the flat spacetime values of the variables and set $\Omega$ as in \eref{ein:omega}, as well as $r_{\!\!\scri}=1$ and $a=1$. \upda{add equivalent plot for the Schwarzschild case!!}}
\label{fr:lightspeeds}
\end{figure}

The lightspeeds of the system are calculated from the line element by imposing $d\bar s^2=0$ and solving for $c=\case{dr}{dt}$. The result is $c_\pm=-\beta^r\pm\alpha\sqrt{\case{\chi}{\gamma_{rr}}}$. There is an extra speed due to the change in the coordinates given by $c_0=-\beta^r$. These three velocities are plotted in \fref{fr:lightspeeds} for flat spacetime and in presence of a Schwarzschild BH. The speeds are proportional to $|\Kc|$. As expected for asymptotic flatness, the behaviour of the speeds in the vicinity of $\scri^+$ does not change if a compact object is present at the origin. There are no negative speeds at $\scri^+$, which is a consequence of the fact that future null infinity is an ingoing null hypersurface in the conformal picture. Similarly, at the horizon of the Schwarzschild BH there are no positive speeds, because all radiation is infalling. 

The eigendecomposition (or characteristic decomposition) of \eref{es:eeqs} is presented in \tref{eigen}, to which the eigenfields $\chi$, $\gamma_{rr}$, $\gamma_{\theta\theta}$, $\alpha$, $\beta^r$ (if evolved) and $\iPhi$ have to be added with zero propagation speed. For generality we have determined the eigenfields and eigenspeeds keeping the angular metric component $\gamma_{\theta\theta}$ as an evolution variable, but it can be eliminated using the freedom of the determinant of the conformal metric. 
Indeed it is only after fixing this degree of freedom that we obtain the standard Z4c/CCZ4 formulation out of our equations.
If $\gamma_{\theta\theta}$ is eliminated, the first eigenfield listed in \tref{eigen} vanishes and the substitution \eref{es:delgtt} has to be applied to the rest of eigenfields. %$\gamma_{\theta\theta}=1/{\sqrt{\gamma_{rr}}}$

\begin{table}[hhh]
\caption{\label{eigen}Eigenfields and eigenspeeds, using $c_0=-\beta^r$ and $c_\pm=-\beta^r\pm\alpha\sqrt{\case{\chi}{\gamma_{rr}}}$.}
\begin{tabular}{@{}llcc}
\hline
System&Eigenfields&Eigenspeeds&at $\scri^+$\\
\hline
GBSSN / \CZ{}&$\frac{\gamma_{rr}'}{\gamma_{rr}}+\frac{2\gamma_{\theta\theta}'}{\gamma_{\theta\theta}}$&0&0\\
GBSSN / \CZ{}&$2\gamma_{rr}\Lambda^r-\frac{\gamma_{rr}'}{\gamma_{rr}}+\frac{\gamma_{\theta\theta}'}{\gamma_{\theta\theta}}-\frac{2\alpha'}{\alpha}+\frac{\chi'}{\chi}+\frac{3}{\sqrt{\gamma_{rr}\chi}}A_{rr}$&$c_-$&0\\
GBSSN / \CZ{}&$2\gamma_{rr}\Lambda^r-\frac{\gamma_{rr}'}{\gamma_{rr}}+\frac{\gamma_{\theta\theta}'}{\gamma_{\theta\theta}}-\frac{2\alpha'}{\alpha}+\frac{\chi'}{\chi}-\frac{3}{\sqrt{\gamma_{rr}\chi}}A_{rr}$&$c_+$&$\frac{2\rscri}{3}|\Kc|$\\ 
GBSSN / \CZ{}&$\frac{\alpha'}{\alpha}-\sqrt{\frac{\gamma_{rr}}{\chi}}K$&$c_-$&0\\
GBSSN / \CZ{}&$\frac{\alpha'}{\alpha}+\sqrt{\frac{\gamma_{rr}}{\chi}}K$&$c_+$&$\frac{2\rscri}{3}|\Kc|$\\
\hline
GBSSN&$\gamma_{rr}\Lambda^r-\frac{\gamma_{rr}'}{2\gamma_{rr}}-\frac{\gamma_{\theta\theta}'}{\gamma_{\theta\theta}}+\frac{2\chi'}{\chi}$&$c_0$&$\frac{\rscri}{3}|\Kc|$\\
\CZ{}&$\gamma_{rr}\Lambda^r-\frac{\gamma_{rr}'}{2\gamma_{rr}}-\frac{\gamma_{\theta\theta}'}{\gamma_{\theta\theta}}+\frac{2\chi'}{\chi}+2\sqrt{\frac{\gamma_{rr}}{\chi}}\Theta$&$c_-$&0\\
\CZ{}&$\gamma_{rr}\Lambda^r-\frac{\gamma_{rr}'}{2\gamma_{rr}}-\frac{\gamma_{\theta\theta}'}{\gamma_{\theta\theta}}+\frac{2\chi'}{\chi}-2\sqrt{\frac{\gamma_{rr}}{\chi}}\Theta$&$c_+$&$\frac{2\rscri}{3}|\Kc|$\\
\hline
G/Z + fixed shift &$2\gamma_{rr}\Lambda^r-\frac{2\alpha'}{\alpha}+\frac{\chi'}{\chi}$&$c_0$&$\frac{\rscri}{3}|\Kc|$\\
G/Z + harmonic shift & $2\gamma_{rr}\Lambda^r-\frac{2\alpha'}{\alpha}+\frac{\chi'}{\chi}+2\sqrt{\frac{\gamma_{rr}}{\chi}}{\beta^r}'$ &$c_-$&$0$\\
G/Z + harmonic shift & $2\gamma_{rr}\Lambda^r-\frac{2\alpha'}{\alpha}+\frac{\chi'}{\chi}-2\sqrt{\frac{\gamma_{rr}}{\chi}}{\beta^r}'$ &$c_+$&$\frac{2\rscri}{3}|\Kc|$\\
\hline
Scalar field&$\iPhi'+\frac{1}{\alpha}\sqrt{\frac{\gamma_{rr}}{\chi}}(\iPi-\beta^r\iPhi')$&$c_-$&0\\
Scalar field&$\iPhi'-\frac{1}{\alpha}\sqrt{\frac{\gamma_{rr}}{\chi}}(\iPi-\beta^r\iPhi')$&$c_+$&$\frac{2\rscri}{3}|\Kc|$\\
\hline
\end{tabular}
\end{table}
As already shown in figure \ref{fr:lightspeeds} and also indicated in the last column of \tref{eigen}, the eigenspeeds at $\scri^+$ are either zero or positive, so that no incoming radiation exists. 

The first group of five eigenspeeds and eigenfields is given by the Einstein equations and the generalized harmonic lapse (chosen as \eref{eg:harmlapse} or \eref{eg:harmderlapseb}) and is common to all combinations (except for the first one that disappears if $\gamma_{\theta\theta}$ is eliminated). 
In the second group the first row corresponds to an evolution without $\Theta$ variable, whereas the other two appear in case of the Z4c system. 
The choice of a fixed shift or the generalized harmonic shift condition (given by \eref{eg:harmdershift} or \eref{eg:harmdershiftb}, as their principal part is the same) determines which eigenquantities to select from the third group. 
The fourth group accounts for the scalar field equations, whose principal part decouples from the rest of the system.

This eigendecomposition can be compared with the ones presented in \cite{Brown:2007nt,Field:2010mn}, making the substitution of $\Lambda^r$ by $\Gamma^r$, rearranging some eigenvectors, omitting the scalar field part and substituting the harmonic lapse by the 1+log slicing condition and the harmonic shift by a Gamma-driver condition. 

The eigendecomposition corresponding to the Gamma-driver shift conditions (described in subsection \ref{sg:gammadriver}) is not shown here because their free parameters $\lambda$ and $\mu$ make the eigenfield expressions very complicated and long. What is important is the fact that evaluating the eigenspeeds at $\scri^+$ allows us to determine the allowed values of the parameters. There will be no negative eigenspeeds (no incoming modes) at $\scri^+$ if $\lambda\,\mu\leq\left(\case{\rscri}{3}\Kc\right)^2$ of the Gamma-driver condition with auxiliary variable $B^r$ \eref{eg:Gammadriver} and if $\lambda\leq\case{1}{12}\left(\rscri\Kc\right)^2$ for the integrated version \eref{eg:integGammadriver}. The reason for keeping the parameter $\mu$ separated from the parameter $\lambda$ when tuning the Gamma-driver shift condition \eref{eg:Gammadriver} is that even if the eigenspeeds depend only on $\lambda\,\mu$, different compatible choices of $\lambda$ and $\mu$ can present a different numerical behaviour. 

The change from conformal $\cK$ to physical $\pK$ or to $\DPK$ only involves a rescaling by $\Omega$ in $K$ in the eigenfields. The same is valid when transforming from $\cT$ to $\pT$. 
The eigendecomposition for \eref{es:pKeeqs} is obtained by substituting $\cK=\case{\pK}{\Omega}$ and $\cT=\case{\pT}{\Omega}$, and the one for  \eref{es:DPKeeqs} by transforming $\cK=\case{\DPK}{\Omega}$ and $\cT=\case{\pT}{\Omega}$. The other terms involved in the transformation \eref{e3:physconfK} become non-principal part terms and therefore do not appear in the eigendecomposition. 

\section{Regularity of the equations at null infinity}\label{cr:regu}

The regularity of the conformal factor terms was already mentioned in section \ref{sg:preferred}, where the preferred conformal gauge was described. As a reminder, the divergent conformal factor terms arising from the transformation of the Ricci or the Einstein tensors due to the conformal metric rescaling, given by \eref{c3:einsteinR} and \eref{c3:einsteinG} respectively, are formally singular at $\scri^+$. Certain relations between the variables have to be satisfied, so that the numerators cancel with the appropriate order of $\Omega$ and in the limit $\Omega\to0$ the divergent terms attain a finite value. These relations between the variables will be referred to as regularity conditions. We will assume that they are valid at $\scri^+$. 

The regularity conditions arising from \eref{c3:einsteinR} (or equivalently \eref{c3:einsteinG}) are summarized as: 
\begin{subequations}
\begin{eqnarray}
\atscrip{\bar{\nabla}^a\Omega\bar{\nabla}_a\Omega}&=&0 , \label{er:condnull}\\
\atscrip{\left(\bar{\nabla}_a\bar{\nabla}_b\,\Omega-\frac{1}{4} \,\bar{g}_{ab}\,\bar{\Box}\,\Omega\right)}&=&0 , \label{er:tf} \\
\left(\atscrip{\lim_{\Omega\to0}\frac{2\bar{\nabla}^a\Omega \bar{\nabla}_a\Omega}{\Omega}-\bar{\Box}\,\Omega\right)} &=& 0 . \label{er:tl}
\end{eqnarray}
\end{subequations}
The first one is obtained by multiplying \eref{c3:einsteinR} by $\Omega^2$ and evaluating it at $\scri^+$. For the other relations, \eref{c3:einsteinR} (or \eref{c3:einsteinG}) is multiplied by $\Omega$ and its trace-free part for \eref{er:tf} and its trace for \eref{er:tl} are evaluated at $\scri^+$.

\subsection{Regularity of the tensorial equations: approach}

The regularity conditions of the tensorial equations \eref{et:tensoreqs} and \eref{et:pKtensoreqs} can be studied by separating the angular components from the radial one. For this, the variables and equations have to be decomposed in [2+1]+1 form. 

The starting point is to write the line element as 
\begin{eqnarray}
ds^2&=&-\alpha^2\,dt^2
+L^2(dr+\beta^r\,dt)(dr+\beta^r\,dt)
\nonumber\\
&&+q_{AB}
(d\theta^A+b^A\,dr+\beta^A\,dt)
(d\theta^B+b^B\,dr+\beta^B\,dt) ,
\end{eqnarray}
where $L^{-2}=\bar\gamma^{ab}\bar D_a r\bar D_b r$ equivalently as with the lapse function. The normal vector to the level sets of $r$ is denoted by $\bar s^a$ and is defined as $\bar s^a=L\bar{D}^ar$. It is related to the two-metric $\bar q_{ab}$ on the spheres orthogonal to these level sets by
\begin{equation}
\bar q_{ab} = \bar\gamma_{ab} - \bar s_a\bar s_b . 
\end{equation}
The conformal spacetime metric can be expressed in terms of the decompositions as
\begin{equation}
\bar g_{ab} = -\bar n_a\bar n_b + \bar\gamma_{ab} = -\bar n_a\bar n_b + \bar s_a\bar s_b + \bar q_{ab} . 
\end{equation}

The coordinate light-speeds of light-rays traveling in the $\pm s^i$ directions are,
\begin{subequations}
\begin{align}
c^r_\pm&=-\beta^r\pm L^{-1}\,\alpha , \\
c^A_\pm&=-\beta^A\mp b^A\,L^{-1}\, \alpha ,
\end{align}
\end{subequations}
radially and in the angular directions respectively. Obviously, $c^r_\pm$ coincide with the lightspeeds $c_\pm$ of the spherically symmetric reduction presented in subsection \ref{sr:eigen}. 

Expressing the regularity condition \eref{er:condnull} in the [2+1]+1 decomposition gives 
\begin{equation}
\atscrip{\left(-(\mathcal{L}_{\bar{n}}\Omega)^2+(\mathcal{L}_{\bar s}\Omega)^2+q^{ab}\bar \nabla_a\Omega\bar \nabla_b\Omega\right)}=0 .
\end{equation}
Assuming that the boundary $\Omega=0$ is also a level-set of the radial coordinate $r=\rscri$, the previous expression becomes simply,
\begin{align}
\atscrip{(\mathcal{L}_{\bar{n}}\Omega)^2}=\atscrip{(\mathcal{L}_{\bar s}\Omega)^2} . \label{er:level}
\end{align}
One of the conditions for the scri-fixing described in \ref{cg:scrifix} was that in our implementation $\Omega$ is time independent. Assuming the level set condition this translates to 
\begin{equation}
\atscrip{\partial_t\Omega}=\atscrip{\left(\alpha\mathcal{L}_{\bar{n}}\Omega+L\beta^r\mathcal{L}_{\bar s}\Omega\right)}=0 . 
\end{equation}

This is the basic picture where the regularity conditions can be derived. The calculation itself is still ongoing work. 

\subsection{Regularity of the spherically symmetric equations}\label{er:Einstenreqs}

The regularity conditions of the spherically symmetric equations can be easily derived by looking at the numerator of terms divided by $\Omega$. The regularity conditions for the evolution system \eref{es:pKeeqs}, the one that uses the physical $\pK$ and is almost the same as the one tested numerically, are the following: 
\begin{subequations}\label{er:regconds}
\begin{itemize}

\item There is only one term with $\Omega^2$ in the denominator:  $\frac{2 \alpha  \pT  \Omega '}{\Omega^2  \gamma _{rr}}$, which appears in \eref{es:pKLambdardot}, and regularity requires that the numerator vanishes at $\scri^+$. This implies that 
\begin{equation}
\atscrip{\pT}=0 . 
\end{equation}   
Using the l'H\^opital rule, the mentioned term is substituted for the rest of the analysis by $\frac{2 \alpha  \pT'}{\Omega  \gamma _{rr}}$. 

\item The rest of the conditions will be given by the terms that are divided by $\Omega$ in the evolution equations. The condition that makes the numerator of $\Omega$ in $\dot\chi$ vanish is
\begin{equation}\label{er:pKreg}
\atscrip{\pK}=\atscrip{ -\frac{3 {\beta^r} \Omega '}{\alpha }} . 
\end{equation}  

\item Before obtaining the regularity conditions for the variables $A_{rr}$ and $\Lambda^r$, it is convenient to do a separate calculation for the GBSSN and \CZ{} equations. 
\begin{itemize}

\item GBSSN: The regularity condition for $A_{rr}$ is computed from the quantities over $\Omega$ in $\dot A_{rr}$ and then substituted into $\dot\Lambda^r$ to obtain the expression for $\Lambda^r$ in an equivalent way. The resulting conditions are: 
\begin{eqnarray}
\atscrip{A_{rr}}^{GBSSN}&=&\atscrip{\left( -\frac{\alpha  \chi  \gamma _{\theta \theta }'}{3 {\beta^r} \gamma _{\theta \theta }}+\frac{2 \alpha  \chi '}{3 {\beta^r}}+\frac{2 \alpha  \chi  \Omega ''}{3 {\beta^r} \Omega '}-\frac{2 \alpha  \chi }{3 {\beta^r} \rscri}-\frac{\alpha  \chi  \gamma _{rr}'}{3 {\beta^r} \gamma _{rr}} \right)}
, \\
\atscrip{\Lambda^r}^{GBSSN}&=& \left(-\frac{4 \pK'}{3 \kappa_1 \gamma _{rr}}+\frac{8 \alpha  \chi  \Omega '}{3 {\beta^r} \kappa_1 \rscri \gamma _{rr}^2}+\frac{4 \alpha  \chi  \Omega ' \gamma _{\theta \theta }'}{3 {\beta^r} \kappa_1 \gamma _{\theta \theta } \gamma _{rr}^2}-\frac{8 \alpha  \chi ' \Omega '}{3 {\beta^r} \kappa_1 \gamma _{rr}^2}-\frac{8 \alpha  \chi  \Omega ''}{3 {\beta^r} \kappa_1 \gamma _{rr}^2}\right.
\nonumber \\ &&
\atscrip{\left.+\frac{4 \alpha  \chi  \Omega ' \gamma _{rr}'}{3 {\beta^r} \kappa_1 \gamma _{rr}^3}+\frac{2}{\rscri \gamma _{\theta \theta }}-\frac{2}{\rscri \gamma _{rr}}-\frac{\gamma _{\theta \theta }'}{\gamma _{\theta \theta } \gamma _{rr}}+\frac{\gamma _{rr}'}{2 \gamma _{rr}^2}\right)} . 
\end{eqnarray}

\item \CZ{}: To calculate the regularity conditions corresponding to the \CZ{} equations, first all $Z_r$ have to be substituted using \eref{es:pKZrdef}. The $A_{rr}$ and $\Lambda^r$ variables involved in the regularity conditions appear in the singular terms in both $\dot A_{rr}$ and $\dot\Lambda^r$, so that the system has to be solved together. The solution yields: 
\begin{eqnarray}
\atscrip{A_{rr}}^{\CZ{}}&=& \left[\frac{1}{8 \alpha ^2 \chi  \Omega '+3 {\beta^r} \gamma _{rr} \left(\alpha  \kappa_1-2 {\beta^r} \Omega '\right)}\left(-\frac{8}{3} \alpha ^2 \chi  \pK' \gamma _{rr}-\frac{2 \alpha ^2 \kappa_1 \chi  \gamma _{rr}}{\rscri}\right.
\right. \nonumber \\ && \left.
-\frac{\alpha ^2 \kappa_1 \chi  \gamma _{rr} \gamma _{\theta \theta }'}{\gamma _{\theta \theta }}+2 \alpha ^2 \kappa_1 \gamma _{rr} \chi '+\frac{2 \alpha ^2 \kappa_1 \chi  \gamma _{rr} \Omega ''}{\Omega '}-\alpha ^2 \kappa_1 \chi  \gamma _{rr}'
\right. \nonumber \\ && \left.
+\frac{4 \alpha  {\beta^r} \chi  \gamma _{rr} \Omega '}{\rscri}+\frac{8}{3} \alpha ^2 \chi  \Theta ' \gamma _{rr}+2 \alpha  {\beta^r} \chi  \Omega ' \gamma _{rr}'-4 \alpha  {\beta^r} \gamma _{rr} \chi ' \Omega '
\right. \nonumber \\ && 
\atscrip{\left.\left.-4 \alpha  {\beta^r} \chi  \gamma _{rr} \Omega ''+\frac{2 \alpha  {\beta^r} \chi  \gamma _{rr} \Omega ' \gamma _{\theta \theta }'}{\gamma _{\theta \theta }}\right)\right]}
, \\
\atscrip{\Lambda^r}^{\CZ{}}&=& \left[\frac{1}{8 \alpha ^2 \chi  \Omega '+3 {\beta^r} \gamma _{rr} \left(\alpha  \kappa_1-2 {\beta^r} \Omega '\right)}\left(4 \alpha  {\beta^r} \Theta '+\frac{6 {\beta^r}^2 \Omega ' \gamma _{\theta \theta }'}{\gamma _{\theta \theta }}-4 \alpha  {\beta^r} \pK'
\right.\right. \nonumber \\ && \left. 
-\frac{3 \alpha  {\beta^r} \kappa_1 \gamma _{\theta \theta }'}{\gamma _{\theta \theta }}-\frac{6 \alpha  {\beta^r} \kappa_1}{\rscri}+\frac{6 \alpha  {\beta^r} \kappa_1 \gamma _{rr}}{\rscri \gamma _{\theta \theta }}+\frac{3 \alpha  {\beta^r} \kappa_1 \gamma _{rr}'}{2 \gamma _{rr}}+\frac{16 \alpha ^2 \chi  \Omega '}{\rscri \gamma _{\theta \theta }}
\right. \nonumber \\ && \left. 
+\frac{12 {\beta^r}^2 \Omega '}{\rscri}-\frac{8 \alpha ^2 \chi  \Omega '}{\rscri \gamma _{rr}}-\frac{12 {\beta^r}^2 \gamma _{rr} \Omega '}{\rscri \gamma _{\theta \theta }}-\frac{4 \alpha ^2 \chi  \Omega ' \gamma _{\theta \theta }'}{\gamma _{\theta \theta } \gamma _{rr}}-\frac{8 \alpha ^2 \chi ' \Omega '}{\gamma _{rr}}
\right. \nonumber \\ && \atscrip{ \left. \left.
-\frac{8 \alpha ^2 \chi  \Omega ''}{\gamma _{rr}}+\frac{8 \alpha ^2 \chi  \Omega ' \gamma _{rr}'}{\gamma _{rr}^2}-\frac{3 {\beta^r}^2 \Omega ' \gamma _{rr}'}{\gamma _{rr}}\right)\right]}
\end{eqnarray}

\end{itemize}

\item The regularity condition required for equations $\dot\pK$ and $\dot\pT$ is the scri-fixing condition:
\begin{equation} \label{er:scrifixreg}
\atscrip{\left(\frac{{\beta^r}^2}{\alpha}-\frac{\alpha\chi}{\gamma_{rr}}\right)}=0 . 
\end{equation}

‪%\item $\left.\K-\frac{3}{a} \ \right|_{\scri}=\left.-\frac{3\beta^r\Omega'}{\alpha}\ \right|_{\scri}=\left.+\frac{3\sqrt{\gamma_{rr}}\Omega'}{\sqrt{\chi}}\ \right|_{\scri}$, obtained from $\dot \chi$ and $\dot \K$ due to the transformation of the trace of the extrinsic curvature. In the second equality we have used the scri-fixing condition. 
%\item $\left.\K' \right|_{\scri}=\left.-\frac{3A_{rr}\Omega'}{\gamma_{rr}} \ \right|_{\scri}$, from $\dot \Lambda^r$. 
%\item $\left.A_{rr}\right|_{\scri}=\left.-\frac{\alpha\chi}{3\beta^r}\left( \frac{2}{r}+\frac{\gamma_{rr}'}{\gamma_{rr}}+\frac{\gamma_{\theta\theta}'}{\gamma_{\theta\theta}}-\frac{2\chi'}{\chi}-\frac{2\Omega''}{\Omega'}\right) \ \right|_{\scri}$, calculated from $\dot A_{rr}$. 

\item The only missing term in the evolution equation of the scalar field auxiliary variable $\dot\iPi$ is satisfied if 
\begin{equation}
\atscrip{\iPi}=0 . 
\end{equation} 

\end{itemize}
\end{subequations}

Only if the relations \eref{er:regconds} are satisfied, the formally singular terms in the equations will attain a regular limit at $\scri^+$. 
The regularity conditions for the system \eref{es:DPKeeqs} can be obtained substituting $\pK$ by $\DPK$ using \eref{ee:DPKdef}, for instance from \eref{er:pKreg}
\begin{equation} \label{er:Kregrel}
\atscrip{\DPK} = -\Kc \atscrip{-\frac{3\beta^r\Omega'}{\alpha}} .
\end{equation}

\subsection{Regularity conditions arising from the gauge conditions} \label{er:gaugereqs}

Extra regularity conditions may arise from the gauge conditions. To illustrate this, we will consider some examples from the actual gauge equations of motion that have been tested numerically and described in section \ref{se:tests}. 
\begin{subequations}\label{er:regcondsgauge}
\begin{itemize}
\item The gauge conditions where the lapse and shift are damped present the following behaviour: 
\begin{itemize}
\item The harmonic slicing condition \eref{eg:harmlapse} with explicit source terms given by \eref{ee:dinalphaalpha} and the 1+log slicing condition \eref{eg:1ploglapse} with \eref{eg:nokval} and the source terms in \eref{ee:tun1plog} are such that the divergent terms in the equations can only cancel appropriately if, after using \eref{er:Kregrel}, the lapse satisfies
\begin{equation} \label{er:fixalpha}
\atscrip{\alpha}=-\frac{\Kc\rscri}{3}. 
\end{equation}
\item The same happens with the damping terms added in the Gamma-driver shift conditions \eref{eg:Gammadriver} and \eref{eg:integGammadriver} with source terms as in \eref{ee:expGammadriver} and \eref{ee:expintegGammadriver} respectively. If $\xi_{\beta^r}\neq0$, then necessarily 
\begin{equation} 
\atscrip{\beta^r}=\frac{\Kc\rscri}{3} . 
\end{equation}
\end{itemize}
The previous regularity conditions on the gauge quantities have the effect of fixing their values at $\scri^+$, as will be described in chapters \ref{c:exper} and \ref{c:results}. This puts further constraints on the regularity conditions derived in the previous subsection, so that \eref{er:scrifixreg} and \eref{er:Kregrel} now reduce to 
\begin{equation}\label{er:simpliregs}
\atscrip{\chi}=\atscrip{\gamma_{rr}} \qquad \textrm{and} \qquad \atscrip{\DPK}=0 , 
\end{equation}
among other simplifications in the regularity conditions \eref{er:regconds}. 

\item The harmonic gauge conditions with source terms calculated from the background conformal metric components \eref{eg:harmtwobsimpl} have no terms divided by $\Omega$. However, in subsection \ref{se:confbg}, I will describe the form of a source function that maintains the numerical simulation stable, namely $\bar F^t_\alpha=\case{\Kc}{\Omega\alpha^3}\left(\hat\alpha^2-\alpha^2\right)$. The regularity condition induced by this term is the same as \eref{er:fixalpha}. No regularity condition appears for the shift - it is allowed to move at $\scri^+$, see \fref{fs:shiftcondab}. 

\item The harmonic gauge conditions with source terms including background physical metric components \eref{eg:harmtwobpsimpl} do include terms that diverge at $\scri^+$. After setting the source functions as indicated in subsection \ref{se:physbg} ($\tilde F^r=0$ and $\tilde F^t = \xi_\alpha\Omega(\alpha-\hat\alpha)$) and using the regularity conditions \eref{er:regconds}, the numerator of the $\Omega$ terms in $\dot\beta^r$ vanishes, while the one in $\dot\alpha$ gives the relation
\begin{equation}
\atscrip{\left(2   {\beta^r}\Omega '-\xi_\alpha\alpha ^3 -\frac{1}{3} \alpha ^2 \xi_\alpha \Kc \rscri-\frac{6   {\beta^r}^2 \gamma _{rr} \Omega '}{\Kc \rscri \gamma _{\theta \theta }}\right)}=0 .  
\end{equation}
In this case neither $\alpha$ nor $\beta^r$ are fixed at $\scri^+$, but their values have to satisfy this relation. 
\end{itemize}
\end{subequations}

{ \ }

The regularity conditions have to be satisfied at $\scri^+$ so that the formally divergent terms take finite limits there. The conditions also have to vanish with the appropriate power of $\Omega$ to avoid the appearance of stiff terms. 
If a staggered grid (see subsection \ref{sn:orinf}) is used, a stable evolution will automatically satisfy the regularity conditions. Nevertheless, in the case of a non-staggered grid, the conditions will have to be explicitly imposed in order to be able to evaluate finite values of the RHSs exactly on $\scri^+$.  

%\subsection{Preferred conformal gauge ???? - how the regularity conditions deal with it}

%the boxomega condition is satisfied with the scri-fixing condition and the constant areal radius condition $\dot{\frac{\gamma_{\theta\theta}}{\chi}}=0$

%the angular part of nablanabla condition is satisfied by the constant areal radius condition $\dot{\frac{\gamma_{\theta\theta}}{\chi}}=0$

%\upda{think about what to put here}

%\section{Exponential growths and stationary solutions ... }

%\section{Free vs constrained evol, first-set-gauge-then-regularize}

%\renewcommand\bibname{{References}}
%\bibliographystyle{../../master/thesis/tocunsrt}
%\bibliography{../articles/hypcomp} 

\chapter{Numerical methods}\label{c:num}

The main ingredients required for the implementation of the equations in the code are: the discretization in space and time of the variables, the approximation of the spatial derivatives in the RHSs, the integration in time of the equations, boundary conditions, dissipation and the calculation of the initial data. The importance of the Courant factor and of convergence checks is also briefly discussed. 

\section{Spatial discretization}

A continuum system of equations has to be discretized in order to be implemented and solved in a numerical code. There are several discretization procedures available for discretizing the spatial part. For smooth fields and simple geometries, excluding fluid dynamics, the pseudo-spectral methods and the finite difference approach are the most popular ones.  

Pseudo-spectral methods, which are a variation of spectral methods, basically consist of writing the solutions to the differential equations as a sum of some given basis functions with time-dependent coefficients. The basis functions are selected depending on the properties of the problem that has to be solved and common options are Fourier series or Chebyshev polynomials. The number of terms in the sum is the same as the number of points in the discrete grid. The spectral method thus uses all the gridpoints in the interpolations, taking a global approach to the system. The coefficient equations are solved using common numerical techniques and then the solution is reconstructed. Pseudo-spectral methods provide exponential convergence when the solution is smooth and the number of points they require is not very large. In spite of these advantages, pseudo-spectral methods are not optimal for our purposes, because their global character is likely to make them more difficult to stabilize, especially in presence of the divergent terms at $\scri^+$ that appear in the equations we consider. 
For a simple description of the pseudo-spectral approach see \cite{Bonazzola1999433}.  % http://www.sciencedirect.com/science/article/pii/S0377042799001673

%Wiki: The idea is to write the solution of the differential equation as a sum of certain "basis functions" (for example, as a Fourier series which is a sum of sinusoids) and then to choose the coefficients in the sum in order to satisfy the differential equation as well as possible. Wiki: When applying spectral methods to time-dependent PDEs, the solution is typically written as a sum of basis functions with time-dependent coefficients; substituting this in the PDE yields a system of ODEs in the coefficients which can be solved using any numerical method for ODEs. 

The finite differences approach does not rely on a basis in the same way as the pseudo-spectral methods do (the solution is assumed to be represented by a polynomial around every gridpoint) and it only uses a certain number of gripoints in the interpolations (local approach). The convergence order is finite, but due to its robustness it will better fulfill our requirements. 

\subsection{Finite differences}\label{a:fd}

The first step is to discretize the values of the variables to be integrated along a given number of points in the coordinates, or more generally, the variables they depend on. Suppose a given function $u(t,x)$ is one of those integration variables. A grid containing a given number of points in $t$ and $x$ is set and the function $u(t,x)$ is evaluated at those points. Assuming that the chosen grid is equidistant (mesh refinement will not be considered here), the discretized quantities are
\begin{subequations}
\begin{eqnarray}
x_i &=& x_0+i\Delta x \ , i = 0, 1, ..., I  , \\
t_n &=& t_0+n\Delta t \ , n = 0, 1, ..., N  , \\
u_i^n &=& u(t_n,x_i) . 
\end{eqnarray}
\end{subequations}

The spatial derivatives of the variables are approximated by a difference quotient. In the context of finite differences a difference quotient is given by a fractional expression that includes a linear combination of the values of the variable at a certain number of close points of the grid (called the stencil) in the numerator and a multiple of the grid-spacing to the same power as the order of the derivative in the denominator. 
In a more intuitive way, it is as if the derivatives were substituted by a discrete version of their definition. For example, a first derivative is defined by
\begin{equation} \frac{\partial u}{\partial x}=\displaystyle\lim_{\Delta x\to0}\frac{u(x+\Delta x)-u(x)}{\Delta x} \end{equation}
and the corresponding finite difference (first order and forward) with $\Delta x$ grid-spacing is
\begin{equation} \left.\frac{\partial u}{\partial x}\right|_i=\frac{u_{i+1}-u_i}{\Delta x} +{\cal O}(\Delta x)\approx\frac{u_{i+1}-u_i}{\Delta x} \ .\end{equation}

Finite differences are approximations to the analytical derivatives, so that they have an associated error proportional to a certain power of the grid-spacing. The value of this power gives the convergence order of the finite differences. The higher it is, the faster the numerical solution will converge to the analytical one when the grid resolution is increased. On the other hand, the higher the convergence order, the more terms the stencil requires, which increases the calculation time. 
%Depending if it is the first, second, third, ... derivative that is being calculated, the powers of the grid-spacing in the denominator of the finite difference will be the first, second, third ... If the order of the derivative is increased, the stencil becomes larger for the same convergence order. 

The finite difference operators can be calculated using Taylor series, Lagrange polynomials or a simple algorithm derived by Fornberg \cite{Fornberg:1998:CWF}. 
%The simplest one uses Taylor expansions of the variables and allows to calculate the error, but it is not practical for large stencils. A straightforward way of obtaining the correct coefficients for the stencil is by deriving the Lagrange polynomial of the appropriate order. Fornberg \cite{fornberg1998calculation}
As an example, the 4th order centered first and second derivatives are given by: 
%( -invect(i+2) + 8.0_wp*invect(i+1) - 8.0_wp*invect(i-1) + invect(i-2) )
% -(invect (-2 + i) - 16*invect (-1 + i) + 30*invect (i) - 16*invect (1 + i) + invect (2 + i))
\begin{subequations}\begin{eqnarray*}
\left.\frac{\partial u}{\partial x}\right|_i&=&\frac{-u_{i+2}+8u_{i+1}-8u_{i-1}+u_{i-2}}{12\,\Delta x} +{\cal O}(\Delta x^4) , \\
\left.\frac{\partial^2 u}{\partial x^2}\right|_i&=&\frac{-u_{i+2}+16u_{i+1}-30u_i+16u_{i-1}-u_{i-2}}{12\,\Delta x^2} +{\cal O}(\Delta x^4) .
\end{eqnarray*}\end{subequations}

The stencils can be centered (taking the same number of points on both sides) or asymmetric. For the same order, the latter have larger errors. Depending on the features of the problem, one or the other may be more convenient. For instance, one-sided stencils may be useful at the boundaries, whereas using one-point off-centered stencils for advection-type derivatives arising from the shift vector is common practice in numerical relativity simulations (see \cite{Chirvasa:2008xx} for the shifted wave equation case and \cite{Husa:2007hp} for moving-puncture BH results). A centered and an one-sided 4th order stencils at the boundaries are shown in \fref{fn:centone} and the one-point off-centered stencil that will be used for the advection terms in tests performed as part of this work is presented in \fref{fn:offcentdiagram}. 
\begin{figure}[htbp]
\center
	\begin{tikzpicture}[scale=0.72]
		\draw (0cm, 0cm) -- (8.5cm, 0cm); \draw[dashed] (8.5cm, 0cm) -- (9.5cm, 0cm); \draw (9.5cm, 0cm) -- (22cm, 0cm);
		\draw (3cm, 70pt) -- (3cm, - 10pt); \draw (3cm, 0cm) node[above=52pt] {$r=0$};
		\draw (19cm, 70pt) -- (19cm, - 10pt); \draw (19cm, 0cm) node[above=52pt] {$r=\scri^+$};
		\foreach \x in {0, 2} \draw (\x cm, 0cm) circle (4.5pt);
		\foreach \x in {4, 6, 8, 10} \fill (\x cm, 0cm) circle (4.5pt);
		\foreach \x in {12, 14, 16, 18} \fill (\x cm, 0cm) circle (4.5pt);
		\foreach \x in {20, 22} \draw (\x cm, 0cm) circle (4.5pt);
		\fill (4cm, 2cm) circle (4.5pt); \foreach \x in {0,2,4,6,8} \draw[dashed] (\x cm,0cm) -- (4cm, 2cm);
		\fill (18cm, 2cm) circle (4.5pt); \foreach \x in {10,12,14,16,18} \draw[dashed] (\x cm,0cm) -- (18cm, 2cm);
		\foreach \x in {6, 8, 10, 12, 14, 16} \fill (\x cm, 2cm) circle (4.5pt);
	\end{tikzpicture}
	\caption{Centered stencil on the left and completely one-sided stencil on the right, both for 4th order finite differences.} \label{fn:centone}
\end{figure}
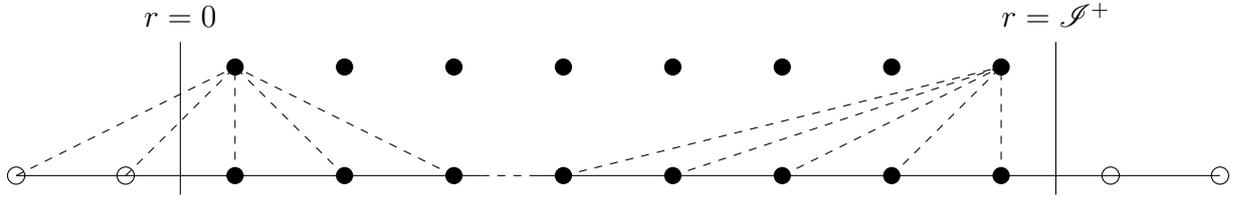
\begin{figure}[htbp]
\center
	\begin{tikzpicture}[scale=0.72]
		\draw (0cm, 0cm) -- (10.5cm, 0cm); \draw[dashed] (10.5cm, 0cm) -- (11.5cm, 0cm); \draw (11.5cm, 0cm) -- (22cm, 0cm);
		\draw (3cm, 70pt) -- (3cm, - 10pt); \draw (3cm, 0cm) node[above=52pt] {$r=0$};
		\draw (19cm, 70pt) -- (19cm, - 10pt); \draw (19cm, 0cm) node[above=52pt] {$r=\scri^+$};
		\foreach \x in {0, 2} \draw (\x cm, 0cm) circle (4.5pt);
		\foreach \x in {4, 6, 8, 10} \fill (\x cm, 0cm) circle (4.5pt);
		\foreach \x in {12, 14, 16, 18} \fill (\x cm, 0cm) circle (4.5pt);
		\foreach \x in {20, 22} \draw (\x cm, 0cm) circle (4.5pt);
		\fill (4cm, 2cm) circle (4.5pt); \foreach \x in {0,2,4,6,8} \draw[dashed] (\x cm,0cm) -- (4cm, 2cm);
		\fill (6cm, 2cm) circle (4.5pt); \foreach \x in {0,2,4,6,8} \draw[dotted] (\x cm,0cm) -- (6cm, 2cm);
		\fill (8cm, 2cm) circle (4.5pt); \foreach \x in {2,4,6,8,10} \draw[loosely dashed] (\x cm,0cm) -- (8cm, 2cm);
		\fill (18cm, 2cm) circle (4.5pt); \foreach \x in {12,14,16,18,20} \draw[dashed] (\x cm,0cm) -- (18cm, 2cm);
		\foreach \x in {10, 12, 14, 16} \fill (\x cm, 2cm) circle (4.5pt);
	\end{tikzpicture}
	\caption{Off-centered stencil used with 4th order finite differences.} \label{fn:offcentdiagram}
\end{figure}
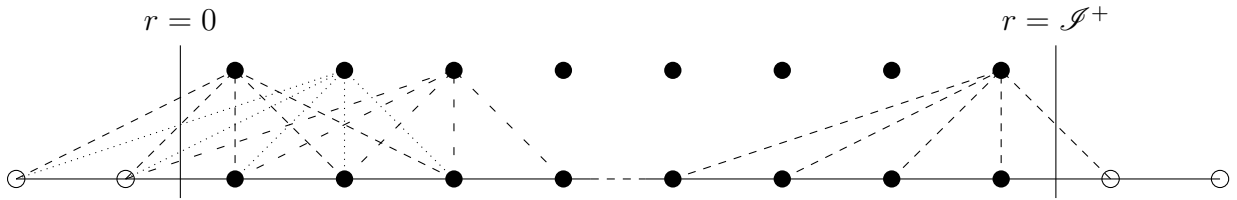

\section{Method of Lines}

The Method of Lines (MoL) technique allows to separate the space and time discretization processes. 
In a system of partial differential equations (PDE)
\begin{equation}\label{en:mol}
\partial_t\vec{u}=\vec f(t,\vec{x},\vec {u},\partial_{\vec{x}}\vec{u},\partial_{\vec{x}}^2\vec{u},...) \ , 
\end{equation}
where the array $\vec f$ in the RHS does not contain any time derivatives, the spatial derivatives are discretized leaving the time dependence continuous. The resulting semi-discrete system is no longer a PDE system, but has been transformed into a system of ordinary differential equations (ODE). The time integration can now be performed using any ODE discretization and integration method. For example, applying the Euler method (first order) to the system \eref{en:mol} gives
\begin{equation}
\vec{u}^{(n+1)}=\vec{u}^{(n)}+\Delta t\vec f(t_n,\vec{x},\vec{u}^{(n)},\partial_{\vec{x}}\vec{u}^{(n)},\partial_{\vec{x}}^2\vec{u}^{(n)},...) \ .
\end{equation}
Generally it is more convenient to use more elaborate methods of higher order, such as the Runge-Kutta (RK) methods, which require some intermediate time-steps. 
Of those, the most common one is the 4th order RK with 3 time-levels. For a differential equation of the form $\partial_tu=\textrm{rhs}(u,t)$ and time-step $\Delta t$, its expression is the following: 
\begin{subequations}\label{a:rk4}
\begin{eqnarray}
k_1 &=& \Delta t\ \textrm{rhs}(u^n,t_n)  , \\
k_2 &=& \Delta t\ \textrm{rhs}(u^n+k_1/2,t_n+\Delta t/2)  , \\
k_3 &=& \Delta t\ \textrm{rhs}(u^n+k_2/2,t_n+\Delta t/2)  , \\
k_4 &=& \Delta t\ \textrm{rhs}(u^n+k_3,t_n+\Delta t)  , \\
u^{n+1} &=& u^n+\frac{1}{6}(k_1+2k_2+2k_3+k_4)  .
\end{eqnarray}
\end{subequations} 
This method is 4th order of convergence. Nevertheless, if the finite difference scheme used for the spatial derivatives has a higher convergence order, the latter can still be achieved for the numerical results by setting a sufficiently small time-step $\Delta t$, because in this way the error of the time integrations becomes negligible. 

\subsection{Explicit and implicit methods}\label{sn:imex}

There are mainly two options to solve numerically a time-dependent problem: using an explicit or an implicit algorithm. An explicit method obtains a solution for later times from calculations that involve the current (and possibly previous) times, conceptually written as 
\begin{equation*}
u^{n+1} = F(u^n) . 
\end{equation*}
In the case of an implicit integration method, the solution at the following time-step is included together with the previous time solutions in a function of the form 
\begin{equation*}
G(u^n, u^{n+1})=0 ,  
\end{equation*}
so that the numerical implementation for solving this implicit equation is more complicated and more expensive computationally. 

The Courant number $C$ is the quotient of the time-step $\Delta t$ and the time interval given by $\case{\Delta x}{v}$, where $v$ is the maximum propagation speed of the system. In the case of explicit time integration algorithms, there is a maximum allowed value of the Courant number that will provide a correct and stable solution. 
This is the Courant-Friedrichs-Lewy (CFL) condition \cite{Courant1928Uber} and it can be written for one and n-dimensional systems of equations as
\begin{equation}\label{en:Courant}
C = \frac{v\,\Delta t}{\Delta x} \leq  C_{max} , \qquad \qquad C = \Delta t \sum_{i=1}^n\frac{v_{i}}{\Delta x_i} \leq C_{max} . 
\end{equation}
The value of $C_{max}$ will depend on the exact integration method chosen. The interpretation of this condition is that the numerical domain of dependence must include the analytical one, so that the information required to integrate to the next time-step is complete. 

The CFL limit can suppose a problem in explicit methods when stiff terms are present in the equations. Stiff terms require very small time-steps to provide a stable evolution. In this case the best option is usually to use an implicit integration scheme, because even if they are more expensive computationally, they do not have a maximum allowed time-step $\Delta t$ and the integration can be performed using larger time intervals. 

Efforts to obtain mixed implicit-explicit (IMEX) methods that combine the best features of both approaches include the PIRK (partially-implicit RK) methods \cite{Montero:2012yr,CorderoCarrion:2012ic}, see also \cite{Ascher:1995:IEM}, which have also been used in three-dimensional simulations with spherical polar coordinates \cite{Baumgarte:2012xy}.

\section{Boundary conditions}\label{sn:bcs}

The spatial integration domain where the variables are discretized and evolved is finite due to the limited memory and computational power of the machines. Due to these restrictions, the n-dimensional grids at whose nodes the variables are evaluated at each time-step only cover a certain interval of the coordinates. 

The finite difference stencils implemented to discretize the spatial derivatives require the values of the variables at the gridpoints around the one where the derivative is calculated. At the outer points of the grid some of these values will be missing. In order to calculate the derivative at the boundaries, the missing information has to be included. This is done by means of boundary conditions imposed on the ghost points, which are extra points that lie outside of the original grid. The number of required gridpoints in the stencil (and thus of ghost points) is determined by the order of the finite differences. For instance, for a 4th order centered stencil, two ghost points have to be set. The boundary conditions contain information about how the variables should behave on the ghost points and will vary depending on the physical properties of the fields at each boundary. In general, finding appropriate boundary conditions is a difficult problem in numerical simulations. 

The use of ghost points and boundary conditions can be avoided by using one-sided stencils at the boundaries. However, the errors associated to one-sided stencils are larger than in the centered case and in some cases special symmetry conditions may have to be explicitly imposed.

\subsection{Boundary conditions for the spherically symmetric code}

In spherical symmetry the only spatial dependence is given by the radial coordinate $r$. The initial data choices described in chapter \ref{c:initial} use a compactified isotropic radius that takes values from the origin $r=0$ to the location of future null infinity $r=\rscri$. The different physics content at both ends of the grid requires a specific treatment for each of them. 

\subsubsection{Boundary conditions at the origin}

The parity (symmetric or antisymmetric behaviour with respect to $r=0$) is a physical property of each of the integration variables and has to be preserved at the origin. It can be easily imposed as a boundary condition by filling the corresponding ghost points (the ones on the left of the integration domain) with either with the mirrored value of the variable with respect to $r=0$ (even parity) or with minus the mirrored values (odd parity). 

Whether a variable has an even (symmetric) or odd (antisymmetric) parity is determined from physical considerations. 
In spherical symmetry scalar quantities will be even, vectors can be odd or even and tensors inherit the parity property from the vectors that compose them. 
The time derivative of a variable does change parity, whereas a spatial derivative does change it from even to odd and vice versa. Similarly, multiplying (or dividing) a variable by an odd power of the radial coordinate also inverts the parity. 
The parity of each on the spherically symmetric variables in equations \eref{es:eeqs} is the following (compare to section 4.2 in \cite{Sorkin:2009bc} and section VI.A in \cite{Alcubierre:2010is}): 
\begin{description}
\item[Even:] condition is $A(-r)=A(r)$ and is satisfied by $\chi, \gamma_{rr}, \gamma_{\theta\theta}, A_{rr}, K, \Theta, \alpha, \iPhi$ and $\iPi$. 
\item[Odd:] condition satisfies $B(-r)=-B(r)$ and the odd variables are $\Lambda^r, \beta^r$ and $B^r$. 
\end{description}
The conformal factor $\Omega$ is even and any rescaling of the variables involving it will not change their parity. On the other hand, $\Omega'$ is odd (this can be checked explicitly in \eref{ein:omega}). The compactification factor $\aconf$ is odd at the origin, as it is proportional to $r$ there. 

\subsubsection{Boundary conditions at the outer boundary}

No information can enter the spacetime through future null infinity because it is an ingoing null hypersurface. This means that if the boundary of the spatial grid is set exactly at $\scri^+$ (and maintained there using the scri-fixing condition, see section \ref{cg:scrifix}), no assumptions regarding the incoming modes have to be encoded in the boundary conditions because these modes do not exist. Also if the values set at the ghost points (which lie on the right of the location of $\scri^+$, outside of the compactified spacetime) were not correctly set, they should not affect the solution in the interior spacetime. This very convenient, as the complicated problem of finding the appropriate boundary conditions is radically simplified. 

In an equivalent way as is done when using excision in BH simulations (putting an inner boundary inside of the BH horizon), in principle we could also go beyond $\scri^+$ in the integration domain and put the outer boundary at a value of the radial coordinate $r>\rscri$. The first drawback is that the conformal factor $\Omega$ becomes negative for $r>\rscri$ and this turns some of the damping terms with $\Omega$ in their denominator into exponentially growing ones, which is very likely to render the simulation unstable. 

A simple and sufficiently good option for the calculation of spatial derivatives in the vicinity of $\scri^+$ is to use one-sided stencils there. Nevertheless, a centered stencil may be preferable in order to decrease the error of the finite differencing (we are actually interested in calculating quantities on $\scri^+$). Besides, the dissipation operator that will be described in section \ref{sn:dissipation} is defined using a centered stencil and will require ghost point information anyway. 

Taking into account the speed profile at $\scri^+$ (all eigenspeeds are outgoing), extrapolation is a good candidate to fill the values of the grid points. This is implemented in the code using the outflow boundary conditions in \cite{Calabrese:2005fp}. 

According to the analysis in \cite{Calabrese:2005fp}, the minimum order of extrapolation is different depending on the kind of variables. For the $u$ variables (the ones that appear spatially derived twice) defined in subsection \ref{sr:1st2nd} the extrapolation order has to be an order higher than the order of the finite differencing, whereas for the $v$ variables (with only first spatial derivatives) it can be the order of the finite differences or an order higher like for $u$. The conditions that are used together with 4th order finite differences are thus 
\begin{subequations}\label{en:outflow}
\begin{eqnarray}
\Delta x^5D_-^5u_{I+1}&=&0 , \\
\Delta x^5D_-^5u_{I+2}&=&0 , \\
\Delta x^4D_-^4v_{I+1}&=&0 , \\
\Delta x^4D_-^4v_{I+2}&=&0 .
\end{eqnarray}
\end{subequations}
The condition is applied on the $I+1$ and $I+2$ locations because a 4th order finite difference stencil at the boundary requires two ghost points. The derivative operator $D_-$ is a backward 1st order finite difference, whereas $D_+$ is the equivalent forward one. They are defined as
\begin{equation}\label{en:Dpm}
D_\pm u_i = \pm\frac{u_{i\pm1}-u_i}{\Delta x} . 
\end{equation}
Expanding the conditions \eref{en:outflow} gives the following values for the variables on the ghost points $I+1$ and $I+2$: 
\begin{subequations}\label{en:outflowexpl}
\begin{eqnarray}
u_{I+1} &=& 5\,u_{I}-10\,u_{I-1}+10\,u_{I-2}-5\,u_{I-3}+u_{I-4} , \\ 
u_{I+2} &=& 5\,u_{I+1}-10\,u_{I}+10\,u_{I-1}-5\,u_{I-2}+u_{I-3} , \\
v_{I+1} &=& 4\,v_{I}-6\,v_{I-1}+4\,v_{I-2}-\,v_{I-3} , \\ 
v_{I+2} &=& 4\,v_{I+1}-6\,v_{I}+4\,v_{I-1}-\,v_{I-2}  .
\end{eqnarray}
\end{subequations}
The ``$u$'' expressions are the ones to impose for the specially symmetric variables $\chi$, $\gamma_{rr}$, $\gamma_{\theta\theta}$, $\alpha$, $\iPhi$ and $\beta^r$, whereas the quantities $A_{rr}$, $K$, $\Lambda^r$, $\Theta$, $\iPi$ and, if present, $B^r$ can use the ``$u$'' or the ``$v$'' expressions. 

\subsubsection{Boundary conditions at the excision boundary}

If excision is used inside of the BH, the treatment of the stencils at the inner boundary $r=r_{exc}<r_{horizon}$ is similar to the one used at $\scri^+$. Provided that appropriate gauge conditions are chosen, no information can enter the domain outside of the horizon from the inside, so that no boundary conditions are required. The derivatives next to the excision boundary can be evaluated using one-sided stencils or using extrapolation for the ghost points at locations $-1$ and $-2$ (for 4th order finite differences) in an equivalent way as in \eref{en:outflow}, 
\begin{subequations}\label{en:excoutflow}
\begin{eqnarray}
\Delta x^5D_+^5u_{-1}&=&0 , \\
\Delta x^5D_+^5u_{-2}&=&0 , \\
\Delta x^4D_+^4v_{-1}&=&0 , \\
\Delta x^4D_+^4v_{-2}&=&0 .
\end{eqnarray}
\end{subequations}
The final expressions are the same as those in \eref{en:outflowexpl}, making the substitutions $I$ by $0$ and $+$ by $-$ and vice versa in the indices. 

\subsection{Treatment of the origin and future null infinity}\label{sn:orinf}

The spherically symmetric equations \eref{es:eeqs}, \eref{es:pKeeqs} and \eref{es:DPKeeqs} are divergent at the origin, $r=0$, and at $\scri^+$, located at $r=\rscri$. There are two main ways to treat these divergences: either avoid the two problematic points using a staggered grid or use special conditions on the variables and take limits in the equations to obtain the corresponding finite behaviour at these two points. 

\subsubsection{Staggered grid}

\begin{figure}[htbp]
\center
	\begin{tikzpicture}[scale=0.87]
		\draw (0cm, 0cm) -- (8.44cm, 0cm);  \draw[dashed] (8.44cm, 0cm) -- (9.56cm, 0cm); \draw (9.56cm, 0cm) -- (18cm, 0cm);
		\draw (3cm, 20pt) -- (3cm, - 70pt);
		\draw (3cm, 0cm) node[above=22pt] {$r=0$};
		\draw (15cm, 20pt) -- (15cm, - 70pt);
		\draw (15cm, 0cm) node[above=22pt] {$r=\rscri$};
		\foreach \x in {0, 2, 16,18} \draw (\x cm, 0cm) circle (4.5pt);
		\foreach \x in {4, 6, 8, 10, 12, 14} \fill (\x cm, 0cm) circle (4.5pt);
		\foreach \x in {0, 2, 4, 6, 12, 14, 16, 18} \draw[thick] (\x cm, -20pt) -- (\x cm, - 25pt);
		\draw (9cm, -22pt) node {Low resolution};%[above=-30pt]
		\foreach \x in {2, 2.67, 15.33, 16} \draw (\x cm, 0cm) circle (3pt);
		\foreach \x in {3.33, 4, 4.67, 5.33, 6, 6.67, 7.33, 8, 10, 10.67, 11.33, 12, 12.67, 13.33, 14, 14.67} \fill (\x cm, 0cm) circle (3pt);
		\foreach \x in {2, 2.62, 3.33, 4, 4.67, 5.33, 6, 6.67, 11.33, 12, 12.67, 13.33, 14, 14.67, 15.33, 16} \draw[thick] (\x cm, -35pt) -- (\x cm, - 40pt);
		\draw (9cm, -37pt) node {Middle resolution};
		\foreach \x in {2.67, 2.89, 15.11, 15.33} \draw (\x cm, 0cm) circle (1.5pt);
		\foreach \x in {3.11, 3.33, 3.56, 3.78, 4, 4.22, 4.44, 4.67, 4.89, 5.11, 5.33, 5.56, 5.78, 6, 6.22, 6.44, 6.67, 6.89, 7.11, 7.33, 7.56, 7.78, 8, 8.22, 8.44, 9.56, 9.78, 10, 10.22, 10.44, 10.67, 10.89, 11.11, 11.33, 11.56, 11.78, 12, 12.22, 12.44, 12.67, 12.89, 13.11, 13.33, 13.56, 13.78, 14, 14.22, 14.44, 14.67, 14.89} \fill (\x cm, 0cm) circle (1.5pt);
		\foreach \x in {2.67, 2.89, 3.11, 3.33, 3.56, 3.78, 4, 4.22, 4.44, 4.67, 4.89, 5.11, 5.33, 5.56, 5.78, 6, 6.22, 6.44, 6.67, 6.89, 11.11, 11.33, 11.56, 11.78, 12, 12.22, 12.44, 12.67, 12.89, 13.11, 13.33, 13.56, 13.78, 14, 14.22, 14.44, 14.67, 14.89, 15.11, 15.33} \draw[thick] (\x cm, -50pt) -- (\x cm, - 55pt);
		\draw (9cm, -52pt) node {High resolution};
	\end{tikzpicture}
	\caption{Disposition of the points in the staggered grid for three different resolutions.}\label{fn:stgrid}
\end{figure}
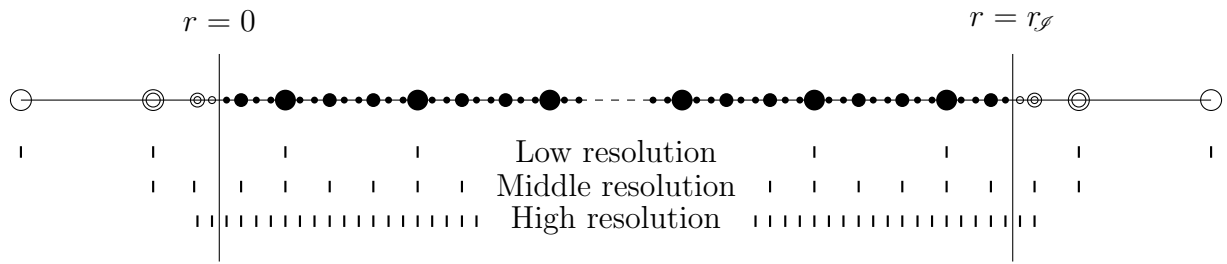

In a staggered grid the origin and $\scri^+$ can be easily avoided: the gridpoints are moved half a spatial step to the right and the rightmost point is eliminated. In this way, the integration domain covers $r\in(0,\rscri)$ and the ghost points lie on the left of $r=0$ and on the right of $r=\rscri$. %(gridpoints used to calculate derivatives and that have to be filled according to the boundary conditions, as their values are not evolved) 
An example of a staggered grid with two ghost points is illustrated by \fref{fn:stgrid}. 

As a curiosity, a comparison between different resolution runs can be performed without the need for interpolation if the increase of resolution is a multiple of three. This is also depicted in \fref{fn:stgrid}, where a third of the gridpoints used for a given resolution coincide with those of a lower one and thus allow for a direct comparison. In practice it may be more convenient to use a smaller factor to increase the resolution (1.5 or 2) and then interpolate in the post-processing of the data.  

The advantages of using a staggered grid is that it is very simple to use and no special treatment is required at the points where the equations diverge. However, as the quantities are not evaluated right on $\scri^+$, extrapolation is required to extract the signal there.

\subsubsection{Non-staggered grid} \label{a:nonstag}

\begin{figure}[htbp]
\center
	\begin{tikzpicture}[scale=0.87]
		\draw (0cm, 0cm) -- (8.44cm, 0cm);  \draw[dashed] (8.44cm, 0cm) -- (9.56cm, 0cm); \draw (9.56cm, 0cm) -- (18cm, 0cm);
		\draw (4cm, 20pt) -- (4cm, - 70pt);
		\draw (4cm, 0cm) node[above=22pt] {$r=0$};
		\draw (14cm, 20pt) -- (14cm, - 70pt);
		\draw (14cm, 0cm) node[above=22pt] {$r=\rscri$};
		\foreach \x in {0, 2, 16,18} \draw (\x cm, 0cm) circle (4.5pt);
		\foreach \x in {4, 6, 8, 10, 12, 14} \fill (\x cm, 0cm) circle (4.5pt);
		\foreach \x in {0, 2, 4, 6, 12, 14, 16, 18} \draw[thick] (\x cm, -20pt) -- (\x cm, - 25pt);
		\draw (9cm, -22pt) node {Low resolution};%[above=-30pt]
		\foreach \x in {2.67, 3.33, 14.67, 15.33} \draw (\x cm, 0cm) circle (3pt);
		\foreach \x in {4, 4.67, 5.33, 6, 6.67, 7.33, 8, 10, 10.67, 11.33, 12, 12.67, 13.33, 14} \fill (\x cm, 0cm) circle (3pt);
		\foreach \x in {2.62, 3.33, 4, 4.67, 5.33, 6, 6.67, 11.33, 12, 12.67, 13.33, 14, 14.67, 15.33} \draw[thick] (\x cm, -35pt) -- (\x cm, - 40pt);
		\draw (9cm, -37pt) node {Middle resolution};
		\foreach \x in {3.56, 3.78, 14.22, 14.44} \draw (\x cm, 0cm) circle (1.5pt);
		\foreach \x in {4, 4.22, 4.44, 4.67, 4.89, 5.11, 5.33, 5.56, 5.78, 6, 6.22, 6.44, 6.67, 6.89, 7.11, 7.33, 7.56, 7.78, 8, 8.22, 8.44, 9.56, 9.78, 10, 10.22, 10.44, 10.67, 10.89, 11.11, 11.33, 11.56, 11.78, 12, 12.22, 12.44, 12.67, 12.89, 13.11, 13.33, 13.56, 13.78, 14} \fill (\x cm, 0cm) circle (1.5pt);
		\foreach \x in { 3.56, 3.78, 4, 4.22, 4.44, 4.67, 4.89, 5.11, 5.33, 5.56, 5.78, 6, 6.22, 6.44, 6.67, 6.89, 11.11, 11.33, 11.56, 11.78, 12, 12.22, 12.44, 12.67, 12.89, 13.11, 13.33, 13.56, 13.78, 14, 14.22, 14.44} \draw[thick] (\x cm, -50pt) -- (\x cm, - 55pt);
		\draw (9cm, -52pt) node {High resolution};
	\end{tikzpicture}
	\caption{Disposition of the points in the non-staggered grid for three resolutions.}\label{fn:nstgrid}
\end{figure}
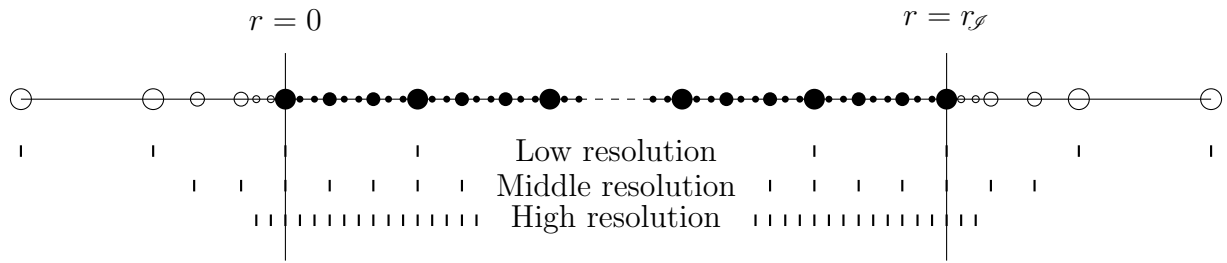

A non-staggered grid is shown in \fref{fn:nstgrid}, together with the higher resolution grids also shown for the staggered case. 

The coordinate locations $r=0$ and $r=\rscri$ are completely different from the physical point of view, so that the approaches used to evaluate the variables and evolution equations at those points will also be different. 

The value of the variables at the origin can be determined easily using parity considerations. It is done as follows, where we use a generic variable denoted by $u$: 
\begin{itemize} 
\item Odd variables: an antisymmetric quantity necessarily becomes zero at $r=0$, so that the condition is simply $u_0^{n+1}=0$. 
\item Even variables: if they are smooth at the origin, they must have a vanishing derivative at $r=0$. The desired expression is obtained by approximating the first radial derivative at the origin with a one-sided backwards finite difference of the desired convergence order and then setting it to zero. The value of the variable at $r=0$ is then isolated and expressed in terms of the other values on its right in the grid. For instance, for 4th order the resulting expression is
\begin{equation*}u_0^{n+1}=\frac{48\,u_1^{n+1}-36\,u_2^{n+1}+16\,u_3^{n+1}-3\,u_4^{n+1}}{25}  . \end{equation*}
This procedure is mentioned in section 5.2 of \cite{Sorkin:2009bc}.  
\end{itemize}

%Another option would be to evaluate the system of equations at $r=0$ and use the l'H\^opital rule to calculate the limit for fractions until no more terms with $\frac{1}{r^n}$ are present. This system then has to be implemented in the code and used only to integrate at $r=0$. 

The value of the variables at $\scri^+$ cannot be determined by parity considerations. In order to make sure that the terms divided by $\Omega$ in the evolution and constraint equations are well behaved, i.e. do not become infinite, the variables have to satisfy certain regularity conditions at $r=\rscri$. This is explained in detail in section \ref{cr:regu}. 
To be able to evaluate the RHSs on $\scri^+$, the l'H\^opital rule has to be applied to the formally divergent terms, so obtaining equations which are regular there. 
%\upda{apply l'H\^opital rule at divergent terms. numerical test?}

\section{Dissipation}\label{sn:dissipation}

The presence of artificial dissipation in numerical simulations has the effect of damping unphysical high frequency oscillations and thus keeps the simulation stable. For this reason adding an extra dissipation term to the equations is common practice. 
A possible way of introducing dissipation is adding Kreiss-Oliger dissipation terms \cite{kreiss1973methods} as used in \cite{Babiuc:2007vr} to the RHS of the evolution equations,
\begin{equation}
\partial_t \boldsymbol{u}\to\partial_t\boldsymbol{u} + Q\boldsymbol{u} \ ,
\end{equation}
where $Q$ is the Kreiss-Oliger dissipation operator of order $2n$ (suitable for a $2n-2$ convergence finite differencing), given by 
\begin{equation}\label{en:KOdiss}
Q=\epsilon\,(-1)^n(\Delta x)^{2n-1}\frac{D_+^nD_-^n}{2^{2n}}   .
\end{equation}
The parameter $\epsilon$ regulates the strength of the dissipation and the derivative operators $D_+$ and $D_-$ are given by \eref{en:Dpm}. 

The convergence order of the finite differencing is unaffected by the addition of Kreiss-Oliger dissipation, because the order of convergence chosen for the dissipation terms is 2 orders higher. This requires an extra ghost point, which is filled using the parity conditions at the origin and using extrapolation at $\scri^+$, as described in section \ref{sn:bcs}. 

\section{Convergence} \label{n:conve}

A well-posed initial value problem needs an appropriate numerical implementation to provide correct results. 

\subsection{Lax equivalence theorem}

One of the ingredients of a successful numerical implementation is consistency. A difference scheme is consistent with the original equations if the difference between both decreases as the spatial and time resolutions increase. This is possible if the truncation error of the discretization scheme is proportional to a certain power of $\Delta x$ and $\Delta t$. 

Another important ingredient is stability. Numerical stability of a numerical algorithm can be studied with a von Neumann analysis. This includes the CFL condition mentioned in section \ref{sn:imex}. In this sense stability can be considered the discrete equivalent to well-posedness. 

The Lax-Richtmyer Equivalence Theorem (\cite{CPA:CPA3160090206}, included as theorem 5.1.4 in \cite{gustafsson1995time}) states for a well-posed problem of a linear system of PDEs that consistency and stability imply convergence. Convergence means that an increase in spatial and time resolution causes that the numerical solution approaches the exact one, thus ensuring that the results are correct.

\subsection{Convergence of the numerical results}\label{sn:conve}

The order of convergence of a discretization method can be calculated from the numerical results even in absence of an exact solution. Suppose a method of convergence order $p$ used with a grid of spacing $h$. The numerical solution is given by $u_h=u_{exact}+\varepsilon\, h^p$, where $u_{exact}$ is the exact solution and $\varepsilon$ is the $h$-independent scaled error. If the resolution is increased by a factor $f$, the new solution will be $u_{fh}=u_{exact}+\varepsilon\, (fh)^p$. Increasing the resolution again by the same factor $f$ gives $u_{f^2h}=u_{exact}+\varepsilon\, (f^2h)^p$. The quotient of the differences between consecutive increases of resolution provides a value that depends only on the known factor $f$  and the order of convergence of the method $p$:
\begin{equation}
\frac{u_{f^2h}-u_{fh}}{u_{fh}-u_h} = \frac{(u_{exact}+\varepsilon\, f^{2p}h^p)-(u_{exact}+\varepsilon \,f^{p}h^p)}{(u_{exact}+\varepsilon\, f^{p}h^p)-(u_{exact}+\varepsilon \,h^p)} = \frac{\varepsilon\, h^p (f^{2p}-f^{p})}{\varepsilon\, h^p(f^{p}-1)} = \frac{f^{p}(f^{p}-1)}{f^{p}-1} = f^p . \label{en:factorp}
\end{equation}
This procedure can be applied to any numerical quantity, for which results at the necessary resolutions exist, to obtain a numerical confirmation of the value of the convergence order. If it is correct, then the numerical implementation of the equations converges appropriately and we know that the solution obtained is the correct one. 

In the case of constraint equations, whose exact value is zero, convergence can also be checked by comparing the different resolution values directly. Denoting the value of the constraint by $u$, we have that $u_h=\varepsilon\, h^p$, $u_{fh}=\varepsilon\, (fh)^p$ and $u_{f^2h}=\varepsilon\, (f^2h)^p$, so that $u_{f^2h}/u_{fh}=u_{fh}/u_h=f^p$.

\section{Numerical calculation of initial data}

%\section{Numerical code used}

The code used to run the simulations presented in this thesis is a one-dimensional spherically symmetric code. 
The time integration by the MoL is a 4th order RK, as described in \eref{a:rk4}. 
Finite differences of orders 2nd, 4th, 6th and 8th are implemented, but for the results shown here 4th order differences were used. 
Kreiss-Oliger dissipation is added to the discretized RHSs as described in section \ref{sn:dissipation}.

\subsection{Calculation of the compactification factor $\aconf$}\label{sn:aconf}

In flat spacetime $\aconf\equiv\Omega$, where there is a simple analytical expression. However, trumpet initial data require the calculation of the critical value of $\Cc$ and the numerical determination of $\aconf$. 

The critical value of $\Cc$ calculated from the values of $M$, $Q$ and $\Kc$ by isolating it from \ref{ein:discrim} is given by an analytical expression only for $Q=0,1$. This analytical expression can acquire an imaginary part due to round-off errors, so that in practice it is more convenient to calculate the critical value of $\Cc$ numerically (for any value of $Q\le M$). In the code this is done by means of a simple bisection method. The bisection method is robust but relatively slow, although the second aspect is not a problem because the $\Cc$ is only determined once in the numerical simulation. 

Once the trumpet value of $\Cc$ is determined, we proceed to construct numerically the profile of the compactification factor $\aconf$. As a first step it is convenient to calculate the value of $R_0$, the double root of \eref{ein:exproots} that corresponds to the choice of critical $\Cc$. This is also done with help of the bisection method. 

\begin{figure}[htbp!!]
\center
	\includegraphics[width=0.75\linewidth]{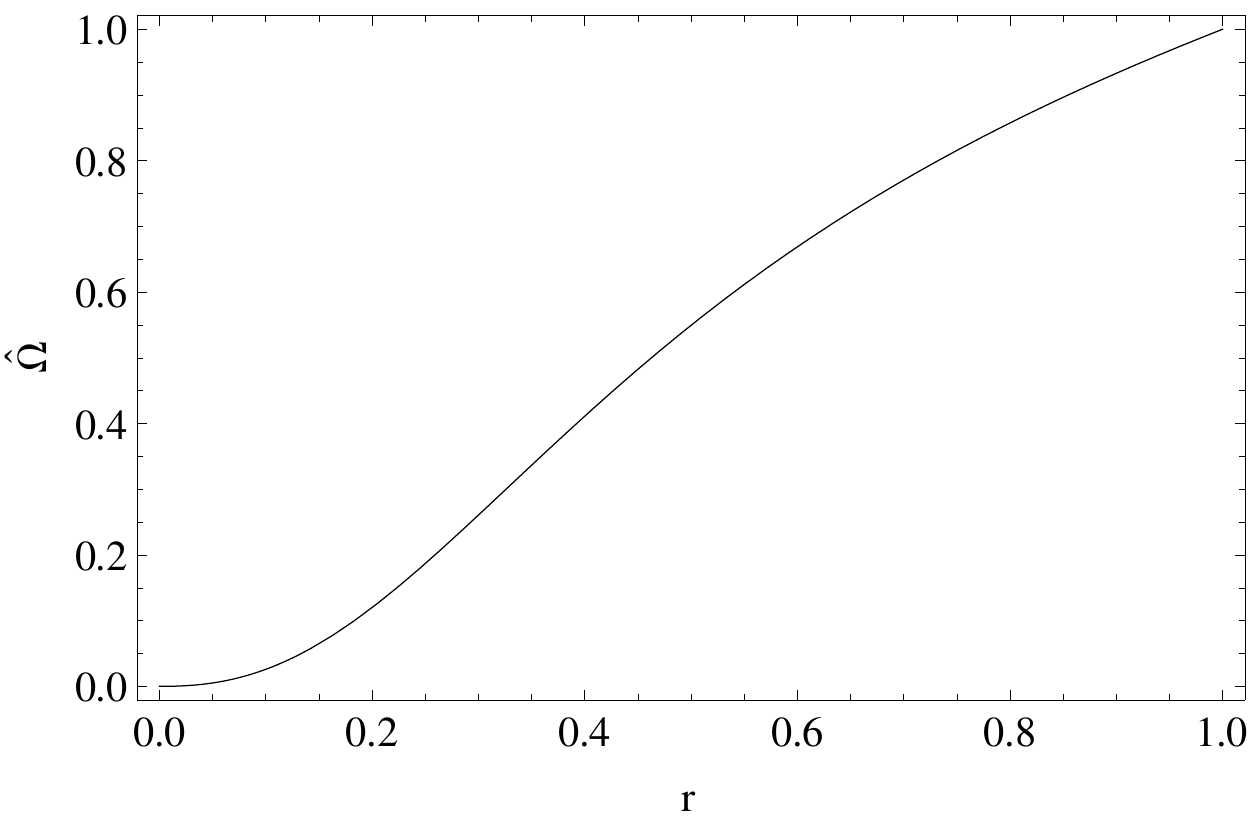}
\caption{Transformed compactification factor $\hat\Omega$ for $M=1$, $Q=0$, $\Kc=-1$ and corresponding critical $\Cc$. Future null infinity is located at $\rscri=1$.}
\label{fn:hatomega}
\end{figure}

The differential equation for $\aconf$ is given by \eref{ein:conflat} substituting the appropriate expression for $A(\case{r}{\aconf})$. We now transform the compactification factor to a new $\hat\Omega$ given by 
\begin{equation}\label{en:hatomega}
\hat\Omega= 1-\frac{R_0}{r}\aconf . 
\end{equation}
Now the differential equation for $\hat\Omega$ is given by
\begin{eqnarray}
%\hat{\Omega }'= \frac{\sqrt{-6 \hat{\Omega }^3 \left(30 \Cc^2+\Cc \Kc R_0^3-3 M R_0^3\right)+9 \hat{\Omega }^2 \left(15 \Cc^2+2 \Cc   \Kc R_0^3+R_0^2 \left(R_0^2-6 M R_0\right)\right)-18 \hat{\Omega } \left(3 \Cc^2+\Cc \Kc R_0^3+R_0^2 \left(R_0^2-3 M R_0\right)\right)+9   \Cc^2 \hat{\Omega }^6-54 \Cc^2 \hat{\Omega }^5+135 \Cc^2 \hat{\Omega }^4+9 \Cc^2+6 \Cc \Kc R_0^3+R_0^2 \left(\Kc^2 R_0^4-18   M R_0+9 R_0^2\right)}}{3 r R_0^2}
\hat{\Omega }'&=& \frac{1}{3 r R_0^2}\left[9   \Cc^2 \hat{\Omega }^6-54 \Cc^2 \hat{\Omega }^5-6 \hat{\Omega }^3 \left(30 \Cc^2+\Cc \Kc R_0^3-3 M R_0^3\right)
\right. \nonumber\\ && \left. 
+135 \Cc^2 \hat{\Omega }^4 +9 \hat{\Omega }^2 \left(15 \Cc^2+2 \Cc   \Kc R_0^3+R_0^2 \left(R_0^2-6 M R_0\right)\right)
\right. \nonumber\\ && \left. 
-18 \hat{\Omega } \left(3 \Cc^2+\Cc \Kc R_0^3+R_0^2 \left(R_0^2-3 M R_0\right)\right)
\right. \nonumber\\ && \left. 
+9 \Cc^2+6 \Cc \Kc R_0^3+R_0^2 \left(\Kc^2 R_0^4-18   M R_0+9 R_0^2\right)\right]^{1/2}
\end{eqnarray}
Note that there is no dependence on $r$ inside of the square root. 
The transformation \eref{en:hatomega} is such that $\hat\Omega$ and $\hat\Omega'$ vanish at $r=0$ and $\hat\Omega$ goes to unity at $\scri^+$. The transformed compactification factor $\hat\Omega$ corresponding to the $\aconf$ that was already presented in \fref{fin:omegas} is shown here in \fref{fn:hatomega} %\upda{figure reference does not work. fix when chapter finished}. putting label after caption fixes it!! 

The quantity $\hat\Omega$ is determined via a shooting and matching approach. First an initial guess for the value of a quantity $A$ is set at the origin as $\hat\Omega(r=\Delta r)= \Delta r^2 A$ and it is integrated along the radial coordinate with a RK of 4th order until $\scri^+$ is reached. The value obtained there is compared to $\hat\Omega(r=\rscri)=1$. Using the bisection method a better initial guess for $A$ is found and the process is iterated until the discrepancy between the calculated value at $\scri^+$ and $\hat\Omega(r=\rscri)=1$ is smaller than a chosen threshold. % need more than quad precision for $\Kc\sim-3$
Now only transforming the obtained profile back to $\aconf$ using \eref{en:hatomega} is left. %\upda{mention that this numerical calculation requires quad precision at least? - done in next chapter} % - putting something at the end of next chapter, may have to be moved here.

There is a limit to the maximum value of $|\Kc|$ that could be set in the BH simulations. The reason is that the just described calculation of the compactification factor $\aconf$ requires more than {\tt double} precision to be calculated numerically. The larger the absolute value of $\Kc$, the larger the numerical precision required and $\Kc<-1.5$ needs more than {\tt quad} precision, the maximum available by the compiler used. 

\subsection{Numerical solution of the spherically symmetric constraint equations}

%The chosen initial data for the scalar field $\tilde\Phi$ is of compact support, so that its value at $\scri^+$ is negligible. Therefore the divergente terms in the constraint equations \eref{ein:constr} will not pose problems provided that they are not evaluated exactly on $r=0$ or $r=\rscri$. 

The momentum constraint \eref{ein:momen} is independent of the quantity $\psi$, so that it can be solved separately. The spacetime at $\scri^+$ is asymptotically flat, which means that the scalar field $\tilde\Phi$ tends to zero when approaching $\scri^+$ ($\bPhi= \frac{\iPhi}{\Omega}$ is finite there). The quantity $A_{rr}$ should therefore take initially its flat spacetime value there ($\atscrip{A_{rr}{}_0}=0$). This implies that the value of $\psi_A$ defined in \eref{ein:constrvars} has to satisfy $\atscrip{\psi_A}=0$. 

We will choose initial data of compact support for $\tilde\Phi$ as described in subsection \ref{sin:iniperturbs}. The parameters will be set in such a way that the perturbation vanishes at the origin and at null infinity. This means that the source term in \eref{ein:momen} corresponding to the scalar field vanishes everywhere where $\iPhi_0'$ is zero, more specifically at $r=0$ and $r=\rscri$. Thus the solution for $\psi_A$ will be zero there. 

\begin{figure}[htbp!!]
\center
\begin{tabular}{@{}c@{}@{}c@{}}
	\includegraphics[width=0.5\linewidth]{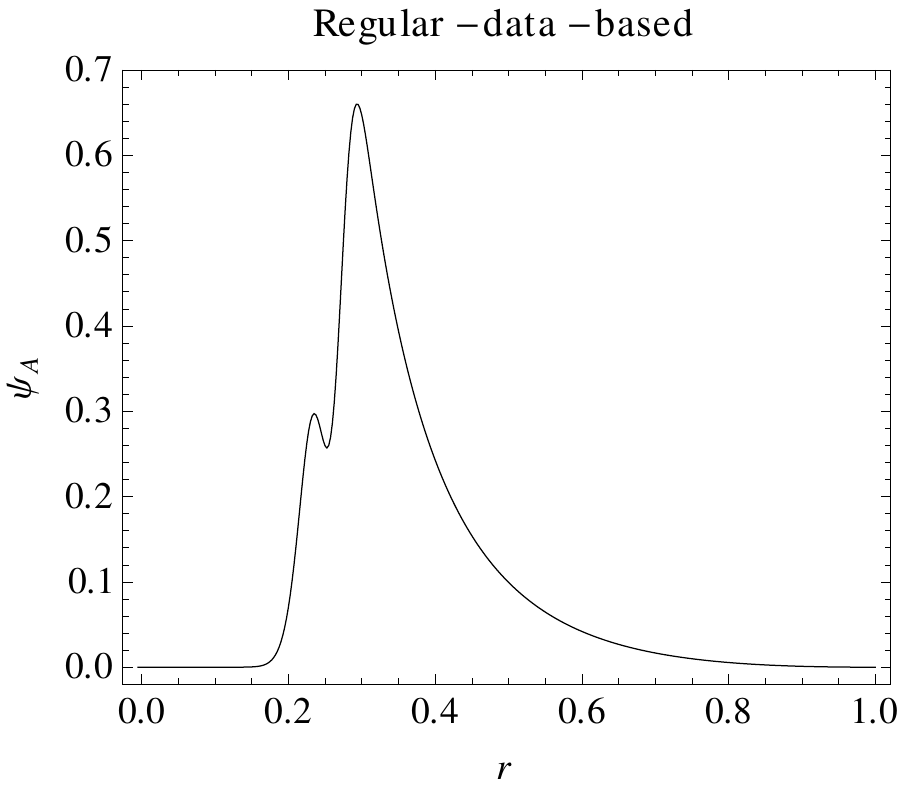}&
	\includegraphics[width=0.5\linewidth]{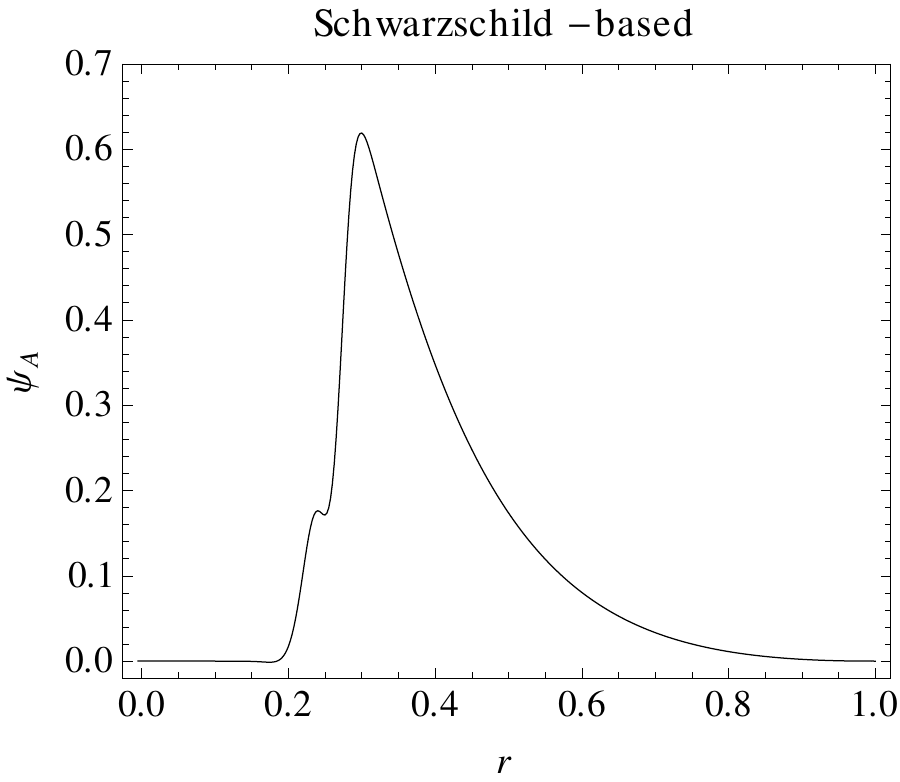}
\end{tabular}
\caption{Examples of solution of $\psi_A$ with $\Kc=-1$ and $sign=1$: on the left the solution for regular data and on the right the one in presence of a BH with  $M=1$, $Q=0$ and critical $\Cc$. If $sign<0$, then $\psi_A$ would take negative values.}
\label{fn:psiA}
\end{figure}

Knowing this we can set $\left.\psi_A\right|_{r=0}=0$ as initial value and integrate towards $\scri^+$ using a RK of 4th order. This method requires calculating the solution at half spatial steps. One possibility to go around this is to integrate first the even gridpoints and then the odd ones, which does not pose any problems given the simplicity of the differential equation. At the end of the integration procedure we obtain that the value of $\psi_A$ in the neighborhood of $\scri^+$ vanishes, as expected. Two example solutions for $\psi_A$ are shown in \fref{fn:psiA}, where $sign=1$ was was chosen to obtain a mostly outgoing scalar field perturbation. The corresponding ingoing versions ($sign=-1$) would have the exact same form below zero, so $sign\to-sign \Rightarrow\psi_A\to-\psi_A$. The parameters are the same ones used in the plots in subsection \ref{sin:inidataplots}.

\begin{figure}[htbp!!]
\center
\begin{tabular}{@{}c@{}@{}c@{}}
	\includegraphics[width=0.5\linewidth]{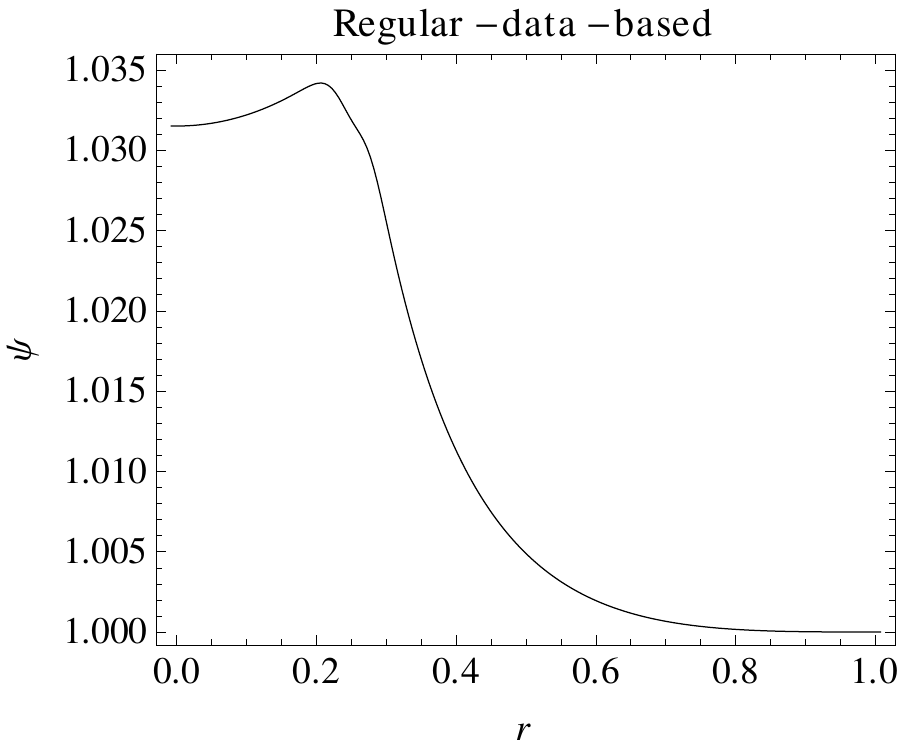}&
	\includegraphics[width=0.5\linewidth]{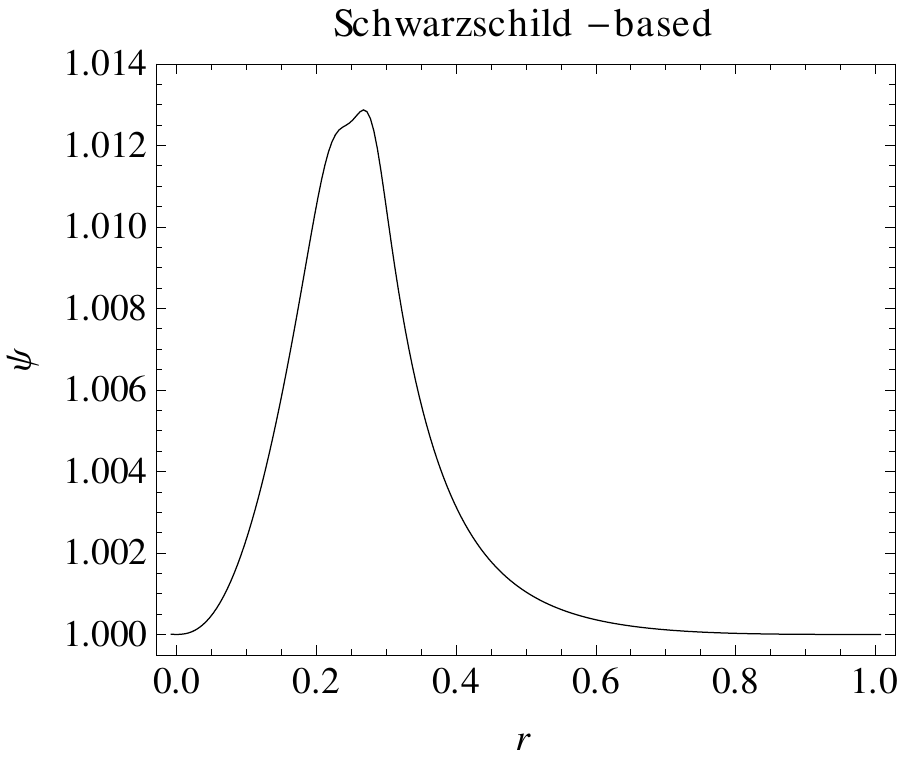}
\end{tabular}
\caption{Examples of solution of $\psi$ with $\Kc=-1$ and $sign=1$: on the left the flat-spacetime-based solution and on the right the one in presence of a BH with  $M=1$, $Q=0$ and critical $\Cc$.}
\label{fn:psi}
\end{figure}

Once the value of $\psi_A$ is known, it is substituted into the Hamiltonian constraint \eref{ein:hamil} and the solution for $\psi$, defined in \eref{ein:constrvars}, can be obtained. This is done via a shooting and matching method. An initial guess is made for the value of $\psi$ at the origin and then the differential equation is integrated using a 4th order RK until $\scri^+$ is reached, where by comparing the result with $\atscrip{\psi}=0$ (asymptotic flatness condition) a new guess is produced. The procedure is repeated until the error in the $\atscrip{\psi}$ obtained is below the chosen threshold. Examples of $\psi$ corresponding to perturbations of regular initial data and a Schwarzschild BH, calculated using the $\psi_A$ profiles in \fref{fn:psiA}, are presented in \fref{fn:psi}. 

%\renewcommand\bibname{{References}}
%\bibliographystyle{../../master/thesis/tocunsrt}
%\bibliography{../articles/hypcomp} 

\chapter{Numerical experiments}\label{c:exper}

\section{Results of a straightforward implementation}

The first numerical test I performed with the Einstein equations expressed in terms of the conformal metric was with the GBSSN evolution system in \eref{es:eeqs}, with a fixed shift, the slicing condition \eref{eg:harmlapse} and with flat spacetime initial data as shown in \fref{fin:Fini}. The spatial grid was staggered, so that the location of $\scri^+$ was avoided. 
The numerical evolution was unstable and it crashed after a few time-steps. With ``crash'' I mean that the variables (at least one of them) acquire a very large absolute value, of the order of $10^{300}$, or they become indeterminate, with output ``NaN'' (not-a-number), in some part of the integration domain. These effects are the result of either a continuum instability of the equations or a numerical instability of the implementation. 

%Most of the instabilities that were observed in this numerical implementation of the spherically symmetric hyperbolic initial value problem presented $\scri^+$-localized high-frequency fluctuations that would cause a fast crash of the simulation. 
%These instabilities quickly propagate to all evolved variables, making it difficult to locate their origin by visual inspection of the variable's data. 

The instability observed in this first numerical test presented high-frequency fluctuations close to $\scri^+$ that would appear after very few time-steps. At some point the values of the variables become very large or NaN at one of the nearest gridpoints to $\scri^+$ and the simulation was considered to have crashed. The ``crash-time'' was not affected by changes in the time-step, but it became smaller with an increase in spatial resolution - meaning that the outermost gridpoint was closer to $\scri^+$. This indicates that the instability is originated by a property of the continuum equations and not by a numerical artifact of the implementation. 
Knowing that the system of equations is complicated and that the CFEs are prone to continuum instabilities \cite{Husa:2005ns,Husa:2002zc}, such a behaviour was expected. Nevertheless, the problem can be analyzed and fixed. 

\section{Steps towards stability}

%difficult to stabilize: non-linear system
%can first test linear system, possibly behaviour of non-linear one will be the same

The Einstein equations form a non-linear system, which makes the study of stability of their implementation especially difficult. If an instability arises in any of the equations, it may contaminate the evolution of the rest of variables very fast due to the non-linear couplings between the equations. This makes the task of finding the origin of the instability very difficult. 

Stability does not only depend on the evolution equations themselves, but also on the background around which the evolution takes place. A stationary background (like flat or Schwarzschild spacetimes) is relevant for our purposes and also a simple choice. The initial data will consist of perturbations of the given stationary solution. 
Choosing a stationary background also allows to study a linearized version of the equations around it. To do so, all variables in the equations are expressed as a Taylor expansion to first order around the given background. An unstable linearized system is unlikely to become stable in its non-linear form, so first trying the linearized system can be a useful approach, because it is a simpler problem to solve. 

%A possible approach is to evolve a linearized version of the equations. To linearize the equations, an appropriate background solution is chosen, e.g. a stationary solution of the equations, and all variables in the equations are expressed as a Taylor expansion to first order around it. 
%The linearized equations are probably easier to stabilize and it is likely that the stabilizing procedures of the linearized equations also apply to the nonlinear ones. 

When studying the equations in search of the origin of the instabilities, it is very helpful to divide the system of equations into smaller parts and analyze those. The smallest division is to consider the single equations, then subsystems of two or three equations will be constructed and analyzed. 

\subsection{Analysis of the single equations}\label{se:single}

The simplest subsystem that can be analyzed are the single equations. For each equation in \eref{es:eeqs} or \eref{es:pKeeqs}, as well as the gauge equations, all quantities except the one that is being evolved are substituted by their stationary solutions. 
For simplicity in the analysis we will suppose that the stationary value of the variable under study, that will be denoted by $A$, is zero. This can be assumed without loss of generality, because with the appropriate variable transformation this can be achieved for any evolved quantity. 
A model equation for $A$ including the relevant terms (it is actually a linearization of the complete equation) is
\begin{equation}\label{ee:modelA}
\dot A = a A' + \left(b+\frac{c}{\Omega}\right)A , %+ \left(d+\frac{e}{\Omega}\right)A^2 + f , 
\end{equation}
where the coefficients $a$, $b$ and $c$ are time-independent functions of the radius. Higher order terms in $A$ may appear in the single equations, but they do not have a relevant effect on stability. This equation serves to model the behaviour of the quantities $A_{rr}$, $\Lambda^r$, $\cK$ or $\pK$ and possibly $\alpha$ and $\beta^r$, which are the ones more likely to pose stability problems. Regular equations of motion (without divergent terms), like the ones for $\gamma_{rr}$ or $\gamma_{\theta\theta}$, should not require any numerical tests. 

The objective of the study is to detect exponential growths. They will arise when the coefficient $\left(b+\frac{c}{\Omega}\right)>0$, so that simple inspection of the values taken by $b$ and $c$ will give us useful information. The advection term with coefficient $a$ only propagates information on the grid and does not play an important role for long-term solutions of the equations where all perturbations have already been radiated away, so it does not need to be considered in the stability analysis. If $c>0$, the growth at $\scri^+$ will be very fast and the simulation is very likely to crash after only a few time-steps.

If $b>0$ (or $c>0$), one of the checks that should be performed on the equation is to see if there is a stable stationary solution of the equation, because $A=0$ turns out not to be one. This can be easily done by plotting $\dot A$ as a function of $A$ (for specific values of the radius and dropping the advection term) keeping all of the non-linear terms. If there is another $A_1\neq0$ where $\dot A(A_1)=0$ and the slope of $\dot A$ is negative there, $A_1$ is a stable solution of the equation.

The mentioned stability conditions on the coefficients are a property of the continuum equations. At the numerical level there are also important stability conditions to satisfy. One is
the value of the Courant factor defined in \eref{en:Courant}, which has to satisfy the CFL limit condition. In presence of stiff terms it may even be smaller than expected. 
The initial perturbations to test the stability of the equations are chosen small enough so that they do not ``break'' the coordinate system. It is useful to first select small perturbation amplitudes to avoid exciting the non-linear effects in the equations. 
%If a perturbation of a stationary solution is present in the initial data, make sure it is not too large to ``break'' the coordinate system. 

%check that there are no exponential growths 
%the solution we expect is the stationary one

%\upda{till here}

\subsection{Analysis of subsystems of equations}

Once the single equations present stable and appropriate numerical behaviour, a hierarchical way of constructing a well-behaved complete system consists of first testing small groups of equations. Each single equation of \eref{es:eeqs}, \eref{es:pKeeqs} or \eref{es:DPKeeqs} by itself is hyperbolic, but not every combination of equations taken from \eref{es:eeqs}, \eref{es:pKeeqs} or \eref{es:DPKeeqs} is. With the purpose of reducing the number of variables as much as possible, it is useful to fix the determinant freedom of the conformal spatial metric and eliminate $\gamma_{\theta\theta}$ in terms of $\gamma_{rr}$ using \eref{es:delgtt}, to solve the vacuum equations (not evolving the scalar field), to use a fixed shift and to consider the GBSSN equations, not evolving the Z4 quantity $\Theta$. 

%For each combination of equations, the hyperbolicity analysis presented in subsection \ref{sr:1st2nd} has to be performed. Only subsystems that have real eigenvalues, a complete set of eigenfield and do not present Jordan-blocks in the Jordan form of their $E$ matrix, defined as proportional to \eref{er:principalm}. 

Another requirement of our setup is that the values of the eigenspeeds have to be larger or equal to zero at $\scri^+$. If a BH is present in the simulated spacetime and excision is being used, then the eigenspeeds have to be negative or vanishing at the horizon. For the possible subsystems that can be constructed from \eref{es:eeqs} (and appropriate lapse condition), the resulting characteristic speeds satisfy these conditions. 

The hyperbolic subsystems of \eref{es:eeqs} or \eref{es:pKeeqs} can be divided into two groups: %, which also have the appropriate eigenspeeds \upda{check!},
\begin{itemize}
\item Subsystems that involve the trace-free part of the extrinsic curvature ($A_{rr}$ in our ansatz) and the difference of contracted connections ($\Lambda^r$): certain combinations of the metric component $\gamma_{rr}$ (and $\gamma_{\theta\theta}$ if not eliminated) with $A_{rr}$ and/or $\Lambda^r$ are hyperbolic. The conformal factor $\chi$ can also be added to any of them without breaking hyperbolicity. The possible hyperbolic combinations are listed in the first part of subsection \ref{se:expsubsys}. 
\item Subsystems that involve the trace of the extrinsic curvature ($K$): the equation of motion of $\cK$ (or equivalently $\pK$) only includes $\alpha''$ as second derivatives of the metric components, so that in our setup it can only couple hyperbolically to the lapse evolution equation. %\upda{Note for me: If the change $K\to K+2\Theta$ is not made, there is also $\chi''$ and the subsystem $\chi K$ is well-behaved.} 
The possible hyperbolic choices are $\alpha K$ and $\alpha\chi K$. 
\end{itemize}

%\upda{well-posed subsystems with the shift? with $\Theta$? }

In the same way as with the single equations, the hyperbolic subsystems have to be numerically checked for exponential growths of any of the variables. % and make sure that the stationary values are the expected ones. 

%Make sure that the chosen equations form a strongly hyperbolic system, that is, check that the matrix \eref{er:principalm} from the system is diagonalizable, its eigenspeeds are all real and a complete set of eigenvectors exist. 
%
%The eigenspeeds at $\scri^+$ and at the horizon (if present) have to be the correct ones: no negatives speeds at $\scri^+$ and no positive ones at the horizon. 

%make sure they are hyperbolic and they have the correct eigenspeeds
%no exponential growths and expected behaviour

\subsection{Tracking down instabilities}

Possibly the more powerful tool to locate the origin of instabilities is by visual inspection of the numerical data. In order to simplify the visualization, it is useful to have very simple stationary values (like 1 or 0) for the variables. This can be easily achieved by simple variable transformations, see next subsection. %The visualization of the variables is much simpler in this way. 
Another aid in tracking down the origin of the instability is plotting the RHSs of the equations in time and search for the first exaggerate growth. 

If an exponential growth has been detected, plotting the data in a logarithmic scale allows to estimate the slope of the resulting straight line. Knowing its value can help identify which term in the equation is causing the exponential growth. 
Another way of trying to find where the instability has its source or checking if an existing guess is correct is to add damping terms of the form $-\lambda A$ or $-\case{\xi}{\Omega}A$ with $\lambda,\xi\geq0$ to the corresponding $\dot A$. Obviously, the solution of the resulting set of equations no longer satisfies the Einstein equations, but the terms added by hand serve as a trick to find and understand the instabilities. 

%for visualization purposes, using variables with simple stationary values is convenient (variable transformations)
%visualize RHSs to detect origin of growths
%plot exponential growths in a logarithmic scale and estimate the slope
%tricks: adding damping terms by hand 

%\upda{check regularity conditions?} 

Checking if the regularity conditions listed in subsections \ref{er:Einstenreqs} and \ref{er:gaugereqs} are satisfied by the numerical data during the evolution can also provide hints about the origin of the instabilities.

\subsection{Variable transformations}

Certain well-chosen variable transformations can help solve some of the continuum instabilities present in the equations. Some transformations change the principal part of the system, while others do not. To see this, consider the $u$ and $v$ variables defined in subsection \ref{sr:1st2nd}. Let us denote a given $u$ variable  as $A_u$ and a given $v$ variable as $B_v$. The possible variable transformations can be classified as follows: 
\begin{itemize}
\item The principal part remains the same after performing transformations of the form $B_v\to B_v+f(u)$, where $f$ is the function that defines the variable transformation and $u$ denotes any $u$-type variables involved in the transformation. %\upda{The transformations $B_v\to f(B_v)$ and $A_u\to f(A_u)$ also maintain the principal part. Do the eigenspeeds change?? And the eigenfields?? Check!! (maybe only the eigenspeeds change ...) (if yes, consider $B_v\to B_v+f(u)$ for the previous trafo and put an extra item explaining $B_v\to f(B_v)$ and $A_u\to f(A_u)$)}
The principal part is left unchanged because the terms arising from the mixing with the $u$ variables are all non-principal part terms. 
\item The principal part is likely to change under transformations of the form $A_u\to f(A_u,u)$, $B_v\to f(B_v,v)$ and $B_v\to f(B_v,u')$, because the same type of variables (in the sense of $u$, $v$ and $w\equiv u'$) are involved and thus the spatial derivatives that form the principal part will also change. However, as these are similarity transformations, even if the eigenvectors of the system may change, the eigenvalues and hyperbolicity properties of the system remain the same. 
%In this case, hyperbolicity of the resulting system has to be checked, as well as made sure that the eigenspeeds are the appropriate ones. 
\end{itemize}

\subsection{Constraint damping}

At the continuum level the constraints are satisfied exactly, so that adding a multiple of them to any evolution equation gives a new system of equations that is exactly equivalent to the original one. In a numerical simulation the constraints are only satisfied to some extent, there is always some numerical error. However, if the value of the constraints converges to zero, this manipulation of the equations is also valid for the evolution equations implemented in the code.  
Such a manipulation is performed for instance in the BSSN formulation, where a multiple of $\mathcal{H}$ is added to $\dot{\bar K}$'s RHS to eliminate the Ricci scalar, while in the Z4 formulation  the same effect is achieved with the mixing of $\bar K$ and $\Theta$ in \eref{et:KTmix}. 

Adding a multiple of a constraint can be useful to eliminate a problematic term in an equation of motion or to enforce the vanishing of the given constraint. The terms with $\kappa_1$ in \eref{c3:einsteinp} are the constraint damping terms \cite{Gundlach:2005eh} of the Z4 formulation and tend to minimize the value of the constraint variables $\Theta$ and $Z_a$.  
It can also help eliminate terms that cause exponential growths. Nevertheless, the resulting system of equations has to be hyperbolic and must have the correct eigenspeeds, properties which may change with the new terms brought in by the constraint expression. 

\subsection{Tuning the gauge conditions and the free parameters}

Provided that they form a hyperbolic system with the Einstein equations and give appropriate eigenspeeds, the gauge conditions for the lapse and shift can be specified freely. Convenient damping terms can be added and the eigenspeeds in regions other than the horizon (if the singularity is excised) and $\scri^+$ can be changed. This introduces more free parameters that will control the strength of the damping, the propagation speeds, etc. 

The parameters $\kappa_1$ and $\kappa_2$ are included in the Einstein equations through the Z4 damping terms. Other parameters can appear from variable transformations or addition of constraints. 
Experimenting with all these free parameters helps to understand the formation of the instabilities or, in case the simulation is well behaved, to improve the accuracy of the numerical results. 

%\upda{The exact expressions used for the source terms and interesting ranges for each of the parameters will be shown in the next section.}

%%%%%%%%%%%%%%%%%%%%%%%%%%%%%%%%%%%%%%%%%%%%%%%%%%%%%%%%%%%%%%%%%%%%%%%%%%%%%%%%%%%%%%%%%%%%%%%%%%%%%%%%%%%%%%%%%%%%%%%%%%%%%%%%%%%%%%%%%%%%%%%%%%%%%%%%%%%%%%%%%%%%%%%%%%%%%%%%%%%%%%

\section{Performed tests} \label{se:tests}

The instability detected in the first numerical test with \eref{es:eeqs} was studied following the steps described in the previous section. 
The main numerical experiments performed will be described now. All of them use a staggered grid as illustrated in \fref{fn:stgrid}. 
Unless otherwise specified, flat spacetime will be considered. 

The main ingredients to obtain a stable numerical evolution were the use of the trace of the physical extrinsic curvature $\pK$ as variable instead of the conformal one and the addition of a constraint damping term to $\dot\Lambda^r$. For the Schwarzschild case, the choice of trumpet initial data was especially convenient. 

\subsection{Single equations}

The stability of a single equation can be tested by setting a small perturbation on the initial stationary data. The initial perturbations tested here were random noise of amplitude $\sim10^{-8}$ or a Gaussian-like perturbation like the one in \eref{initialgaussian} with amplitude $A\sim10^{-3}$. In both cases, the perturbations will be propagated or damped away if the system is numerically stable. 

Each of \eref{es:eeqs} by themselves, without being coupled to the rest of the system and using flat background data, reduce to
\begin{subequations}
{\small
\begin{eqnarray}
\dot{\chi }&=& \frac{1}{3} \Kc r \chi '
 , \quad
\dot{\gamma _{rr}}= \frac{2}{9} \Kc r \gamma _{rr}'
 , \quad
\dot{\gamma _{\theta \theta }}= \frac{1}{9} \Kc r \gamma _{\theta \theta }'
, \quad
 \dot{\Lambda ^r}= \frac{1}{3} \Kc r \left(\Lambda ^r\right)'
 , \\
\dot{A_{rr}}&=& -\frac{2}{3} A_{rr}^2 \sqrt{\Kc{}^2 r^2+9 \Omega ^2}+\frac{1}{9} \Kc A_{rr} \left(\frac{2 \sqrt{\Kc{}^2 r^2+9 \Omega ^2}}{\Omega }+3\right)+\frac{1}{3} \Kc r A_{rr}'
 , \qquad \  \\
\dot{\cK}&=& \frac{1}{9}\cK^2 \sqrt{\Kc{}^2 r^2+9 \Omega ^2}+\frac{1}{3} \Kc r\cK'+\frac{K \Kc \left(\sqrt{\Kc{}^2 r^2+9 \Omega ^2}-3 \Omega \right)}{9 \Omega }
\nonumber \\ &&
+\frac{1}{3} \Kc{}^2 \left(-\frac{9 \Omega }{\Kc{}^2 r^2+9 \Omega ^2}+\frac{3}{\sqrt{\Kc{}^2 r^2+9 \Omega ^2}}-\frac{1}{\Omega }\right)
 , %\\
 \\
\dot{\Theta }&=& -\frac{\Theta  \left(\kappa_1 (\kappa_2+2) \sqrt{\Kc{}^2 r^2+9 \Omega ^2}+\Kc \Omega  (3 C_{Z4c}-4)\right)}{3 \Omega }
\nonumber \\ &&
+\frac{2}{9} \Theta ^2  (2-3 C_{Z4c}) \sqrt{\Kc{}^2 r^2+9 \Omega ^2}+\frac{1}{3} \Kc r \Theta ' . 
\end{eqnarray}
}%
Numerically they are all well-behaved: they show no exponential growths and their hyperboloidal values of flat spacetime are the stationary solution of the system. Actually, the first term in \eref{et:pKL}'s RHS is added to cancel a potential exponential growth, so that the spherically component equation is well-behaved. 

The slicing condition, whose functions $K_0$ and $L_0$ as appeared in \eref{eg:harmlapse} have been determined under the conditions that the coefficient in front of $-\alpha^2$ be the freely specifiable parameter $\xi_\alpha$ and $\dot\alpha(\alpha_0)=0$, is given as a single equation by
{\small
\begin{eqnarray}
\dot\alpha = -\xi_\alpha \alpha ^2 +\frac{1}{9} \left(\Kc \left(\xi_\alpha \Kc r^2-\sqrt{\Kc{}^2 r^2+9 \Omega ^2}\right)+9 \xi_\alpha \Omega ^2+3 \Kc \Omega \right)
%\nonumber \\ &&
+\frac{1}{3} \Kc r \alpha ' . \quad  
\end{eqnarray}
}%
\end{subequations}
Provided that $\xi_\alpha\ge0$, the numerical simulation does not crash and the expected stationary value ($\alpha$'s flat spacetime profile) is reached. 

\subsection{Subsystems}\label{se:expsubsys}

%When testing the possible hyperbolic subsystems, two difficulties arised: on the one hand and while $\gamma_{rr}A_{rr}$ and $\gamma_{rr}\Lambda^r$ were numerically stable, the combination \upda{$\gamma_{rr}A_{rr}\Lambda^r$ was not}, and on the other hand, the subsystem $\alpha K$ was also crashing. 

\subsubsection{Involving $\gamma_{rr}$, ($\gamma_{\theta\theta}$,) $A_{rr}$, $\Lambda^r$ (and $\chi$)}

The possible hyperbolic combinations are: $\gamma_{rr}A_{rr}$, $\gamma_{rr}\Lambda^r$ and the same ones also including $\gamma_{\theta\theta}$; also $\gamma_{rr}\gamma_{\theta\theta}A_{rr}\Lambda^r$ and, if the metric component $\gamma_{\theta\theta}$ has been previously eliminated from the evolution system using the substitution \eref{es:delgtt}, the new $\gamma_{rr}A_{rr}$ and $\gamma_{rr}A_{rr}\Lambda^r$ as well. 
As an example I explicitly write the last subsystem, $\gamma_{rr}A_{rr}\Lambda^r$ with eliminated $\gamma_{\theta\theta}$: 
{\small
\begin{subequations}
\begin{eqnarray}
\dot{\gamma _{rr}} &=& \frac{1}{3} \Kc r \gamma _{rr}'-\frac{2}{3} A_{rr} \sqrt{\Kc{}^2 r^2+9 \Omega ^2}
, \\
\dot{A_{rr}}&=& -\frac{2 A_{rr}^2 \sqrt{\Kc{}^2 r^2+9 \Omega ^2}}{3 \gamma _{rr}}+\frac{2 \Kc A_{rr} \sqrt{\Kc{}^2 r^2+9 \Omega ^2}}{9 \Omega }+\frac{1}{3} \Kc r A_{rr}'
\nonumber \\ &&
+\frac{\Kc A_{rr}}{3}+\frac{1}{6} \Lambda ^r \sqrt{\Kc{}^2 r^2+9 \Omega ^2} \gamma _{rr}'+\frac{2}{9} \gamma _{rr} \sqrt{\Kc{}^2 r^2+9 \Omega ^2} \left(\Lambda ^r\right)'
\nonumber \\ &&
-\frac{2 \Lambda ^r \gamma _{rr} \sqrt{\Kc{}^2 r^2+9 \Omega ^2}}{9 r}+\frac{7 \sqrt{\Kc{}^2 r^2+9 \Omega ^2} \left(\gamma _{rr}'\right){}^2}{36 \gamma _{rr}^2}
\nonumber \\ &&
-\frac{5 \sqrt{\gamma _{rr}} \sqrt{\Kc{}^2 r^2+9 \Omega ^2} \gamma _{rr}'}{9 r}+\frac{\sqrt{\Kc{}^2 r^2+9 \Omega ^2} \gamma _{rr}'}{6 r \gamma _{rr}}-\frac{\sqrt{\Kc{}^2 r^2+9 \Omega ^2} \gamma _{rr}''}{6 \gamma _{rr}}
\nonumber \\ &&
+\frac{2 \gamma _{rr}^{3/2} \sqrt{\Kc{}^2 r^2+9 \Omega ^2}}{3 r^2}-\frac{2 \sqrt{\Kc{}^2 r^2+9 \Omega ^2}}{3 r^2}+\frac{\Kc{}^2 r \gamma _{rr}'}{27 \Omega  \gamma _{rr}}+\frac{\Omega  \gamma _{rr}'}{6 r \gamma _{rr}}
, \\
\dot{\Lambda ^r}&=&\frac{A_{rr} \sqrt{\Kc{}^2 r^2+9 \Omega ^2} \gamma _{rr}'}{6 \gamma _{rr}^3}-\frac{2 A_{rr} \sqrt{\Kc{}^2 r^2+9 \Omega ^2}}{3 r \sqrt{\gamma _{rr}}}-\frac{4 A_{rr} \sqrt{\Kc{}^2 r^2+9 \Omega ^2}}{3 r \gamma _{rr}^2}
\nonumber \\ &&
+\frac{4 \Kc{}^2 r A_{rr}}{9 \Omega  \gamma _{rr}^2}+\frac{6 \Omega  A_{rr}}{r \gamma _{rr}^2}+\frac{1}{3} \Kc r {\Lambda ^r}'+\frac{2 \Kc \sqrt{\gamma _{rr}}}{3 r}-\frac{2 \Kc}{3 r \gamma _{rr}}+\frac{\Kc \gamma _{rr}'}{3 \gamma _{rr}^2} . \qquad \ \ 
\end{eqnarray}
\end{subequations}
}%
The numerical behaviour of these subsystems is not as good as expected if a centered stencil is used in the neighborhood of $\scri^+$: at some point the values of the quantities separate from the stationary ones, but this difference does not continue growing in time, so that the system does not crash. However, if an one-sided finite difference stencil is used at $\scri^+$ or no dissipation is set, the described effect does not appear. 

The $\chi$ evolution equation can be added to the listed subsystems, providing another hyperbolic system with an equivalent numerical behaviour. 

\subsubsection{Involving $\alpha$, $K$ (and $\chi$)}

The first test with the subsystem $\alpha K$ was performed using the harmonic lapse \eref{eg:harmlapse} and the evolution equation for the conformal trace of the extrinsic curvature. The exact expressions are: 
{\small
\begin{subequations}
\begin{eqnarray}
\dot \alpha &=&  -\alpha ^2 \xi_\alpha+\frac{1}{9} \xi_\alpha \Kc{}^2 r^2+\xi_\alpha \Omega ^2-\alpha ^2\cK-\frac{\Kc{}^3 r^2}{27 \Omega }+\frac{3 \alpha ^2 \Kc r \Omega '}{\Omega  \sqrt{\Kc{}^2 r^2+9 \Omega ^2}}
\nonumber \\ &&
-\frac{\Kc r \Omega ' \sqrt{\Kc{}^2 r^2+9 \Omega ^2}}{9 \Omega }+\frac{\alpha ^2 \Kc}{\Omega }+\frac{1}{3} \Kc r \alpha '
, \\ 
\dot\cK &=& -\frac{\Kc r \dot{\alpha } \Omega '}{\alpha ^2 \Omega }-\alpha ''+\frac{3 \alpha ' \Omega '}{\Omega }+\frac{\alpha  \Omega ''}{\Omega }-\frac{3 \alpha  \left(\Omega '\right)^2}{\Omega ^2}+\frac{\alpha \cK^2}{3}+\frac{1}{3} \Kc r\cK'-\frac{K \Kc r \Omega '}{3 \Omega }
\nonumber \\ &&
+\frac{\Kc{}^2 r^2 \alpha ' \Omega '}{3 \alpha ^2 \Omega }-\frac{\Kc{}^2 r^2 \Omega ''}{3 \alpha  \Omega }+\frac{\Kc{}^2 r^2 \left(\Omega '\right)^2}{3 \alpha  \Omega ^2}-\frac{\Kc{}^2 r \Omega '}{3 \alpha  \Omega }-\frac{2 \alpha '}{r}+\frac{2 \alpha  \Omega '}{r \Omega} . \qquad \ \label{ee:acKK}
\end{eqnarray}
\end{subequations}
}%
They were numerically unstable and varying the free parameter $\xi_\alpha$ in $\dot\alpha$ did not help. The problem is that there is an exponential growth in the $\cK$ variable. In its RHS in \eref{ee:acKK} there is a $-\case{\beta^r\Omega'}{\Omega}K=-\case{K \Kc r \Omega '}{3 \Omega }$ term that damps the behaviour of $\cK$ to the correct solution, but when $\dot\alpha$ is substituted there, the coefficient transforms to $\case{2\beta^r\Omega'}{\Omega}K=\case{2K \Kc r \Omega '}{3 \Omega }$, which causes an exponential growth (note that $\Kc<0$, $\Omega'<0$ and $\Omega>0$) in $\cK$ at $\scri^+$. 

This can be solved by means of a variable transformation in $K$ that takes away the $\dot \alpha$ term from $\dot\cK$'s RHS. Inspired in the relation between the traces of the conformal and physical extrinsic curvatures \eref{e3:physconfK}, a possible variable transformation is
\begin{equation}\label{ee:oldK}
\odK =  K-\left(\frac{\tilde K_0}{\Omega}+\frac{3\,\beta^r\Omega'}{\Omega\,\alpha}\right), % from old version of paper1
\end{equation}
where we set $\tilde K_0=\Kc$, the flat spacetime value of the trace of the physical extrinsic curvature. Now the flat value of the new variable $\odK$ is simply zero. This transformation can be regarded as subtracting its flat spacetime value from the variable, but keeping the dependence on $\alpha$ and $\beta^r$. The resulting system is
{\small
\begin{subequations}
\begin{eqnarray}
\dot \alpha &=&  -\frac{1}{27} \xi_\alpha \Kc{}^3 r^2+\frac{1}{3} \alpha ^2 \xi_\alpha \Kc-\frac{1}{3} \xi_\alpha \Kc \Omega ^2-\alpha ^2 \odK-\frac{\Kc{}^3 r^2}{27 \Omega }
\nonumber \\ &&
+\frac{2 \Kc r \Omega  \Omega '}{\sqrt{\Kc{}^2 r^2+9 \Omega ^2}}+\frac{2 \Kc{}^3 r^3 \Omega '}{9 \Omega  \sqrt{\Kc{}^2 r^2+9 \Omega ^2}}+\frac{1}{3} \Kc r \alpha '-\frac{\alpha  \Kc r \Omega '}{\Omega }
\label{ee:oldaKalpha}, \quad \\ 
\dot\odK &=& -\alpha ''+\frac{3 \alpha ' \Omega '}{\Omega }+\frac{\alpha  \Omega ''}{\Omega }-\frac{3 \alpha  \left(\Omega '\right)^2}{\Omega ^2}+\frac{\alpha  \odK^2}{3}+\frac{1}{3} \Kc r \odK'+\frac{2 \alpha  \odK \Kc}{3 \Omega }
\nonumber \\ &&
+\frac{\odK \Kc r \Omega '}{3 \Omega }+\frac{\alpha  \Kc{}^2}{3 \Omega ^2}-\frac{2 \alpha '}{r}+\frac{2 \alpha  \Omega '}{r \Omega } . 
\end{eqnarray}
\end{subequations}
}%
With this change the subsystem $\alpha \odK$ was finally stable with a choice of parameter $\xi_\alpha\ge1$. If $\xi_\alpha=0$, slow fluctuations, which appear at $\scri^+$ and propagate inwards, start growing in amplitude and in the end cause the simulation to crash.
Apart from eliminating the exponential growth of the $\cK$ variable, the effect of the transformation \eref{ee:oldK} onto the slicing condition is the appearance of terms that include a conformal factor $\Omega$ in their denominator and thus diverge at $\scri^+$. The regularity condition that ensures that a finite limit for the divergent terms at $\scri^+$ is attained is $\atscrip{\alpha}=-\case{\rscri\Kc}{3}$ \eref{er:fixalpha}, which coincides with the flat stationary value of the lapse there. This means that the equations ``force'' a certain value of $\alpha$ at null infinity, so that during the evolution any perturbation that may have affected the lapse is damped away before reaching $\scri^+$. Note that this ``$\alpha$-fixing'' effect is not the result of explicitly imposing any boundary condition. The last term in \eref{ee:oldaKalpha} is mainly the one responsible for the damping of the perturbations in $\alpha$ as they propagate towards $\scri^+$. 
%\upda{Describe effect of ``$\alpha$-fixing'' and damping terms for $\alpha$.} %$\xi_\alpha\in[1,100 according to a simulation]$.

These results motivate the use of the trace of the physical extrinsic curvature instead of the conformal one as evolution variable. %\upda{check Moncrief and Rinne} 
The set of evolution equations that include $\pK$ are presented in \eref{et:pKtensoreqs} and \eref{es:pKeeqs}. In this choice of variables, the constraint equations \eref{et:pKtensorcqs} and \eref{es:pKceqs} in the vacuum are independent of the gauge variables $\alpha$ and $\beta^r$. For the simulations performed here, instead of choosing the trace of the physical extrinsic curvature $\pK$ as a variable, for simplicity we will use its variation with respect to the stationary value introduced in \eref{ee:DPKdef}, that is
\begin{equation}
\DPK = \pK-\Kc . 
\end{equation}
The subsystem $\alpha \DPK$ is given by 
{\small
\begin{eqnarray}\label{ee:finalpha}
\dot \alpha &=&  -\alpha ^2 \xi_\alpha+\frac{1}{9} \xi_\alpha \Kc{}^2 r^2+\xi_\alpha \Omega ^2-\frac{\alpha ^2\DPK}{\Omega }-\frac{\Kc{}^3 r^2}{27 \Omega }+\frac{2 \Kc r \Omega  \Omega '}{\sqrt{\Kc{}^2 r^2+9 \Omega ^2}}
\nonumber \\ &&
+\frac{2 \Kc{}^3 r^3 \Omega '}{9 \Omega  \sqrt{\Kc{}^2 r^2+9 \Omega ^2}}+\frac{1}{3} \Kc r \alpha '-\frac{\alpha  \Kc r \Omega '}{\Omega } \label{ee:dinalphaalpha}
, \\
\dot\DPK &=& -\Omega  \alpha ''+3 \alpha ' \Omega '+\alpha  \Omega ''-\frac{3 \alpha  \left(\Omega '\right)^2}{\Omega }+\frac{\alpha \DPK^2}{3 \Omega }+\frac{1}{3} \Kc r\DPK'+\frac{2 \alpha \DPK \Kc}{3 \Omega }
\nonumber \\ &&
+\frac{\alpha  \Kc{}^2}{3 \Omega }-\frac{2 \Omega  \alpha '}{r}+\frac{2 \alpha  \Omega '}{r} . 
\end{eqnarray}
}%
While there are still terms that diverge at $\scri^+$, some others become degenerate there. This has posed no problems at the numerical level so far. 
Adding $\chi$ to the $\alpha \DPK$ also gives a well-behaved system. 

%using physical K (transforming in lapse cond) ok: motivation for using \eref{es:pKeeqs}

%list of subsystems tested with results: 
%- unsuccessful conformal K+alpha, explanation
%- successful physical K +alpha, with coealphah and old damping terms. 

\subsection{Complete system: GBSSN}\label{se:gbssnstabi}

The simplest complete evolution system of the Einstein equation for the rescaled metric is obtained by putting together the two subsystems described in the previous subsection. The evolved quantities are $\chi$, $\gamma_{rr}$, $A_{rr}$, $\DPK$, $\Lambda^r$ and $\alpha$ (the shift $\beta^r$ is fixed and $\gamma_{\theta\theta}$ has been eliminated in terms of $\gamma_{rr}$). 

Even when the two subsystems were stable, the complete system turned out not to be. A useful observation is that many of the equations include a damping term of the form $\dot A=\frac{c}{\Omega}A$ in their RHSs, with $A$ some evolution variable and $c$ a coefficient that depends on other evolution variables and on the radial coordinate. In \tref{te:damp} the value of these $c$s is shown for the equations in the complete system, where the flat spacetime values of the variables in each $c$ has been substituted and they have been evaluated at $\scri^+$ ($r=\rscri$). A heuristic requirement for stability is that these coefficients of the divergent terms at $\scri^+$ have to be negative everywhere, as $\Omega$ is always positive in the physical domain. 
\begin{table}[hhh]
\center
\caption{\label{te:damp} Values at $\scri^+$ of the coefficients over $\Omega$ that precede the variables.}
\begin{tabular}{ccccccc}
\hline
$\dot \chi$&$\dot \gamma_{rr}$&$\dot A_{rr}$&$\dot{\DPK}$&$\dot \Lambda^r$&$\dot \alpha$\\
\hline
0&0&$-2 \rscri\left(\frac{\Kc}{3}\right)^2$&$-2 \rscri\left(\frac{\Kc}{3}\right)^2$&0&$-\frac{\rscri\Kc{}^2}{3}$\\
\hline
\end{tabular}
\end{table}

The evolution equation of $\Lambda^r$ does not have a damping term like those present in $\dot A_{rr}$ and $\dot\DPK$. Given that they are of the same type of $v$ variables, it may indeed be required. The simplest way to include a damping term of the form $-\case{\xi_{\Lambda^r}}{\Omega}\Lambda^r$ is by adding the constraint \eref{es:pKZrdef}'s RHS multiplied by $-\case{2\xi_{\Lambda^r}}{\gamma_{rr}\Omega}$ to $\dot\Lambda^r$'s RHS. 
Evolving the complete system with this constraint damping term with $\xi_{\Lambda^r}\in[2,4]$ finally gave a stable numerical evolution. 

There is a more natural way of obtaining the desired damping term in $\dot \Lambda^r$. It is done by keeping the Z4 damping term in $\Lambda^r$'s equation of motion even when the GBSSN formulation is chosen: 
\begin{equation}\label{lambdarterm}
\dot \Lambda^r=\textrm{GBSSN-RHS} - \frac{2\kappa_1 \alpha Z_r}{\gamma_{rr}\Omega}, \quad \textrm{substituting} \  Z_{r}=\frac{1}{2}\gamma_{rr}\Lambda^r+\frac{1}{r}-\frac{\gamma_{rr}}{\gamma_{\theta\theta}r}-\frac{\gamma_{rr}'}{4\gamma_{rr}}+\frac{\gamma_{\theta\theta}'}{2\gamma_{\theta\theta}}  .
\end{equation}
In \eref{es:pKLambdardot} and \eref{es:DPKLambdardot} the $Z_r$ in this term has already been substituted. Using the GBSSN system with this extra damping term would be called ``Z3 damping'' in \cite{Gundlach:2005eh}. 

With $\alpha$ and $\beta^r$ ``fixed'' at $\scri^+$, the scri-fixing condition implies that $\atscrip{\chi}=\atscrip{\gamma_{rr}}$ (see subsection \ref{er:gaugereqs}), and this is exactly the behaviour observed at future null infinity. This also requires that $\atscrip{\K}=0$, as can be seen in the simulation results shown in \fref{fs:shiftcondKL}. 

\subsection{Complete system: Z4c}

Adding the $\cT$ evolution equation and the remaining $Z_r$ terms to the system did not require any extra work to obtain a successful simulation. The equation of motion $\dot \cT$ already has an appropriate damping term like those presented in \tref{te:damp}, namely $\kappa_1(2+\kappa_2)\rscri\case{\Kc}{3}$. 

For similarity with the choice of the trace of the physical extrinsic curvature as evolution variable, the $\Theta$ variable will also be transformed using the same $\Omega$ factor: 
\begin{equation}
\pT=\Omega\cT . 
\end{equation}
%mention trafo for $\Theta$, for similarity with $K$
The sets of equations \eref{es:pKeeqs}, \eref{es:pKceqs}, \eref{es:DPKeeqs} and \eref{es:DPKceqs} were already expressed in terms of this physical $\pT$. 

%\upda{mention and describe equations when I also evolved $Z_r$?}

\subsection{Massless scalar field}

%The scalar field on a compactified hyperboloidal background in known to be well-behaved \upda{references?, \cite{masterAlex}}. 
When the scalar field equations were evolved coupled to the Einstein equations, no new instabilities appeared. 
Although it does not necessarily provide better numerical results, it is useful to evolve the fields $\bPhi$ and $\bPi$ defined in \eref{esca:tildebarphi}, because they do not vanish at $\scri^+$ and they are more convenient for visualization and data treatment purposes. The values of $\bPhi$ can be obtained from $\iPhi$ by dividing its numerical values by $\Omega$ and substituting the analytical expression \eref{ein:omega}. 
%useful to use trafo \eref{esca:tildebarphi} and evolve $\bPhi$ and $\bPi$, not zero at $\scri^+$, easier to visualize and check convergence there. 

\subsection{Evolved shift}\label{se:evolshift}

The first test with an evolved shift condition was using the generalized Gamma-driver condition as defined in \eref{eg:Gammadriver}. The implemented expressions including the explicit source function terms are 
%\upda{add equation with exact source terms}. 
{\small
\begin{subequations}\label{ee:expGammadriver}
\begin{eqnarray}
\dot \beta^r & = & \beta^r{\beta^r}'+\frac{3}{4} \mu B^r  -\frac{\Kc^2 r}{9} -\frac{\xi_{\beta^r}}{\Omega}\left(\frac{\Kc}{3}\right)^2\left(\beta^r - \frac{\Kc r}{3}\right) , \\ 
\dot B^r & = & \beta^r {B^r}'-\eta B^r +\lambda\left(\dot \Lambda^r-\beta^r{\Lambda^r}'\right)  . 
\end{eqnarray}	
\end{subequations}
}%

The parameter choices that gave a stable numerical evolution were $\xi_{\beta^r}=5$, $\eta=0.1$, and for the choice $\Kc=-3$ (in flat space time), the values $\lambda=3/4$ and $\mu=1$ give the correct eigenspeeds. The damping term with $\xi_{\beta^r}$ was added to keep the value of $\beta^r$ fixed at $\scri^+$ and with the choice of slicing condition \eref{ee:finalpha} the evolution was not stable without a minimum value of $\xi_{\beta^r}$. %\upda{The behaviour at $\scri^+$ when using a fixed and Gamma-driver damped condition is the same. False!! It does have an effect!}
A consequence of the damping effect in $\dot\alpha$ and $\dot\beta^r$ at $\scri^+$ is that, according to the regularity condition  \eref{er:simpliregs}, the quantity $\DPK$ is not allowed to take values different from zero there. This is indeed the observed numerical behaviour, see figures \ref{fs:shiftcondab} and \ref{fs:shiftcondKL}.  

The generalized Gamma-driver condition in its integrated form \eref{eg:integGammadriver} was implemented numerically for the flat spacetime case as 
{\small
\begin{equation}\label{ee:expintegGammadriver}
\dot \beta^r = \beta^r{\beta^r}' + \lambda \, \Lambda^r - \eta \left(\beta^r-\frac{\Kc r}{3}\right) -\frac{\Kc^2 r}{9} -\frac{\xi_{\beta^r}}{\Omega}\left(\frac{\Kc}{3}\right)^2\left(\beta^r-\frac{\Kc r}{3}\right) . 
\end{equation}
}%
The choices $\xi_{\beta^r}=5$, $\eta=0.1$ and $\lambda=3/4$ provide a stable evolution for $\Kc=-3$ with regular initial data. 

\subsection{Schwarzschild initial data}\label{se:schwini}

For the first tests with Schwarzschild initial data on the hyperboloidal slice, a non-critical value of $\Cc$ was chosen, as had been done in chapter 2 of \cite{Zenginoglu:2007it}. It was large enough to satisfy condition \eref{ein:BHintersect}, which translates to $\Cc>-\case{8}{3}\Kc M^3$ in the Schwarzschild case and implies that the hyperboloidal slices intersected the BH horizon.

\subsubsection{Non-critical $\Cc$}

Together with a non-critical value of $\Cc$ the compactification factor $\aconf$ was chosen to be the flat spacetime one \eref{ein:aconfflat}. With this choice the initial values of some of the evolution variables ($A_{rr}$, $\alpha$ and $\beta^r$) diverge at the singularity, which is mapped to $r=0$ when a non-critical value of $\Cc$ is chosen. This forces the use of excision inside of the BH horizon to avoid the divergent behaviour of the variables in the integration domain. 

The evolution of the Einstein equations with these initial data showed an exponential growth, originated at the horizon and propagating outward, and that in the end rendered the simulation unstable. %\upda{This instability was not as fast as the ones observed at $\scri^+$ when the complete system was unstable and also was not accompanied by high-frequency oscillations. Check!} 
When the equations were inspected one by one, the exponential growth was located in the $\dot A_{rr}$ evolution equation. Plotting the numerical behaviour of $A_{rr}$ over time in a logarithmic scale revealed that the slope coincided with the coefficient $b$ of a term of the form $b A_{rr}$ present in $\dot A_{rr}$. If $\dot A_{rr}$ was evolved as a single equation, at some time the exponential growth stopped and the value of $A_{rr}$ stayed constant. %\upda{put plot? - will have to look for it ...} 
The explanation for this behaviour was that the $\dot A_{rr}$ equation had two solutions: the one calculated from the metric initial data \eref{ein:metrinia} using \eref{ein:Acurv0} and another one which diverges at $\scri^+$. It turned out that, at least in some part of the integration domain, the second solution was the stable stationary one, whereas the first one (the one expected by the rest of evolution equations) was an unstable stationary solution. Here ``unstable'' means that a small perturbation of the solution would make the value of $A_{rr}$ drift away from it. 

Several variable transformations involving $A_{rr}$ were tested to try to obtain a consistent stationary solution for $\dot A_{rr}$ and the rest of equations of motion, but none succeeded. When the critical $\Cc$ value of the hyperboloidal trumpet was chosen, together with the compactification factor $\aconf$ calculated from condition \eref{ein:conflat}, this large growth in the $A_{rr}$ variable no longer appeared. 

\subsubsection{Critical $\Cc$ (trumpet initial data)}

The choice of critical $\Cc$ and the corresponding $\aconf$ that makes the initial radius isotropic gives finite initial data for all variables in the range $r\in[0,\rscri]$, so that the use of excision is not required. 

Evolving the hyperboloidal trumpet initial data provides the expected results. The only drawback observed so far is a slow drift that appears for all variables and takes them away from their stationary value (it can be seen in an especially clear way in $\Lambda^r$). In spite of this slow drifting motion of the variables, they converge at the appropriate rate. However, this drifting does not converge away. 
If the shift is fixed, a slowly growing bulk instability (possibly an amplified version of the drift) appears and makes the simulation crash around a time $t\sim50-100$ for $\Kc=-1$. The region where it crashes varies depending on the chosen value for $\Kc$.  

A possibly similar drift was detected 
%in other simulations including strong fields, e.g. \cite{Hannam:2008sg}, and 
in the preliminary results in spherical symmetry of the hyperboloidal initial value problem presented in \cite{Zenginoglu:2007it} (see p.47-48), which were crashing due to a bulk instability: depending on the $\Kc$ value, the errors and the crash would appear closer to $\scri^+$ (larger $\Kc$) or in the interior of the domain (smaller $\Kc$). A similar effect has also been detected in the numerical simulations performed for this thesis.  %\upda{describe more this drift, perform convergence test.} % cited by An\i l when talking about the effect that $\Kc$ has on the point where the code crashes \cite{Frauendiener:2004bj}

The origin of the slow drift seems to be due to the fact that the initial data are an unstable stationary solution of the system (probably a similar effect to the growth observed in the non-critical $\Cc$ case, but much smaller), and the drift is the movement of the variables towards their stable stationary solution. The results shown in subsections \ref{se:onlySchw} and \ref{ss:large} of the next chapter seem to point in this direction.  

For small initial perturbations, the effect of the drift can be neglected for a long time ($t\sim500,1000$) and it can even be decreased by enlarging the eigenspeeds introduced by the shift evolution equation in the interior of the integration domain. For instance, this can be done by substituting the constant value of $\mu$ (or $\lambda$) by a $r$-dependent function that takes the appropriate value at $\scri^+$ and grows for $r<\rscri$, e.g. $\mu=0.15+4(\rscri^2-r^2)$ for $\Kc=-1$. Using this tuning of the shift eigenspeeds, the drift can be minimized long enough to be able to observe the tail of the scalar field. As the stable stationary solution of the Schwarzschild spacetime is not yet known, this is the most convenient way I found to study the behaviour of the scalar field.

\subsection{Tuned 1+log slicing condition}

The presence of a BH in the integration domain recommends the implementation of a singularity avoidant slicing condition, such as the 1+log condition. The evolved condition \eref{eg:1ploglapse} with explicit source terms, calculated such that a damping term of the form $\case{\Kc\,\alpha}{3\Omega}$ with tunable strength parameter $\xi _{1+\log }$ is present, is given by
{\small
\begin{eqnarray}
\dot\alpha &=& \frac{2 \Cc^3 \Omega  \aconf^7}{r^7}-\frac{3 \Cc^3 \aconf^8 \Omega '}{r^4 \rootp}+\frac{\Cc^2 \Kc \Omega  \aconf^4}{r^4}-\frac{\Cc^2 \Kc \aconf^5 \xi _{1+\log }}{r^2 \rootp}
\nonumber \\ && 
-\frac{3 \Cc^2 \Kc \aconf^5 \Omega '}{r \rootp}-\frac{\Cc \Kc^2 r^2 \aconf^2 \Omega '}{\rootp}-\frac{2 \Cc \Kc^2 r \aconf^2 \xi _{1+\log }}{3 \rootp}-\frac{\Cc M \Omega  \aconf^4}{r^4}
\nonumber \\ && 
+\frac{6 \Cc M \aconf^5 \Omega '}{r \rootp}+\frac{3 \Cc n \aconf^3 \Omega '}{r^2 \Omega }-\frac{3 \Cc \aconf^4 \Omega '}{\rootp}-\frac{\Kc^3 r^5 \Omega '}{9 \rootp \aconf}-\frac{\Kc^3 r^4 \xi _{1+\log }}{9 \rootp \aconf}
\nonumber \\ && 
-\frac{\Kc^3 r^2 \Omega }{27 \aconf^2}+\frac{2 \Kc M r^2 \aconf^2 \Omega '}{\rootp}+\frac{2 \Kc M r \aconf^2 \xi _{1+\log }}{\rootp}-\frac{\Kc M \Omega  \aconf}{3 r} -\frac{3 {\beta^r} n \Omega '}{\Omega }
\nonumber \\ && 
-\frac{\Kc r^3 \aconf \Omega '}{\rootp}-\frac{\Kc r^2 \aconf \xi _{1+\log }}{\rootp}+{\beta^r} \alpha '-\frac{\alpha  K n}{\Omega }+\frac{\alpha  \Kc \xi _{1+\log }}{3 \Omega }+\frac{\Kc n r \Omega '}{\Omega } , \label{ee:tun1plog}
\end{eqnarray}
}%
where $\rootp=\sqrt{9 \Cc^2 \aconf^6+6 r^3 \aconf^3 (\Cc \Kc-3 M)+9 r^4 \aconf^2+\Kc^2 r^6}$. The source terms are written in such a way that both trumpet initial data or flat spacetime initial data (with $M=0$, $\Cc=0$ and $\aconf=\Omega$) can be set. The value of $n$ that makes sure that all characteristic speeds at $\scri^+$ are positive is given by \eref{eg:nokval} and the value of the damping parameter, tuned experimentally, that provides a stable numerical behaviour is $\xi _{1+\log }=2$. 

%dissipation tuning
The fact that the amount of dissipation needed in the BH interior was smaller than in the region close to $\scri^+$ had already been realized in the tests performed with the harmonic slicing condition. Nevertheless, when using the 1+log condition with BH initial data this effect was more accentuated and the simulations would crash at an early time in the interior of the BH. The use of excision is not appropriate with the tuned 1+log slicing condition \eref{ee:tun1plog}, because the value of $n$ given by \eref{eg:nokval} makes sure that the characteristic speeds are all zero or positive at $\scri^+$, but it does not make them zero or negative at the horizon. 
A possible way of tuning the dissipation applied to the equations is by giving a radial dependence to the parameter $\epsilon$ in the dissipation operator \eref{en:KOdiss}. Stable results were obtained for some tests with the simple substitution 
\begin{equation}\label{ee:tunediss}
\epsilon = \epsilon_0+r\epsilon_1 , 
\end{equation}
choosing $\epsilon_0=0$ and $\epsilon_1$ to be the same value as the one appropriate for the former $\epsilon$ at $\scri^+$, so $\epsilon_1=0.5$. 

%The 1+log condition is the most suitable slicing condition to simulate a scalar field that collapses into a BH. An example of a collapse was 

\subsection{Gauge source functions of a conformal background metric}\label{se:confbg}

The use of the gauge conditions calculated from \eref{eg:gharmb} has some advantages, but unfortunately the preferred conformal gauge is not achieved. The equations \eref{eg:harmtwobsimpl} cannot be directly implemented in the code because the lapse condition \eref{eg:harmderlapsebsimpl} presents an exponential growth (the one that the function $K_0$ suppresses in \eref{eg:bonamasso} and derived equations). This lapse equation of motion can be made well-behaved by substituting $\bar F^t\equiv \bar F^t_\alpha=\case{\Kc}{\Omega\alpha^3}\left(\hat\alpha^2-\alpha^2\right)$ only in $\dot\alpha$. In $\dot\beta^r$ we set $\bar F^t\equiv \bar F^t_{\beta^r}=0$ and $\bar F^r=0$, because setting $\dot\alpha$'s value of $\bar F^t_\alpha$ instead of $\bar F^t_{\beta^r}=0$ leads to an unstable numerical behaviour. Using the evolution equations in terms of $\DPK$ to substitute $\dot\chi$, $\dot\gamma_{rr}$ and $\dot\gamma_{\theta\theta}$ yields
{\small
\begin{subequations}\label{ee:improvedbg}
\begin{eqnarray}
\dot\alpha &=& {\beta^r} \alpha '-\frac{3 \alpha  {\beta^r} \Omega '}{\Omega }-\frac{2 \alpha  {\beta^r} \hat{\alpha }'}{\hat{\alpha }}-\frac{\alpha ^2 \DPK}{\Omega }-\frac{\hat{\alpha }^2 \Kc}{\Omega }-\frac{\alpha ^3 \chi  \hat{\beta ^r} \hat{\chi }'}{\hat{\alpha }^2 \hat{\chi }^2 \gamma _{\theta \theta }}+\frac{2 \alpha ^3 \chi  \hat{\beta ^r}}{\hat{\alpha }^2 r \hat{\chi } \gamma _{\theta \theta }}-\frac{\alpha  {\beta^r}^2 \hat{\beta ^r}'}{\hat{\alpha }^2 \hat{\chi }}
\nonumber \\ && 
+\frac{\alpha  {\beta^r}^2 \hat{\beta ^r} \hat{\chi }'}{2 \hat{\alpha }^2 \hat{\chi }^2}+\frac{2 \alpha  {\beta^r} \hat{\beta ^r} \hat{\beta ^r}'}{\hat{\alpha }^2 \hat{\chi }}-\frac{\alpha  {\beta^r} \hat{\beta ^r}^2 \hat{\chi }'}{\hat{\alpha }^2 \hat{\chi }^2}-\frac{\alpha  \hat{\beta ^r}^2 \hat{\beta ^r}'}{\hat{\alpha }^2 \hat{\chi }}+\frac{\alpha  \hat{\beta ^r}^3 \hat{\chi }'}{2 \hat{\alpha }^2 \hat{\chi }^2}+\frac{\alpha  \hat{\alpha }' \hat{\beta ^r}}{\hat{\alpha }}+\frac{\alpha ^3 \chi  \hat{\beta ^r}'}{\hat{\alpha }^2 \hat{\chi } \gamma _{rr}}
\nonumber \\ && 
-\frac{\alpha ^3 \chi  \hat{\beta ^r} \hat{\chi }'}{2 \hat{\alpha }^2 \hat{\chi }^2 \gamma _{rr}}-\frac{2 \alpha ^2 \Theta}{\Omega }
, \\
\dot \beta^r &=& -\frac{\alpha ^2 \chi  \hat{\chi }'}{\hat{\chi } \gamma _{\theta \theta }}-\frac{2 {\beta^r}^2 \hat{\alpha }'}{\hat{\alpha }}+\hat{\alpha } \hat{\chi } \hat{\alpha }'-\frac{{\beta^r}^2 \hat{\chi }'}{2 \hat{\chi }}+{\beta^r} {\beta^r}'-\frac{\alpha ^2 {\beta^r} \chi  \hat{\beta ^r} \hat{\chi }'}{\hat{\alpha }^2 \hat{\chi }^2 \gamma _{\theta \theta }}+\frac{2 \alpha ^2 {\beta^r} \chi  \hat{\beta ^r}}{\hat{\alpha }^2 r \hat{\chi } \gamma _{\theta \theta }}+\frac{\alpha ^2 \chi  \hat{\beta ^r}^2 \hat{\chi }'}{\hat{\alpha }^2 \hat{\chi }^2 \gamma _{\theta \theta }}
\nonumber \\ && 
-\frac{2 \alpha ^2 \chi  \hat{\beta ^r}^2}{\hat{\alpha }^2 r \hat{\chi } \gamma _{\theta \theta }}+\alpha ^2 \chi  \Lambda ^r-\frac{{\beta^r}^3 \hat{\beta ^r}'}{\hat{\alpha }^2 \hat{\chi }}+\frac{{\beta^r}^3 \hat{\beta ^r} \hat{\chi }'}{2 \hat{\alpha }^2 \hat{\chi }^2}+\frac{3 {\beta^r}^2 \hat{\beta ^r} \hat{\beta ^r}'}{\hat{\alpha }^2 \hat{\chi }}-\frac{3 {\beta^r}^2 \hat{\beta ^r}^2 \hat{\chi }'}{2 \hat{\alpha }^2 \hat{\chi }^2}-\frac{3 {\beta^r} \hat{\beta ^r}^2 \hat{\beta ^r}'}{\hat{\alpha }^2 \hat{\chi }}
\nonumber \\ && 
+\frac{3 {\beta^r} \hat{\beta ^r}^3 \hat{\chi }'}{2 \hat{\alpha }^2 \hat{\chi }^2}+\frac{3 {\beta^r} \hat{\alpha }' \hat{\beta ^r}}{\hat{\alpha }}+\frac{\hat{\beta ^r}^3 \hat{\beta ^r}'}{\hat{\alpha }^2 \hat{\chi }}-\frac{\hat{\beta ^r}^4 \hat{\chi }'}{2 \hat{\alpha }^2 \hat{\chi }^2}-\frac{\hat{\alpha }' \hat{\beta ^r}^2}{\hat{\alpha }}+\frac{\hat{\beta ^r}^2 \hat{\chi }'}{2 \hat{\chi }}-\hat{\beta ^r} \hat{\beta ^r}'+\frac{\alpha ^2 {\beta^r} \chi  \hat{\beta ^r}'}{\hat{\alpha }^2 \hat{\chi } \gamma _{rr}}
\nonumber \\ && 
-\frac{\alpha ^2 {\beta^r} \chi  \hat{\beta ^r} \hat{\chi }'}{2 \hat{\alpha }^2 \hat{\chi }^2 \gamma _{rr}}-\frac{\alpha ^2 \chi  \hat{\beta ^r} \hat{\beta ^r}'}{\hat{\alpha }^2 \hat{\chi } \gamma _{rr}}+\frac{\alpha ^2 \chi  \hat{\beta ^r}^2 \hat{\chi }'}{2 \hat{\alpha }^2 \hat{\chi }^2 \gamma _{rr}}+\frac{\alpha ^2 \chi '}{2 \gamma _{rr}}+\frac{\alpha ^2 \chi  \hat{\chi }'}{2 \hat{\chi } \gamma _{rr}}-\frac{\alpha  \chi  \alpha '}{\gamma _{rr}} . \label{ee:bgshift}
\end{eqnarray}	
\end{subequations}
}%
However, the choice of this source function for $\bar F_\alpha^t$ in $\dot\alpha$ introduces the extra regularity condition $\atscrip{\alpha}=\atscrip{\hat\alpha}$, fixing the value of the lapse at future null infinity. The shift evolution equation \eref{eg:harmdershiftb} has no damping terms for $\beta^r$.  
This means that only $\alpha$ has a ``fixed'' value at $\scri^+$; $\chi$, $\gamma_{rr}$ and $\beta^r$ are allowed to vary their values there provided that the scri-fixing condition \eref{eg:scrifixspher} is satisfied. This is exactly the numerical behaviour observed, including that now $\DPK$ is allowed to be non-zero at future null infinity (compare to the explanation in subsection \ref{se:evolshift} and figures \ref{fs:shiftcondab} and \ref{fs:shiftcondKL}). %\upda{relate to the reg conds?}

%The difference \upda{$\atscrip{coef\case{\alpha_0-\alpha}{\Omega}}$} is exactly the amount by which the preferred conformal gauge condition is not satisfied at $\scri^+$. 
With the selected values for the source functions, the amount by which the preferred conformal gauge condition is not satisfied at $\scri^+$ is $\atscrip{F^t_{\alpha}\beta^r\Omega'} = \atscrip{\case{\Kc}{\Omega\alpha^3}\left(\hat\alpha^2-\alpha^2\right)\beta^r\Omega'}$. In a numerical simulation this quantity does attain a finite value at $\scri^+$, but it is only zero when the variables are at their stationary values. 

A small drawback of this choice of gauge conditions is that the drift that appears in simulations with BH trumpet initial data is more pronounced. The freedom of the parameters $\lambda$ and $\mu$ present in the Gamma-driver shift condition \eref{eg:Gammadriver} is not present here (all terms are directly derived from \eref{eg:gharmb}), so that in principle the eigenspeeds cannot be tuned. One can however manually modify the coefficient of the $\alpha^2\chi\Lambda^r$ term in \eref{ee:bgshift} to increase its value for $r<\rscri$ and in this way slow down the drift's growth. 
%\upda{understand numerically!}

\subsection{Gauge source functions of a physical background metric}\label{se:physbg}

A direct implementation of \eref{eg:harmtwobsimpl} with vanishing source functions gives an unstable simulation. With $\Kc=-3$ it crashes almost instantaneously, but with $\Kc>-2$ it lasts longer and the instability can be studied. Even if for some parameter choices the evolution can last long enough to see how the reflected initial perturbation pulses leave the domain, it will crash at some point: for instance, a test with GBSSN crashed at $t\approx3.7$ and the same setup with some configurations of Z4c would last until around $t=7.5$. 

The choice of source functions $\tilde F^r=0$ and $\tilde F^t = \xi_\alpha\Omega(\alpha-\hat\alpha)$ (used with $ \xi_\alpha=2$) finally gave a stable evolution (at least for some configurations with regular initial data), with the complete equations:
{\small
\begin{subequations}\label{ee:improvedpbg}
\begin{eqnarray}
\dot\alpha &=& {\beta^r} \alpha '+\frac{\alpha  {\beta^r} \Omega '}{\Omega }-\frac{2 \alpha  {\beta^r} \hat{\alpha }'}{\hat{\alpha }}-\frac{\alpha ^4 \xi_\alpha}{\Omega }+\frac{\hat{\alpha } \alpha ^3 \xi_\alpha}{\Omega }-\frac{\alpha ^2 \DPK}{\Omega }-\frac{\alpha ^2 \Kc}{\Omega }-\frac{2 \alpha ^3 \chi  \hat{\beta ^r} \Omega '}{\hat{\alpha }^2 \hat{\chi } \Omega  \gamma _{\theta \theta }}
\nonumber \\ && 
-\frac{\alpha ^3 \chi  \hat{\beta ^r} \hat{\chi }'}{\hat{\alpha }^2 \hat{\chi }^2 \gamma _{\theta \theta }}+\frac{2 \alpha ^3 \chi  \hat{\beta ^r}}{\hat{\alpha }^2 r \hat{\chi } \gamma _{\theta \theta }}+\frac{\alpha  {\beta^r}^2 \hat{\beta ^r} \Omega '}{\hat{\alpha }^2 \hat{\chi } \Omega }-\frac{\alpha  {\beta^r}^2 \hat{\beta ^r}'}{\hat{\alpha }^2 \hat{\chi }}+\frac{\alpha  {\beta^r}^2 \hat{\beta ^r} \hat{\chi }'}{2 \hat{\alpha }^2 \hat{\chi }^2}-\frac{2 \alpha  {\beta^r} \hat{\beta ^r}^2 \Omega '}{\hat{\alpha }^2 \hat{\chi } \Omega }
\nonumber \\ && 
+\frac{2 \alpha  {\beta^r} \hat{\beta ^r} \hat{\beta ^r}'}{\hat{\alpha }^2 \hat{\chi }}-\frac{\alpha  {\beta^r} \hat{\beta ^r}^2 \hat{\chi }'}{\hat{\alpha }^2 \hat{\chi }^2}+\frac{\alpha  \hat{\beta ^r}^3 \Omega '}{\hat{\alpha }^2 \hat{\chi } \Omega }-\frac{\alpha  \hat{\beta ^r}^2 \hat{\beta ^r}'}{\hat{\alpha }^2 \hat{\chi }}+\frac{\alpha  \hat{\beta ^r}^3 \hat{\chi }'}{2 \hat{\alpha }^2 \hat{\chi }^2}-\frac{\alpha  \hat{\beta ^r} \Omega '}{\Omega }+\frac{\alpha  \hat{\alpha }' \hat{\beta ^r}}{\hat{\alpha }}
\nonumber \\ && 
-\frac{\alpha ^3 \chi  \hat{\beta ^r} \Omega '}{\hat{\alpha }^2 \hat{\chi } \Omega  \gamma _{rr}}+\frac{\alpha ^3 \chi  \hat{\beta ^r}'}{\hat{\alpha }^2 \hat{\chi } \gamma _{rr}}-\frac{\alpha ^3 \chi  \hat{\beta ^r} \hat{\chi }'}{2 \hat{\alpha }^2 \hat{\chi }^2 \gamma _{rr}}-\frac{2 \alpha ^2 \Theta}{\Omega }
, \\
\dot \beta^r &=& \alpha ^2 \chi  \Lambda ^r+\frac{\xi_\alpha \alpha ^2 {\beta^r} \hat{\alpha }}{\Omega }-\frac{\alpha  \chi  \alpha '}{\gamma _{rr}}+{\beta^r} {\beta^r}'+\frac{\alpha ^2 \chi '}{2 \gamma _{rr}}+\frac{2 \hat{\beta ^r}^2 \Omega '}{\Omega }-\frac{3 {\beta^r} \hat{\beta ^r} \Omega '}{\Omega }-\frac{\hat{\alpha }^2 \hat{\chi } \Omega '}{\Omega }+\frac{{\beta^r}^2 \Omega '}{\Omega }
\nonumber \\ && 
-\frac{\hat{\beta ^r}^4 \Omega '}{\Omega  \hat{\alpha }^2 \hat{\chi }}+\frac{3 {\beta^r} \hat{\beta ^r}^3 \Omega '}{\Omega  \hat{\alpha }^2 \hat{\chi }}-\frac{3 {\beta^r}^2 \hat{\beta ^r}^2 \Omega '}{\Omega  \hat{\alpha }^2 \hat{\chi }}+\frac{{\beta^r}^3 \hat{\beta ^r} \Omega '}{\Omega  \hat{\alpha }^2 \hat{\chi }}+\frac{3 \alpha ^2 \chi  \Omega '}{\Omega  \gamma _{rr}}+\frac{\alpha ^2 \chi  \hat{\beta ^r}^2 \Omega '}{\Omega  \hat{\alpha }^2 \hat{\chi } \gamma _{rr}}-\frac{\alpha ^2 {\beta^r} \chi  \hat{\beta ^r} \Omega '}{\Omega  \hat{\alpha }^2 \hat{\chi } \gamma _{rr}}
\nonumber \\ && 
-\frac{2 \alpha ^2 \chi  \Omega '}{\Omega  \gamma _{\theta \theta }}+\frac{2 \alpha ^2 \chi  \hat{\beta ^r}^2 \Omega '}{\Omega  \hat{\alpha }^2 \hat{\chi } \gamma _{\theta \theta }}-\frac{2 \alpha ^2 {\beta^r} \chi  \hat{\beta ^r} \Omega '}{\Omega  \hat{\alpha }^2 \hat{\chi } \gamma _{\theta \theta }}-\frac{\hat{\beta ^r}^2 \hat{\alpha }'}{\hat{\alpha }}+\frac{3 {\beta^r} \hat{\beta ^r} \hat{\alpha }'}{\hat{\alpha }}+\hat{\alpha } \hat{\chi } \hat{\alpha }'-\frac{2 {\beta^r}^2 \hat{\alpha }'}{\hat{\alpha }}-\hat{\beta ^r} \hat{\beta ^r}'
\nonumber \\ && 
+\frac{\hat{\beta ^r}^3 \hat{\beta ^r}'}{\hat{\alpha }^2 \hat{\chi }}-\frac{3 {\beta^r} \hat{\beta ^r}^2 \hat{\beta ^r}'}{\hat{\alpha }^2 \hat{\chi }}+\frac{3 {\beta^r}^2 \hat{\beta ^r} \hat{\beta ^r}'}{\hat{\alpha }^2 \hat{\chi }}-\frac{{\beta^r}^3 \hat{\beta ^r}'}{\hat{\alpha }^2 \hat{\chi }}-\frac{\alpha ^2 \chi  \hat{\beta ^r} \hat{\beta ^r}'}{\hat{\alpha }^2 \hat{\chi } \gamma _{rr}}+\frac{\alpha ^2 {\beta^r} \chi  \hat{\beta ^r}'}{\hat{\alpha }^2 \hat{\chi } \gamma _{rr}}-\frac{{\beta^r}^2 \hat{\chi }'}{2 \hat{\chi }}
\nonumber \\ && 
+\frac{\hat{\beta ^r}^2 \hat{\chi }'}{2 \hat{\chi }}+\frac{\alpha ^2 \chi  \hat{\chi }'}{2 \hat{\chi } \gamma _{rr}}+\frac{\alpha ^2 \chi  \hat{\beta ^r}^2 \hat{\chi }'}{2 \hat{\alpha }^2 \hat{\chi }^2 \gamma _{rr}}-\frac{\alpha ^2 {\beta^r} \chi  \hat{\beta ^r} \hat{\chi }'}{2 \hat{\alpha }^2 \hat{\chi }^2 \gamma _{rr}}-\frac{\alpha ^2 \chi  \hat{\chi }'}{\hat{\chi } \gamma _{\theta \theta }}+\frac{\alpha ^2 \chi  \hat{\beta ^r}^2 \hat{\chi }'}{\hat{\alpha }^2 \hat{\chi }^2 \gamma _{\theta \theta }}-\frac{\alpha ^2 {\beta^r} \chi  \hat{\beta ^r} \hat{\chi }'}{\hat{\alpha }^2 \hat{\chi }^2 \gamma _{\theta \theta }}
\nonumber \\ && 
-\frac{\hat{\beta ^r}^4 \hat{\chi }'}{2 \hat{\alpha }^2 \hat{\chi }^2}+\frac{3 {\beta^r} \hat{\beta ^r}^3 \hat{\chi }'}{2 \hat{\alpha }^2 \hat{\chi }^2}-\frac{3 {\beta^r}^2 \hat{\beta ^r}^2 \hat{\chi }'}{2 \hat{\alpha }^2 \hat{\chi }^2}+\frac{{\beta^r}^3 \hat{\beta ^r} \hat{\chi }'}{2 \hat{\alpha }^2 \hat{\chi }^2}-\frac{\xi_\alpha \alpha ^3 {\beta^r}}{\Omega }-\frac{2 \alpha ^2 \chi  \hat{\beta ^r}^2}{r \hat{\alpha }^2 \hat{\chi } \gamma _{\theta \theta }}+\frac{2 \alpha ^2 {\beta^r} \chi  \hat{\beta ^r}}{r \hat{\alpha }^2 \hat{\chi } \gamma _{\theta \theta }} . \qquad \quad \  \label{ee:pbgshift}
\end{eqnarray}	
\end{subequations}
}%

This system has not been tested in much detail yet, but it seems to only be stable with $\Kc>-2$. An important advantage of this gauge condition and the way it is set is that, at least for the \CZ{} ($C_{Z4c}=0$) case, the preferred conformal gauge is satisfied. It has been numerically checked that $\atscrip{\bar\Box\Omega}=0$. 
%ok with Z4 ($C_{Z4c}=0$), preferred conformal gauge satisfied

There is still much to understand and improve about this condition. For instance and unlike the \CZ{} ($C_{Z4c}=0$) case, the GBSSN one does not converge appropriately and the preferred conformal gauge is not satisfied. 
Also in the few tests that have been performed with Schwarzschild initial data, the simulations performed to obtain the convergence of the scalar field at $\scri^+$ showed bad convergence or were unstable.

\section{Conclusions}

The main changes that have been performed to the rescaled GBSSN and $\CZ{}$ spherically symmetric equations are: using the trace of the physical extrinsic curvature $\pK$ (or even better $\DPK$) as variable instead of the conformal one $K$, adding a constraint damping term to the equation of motion of $\Lambda^r$ and setting appropriate source terms in the gauge conditions. 
How the latter are chosen is of great importance, because the gauge conditions are determinant in the evolution of the complete system. The gauge equations of motion presented here allow for a stable (at least for a reasonably long time in the strong field case) evolution of the Einstein equation, but there is clearly much room for improvement.

\subsection{Parameter ranges}

Possibly the most important parameter in the equations is $\kappa_1$. It does not only control the damping of the Z4 quantities, but also gives the necessary constraint damping to keep the evolution stable. In the GBSSN case, in general $\kappa_1\in[1.4,4.5]$ allows for a stable evolution and the default choice is $\kappa_1=1.5$. %\upda{$\kappa_1=1.85$ required in some cases??} 
The value of $\kappa_1$ depends on $\Kc$; if $|\Kc|$ is increased, the value of $\kappa_1$ that will provide a stable resolution is larger, and vice versa.
The \CZ{} equations seem to prefer a smaller value $\kappa_1\sim1$. This value is relatively close to the one used in CCZ4 by \cite{Alic:2011gg}, $\kappa_1\sim1$, but is considerably bigger than the one suggested in Z4c by \cite{Weyhausen:2011cg}, which is $\kappa_1\sim0.02$ also in spherical symmetry. The default choice for $\kappa_2$ is the commonly chosen $\kappa_2=0$, which has worked fine for all the performed tests. 
%The choice for $\kappa_2$ is \upda{$\kappa_2=0$ when dropping the non-principal part Z4 terms as in the Z4c formalism in \cite{Bernuzzi:2009ex,Weyhausen:2011cg} and $\kappa_2=1$ when all terms are kept. Check! - now $\kappa_2=0$ seems a good choice always, as the damping in $\Theta$ is just given by a term proportional to $\kappa_1$}

Certain choices of gauge conditions include some parameters that control the damping of the lapse and shift. 
%\begin{itemize}
%\item 
The generalized harmonic lapse condition as in \eref{ee:finalpha} is stable for the range $\xi_\alpha\in[1,12]$ in flat spacetime with $\Kc=-3$. The larger $\xi_\alpha$, the faster the initial perturbations are suppressed and the earlier the variables reach their stationary values. 
%\item 
The damping parameter of the tuned 1+log lapse condition \eref{ee:tun1plog} had to be tuned between the values $\xi_{1+log}\in[2,5]$ to give stable evolutions. The choice $\xi_{1+log}=2$ was appropriate for simulations with regular data and Schwarzschild trumpet data, but the choice of a larger value $\xi_{1+log}=5$ was necessary to prevent the code from crashing at high grid resolutions in the case of a collapsing scalar field perturbation. 
%\item 
The Gamma-driver shift conditions as in \eref{ee:expGammadriver} and in \eref{ee:expintegGammadriver} are stable with $\xi_{\beta^r}=5$ for flat spacetime with $\Kc=-3$. The case with BH initial data with $\Kc=-1$ requires more complicated source conditions, calculated from the BH initial data values for the variables, but the choice $\xi_{\beta^r}=5$ is still valid. What has to be rescaled when changing the value of $\Kc$ are the parameters $\lambda$ and $\mu$, so that the eigenspeeds at $\scri^+$ are all outgoing. The value of $\eta$ is in general chosen as small as possible to damp as much as possible the behaviour of $\Lambda^r$, so that $\eta=0,0.1$ were common choices in the simulations performed here. 
%\end{itemize}

The dissipation parameter $\epsilon$ included in \eref{en:KOdiss} is chosen to be $\epsilon=0.5$ for $\Kc=-3$, although a smaller value can be chosen for a smaller $\Kc$, for instance $\epsilon=0.05$ is enough to maintain a simulation with $\Kc=-0.8$ and $\kappa_1=0.5$ stable. % Have to make more specific or leave away: The usual value of the dissipation required here is larger than the ones commonly used. %\upda{check bibliography!} 
%$\epsilon=0.05$, a simulation with $a=3.75$ and $\kappa_1=0.5$ is stable

Without restricting generality, the mass of the BH (if present) can be set to unity and this is done here. In this work the parameter $\Kc$ is chosen to be $\Kc=-3$ for flat spacetime and $\Kc=-1$ for the Schwarzschild BH case. The reason for using a smaller value of $|\Kc|$ in the BH simulations is due to precision limitations of the compiler when calculating the compactification factor $\aconf$, as was described in subsection \ref{sn:aconf}. 
In general the simulations are well-behaved for a $\Kc$ as large as $\Kc=-3 \cdot 10^{-3}$ using $\kappa_1=0.003$. The lower limit achieved (with harmonic lapse and Gamma-driver) with GBSSN was $\Kc=-3.75$ and $\kappa_1=3$, but the only successful example with the \CZ{} equations was with $\Kc=-3.41$, $\kappa_1=1$ and only for the case $C_{Z4c}=1$. The exact values may depend on minor details of the setup (such as the spatial resolution), so that the limits given should not be taken as rigorous constraints, but as restricted examples. 
The gauge conditions \eref{ee:improvedpbg} would only not crash instantaneously if $\Kc>-2$. 

%comment on maximum Courant factor = 0.2 (for Kcmc=-3)
For the choice of $\Kc=-3$, the maximum value allowed for the Courant factor is 0.200 for the \CZ{} equations when $C_{Z4c}=0$  (although 0.224 is possible in this case for $\kappa_2=-1$), 0.224 for \CZ{}  with $C_{Z4c}=1$, and 0.252 for the GBSSN  system. For a smaller absolute value of $\Kc$, the Courant factor can be larger, e.g. 5 for $\Kc=-3 \cdot 10^{-3}$.

\subsection{Difference between formulations}

The comparison of the GBSSN and \CZ{} systems used on the hyperboloidal initial value problem is not among the aims of this work, but it is well worth mentioning some of their differences in behaviour. In most cases these differences are small, because the three possible evolution systems available (GBSSN, \CZ{} ($C_{Z4c}=1$) and \CZ{} ($C_{Z4c}=0$)) perform quite similarly. 

It was already mentioned that in some numerical tests performed a smaller value of $\kappa_1$ was required by the \CZ{} equations. The results in convergence tests vary between different systems and different parameters (damping parameters, dissipation, etc.) may have to be used to obtain optimal results. 
The convergence plots in \fref{fs:convflathscri} show a comparison between the formulations. %\upda{From paper1: As an example, the data in \fref{phiscri} belong to a simulation with \CZ{} ($C_{Z4c}=1$); the same simulation carried out with \CZ{} ($C_{Z4c}=0$) showed a convergence which was practically the same, whereas that of the equivalent GBSSN simulation was considerably less accurate. }
It indicates that the errors  in the \CZ{} case are smaller and in most cases its convergence is also better. 

\subsection{Effect of numerical treatment}

%one-sided stencils at $\scri^+$: useful at first, easy to implement if extrapolation not working

If the appropriate extrapolation at the outer boundary is not implemented, a simple choice to treat the derivatives there is to use one-sided stencils at $\scri^+$. %\upda{The stability behaviour and results are very similar. Put plot in next chapter?}

The appropriate extrapolation order at the outer boundary corresponding to $n$th order finite differences is $(n+1)$th order extrapolation for the $u$ variables and $n$th or $(n+1)$th order for the $v$ variables. Using higher or lower extrapolation orders is not recommended, according to my observations of the numerical behaviour close to the boundary. %\upda{put plots to explain?}

The use of off-centered stencils in the derivatives of the advection terms requires less dissipation than the centered stencils case to obtain a stable numerical evolution. In most cases the results of both options are quite similar, see \fref{fs:convflathscri}. However, when checking the convergence of the small initial scalar perturbation of a Schwarzschild BH, the convergence was better in the off-centered case, see \fref{fs:tailsconv}. 

%\upda{From blog: I also ran the simulations using off-centered stencils in the advection terms. The 4th order off-sided stencil was implemented as illustrated by the diagram in figure \ref{offcentdiagram}.  The statement that using off-centered stencils in the advection terms still holds. I ran convergence tests using $\epsilon=0.1,0.25,0.5$ and there is almost no difference between them (the smaller the dissipation, the slightly better). The convergence results, compared to the centered case in blue, are shown in figure \ref{offcentered}. In general the errors(especially for the $C_{Z4c}=1$ (zz) case) are smaller, but the coincidence between the curves is slightly worse. Leave for further work. }

\subsection{Other observations}

Choosing $\iPhi$ and $\iPi$ or $\bPhi$ and $\bPi$ for the scalar field variables does not make a real difference in the numerical results. For convenience, $\bPhi$ and $\bPi$ are preferred, because they do not vanish at $\scri^+$. %\upda{Results with other option??}

If a transformation to evolution variables whose stationary values are $1$ or $0$ is performed, the radial dependence of the stationary solutions is put into the equations, so that they look considerably more complicated. In any case, the numerical behaviour of the results in terms of stability or accuracy in the convergence results seems to be unaffected by this kind of transformation.

\chapter{Results}\label{c:results}

%\cite{Vano-Vinuales:2014koa}

In this chapter I will show some of the results that have been obtained with the formulations, initial data and gauge conditions described previously.

In the equations used, the variable $\gamma_{\theta\theta}$ has been eliminated in terms of $\gamma_{rr}$ as in \eref{es:delgtt}. The common parameter choices are $\kappa_1=1.5$, $\kappa_2=0$ and $\rscri=1$. In the case of Schwarzschild initial data, $M=1$, $\Kc=-1$ and the value of $\Cc=3.11$ is the critical one.

Regarding the numerical setup, some choices are common to all simulations presented here. The spatial grid is staggered, so it takes the form of \fref{fn:stgrid}. The Courant factor is chosen to be $0.2$ and the finite differences used are 4th order, so that the convergence order is 4th order. For convergence runs, the factor by which spatial and time resolution is increased is always 1.5. The finite difference stencil at the boundaries are centered ones: the ghost points at the inner boundary are filled in according to the parity conditions of the variables and at the outer boundary  5th order extrapolation is used for all variables.

\section{Regular initial data}

This section's results have flat spacetime as stationary state, except the collapse example in subsection \ref{ss:col}. Flat spacetime stationary values were shown in \fref{fin:Fini}.

\subsection{Gauge waves in flat spacetime}

\begin{figure}[htbp!!]
\center
\vspace{-2ex}
 \begin{tikzpicture}[scale=1.5]\draw (-1cm,0cm) node {};
		\draw (0cm, 0cm) node {$\chi$}; \draw (0.3cm, 0cm) -- (1cm, 0cm);
		\draw (1.5cm, 0cm) node {$\gamma_{rr}$}; \draw [dashed] (1.8cm, 0cm) -- (2.5cm, 0cm);
		\draw (3cm, 0cm) node {$A_{rr}$}; \draw [dotted] (3.3cm, 0cm) -- (4cm, 0cm);
		\draw (4.5cm, 0cm) node {$\DPK$}; \draw [dash pattern= on 4pt off 2pt on 1pt off 2pt] (4.8cm, 0cm) -- (5.5cm, 0cm);
		\draw (6cm, 0cm) node {$\Lambda^r$}; \draw [dash pattern= on 8pt off 2pt] (6.3cm, 0cm) -- (7cm, 0cm);
		\draw (7.5cm, 0cm) node {$\alpha$}; \draw [dash pattern= on 6pt off 2pt on 1pt off 2pt on 1pt off 2pt] (7.8cm, 0cm) -- (8.5cm, 0cm);
	\end{tikzpicture}
\\
\begin{tabular}{ m{0.5\linewidth}@{} @{}m{0.5\linewidth}@{} }
\mbox{\includegraphics[width=1\linewidth]{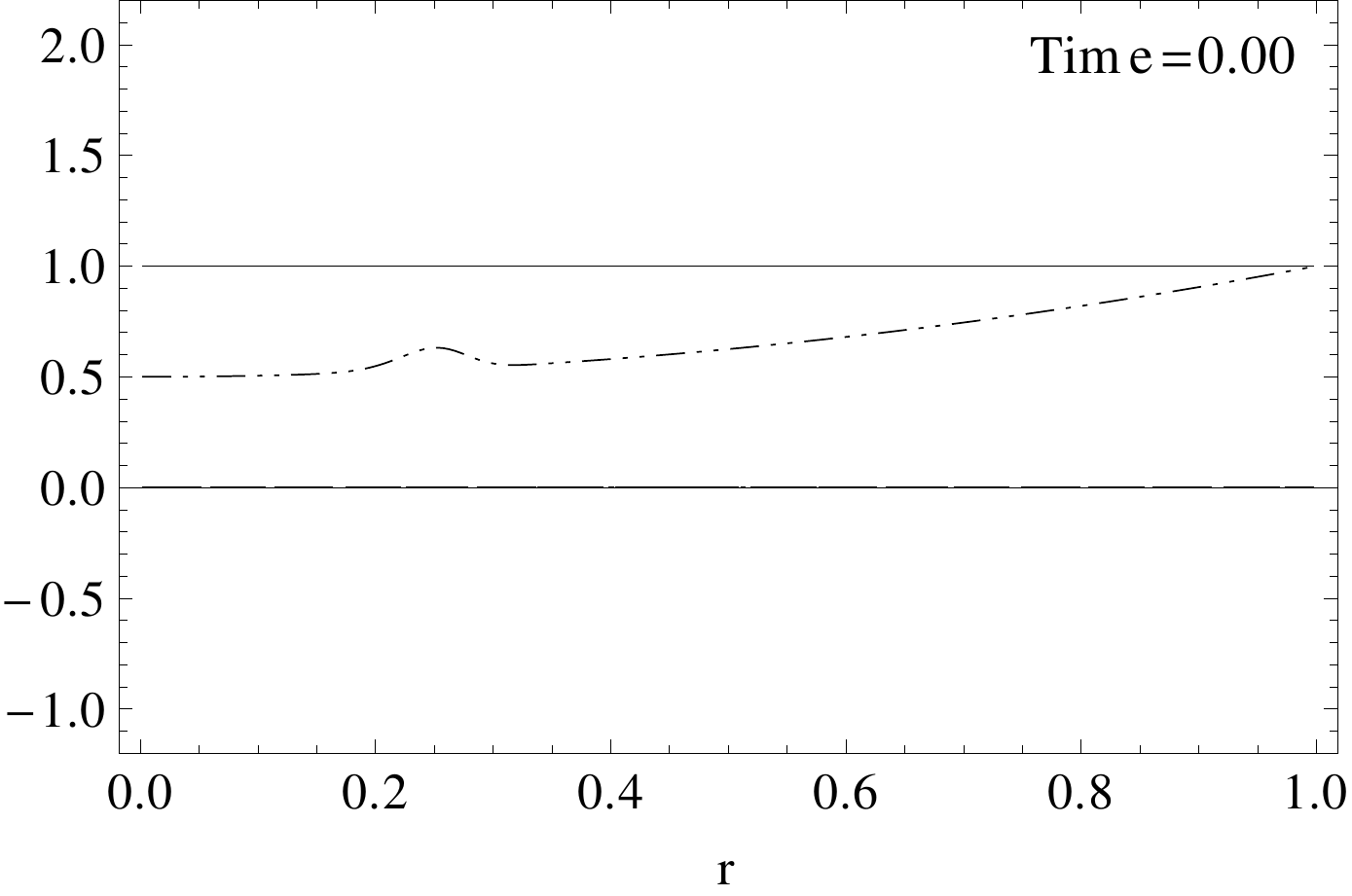}}&
\hspace{-0.8ex} \mbox{\includegraphics[width=1\linewidth]{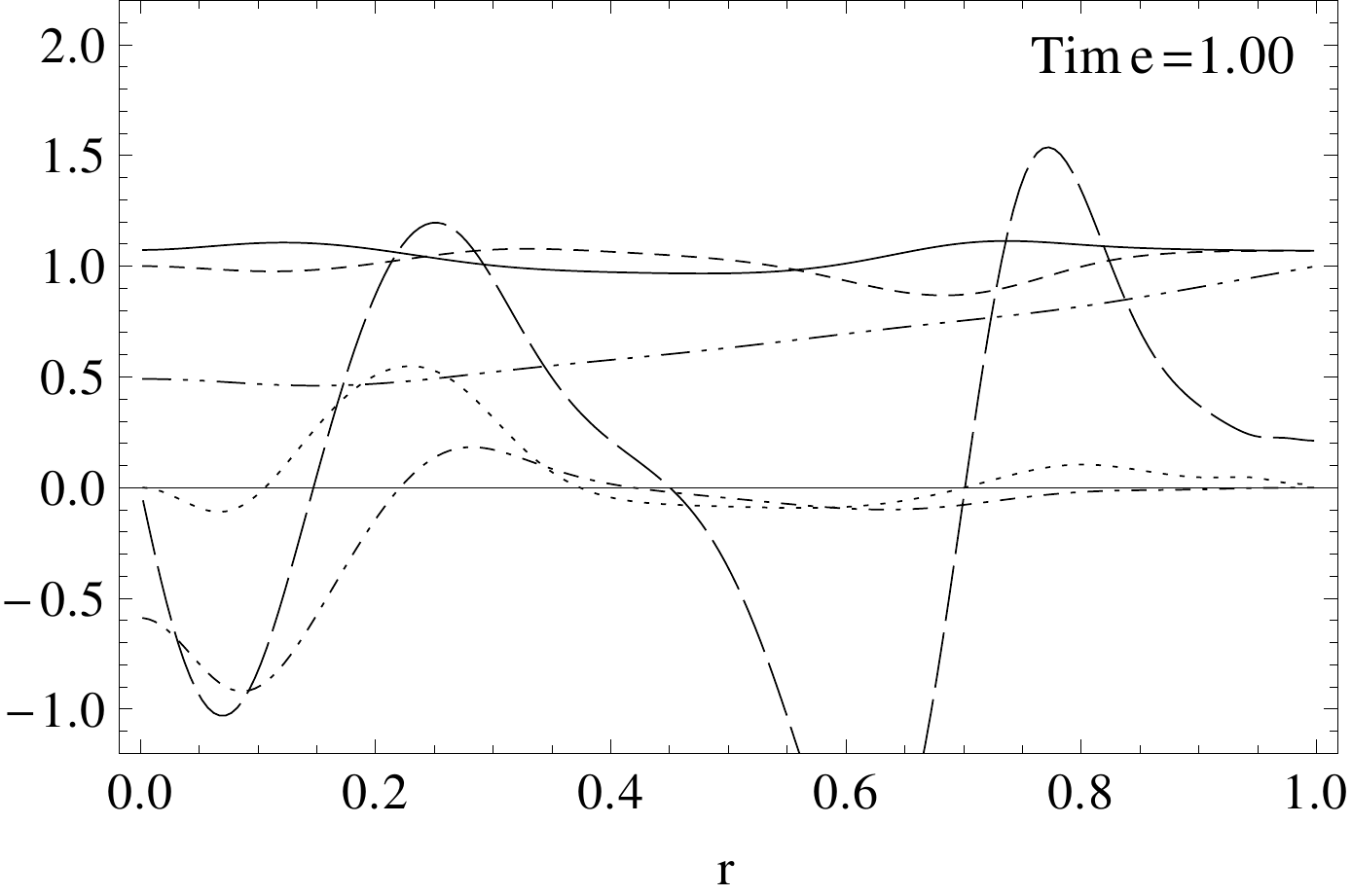}}\\
\vspace{-5.5ex} \mbox{\includegraphics[width=1\linewidth]{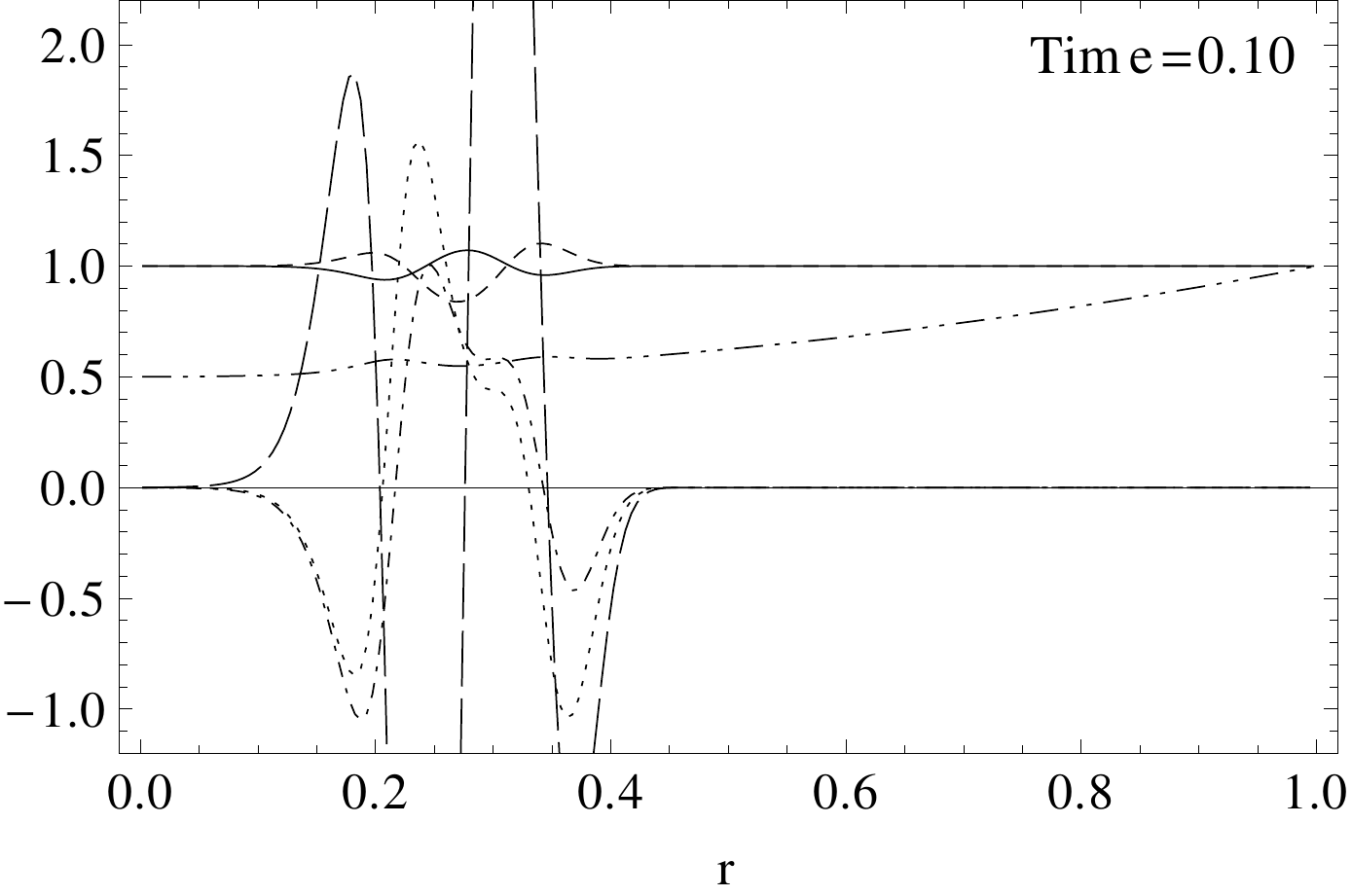}}&
\vspace{-5.5ex} \hspace{-0.8ex} \mbox{\includegraphics[width=1\linewidth]{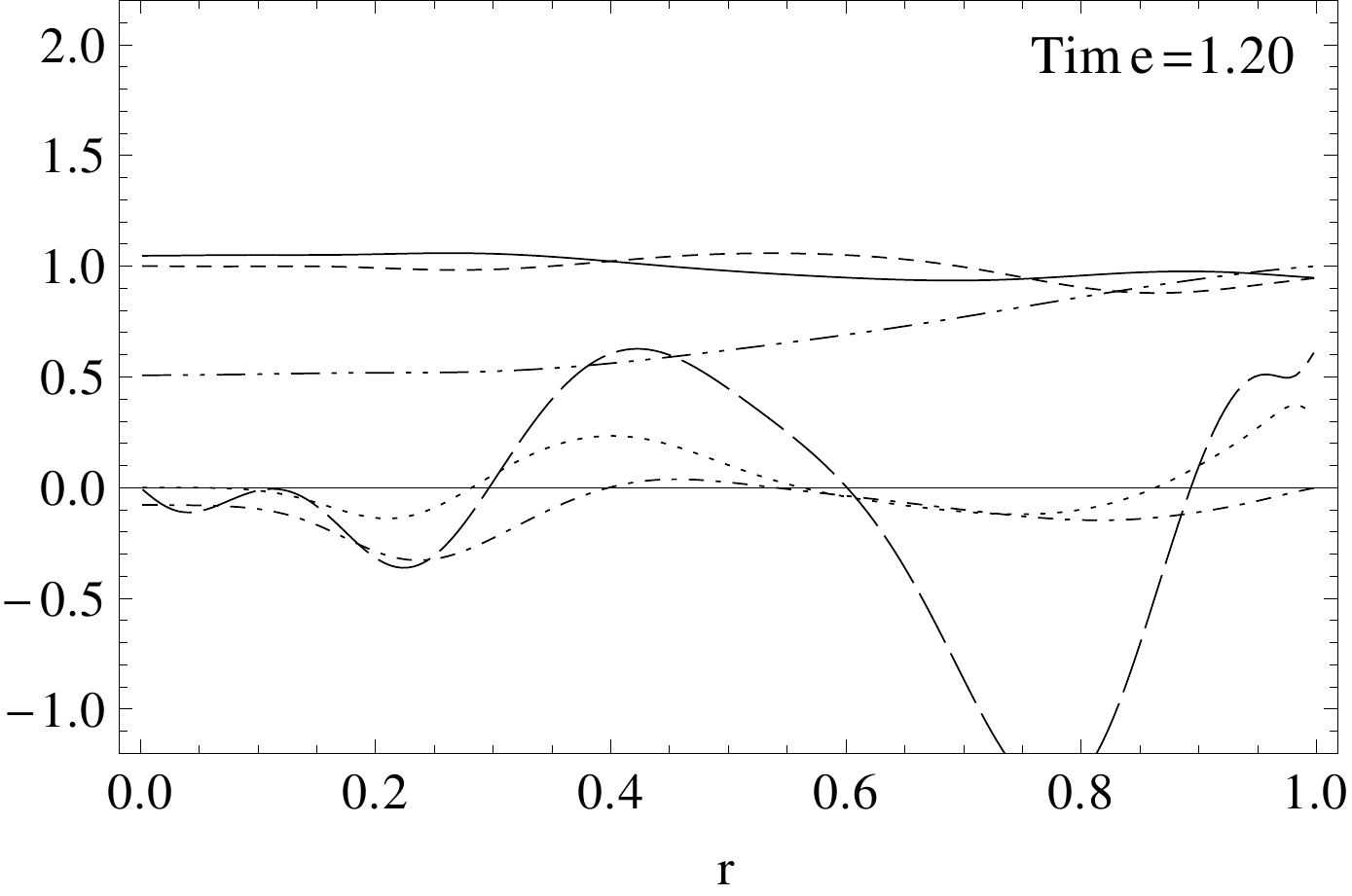}}\\
\vspace{-5.5ex} \mbox{\includegraphics[width=1\linewidth]{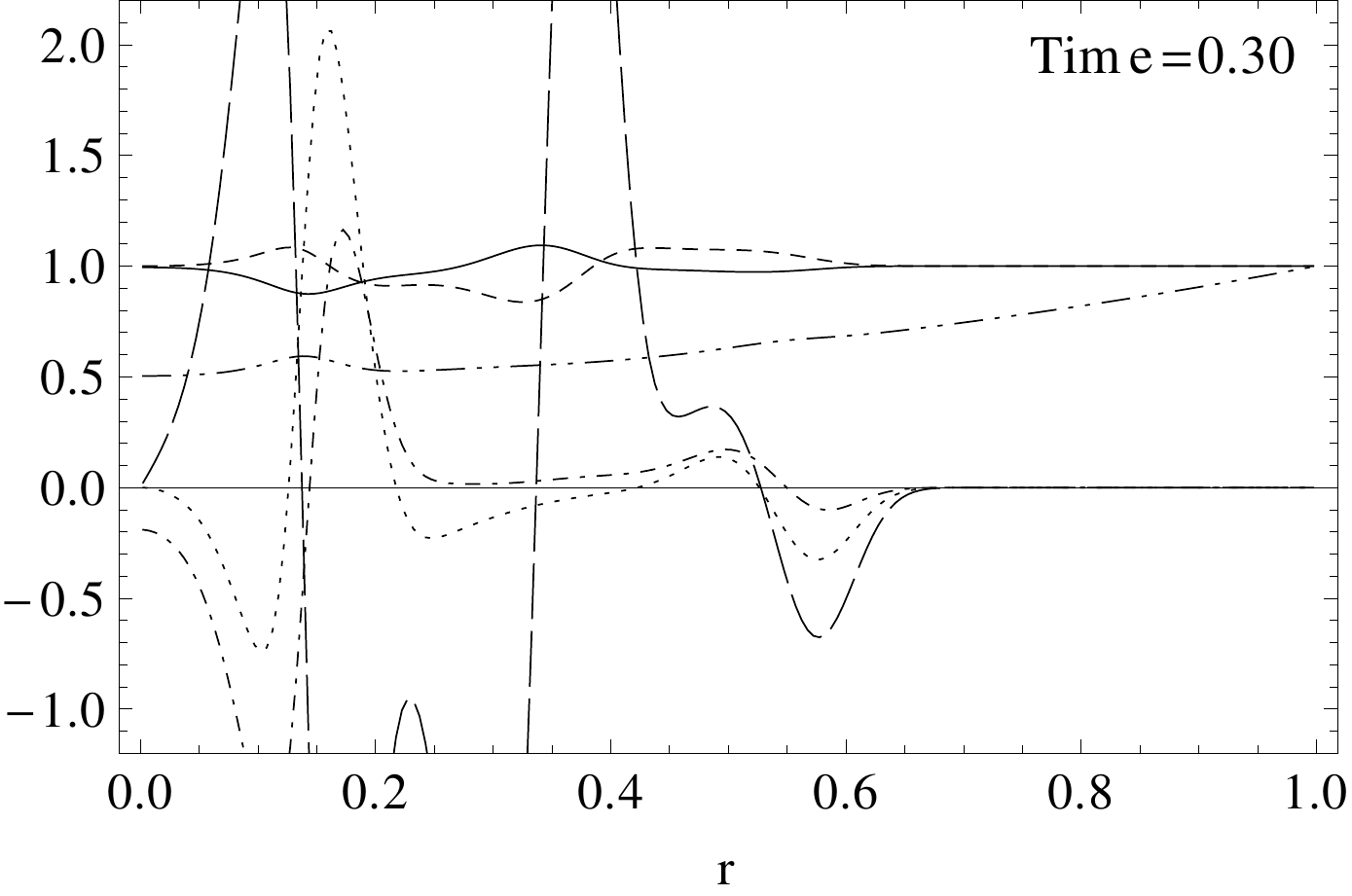}}&
\vspace{-5.5ex} \hspace{-0.8ex} \mbox{\includegraphics[width=1\linewidth]{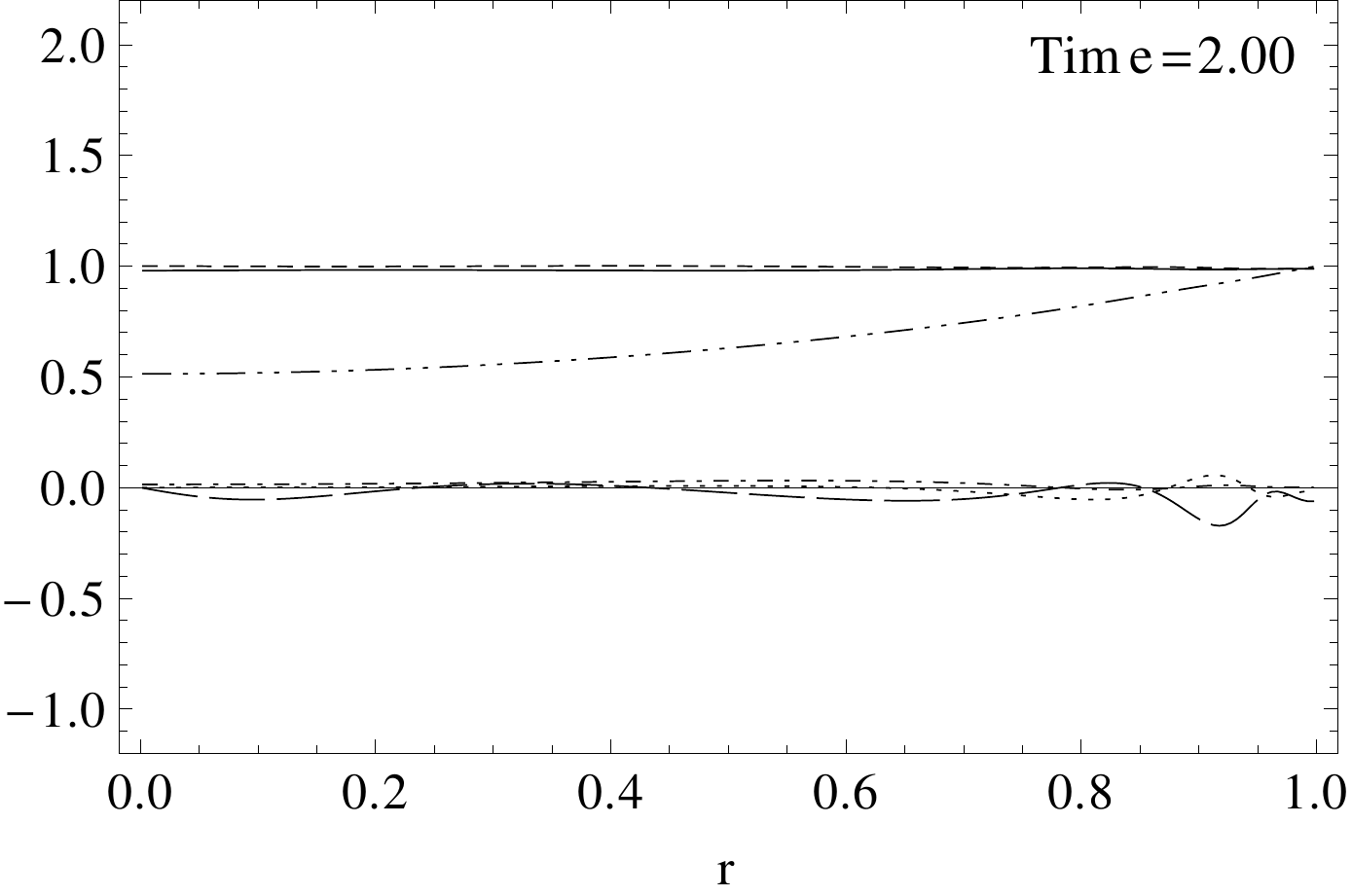}}\\
\vspace{-5.5ex} \mbox{\includegraphics[width=1\linewidth]{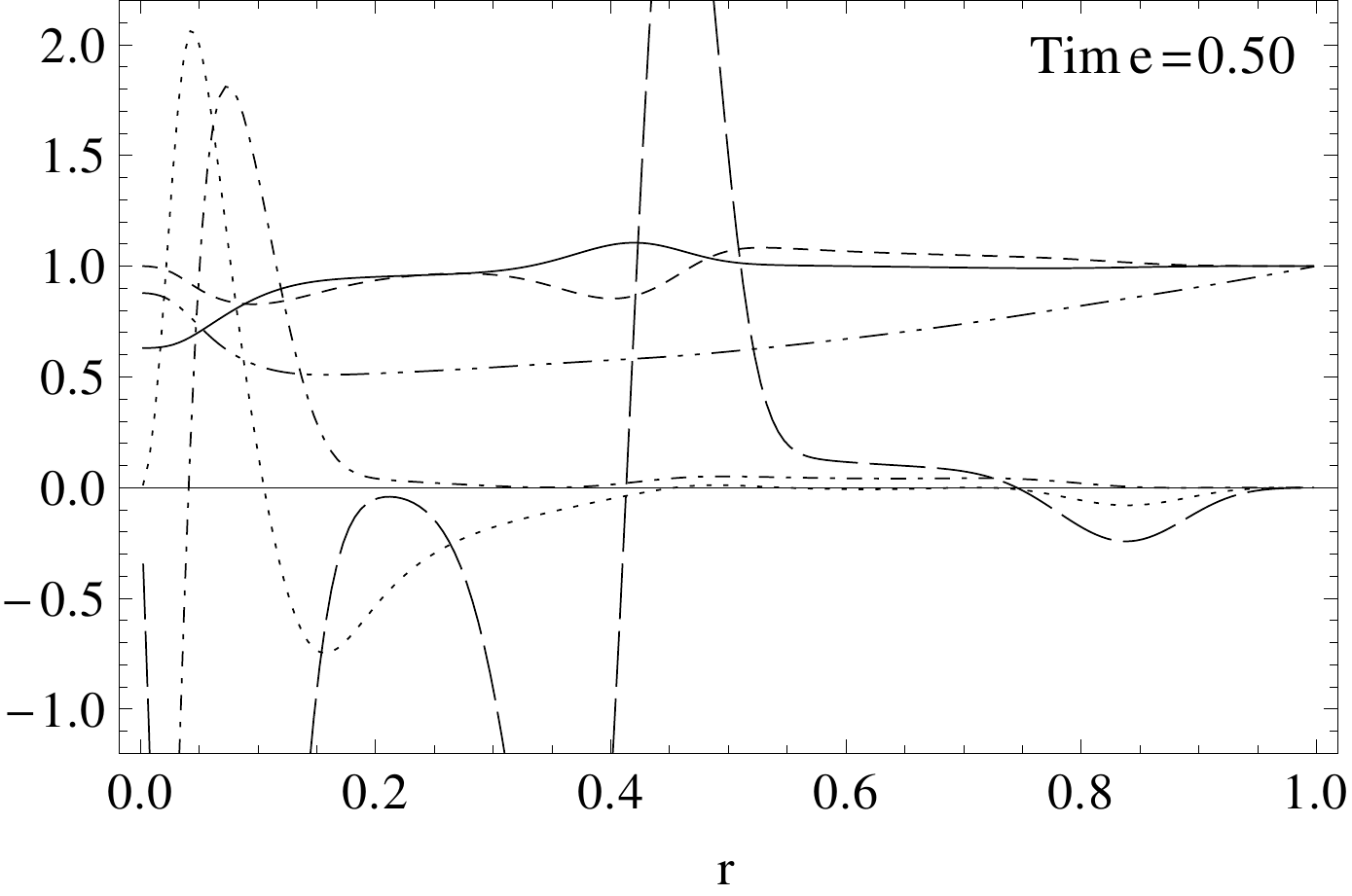}}&
\vspace{-5.5ex} \hspace{-0.8ex} \mbox{\includegraphics[width=1\linewidth]{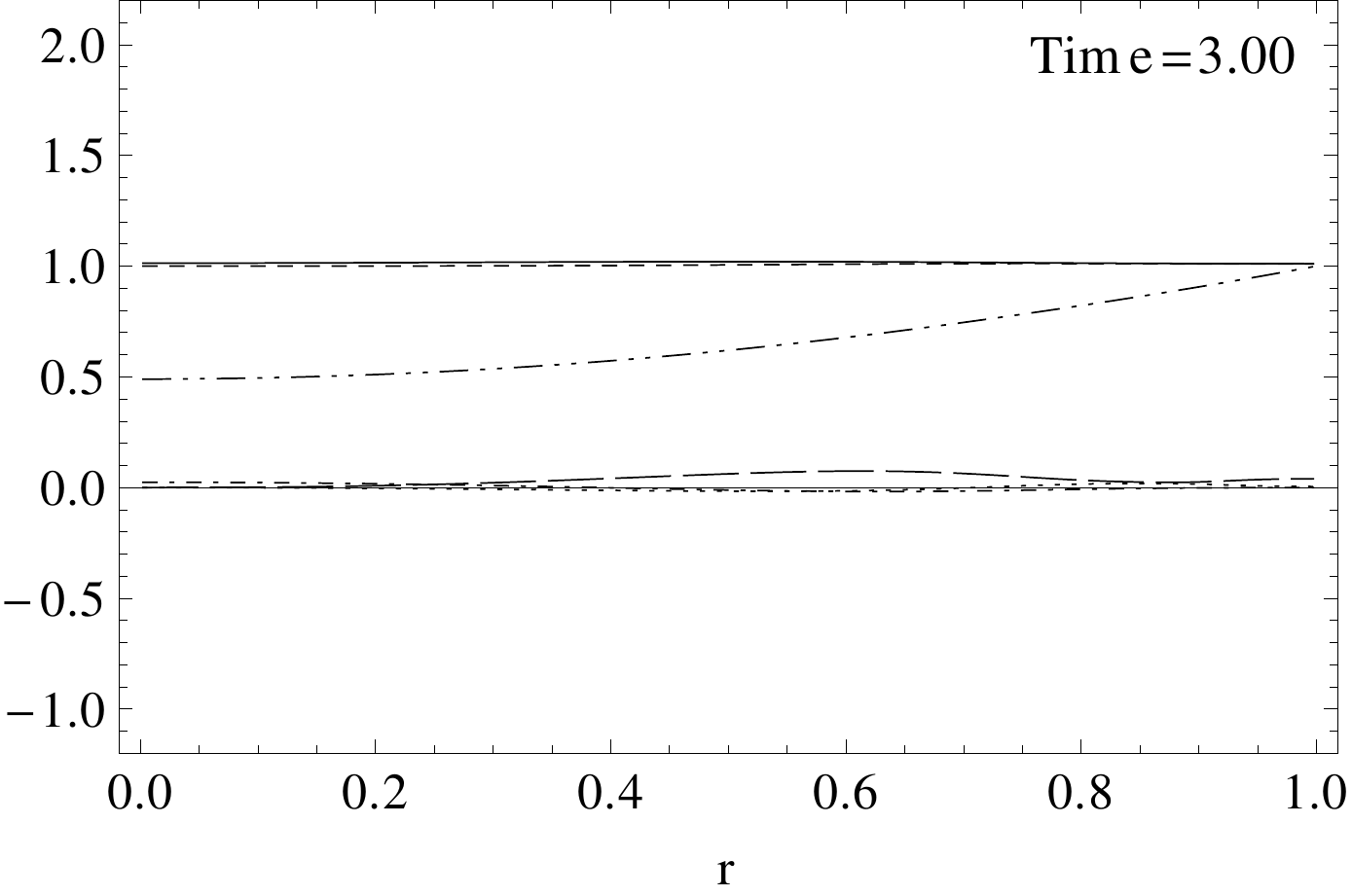}}\\
\vspace{-5.5ex} \mbox{\includegraphics[width=1\linewidth]{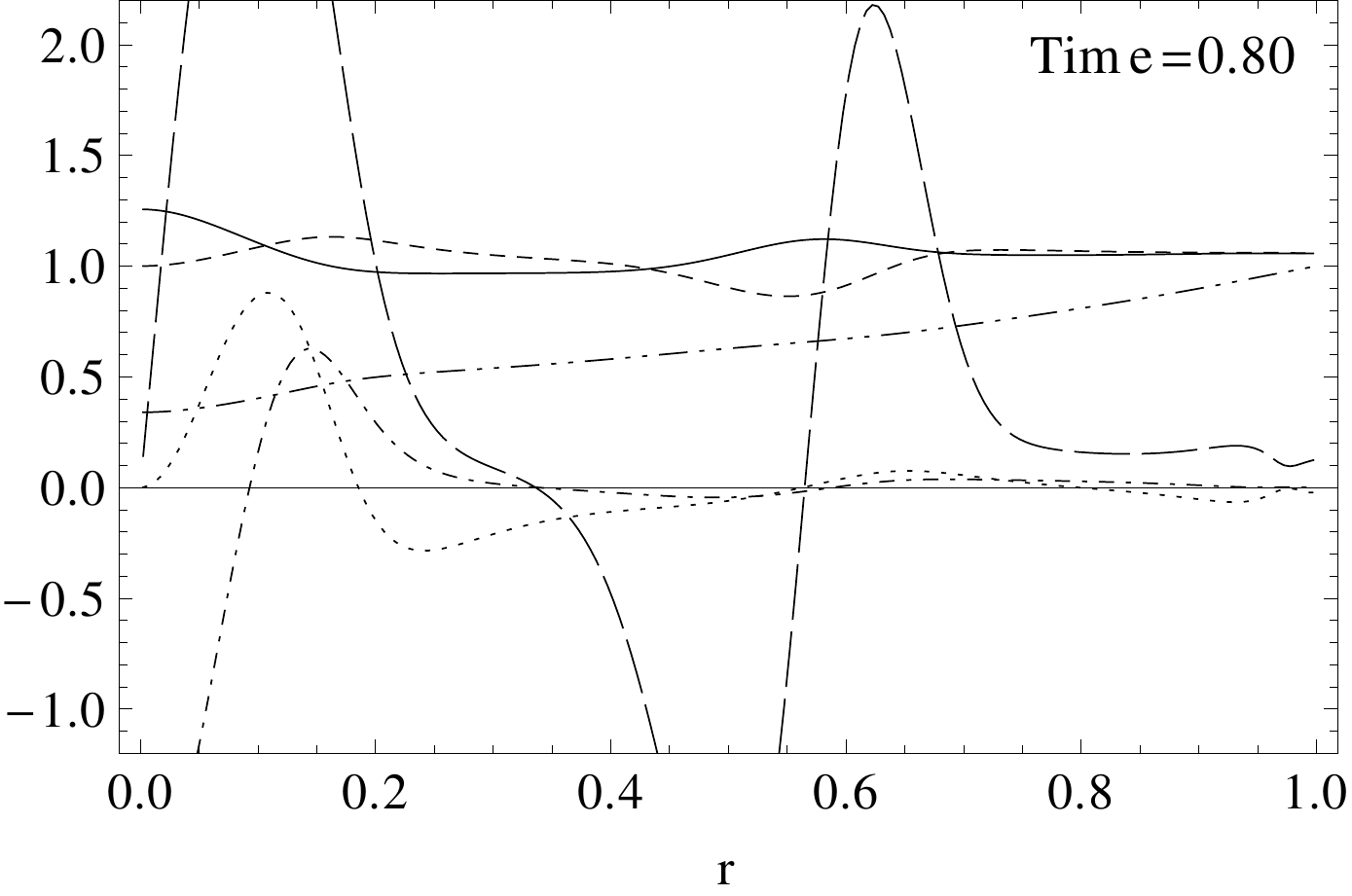}}&
\vspace{-5.5ex} \hspace{-0.8ex} \mbox{\includegraphics[width=1\linewidth]{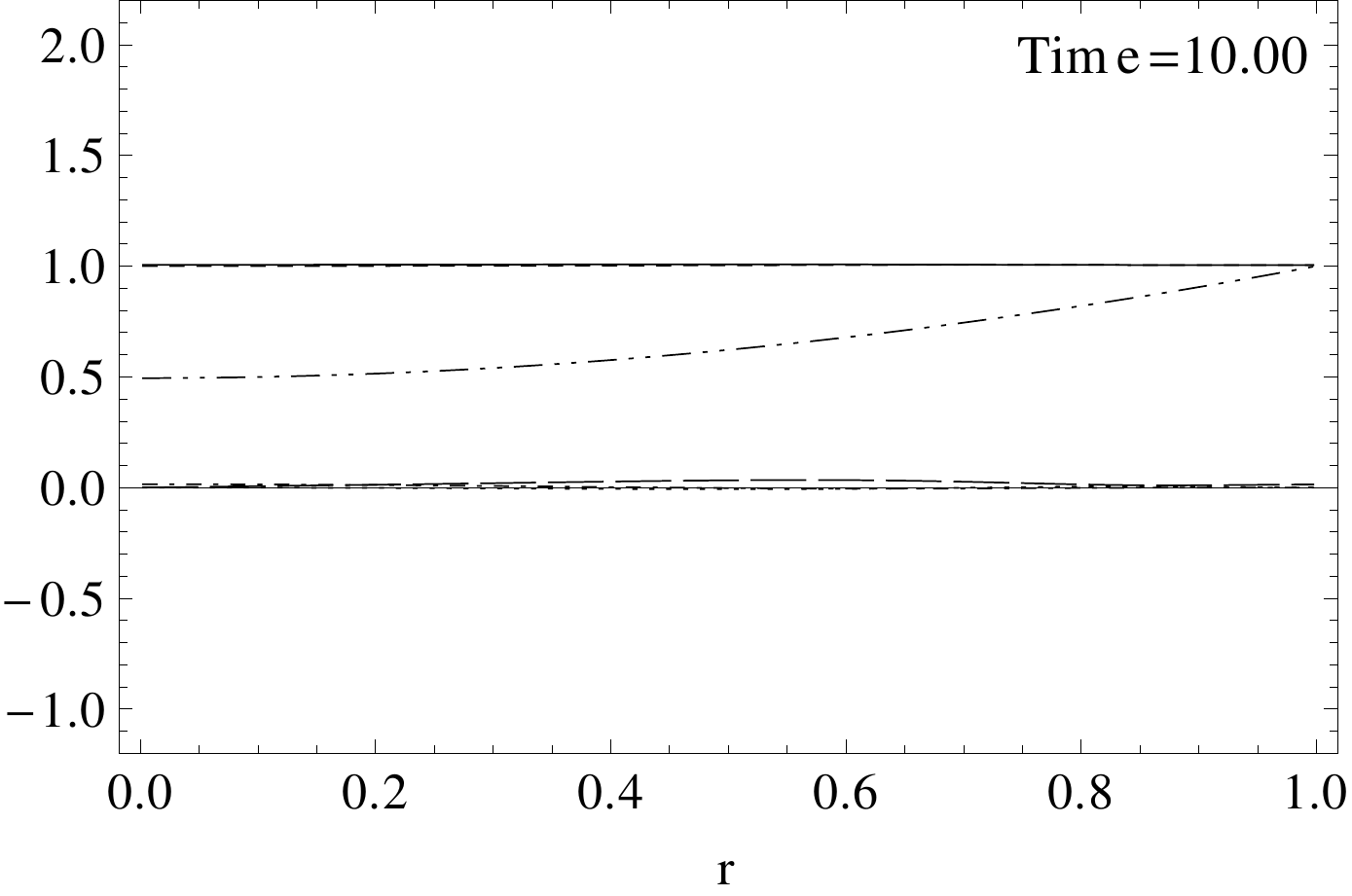}}
\end{tabular}
\vspace{-2ex}
\caption{Evolution of the variables (gauge waves) in flat spacetime.}
\label{fs:allevolgauge}
\end{figure}
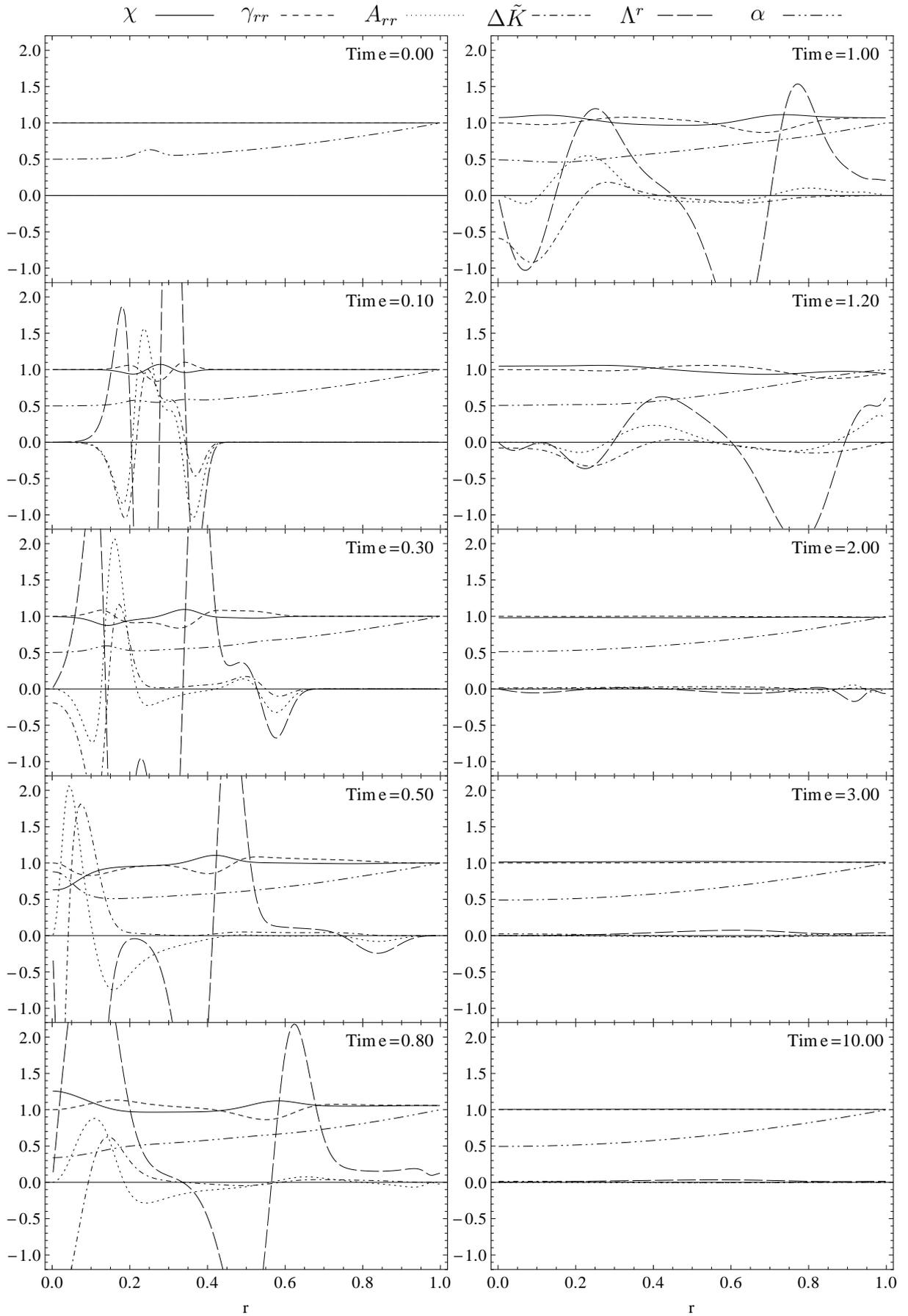

The simulation shown in \fref{fs:allevolgauge} was performed with 200 gridpoints and $\Delta t=0.001$, using initial perturbation parameters $A_\alpha=0.1$, $\sigma=0.1$ and $c=0.25$ and the parameter choice $\Kc=-3$. The shift is fixed and the GBSSN equations with the harmonic slicing condition \eref{ee:finalpha} are used, so all of the evolved quantities are plotted. This is actually the smallest system of the complete Einstein equations that can be evolved. The state of the system is shown at 10 different times. The initial perturbation of the lapse affects the rest of the variables; it splits into an ingoing part and an outgoing part. The latter moves towards $\scri^+$ and leaves the domain. The ingoing perturbation is first reflected at the origin and then the reflection propagates to the right and leaves the domain through $\scri^+$. Due to the fact that the shift is fixed and a regularity condition forces $\atscrip{\alpha}=-\case{\Kc\rscri}{3}=1$, the conditions \eref{er:simpliregs} must hold and are indeed satisfied by the numerical data. The quantities $A_{rr}$ and $\Lambda^r$ are however allowed to take non-vanishing values at $\scri^+$.
This is the same simulation as shown in figure 3 in \cite{Vano-Vinuales:2014koa}.

% Gauge waves

\subsection{Scalar field perturbation on regular initial data}

The simulation shown in figures \ref{fs:allevolscacent1} and \ref{fs:allevolscacent2} uses the same evolution equations as the one in \fref{fs:allevolgauge}, but here they are coupled to the scalar field in its rescaled form ($\bPhi$ and $\bPi$). The initial time-symmetric perturbation on the physical scalar field $\iPhi$ has parameters $A_\Phi=0.058$, $\sigma=0.1$ and $c=0.25$. It corresponds to a total Misner-Sharp mass of $M_{MS}=0.1057$ (see \fref{es:msmflat}). This is actually a quite large initial perturbation, as the same setup with $A_\Phi=0.060$ ($M_{MS}=0.1132$, only a 7.1\% increase in the total mass) is enough to form a BH at the origin. % (M_{0.060}/M_{0.058}-1)*100
The effect of the initial perturbation on the conformal factor $\chi$ can be seen in the top-left plot of \fref{fs:allevolscacent1}; this was already shown in \fref{fin:Finic}.

The initial pulse in $\bPhi$ splits into two approximately equal parts - this is an effect of the initial data being time symmetric. The outgoing one propagates towards $\scri^+$ and leaves the domain and the ingoing one is reflected at $r=0$ and then continues until it crosses $\scri^+$. Note that the relaxation of the variables evolved in the Einstein equations to their final stationary values continues after the scalar field perturbation has already left the domain.

\begin{figure}[htbp!!]
\center
	\includegraphics[width=0.7\linewidth]{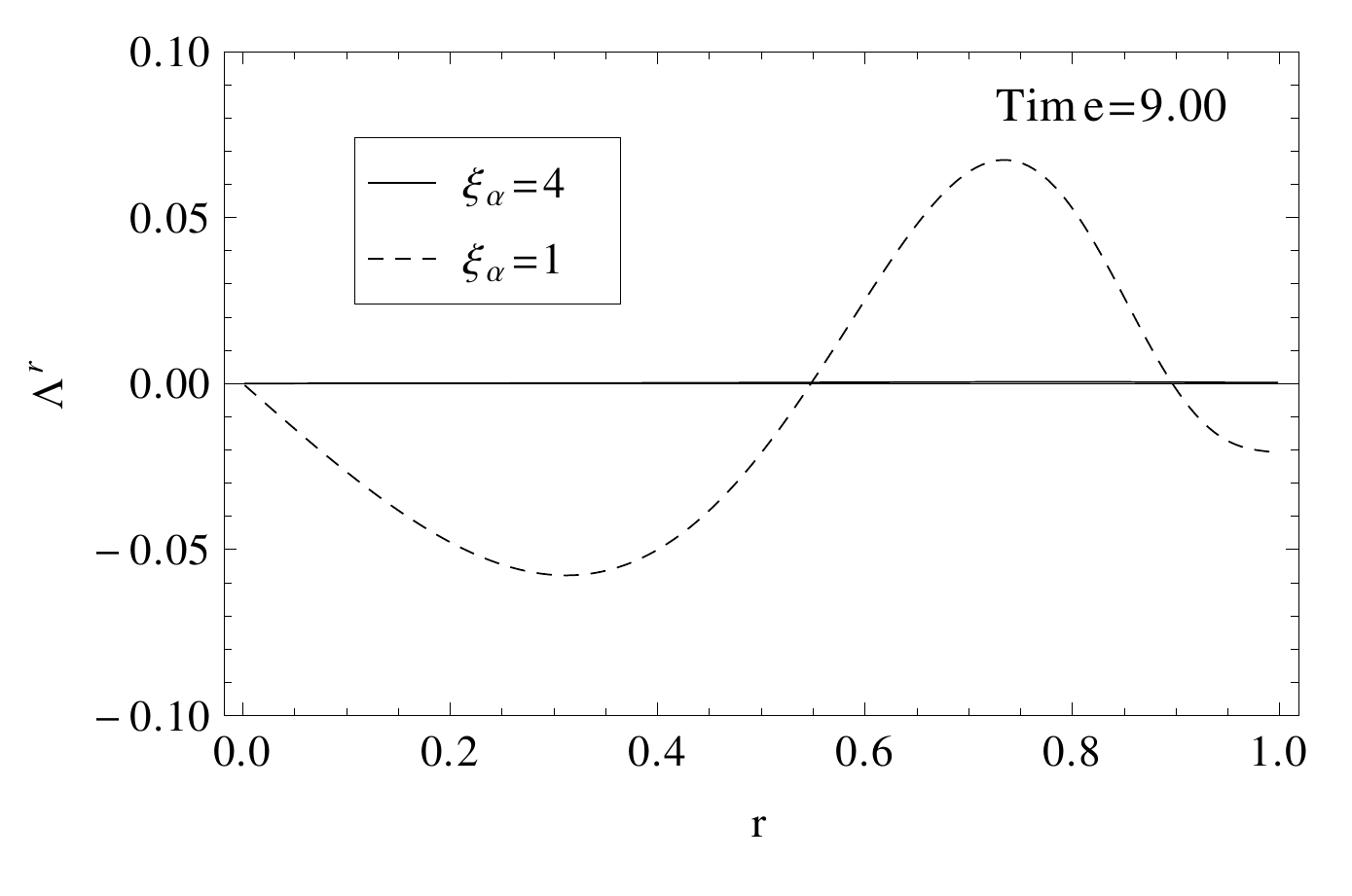}
\caption{The effect of the parameter $\xi_\alpha$ on the behaviour of $\Lambda^r$ for the simulation shown in figures \ref{fs:allevolscacent1} and \ref{fs:allevolscacent2}: the larger $\xi_\alpha$, the faster the metric quantities reach their stationary values.}\label{fs:scaLambda}
\end{figure}

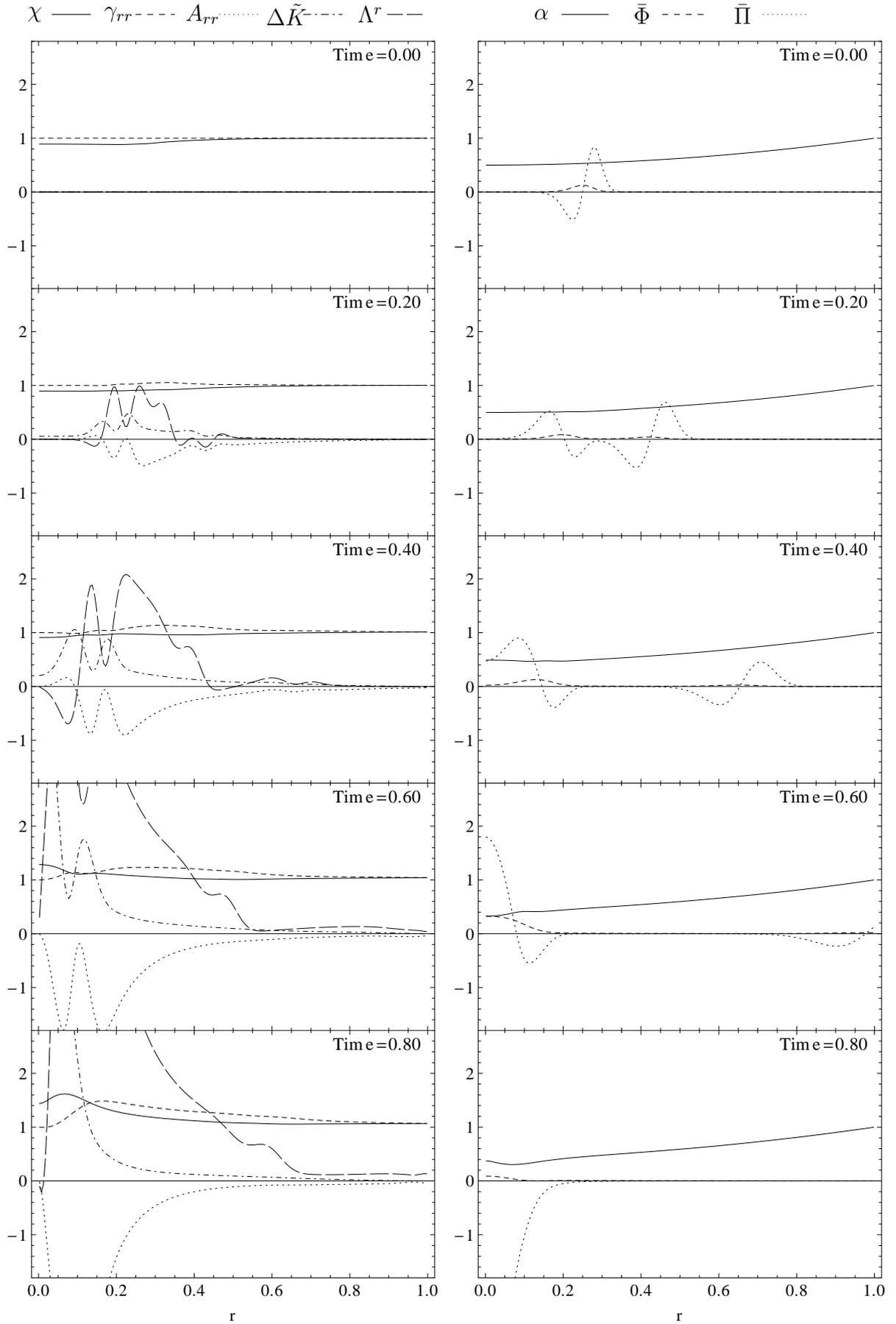
\begin{figure}[htbp!!]
\center
\vspace{-2ex}
\begin{tabular}{ m{0.5\linewidth}@{} @{}m{0.5\linewidth}@{} }
\begin{tikzpicture}[scale=1.0] \draw (-0.5cm,0cm) node {};
		\draw (0cm, 0cm) node {$\chi$}; \draw (0.3cm, 0cm) -- (1cm, 0cm);
		\draw (1.5cm, 0cm) node {$\gamma_{rr}$}; \draw [dashed] (1.8cm, 0cm) -- (2.5cm, 0cm);
		\draw (3cm, 0cm) node {$A_{rr}$}; \draw [dotted] (3.3cm, 0cm) -- (4cm, 0cm);
		\draw (4.5cm, 0cm) node {$\DPK$}; \draw [dash pattern= on 4pt off 2pt on 1pt off 2pt] (4.8cm, 0cm) -- (5.5cm, 0cm);
		\draw (6cm, 0cm) node {$\Lambda^r$}; \draw [dash pattern= on 8pt off 2pt] (6.3cm, 0cm) -- (7cm, 0cm);
	\end{tikzpicture}
&
\begin{tikzpicture}[scale=1.2] \draw (-1.3cm,0cm) node {};
		\draw (0cm, 0cm) node {$\alpha$}; \draw (0.3cm, 0cm) -- (1cm, 0cm);
		\draw (1.5cm, 0cm) node {$\bPhi$}; \draw [dashed] (1.8cm, 0cm) -- (2.5cm, 0cm);
		\draw (3cm, 0cm) node {$\bPi$}; \draw [dotted] (3.3cm, 0cm) -- (4cm, 0cm);
	\end{tikzpicture}
\\
\mbox{\includegraphics[width=1\linewidth]{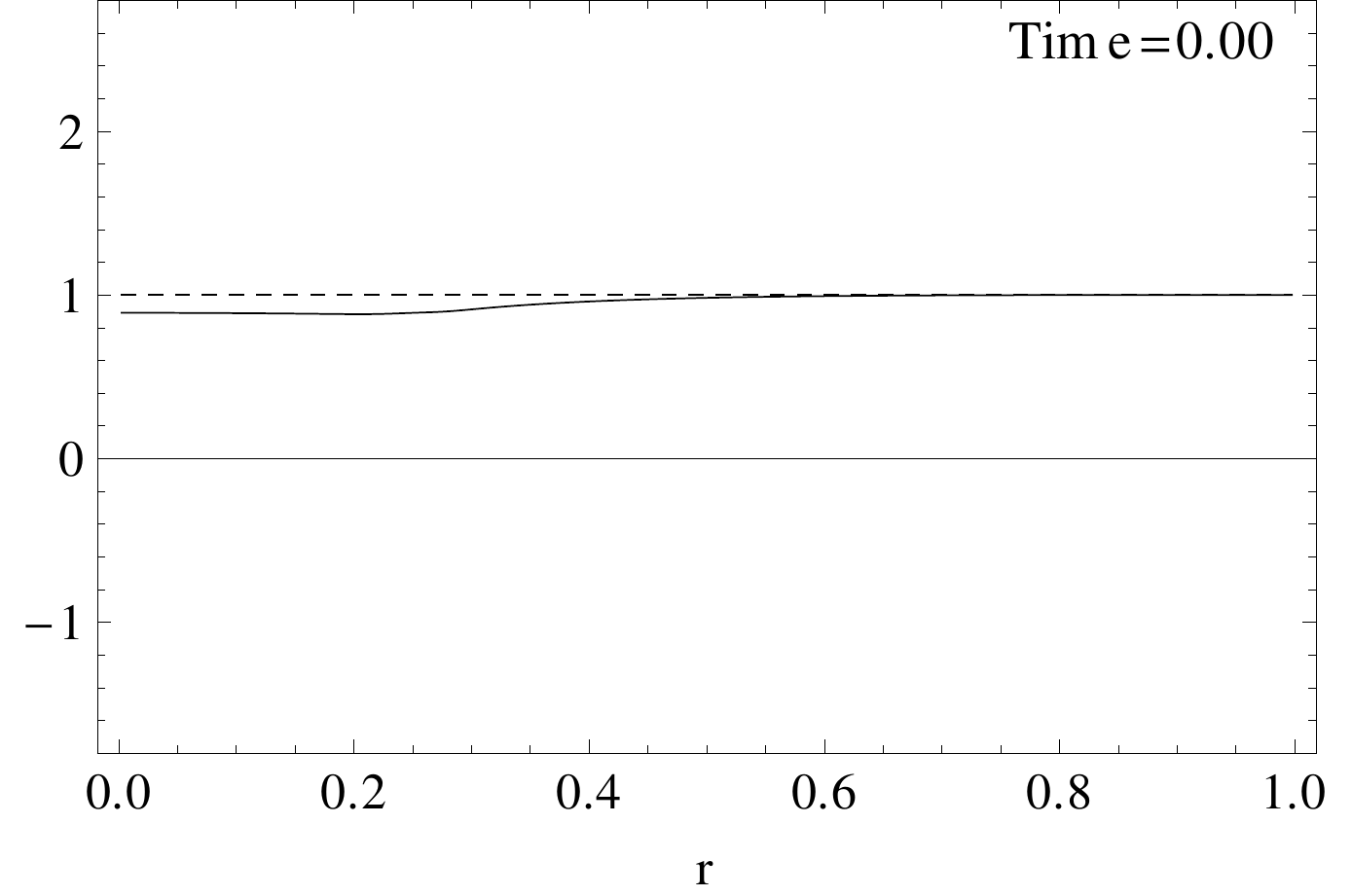}}&
\hspace{-0.8ex} \mbox{\includegraphics[width=1\linewidth]{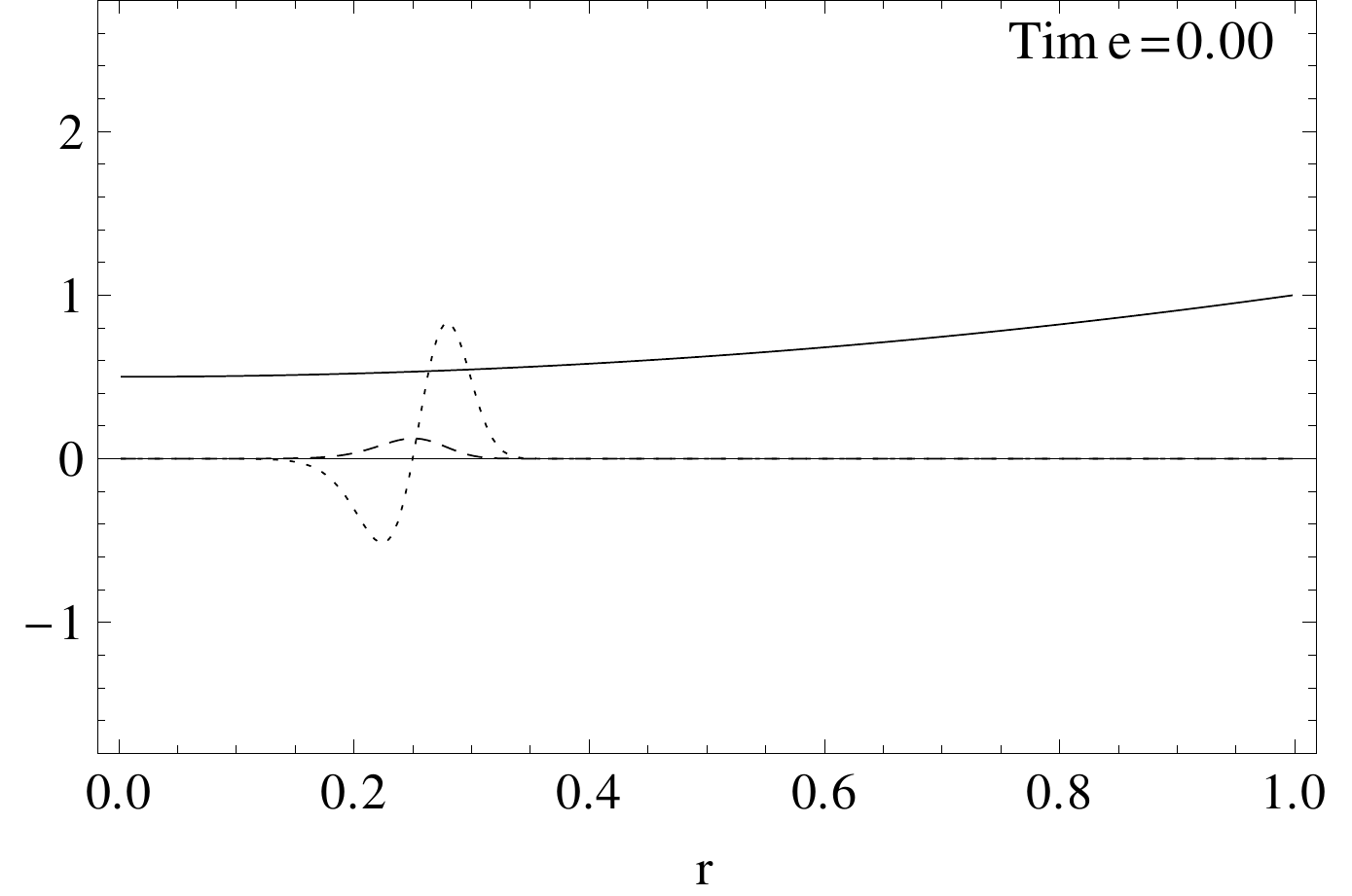}}\\
\vspace{-5.5ex} \mbox{\includegraphics[width=1\linewidth]{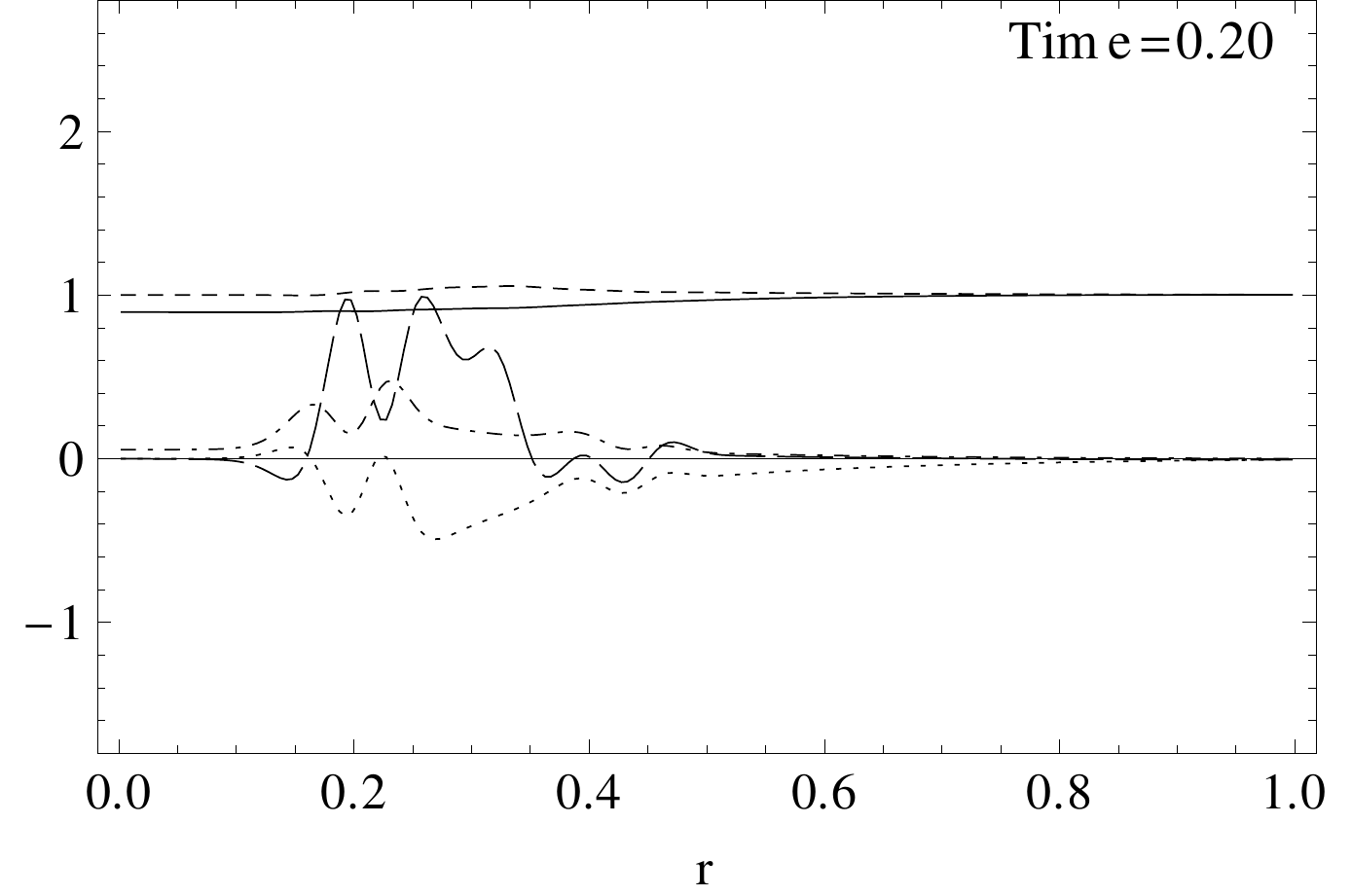}}&
\vspace{-5.5ex} \hspace{-0.8ex} \mbox{\includegraphics[width=1\linewidth]{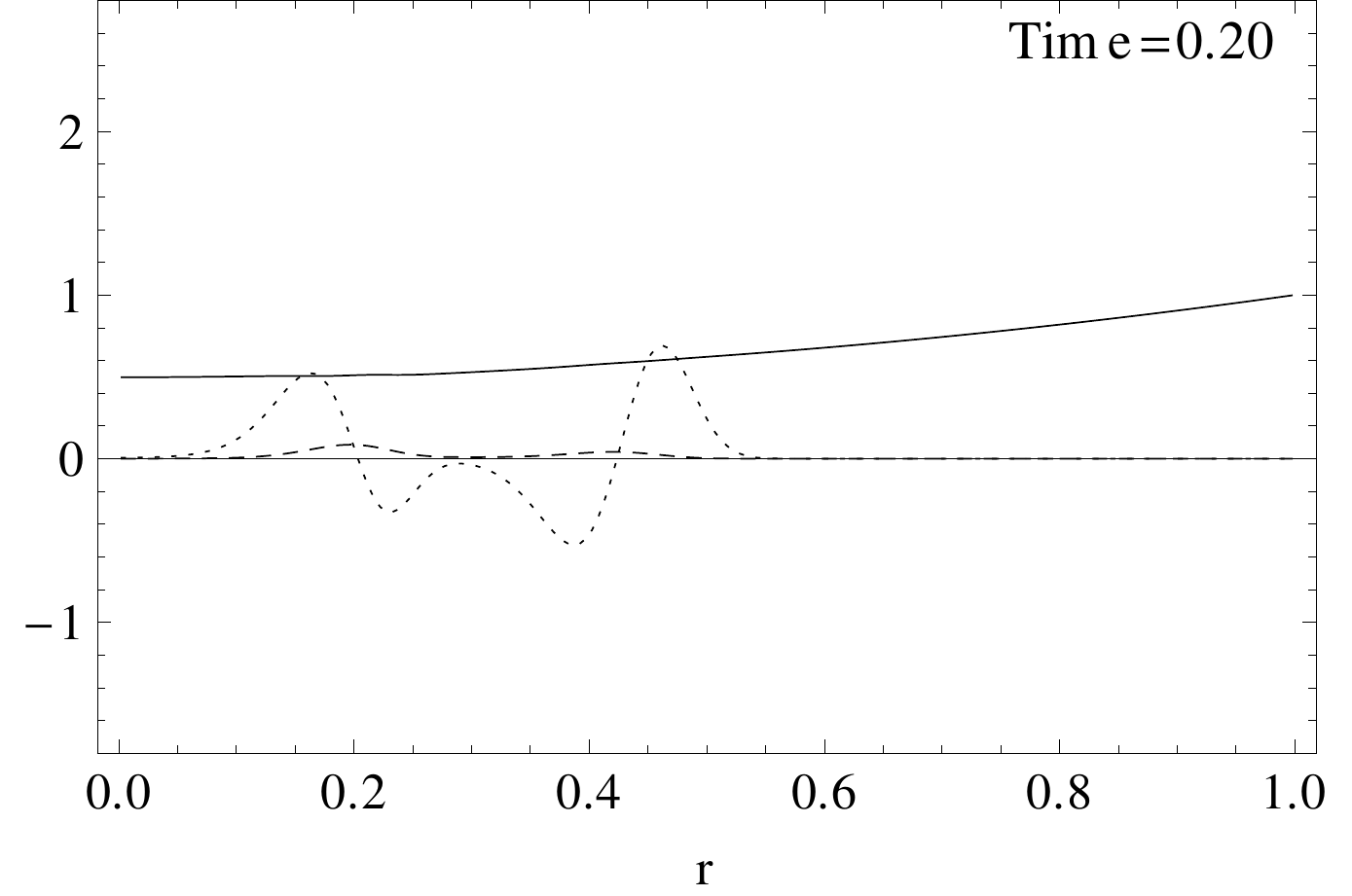}}\\
\vspace{-5.5ex} \mbox{\includegraphics[width=1\linewidth]{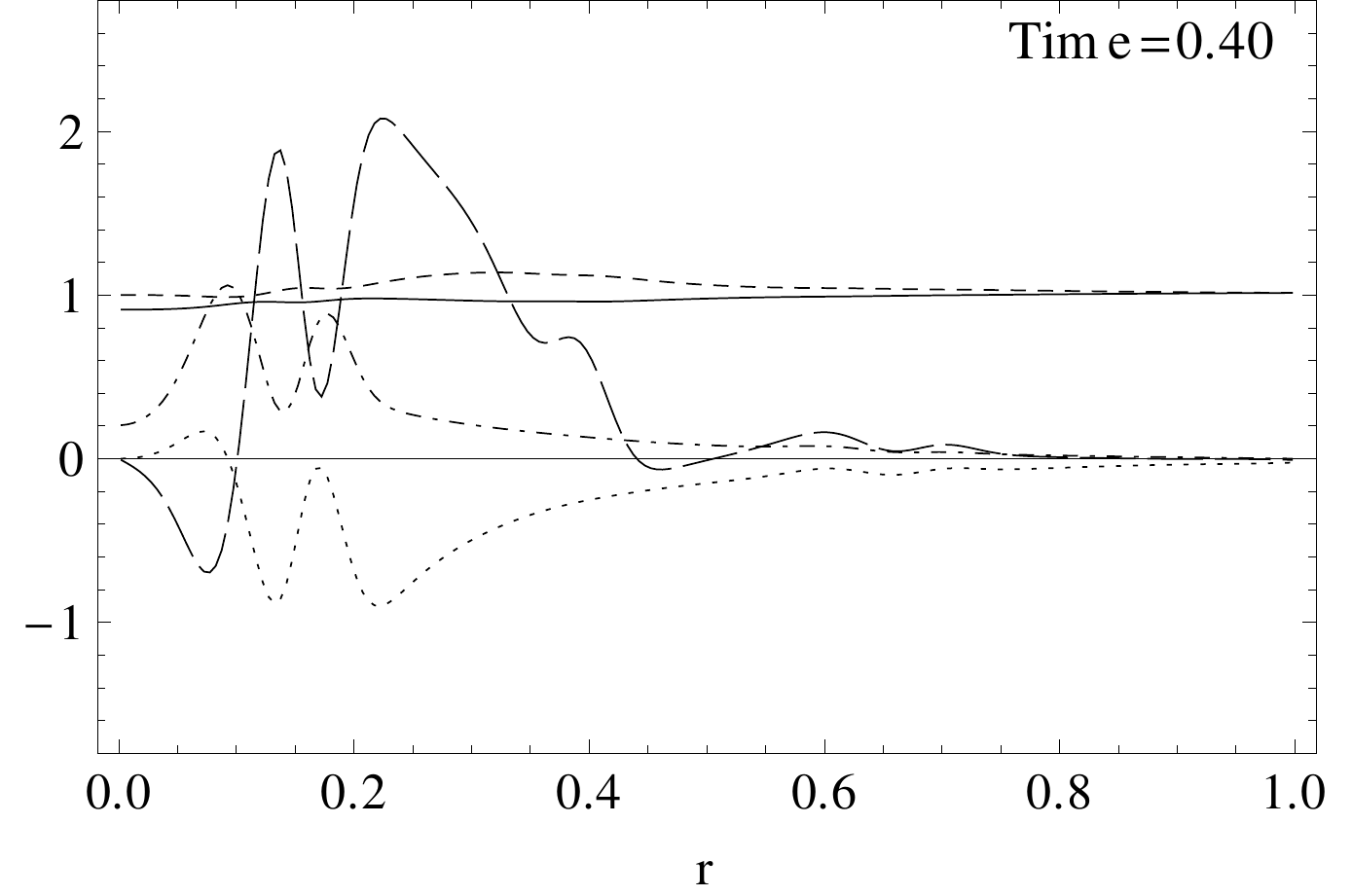}}&
\vspace{-5.5ex} \hspace{-0.8ex} \mbox{\includegraphics[width=1\linewidth]{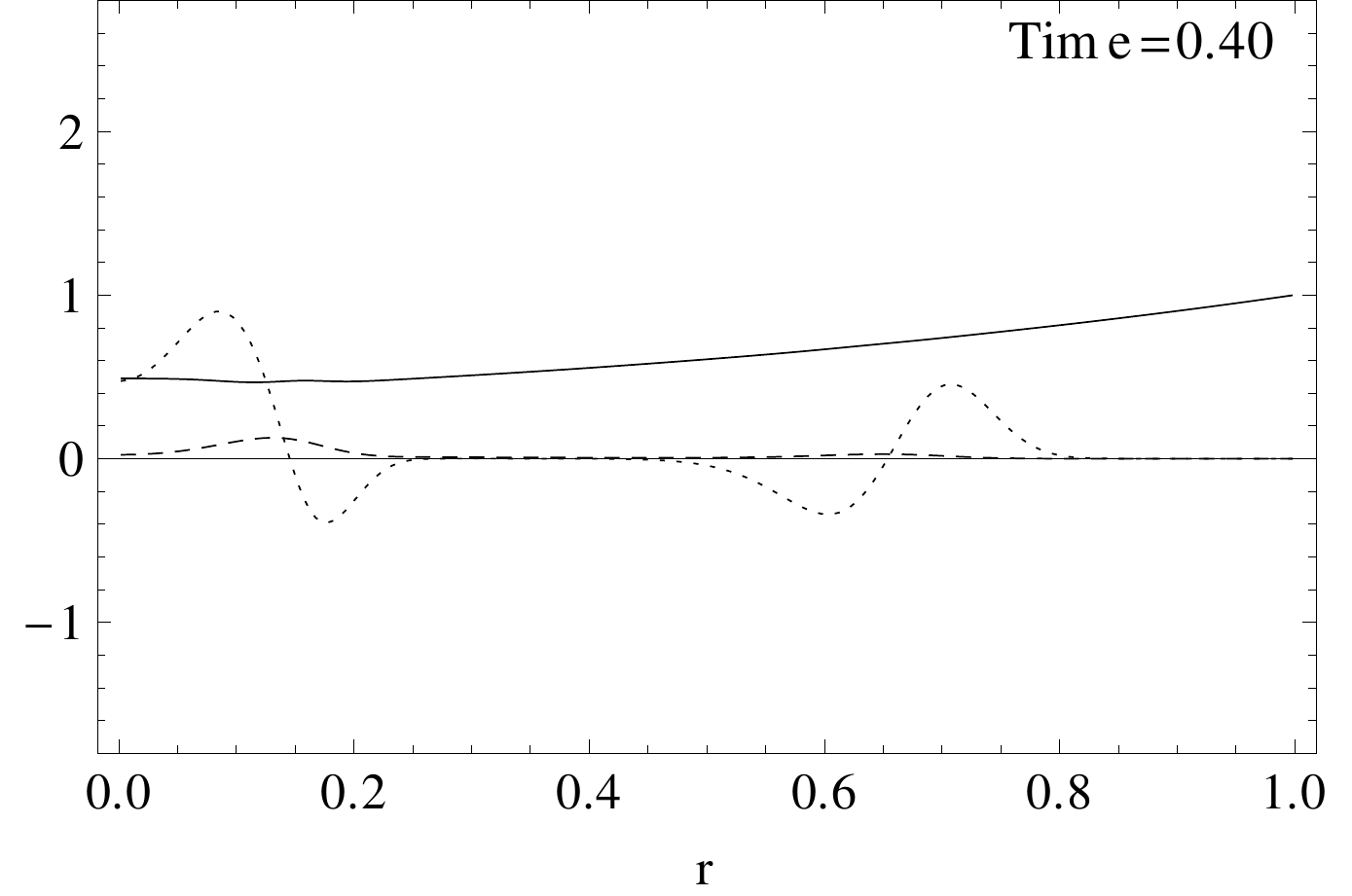}}\\
\vspace{-5.5ex} \mbox{\includegraphics[width=1\linewidth]{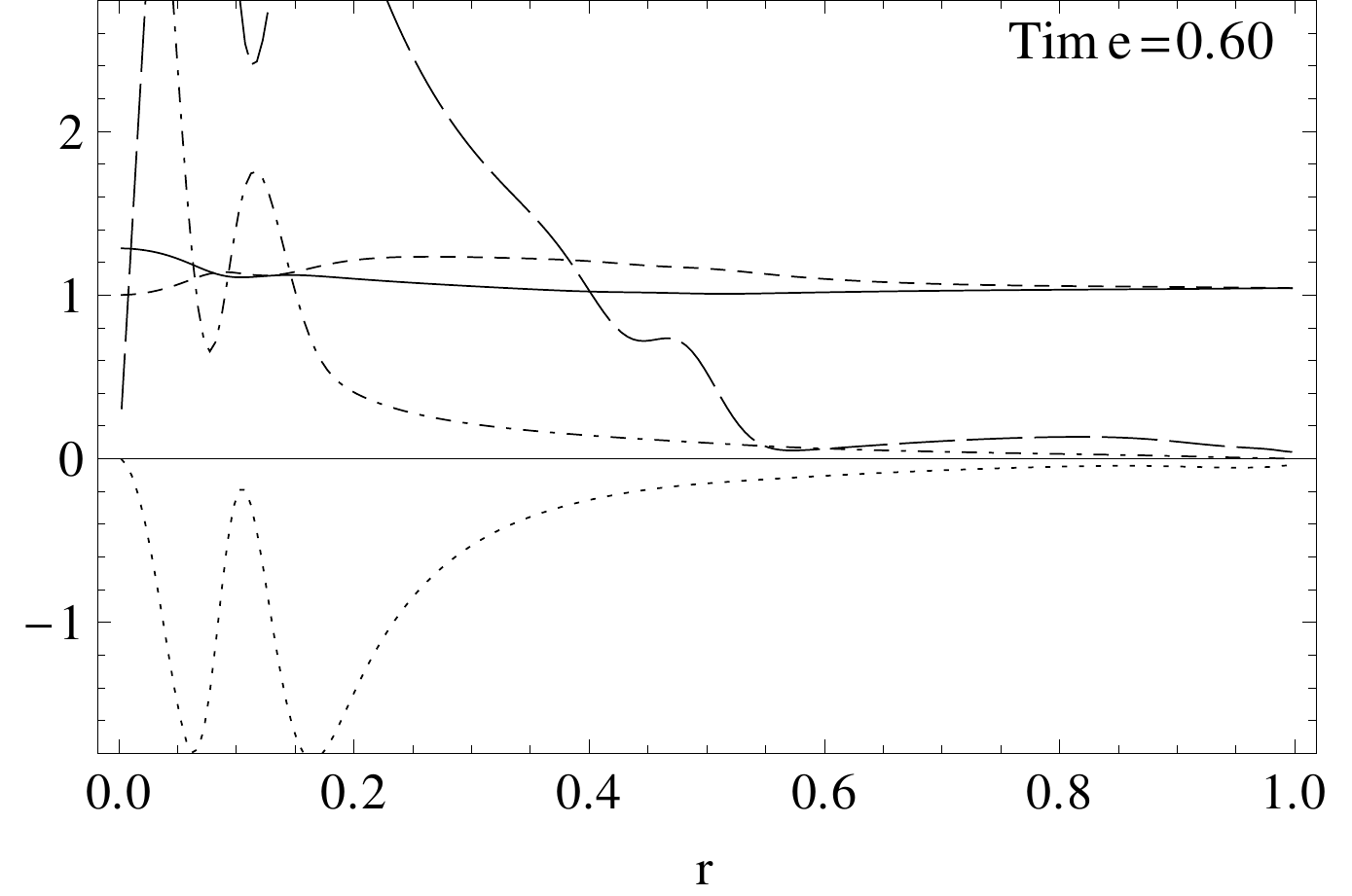}}&
\vspace{-5.5ex} \hspace{-0.8ex} \mbox{\includegraphics[width=1\linewidth]{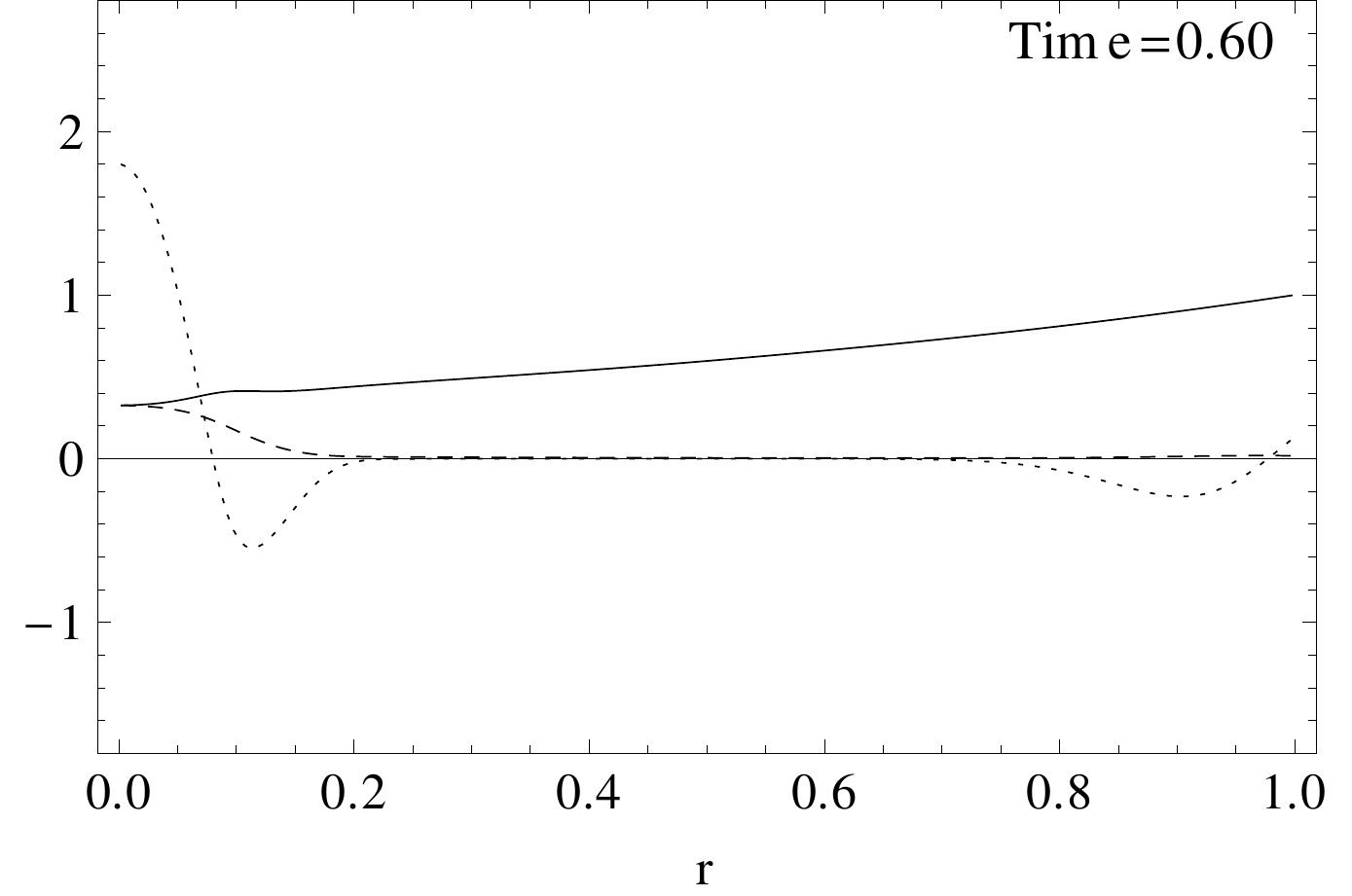}}\\
\vspace{-5.5ex} \mbox{\includegraphics[width=1\linewidth]{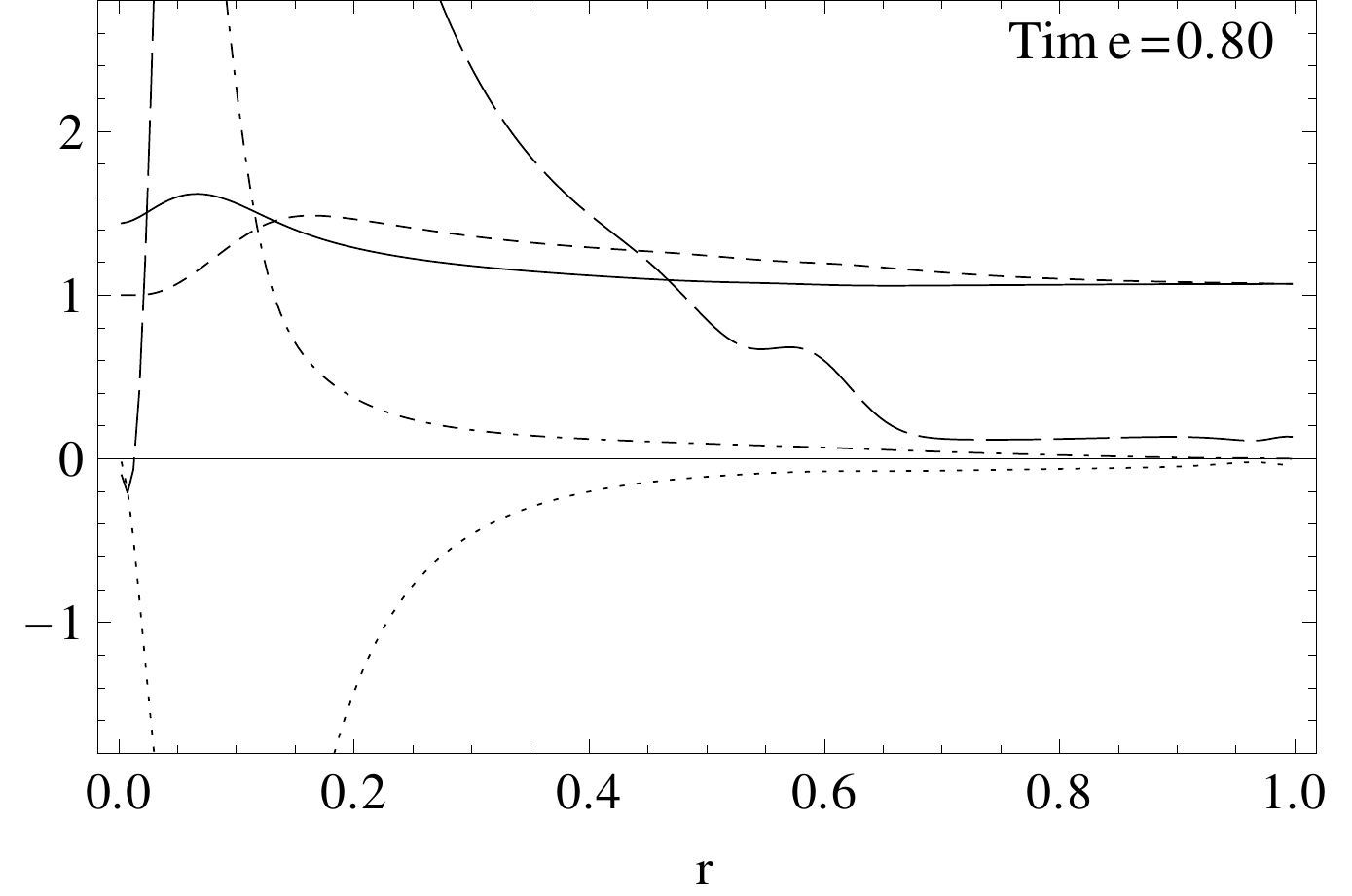}}&
\vspace{-5.5ex} \hspace{-0.8ex} \mbox{\includegraphics[width=1\linewidth]{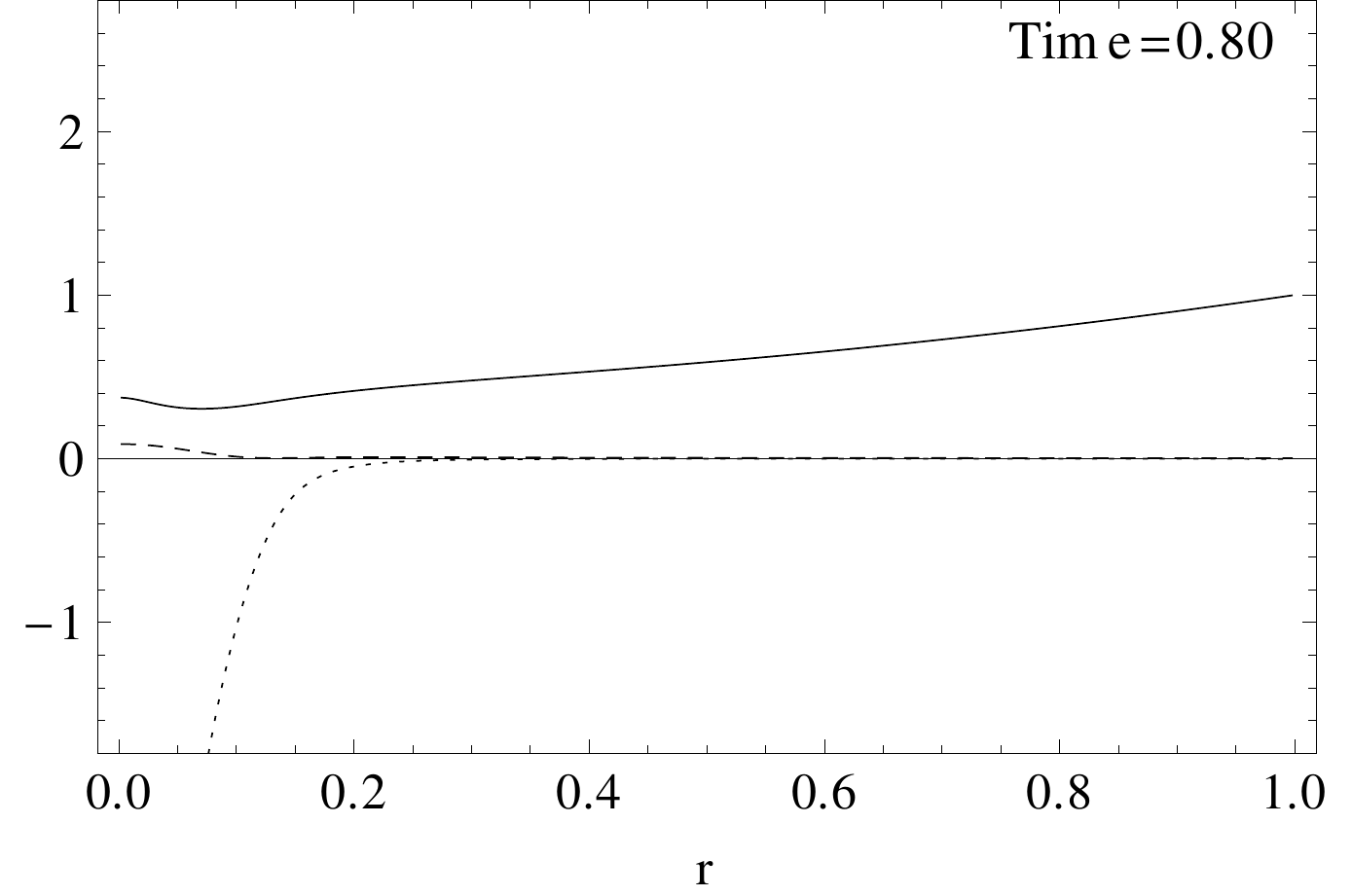}}
\end{tabular}
\vspace{-2ex}
\caption{Evolution of the variables (scalar field): $\chi$, $\gamma_{rr}$, $A_{rr}$, $\DPK$ and $\Lambda^r$ on the left and $\alpha$, $\bPhi={\iPhi}/{\Omega}$ and $\bPi={\iPi}/{\Omega}$ on the right.} %(defined as in \eref{esca:tildebarphi})
\label{fs:allevolscacent1}
\end{figure}

\begin{figure}[htbp!!]
\center
\vspace{-2ex}
\begin{tabular}{ m{0.5\linewidth}@{} @{}m{0.5\linewidth}@{} }
\begin{tikzpicture}[scale=1.0] \draw (-0.5cm,0cm) node {};
		\draw (0cm, 0cm) node {$\chi$}; \draw (0.3cm, 0cm) -- (1cm, 0cm);
		\draw (1.5cm, 0cm) node {$\gamma_{rr}$}; \draw [dashed] (1.8cm, 0cm) -- (2.5cm, 0cm);
		\draw (3cm, 0cm) node {$A_{rr}$}; \draw [dotted] (3.3cm, 0cm) -- (4cm, 0cm);
		\draw (4.5cm, 0cm) node {$\DPK$}; \draw [dash pattern= on 4pt off 2pt on 1pt off 2pt] (4.8cm, 0cm) -- (5.5cm, 0cm);
		\draw (6cm, 0cm) node {$\Lambda^r$}; \draw [dash pattern= on 8pt off 2pt] (6.3cm, 0cm) -- (7cm, 0cm);
	\end{tikzpicture}
&
\begin{tikzpicture}[scale=1.2] \draw (-1.3cm,0cm) node {};
		\draw (0cm, 0cm) node {$\alpha$}; \draw (0.3cm, 0cm) -- (1cm, 0cm);
		\draw (1.5cm, 0cm) node {$\bPhi$}; \draw [dashed] (1.8cm, 0cm) -- (2.5cm, 0cm);
		\draw (3cm, 0cm) node {$\bPi$}; \draw [dotted] (3.3cm, 0cm) -- (4cm, 0cm);
	\end{tikzpicture}
\\
\mbox{\includegraphics[width=1\linewidth]{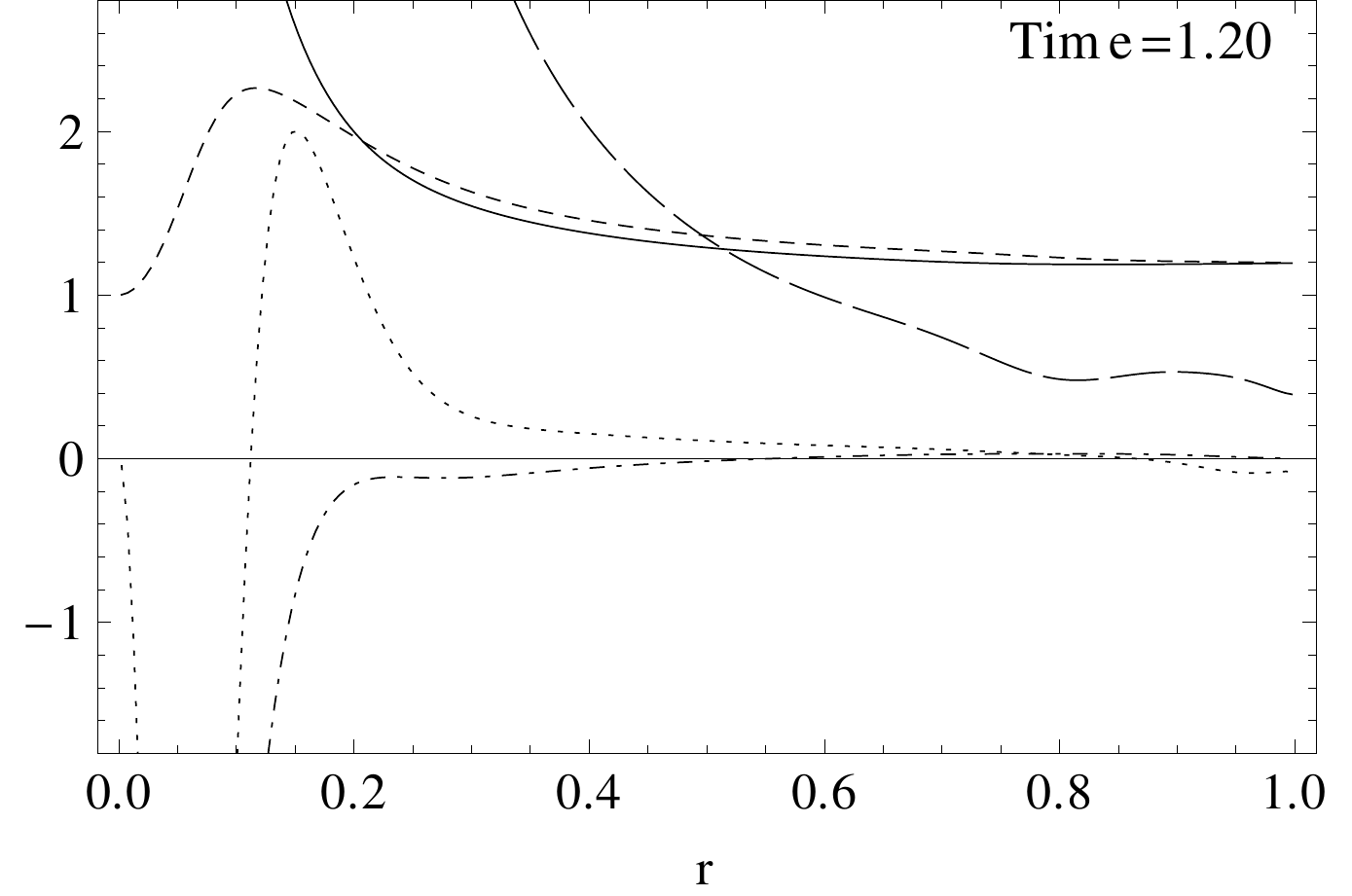}}&
\hspace{-0.8ex} \mbox{\includegraphics[width=1\linewidth]{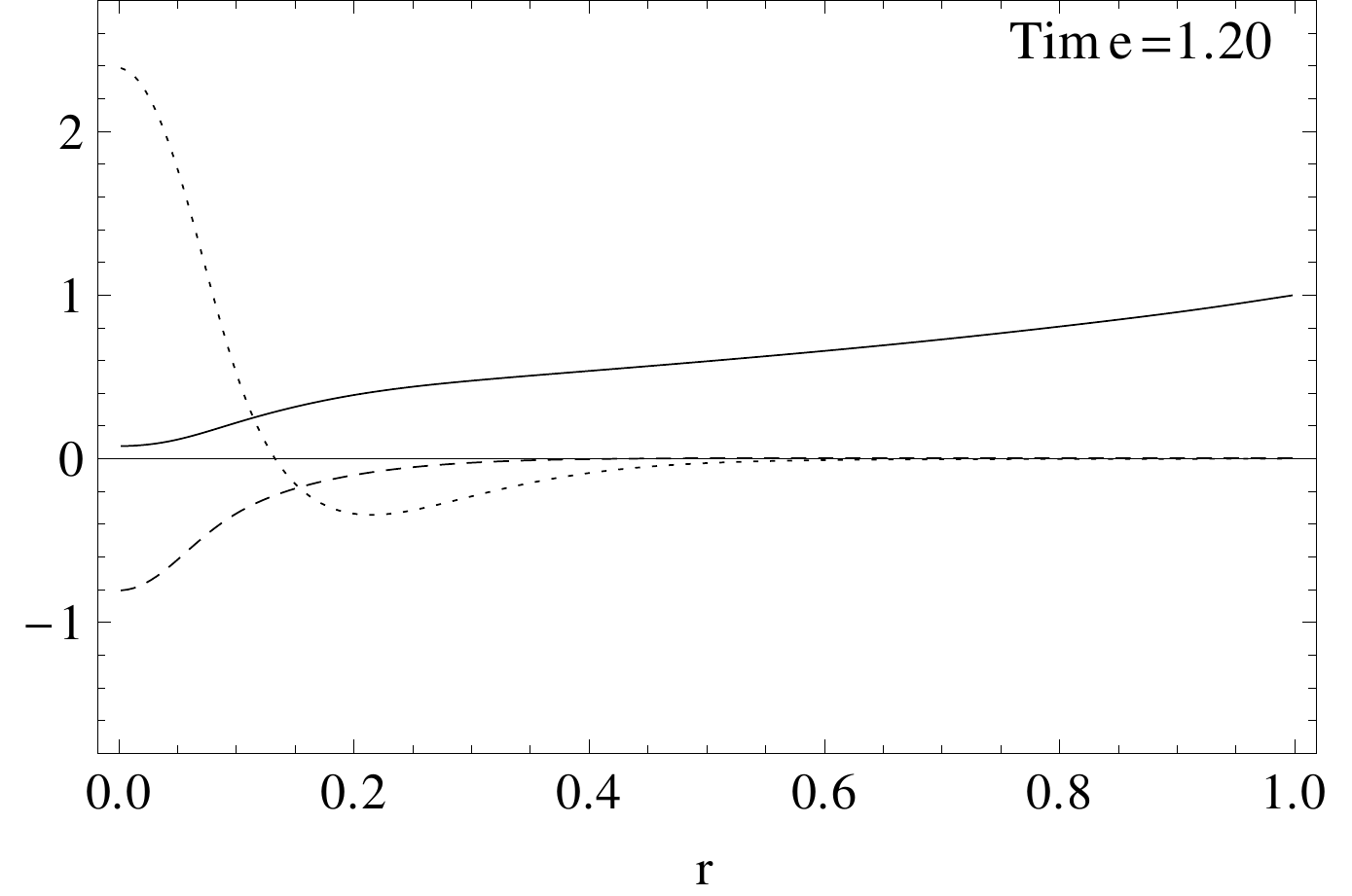}}\\
\vspace{-5.5ex} \mbox{\includegraphics[width=1\linewidth]{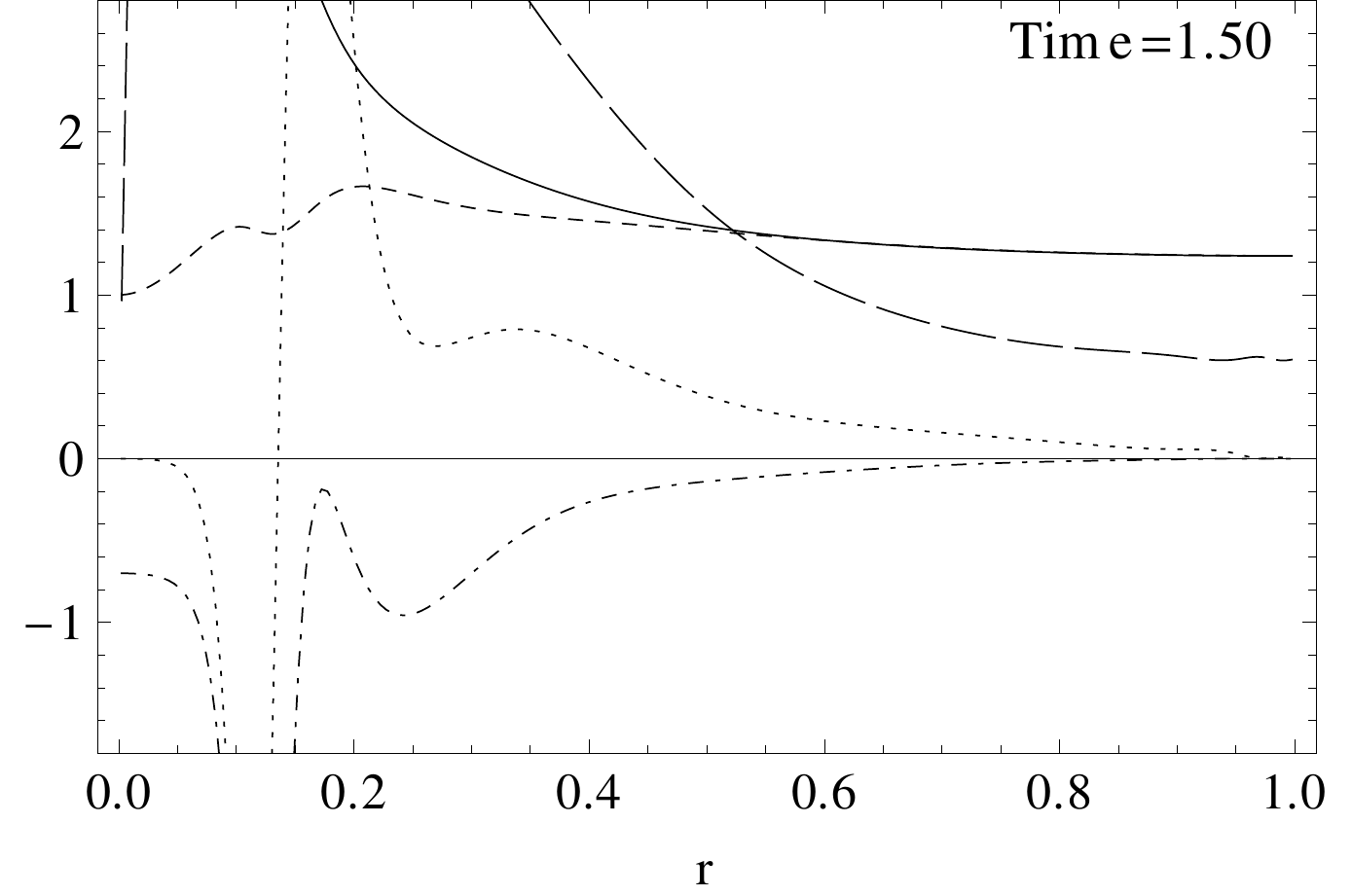}}&
\vspace{-5.5ex} \hspace{-0.8ex} \mbox{\includegraphics[width=1\linewidth]{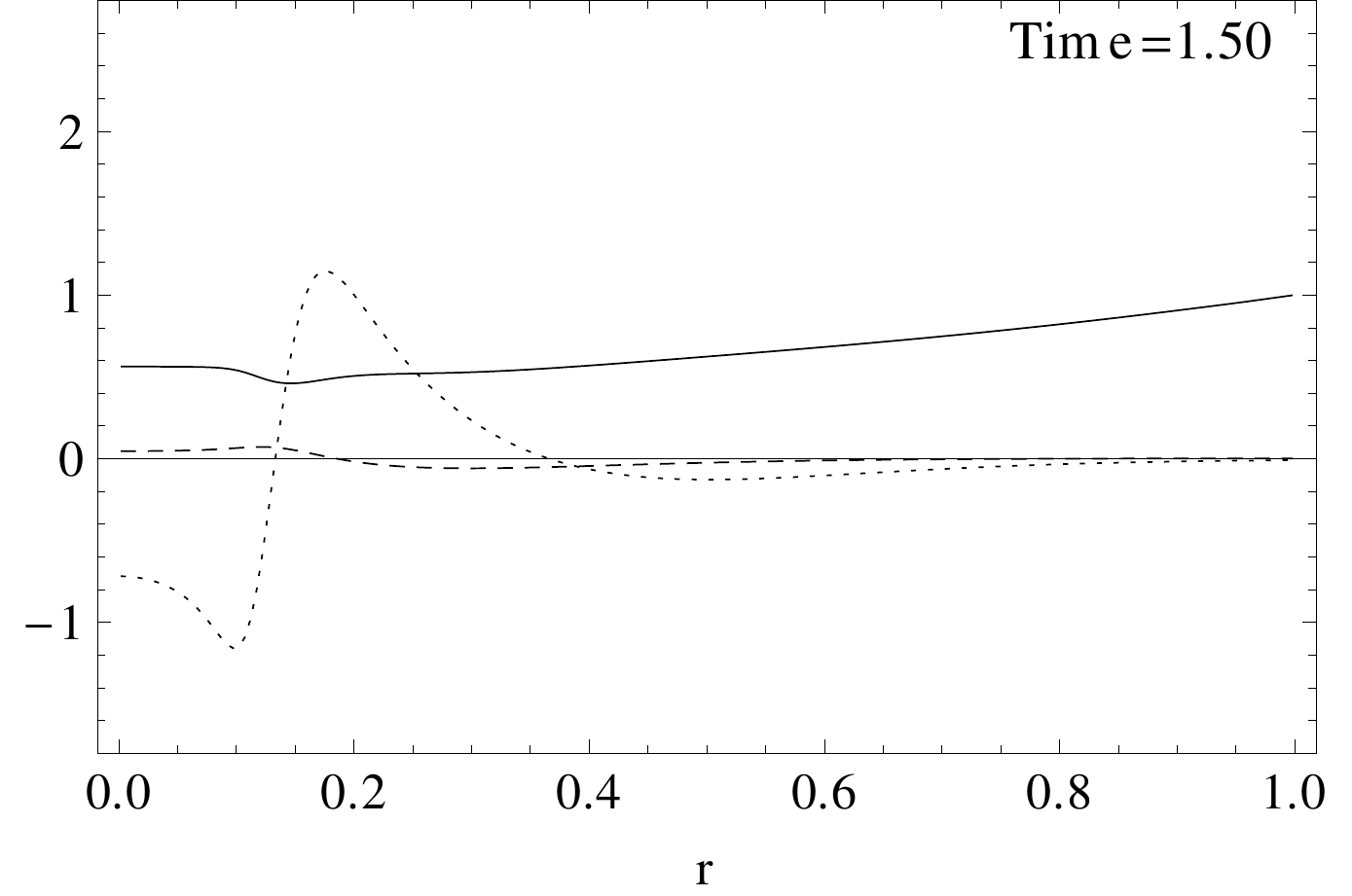}}\\
\vspace{-5.5ex} \mbox{\includegraphics[width=1\linewidth]{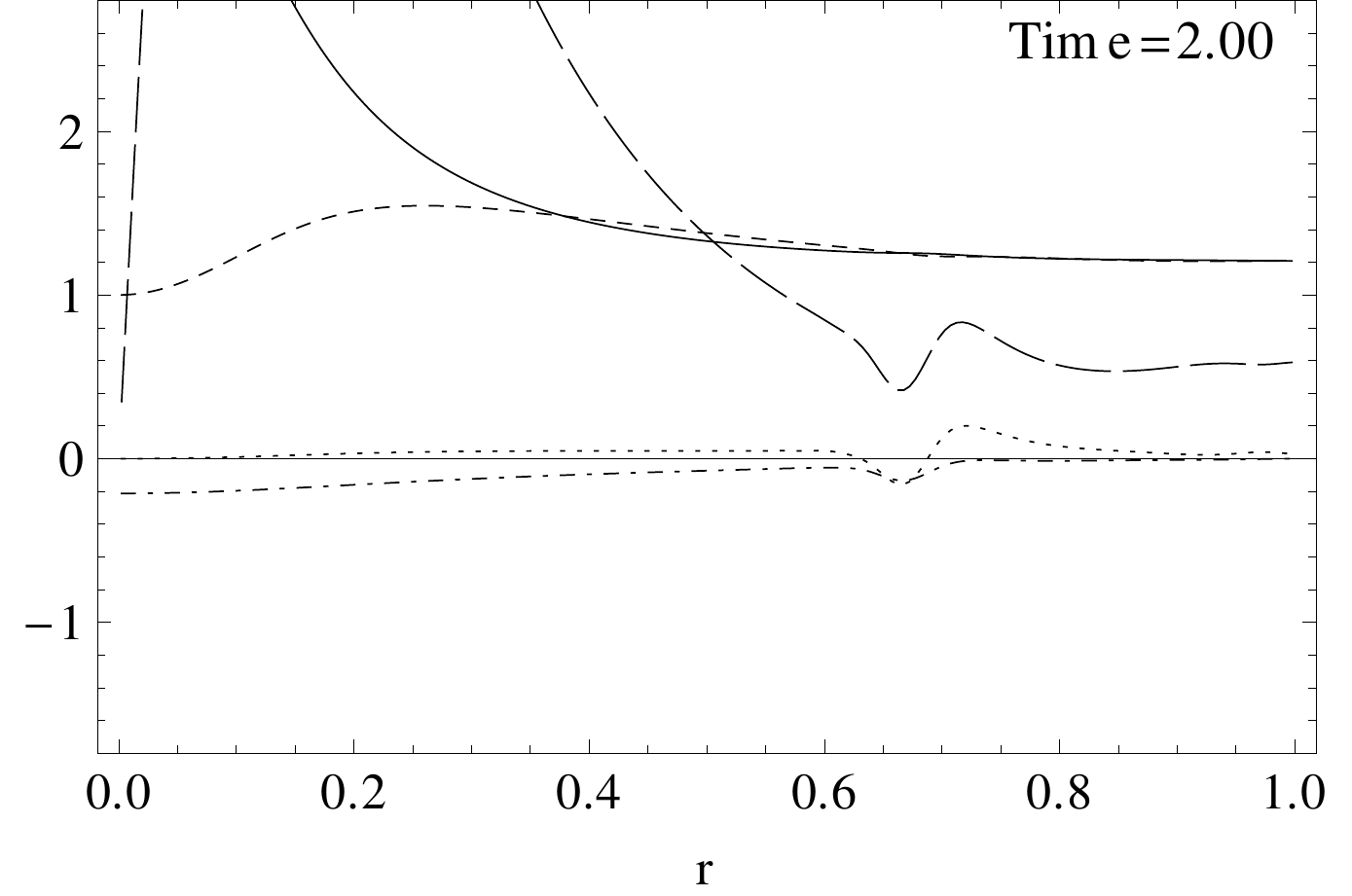}}&
\vspace{-5.5ex} \hspace{-0.8ex} \mbox{\includegraphics[width=1\linewidth]{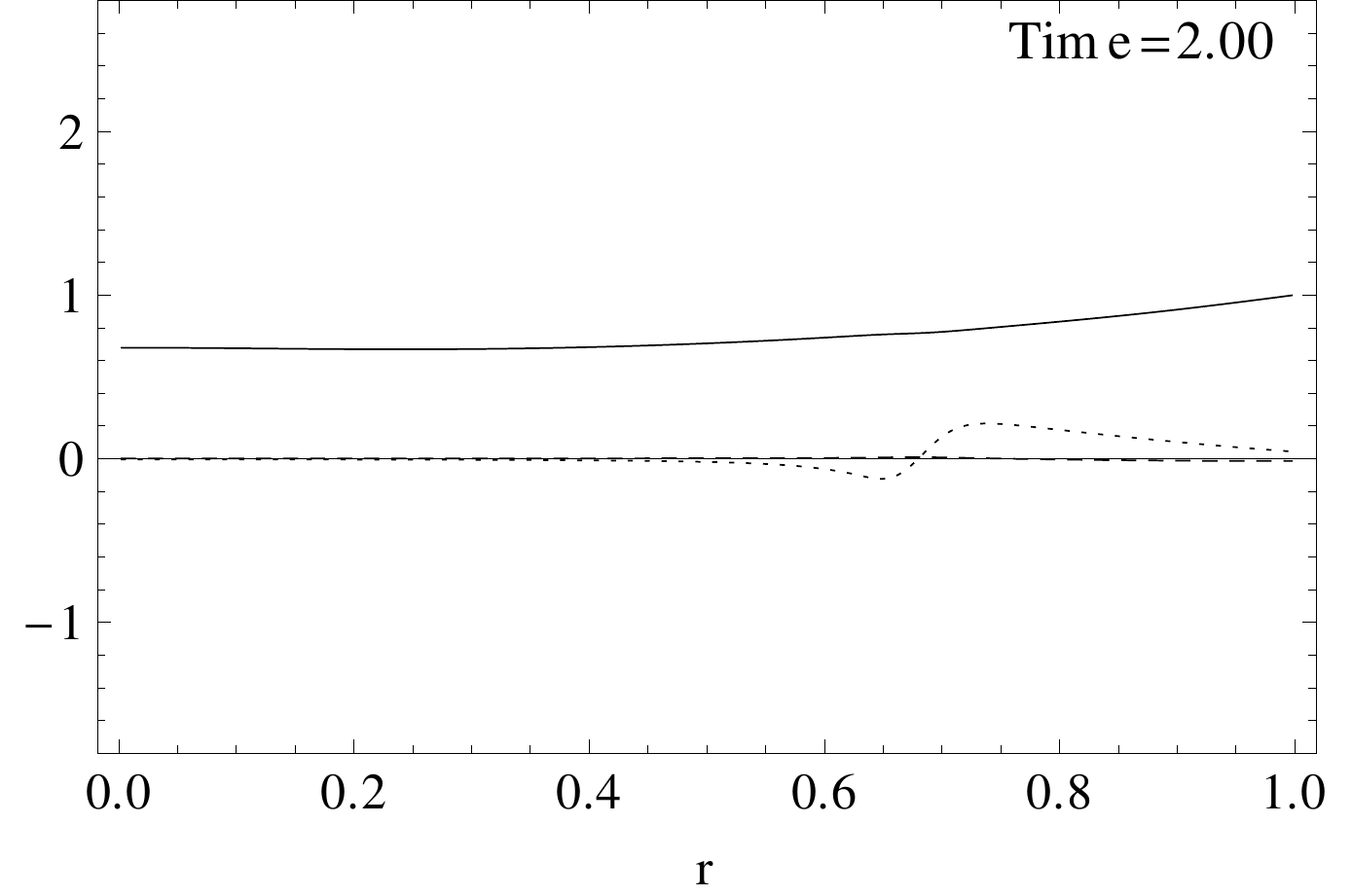}}\\
\vspace{-5.5ex} \mbox{\includegraphics[width=1\linewidth]{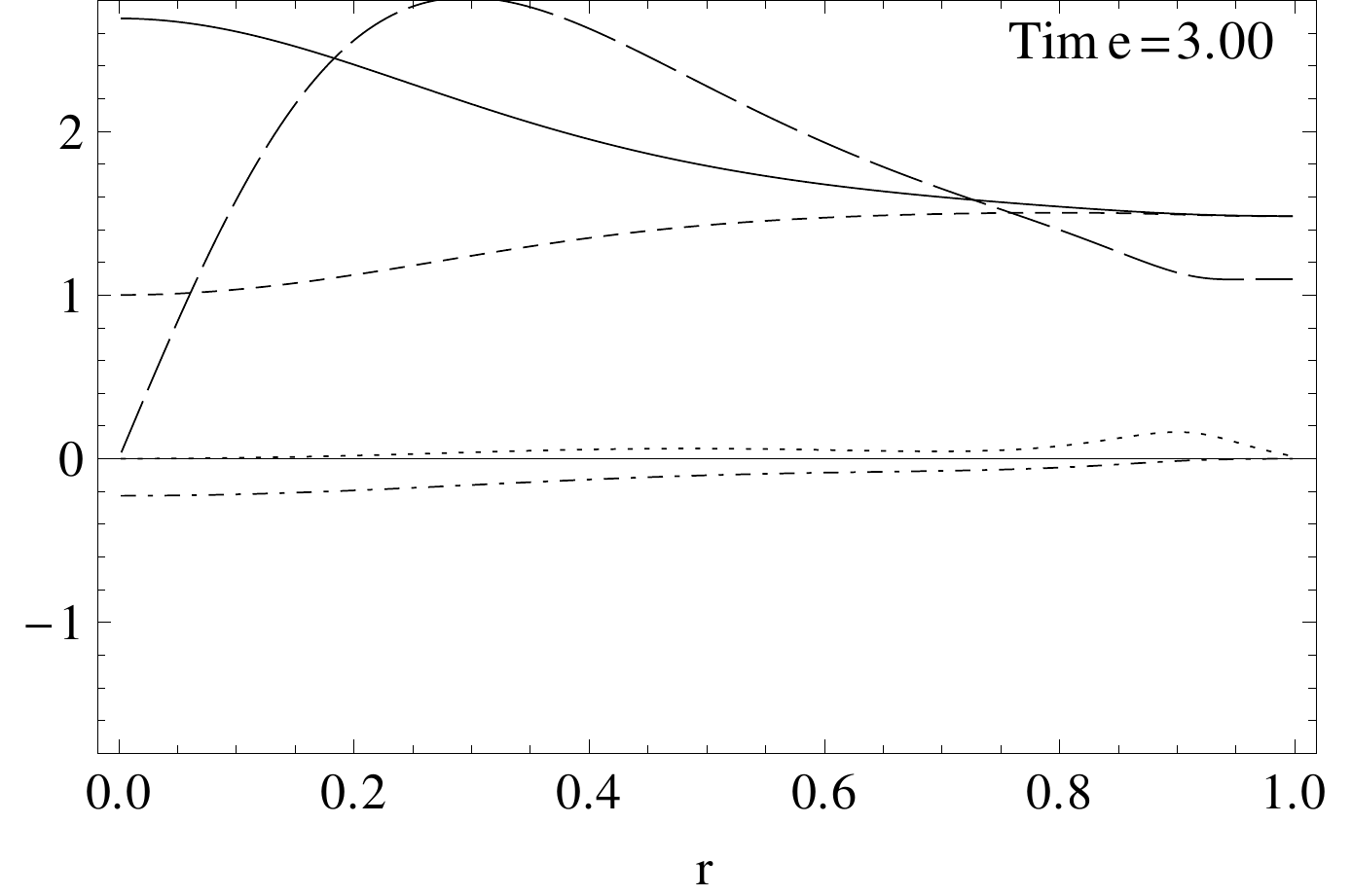}}&
\vspace{-5.5ex} \hspace{-0.8ex} \mbox{\includegraphics[width=1\linewidth]{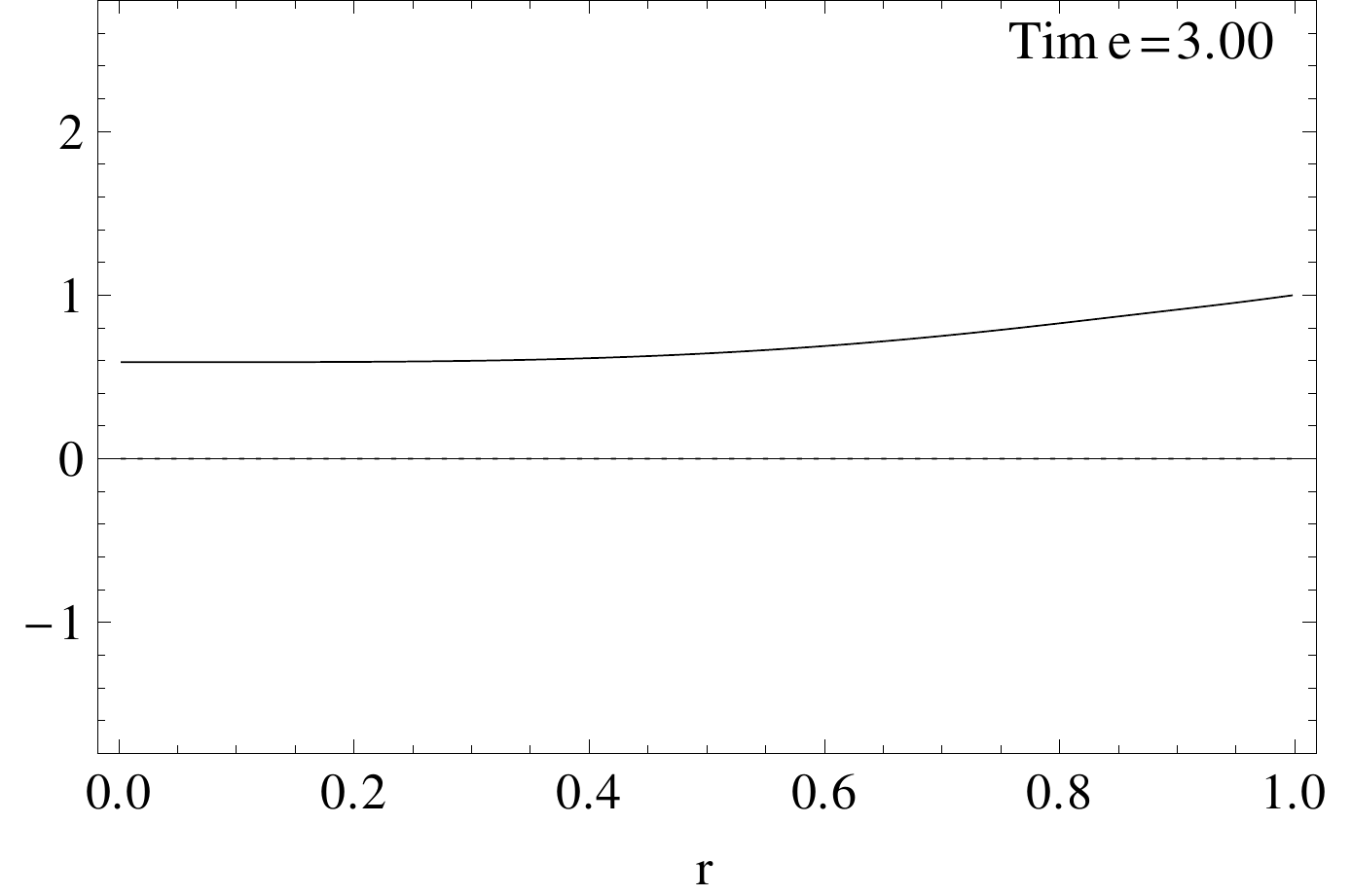}}\\
\vspace{-5.5ex} \mbox{\includegraphics[width=1\linewidth]{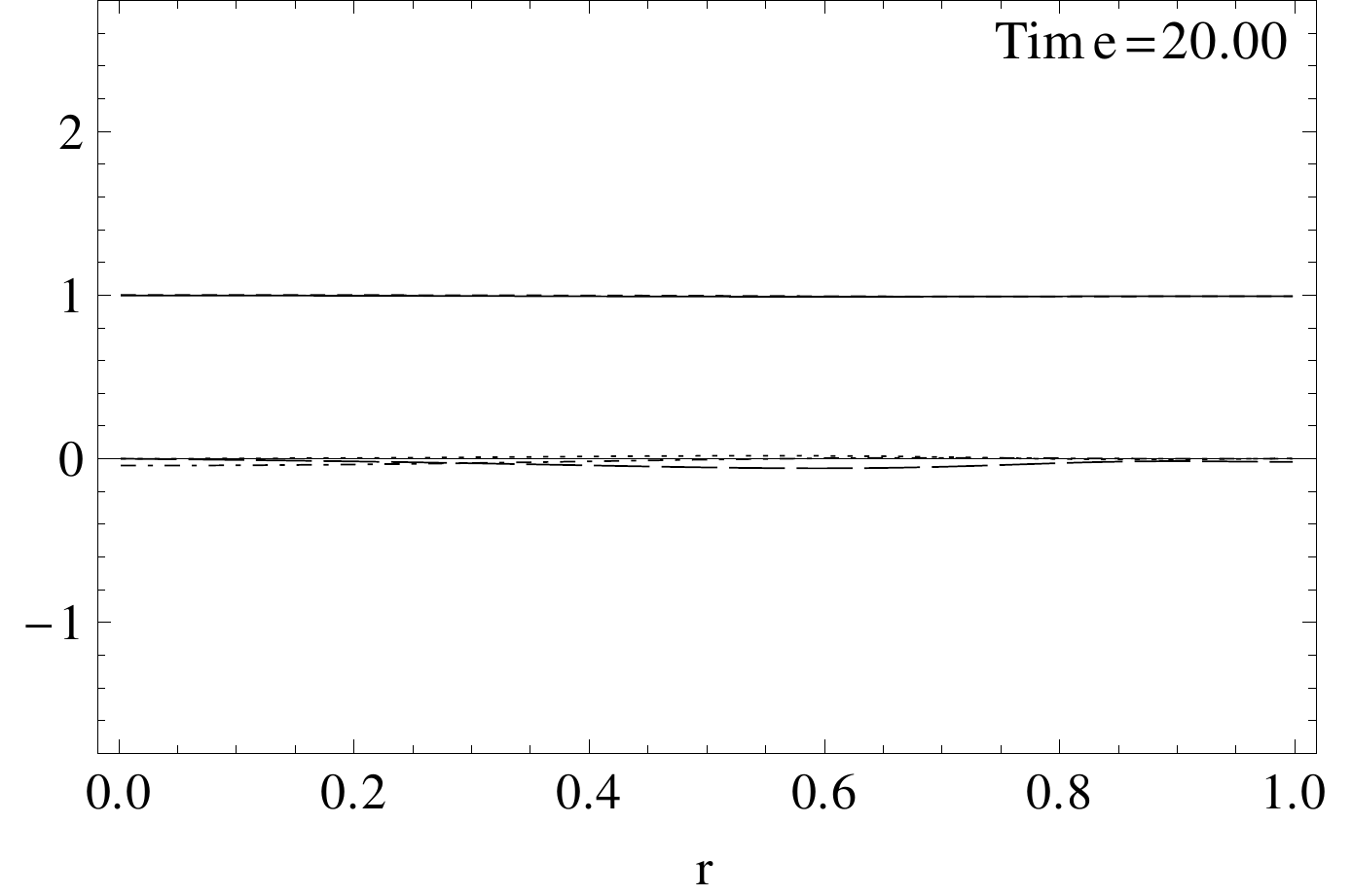}}&
\vspace{-5.5ex} \hspace{-0.8ex} \mbox{\includegraphics[width=1\linewidth]{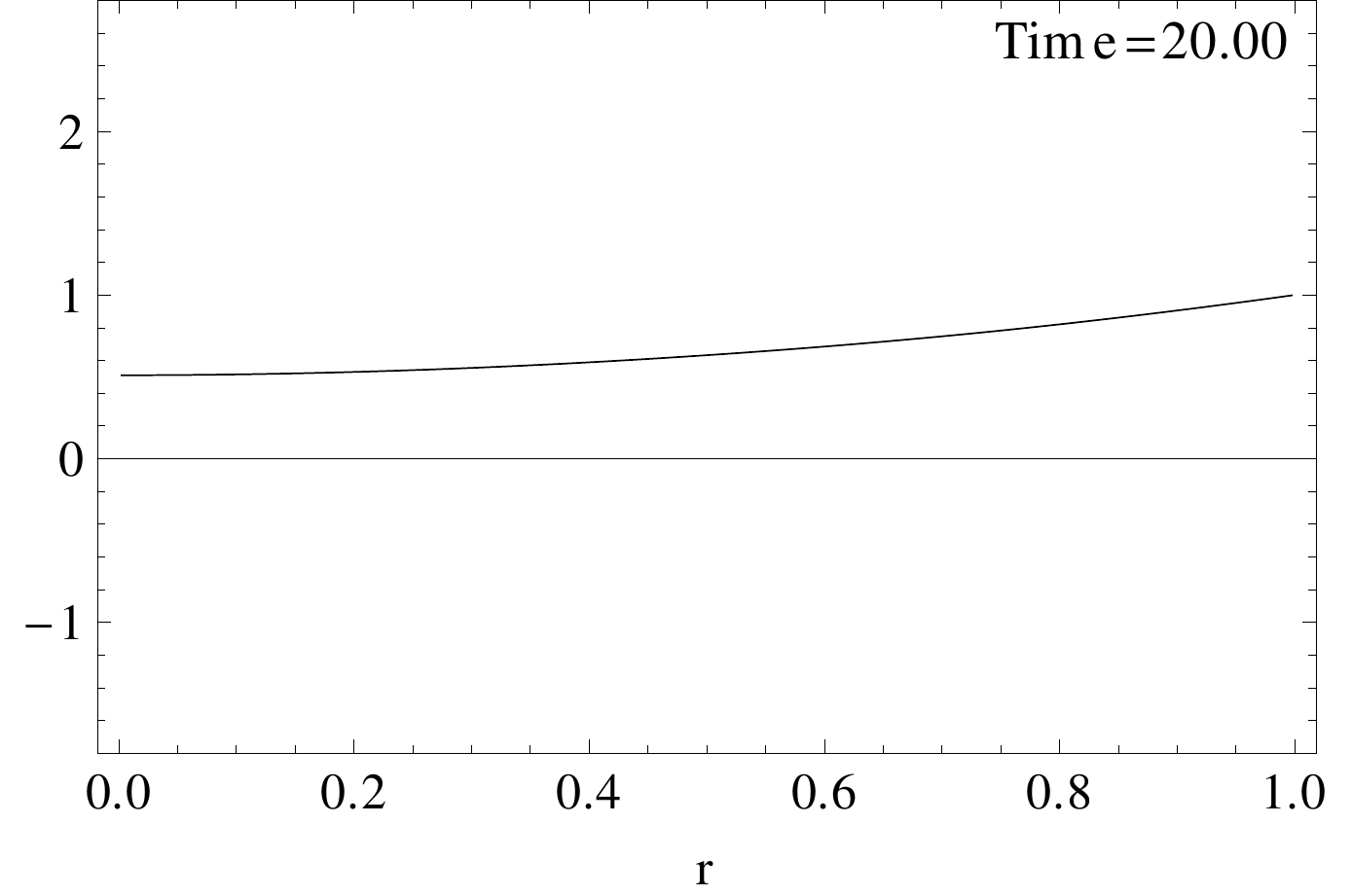}}
\end{tabular}
\vspace{-2ex}
\caption{Evolution of the variables (scalar field): continuation.}
\label{fs:allevolscacent2}
\end{figure}

\begin{figure}[htbp!!]
\center
	\includegraphics[width=0.99\linewidth]{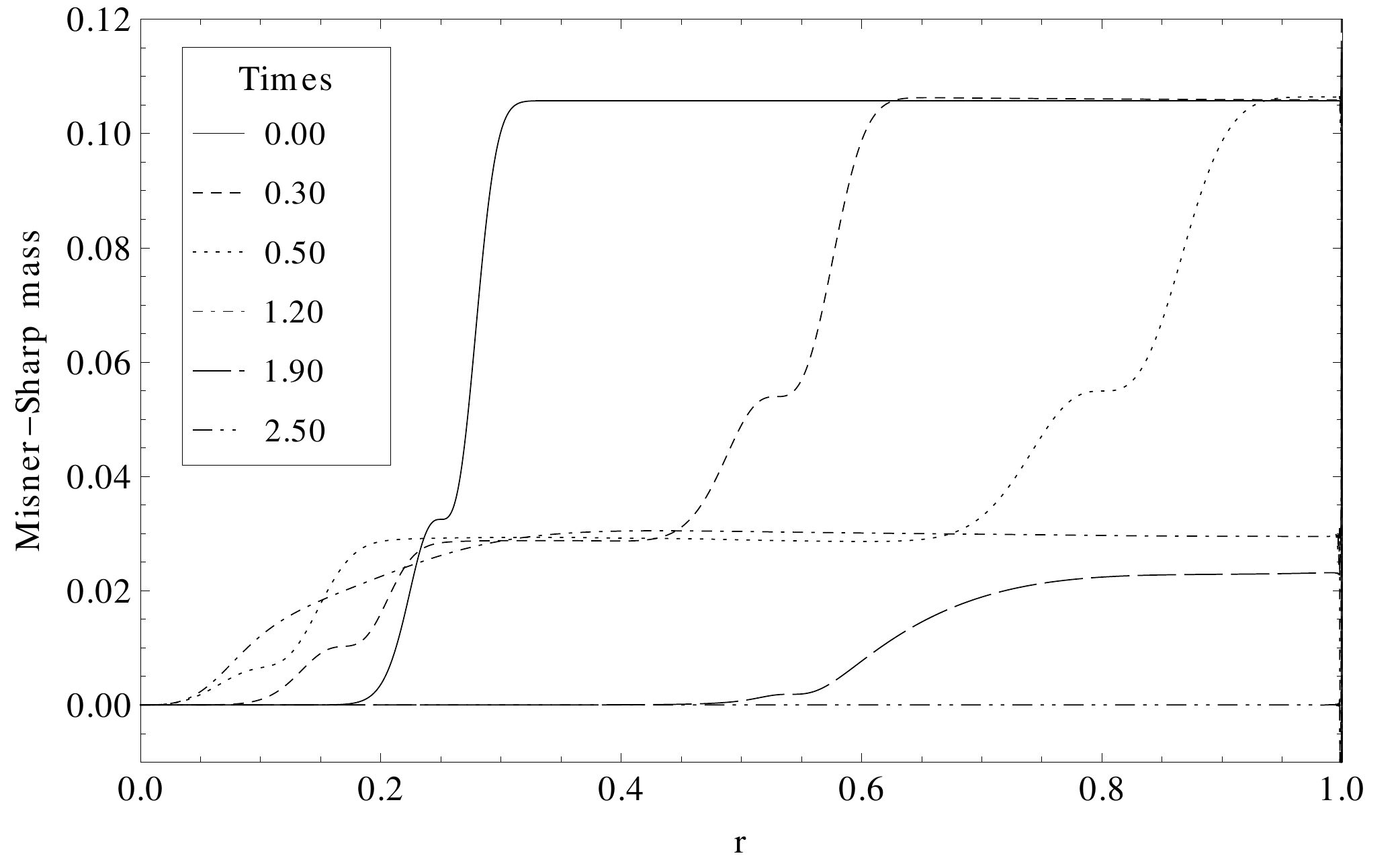} \vspace{-2ex}
\caption{Misner-Sharp mass corresponding to the simulations of figures \ref{fs:allevolscacent1} and \ref{fs:allevolscacent2}.}\label{es:msmflat}
\end{figure}

For the chosen value of the parameter $\xi_\alpha=1$, some fluctuations originated at $\scri^+$ appear during this relaxation time and are finally damped away so that the final state is a stable stationary one. A larger value of  $\xi_\alpha$ will damp these fluctuations more efficiently and the stationary state will be achieved earlier. To illustrate this effect, the $\Lambda^r$ quantity of two simulations with different values of $\xi_\alpha$ is plotted in \fref{fs:scaLambda} at a time where the relaxation phase in \fref{fs:allevolscacent2}'s simulation is taking place: the fluctuations corresponding to the $\xi_\alpha=1$ case are still present, while the $\Lambda^r$ from the $\xi_\alpha=4$ case has already reached its final stationary value.
%no preferred conformal gauge

% Scalar field

% Scalar field - continuation

The value of the Misner-Sharp mass as expressed in \eref{em:msmassexpl} is plotted in \fref{es:msmflat} at some given times. The variation of $M_{MS}$ at the initial time is located around the position of the initial scalar field perturbation. As the pulses propagate through the domain, the parts where $M_{MS}$ varies move with them: for instance, at $t=0.5$ the outgoing pulse is still inside of the domain (compare with \fref{fs:allevolscacent1}), while at $t=1.2$ it has already left - the total mass of the system has become smaller by the amount of energy carried away by the outgoing scalar field pulse. The equivalent effect takes place when the reflected pulse leaves through $\scri^+$.

\begin{figure}[htbp!!]
\center
\begin{tabular}{ m{0.5\linewidth}@{} @{}m{0.5\linewidth}@{} }
\hspace{-2ex} \vspace{-3ex} \mbox{\includegraphics[width=1.08\linewidth]{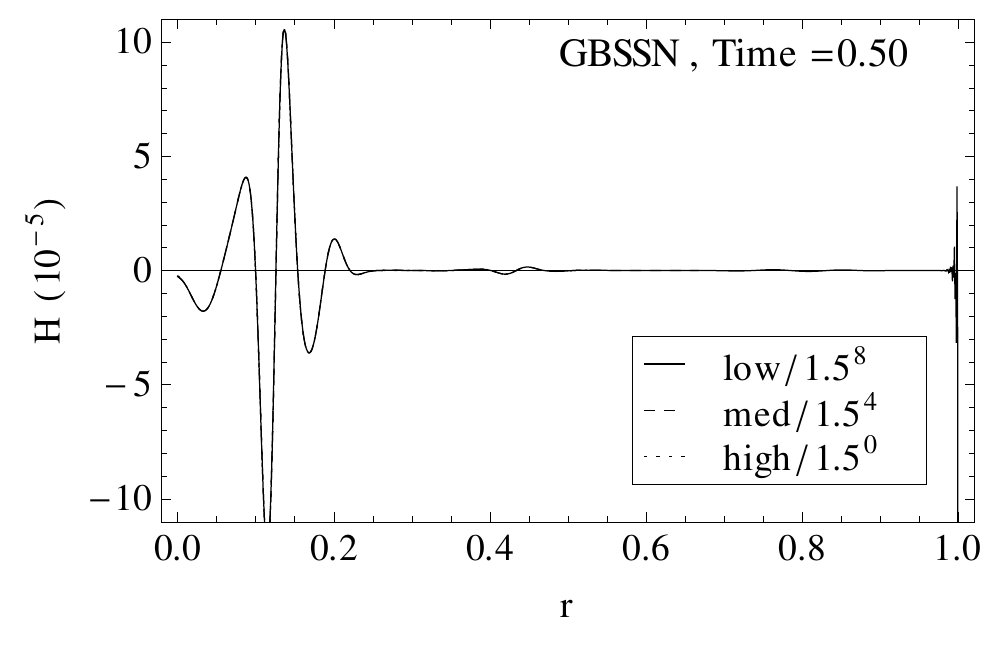}}&
\hspace{-1ex} \vspace{-3ex} \mbox{\includegraphics[width=1.08\linewidth]{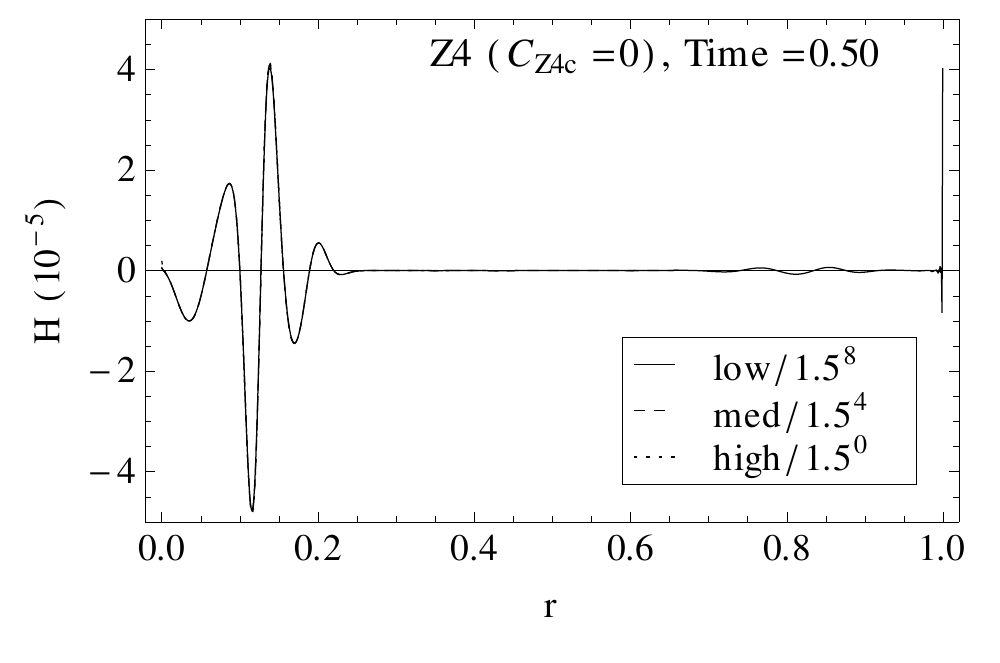}}\\
\hspace{-2ex} \vspace{-2ex} \mbox{\includegraphics[width=1.08\linewidth]{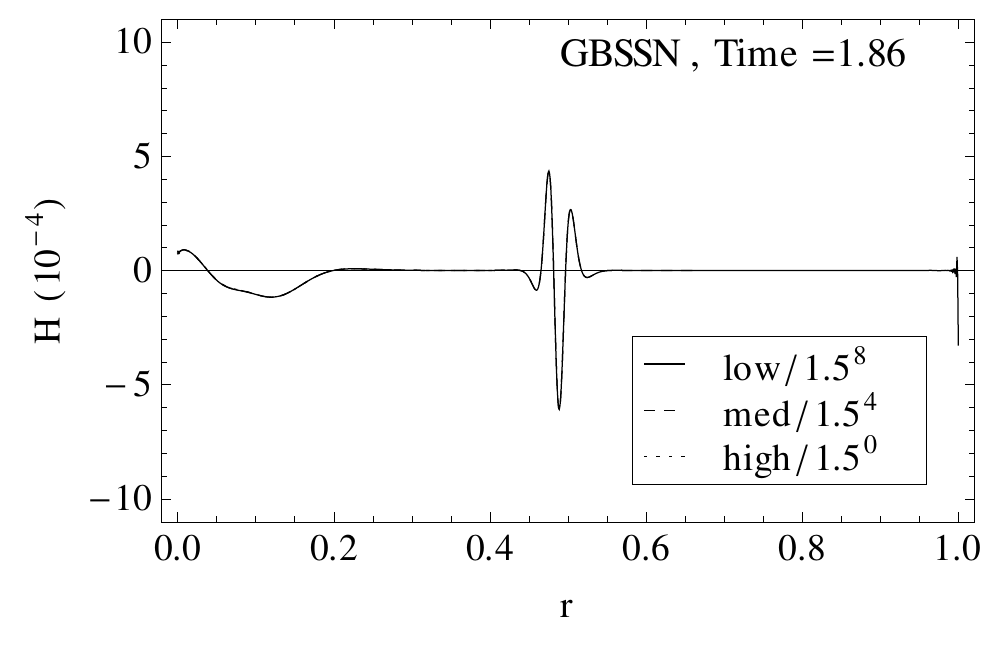}}&
\hspace{-1ex} \vspace{-2ex} \mbox{\includegraphics[width=1.08\linewidth]{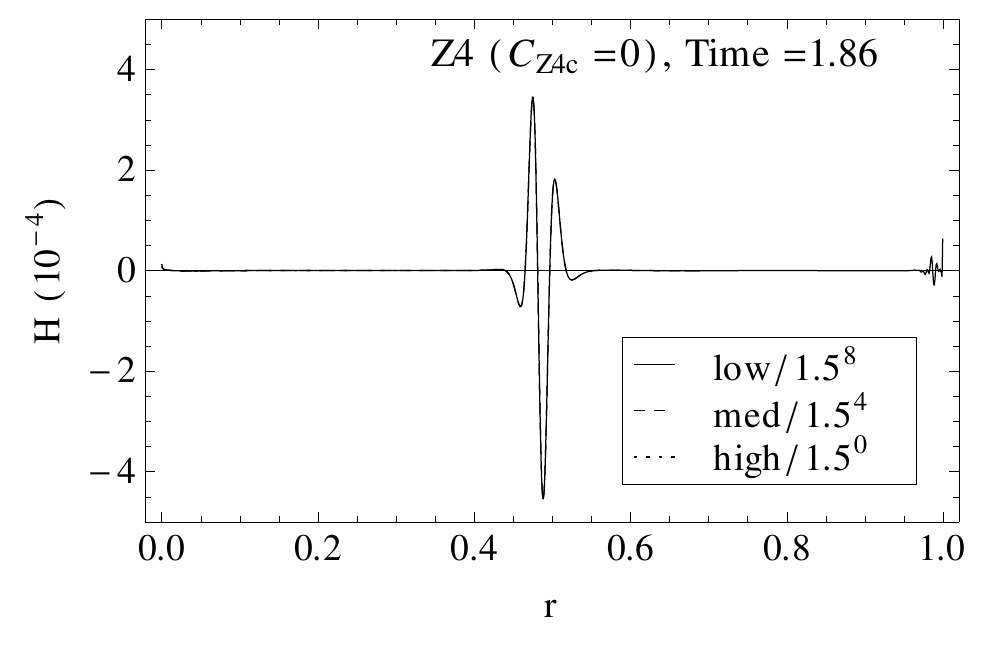}}
\end{tabular}
\vspace{-3ex}
\caption{Point-wise convergence of the Hamiltonian constraint for the GBSSN (left) and Z4c ($C_{Z4c}=0$) (right) formulations at two different times.}
\label{fs:convflatconstrH}
%\end{figure}
%\begin{figure}[htbp!!]
\center
\begin{tabular}{ m{0.5\linewidth}@{} @{}m{0.5\linewidth}@{} }
\hspace{-2ex} \vspace{-3ex} \mbox{\includegraphics[width=1.08\linewidth]{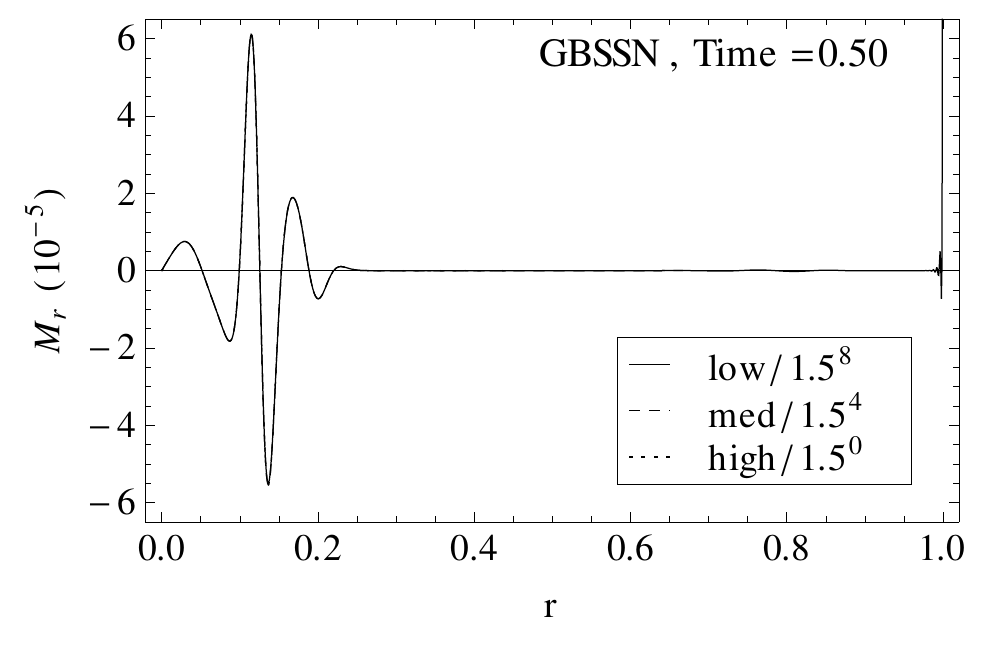}}&
\hspace{-1ex} \vspace{-3ex} \mbox{\includegraphics[width=1.08\linewidth]{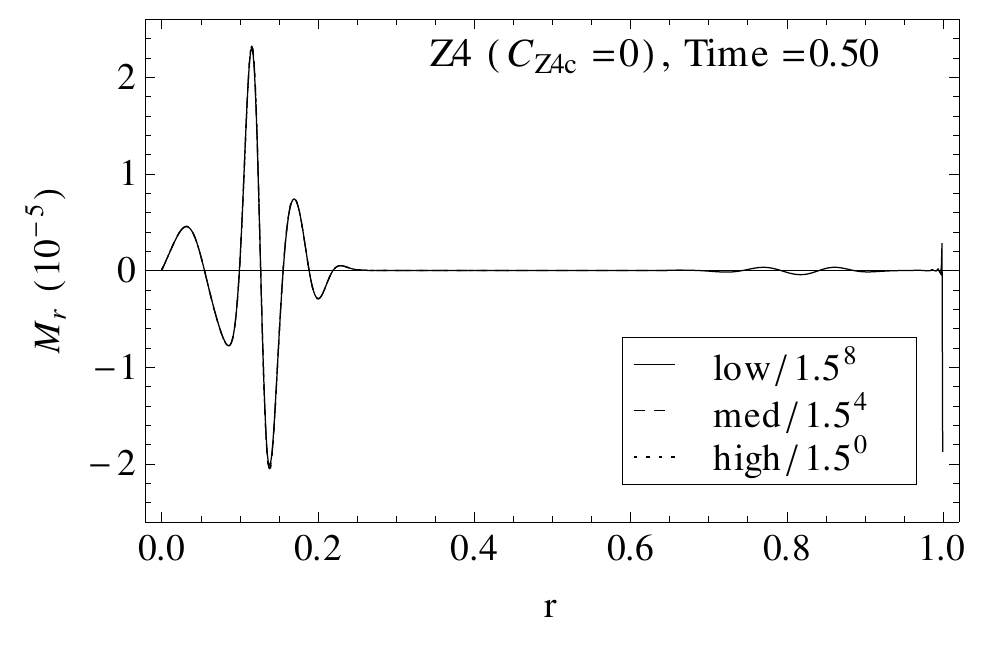}}\\
\hspace{-2ex} \vspace{-2ex} \mbox{\includegraphics[width=1.08\linewidth]{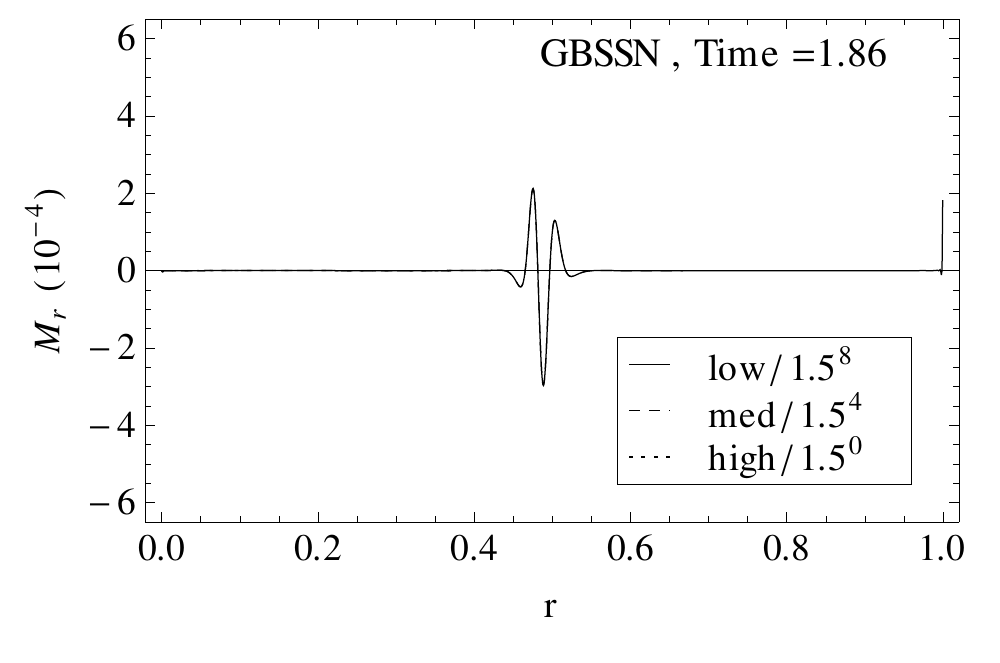}}&
\hspace{-1ex} \vspace{-2ex} \mbox{\includegraphics[width=1.08\linewidth]{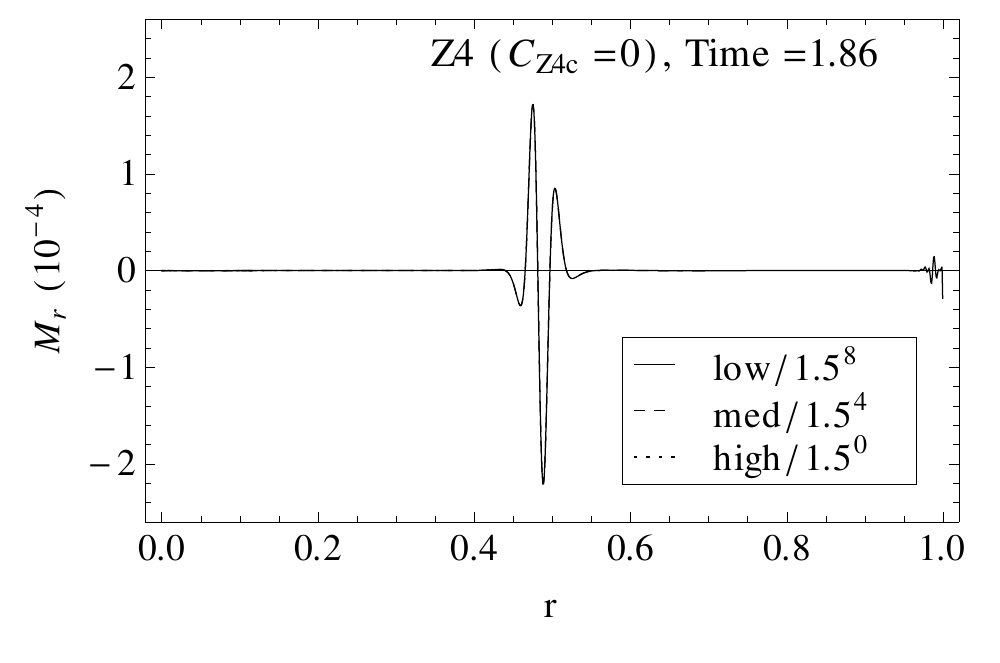}}
\end{tabular}
\vspace{-3ex}
\caption{Point-wise convergence of the momentum constraint for the GBSSN (left) and Z4c ($C_{Z4c}=0$) (right) formulations at two different times.}
\label{fs:convflatconstrM}
\end{figure}

Figures \ref{fs:convflatconstrH} and \ref{fs:convflatconstrM} respectively show the convergence (checked as indicated in subsection \ref{sn:conve}) of the Hamiltonian and momentum constraints at two different times (before and after the first pulse has left) for the GBSSN and Z4c ($C_{Z4c}=0$) equations. The lowest resolution had 1200 gridpoints and $\Delta t=9\cdot10^{-5}$. The 4th order convergence in the interior of the domain is achieved nicely, the three curves lie exactly on top of each other. Some non-converging noise can be spotted on the closest points to $\scri^+$, but this is something that does not affect the evolved variables.
The errors of the constraints are larger for the GBSSN formulation. As indicated by the axes labels, the vertical axes have been rescaled by $10^5$ or $10^4$.
These simulations used centered stencils for the advection terms. The difference in the same formulation using centered and off-centered stencils (see \fref{fn:offcentdiagram}) in the advection terms is very small.
% Convergence of constraints

\begin{figure}[htbp!!]
\center
\begin{tabular}{ m{0.5\linewidth}@{} @{}m{0.5\linewidth}@{} }
\hspace{-2ex} \mbox{\includegraphics[width=1.08\linewidth]{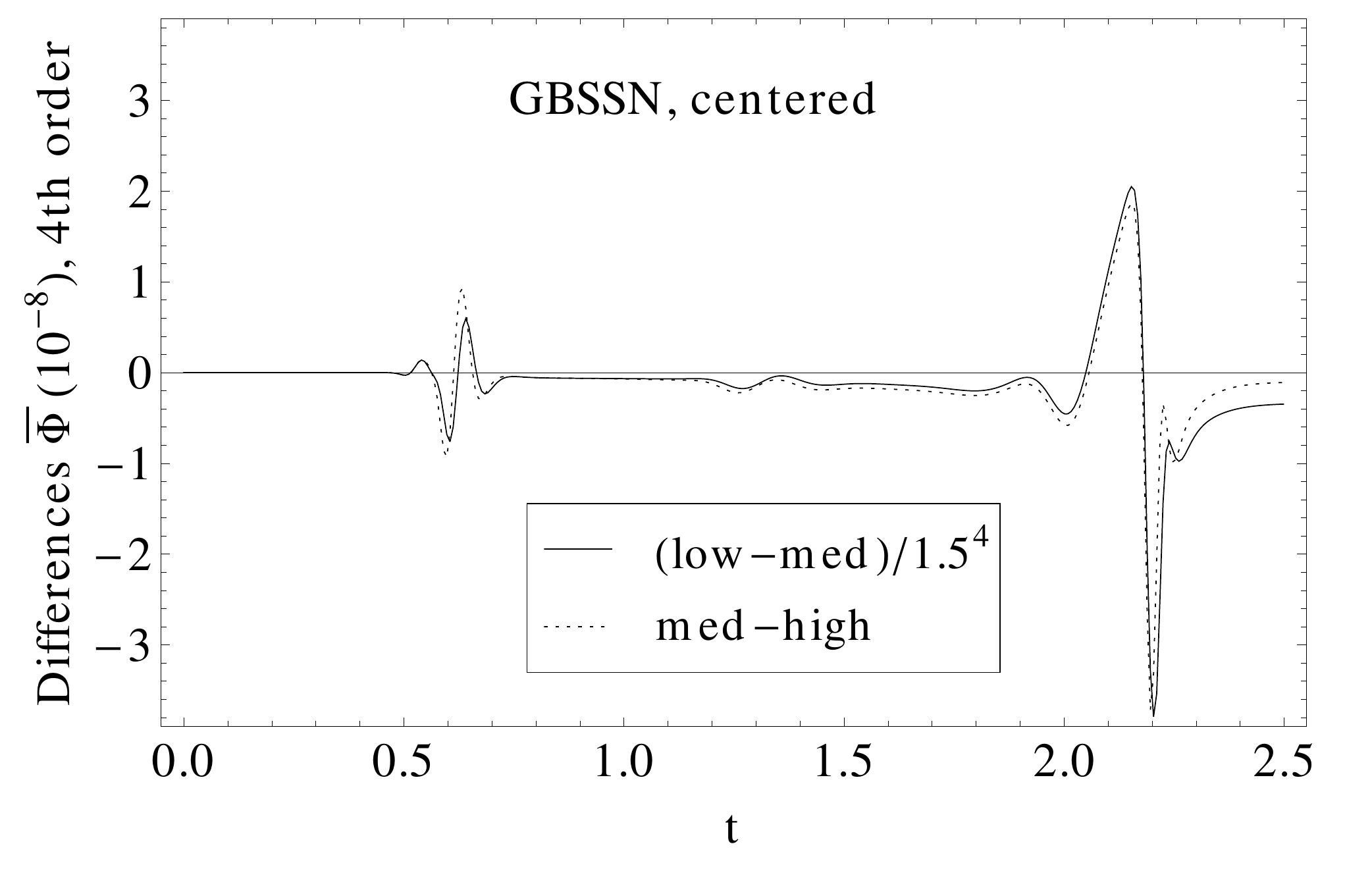}}&
\hspace{-0ex} \mbox{\includegraphics[width=1.08\linewidth]{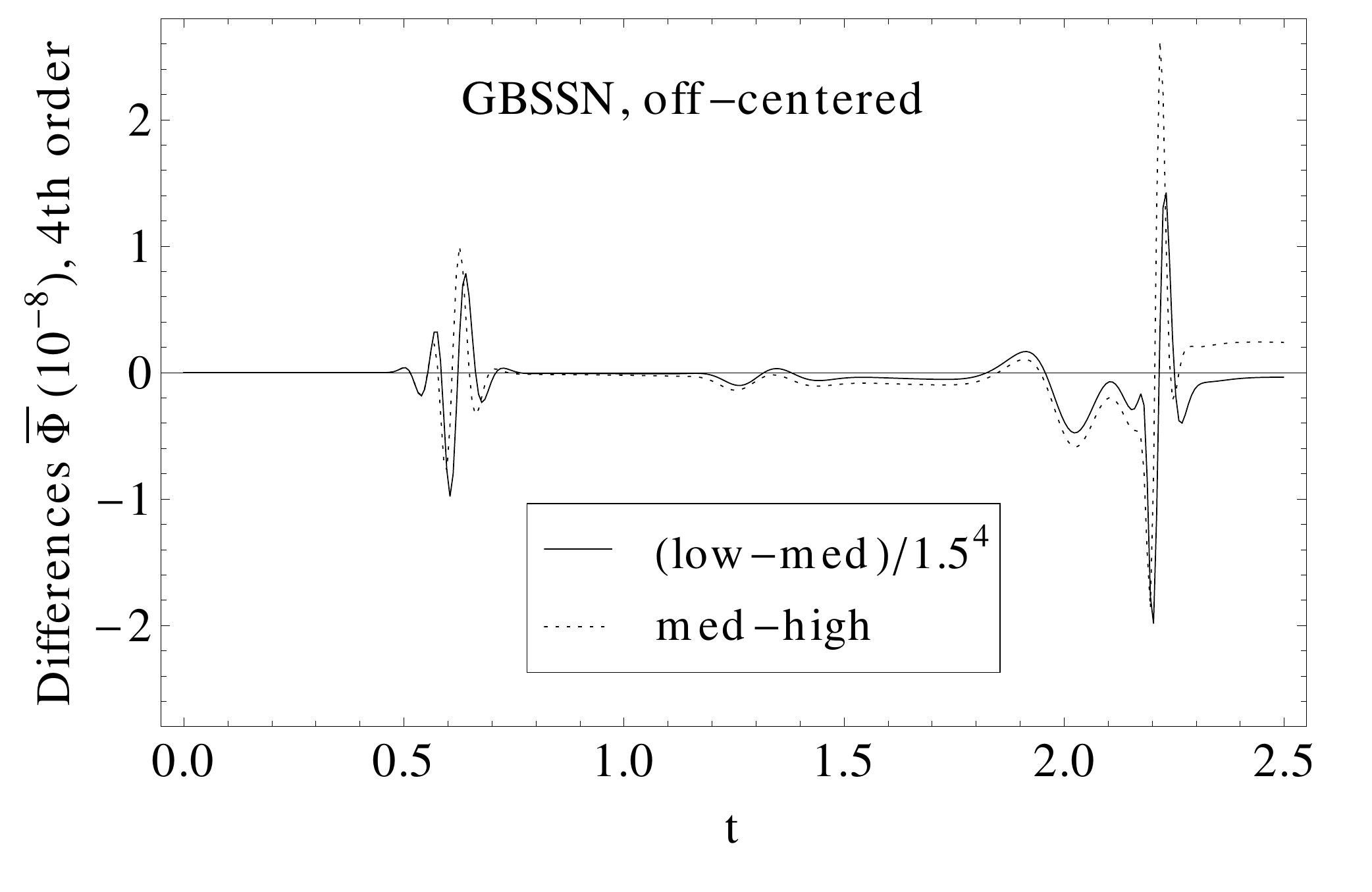}}\\
\hspace{-3ex} \mbox{\includegraphics[width=1.08\linewidth]{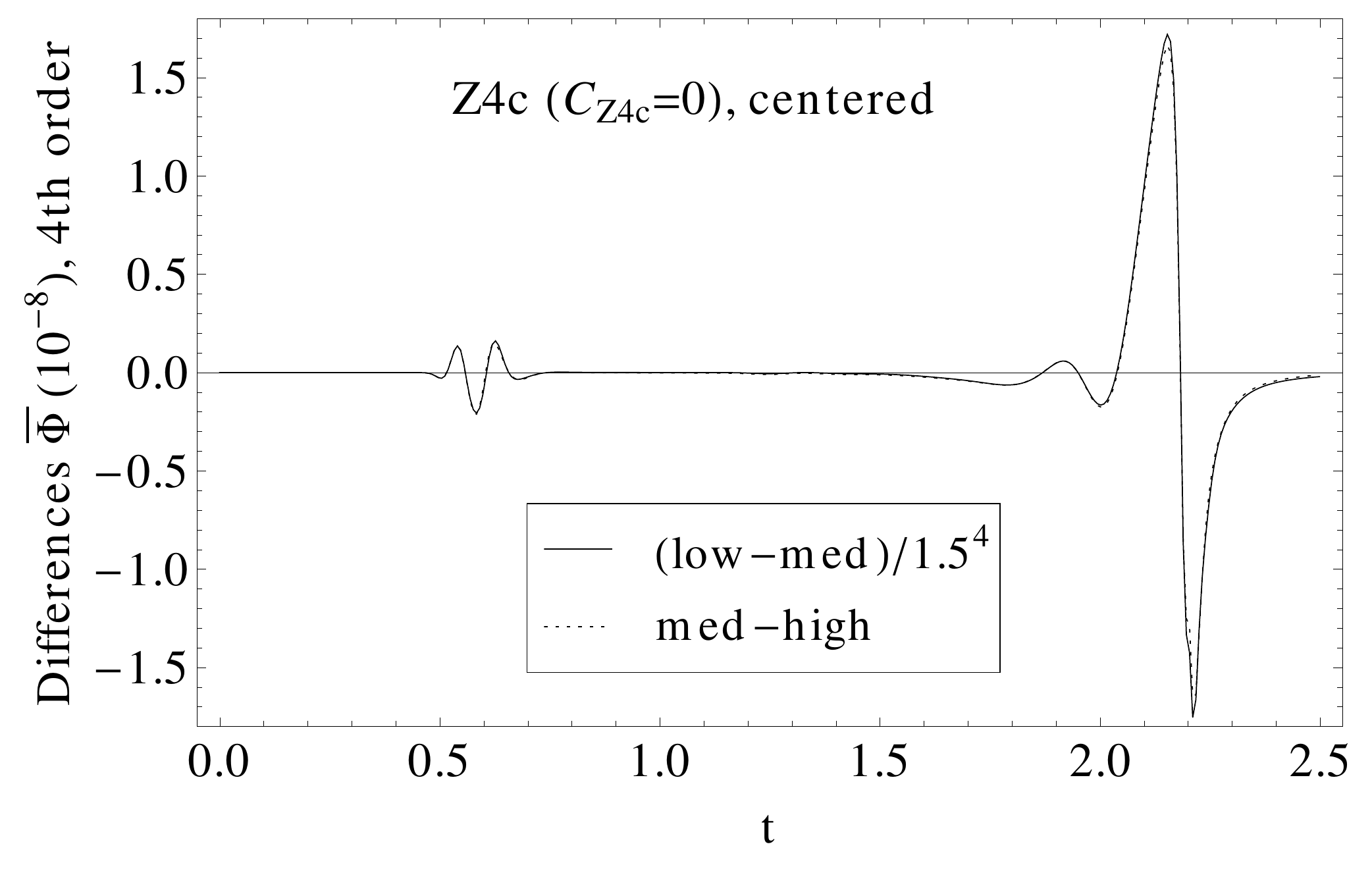}}&
\hspace{-0ex} \mbox{\includegraphics[width=1.08\linewidth]{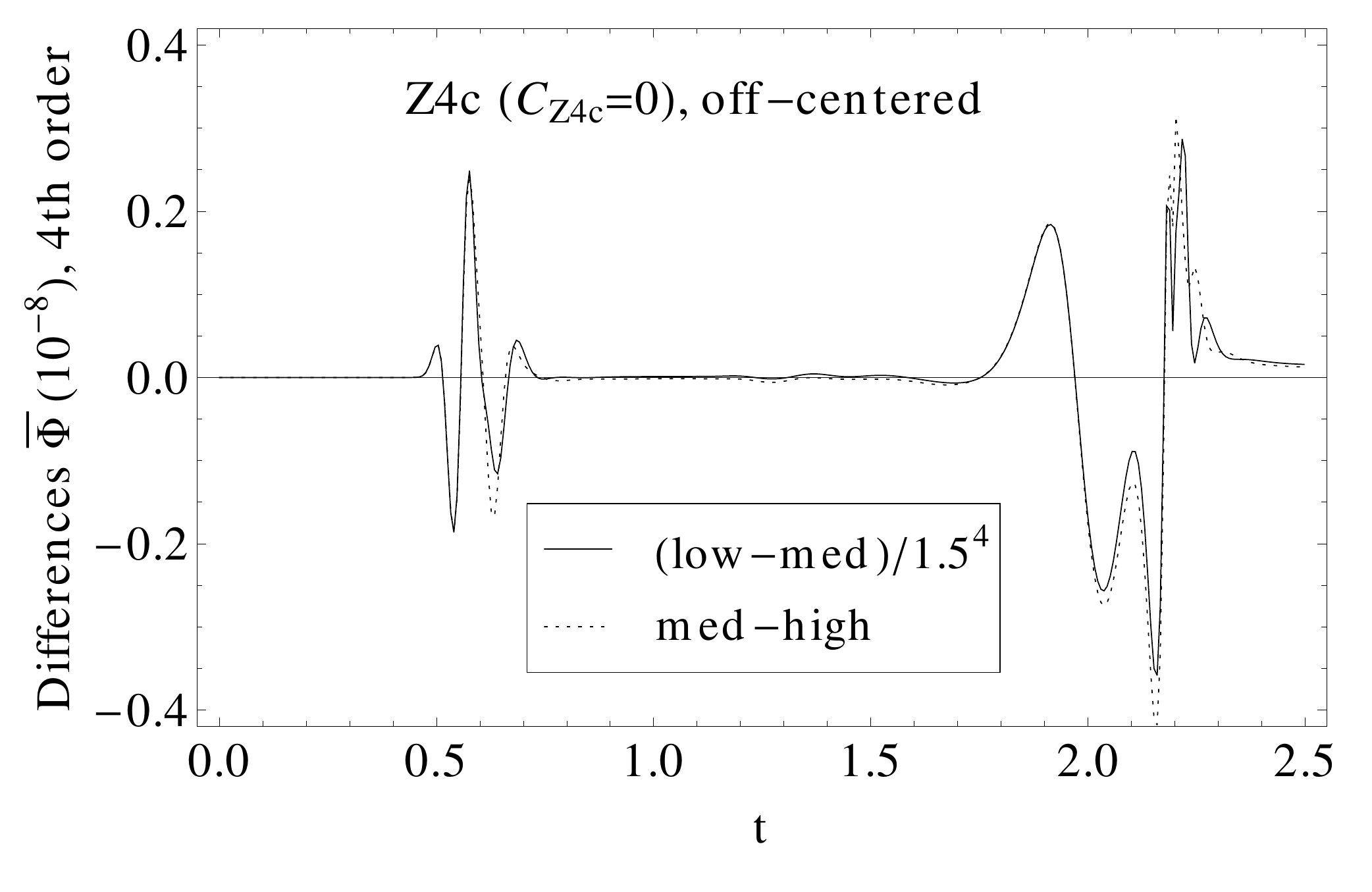}}\\
\hspace{-2ex} \mbox{\includegraphics[width=1.08\linewidth]{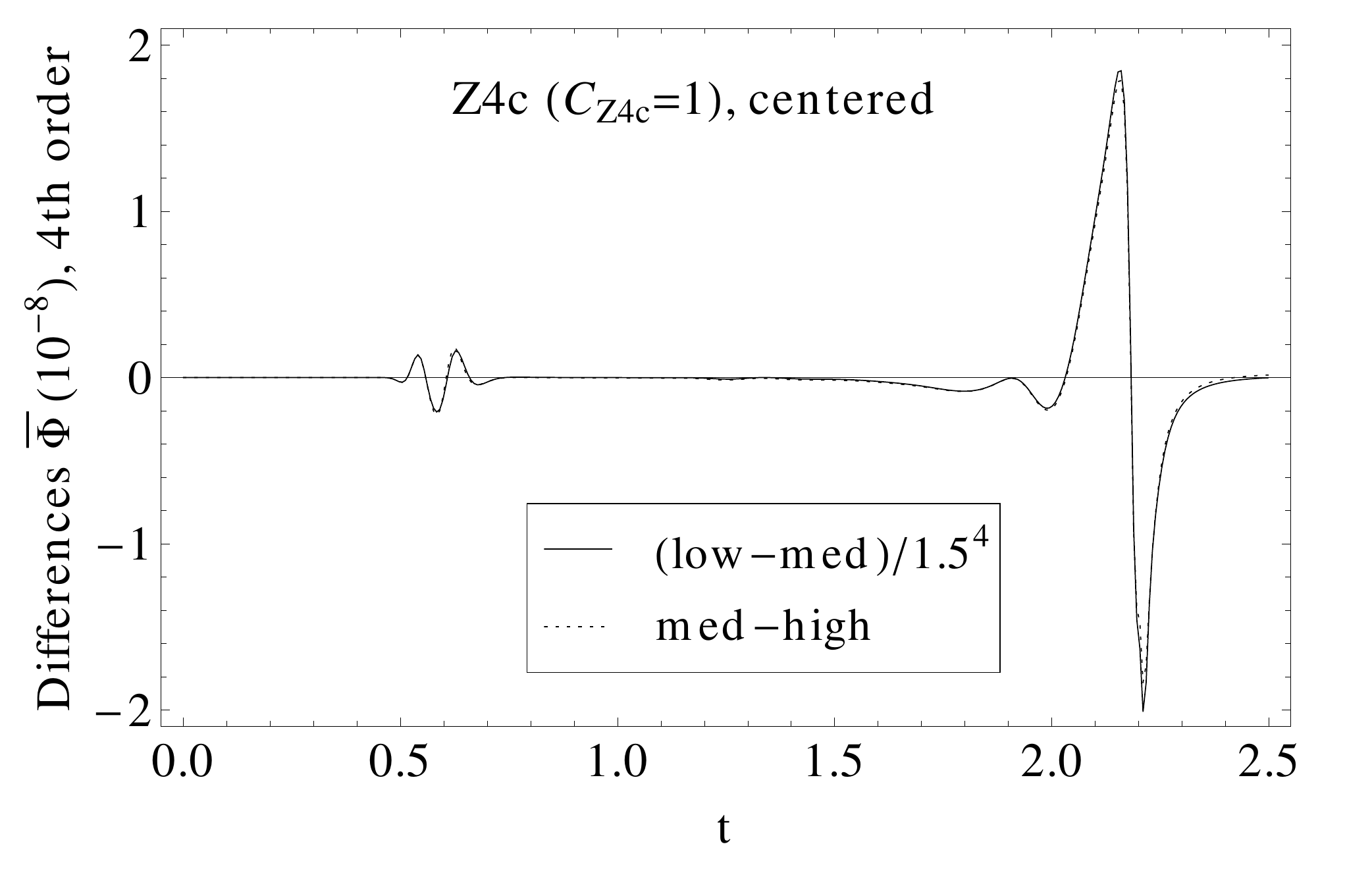}}&
\hspace{-1ex} \mbox{\includegraphics[width=1.08\linewidth]{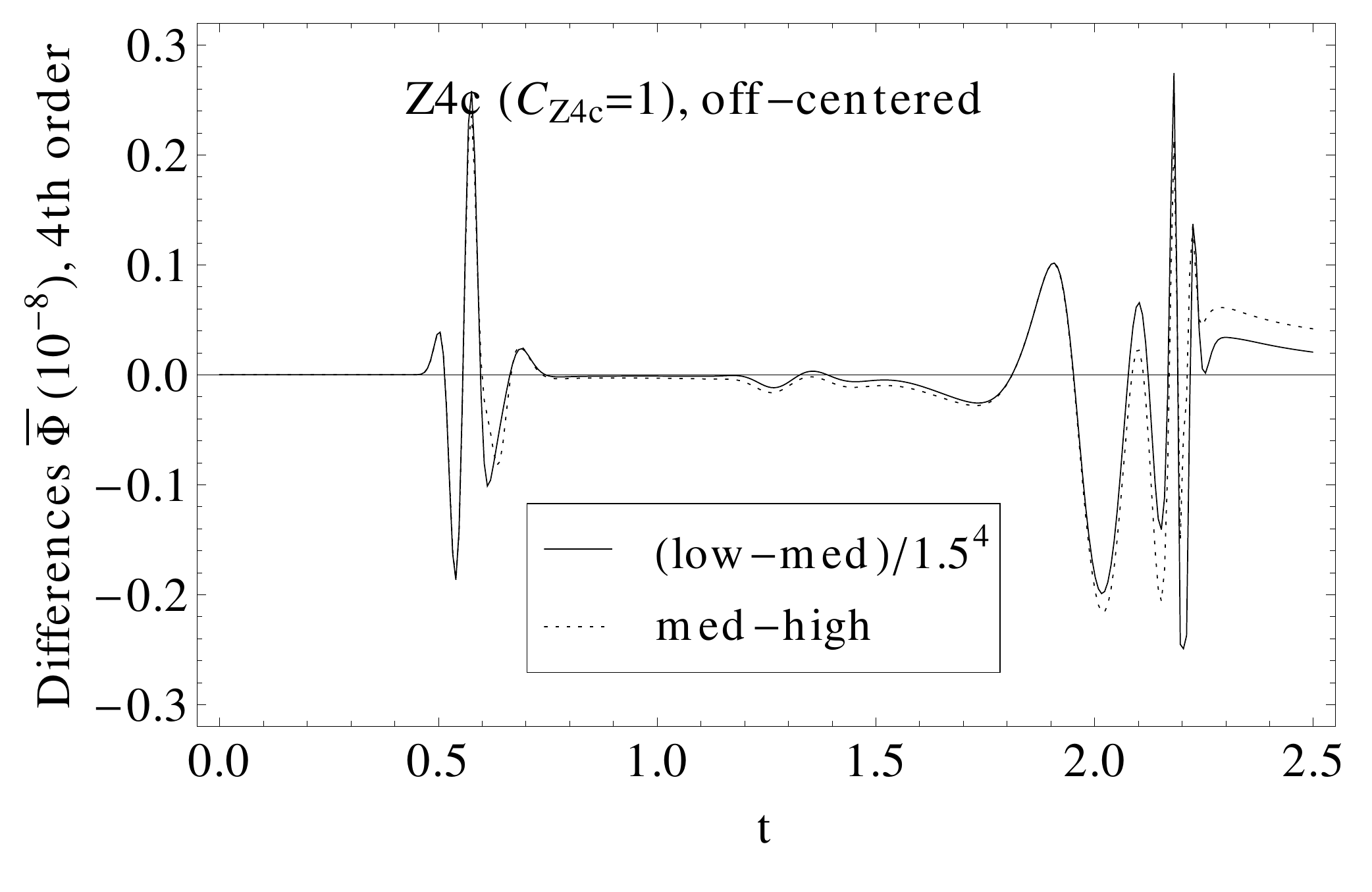}}
\end{tabular}
\vspace{-2ex}
\caption{Convergence of the rescaled scalar function $\bPhi$ at $\scri^+$ over time for different configurations: on the left the stencils of the advection terms were centered (using $\epsilon=0.5$ of dissipation), while on the right they were off-centered (with $\epsilon=0.25$). Note the changes in the vertical axis between the cases (as indicated, it is rescaled by $10^{8}$).}
\label{fs:convflathscri}
\end{figure}

Convergence of the rescaled scalar function $\bPhi$ at $\scri^+$ is presented in \fref{fs:convflathscri} for the three formulations GBSSN, \CZ{} ($C_{Z4c}=0$) and \CZ{} ($C_{Z4c}=1$) and using centered and off-centered stencils in the advection terms for each of them. The used grid is staggered, so that the values of $\bPhi$ at $\scri^+$ have been obtained using 4th order extrapolation. The amplitude in the errors for the  off-centered cases is considerably smaller, but the coincidence between the curves is not as good as in the centered case.
Both in terms of error amplitude and convergence, \CZ{} performs better than GBSSN.
%\upda{comment that off-centered stencils in the advection terms give smaller errors, but they do not coincide as nicely. In terms of error and convergence, \CZ{} performs better than GBSSN. }
% Convergence

The results presented so far have been obtained in simulations where the shift was fixed in time. It is interesting to study the behaviour of the evolution variables at $\scri^+$ under different gauge conditions (especially the ones regarding the shift), as they deliver important information about the appropriate numerical treatment of $\scri^+$.

The considered shift conditions are: the fixed shift, the Gamma-driver condition \eref{ee:expGammadriver} and the integrated Gamma-driver \eref{ee:expintegGammadriver}. Then also the pair of slicing and shift conditions given by \eref{ee:improvedbg} (harmonic condition with conformal background metric source terms) and \eref{ee:improvedpbg} (harmonic condition with physical background metric source terms) have been tested.
The parameter choices for each setup (all use \CZ{} ($C_{Z4c}=0$) and $\Kc=-2$) are: fixed shift, $\xi_\alpha=1$; Gamma-driver, $\xi_\alpha=1$, $\xi_{\beta^r}=5$, $\mu=1$ and $\lambda=\case{3}{4}$; integrated Gamma-driver, $\xi_\alpha=1$, $\xi_{\beta^r}=5$ and $\lambda=\case{3}{4}$; conformal harmonic with background sources has no parameters; and physical harmonic with background sources, $\xi_\alpha=2$.

\begin{figure}[htbp!!]
\center
\vspace{-2.5ex}
 \begin{tikzpicture}[scale=2.0]\draw (-1cm,0cm) node {};
		\draw (0cm, 0cm) node {\small Fixed shift}; \draw (0.5cm, 0cm) -- (0.8cm, 0cm);
		\draw (1.5cm, 0cm) node {\small Conf.har.bg.}; \draw [dashed] (2cm, 0cm) -- (2.3cm, 0cm);
		\draw (3cm, 0cm) node {\small Phys.har.bg.}; \draw [dotted] (3.5cm, 0cm) -- (3.8cm, 0cm);
		\draw (4.5cm, 0cm) node {\small Int.$\Gamma$-driver}; \draw [dash pattern= on 4pt off 2pt on 1pt off 2pt] (5cm, 0cm) -- (5.3cm, 0cm);
		\draw (6cm, 0cm) node {\small $\Gamma$-driver}; \draw [dash pattern= on 8pt off 2pt] (6.5cm, 0cm) -- (6.8cm, 0cm);
	%	\draw (7.5cm, 0cm) node {$\alpha$}; \draw [dash pattern= on 6pt off 2pt on 1pt off 2pt on 1pt off 2pt] (7.8cm, 0cm) -- (8.5cm, 0cm);
	\end{tikzpicture}
\\
\begin{tabular}{ m{0.5\linewidth}@{} @{}m{0.5\linewidth}@{} }
\mbox{\includegraphics[width=1\linewidth]{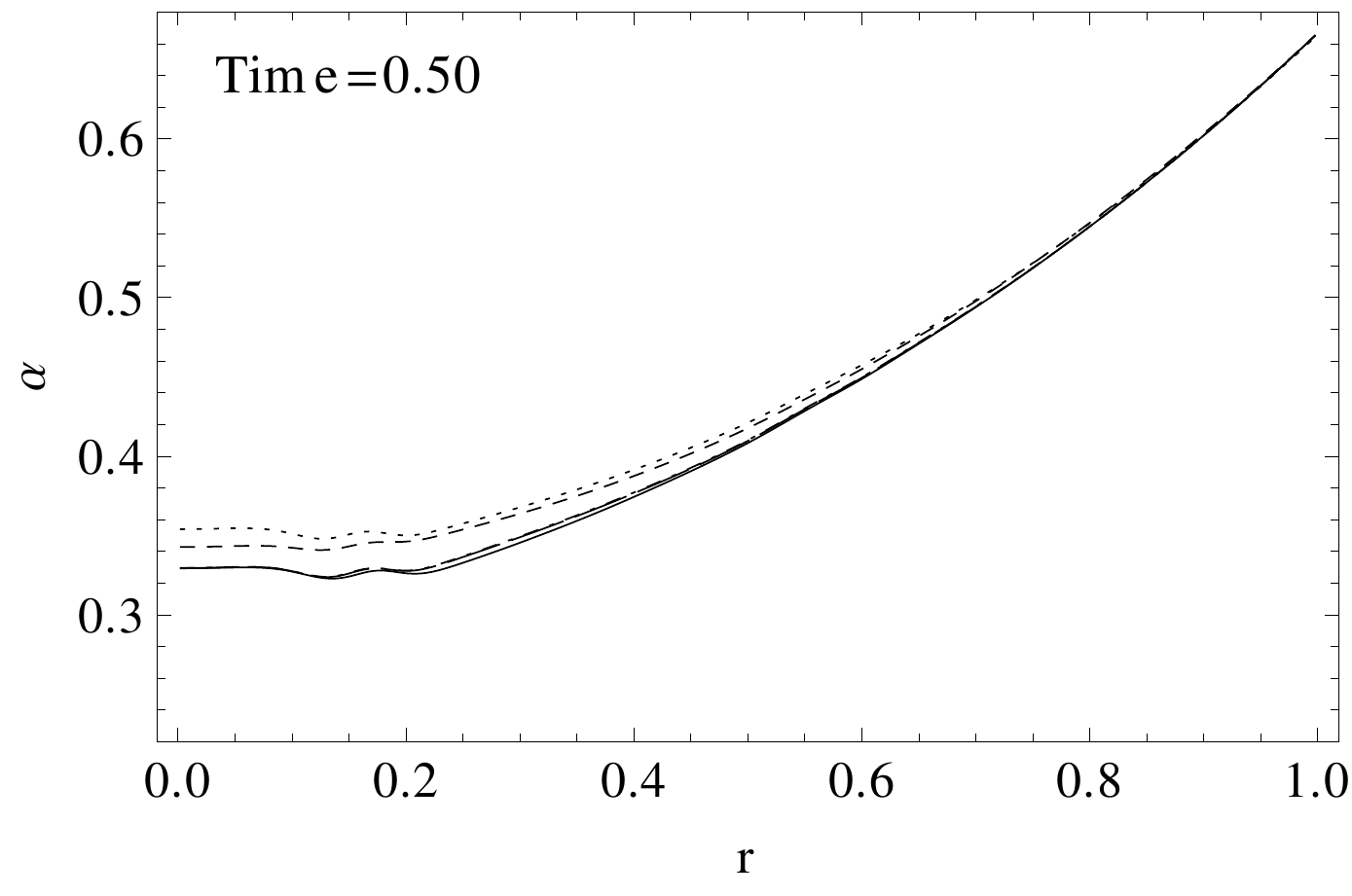}}&
\hspace{-0.6ex} \mbox{\includegraphics[width=1.06\linewidth]{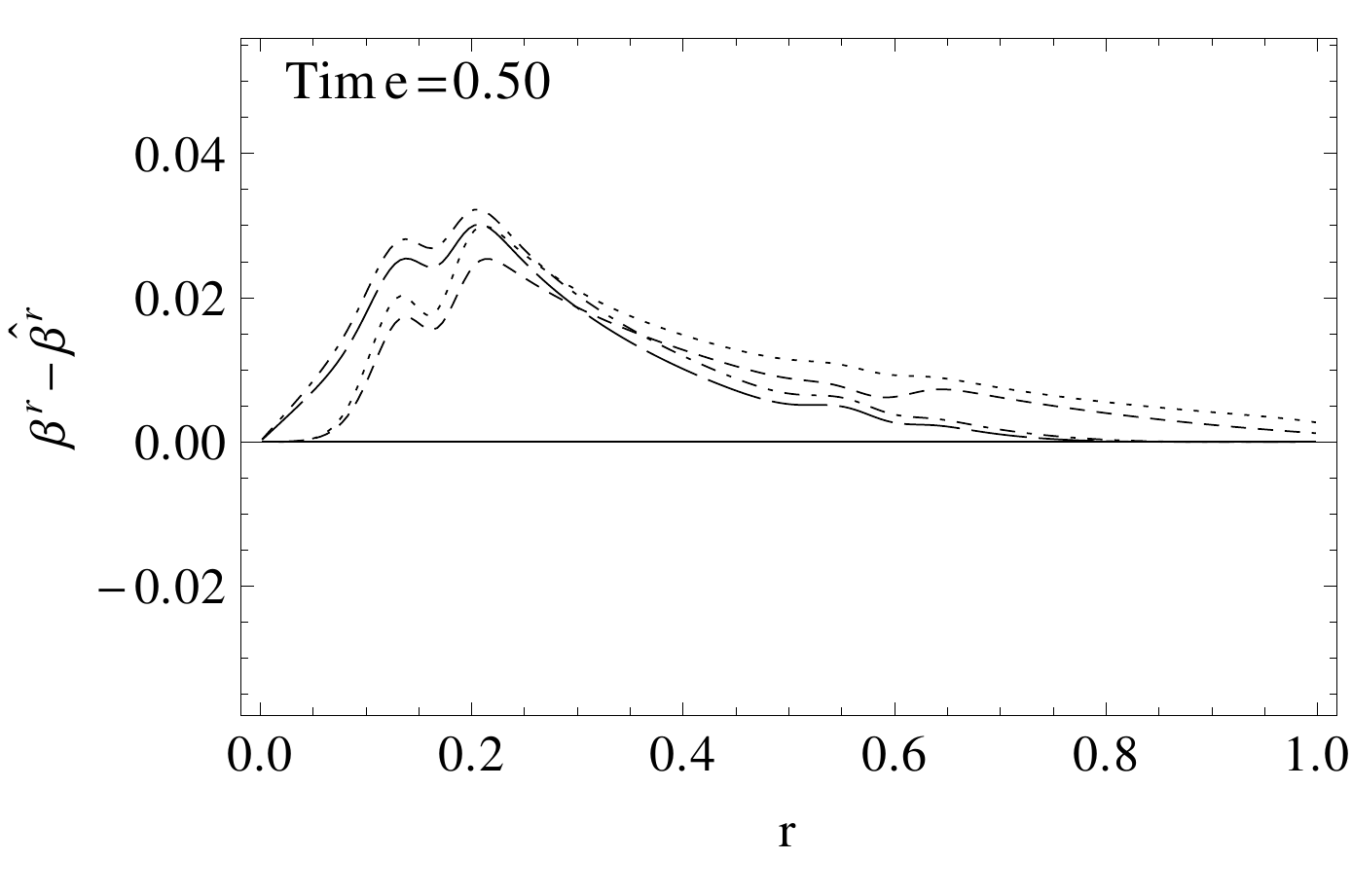}}\\
\vspace{-7.0ex} \mbox{\includegraphics[width=1\linewidth]{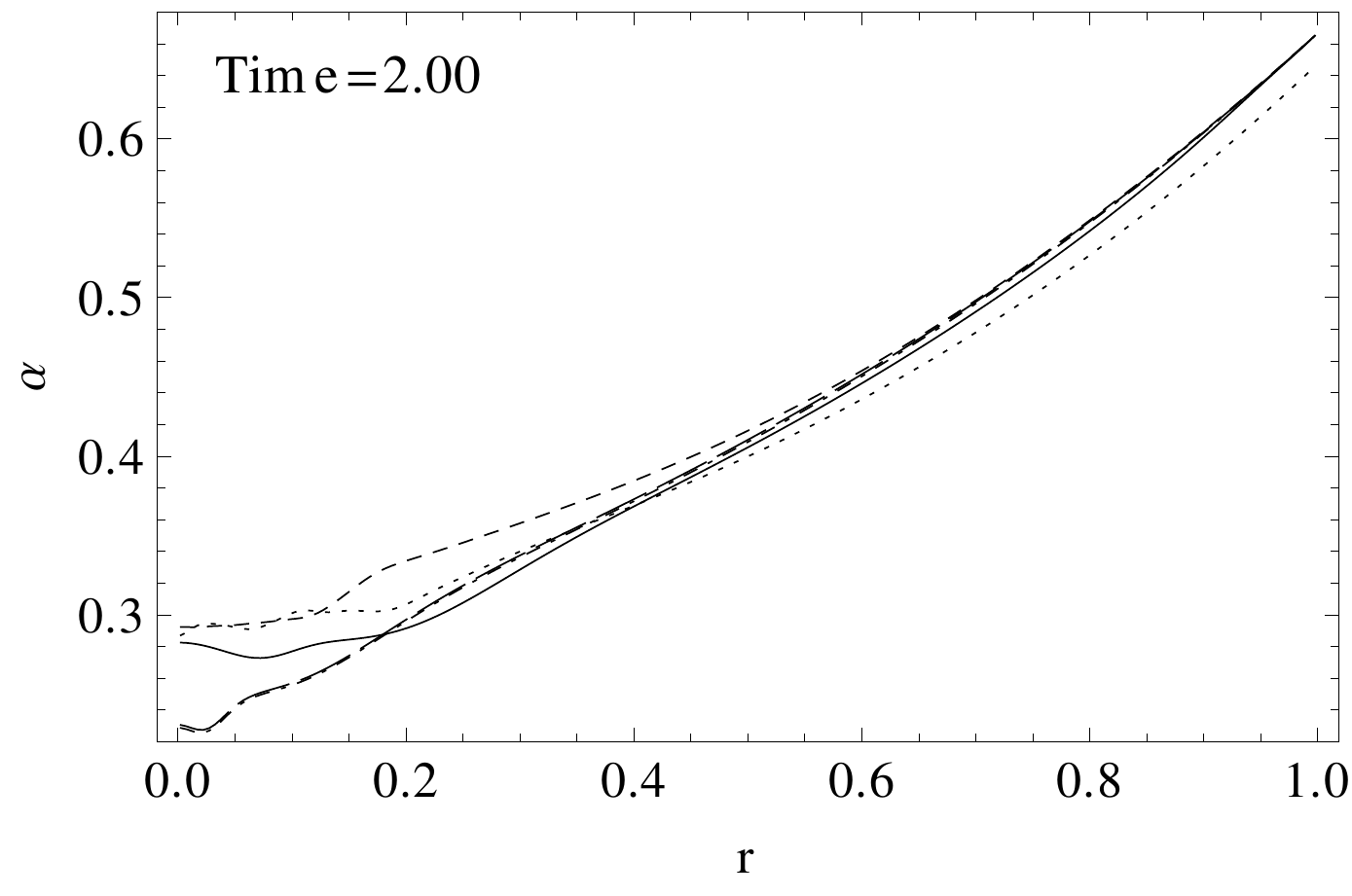}}&
\vspace{-7.0ex} \hspace{-0.6ex} \mbox{\includegraphics[width=1.06\linewidth]{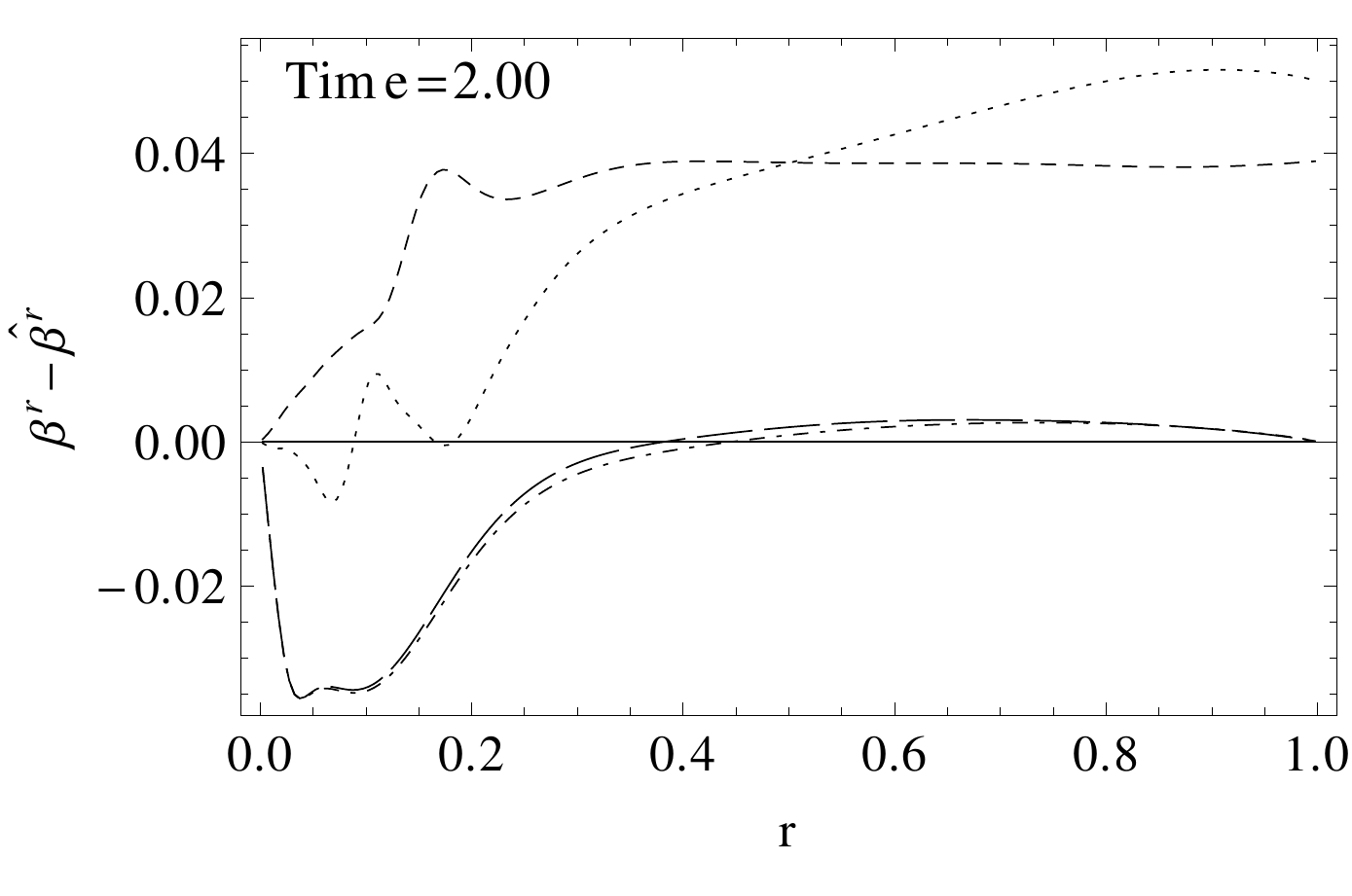}}\\
\vspace{-7.0ex} \mbox{\includegraphics[width=1\linewidth]{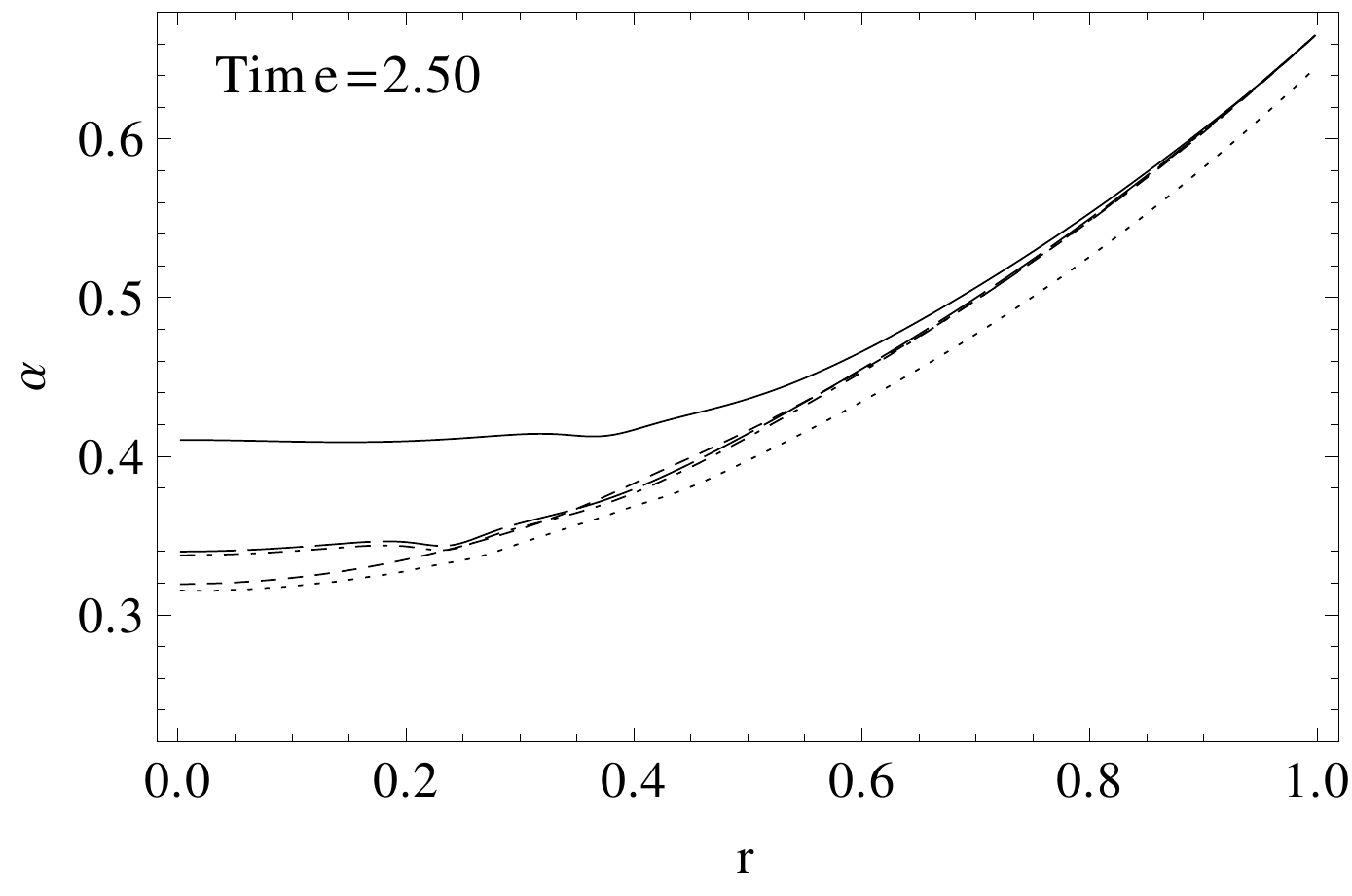}}&
\vspace{-7.0ex} \hspace{-0.6ex} \mbox{\includegraphics[width=1.06\linewidth]{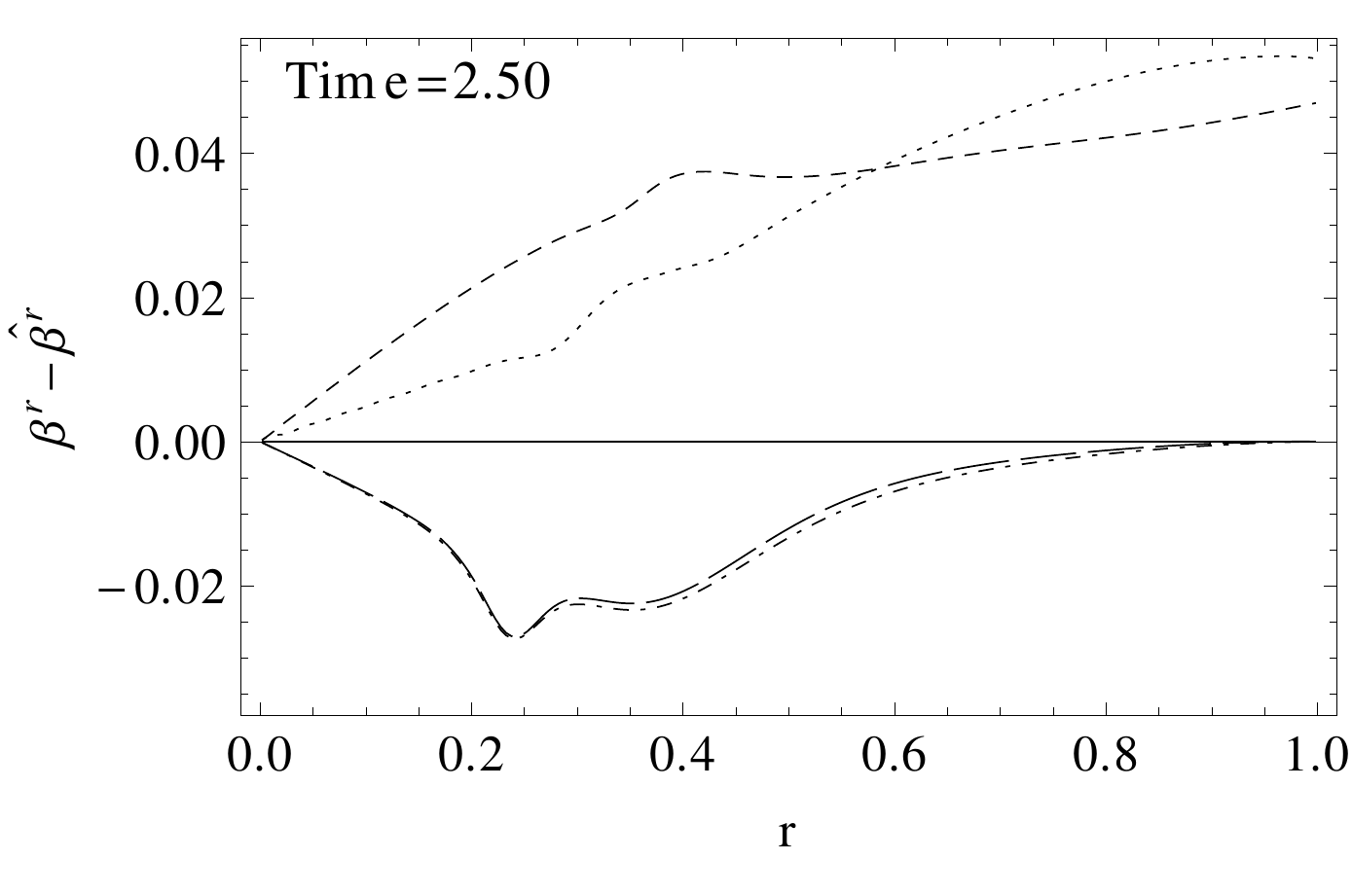}}\\
\vspace{-7.0ex} \mbox{\includegraphics[width=1\linewidth]{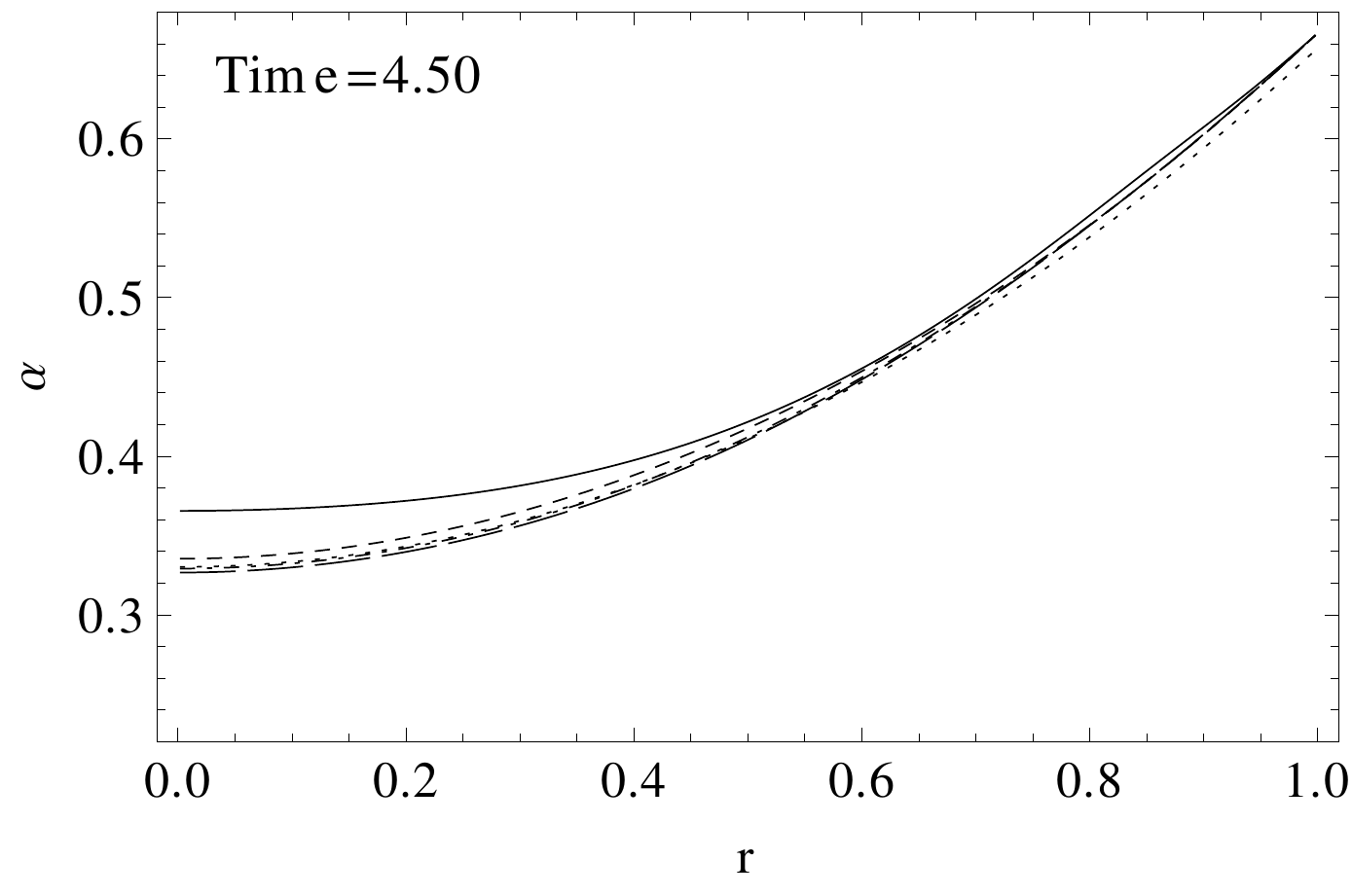}}&
\vspace{-7.0ex} \hspace{-0.6ex} \mbox{\includegraphics[width=1.06\linewidth]{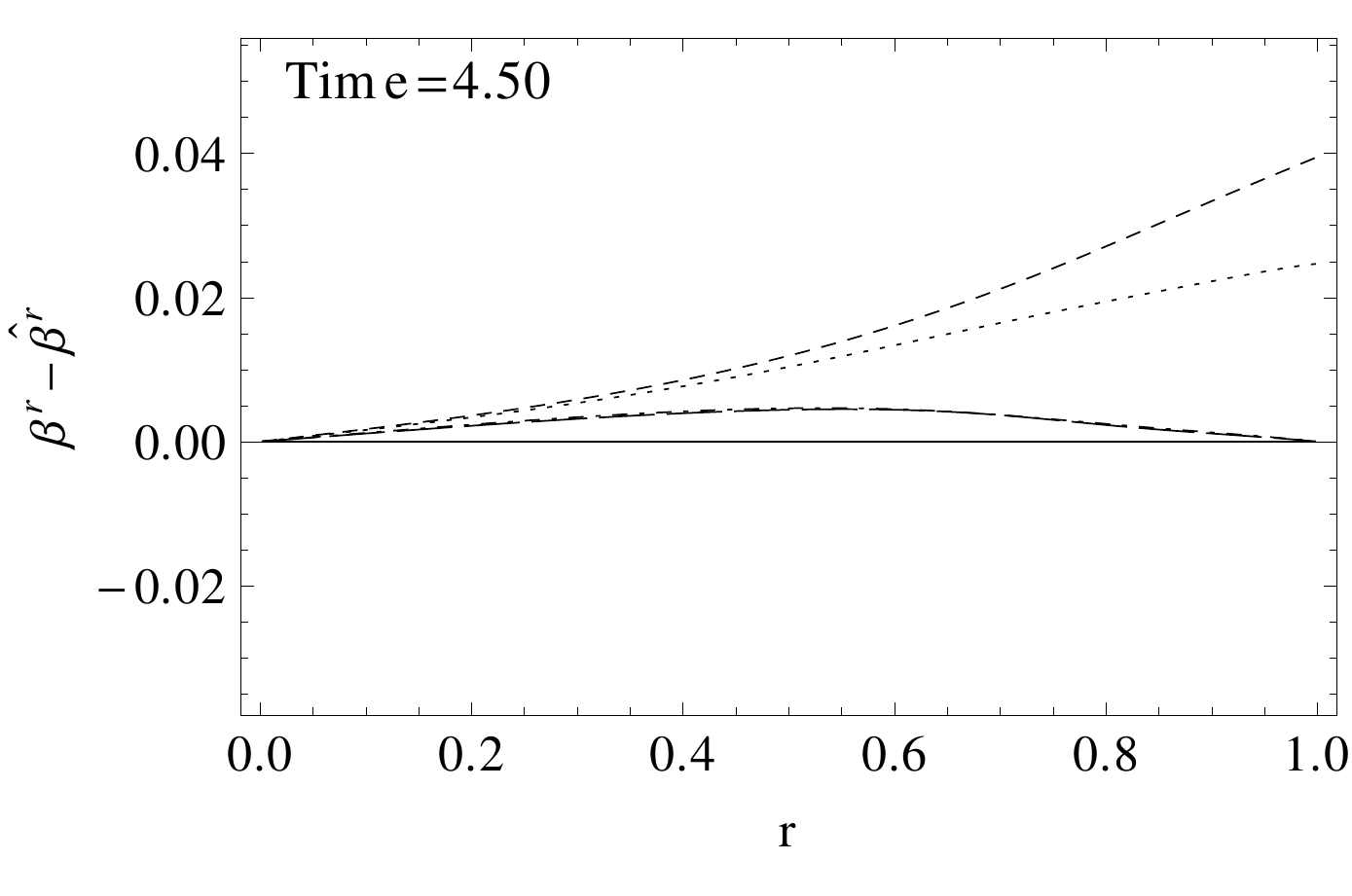}}\\
\vspace{-7.0ex} \mbox{\includegraphics[width=1\linewidth]{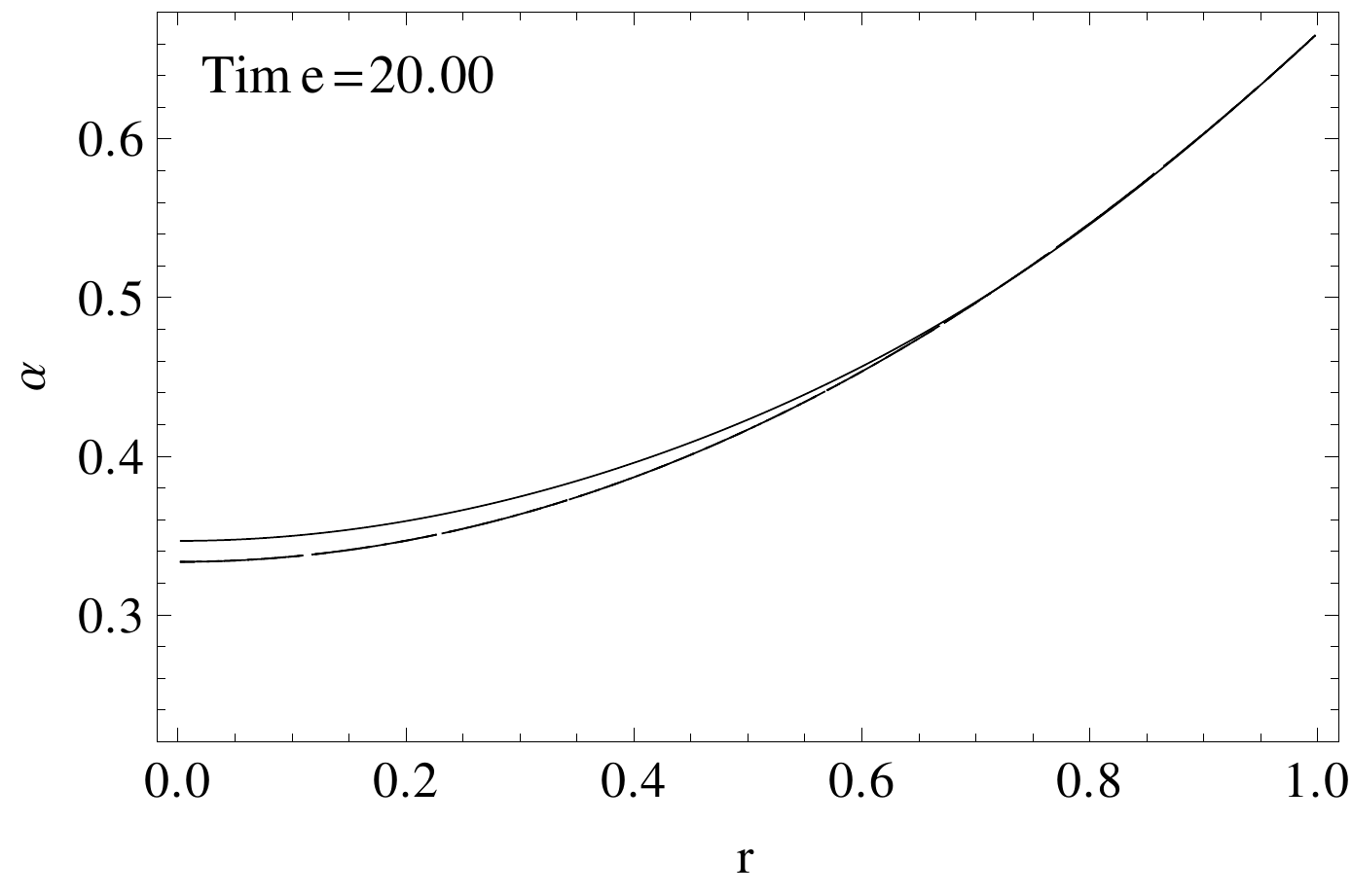}}&
\vspace{-7.0ex} \hspace{-0.6ex} \mbox{\includegraphics[width=1.06\linewidth]{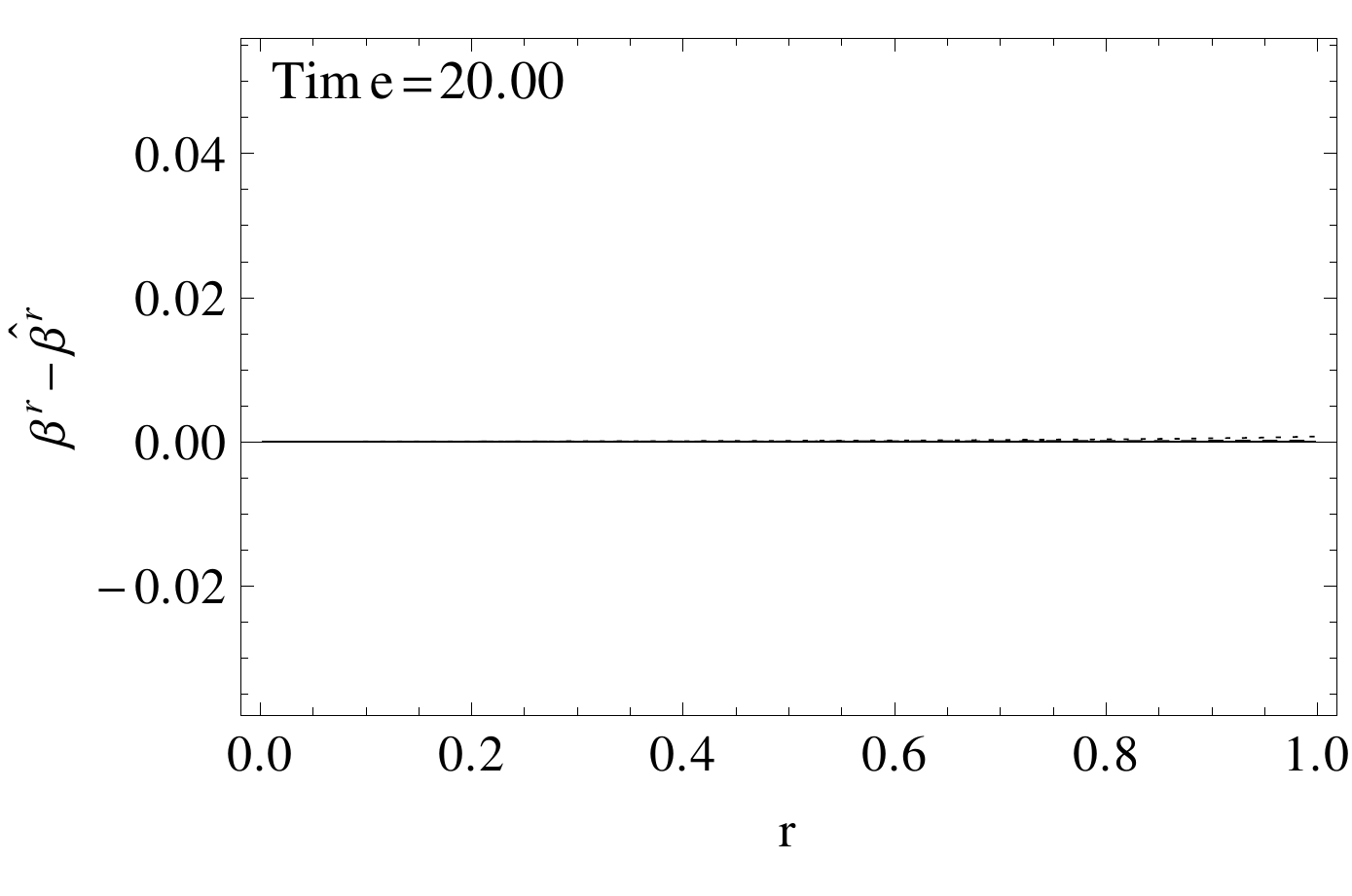}}
\end{tabular}
\vspace{-3ex}
\caption{Evolution of $\alpha$ and $\beta^r$. The auxiliary variable $B^r$ used in the Gamma-driver shift is not shown because its behaviour is very similar to $\Lambda^r$ of the Gamma-driver case.}
\label{fs:shiftcondab}
\end{figure}
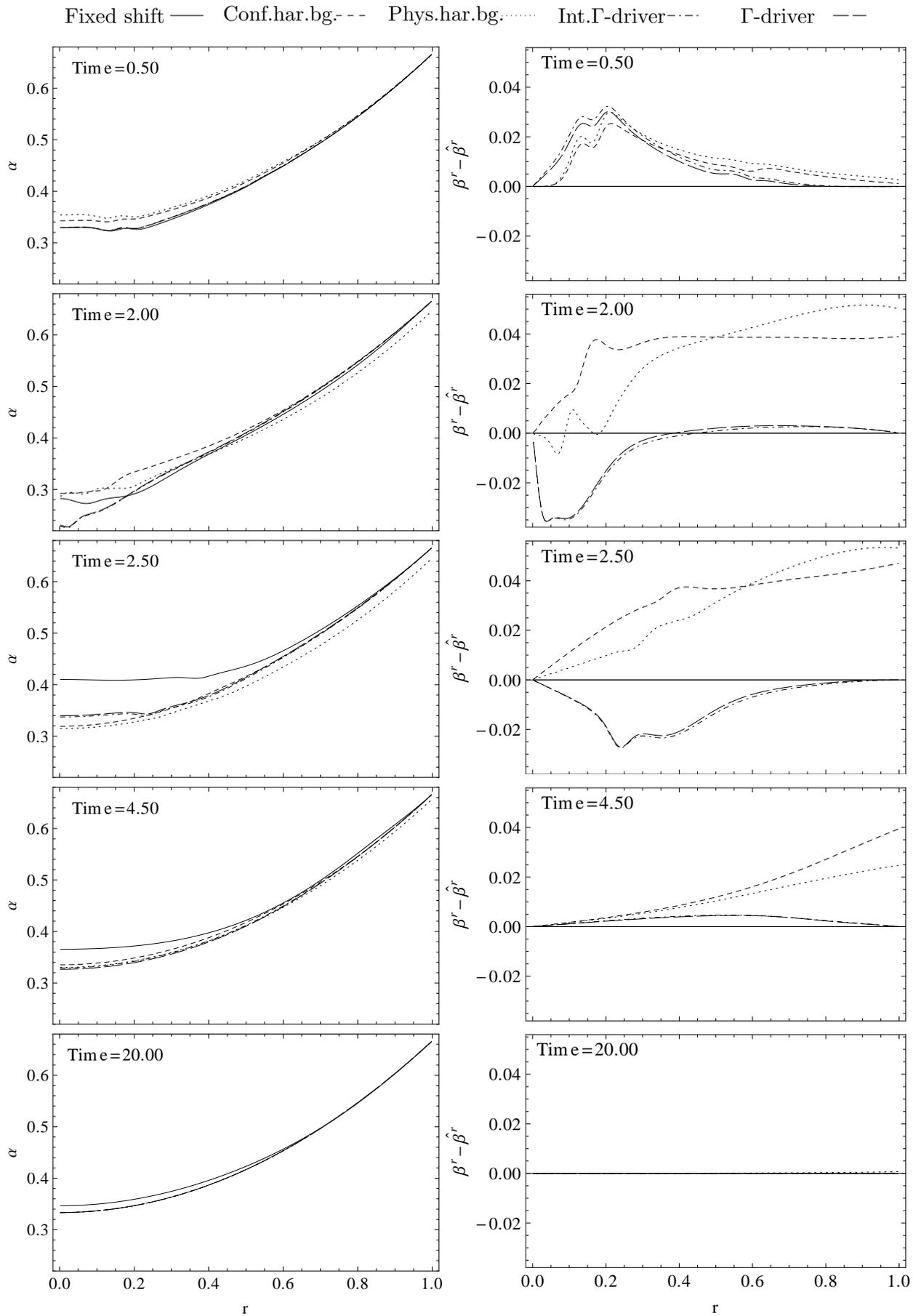

\begin{figure}[htbp!!]
\center
\vspace{-2ex}
 \begin{tikzpicture}[scale=2.0]\draw (-1cm,0cm) node {};
		\draw (0cm, 0cm) node {\small Fixed shift}; \draw (0.5cm, 0cm) -- (0.8cm, 0cm);
		\draw (1.5cm, 0cm) node {\small Conf.har.bg.}; \draw [dashed] (2cm, 0cm) -- (2.3cm, 0cm);
		\draw (3cm, 0cm) node {\small Phys.har.bg.}; \draw [dotted] (3.5cm, 0cm) -- (3.8cm, 0cm);
		\draw (4.5cm, 0cm) node {\small Int.$\Gamma$-driver}; \draw [dash pattern= on 4pt off 2pt on 1pt off 2pt] (5cm, 0cm) -- (5.3cm, 0cm);
		\draw (6cm, 0cm) node {\small $\Gamma$-driver}; \draw [dash pattern= on 8pt off 2pt] (6.5cm, 0cm) -- (6.8cm, 0cm);
	%	\draw (7.5cm, 0cm) node {$\alpha$}; \draw [dash pattern= on 6pt off 2pt on 1pt off 2pt on 1pt off 2pt] (7.8cm, 0cm) -- (8.5cm, 0cm);
	\end{tikzpicture}
\\
\begin{tabular}{ m{0.5\linewidth}@{} @{}m{0.5\linewidth}@{} }
\mbox{\includegraphics[width=1\linewidth]{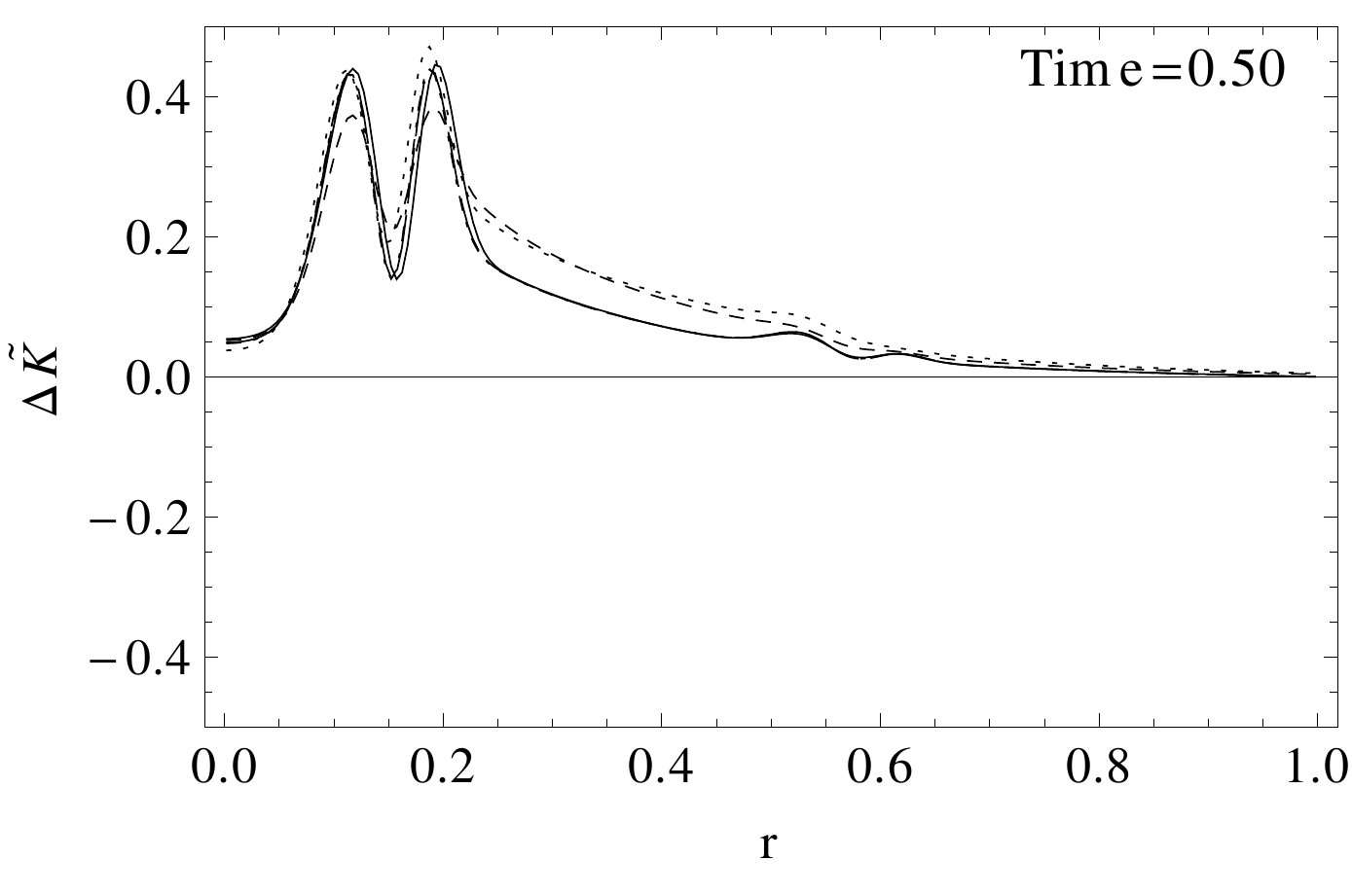}}&
\hspace{-0.6ex} \mbox{\includegraphics[width=1\linewidth]{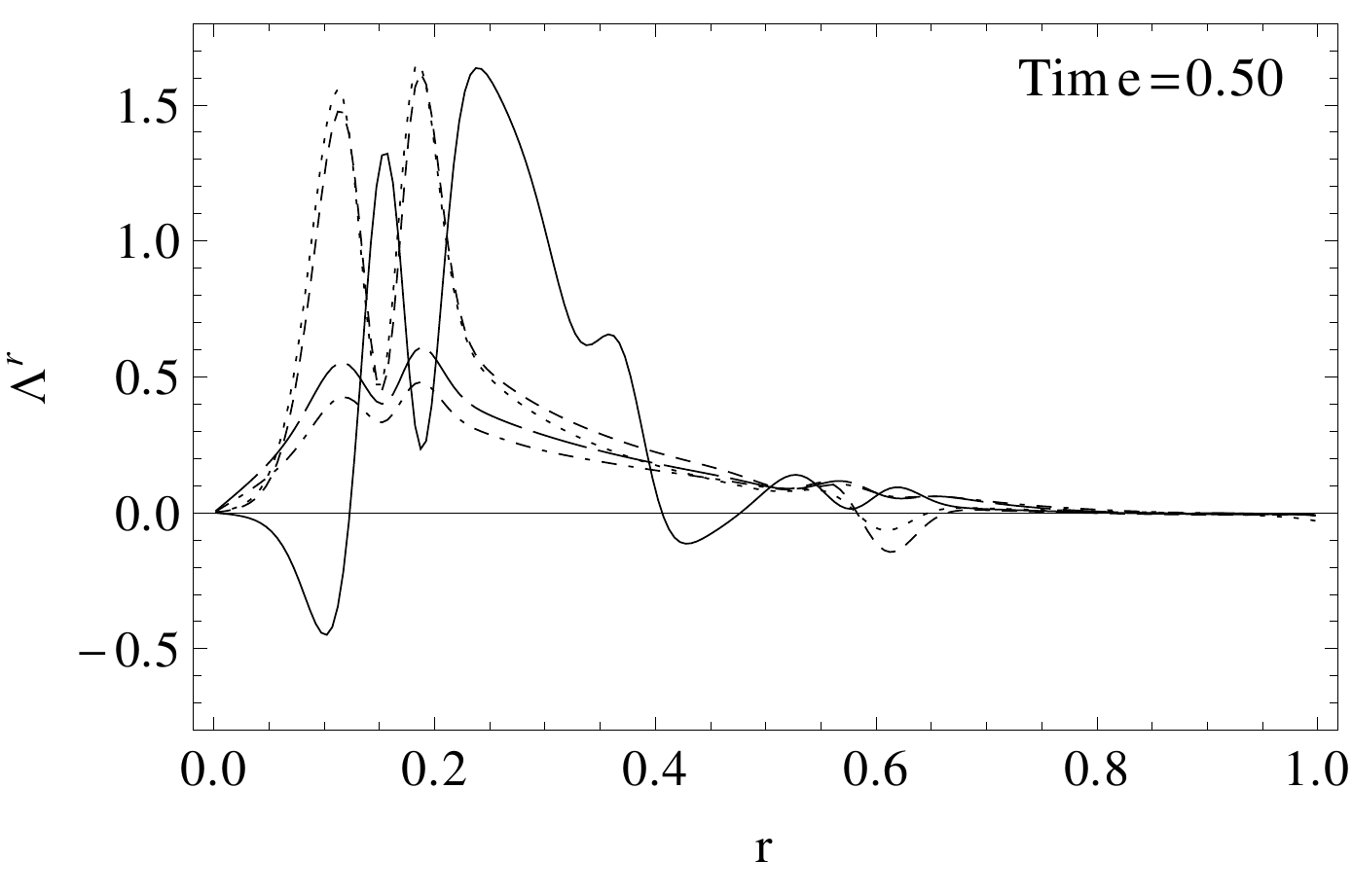}}\\
\vspace{-5.9ex} \mbox{\includegraphics[width=1\linewidth]{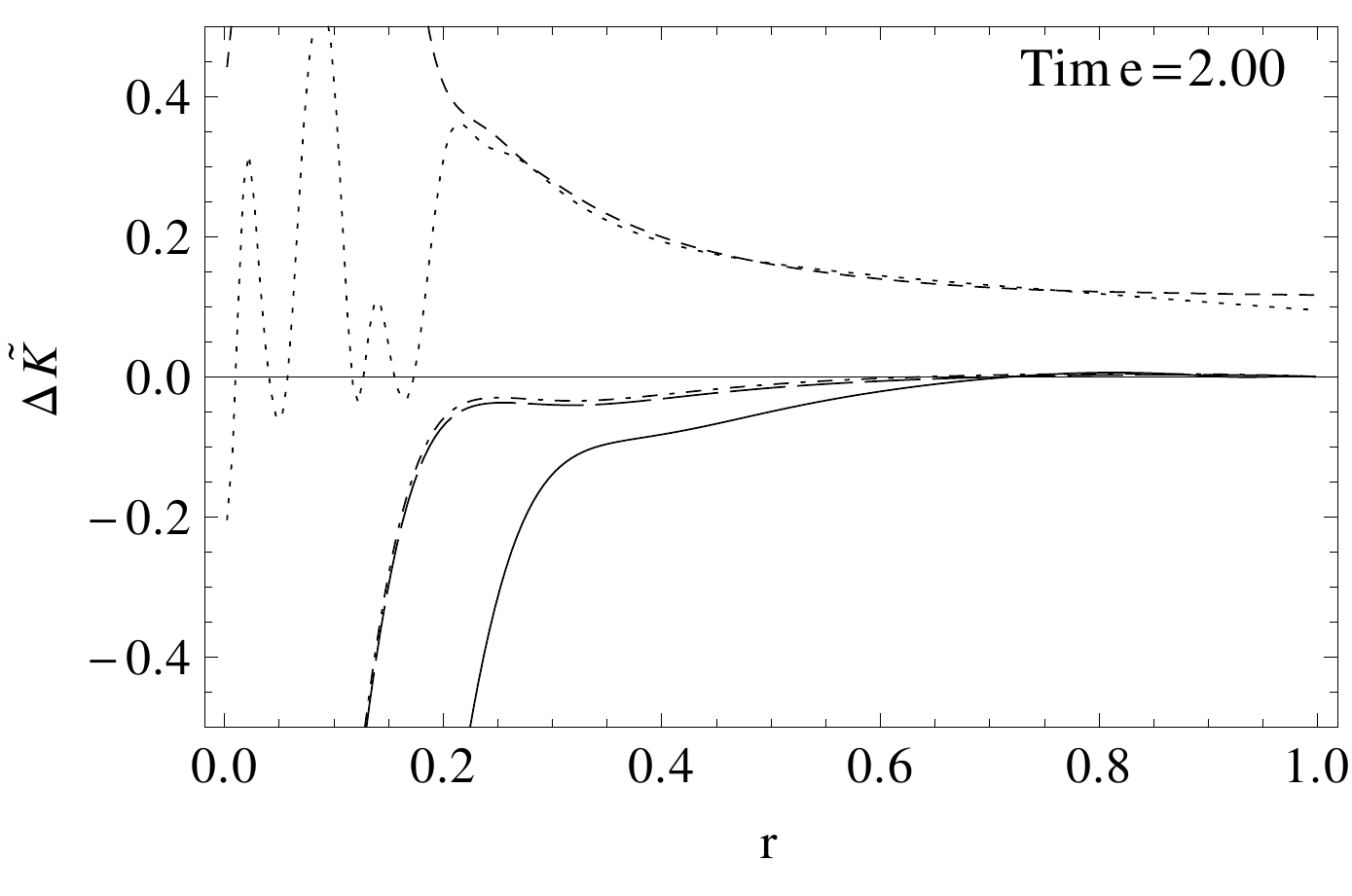}}&
\vspace{-5.9ex} \hspace{-0.6ex} \mbox{\includegraphics[width=1\linewidth]{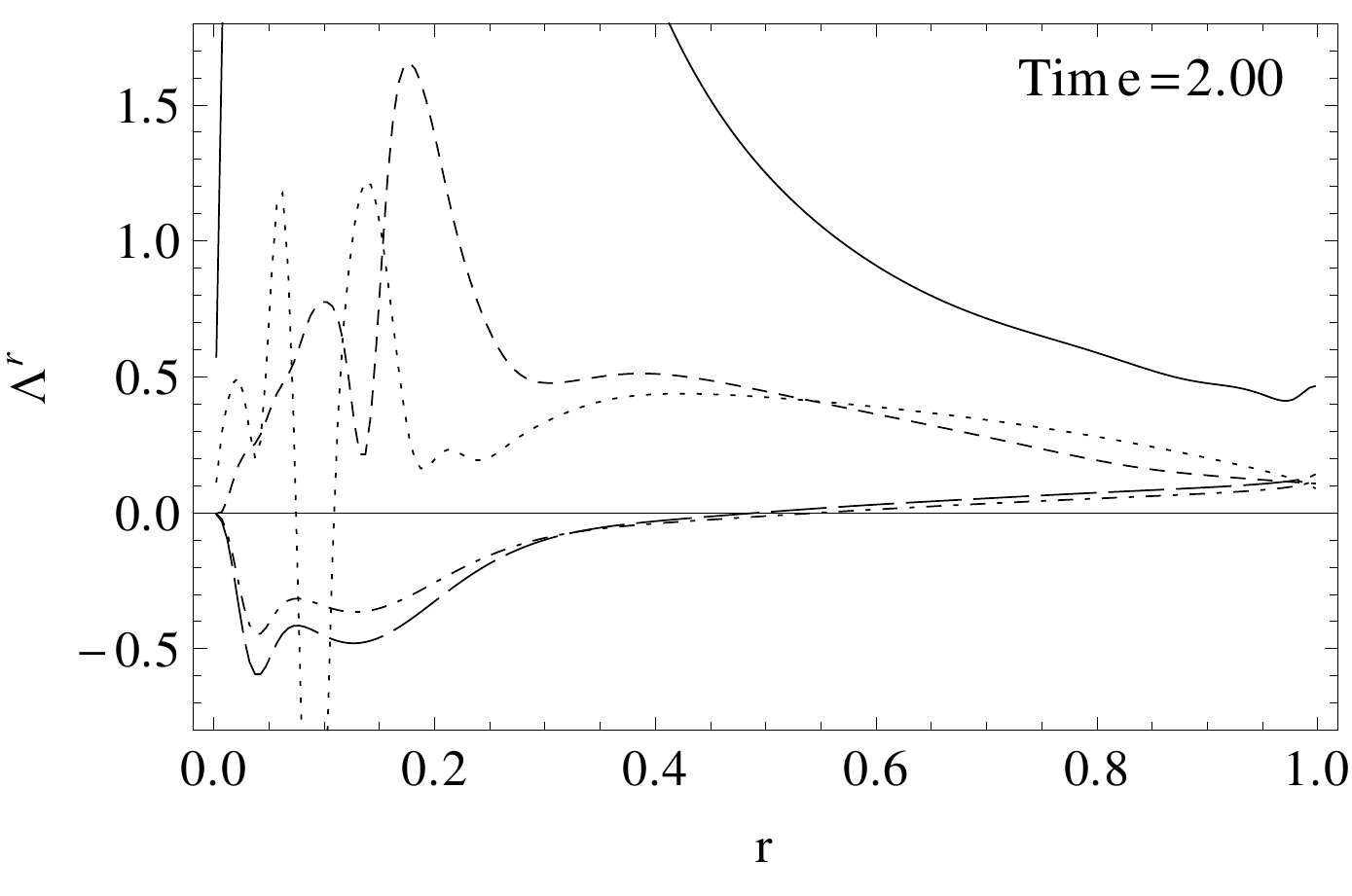}}\\
\vspace{-5.9ex} \mbox{\includegraphics[width=1\linewidth]{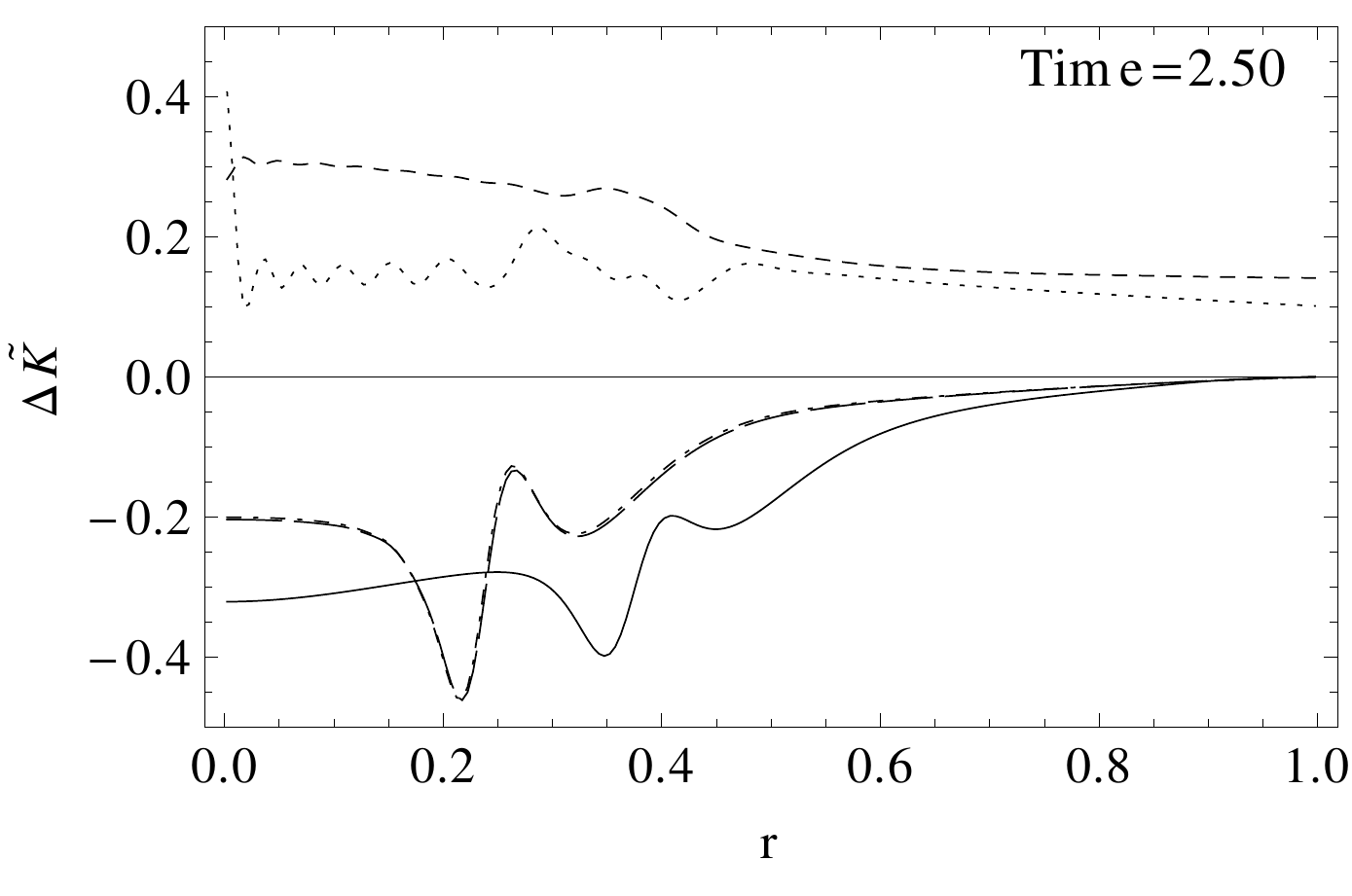}}&
\vspace{-5.9ex} \hspace{-0.6ex} \mbox{\includegraphics[width=1\linewidth]{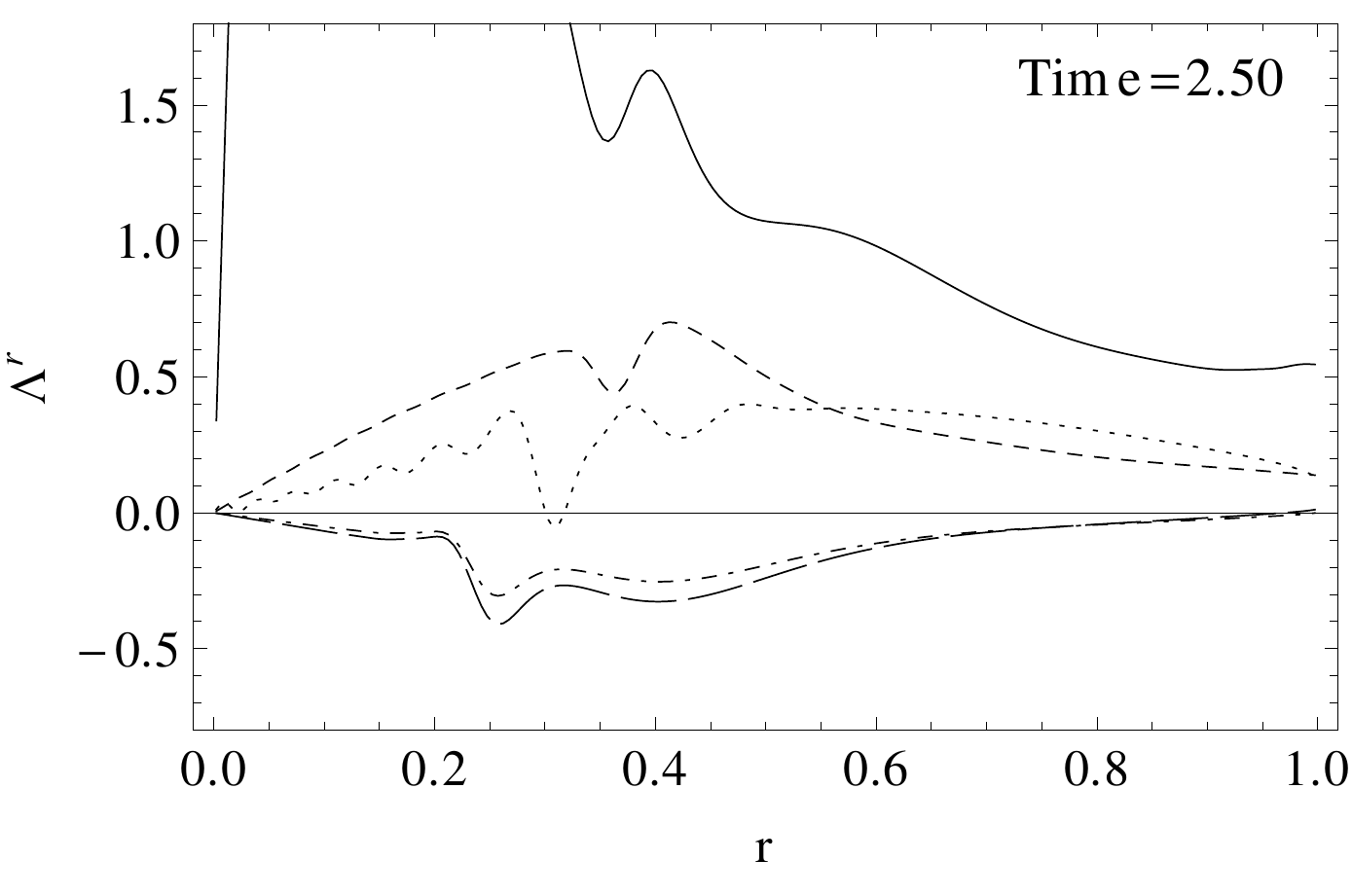}}\\
\vspace{-5.9ex} \mbox{\includegraphics[width=1\linewidth]{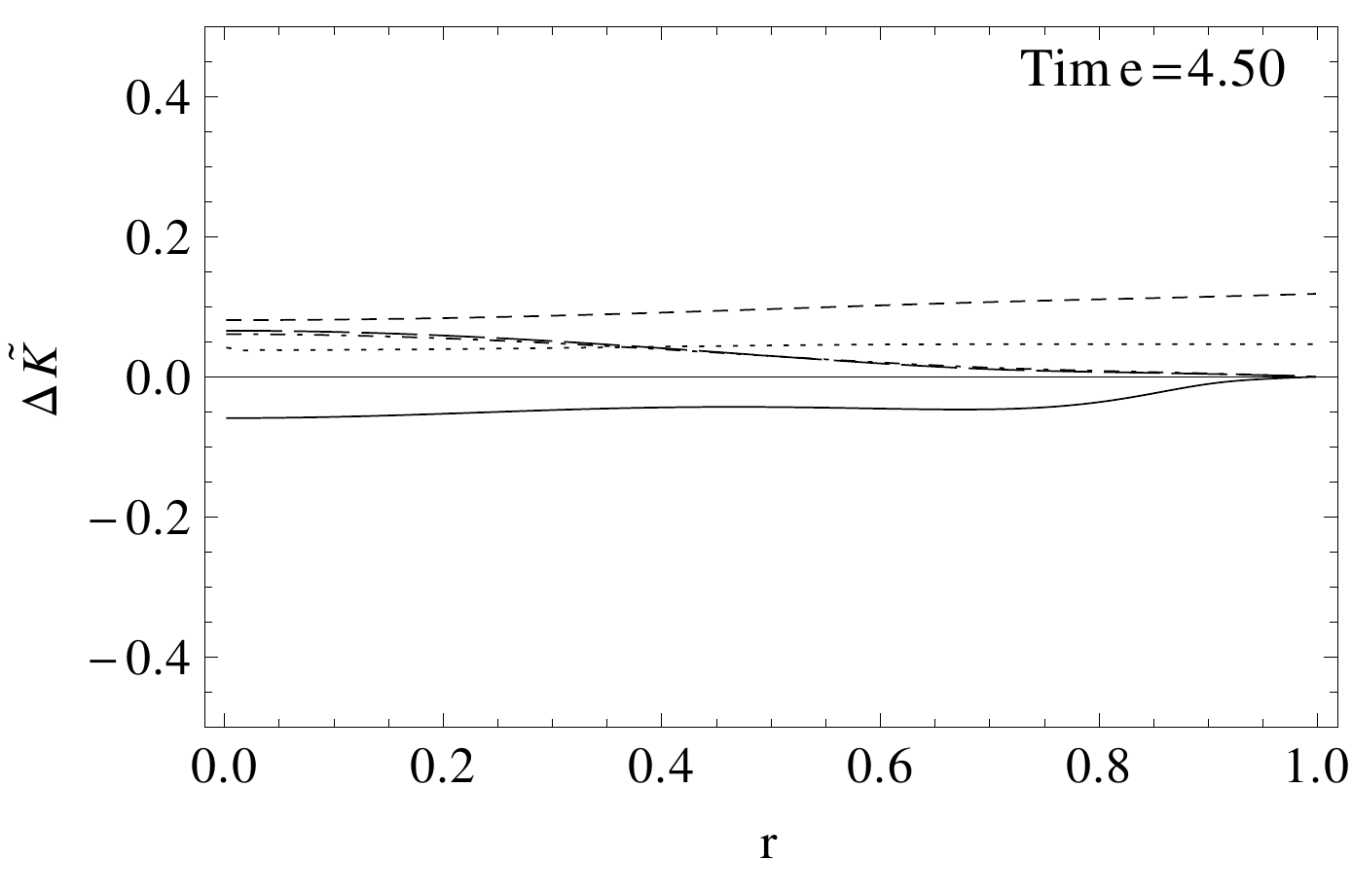}}&
\vspace{-5.9ex} \hspace{-0.6ex} \mbox{\includegraphics[width=1\linewidth]{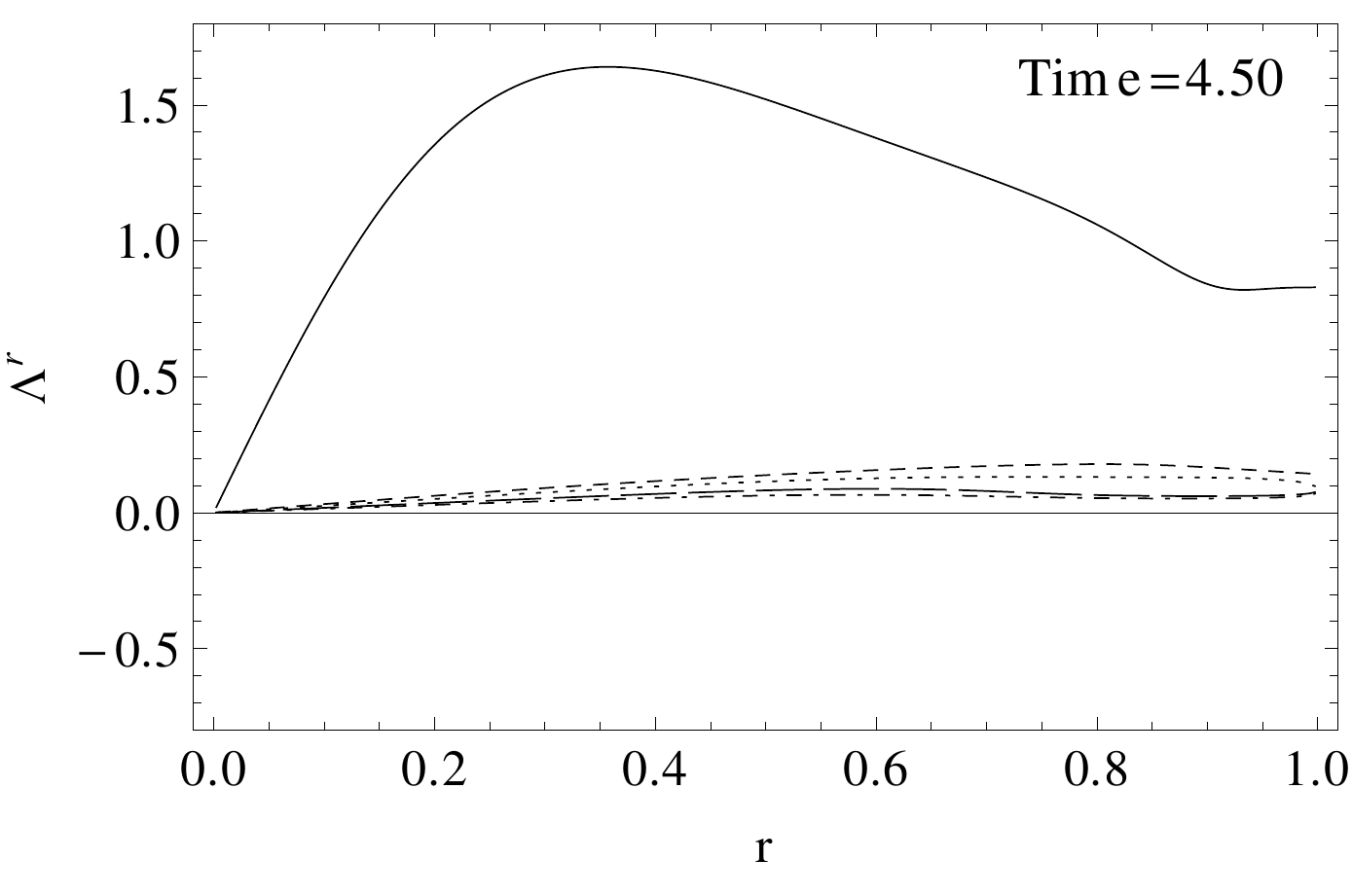}}\\
\vspace{-5.9ex} \mbox{\includegraphics[width=1\linewidth]{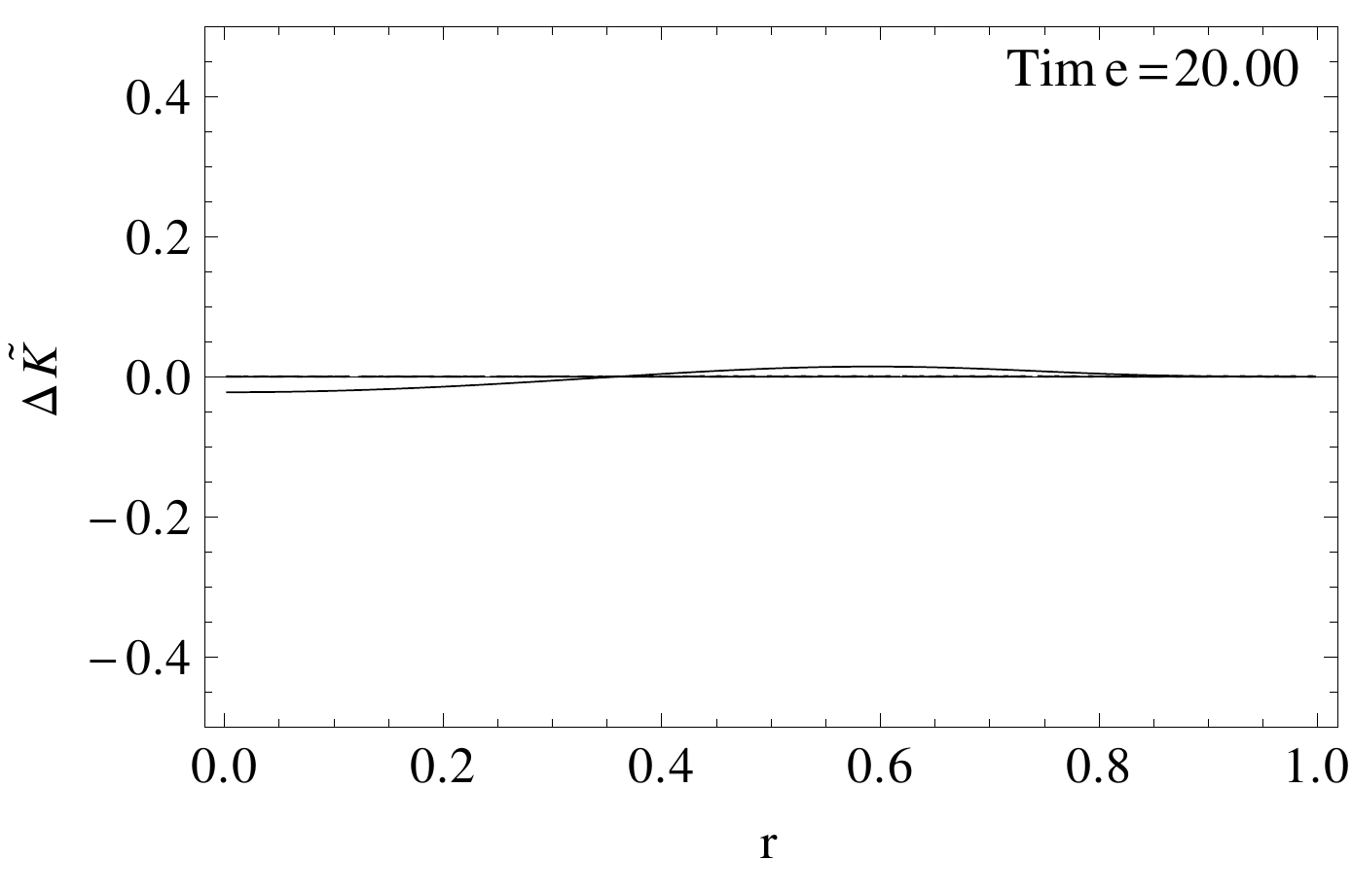}}&
\vspace{-5.9ex} \hspace{-0.6ex} \mbox{\includegraphics[width=1\linewidth]{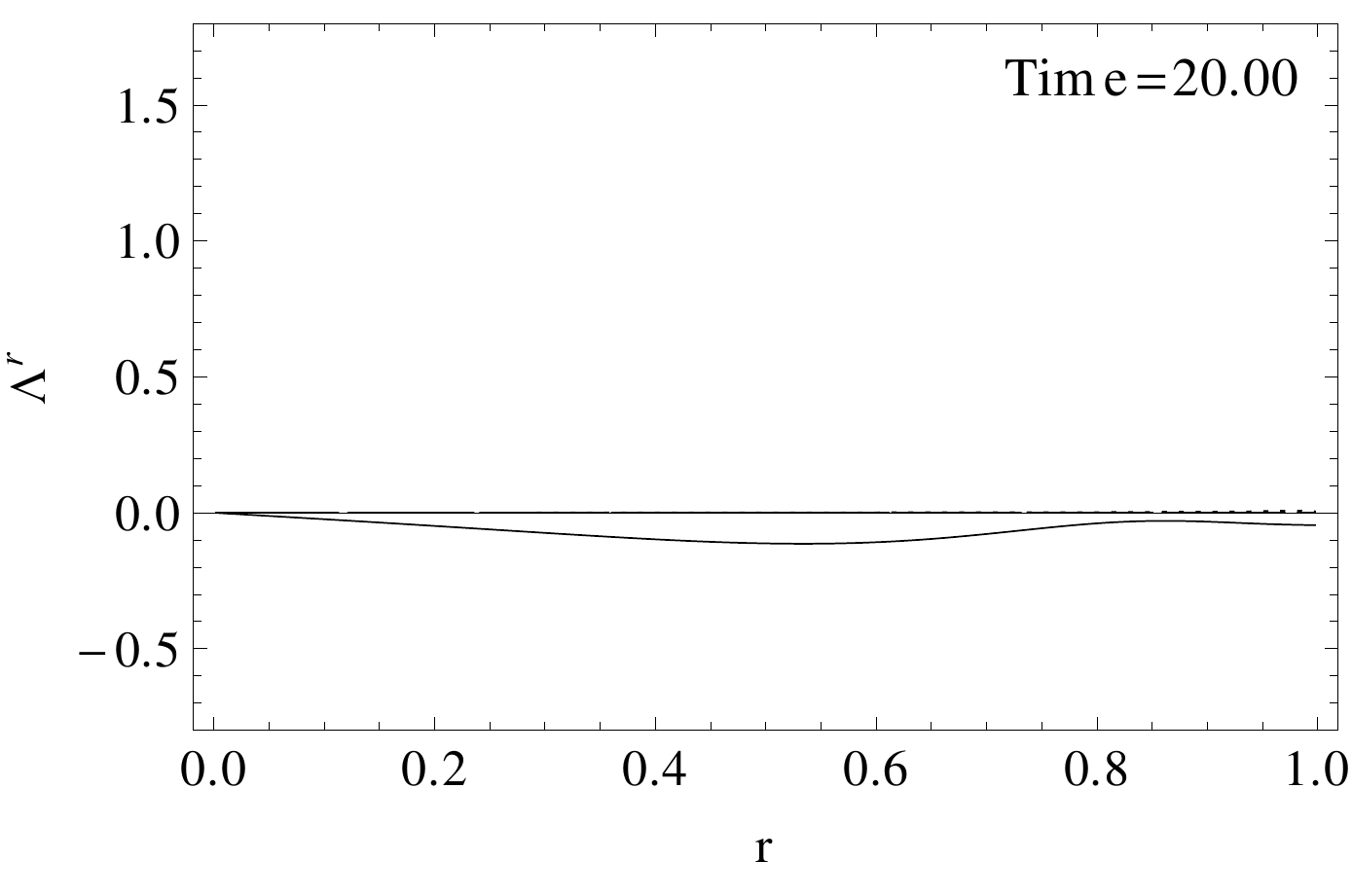}}
\end{tabular}
\vspace{-2ex}
\caption{Evolution of $\DPK$ and $\Lambda^r$.}
\label{fs:shiftcondKL}
\end{figure}

Instead of showing the behaviour of all the variables, I selected the ones I think are more interesting to compare, namely the gauge variables $\alpha$ and $\beta^r$, $\DPK$ and $\Lambda^r$. The shift auxiliary variable $B^r$ that appears in \eref{ee:expGammadriver} is not plotted, because its behaviour is very similar to that of $\Lambda^r$. Instead of plotting $\beta^r$, what is shown is its variation with respect to its stationary value for flat spacetime, so $\beta^r-\hat\beta^r\equiv\beta^r-\case{\Kc r}{3}$.
The comparison between the variables is shown in figures \ref{fs:shiftcondab} and \ref{fs:shiftcondKL}. The initial data of these simulations is exactly the same as in \fref{fin:Finic} or in \fref{fs:allevolscacent1}.

Regarding $\alpha$, the main point to mention is that the only case where $\atscrip{\alpha}$ is not fixed is for the physical harmonic with background source conditions. This relates to the regularity conditions described in subsection \ref{er:gaugereqs}. The previously mentioned fluctuations that appear for the fixed shift case can also be seen here in the solid line, which has not yet settled to its stationary value at $t=20$.

The two Gamma-driver gauge conditions, which also show a very similar behaviour during the evolution, are the only ones whose value at $\scri^+$ does not move in time, consequence of the damping terms with $\xi_{\beta^r}\neq0$ that are needed to keep the numerical simulation stable. The other two configurations allow the shift to move at $\scri^+$, always satisfying the regularity conditions.

If at least $\atscrip{\alpha}$ or $\atscri{\beta^r}$ is not fixed at $\scri^+$, then according to the regularity condition \eref{er:Kregrel} the variable $\DPK$ is allowed to take non-zero values at $\scri^+$. The latter is exactly what happens with the conformal and physical harmonic with background source conditions. Close to the origin, the physical condition shows a less smooth behaviour.

For the quantity $\Lambda^r$, the most relevant difference between the different gauge conditions is that in the fixed shift case, $\Lambda^r$ separates more from its stationary value and also needs more time to go back to it again.

% Comparison of shift conditions

{ \ }

\begin{figure}[htbp!!]
\center
\mbox{\includegraphics[width=0.8\linewidth]{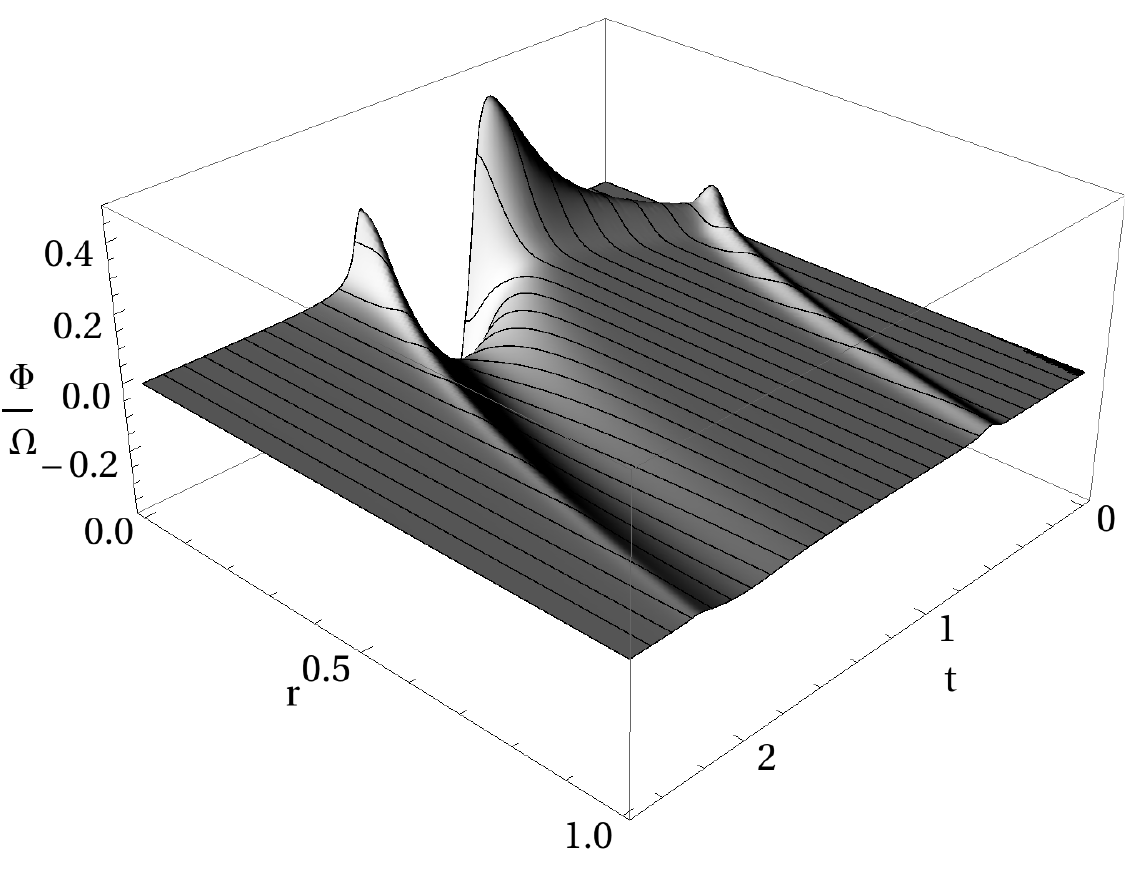}}\vspace{-2ex}
\caption{Behaviour of $\bPhi$ over time and compactified spatial coordinate.}
\label{fs:phirt}
\end{figure}

Figure \ref{fs:phirt} shows the behaviour of the rescaled scalar field $\bPhi$ in the spacetime. The splitting of the initial perturbation and the reflection of the ingoing pulse at the origin can be clearly seen. An important point is that the propagation speed remains finite and non-zero as the pulses approach $\scri^+$, as can be seen in the outgoing pulses.

\begin{figure}[htbp!!]
\center\vspace{-1ex}
\begin{tabular}{@{}c@{}@{}c@{}}
\mbox{\includegraphics[width=0.83\linewidth]{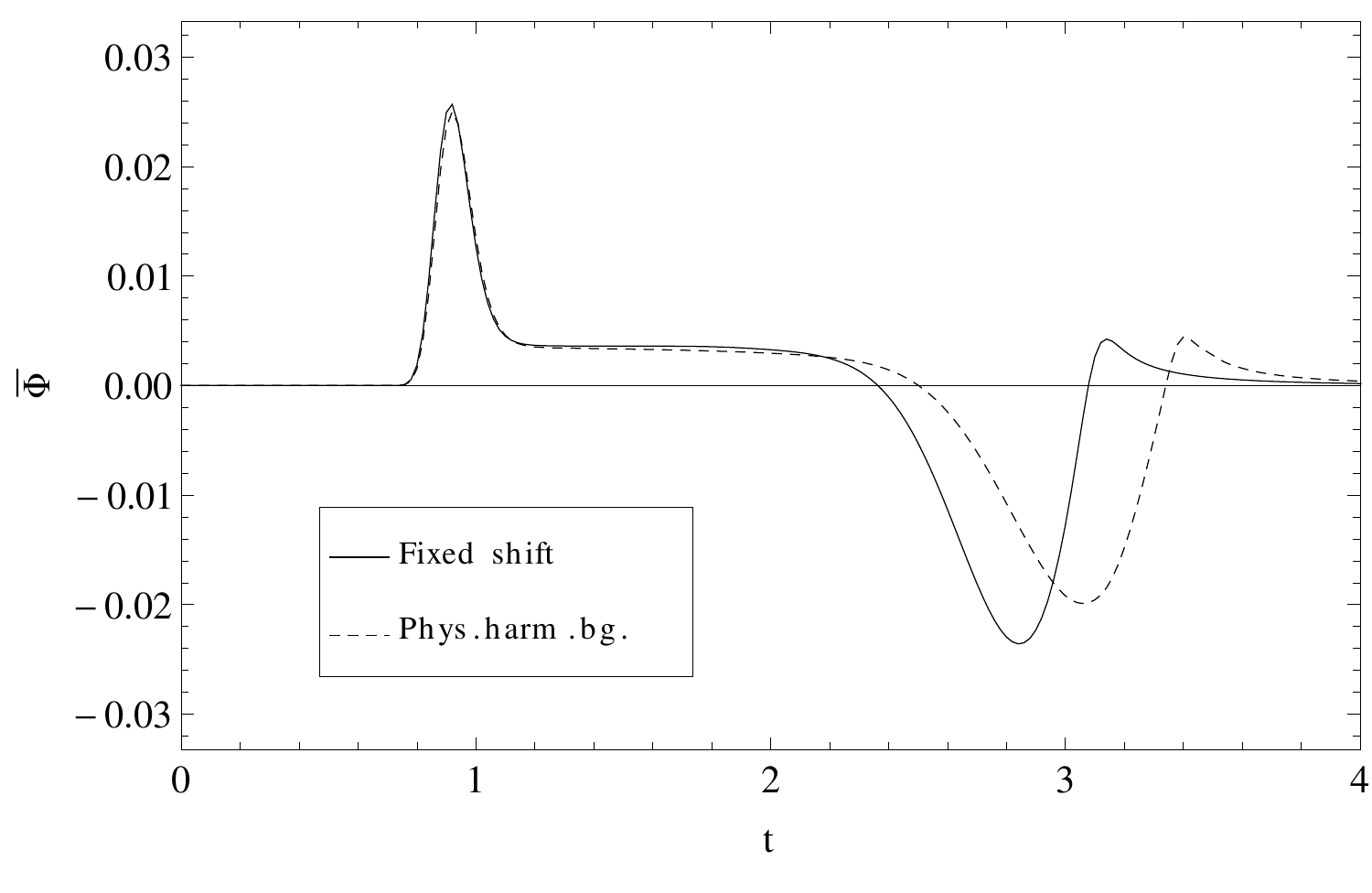}}
\end{tabular}\vspace{-4ex}
\caption{Signal of $\bPhi$ over time at $\scri^+$. The signal with the solid line corresponds to a simulation without preferred conformal gauge, while in the simulation for the dashed line the preferred conformal gauge was satisfied.}
\label{fs:phiscri}
\end{figure}

The signal of the scalar field at $\scri^+$ (calculated using 4th order extrapolation) is shown in \fref{fs:phiscri}. In the previously presented evolution with fixed shift, the preferred conformal gauge \eref{eg:condother} is not satisfied. This means that our time coordinate is not affinely parametrized and the signal we extract at $\scri^+$ will look deformed. This is the case of the solid line in \fref{fs:phiscri}. The dashed one shows the value at $\scri^+$ of the scalar field from a simulation with the harmonic gauge conditions with physical background source terms \eref{ee:improvedpbg}, which indeed does satisfy the preferred conformal gauge (at least for the choice of \CZ{} ($C_{Z4c}=0$)) - in order to on top obtain an affine time, the variation of the gauge variables at $\scri^+$ has to be taken into account. The difference between both signals not only appears in the rescaling of the time, but also in the amplitude of the signal.

\subsection{Collapse of the scalar field perturbation into a black hole}\label{ss:col}

\begin{figure}[htbp!!]
\center\vspace{-1.5ex}
\mbox{\includegraphics[width=0.65\linewidth]{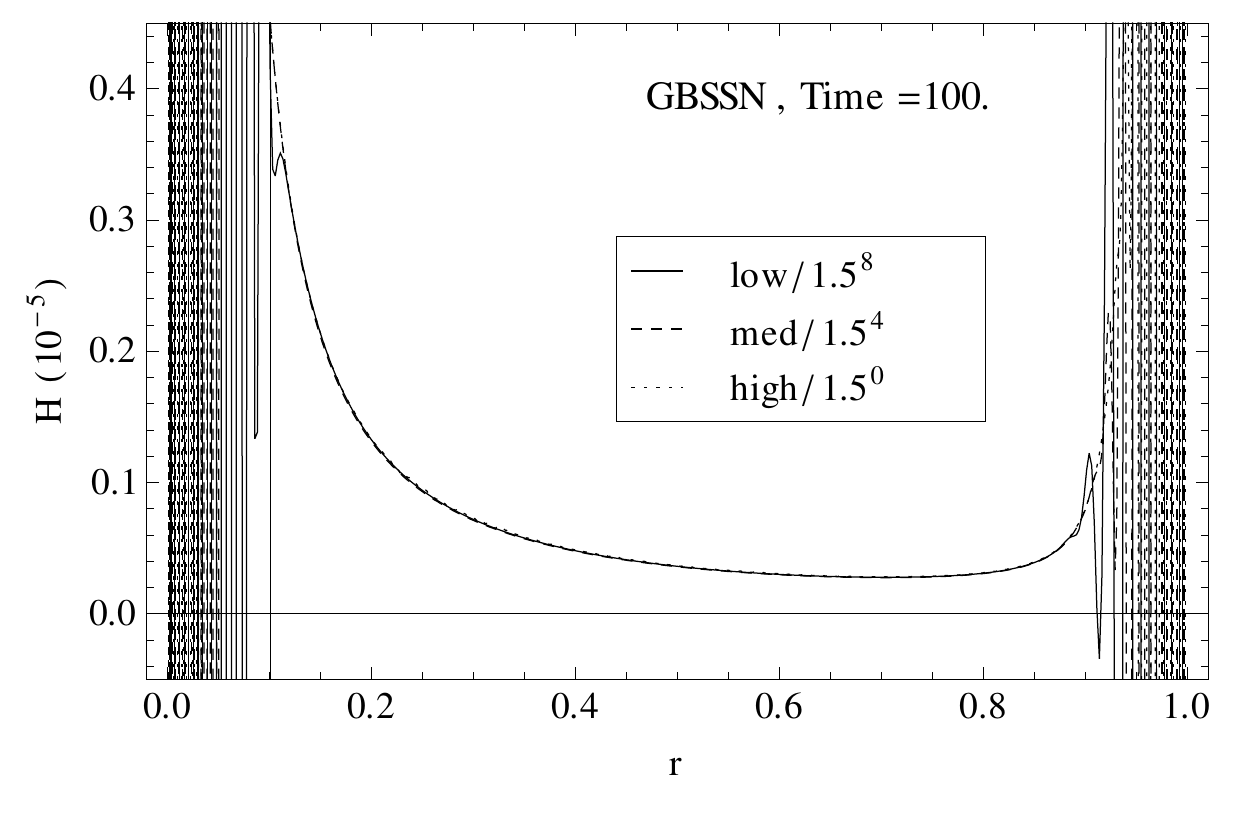}}\vspace{-3.7ex}
\caption{Convergence of the Hamiltonian constraint at a time where the stable stationary end state of the collapse has already been reached.}
\label{fs:colH}
\end{figure}

\begin{figure}[htbp!!]
\center
\vspace{-2ex}
\begin{tabular}{ m{0.5\linewidth}@{} @{}m{0.5\linewidth}@{} }
\begin{tikzpicture}[scale=1.0] \draw (-0.5cm,0cm) node {};
		\draw (0cm, 0cm) node {$\chi$}; \draw (0.3cm, 0cm) -- (1cm, 0cm);
		\draw (1.5cm, 0cm) node {$\gamma_{rr}$}; \draw [dashed] (1.8cm, 0cm) -- (2.5cm, 0cm);
		\draw (3cm, 0cm) node {$A_{rr}$}; \draw [dotted] (3.3cm, 0cm) -- (4cm, 0cm);
		\draw (4.5cm, 0cm) node {$\DPK$}; \draw [dash pattern= on 4pt off 2pt on 1pt off 2pt] (4.8cm, 0cm) -- (5.5cm, 0cm);
		\draw (6cm, 0cm) node {$\Lambda^r$}; \draw [dash pattern= on 8pt off 2pt] (6.3cm, 0cm) -- (7cm, 0cm);
	\end{tikzpicture}
&
\begin{tikzpicture}[scale=1.0] \draw (-1.3cm,0cm) node {};
		\draw (0cm, 0cm) node {$\alpha$}; \draw (0.3cm, 0cm) -- (1cm, 0cm);
		\draw (1.5cm, 0cm) node {$\beta^r$}; \draw [dashed] (1.8cm, 0cm) -- (2.5cm, 0cm);
		\draw (3cm, 0cm) node {$\bPhi$}; \draw [dotted] (3.3cm, 0cm) -- (4cm, 0cm);
		\draw (4.5cm, 0cm) node {$\bPi$}; \draw [dash pattern= on 4pt off 2pt on 1pt off 2pt] (4.8cm, 0cm) -- (5.5cm, 0cm);
	\end{tikzpicture}
\\
\mbox{\includegraphics[width=1\linewidth]{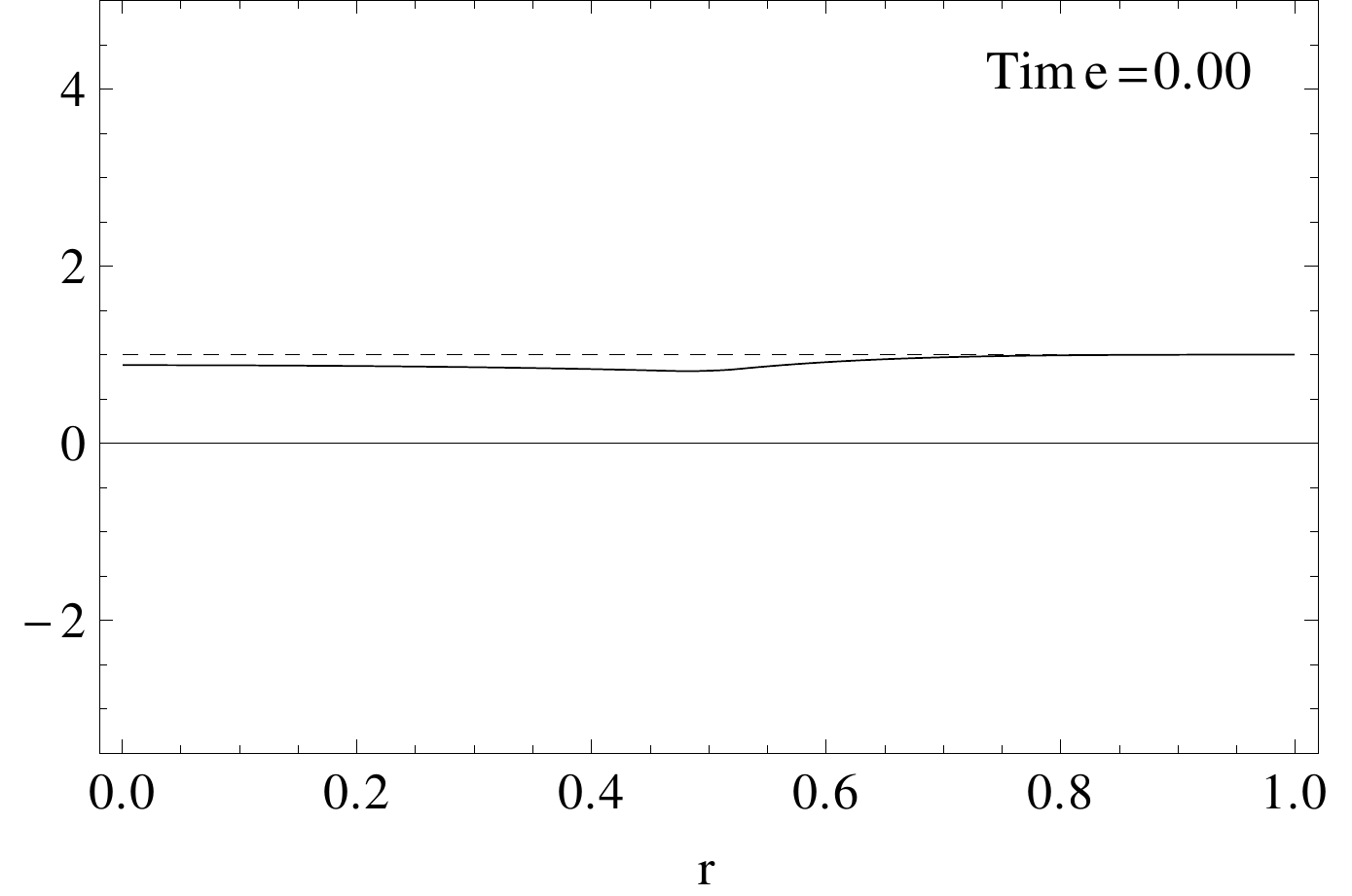}}&
\hspace{-0.8ex} \mbox{\includegraphics[width=1\linewidth]{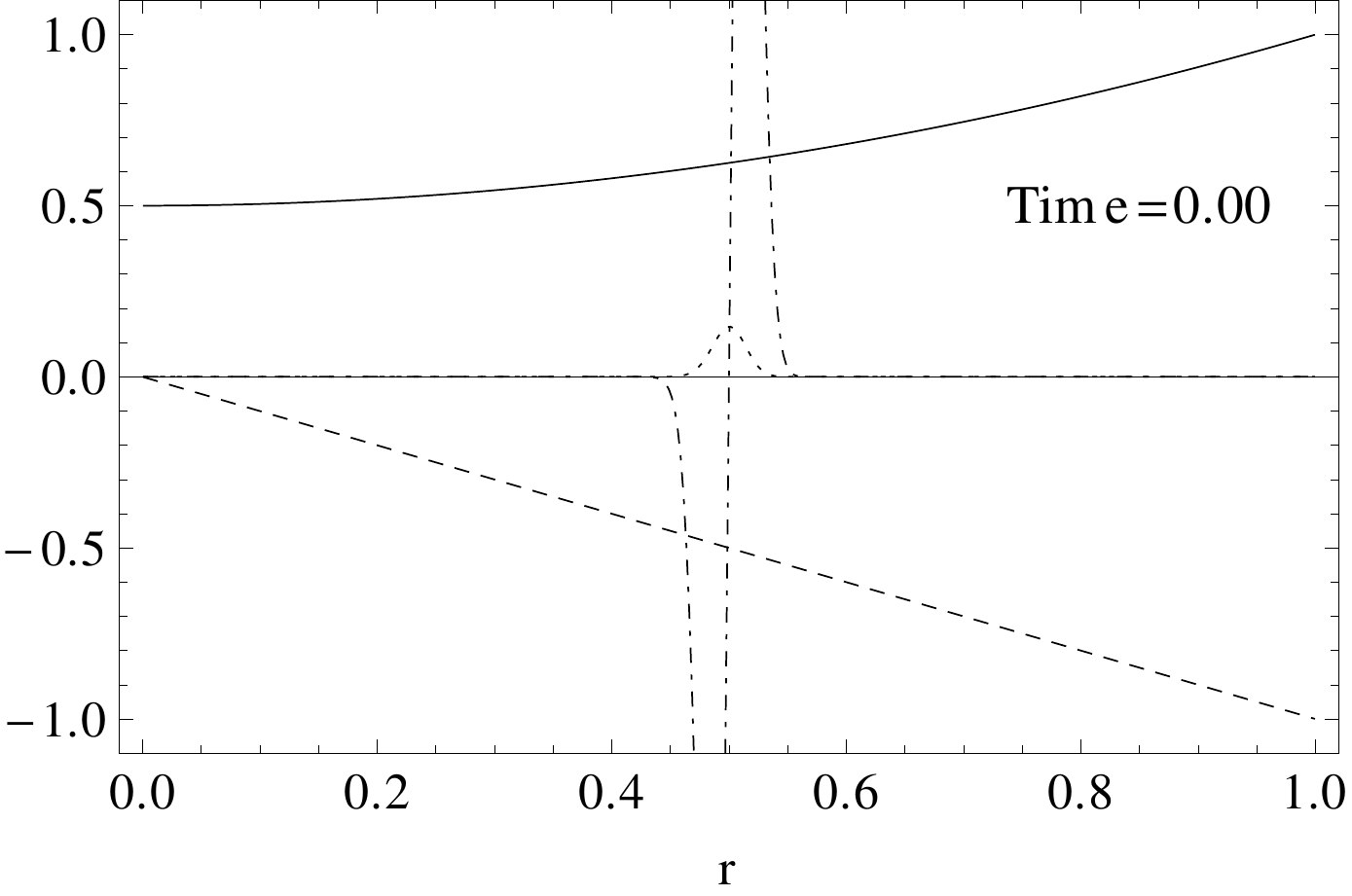}}\\
\vspace{-5.5ex} \mbox{\includegraphics[width=1\linewidth]{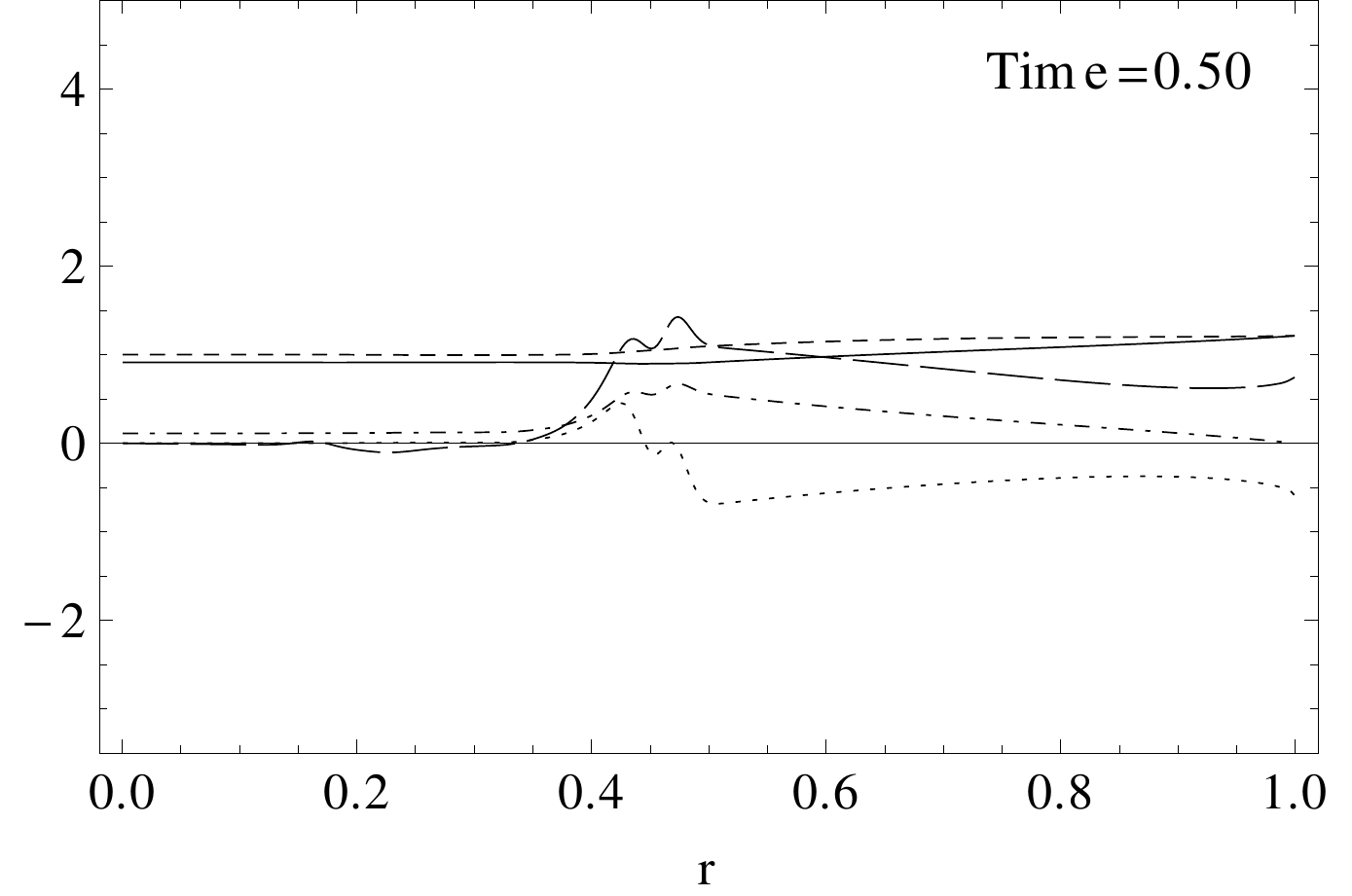}}&
\vspace{-5.5ex} \hspace{-0.8ex} \mbox{\includegraphics[width=1\linewidth]{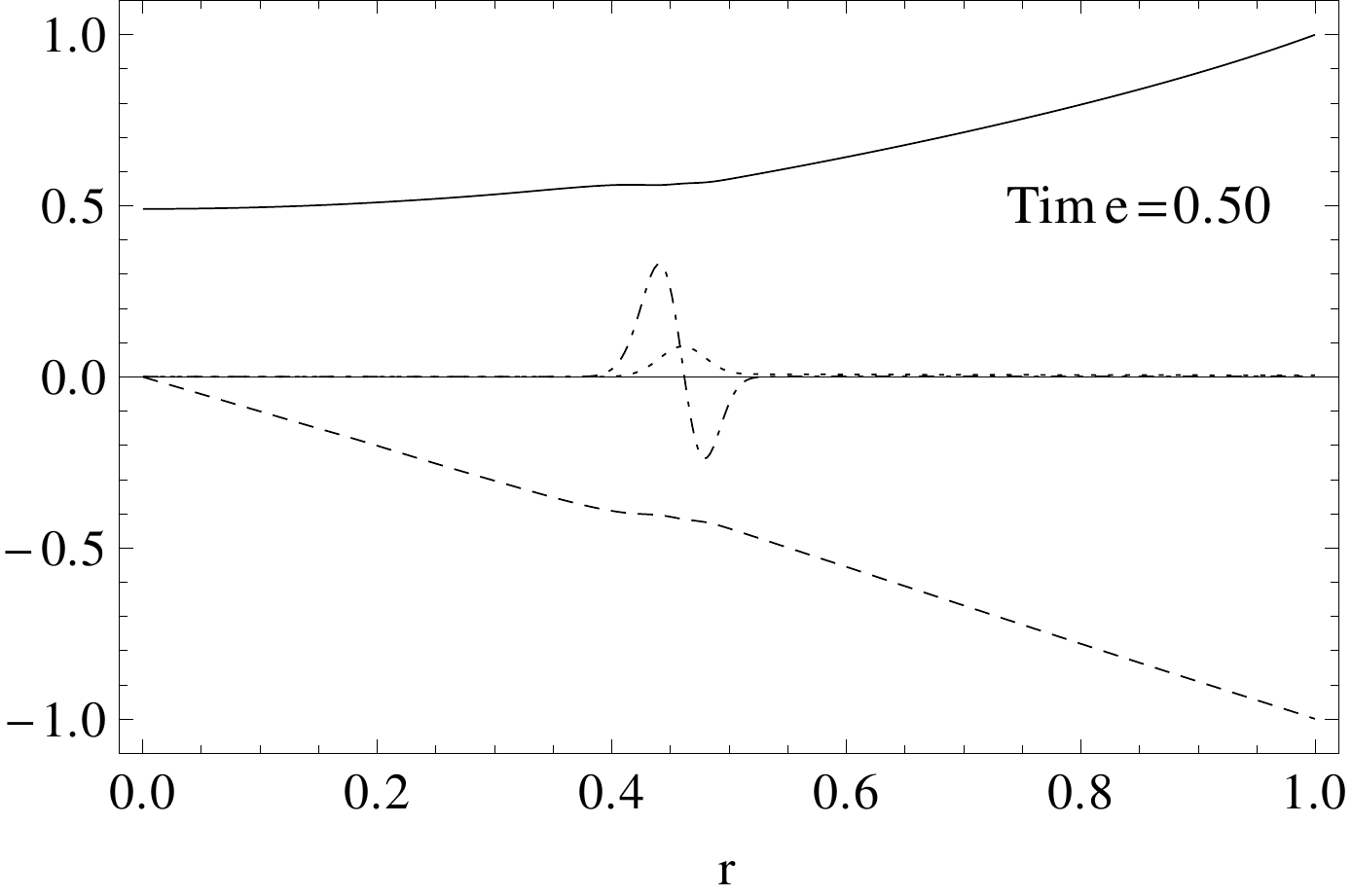}}\\
\vspace{-5.5ex} \mbox{\includegraphics[width=1\linewidth]{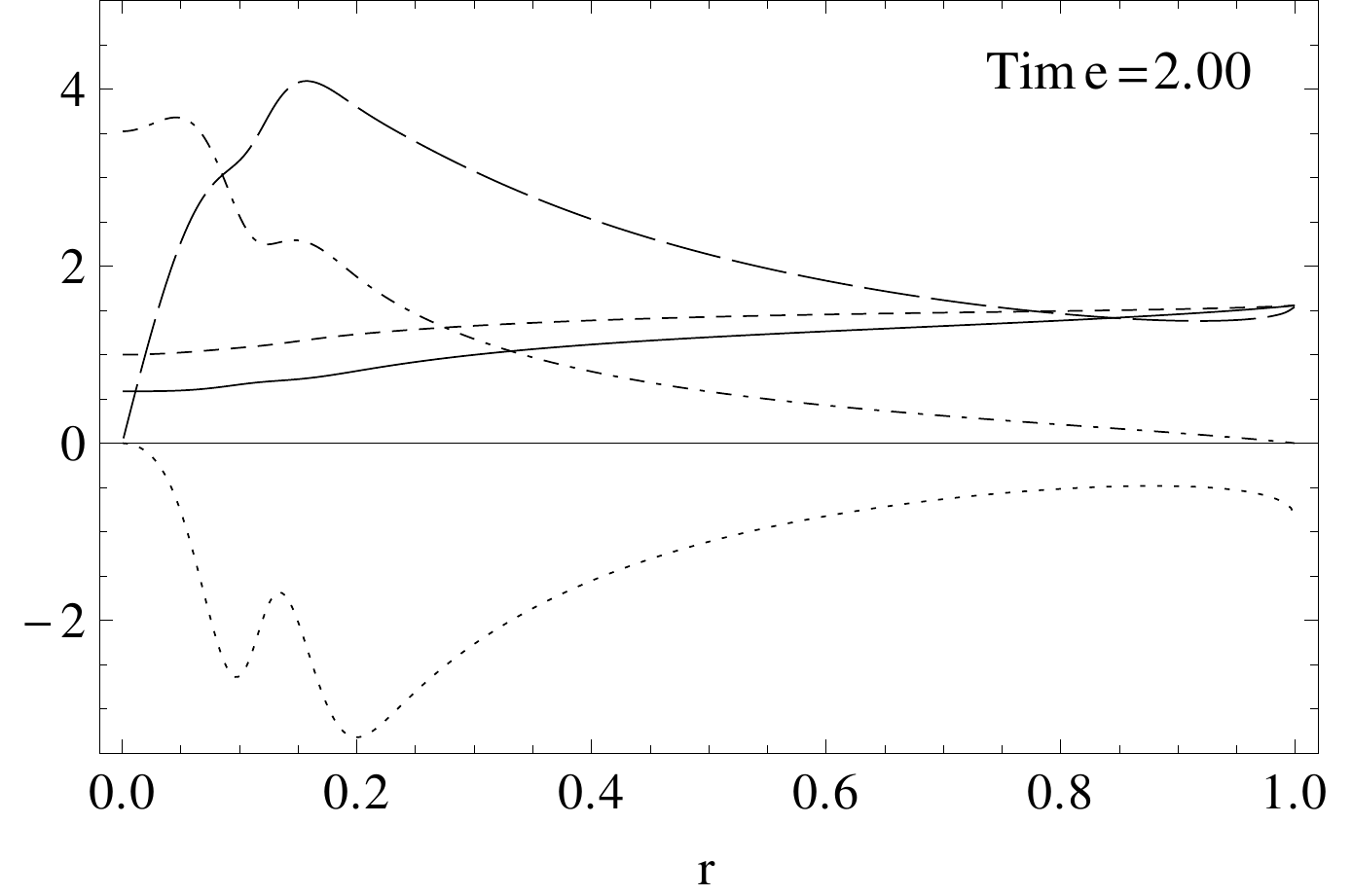}}&
\vspace{-5.5ex} \hspace{-0.8ex} \mbox{\includegraphics[width=1\linewidth]{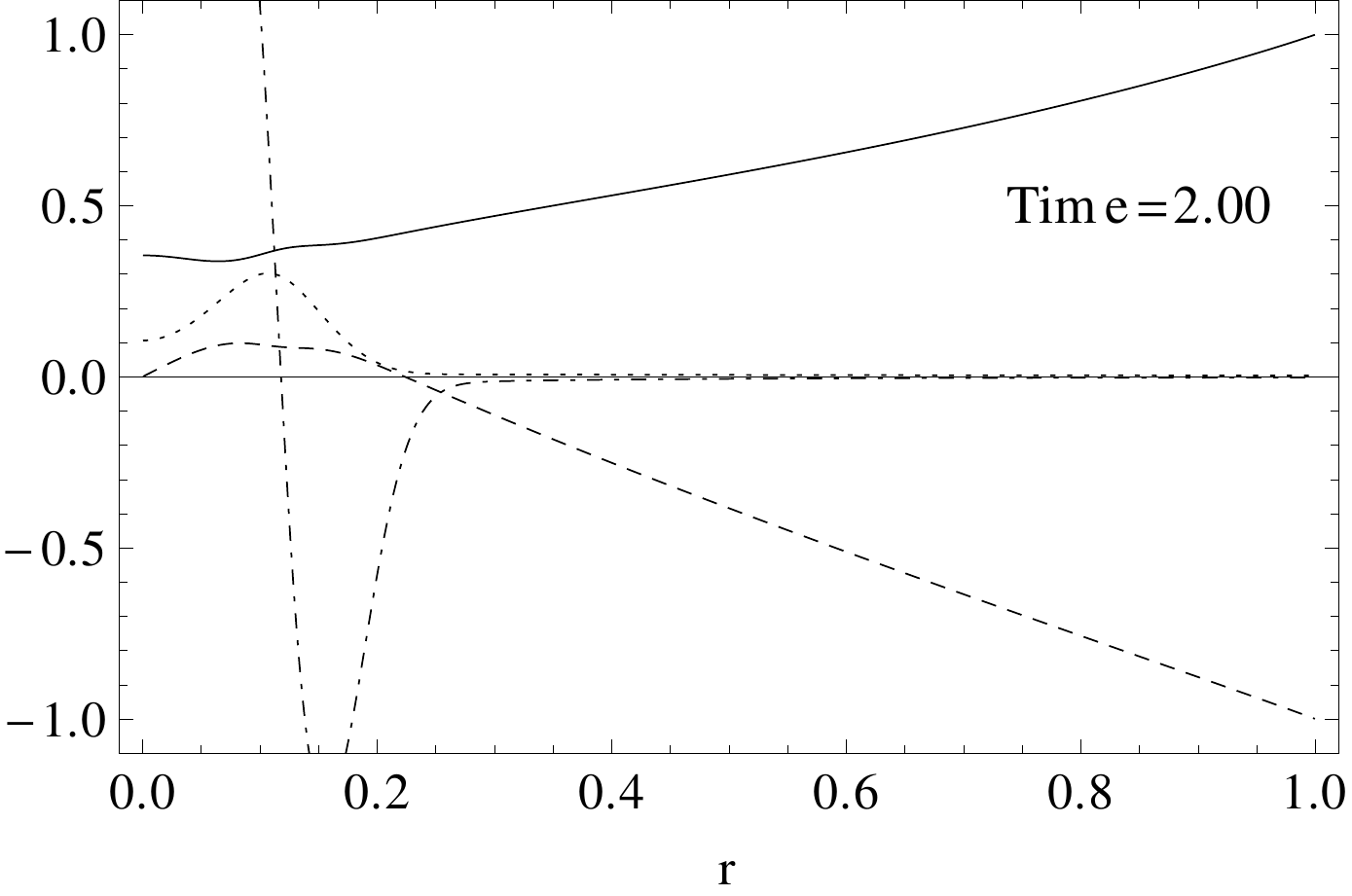}}\\
\vspace{-5.5ex} \mbox{\includegraphics[width=1\linewidth]{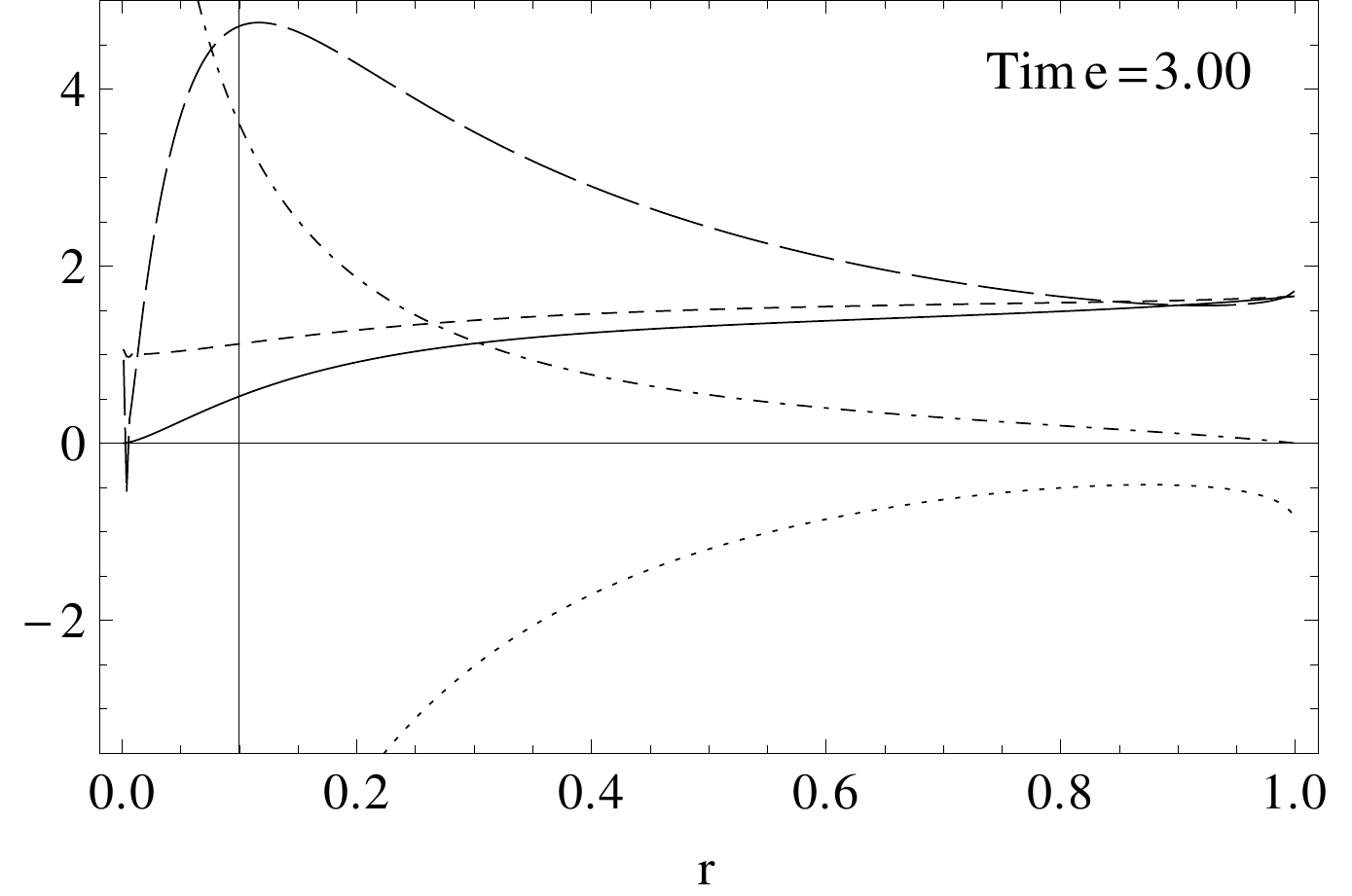}}&
\vspace{-5.5ex} \hspace{-0.8ex} \mbox{\includegraphics[width=1\linewidth]{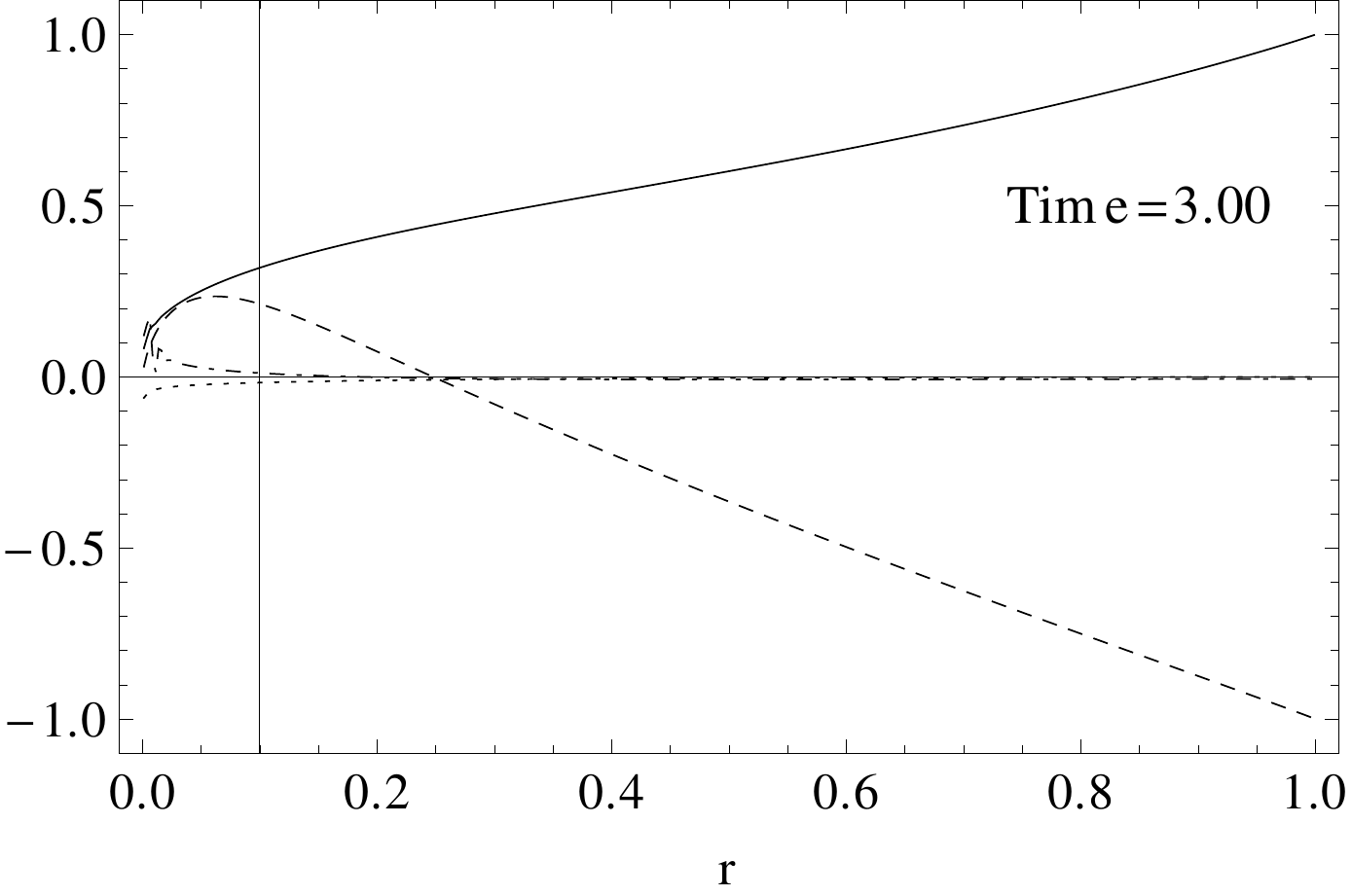}}\\
\vspace{-5.5ex} \mbox{\includegraphics[width=1\linewidth]{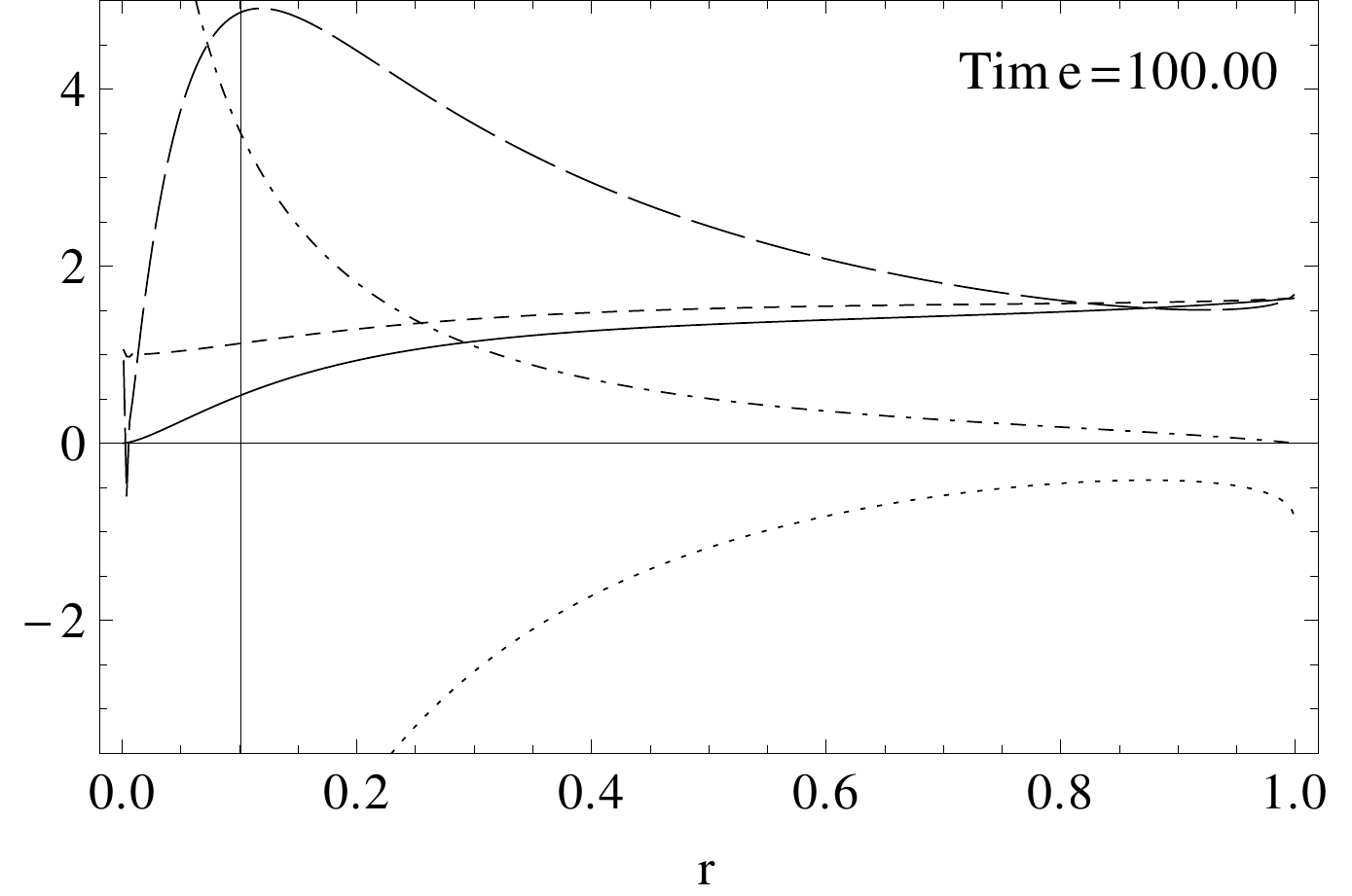}}&
\vspace{-5.5ex} \hspace{-0.8ex} \mbox{\includegraphics[width=1\linewidth]{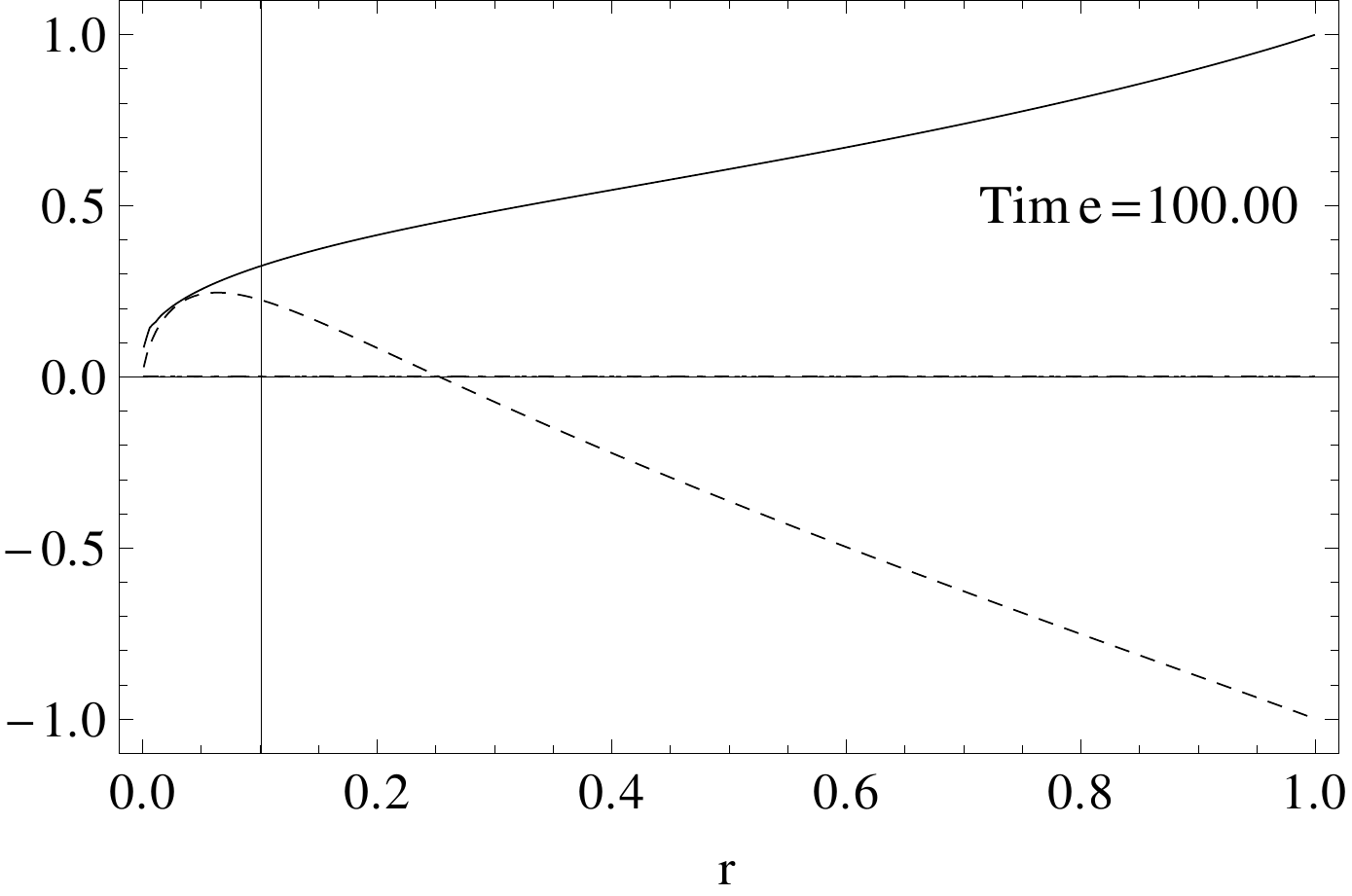}}
\end{tabular}
\vspace{-2ex}
\caption{Evolution of the variables in a BH collapse of the scalar field perturbation. The vertical line denotes the position of the apparent horizon.}
\label{fs:collapse}
\end{figure}
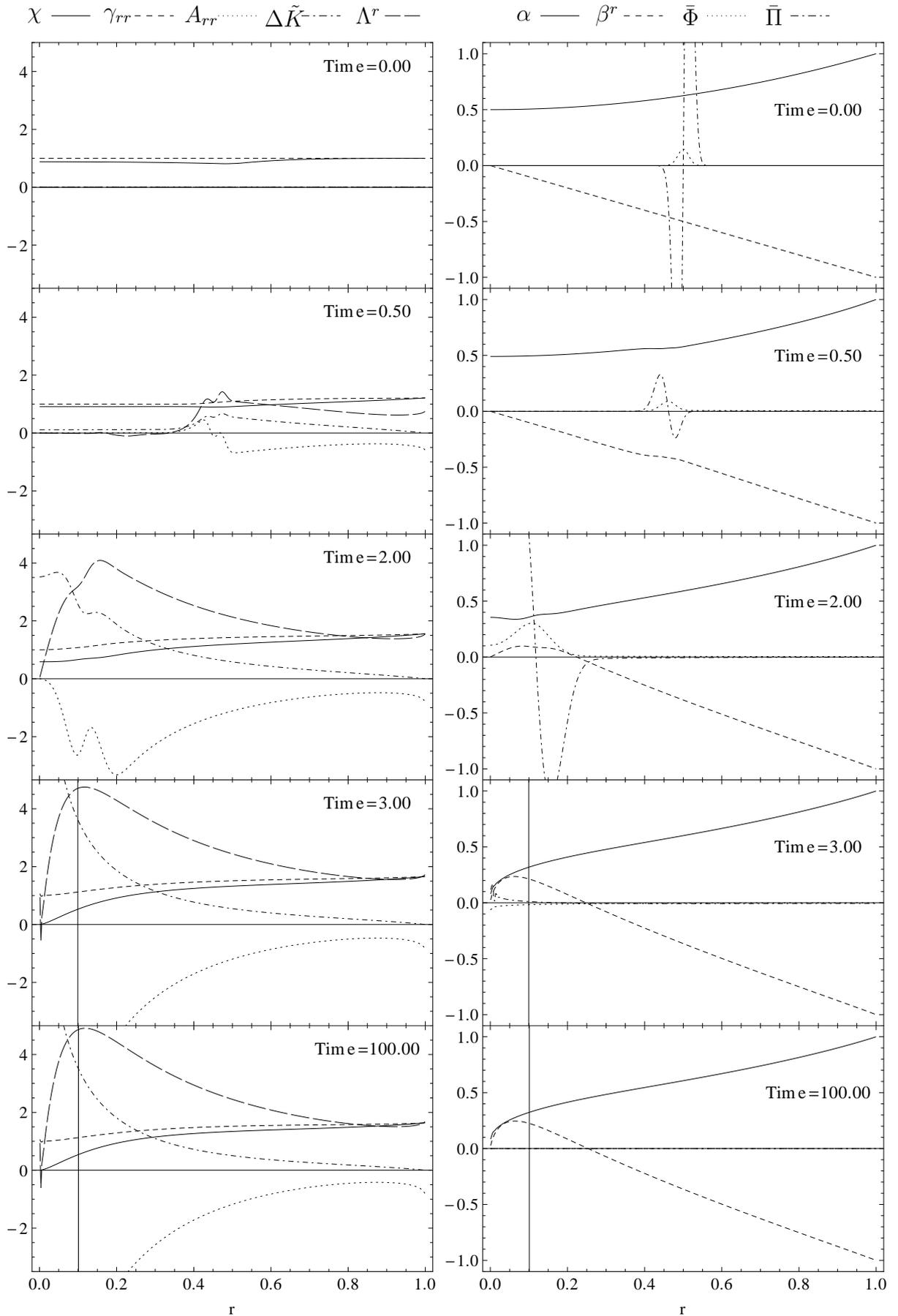

A simulation of the collapse of a scalar field perturbation to a BH has been performed using the GBSSN system with the tuned 1+log \eref{ee:tun1plog} and the integrated Gamma-driver \eref{ee:expintegGammadriver} conditions. The convergence of the Hamiltonian constraint once the stable stationary end state has long been reached is shown in \fref{fs:colH}, while the evolution of the variables is presented in \fref{fs:collapse}.
The parameter choices were $\Kc=-3$, $\kappa_1=1.5$, $\xi_{1+log}=5$ (as with the normally used here $\xi_{1+log}=2$ some fluctuations coming from $\scri^+$ appeared in the variables and could even make the simulation crash), $\xi_{\beta^r}=5$, $\lambda=0.75$, $\eta=0.1$ and $\epsilon=0.75$. % ($\epsilon=0.5$ was enough for the lowest resolutions with 400 gridpoints ($\Delta t=0.0005$) and 600 gridpoints, but the highest resolution one with 900 points was crashing at the origin and required more dissipation - actually with coealphal=5 also for 400 points $\epsilon=0.75$ was needed). % $\cL=0$
%Figure \ref{fs:collapse} shows the collapse to a BH of a massless scalar field perturbation. The GBSSN system with 1+log ($\xi_{1+log}=2$) and integrated Gamma-driver ($\lambda=0.75$ and $\eta=0.1$) was used in the simulation. The choices were $\Kc=-3$, $\kappa_1=1.5$, $\epsilon=0.5$, 200 gridpoints and $\Delta t=0.001$. % $\cL=4$
The initial profile of the scalar field perturbation was time symmetric and used $A_{\Phi}=0.055$, $\sigma=0.1$ and $c=0.5$.
The formation of the apparent horizon of the BH takes place at $t\approx2.28$. The mass of the final BH is $M\approx0.148$ (its profile varies up to 12\% with the radial coordinate) and its apparent horizon is located at $r\approx0.10$ (at the creation of the BH it is at about an 8\% smaller radius); its radial coordinate location is indicated in the two last rows of \fref{fs:collapse} and in \fref{fs:colH} by a vertical solid line. The collapse shown here is above the critical case \cite{Choptuik:1992jv}, whose amplitude would be $A_\Phi\sim0.028$.

The stationary end state is reached at $t\approx50$ and after that the evolution variables and the constraints remain static (this has been checked until $t=500$). The convergence example shown in \fref{fs:colH} shows good coincidence in the interior region, although close to the origin and $\scri^+$ the profile is far from smooth. Almost all of the non-converging noisy part at the origin is located inside of the BH's horizon.
The similar effect that appears at $\scri^+$ is also static and is likely to be caused by the small errors at the variables near $\scri^+$, exaggerated by the divergent terms in the constraint equations.
% Convergence of Hamiltonian constraint at stationary end state of collapse

The profiles of the stationary state shown in the last row of \fref{fs:collapse} vary a little depending on the choice of parameters (like $\xi_{1+log}$, $\xi_{\beta^r}$ or $\eta$), but in spite of these differences, the stationary profiles clearly differ from the CMC trumpet data in \fref{fin:Bini} (where $\Kc=-1$ is used instead of $-3$): $\DPK$ and $\Lambda^r$ do not vanish, the gauge variables go to zero with a steeper slope at the origin (compare to \fref{fin:alphabetar}) and $\chi$ and $\gamma_{rr}$ are not unity at $\scri^+$, among other differences.
Here the source functions of the gauge conditions are calculated from the flat spacetime stationary values of the quantities, so that the simulation is not expected to arrive at a CMC trumpet stationary state.

%also $\Kc=-3$ instead of $-1$ as in \fref{fin:Bini}
%comparison of Kcmc and lambdar for BH evols in next section

% Scalar field collapse

%\upda{The ripples in $\DPK$ and $\Lambda^r$'s profile at times $t=2$ and $t=3$ cannot be prevented from appearing even with more dissipation. In this simulation with 200 gridpoints they are damped away, but the same wimulation with 400 points (and $\Delta t=0.0005$) was crashing due to a growth of this feature.} $\xi_{1+log}=5$ ripples cannot be seen, with $\epsilon=0.75$ can run with 900 points.

\section{Strong field initial data}
\vspace{-1ex}
\subsection{Unperturbed Schwarzschild initial data}\label{se:onlySchw}

\begin{figure}[htbp!!]
\center
\vspace{-2ex}
\begin{tabular}{ m{0.5\linewidth}@{} @{}m{0.5\linewidth}@{} }
\begin{tikzpicture}[scale=1.1] %\draw (-0.5cm,0cm) node {};
		\draw (0cm, 0cm) node {$\Delta\chi$}; \draw (0.4cm, 0cm) -- (1cm, 0cm);
		\draw (1.5cm, 0cm) node {$\Delta\gamma_{rr}$}; \draw [dashed] (1.9cm, 0cm) -- (2.5cm, 0cm);
		\draw (3cm, 0cm) node {$\Delta A_{rr}$}; \draw [dotted] (3.4cm, 0cm) -- (4cm, 0cm);
		\draw (4.5cm, 0cm) node {$\Delta\DPK$}; \draw [dash pattern= on 4pt off 2pt on 1pt off 2pt] (4.9cm, 0cm) -- (5.5cm, 0cm);
		\draw (6cm, 0cm) node {$\Delta\Lambda^r$}; \draw [dash pattern= on 8pt off 2pt] (6.4cm, 0cm) -- (7cm, 0cm);
	\end{tikzpicture}
&
\begin{tikzpicture}[scale=1.1] \draw (-1.3cm,0cm) node {};
		\draw (0cm, 0cm) node {$\Delta\alpha$}; \draw (0.4cm, 0cm) -- (1cm, 0cm);
		\draw (1.5cm, 0cm) node {$\Delta\beta^r$}; \draw [dashed] (1.9cm, 0cm) -- (2.5cm, 0cm);
		\draw (3cm, 0cm) node {$\Delta\bPhi$}; \draw [dotted] (3.4cm, 0cm) -- (4cm, 0cm);
		\draw (4.5cm, 0cm) node {$\Delta\bPi$}; \draw [dash pattern= on 4pt off 2pt on 1pt off 2pt] (4.9cm, 0cm) -- (5.5cm, 0cm);
	\end{tikzpicture}
\\
%\hspace{-0.8ex}\mbox{\includegraphics[width=1\linewidth]{figures/schwl_1.pdf}}&
%\hspace{-1.6ex} \mbox{\includegraphics[width=1\linewidth]{figures/schwr_1.pdf}}\\
\vspace{-1.5ex} \hspace{-0.8ex}\mbox{\includegraphics[width=1\linewidth]{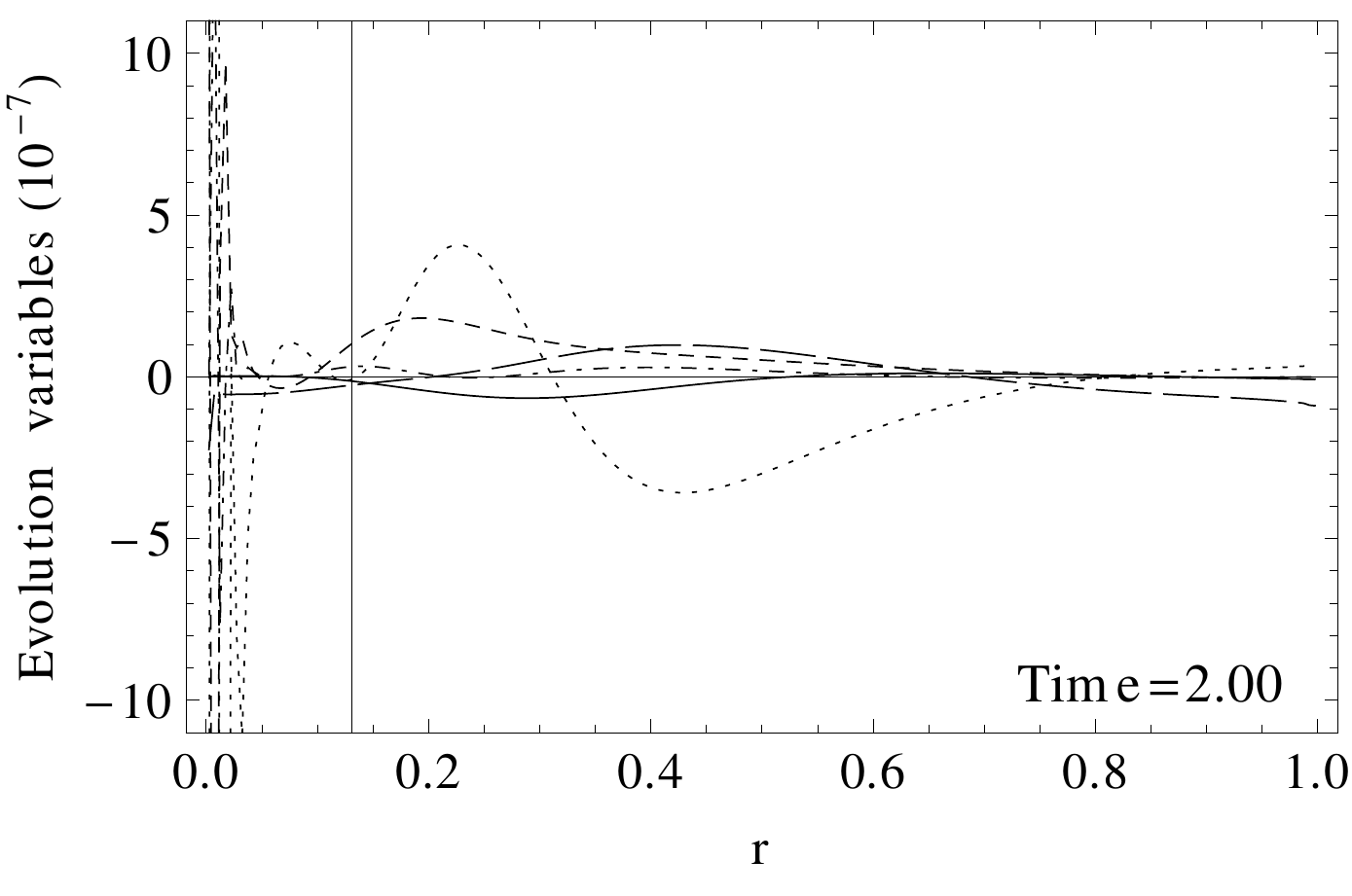}}&
\vspace{-1.5ex} \hspace{-1.6ex} \mbox{\includegraphics[width=1\linewidth]{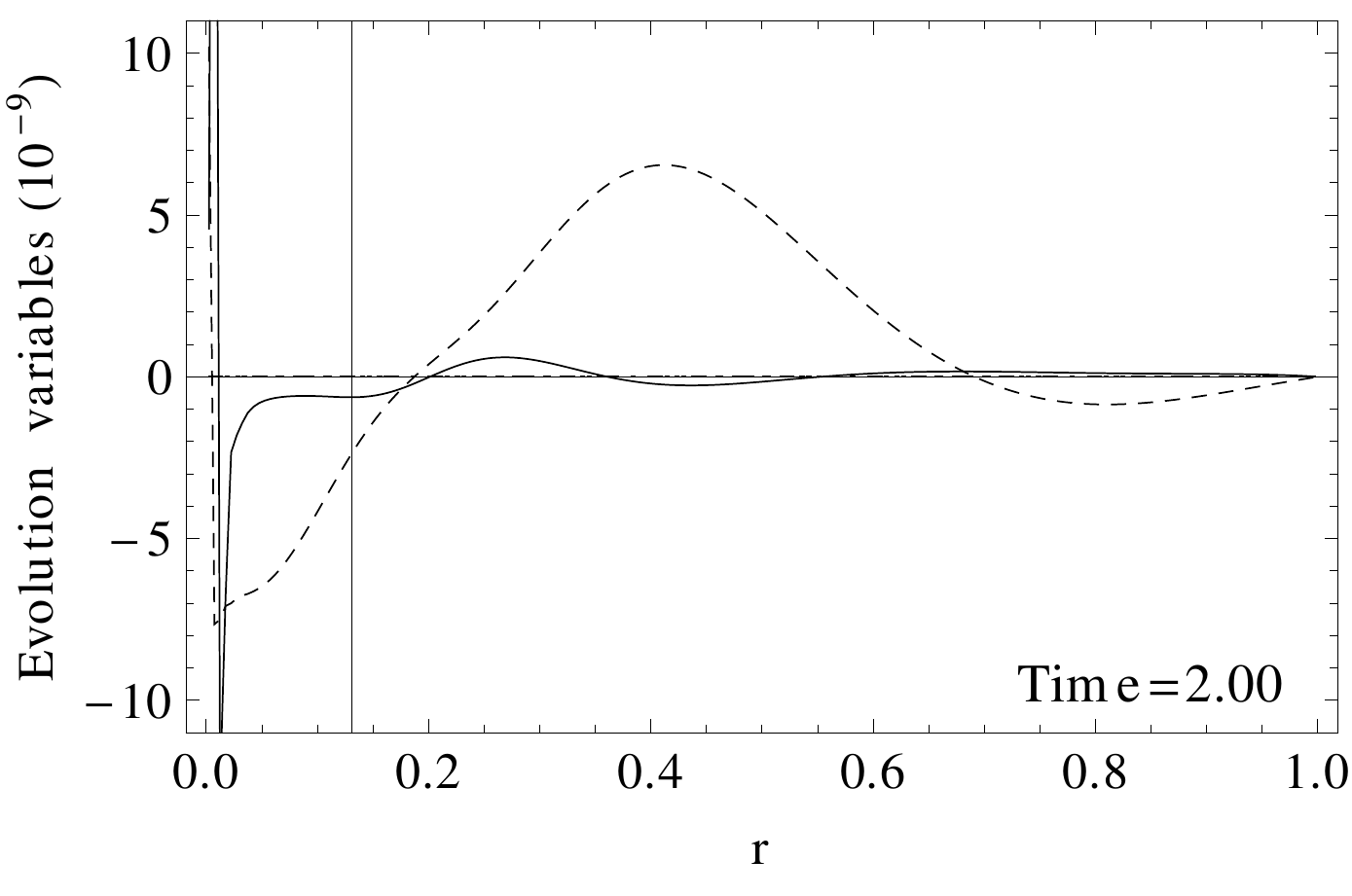}}\\
%\vspace{-6.ex} \hspace{-0.8ex}\mbox{\includegraphics[width=1\linewidth]{figures/schwl_3.pdf}}&
%\vspace{-6.ex} \hspace{-1.6ex} \mbox{\includegraphics[width=1\linewidth]{figures/schwr_3.pdf}}\\
%\vspace{-6.ex} \hspace{-0.8ex}\mbox{\includegraphics[width=1\linewidth]{figures/schwl_4.pdf}}&
%\vspace{-6.ex} \hspace{-1.6ex} \mbox{\includegraphics[width=1\linewidth]{figures/schwr_4.pdf}}\\
\vspace{-6.ex} \hspace{-0.8ex}\mbox{\includegraphics[width=1\linewidth]{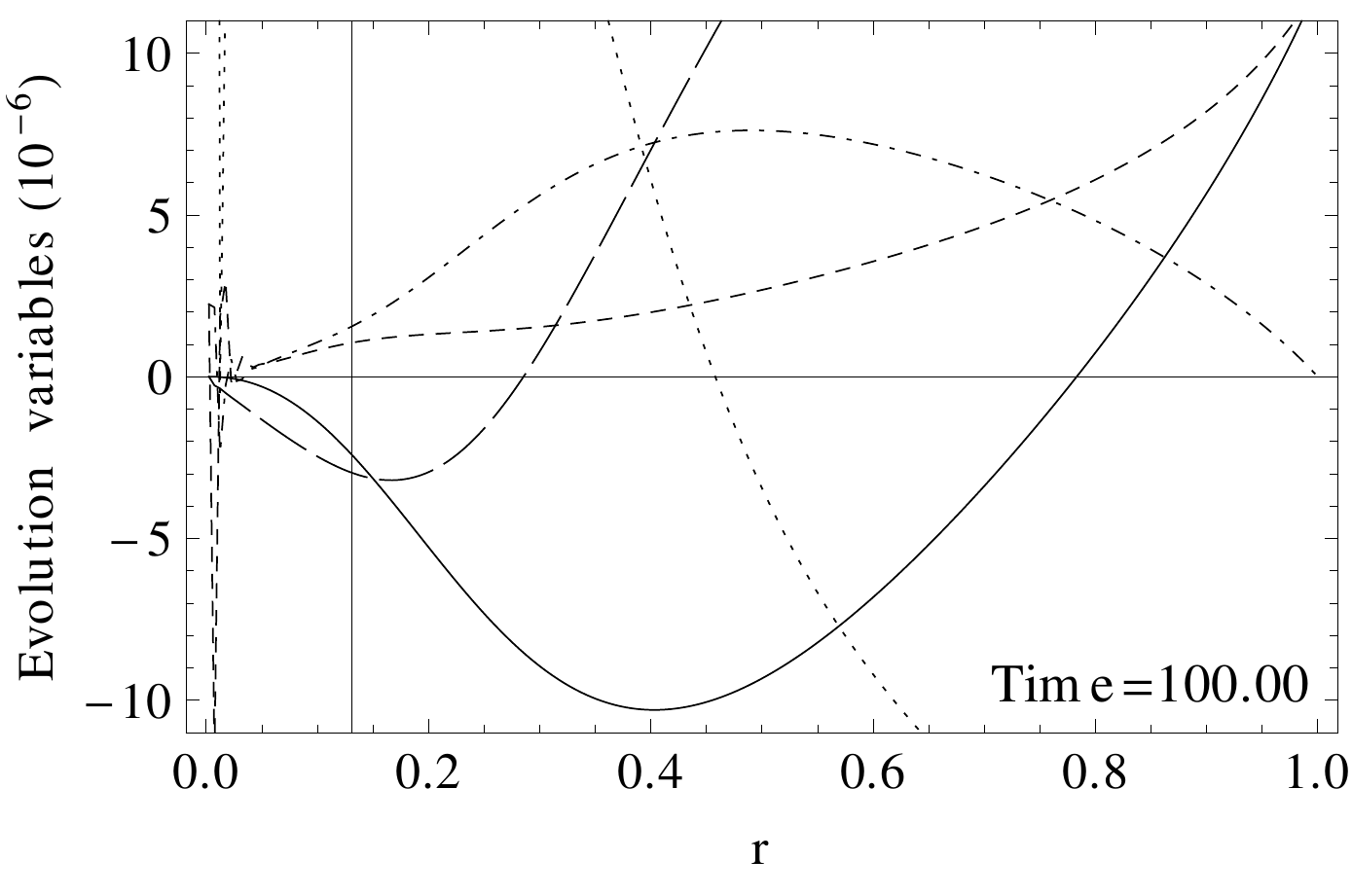}}&
\vspace{-6.ex} \hspace{-1.6ex} \mbox{\includegraphics[width=1\linewidth]{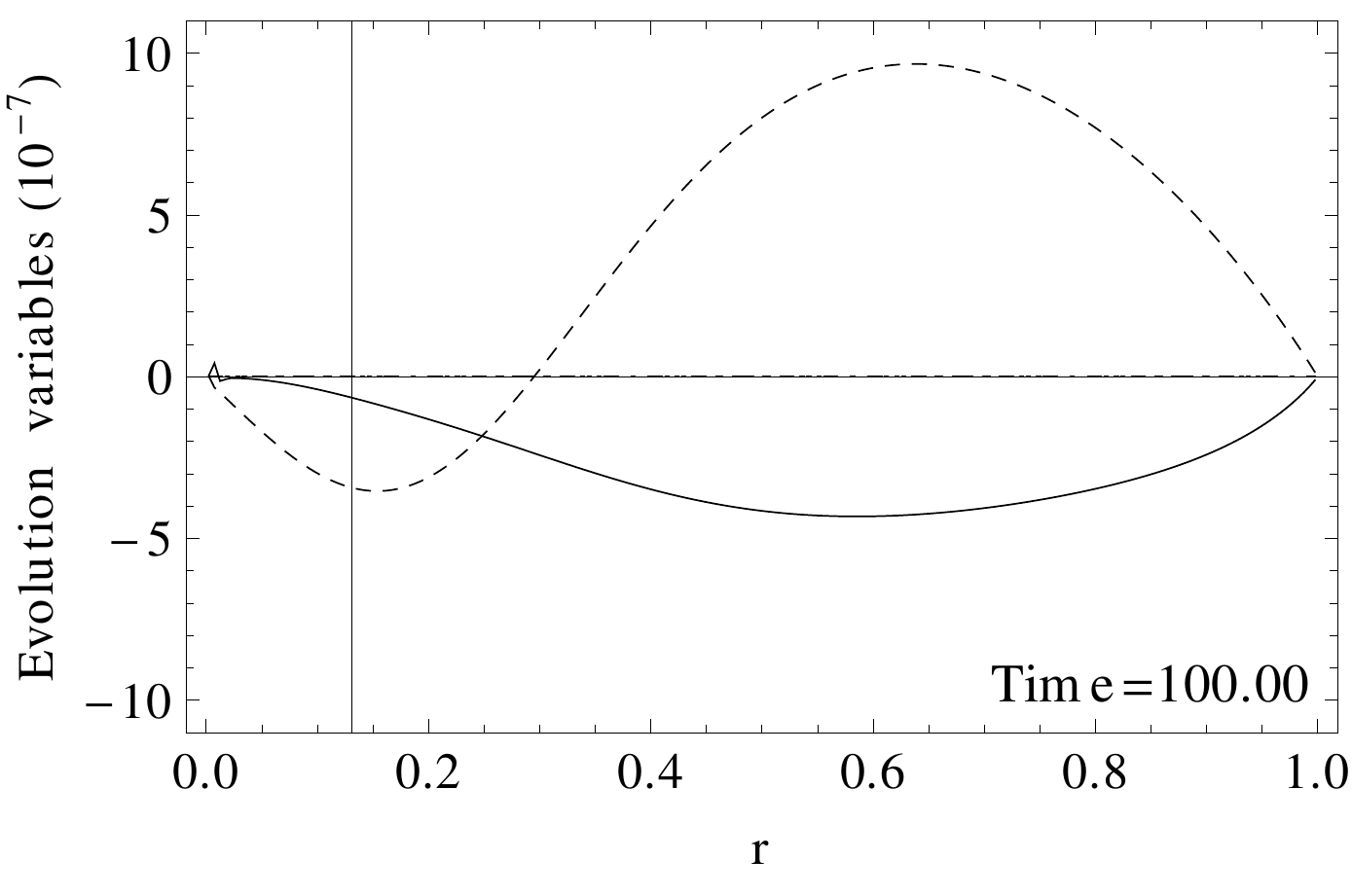}}
\end{tabular}
\vspace{-4ex}
\caption{Variation of the variables ($\Delta X\equiv X-X_0$) in a simulation with unperturbed Schwarzschild CMC trumpet initial data. The vertical line at $r\approx0.13$ locates the horizon.}
\label{fs:onlyschw}
\end{figure}
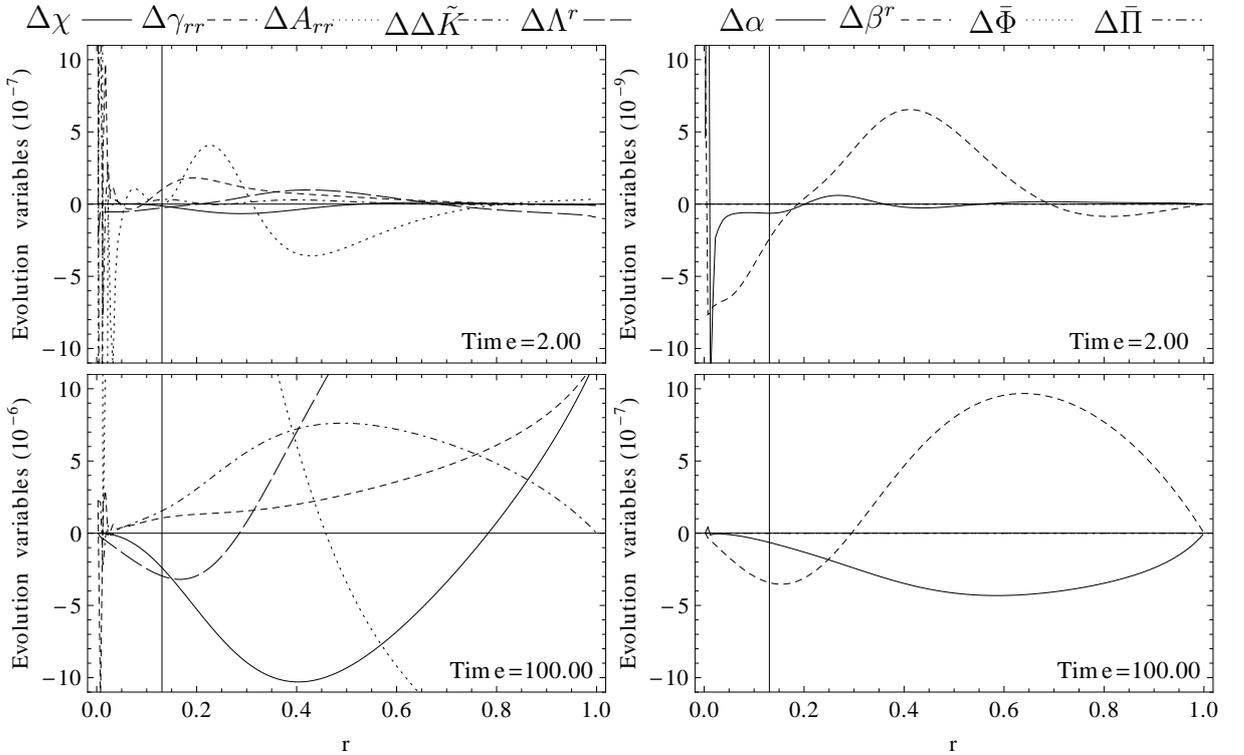

When evolving unperturbed Schwarzschild CMC trumpet initial data, like the values presented in \fref{fin:Bini}, the variables were drifting away from their initial values. This drift is illustrated by the evolution plots in \fref{fs:onlyschw}, from a simulation with the GBSSN equations, tuned 1+log and integrated Gamma-driver. In the same way as was described for  figures \ref{fs:shiftcondab} and \ref{fs:shiftcondKL}, the regularity conditions \eref{er:regcondsgauge} that arise from the gauge conditions imply that the values of $\alpha$, $\beta^r$ and $\DPK$ remain fixed at $\scri^+$ and that the values of $\atscrip{\chi}=\atscrip{\gamma_{rr}}$. Apart from these fixed values and except for the scalar field variables $\bPhi$ and $\bPi$, which do not move away from their initial vanishing value, the rest of the evolution equations are slowly drifting away from their values. Note the rescaling of the vertical axis in the plots; the largest growth is of the order of $10^{-5}$ at $t=100$.

% only Schwarzschild

\begin{figure}[h!!]
\center
\begin{tabular}{ m{0.5\linewidth}@{} @{}m{0.5\linewidth}@{} }
\hspace{-4ex} \mbox{\includegraphics[width=1.15\linewidth]{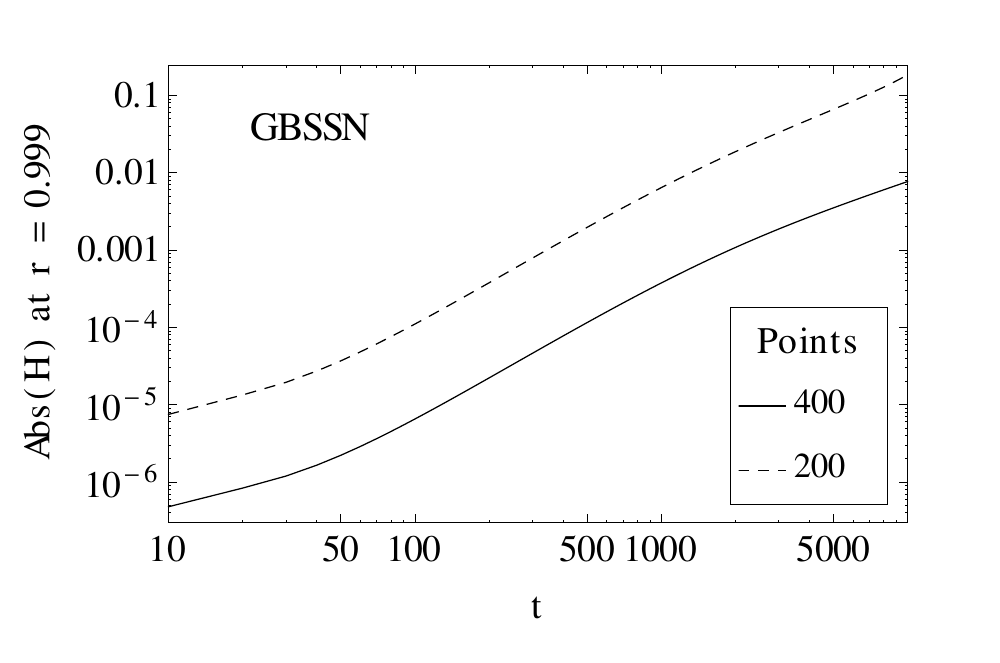}}&
\hspace{-2.5ex} \mbox{\includegraphics[width=1.15\linewidth]{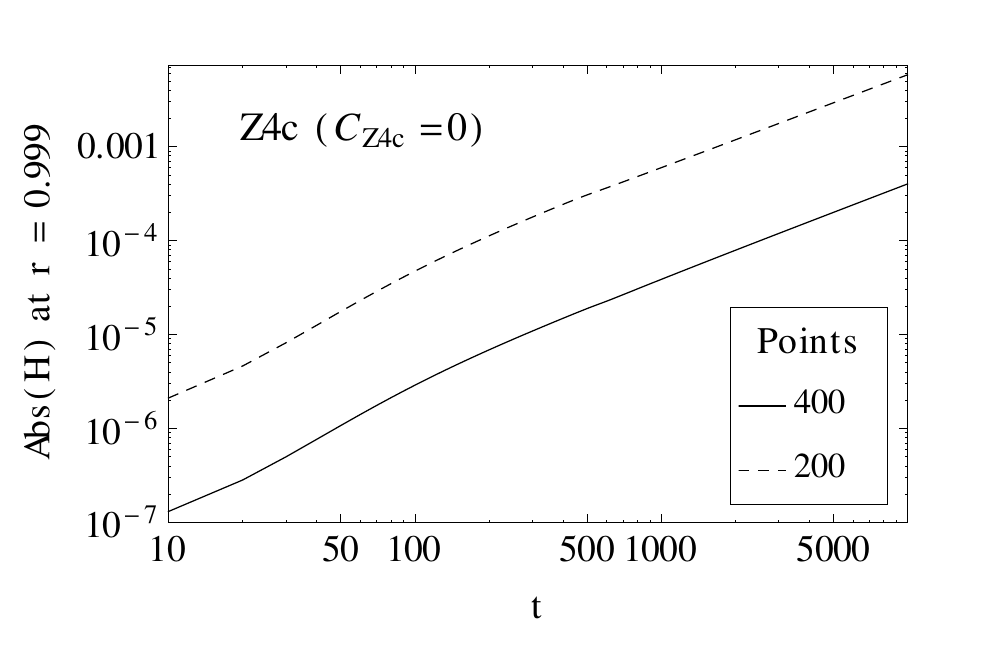}}
\end{tabular}
\vspace{-3ex}
\caption{Logarithmic plot of the drift in the $\Lambda^r$ variable for unperturbed Schwarzschild initial data with two different resolutions.}
\label{fs:Schw}
\end{figure}

In order to better understand how the drift of the variables behaves in time, the value of $\Lambda^r$ at the closest gridpoint to $\scri^+$ has been plotted in \fref{fs:Schw} for two different resolutions (200 and 400 points, with the time-step rescaled accordingly). The evolution was performed with the GBSSN and the \CZ{} ($C_{Z4c}=0$) systems, together with the tuned 1+log ($\xi_{1+log}=2$) and the integrated Gamma-driver ($\xi_{\beta^r}=5$ and $\lambda=0.0833$) and a dissipation of $\epsilon=0.5r$.
The difference in amplitude of the growth between the two different resolutions in a same formulation is about an order of magnitude. The initial amplitude is larger for the GBSSN system and also its growth is faster (the slope in the plot is steeper). Even at an evolution time of $t=10000$ the drift does not seem to stop, but the values of the variables continue drifting away. It is not clear, whether a stable end state will be reached by the variables or the drift will continue increasing until the simulation crashes.
% Long Schwarzschild

\begin{figure}[htbp!!]
\center
\mbox{\includegraphics[width=0.85\linewidth]{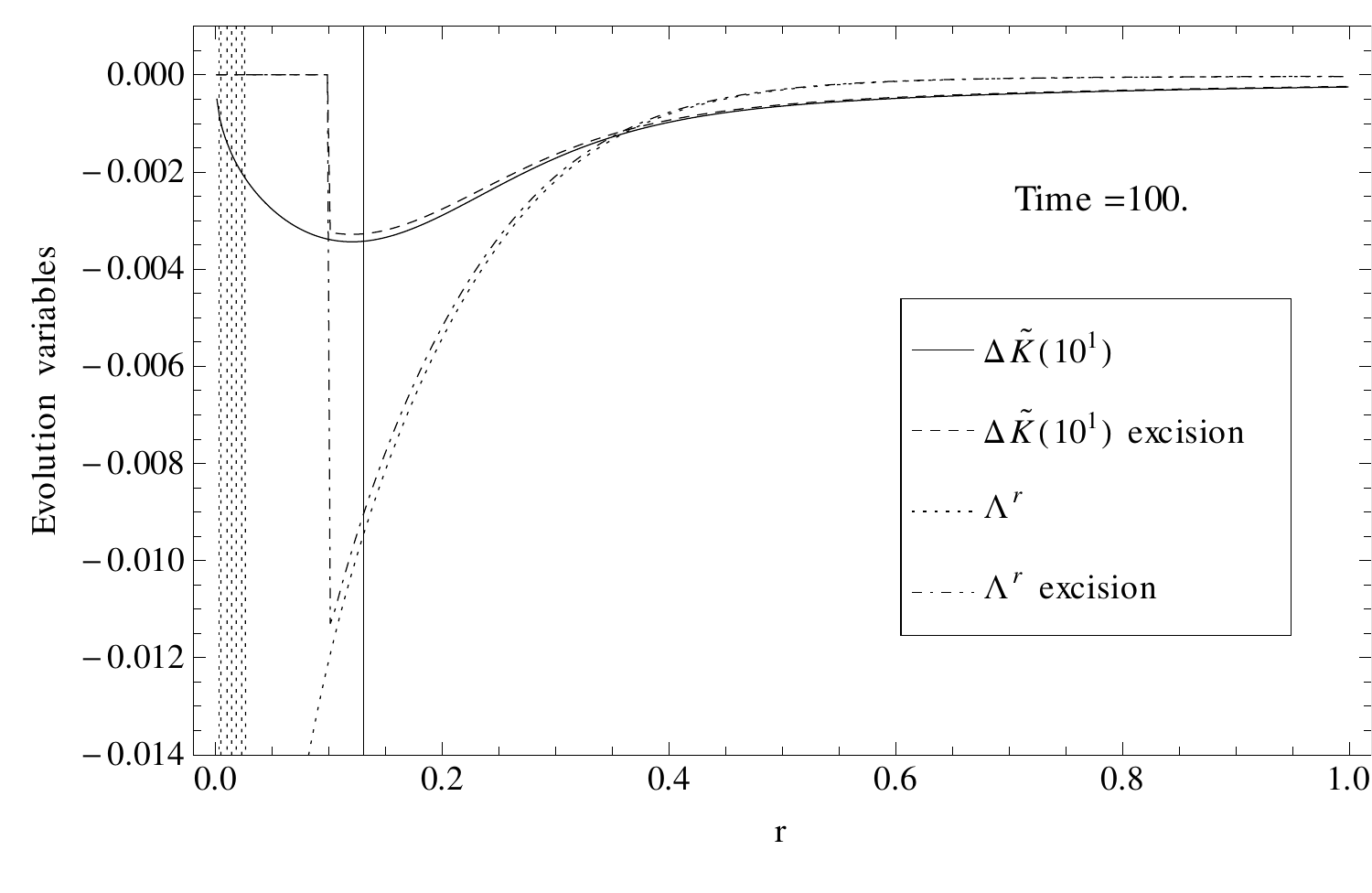}}\vspace{-4ex}
\caption{Values of $\DPK$ and $\Lambda^r$ at $t=100$, evolved with \CZ{} ($C_{Z4c}=0$) and the physical harmonic gauge \eref{ee:improvedpbg}. The variables in the evolution with excision were only evolved for $r\in(0.10,\rscri)$. The vertical line indicates the position of the horizon.}
\label{fs:excSchw}
\end{figure}

To make sure the drift is not due to errors inside of the BH that affect the outer spacetime, another simulation with \CZ{} ($C_{Z4c}=0$) and harmonic gauge conditions with physical background metric source functions was performed with and without excision. These gauge conditions were chosen, because their characteristic speeds at the horizon are negative or zero, as is required for excision.
The comparison of excised and non-excised simulations for the variables $\DPK$ and $\Lambda^r$ is shown in \fref{fs:excSchw} at $t=100$. %, the ones where this effect can be more clearly appreciated,
The difference between the values of $\DPK$ and $\Lambda^r$ for the excised and non-excised simulations are of only 5\% at the excision boundary, so that the coincidence of the curves is quite good. We can conclude that no instabilities arise from the BH interior and that the drift does not arise as a consequence of the singular behaviour at the origin.
% Excision only Schwarzschild

Comparing the CMC trumpet initial data (\fref{fin:Bini}) used here and the end state of the collapse simulation of \fref{fs:collapse}, the most natural conclusion is that the Schwarzschild CMC trumpet is an unstable stationary solution of the Einstein equations and this is why the variables slowly drift away from their initial values.

How the drift that causes the simulations to finally crash can be minimized up to some extent was briefly described in subsection \ref{se:schwini}.
The trick consists of increasing the eigenspeed associated with the shift evolution equation by adding to it a positive function that vanishes at $\scri^+$ and thus leaves the eigenspeeds there unchanged. This increase takes the following form for each of the evolved shift conditions: Gamma-driver, $\mu\to\mu+\cL(\rscri^2-r^2)$; integrated Gamma-driver, $\lambda\to\lambda+\cL(\rscri^2-r^2)$; conformal and physical harmonic gauges with background sources, $\alpha^2\chi\Lambda^r\to\alpha^2\chi\Lambda^r\left[1+\cL(\rscri^2-r^2)\right]$.
Unless otherwise specified, the simulations presented here used $\cL=4$.

The harmonic slicing condition is not the most appropriate one to use in the presence of BH due to its only marginally singularity avoidance property. I performed some preliminary tests matching 1+log and harmonic slicings, but the results regarding stability for the convergence runs were not considerably better than using the harmonic slicing everywhere, so that the latter has become the default choice for the simulations presented here. To improve stability in the region close to the origin, a term of the form $\pphi(\hat\alpha-\alpha)$ is added to $\dot\alpha$. The default choice here is $\pphi=1$.

\subsection{Gauge waves in the Schwarzschild spacetime}

\begin{figure}[htbp!!]
\center
	\includegraphics[width=0.65\linewidth]{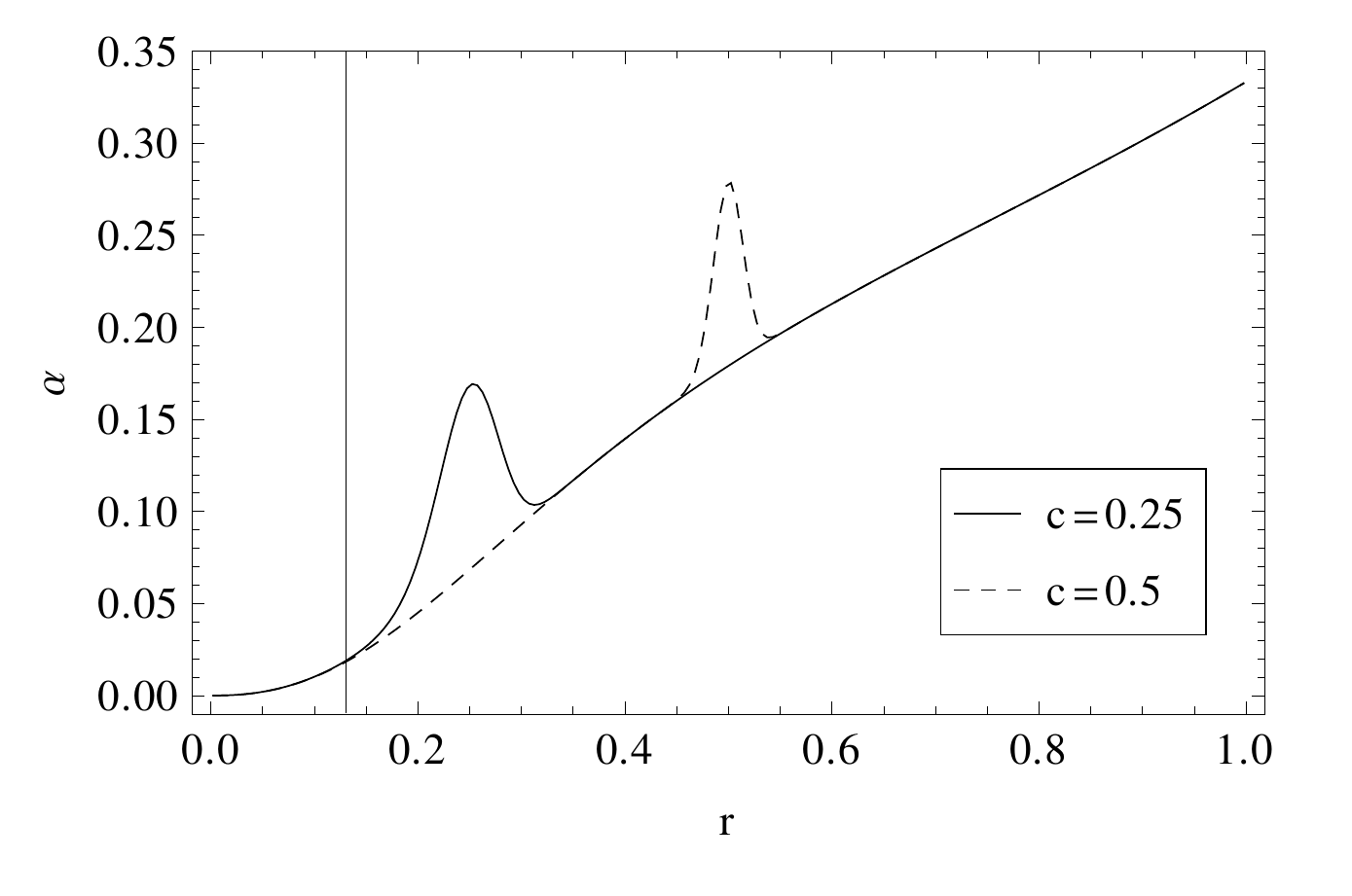} \vspace{-4ex}
\caption{Initial perturbations in the lapse for the gauge waves simulations with Schwarzschild initial data shown in \fref{fs:bhgauge} (case $c=0.25$) and \eref{fs:bhgaugec05} (case $c=0.5$).}\label{fs:bhgaugealpha}
\end{figure}

\begin{figure}[htbp!!]
\center
\vspace{-2ex}
 \begin{tikzpicture}[scale=1.5]\draw (-1cm,0cm) node {};
		\draw (0cm, 0cm) node {$\chi$}; \draw (0.3cm, 0cm) -- (1cm, 0cm);
		\draw (1.5cm, 0cm) node {$\gamma_{rr}$}; \draw [dashed] (1.8cm, 0cm) -- (2.5cm, 0cm);
		\draw (3cm, 0cm) node {$\DPK$}; \draw [dotted] (3.3cm, 0cm) -- (4cm, 0cm);
		\draw (4.5cm, 0cm) node {$\Lambda^r$}; \draw [dash pattern= on 4pt off 2pt on 1pt off 2pt] (4.8cm, 0cm) -- (5.5cm, 0cm);
		\draw (6cm, 0cm) node {$\alpha$}; \draw [dash pattern= on 8pt off 2pt] (6.3cm, 0cm) -- (7cm, 0cm);
		\draw (7.5cm, 0cm) node {$\beta^r$}; \draw [dash pattern= on 6pt off 2pt on 1pt off 2pt on 1pt off 2pt] (7.8cm, 0cm) -- (8.5cm, 0cm);
	\end{tikzpicture}
\\
\begin{tabular}{ m{0.5\linewidth}@{} @{}m{0.5\linewidth}@{} }
\mbox{\includegraphics[width=1\linewidth]{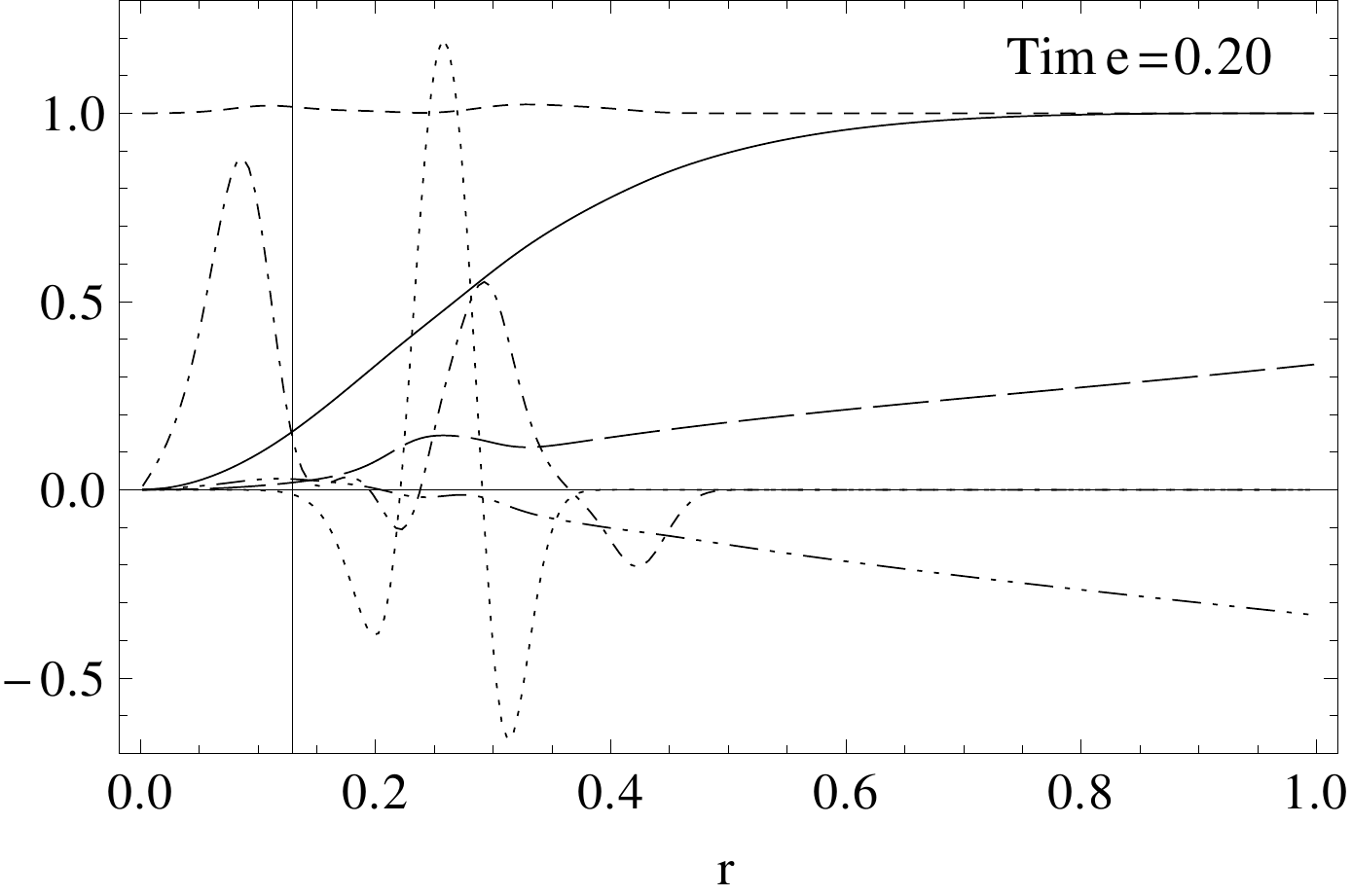}}&
\hspace{-0.8ex} \mbox{\includegraphics[width=1\linewidth]{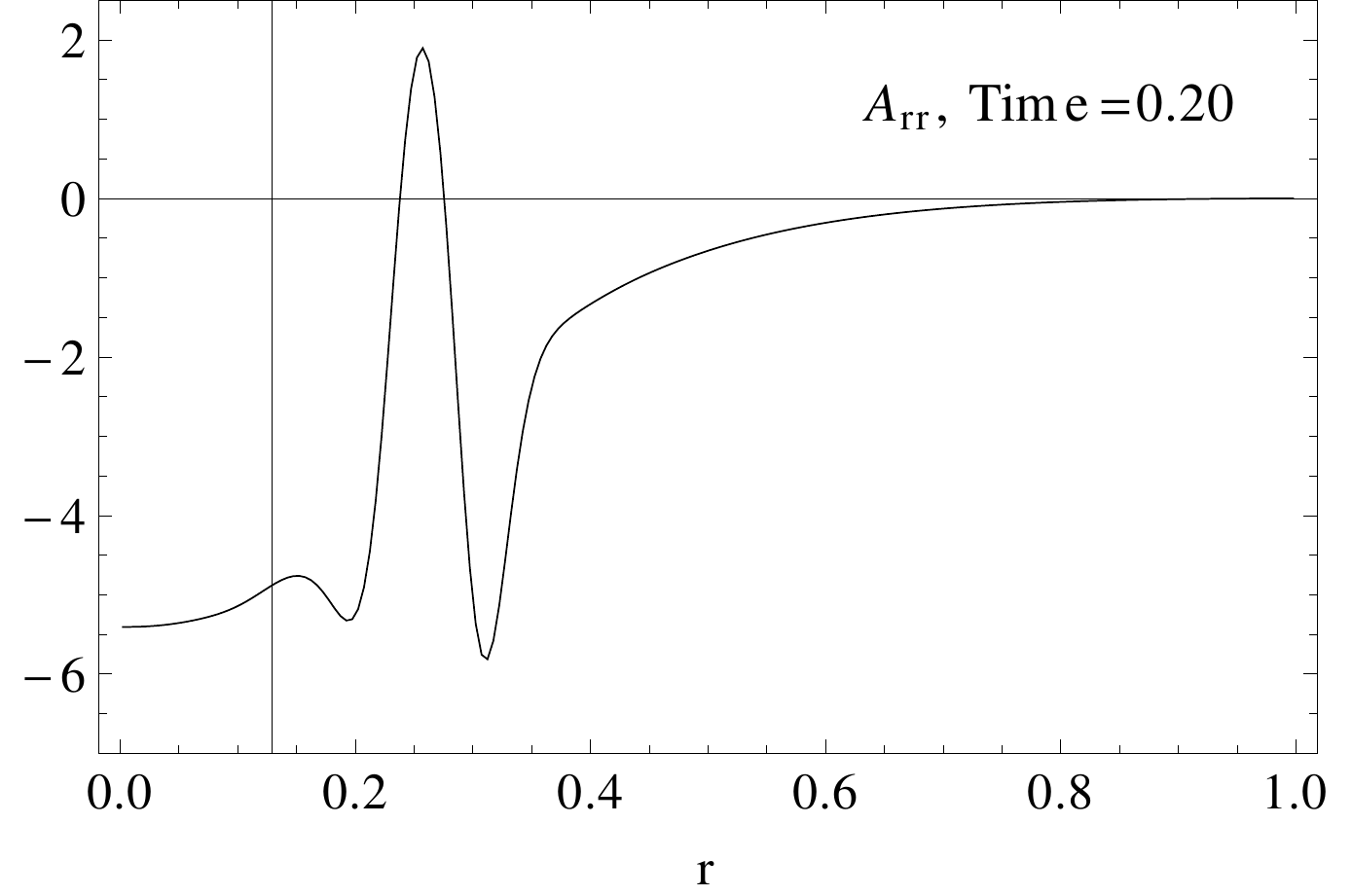}}\\
\vspace{-5.5ex} \mbox{\includegraphics[width=1\linewidth]{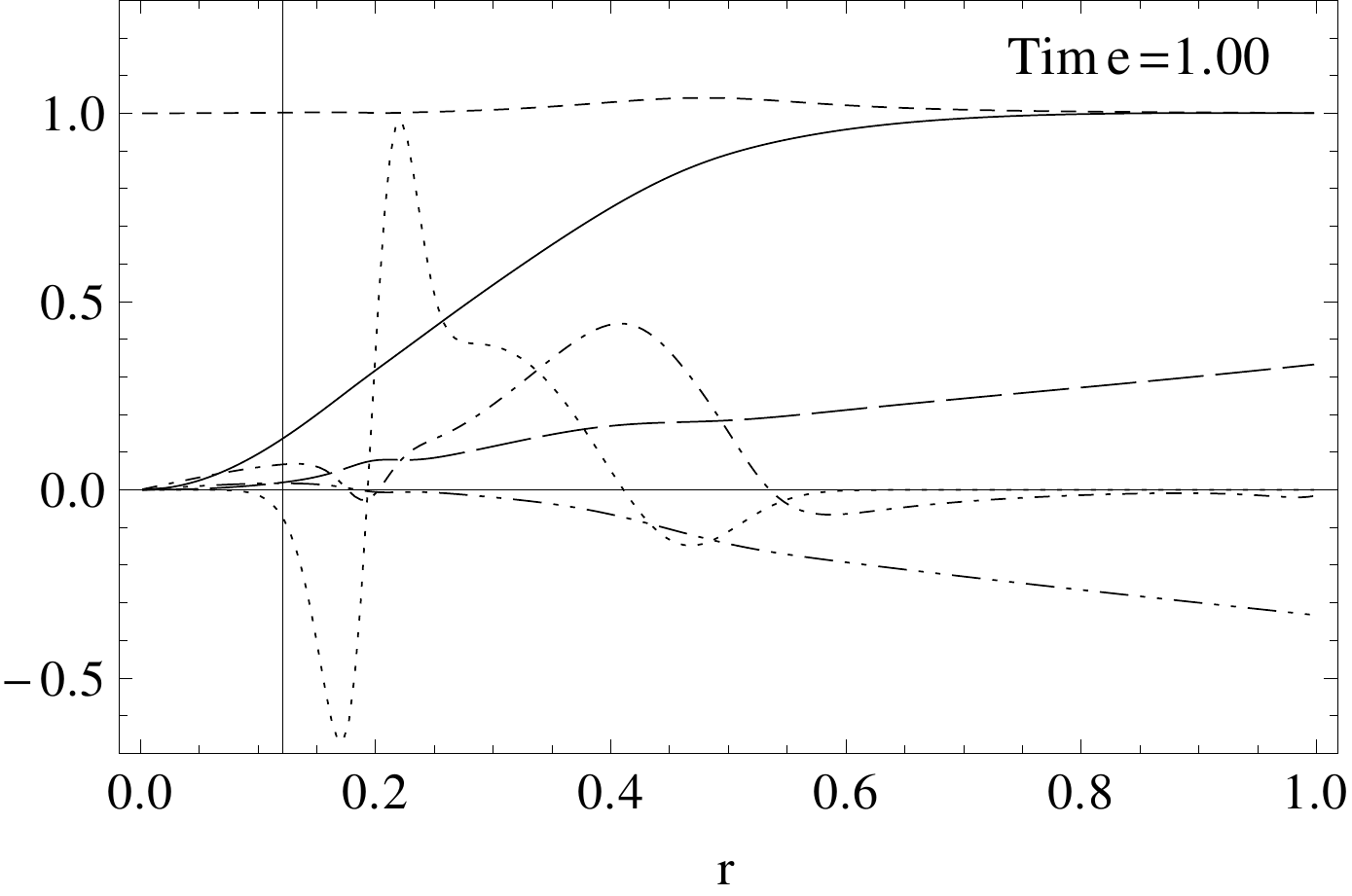}}&
\vspace{-5.5ex} \hspace{-0.8ex} \mbox{\includegraphics[width=1\linewidth]{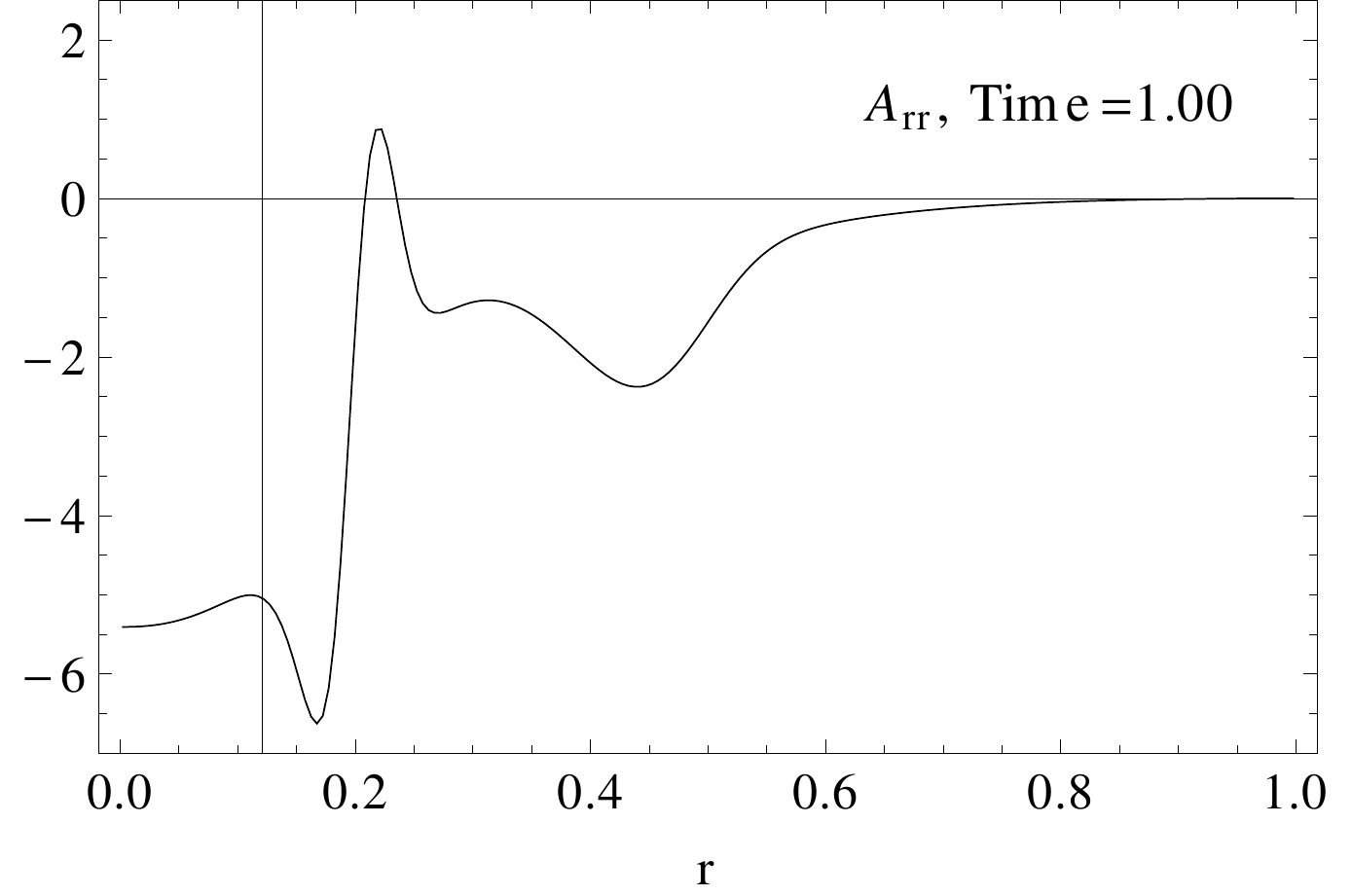}}\\
\vspace{-5.5ex} \mbox{\includegraphics[width=1\linewidth]{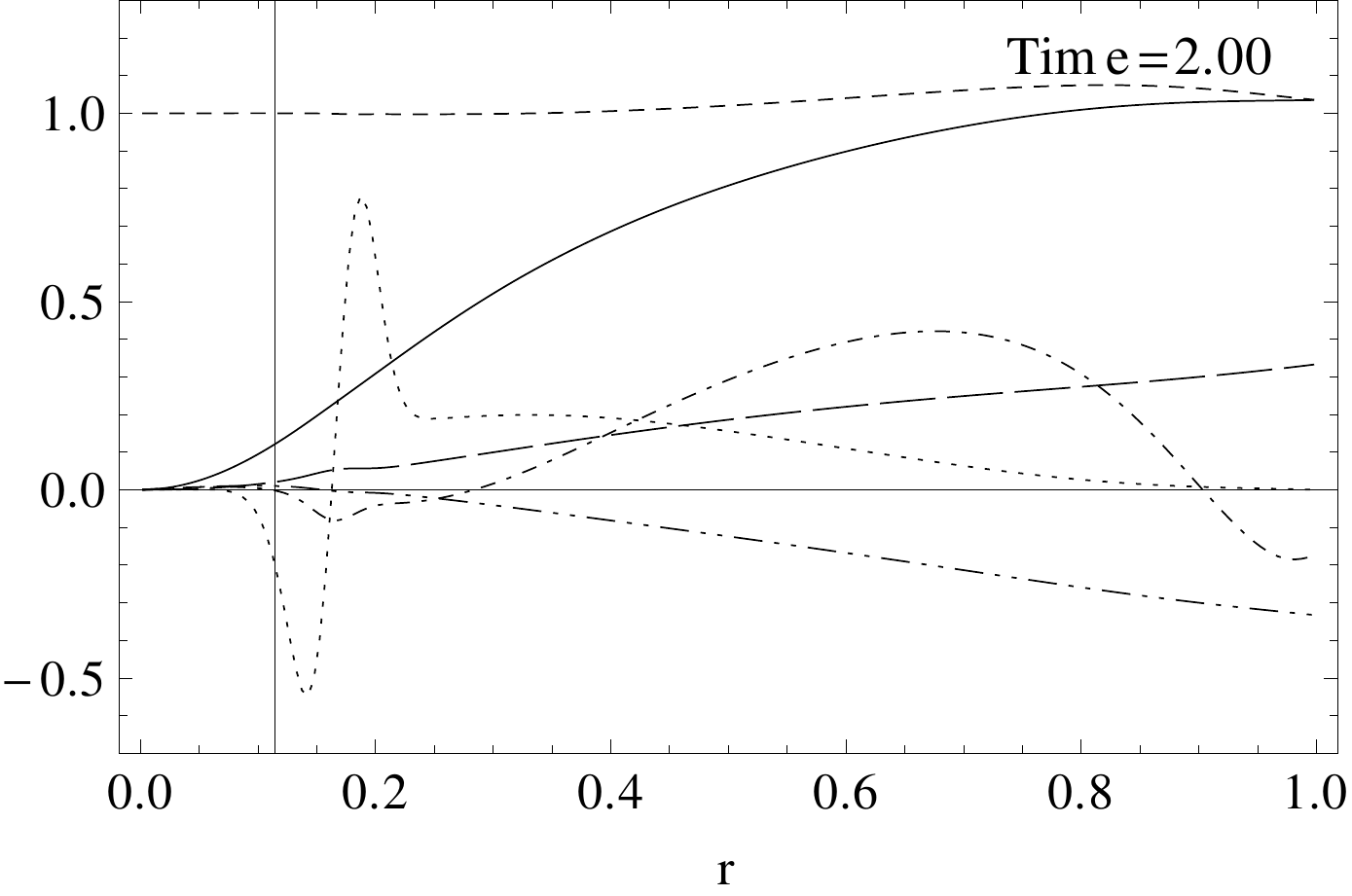}}&
\vspace{-5.5ex} \hspace{-0.8ex} \mbox{\includegraphics[width=1\linewidth]{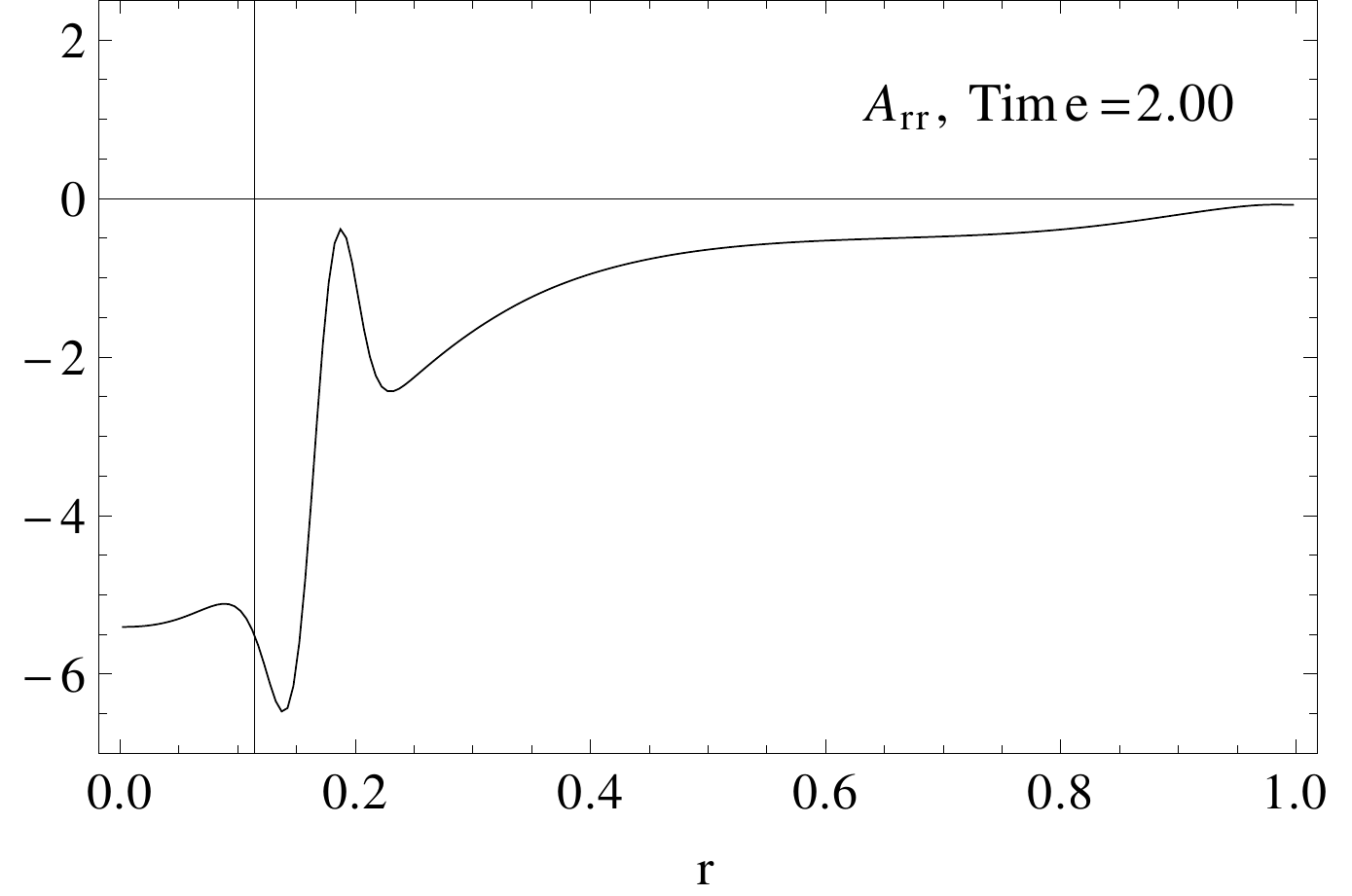}}\\
\vspace{-5.5ex} \mbox{\includegraphics[width=1\linewidth]{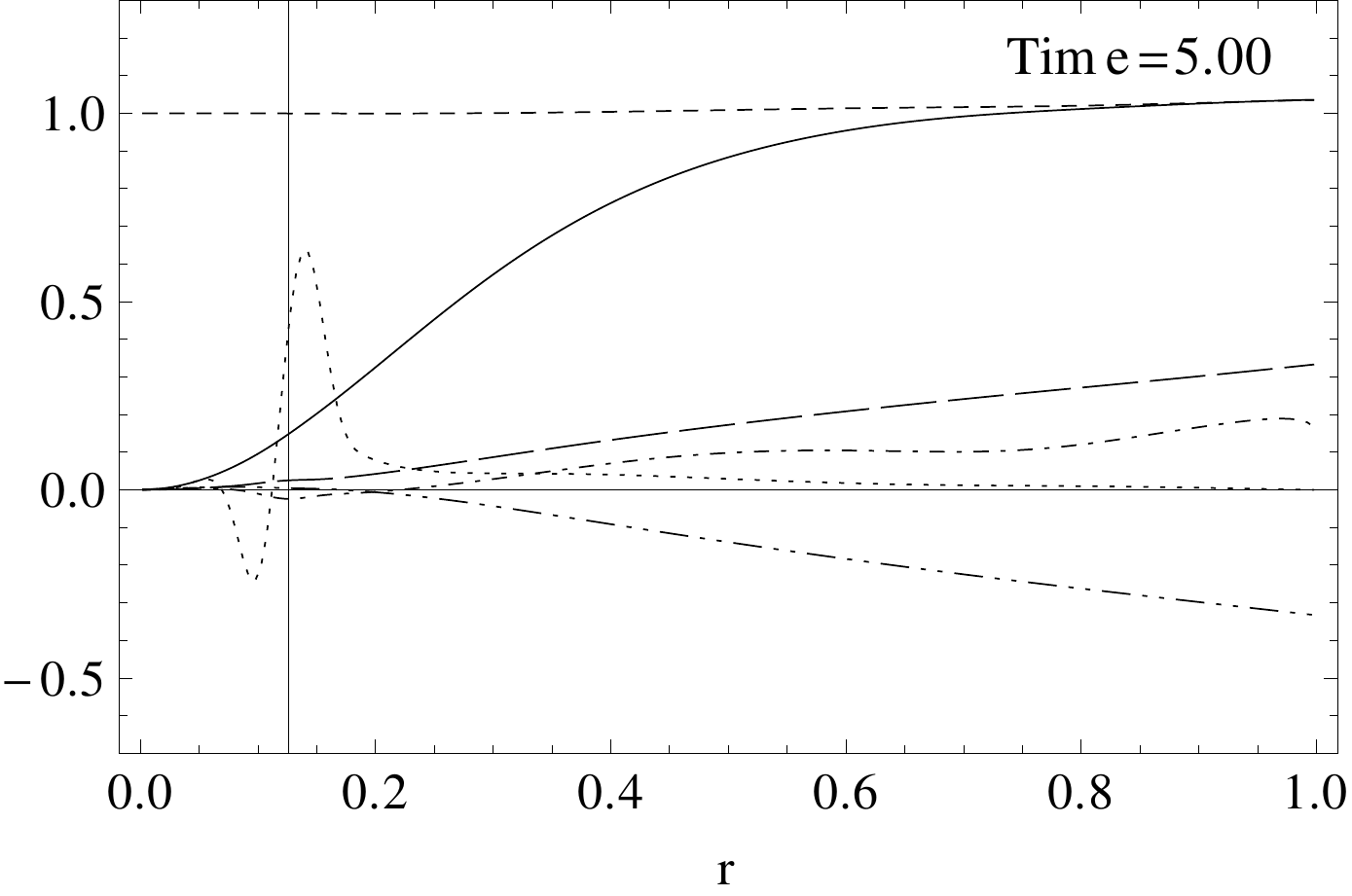}}&
\vspace{-5.5ex} \hspace{-0.8ex} \mbox{\includegraphics[width=1\linewidth]{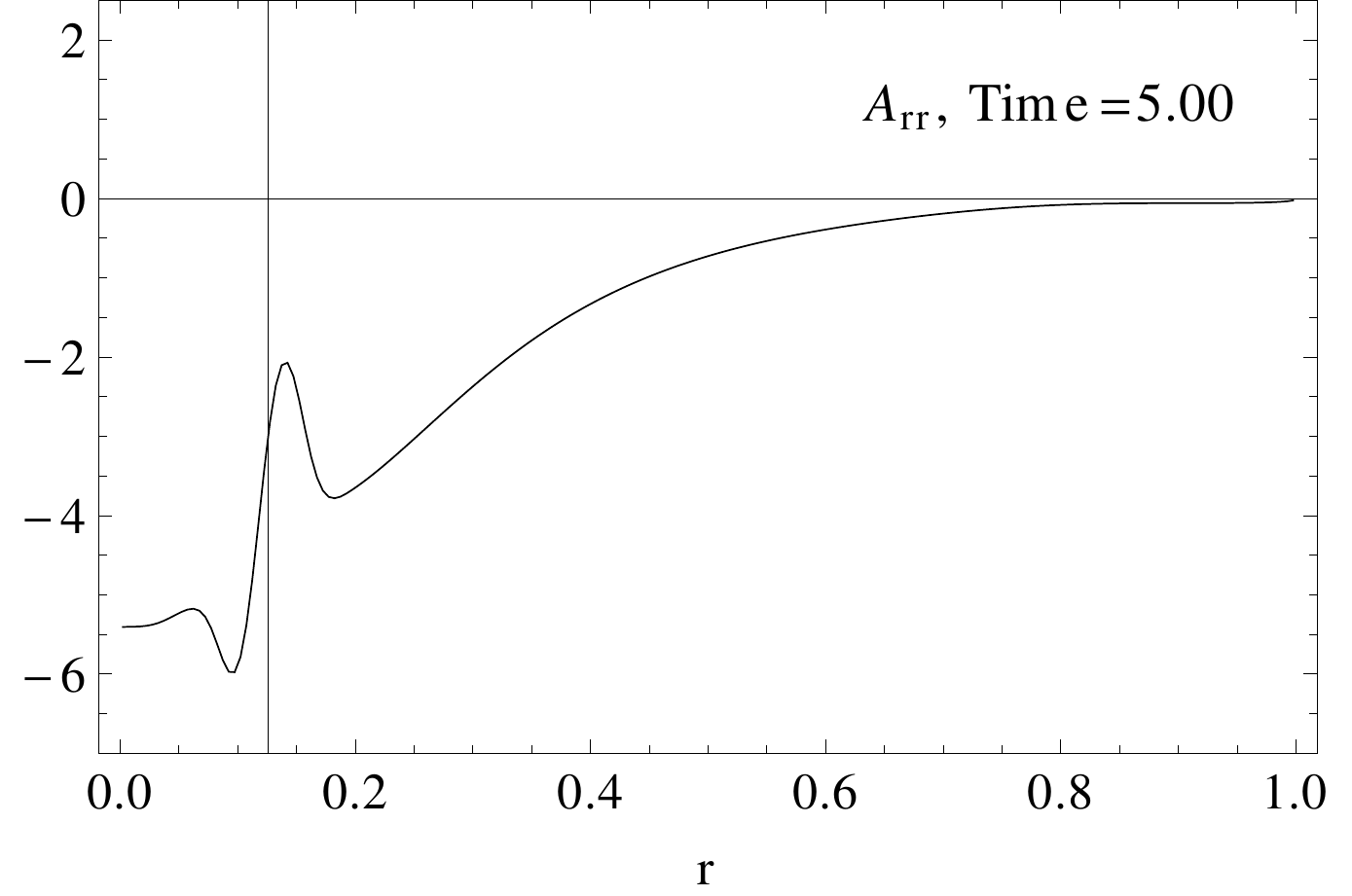}}\\
\vspace{-5.5ex} \mbox{\includegraphics[width=1\linewidth]{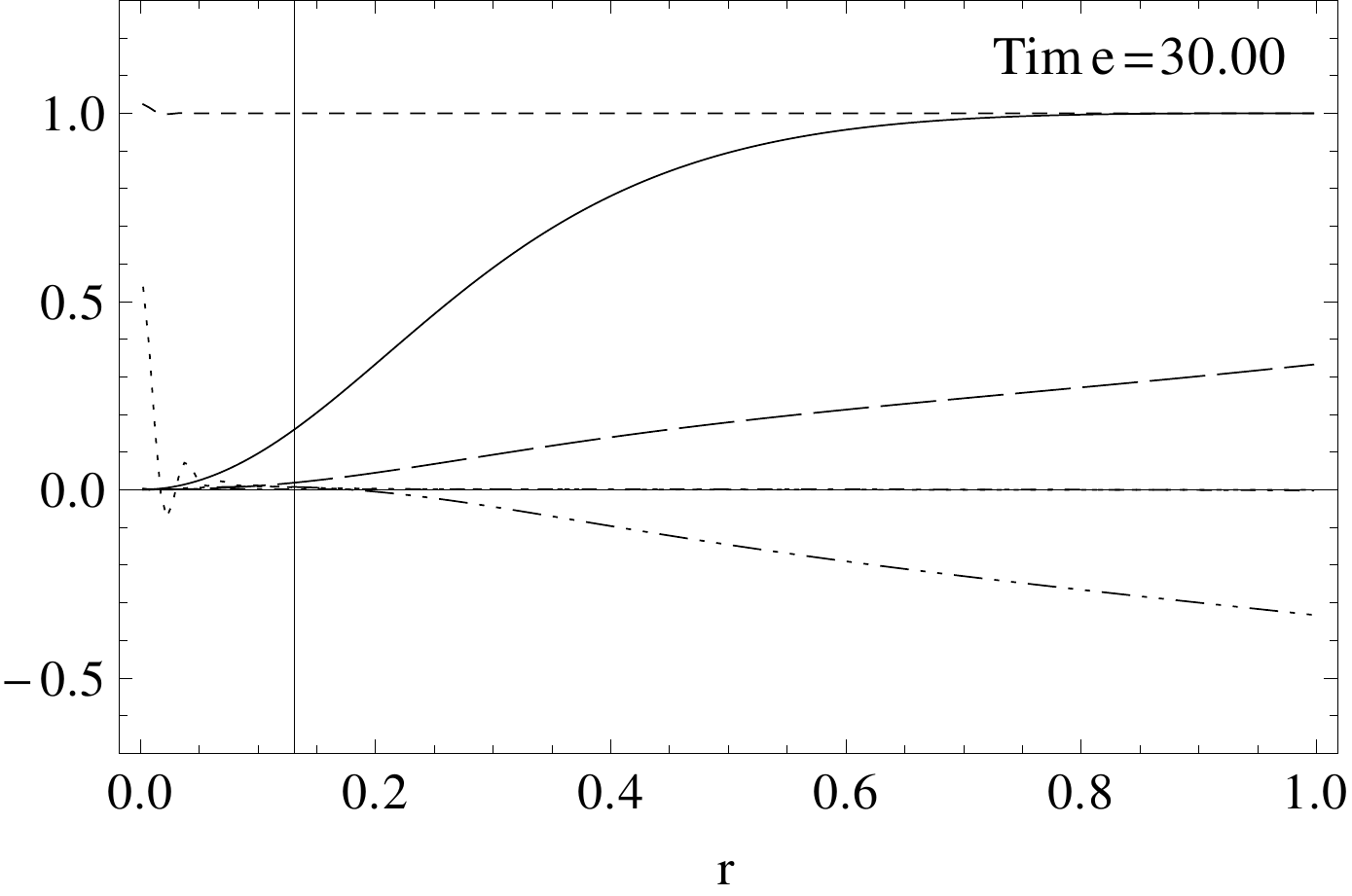}}&
\vspace{-5.5ex} \hspace{-0.8ex} \mbox{\includegraphics[width=1\linewidth]{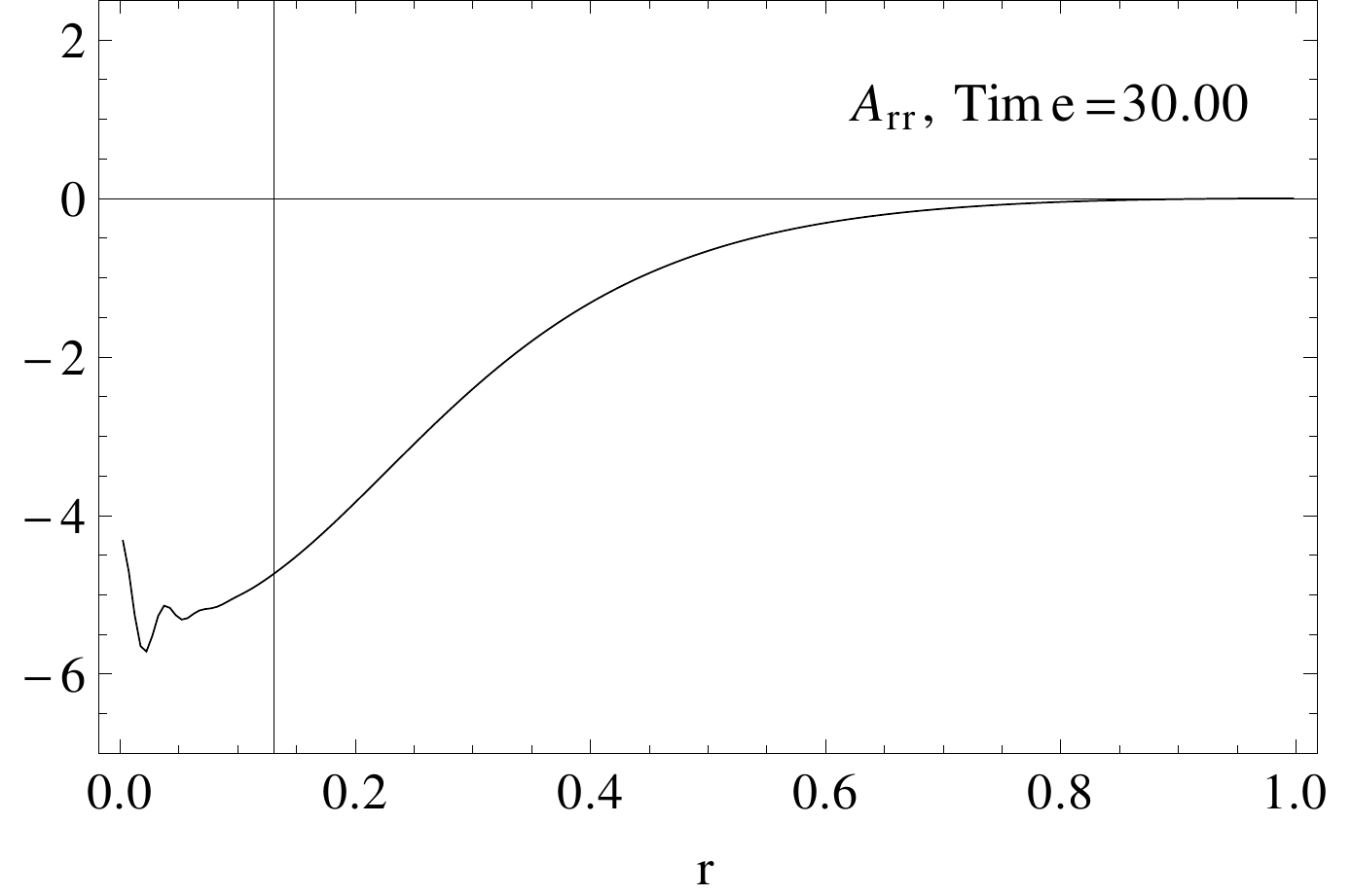}}
\end{tabular}
\vspace{-2ex}
\caption{Evolution of the variables under an initial perturbation of $\alpha$ at $r=0.25$.}
\label{fs:bhgauge}
\end{figure}

\begin{figure}[htbp!!]
\center
\vspace{-2ex}
 \begin{tikzpicture}[scale=1.5]\draw (-1cm,0cm) node {};
		\draw (0cm, 0cm) node {$\chi$}; \draw (0.3cm, 0cm) -- (1cm, 0cm);
		\draw (1.5cm, 0cm) node {$\gamma_{rr}$}; \draw [dashed] (1.8cm, 0cm) -- (2.5cm, 0cm);
		\draw (3cm, 0cm) node {$\DPK$}; \draw [dotted] (3.3cm, 0cm) -- (4cm, 0cm);
		\draw (4.5cm, 0cm) node {$\Lambda^r$}; \draw [dash pattern= on 4pt off 2pt on 1pt off 2pt] (4.8cm, 0cm) -- (5.5cm, 0cm);
		\draw (6cm, 0cm) node {$\alpha$}; \draw [dash pattern= on 8pt off 2pt] (6.3cm, 0cm) -- (7cm, 0cm);
		\draw (7.5cm, 0cm) node {$\beta^r$}; \draw [dash pattern= on 6pt off 2pt on 1pt off 2pt on 1pt off 2pt] (7.8cm, 0cm) -- (8.5cm, 0cm);
	\end{tikzpicture}
\\
\begin{tabular}{ m{0.5\linewidth}@{} @{}m{0.5\linewidth}@{} }
\mbox{\includegraphics[width=1\linewidth]{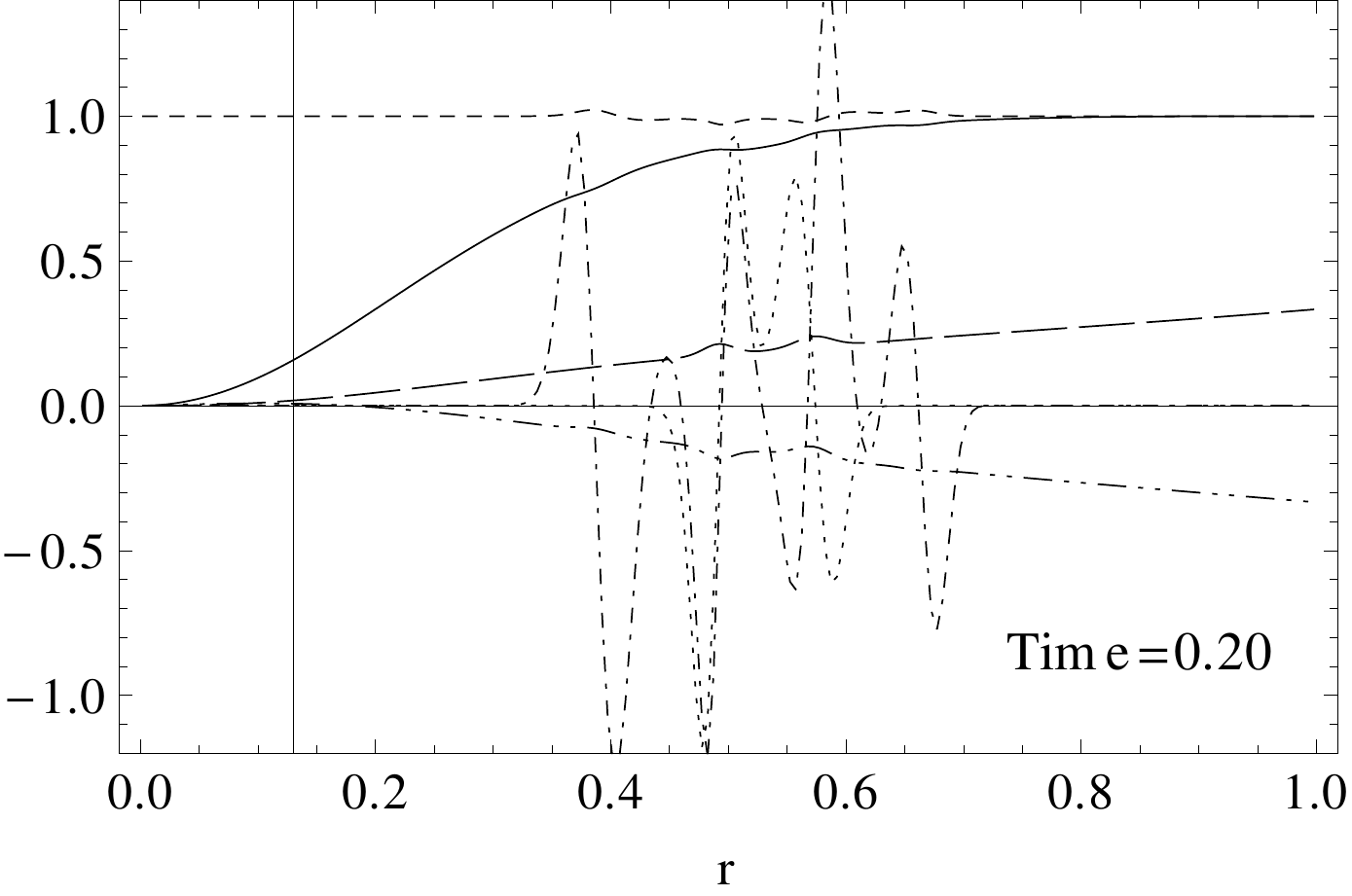}}&
\hspace{-0.8ex} \mbox{\includegraphics[width=1\linewidth]{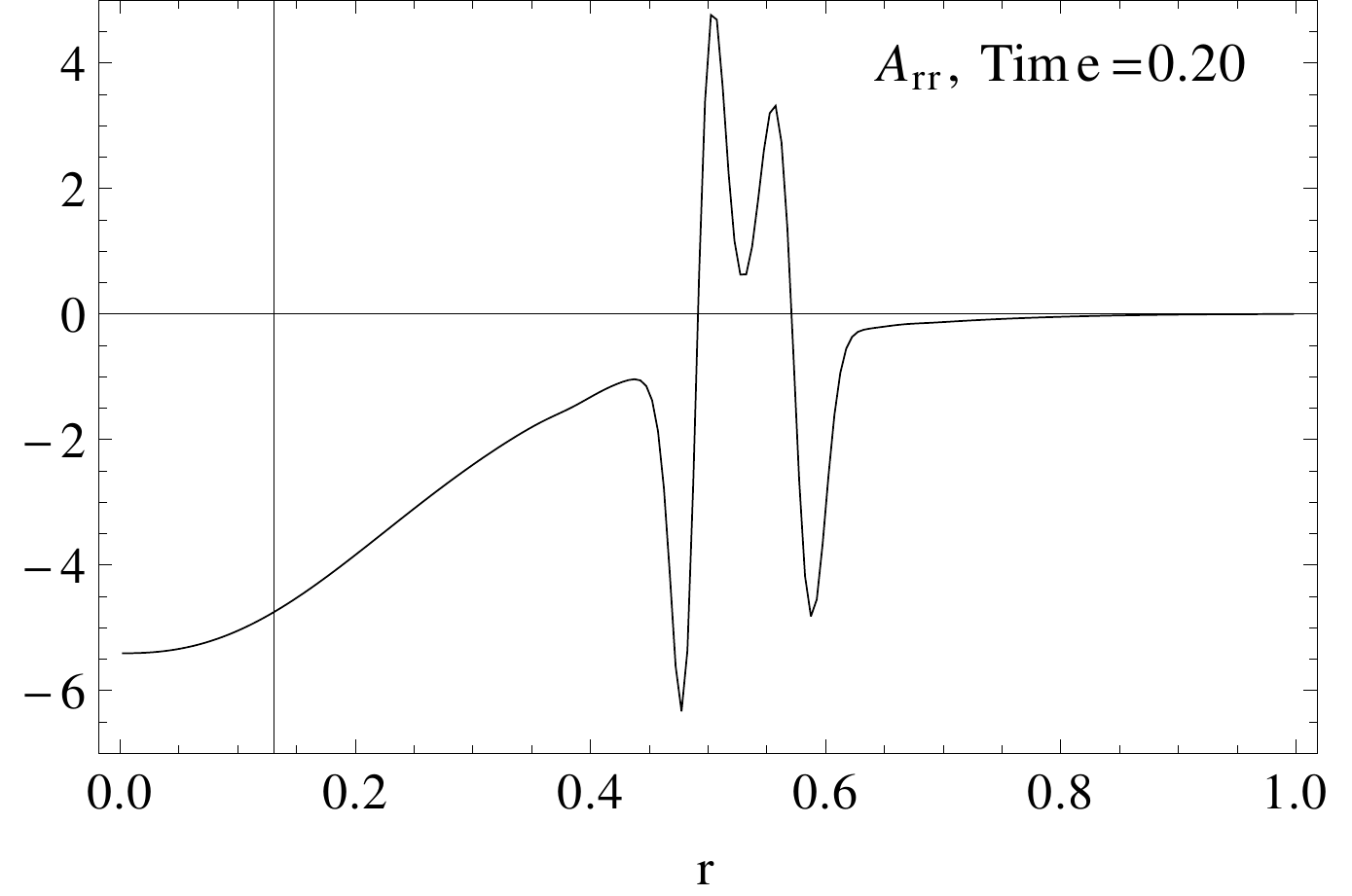}}\\
\vspace{-5.5ex} \mbox{\includegraphics[width=1\linewidth]{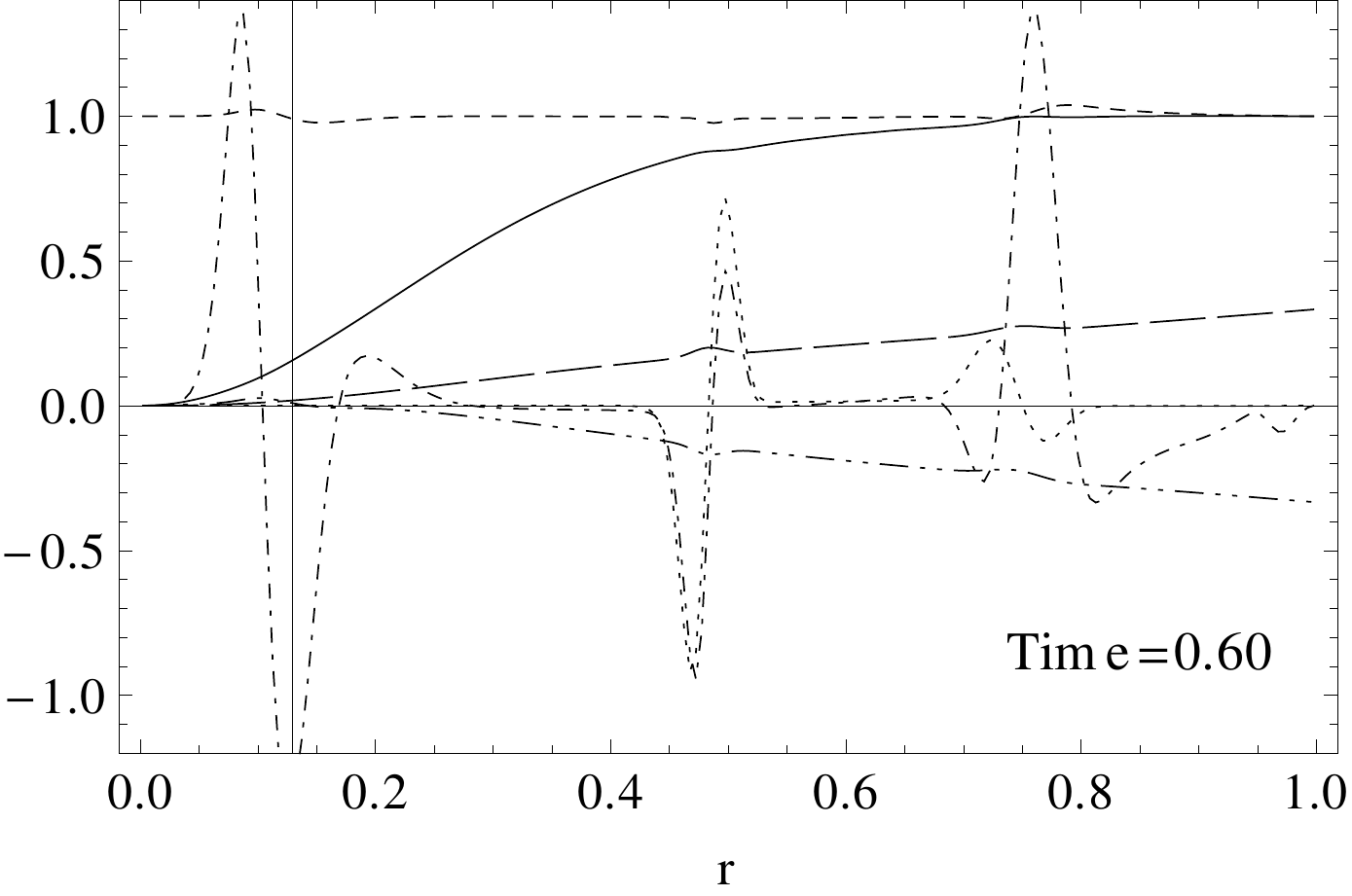}}&
\vspace{-5.5ex} \hspace{-0.8ex} \mbox{\includegraphics[width=1\linewidth]{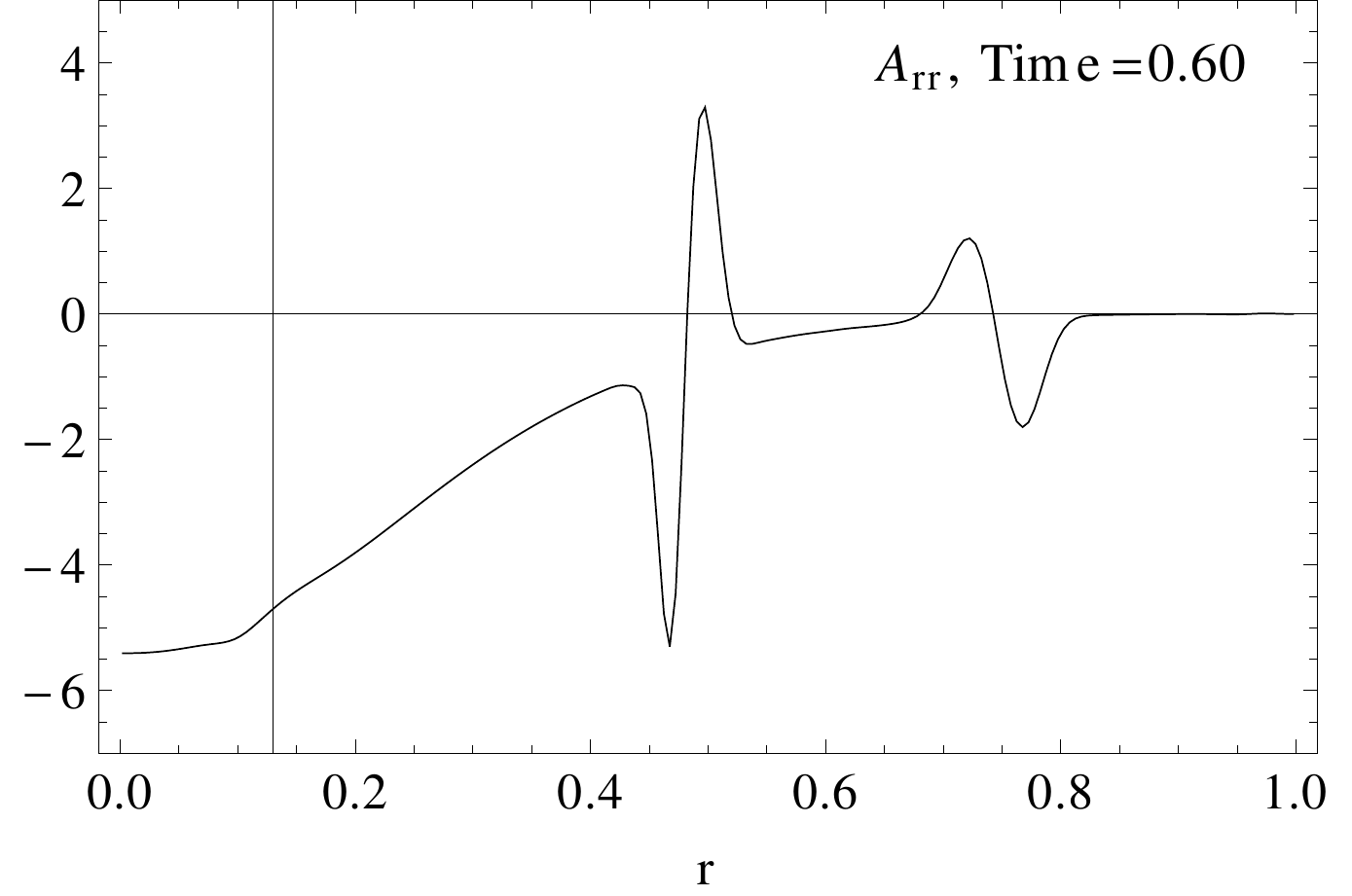}}\\
\vspace{-5.5ex} \mbox{\includegraphics[width=1\linewidth]{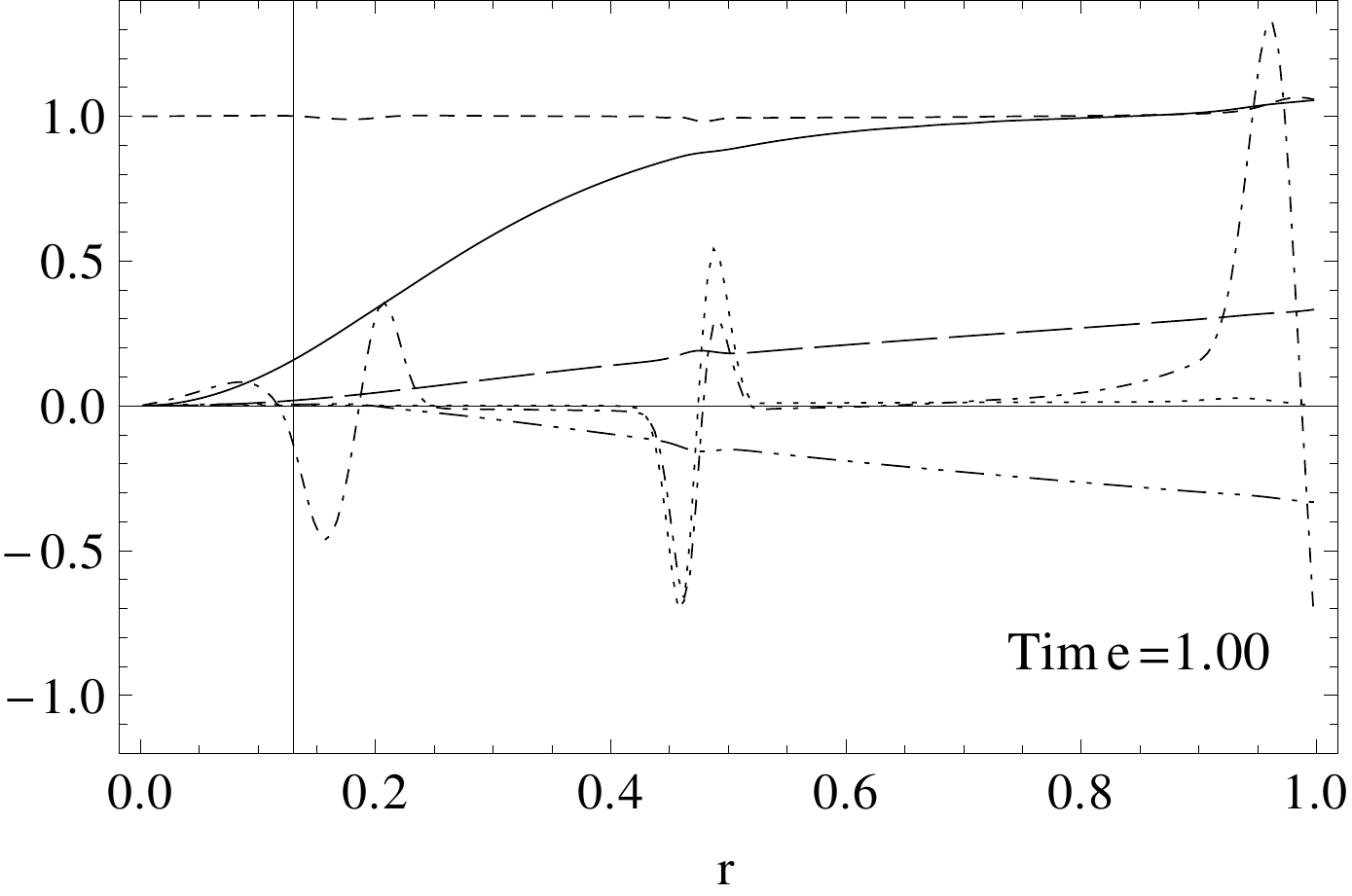}}&
\vspace{-5.5ex} \hspace{-0.8ex} \mbox{\includegraphics[width=1\linewidth]{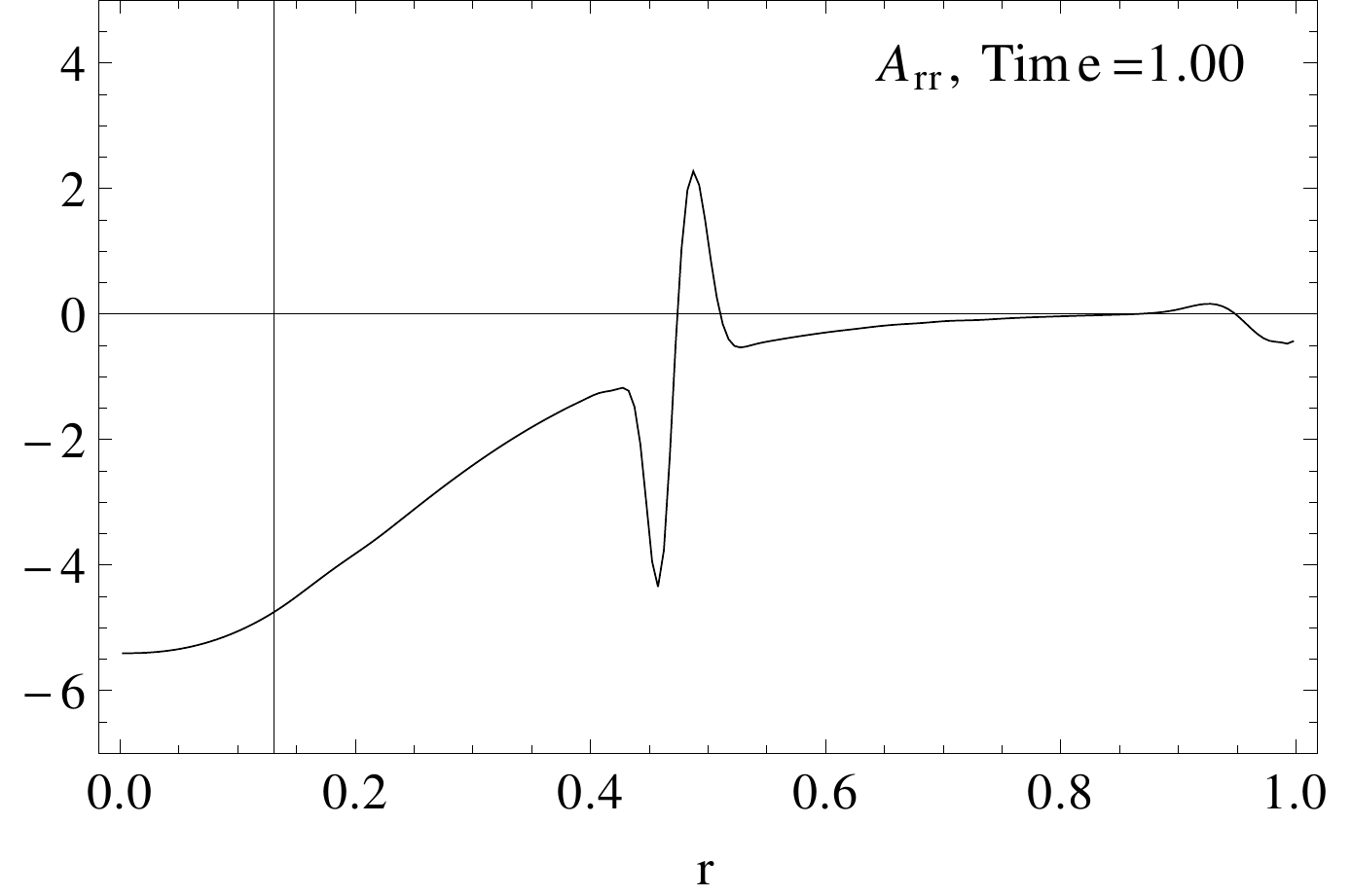}}\\
\vspace{-5.5ex} \mbox{\includegraphics[width=1\linewidth]{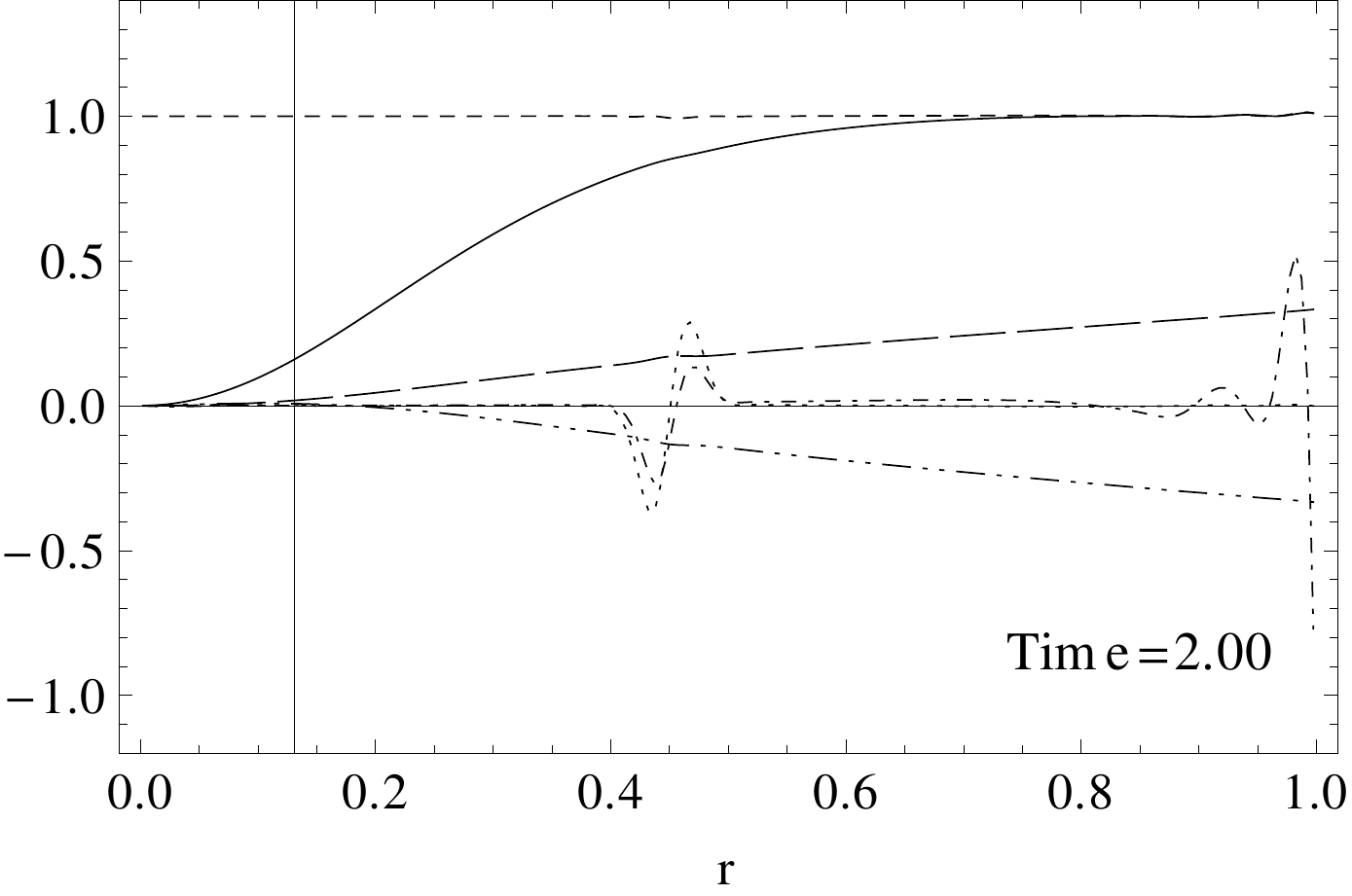}}&
\vspace{-5.5ex} \hspace{-0.8ex} \mbox{\includegraphics[width=1\linewidth]{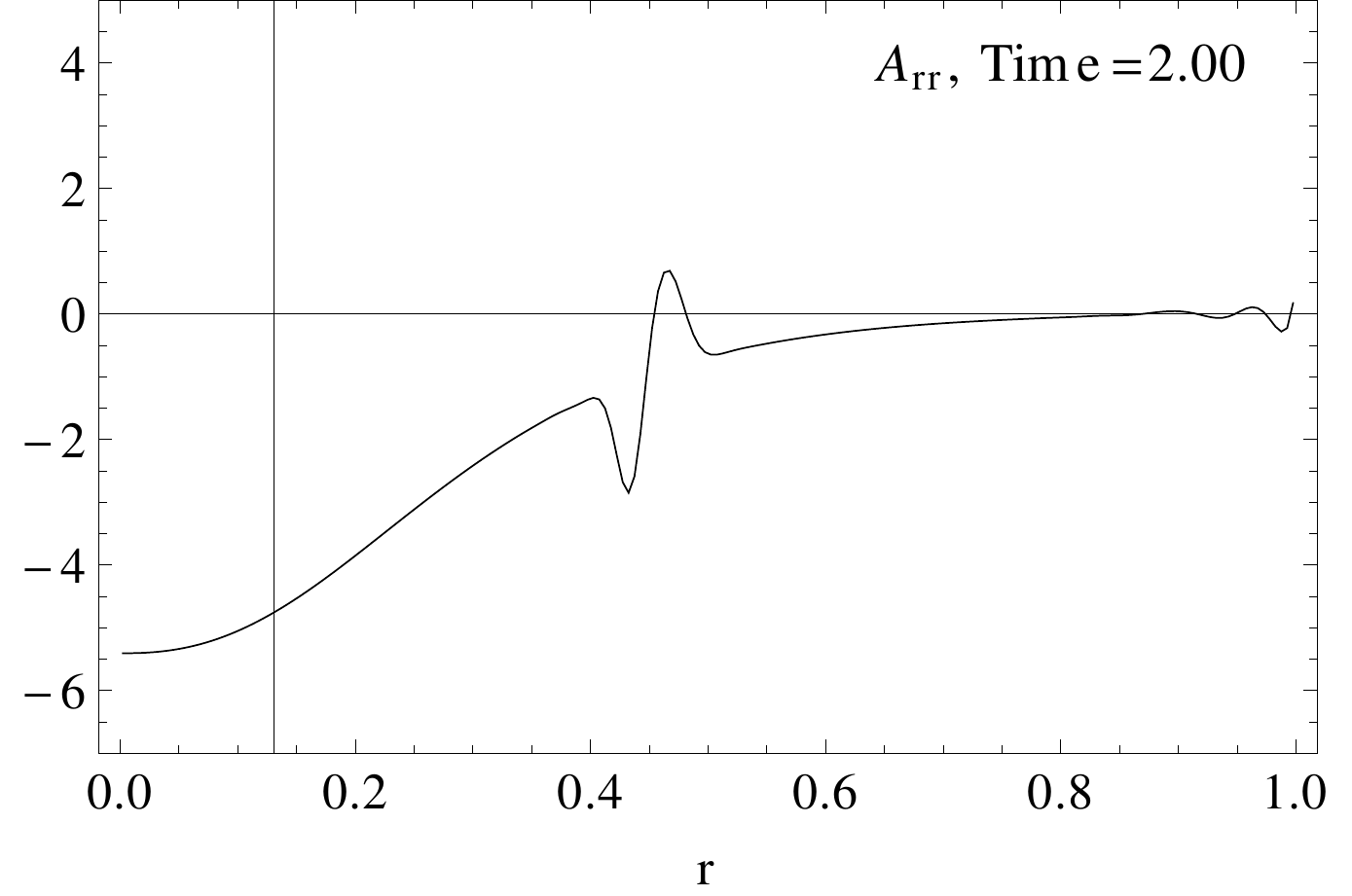}}\\
\vspace{-5.5ex} \mbox{\includegraphics[width=1\linewidth]{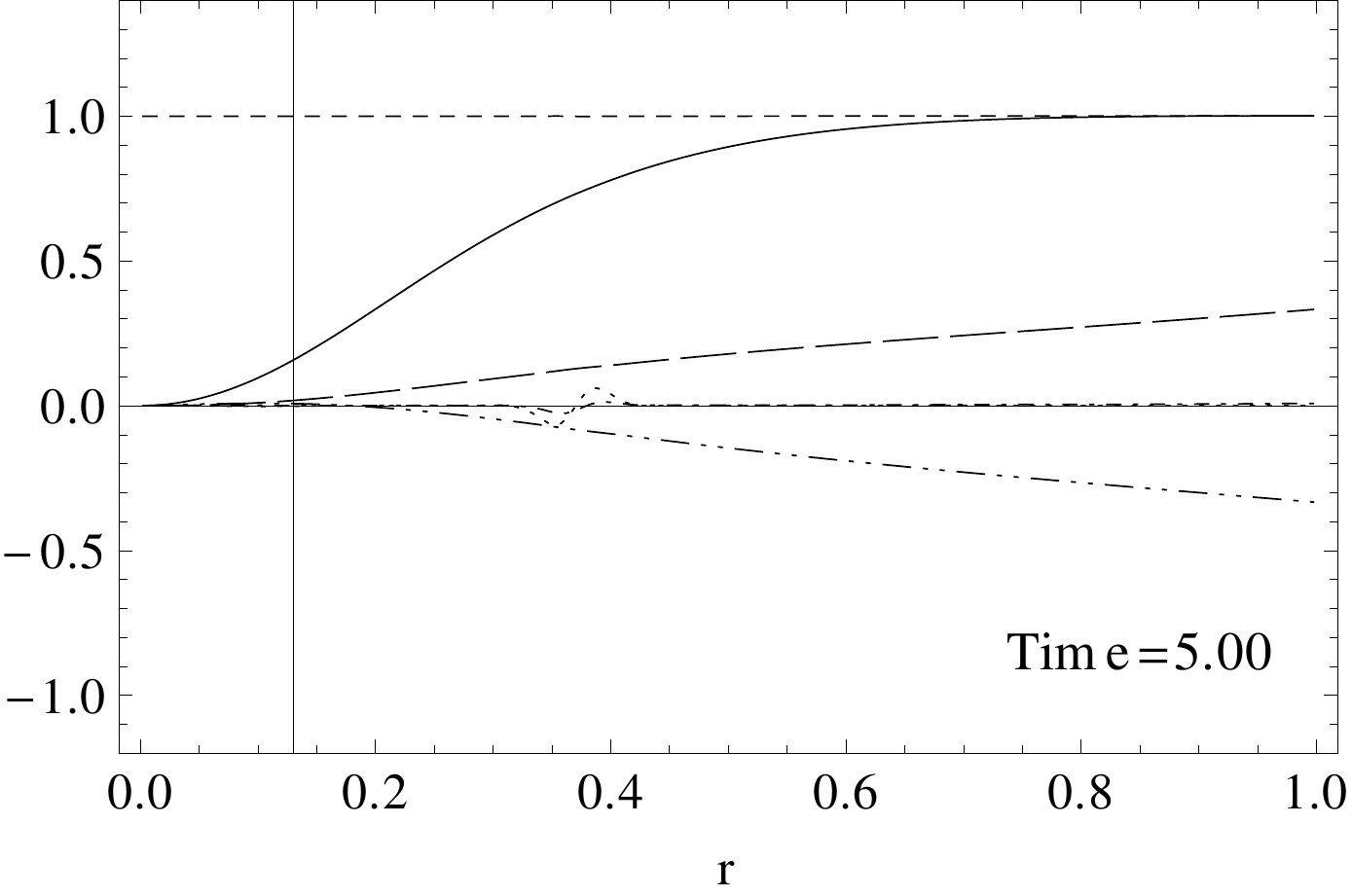}}&
\vspace{-5.5ex} \hspace{-0.8ex} \mbox{\includegraphics[width=1\linewidth]{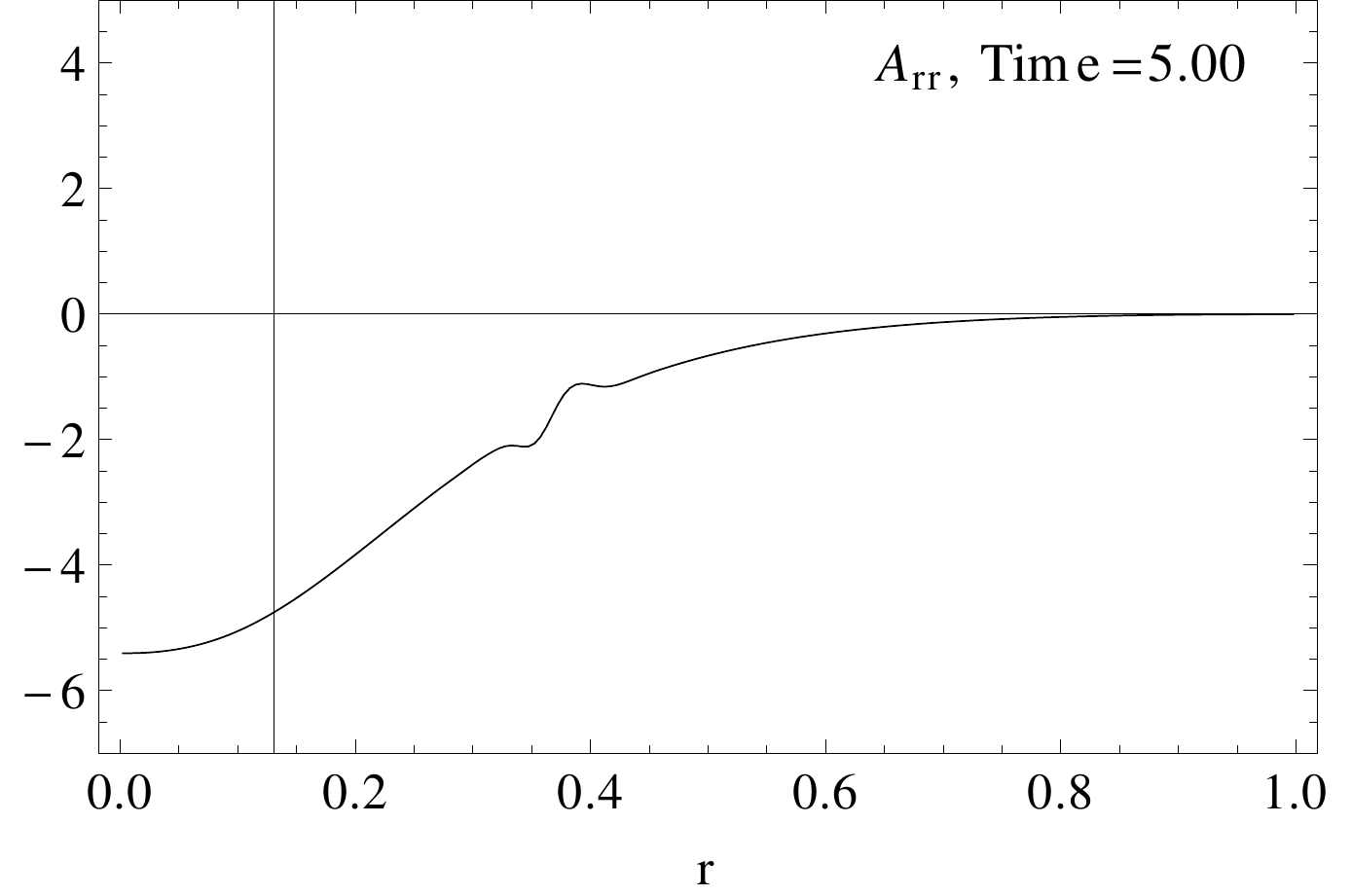}}
\end{tabular}
\vspace{-2ex}
\caption{Evolution of the variables under an initial perturbation of $\alpha$ at $r=0.5$.}
\label{fs:bhgaugec05}
\end{figure}
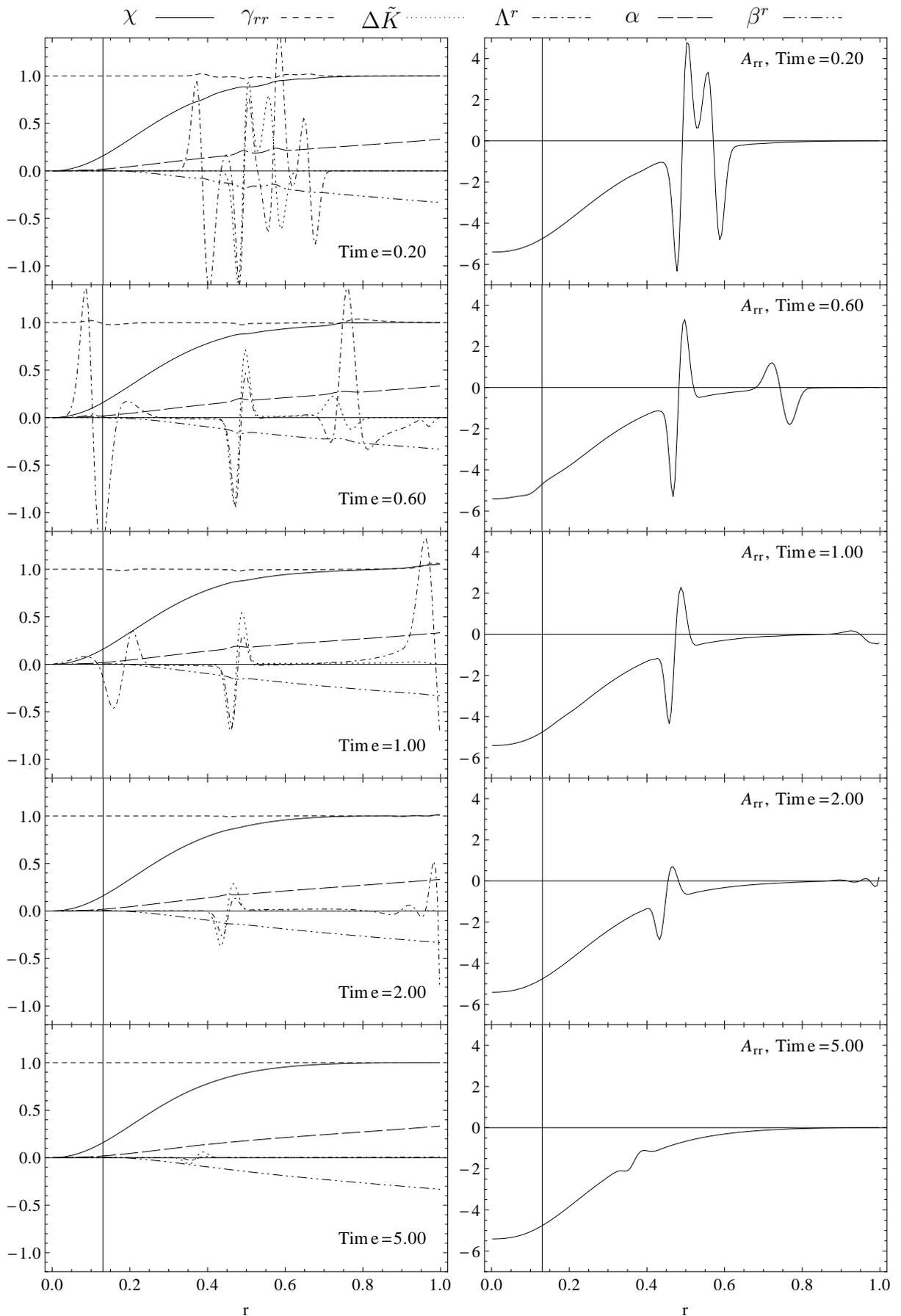

Figures \ref{fs:bhgauge} and \ref{fs:bhgaugec05} show two examples of gauge waves evolution with Schwarzschild initial data. The evolution system consisted of the GBSSN equations with harmonic slicing and the integrated Gamma-driver shift condition, with $\lambda=\case{1}{12}\approx0.083$. The initial data are those in \fref{fin:Bini} and the initial perturbations on $\alpha$ are shown in \ref{fs:bhgaugealpha}: their amplitude ($A_\alpha=0.1$) and width ($\sigma=0.1$) were the same, but one of them was centered in $c=0.25$ and the other at $r=0.5$. As their location is given on the compactified coordinate, the effect of each initial perturbations can be quite different.
%initial data given on the compactified coordinate

% Gauge waves

% Gauge waves: initial centered

\subsection{Small scalar field perturbation on Schwarzschild spacetime}

%Unfortunately, I have not managed to evolve a scalar field with initial perturbation $A_{\Phi}\geq10^{-2}$ with the current setup and the developed gauge conditions. The code would crash at the origin or at some other point in the domain as a consequence of the drift, which is faster the larger the initial perturbation.
The larger the initial amplitude of the scalar field perturbation, the faster the drift experienced by the variables. This drift was first described in subsections \ref{se:schwini} and \ref{se:onlySchw}, and its dependence on the initial amplitude is shown in \fref{fs:drift}. For amplitudes of the order $10^{-2}$ and for some choices of the gauge conditions, in the worst case the numerical evolution can crash at a time as early as $t=30$, which does not even allow to see the dissipation of the scalar field inside of the BH. For this reason we will now concentrate on studying small scalar field perturbations and will leave the drift effect for future work.

\begin{figure}[htbp!!]
\center\vspace{-1.5ex}
\begin{tabular}{@{}c@{}@{}c@{}}
\mbox{\includegraphics[width=0.5\linewidth]{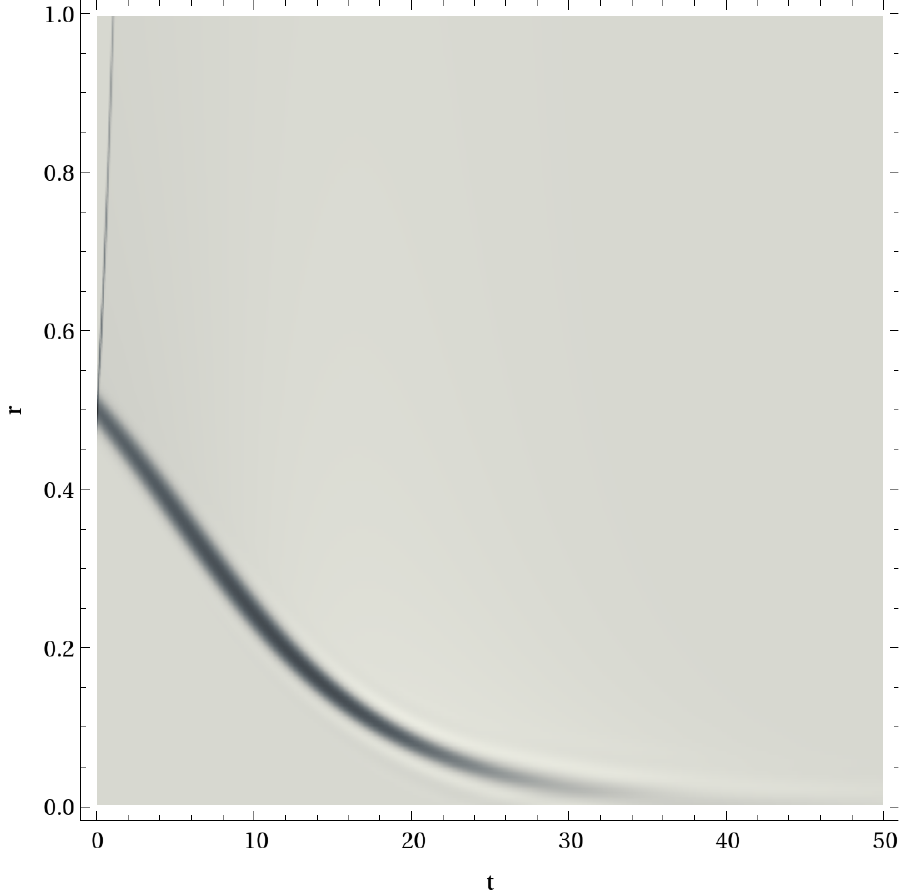}}&
\mbox{\includegraphics[width=0.5\linewidth]{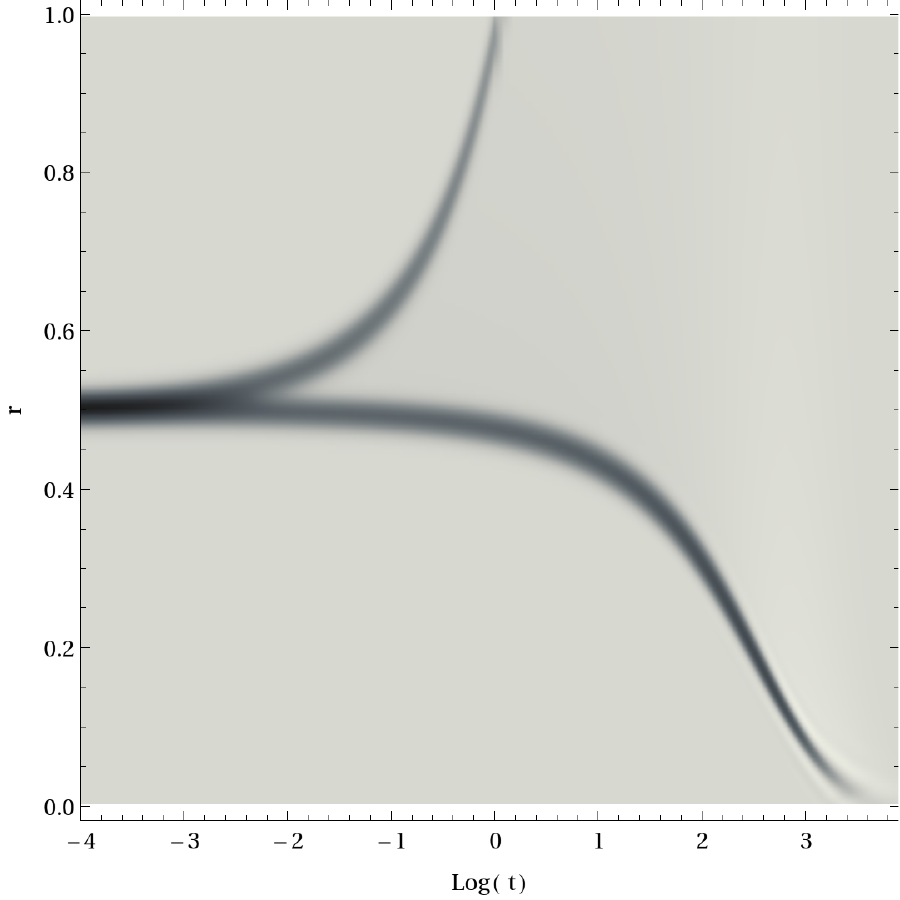}}
\end{tabular}\vspace{-2ex}
\caption{Evolution of the scalar field, in grayscale its amplitude with its maximum value in black. The plot on the left shows clearly the ingoing wave pulse, which approaches the origin with a speed that tends to zero and is damped away in the end, while the outgoing one is represented by a very thin line. The plot on the right presents the same data as a function of the logarithm of time, to be able to clearly distinguish the outgoing wave pulse which leaves through $\scri^+$ (at $r=1$) at a very early time ($t\sim1$ or $\log(t)\sim0$).}
\label{fs:phirtlog}
\end{figure}

\begin{figure}[htbp!!]
\center
\vspace{-2.5ex}
 \begin{tikzpicture}[scale=2.0]\draw (-1cm,0cm) node {};
		\draw (-0.3cm, 0cm) node {\small Time symmetric}; \draw (0.5cm, 0cm) -- (0.8cm, 0cm);
		\draw (1.6cm, 0cm) node {\small Ingoing}; \draw [dashed] (2cm, 0cm) -- (2.3cm, 0cm);
		\draw (3.0cm, 0cm) node {\small Outgoing}; \draw [dotted] (3.5cm, 0cm) -- (3.8cm, 0cm);
	\end{tikzpicture}
\\
\begin{tabular}{ m{0.5\linewidth}@{} @{}m{0.5\linewidth}@{} }
\mbox{\includegraphics[width=1\linewidth]{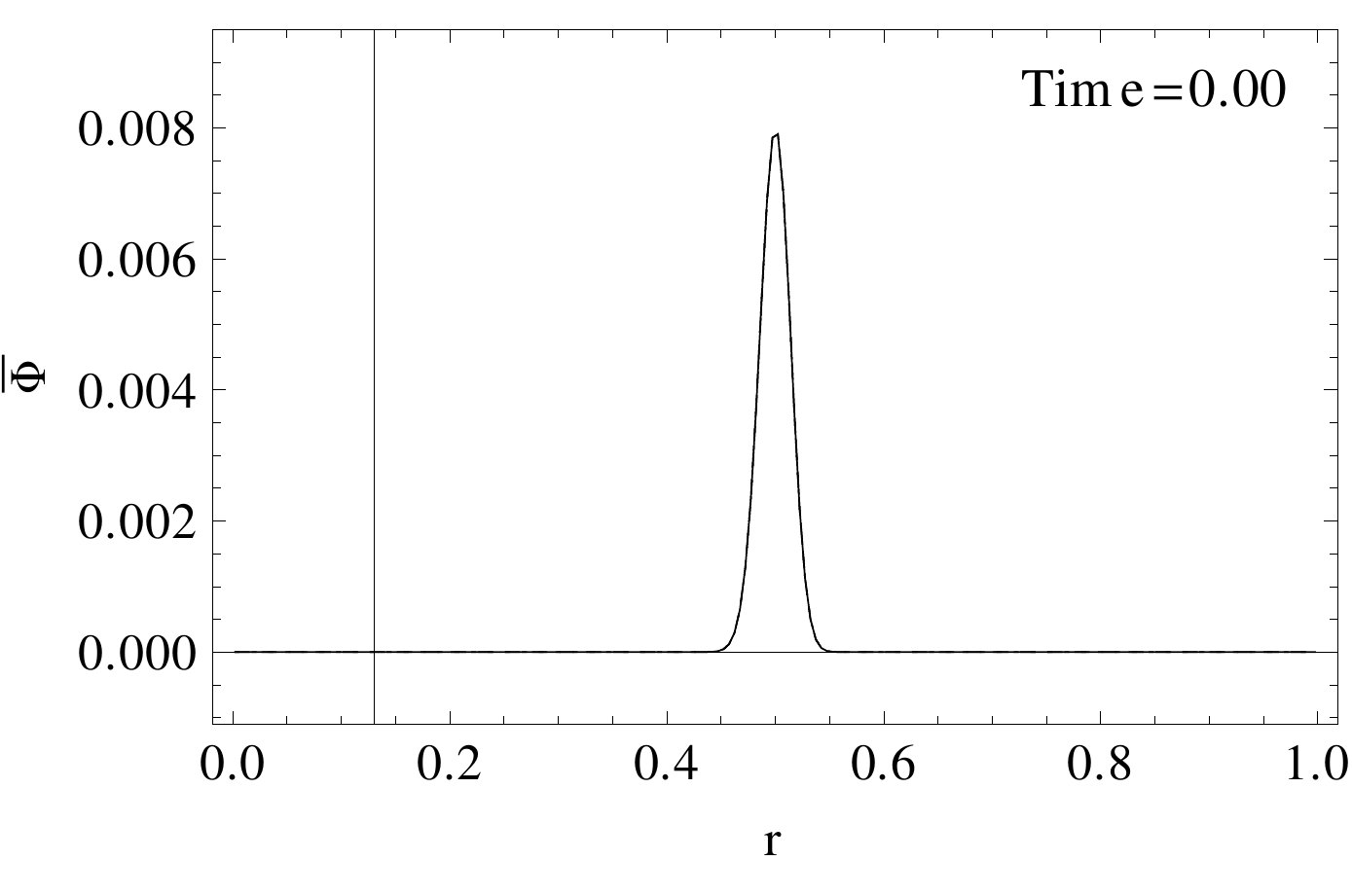}}&
\hspace{-0.6ex} \mbox{\includegraphics[width=1.0\linewidth]{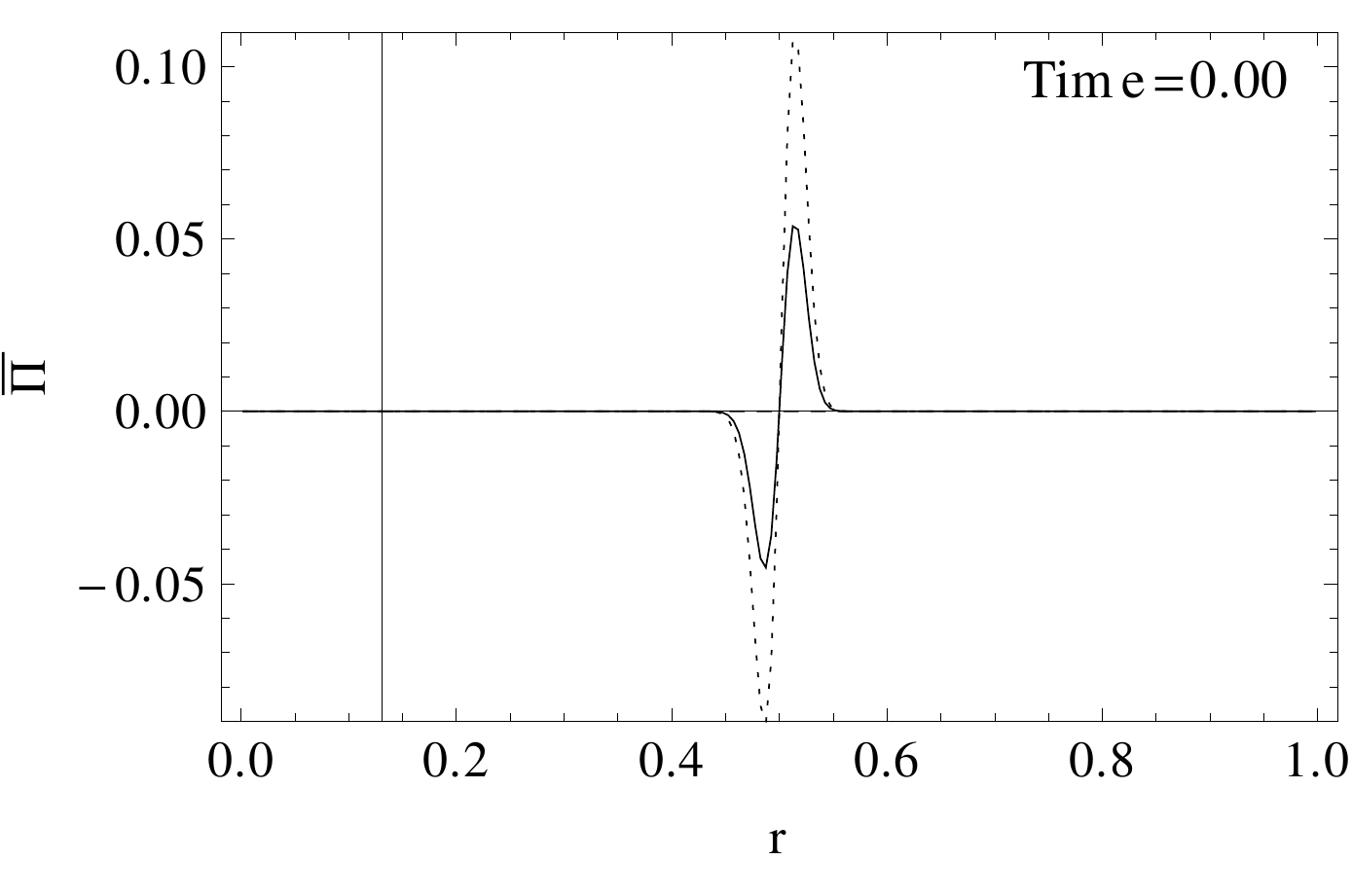}}\\
\vspace{-6.0ex} \mbox{\includegraphics[width=1\linewidth]{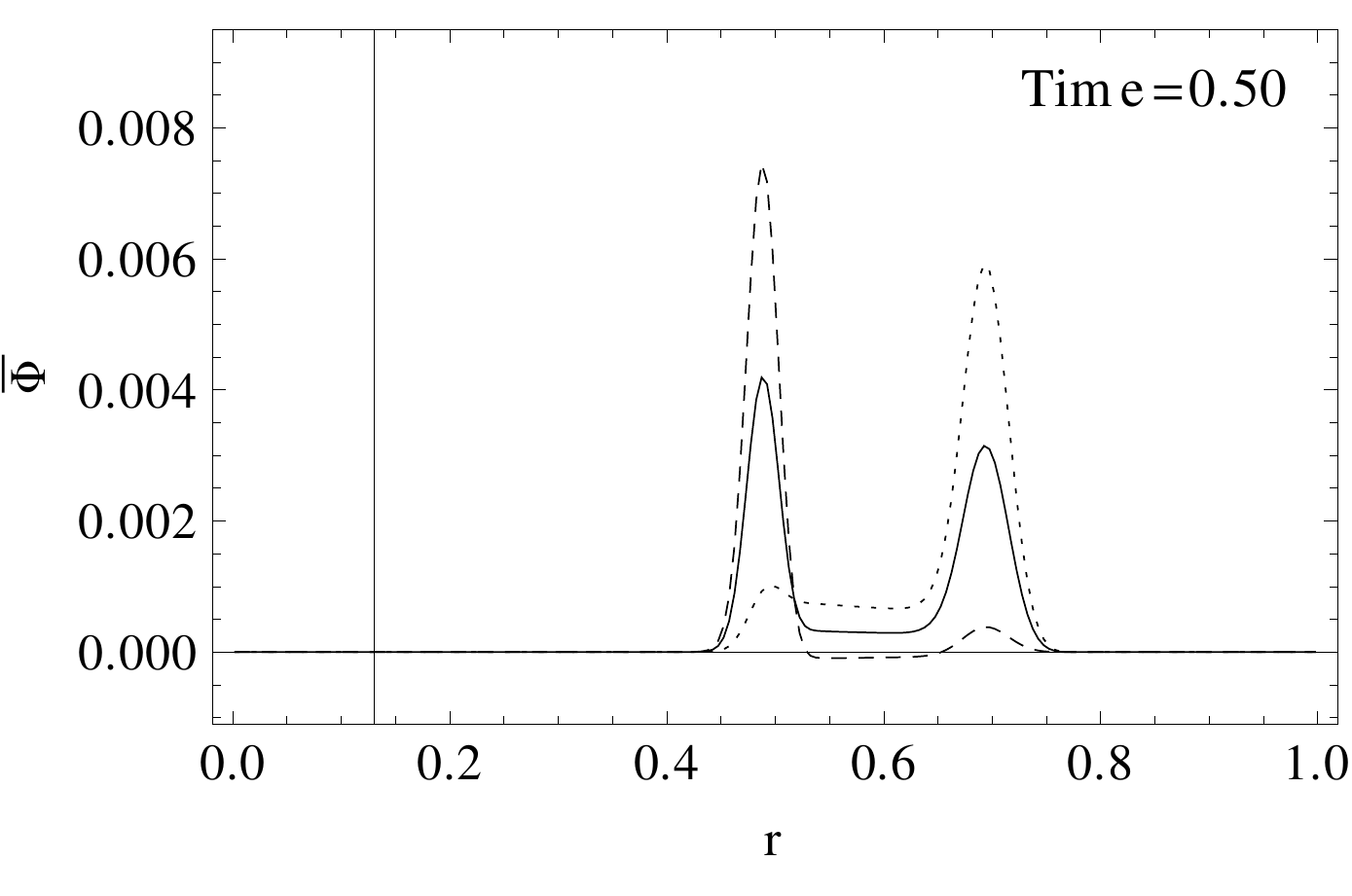}}&
\vspace{-6.0ex} \hspace{-0.6ex} \mbox{\includegraphics[width=1.0\linewidth]{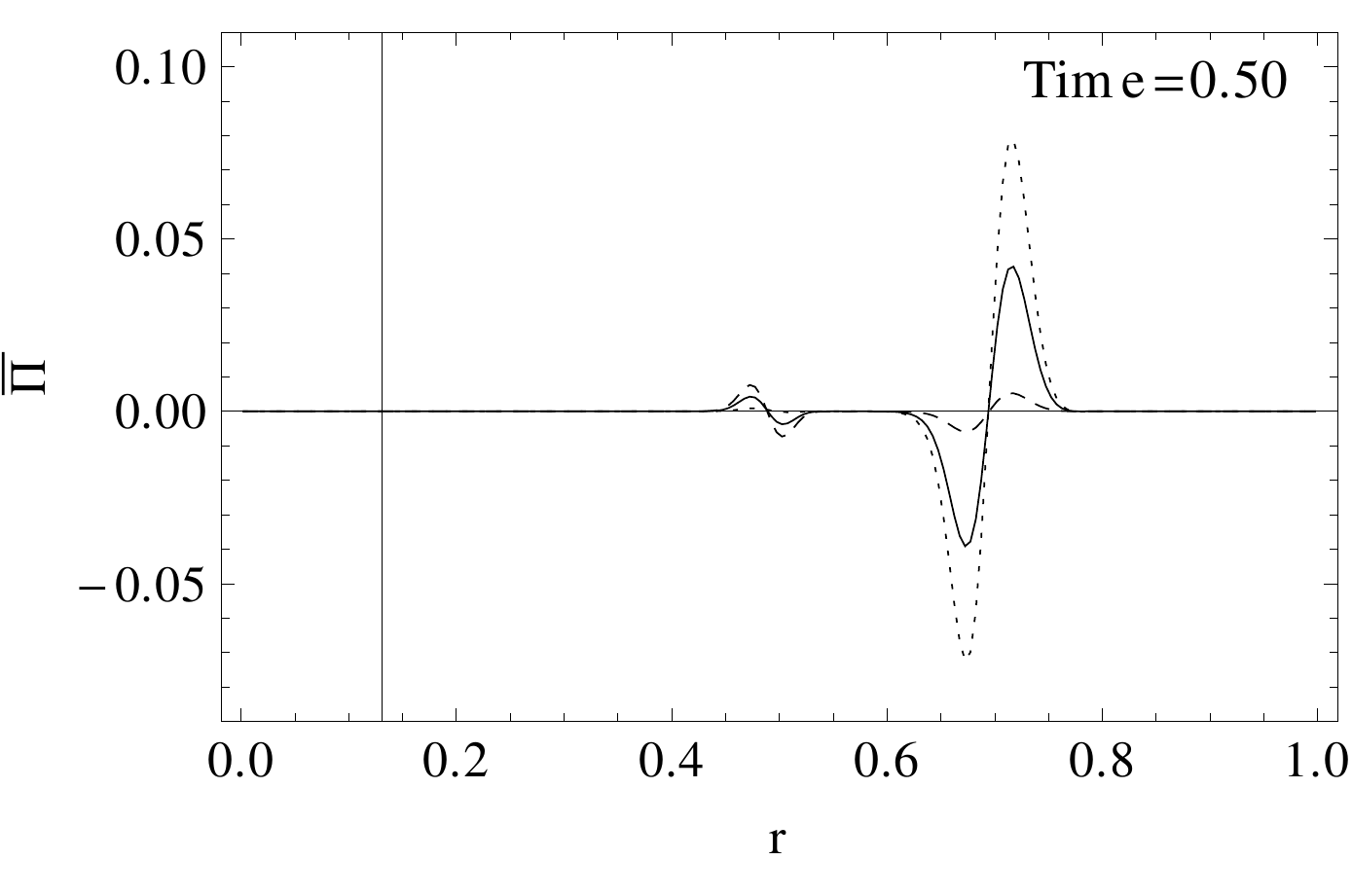}}\\
\vspace{-6.0ex} \mbox{\includegraphics[width=1\linewidth]{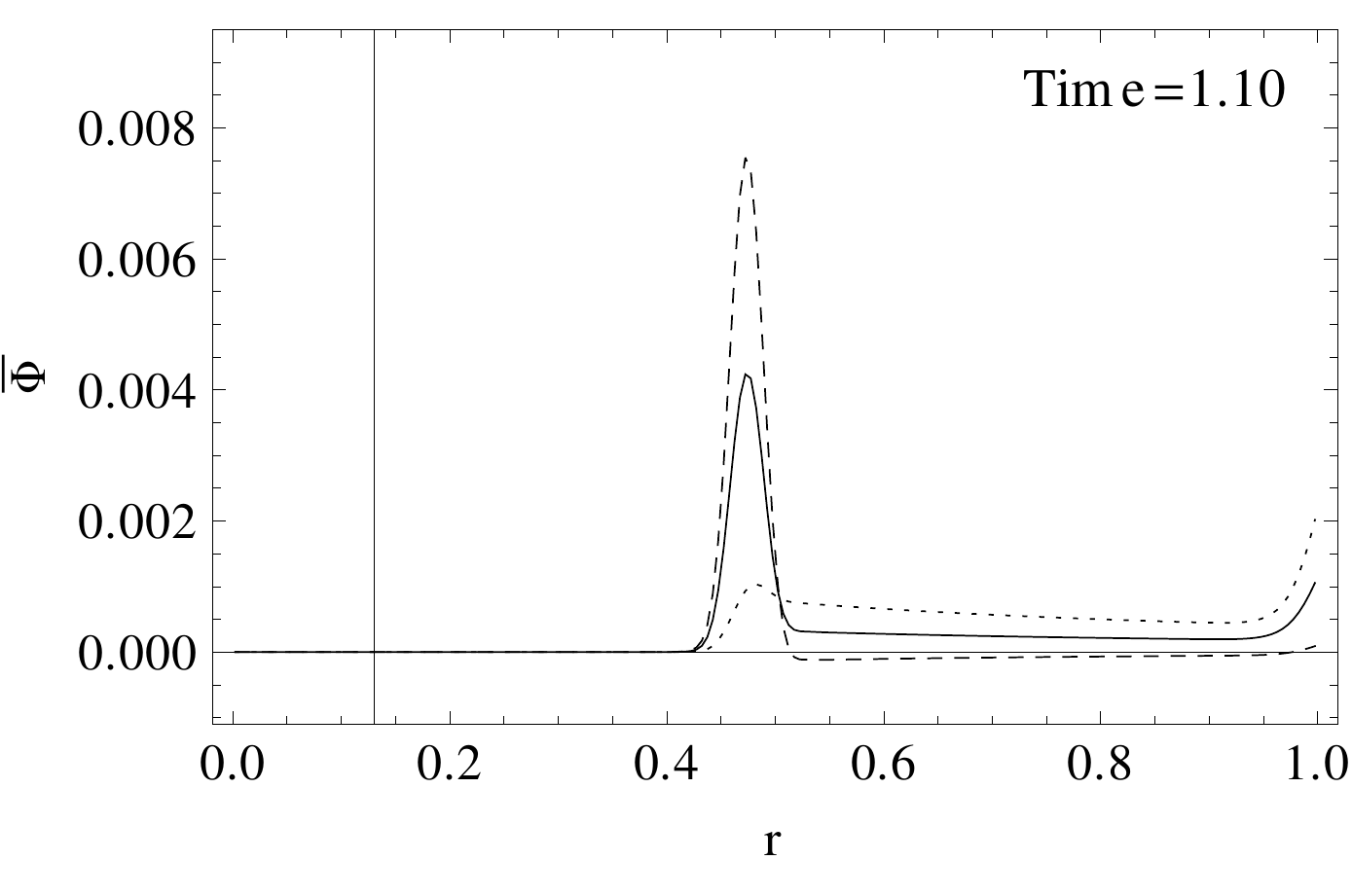}}&
\vspace{-6.0ex} \hspace{-0.6ex} \mbox{\includegraphics[width=1.0\linewidth]{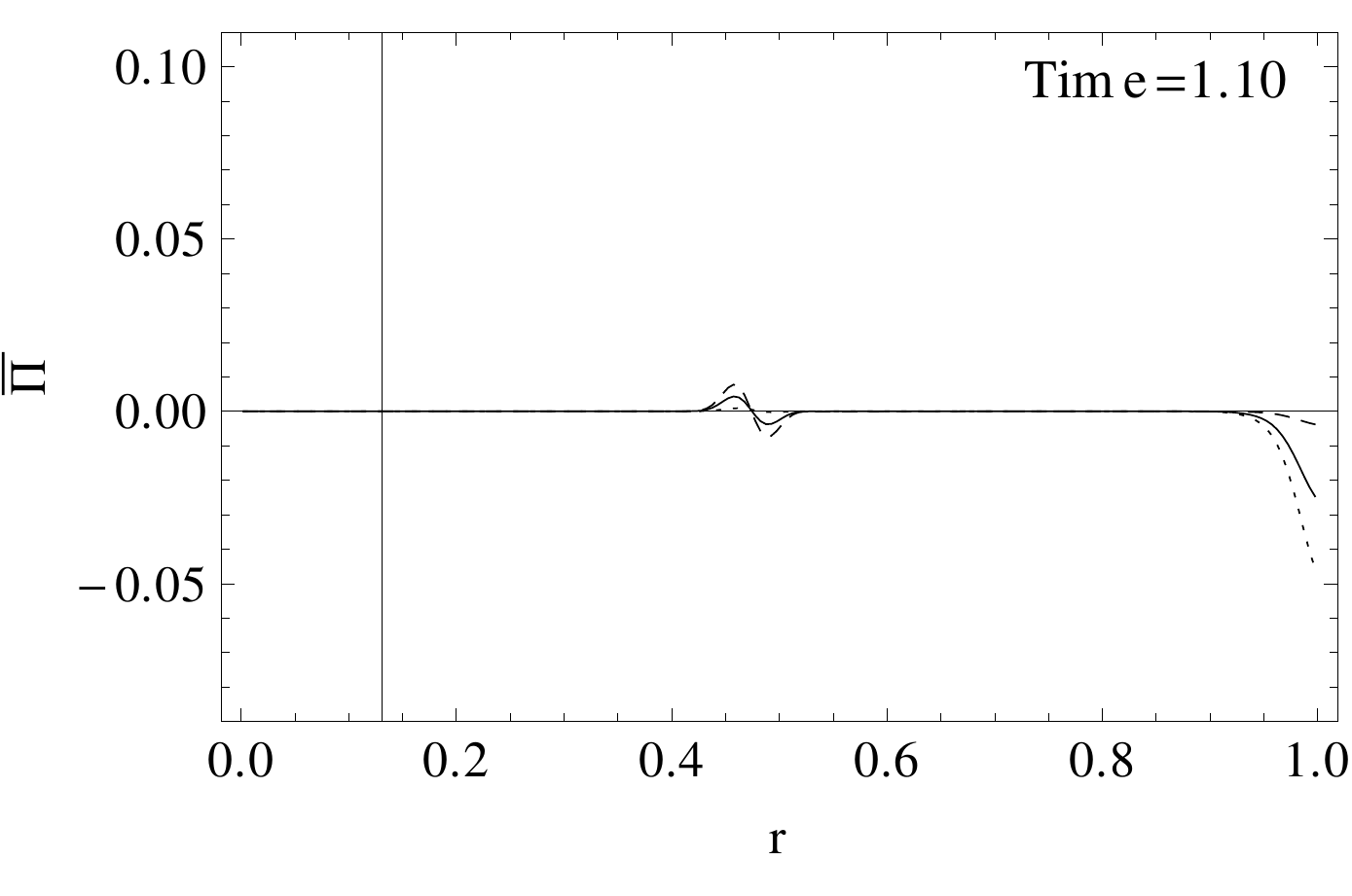}}\\
\vspace{-6.0ex} \mbox{\includegraphics[width=1\linewidth]{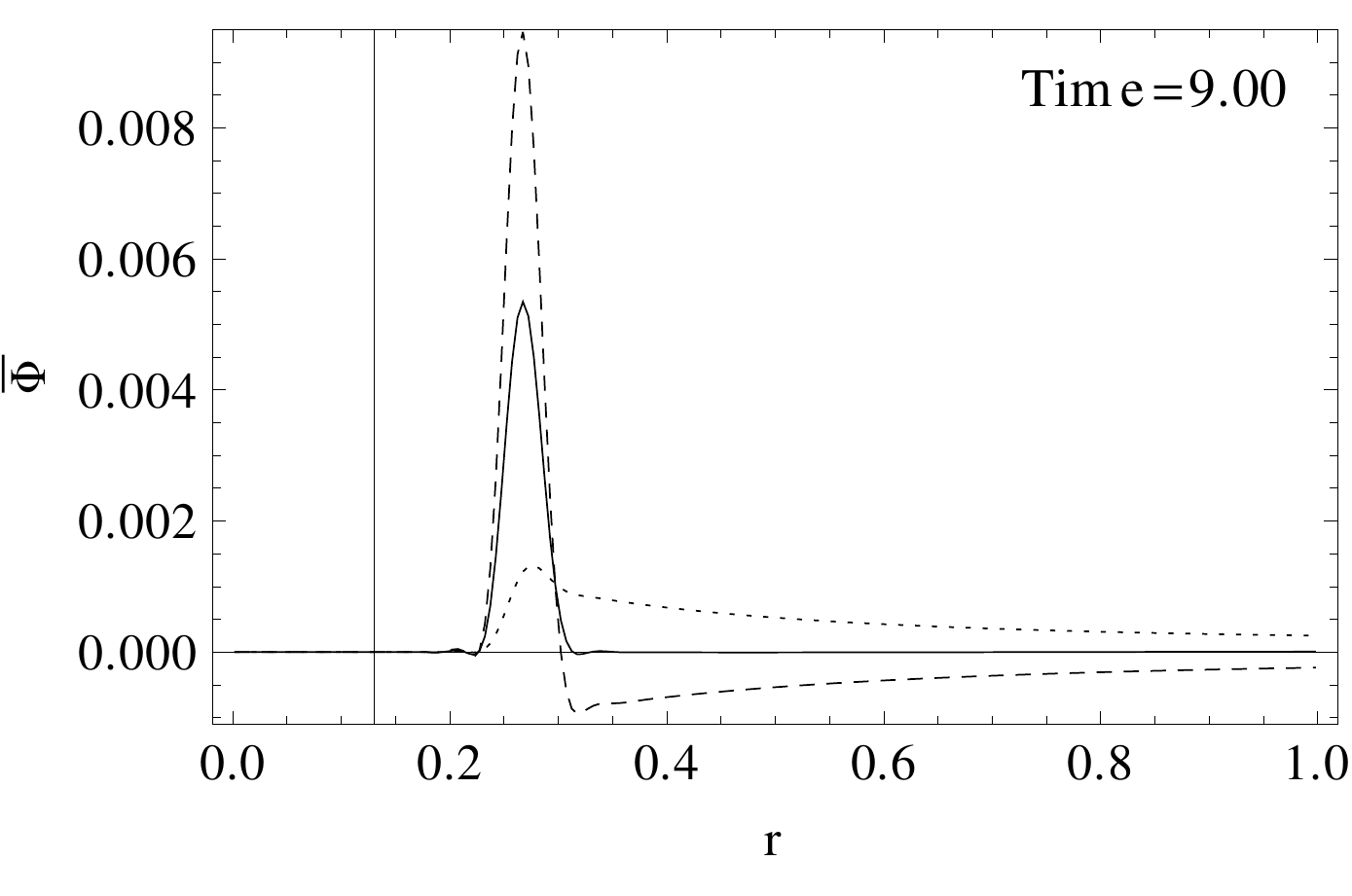}}&
\vspace{-6.0ex} \hspace{-0.6ex} \mbox{\includegraphics[width=1.0\linewidth]{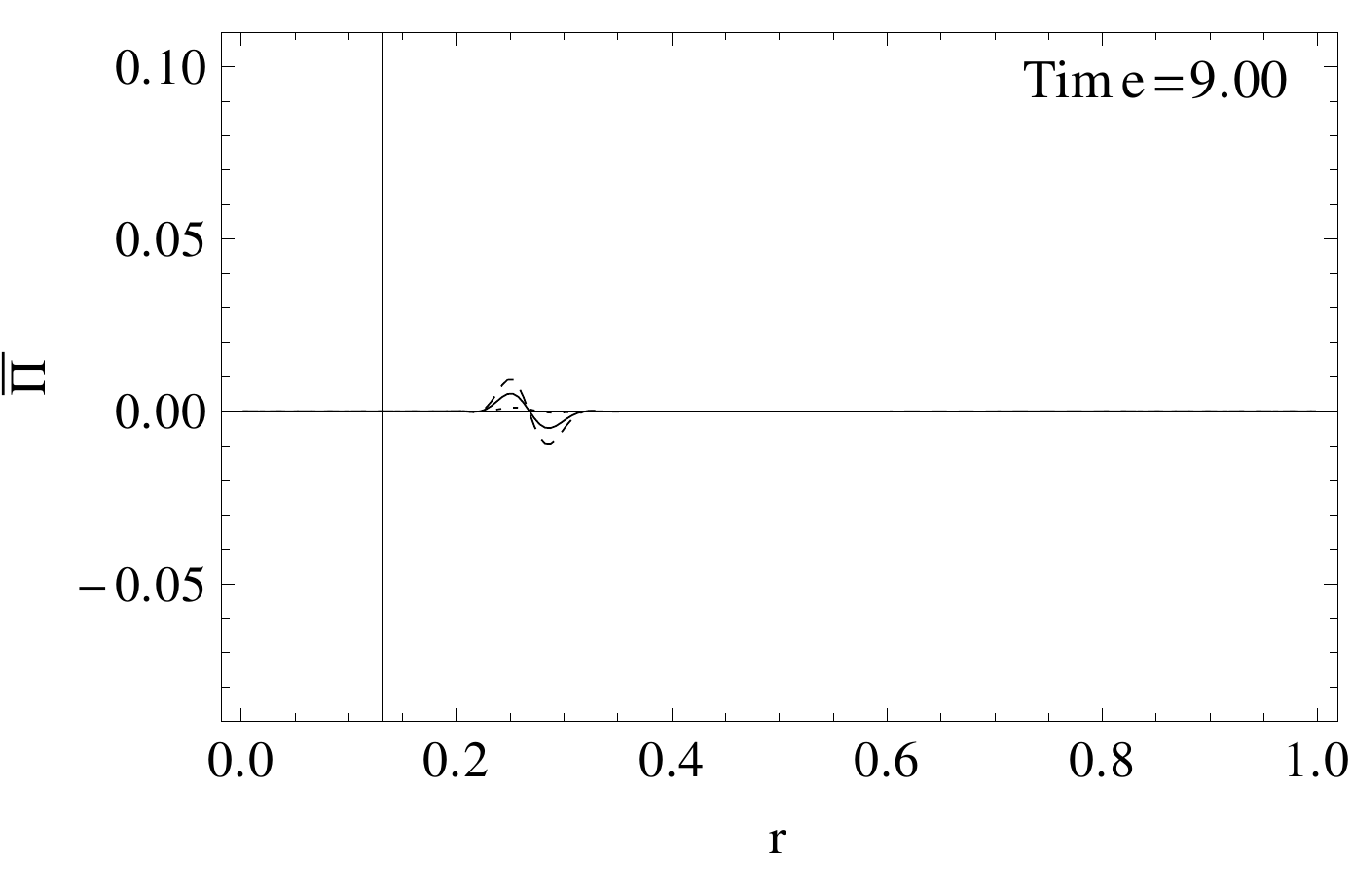}}\\
\vspace{-6.0ex} \mbox{\includegraphics[width=1\linewidth]{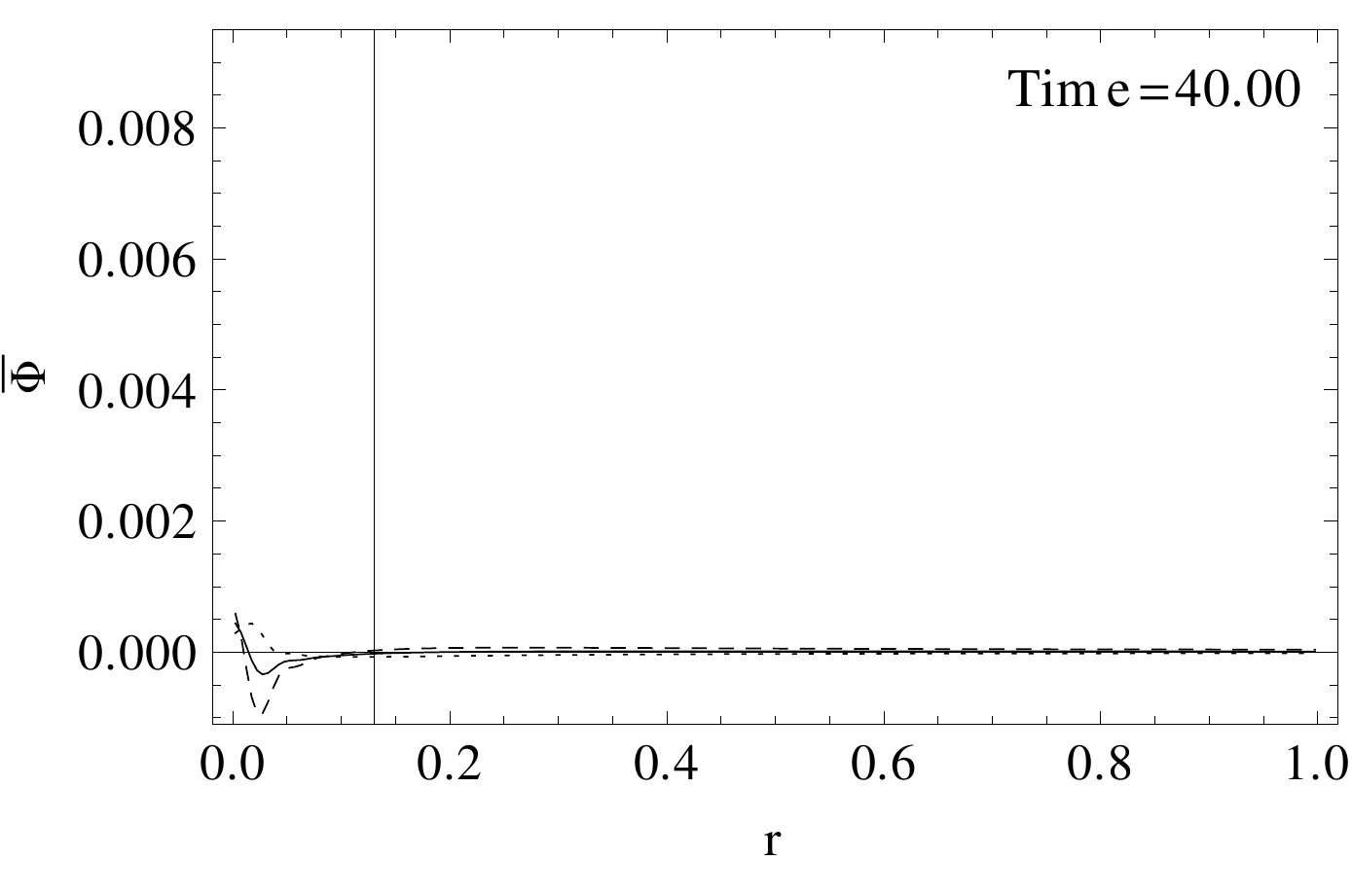}}&
\vspace{-6.0ex} \hspace{-0.6ex} \mbox{\includegraphics[width=1.0\linewidth]{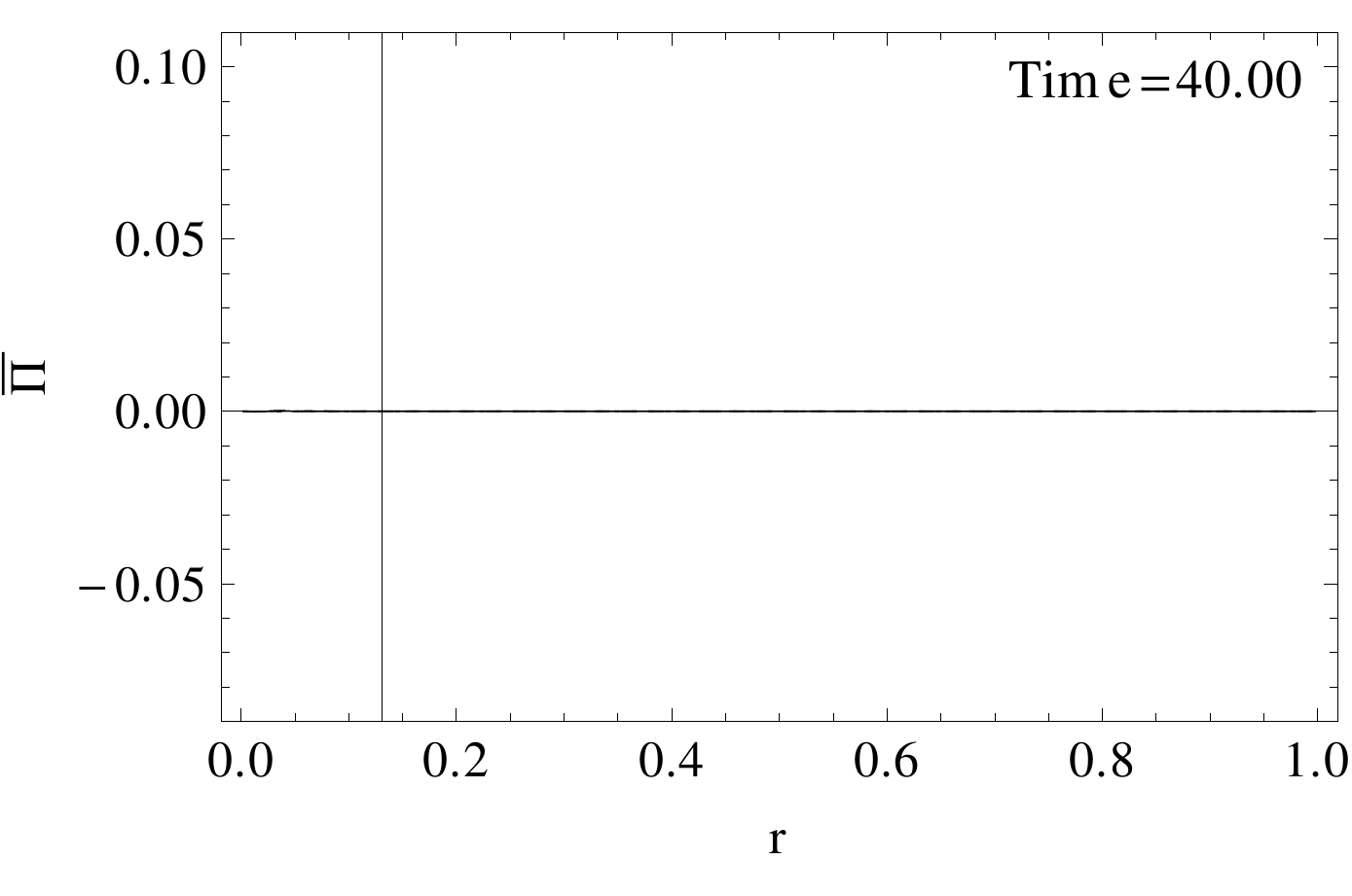}}
\end{tabular}
\vspace{-3ex}
\caption{Rescaled scalar field $\bPhi$ and its auxiliary variable $\bPhi$ for an evolution with 200 gridpoints and initial $A_{\Phi}=10^{-3}$ for time symmetric, ingoing and outgoing initial data. The rest of variables are not shown, because the perturbation is small enough to leave them visually unperturbed.}
\label{fs:phibhevol}
\end{figure}
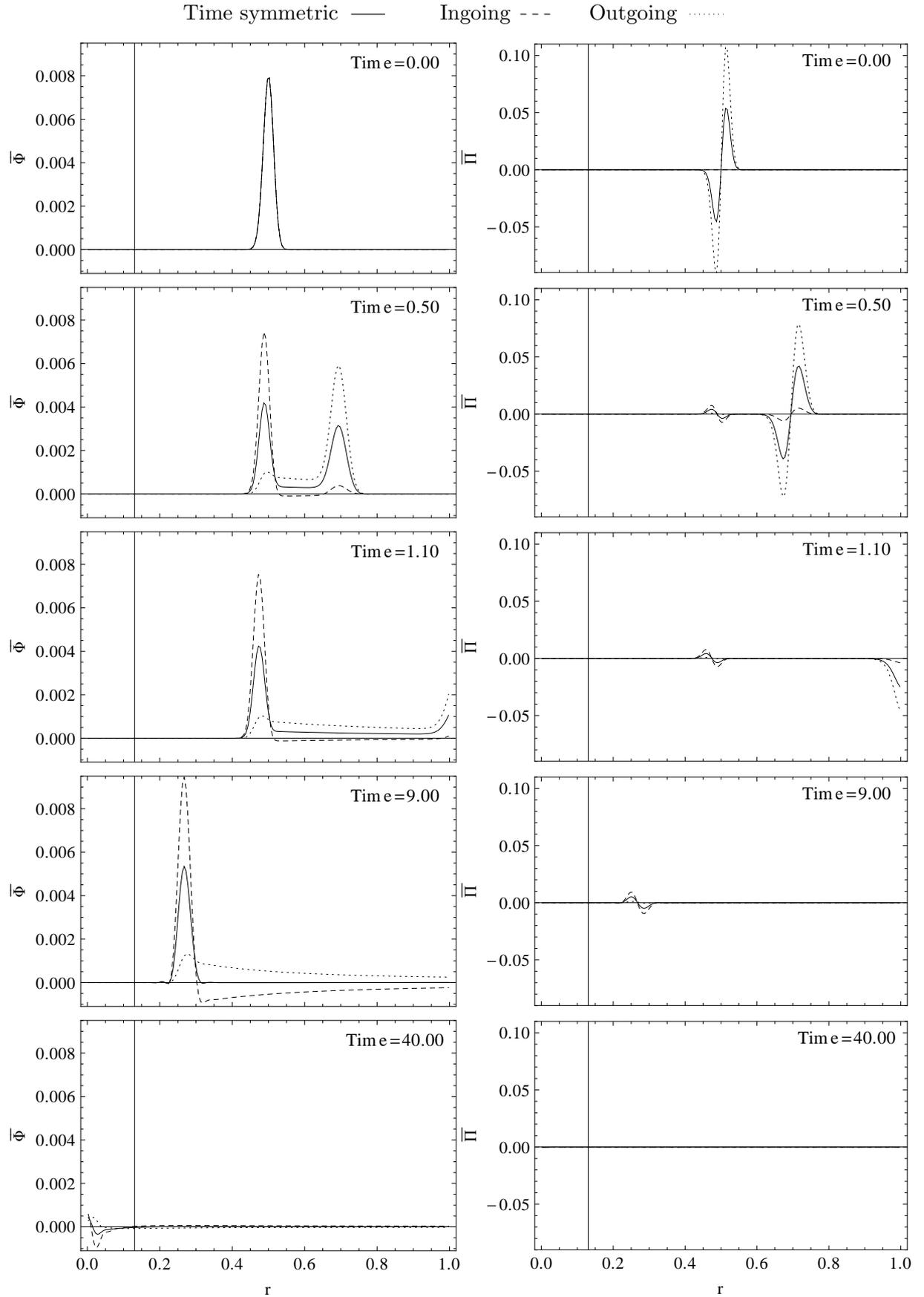

The simulations whose scalar field is shown in figures \ref{fs:phirtlog} and \ref{fs:phibhevol} use an initial amplitude of the scalar field of $A_{\Phi}=10^{-3}$, a width of $\sigma=0.1$ and are centered at $r=c=0.5$. The initial data of the simulation in \fref{fs:phirtlog} and in the solid lines in \fref{fs:phibhevol} are time symmetric, which means that the scalar field perturbation is in- and outgoing in equal parts. The dotted and dashed lines in \fref{fs:phibhevol} represent mostly ingoing and mostly outgoing initial data (the difference can be seen in the value of $\bPi$ in the top-right plot in \fref{fs:phibhevol} and in the amplitude of the pulses of $\bPhi$ at later times).

The behaviour of the scalar field is the following: the outgoing pulse in which the initial perturbation splits moves towards $\scri^+$ and leaves the domain (this is represented by the line going upwards in \fref{fs:phirtlog}). The ingoing pulse propagates towards the origin and its speed decreases as it comes closer to the trumpet. Once it is inside of the horizon (located at $r_{Schw}\approx0.13$), it is slowly dissipated away.

Given the small amplitude of the perturbation, the change in the mass and the apparent horizon of the BH can be neglected: for instance, the largest change in the horizon location (corresponding to the mostly ingoing scalar field) was approximately 0.15\%.

% Scalar field in BH evol

\subsection{Power-law decay tails}

A scalar field perturbation of strong field initial data is expected to decay at late times with a power-law tail of the form \cite{PhysRevD.5.2419}
\begin{equation}
\lim_{t \to +\infty}\Phi(t,r) \propto t^{p}.
\end{equation}
Analytical calculations have determined that for spherical scalar perturbations the decay rate $p$ is $p=-3$ along timelike surfaces \cite{PhysRevD.5.2419} and $p=-2$ along null surfaces ($\scri^+$) \cite{Bonnor66,PhysRevD.49.883}.
In our simulations we found a value of $p\sim -2.08\pm0.09$ for the scalar field on $\scri^+$ \cite{Vano-Vinuales:2014ada}.

\begin{figure}[htbp!!]
\center
	\includegraphics[width=1.02\linewidth]{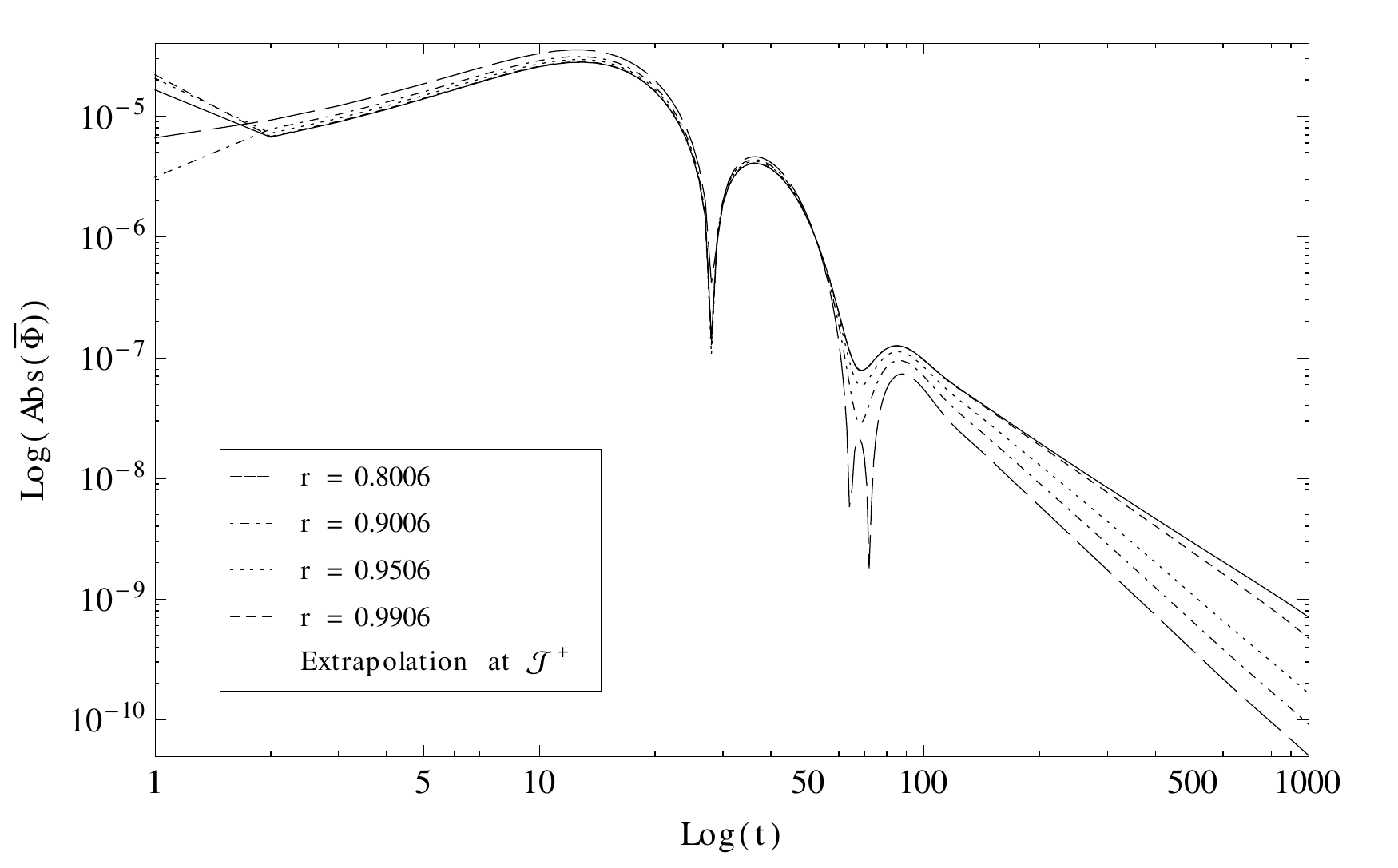}
\caption{Values of the rescaled scalar field at some values of $r$ and extrapolated to $r=r_{\!\!\scri}$ as a function of time.}\label{fs:phitails}
\end{figure}

\begin{figure}[htbp!!]
\center
%\begin{tabular}{ m{0.5\linewidth}@{} @{}m{0.5\linewidth}@{} }
%\hspace{-2ex} \mbox{\includegraphics[width=1.08\linewidth]{figures/tailsr.pdf}}&
%\hspace{-1ex} \mbox{\includegraphics[width=1.08\linewidth]{figures/tailsrl.pdf}}
\mbox{\includegraphics[width=0.8\linewidth]{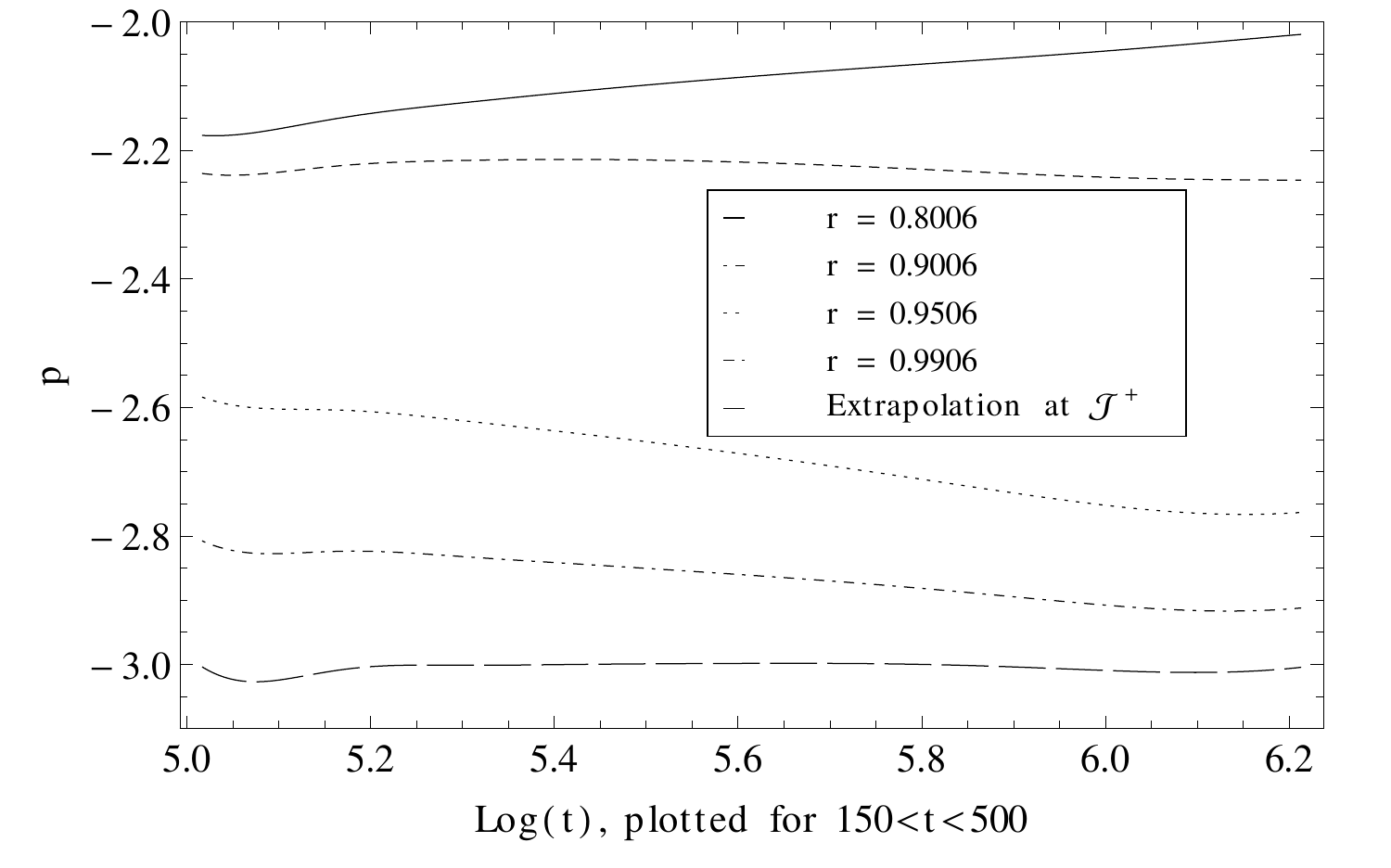}}\vspace{-2ex}
%\end{tabular}
\caption{Slopes of the scalar field tails shown in \fref{fs:phitails} over $\log(t)$ for $150<t<500$. The exponent $p$ of the tails is $p\sim-3$ up to $r\sim 0.95$ and $p\sim-2$ at $\scri^+$.}
\label{fs:tailsr}
\end{figure}

An example of a scalar field perturbation with initial $A_\Phi=10^{-4}$ in a simulation with 800 gridpoints is shown in \fref{fs:phitails}. The \CZ{} ($C_{Z4c}=0$) system with harmonic slicing \eref{eg:harmlapse} and Gamma-driver \eref{eg:Gammadriver} was used. The rescaled scalar field at some selected values of the radial coordinate, as well as extrapolated to $\scri^+$, is plotted logarithmically over time, so that the decay tails can be appreciated. Although the difference in the slope of the tails is clearly visible, it can be better understood by looking at \fref{fs:tailsr}, where the value of the slopes is plotted over $\log(t)$ for $150<t<500$. For $t<150$ the tail regime has not yet been reached, while after $t=500$ the value of the scalar field starts to be affected by the variable's drift and the loss of convergence at $\scri^+$.

% Tails
The time coordinate used in the code coincides quite well with the Bondi time for the tails, because in the tail regime the variables are very close to their CMC values (not taking into account the drift), which satisfy the preferred conformal gauge.

According to \fref{fs:tailsr}, the change from $p=-3$ (along timelike slices) to $p=-2$ (along null slices) takes place continuously approximately in the outermost $0.2$ fraction of the compactified radial coordinate. The lines shown present some variations, due to the limited resolution of the simulation, and the value of $p$ in the extrapolation at $\scri^+$ approaches $-2$ after some time.
% Slopes of tails

The following figures \ref{fs:driftvars} and \ref{fs:Hgztail} exemplify the behaviour of the rest of the system at the times when the scalar field tail is observed. The profiles in \fref{fs:driftvars} allow to compare the variation of some of the evolution variables at late times. It is similar to the drift presented in \fref{fs:onlyschw} (for unperturbed Schwarzschild initial data), because as is shown in \fref{fs:drift} in the next subsection, at the late times considered here an initial amplitude of $A_\Phi\sim10^{-4}$ does not make much difference from the unperturbed case in the behaviour of the variables other than the scalar field ones.
% Late variables for tails
\begin{figure}[htbp!!]
\center
%\mbox{\includegraphics[width=1.0\linewidth]{figures/driftvars.pdf}}
\begin{tabular}{@{}c@{}@{}c@{}}
\hspace{-3ex} \mbox{\includegraphics[width=0.55\linewidth]{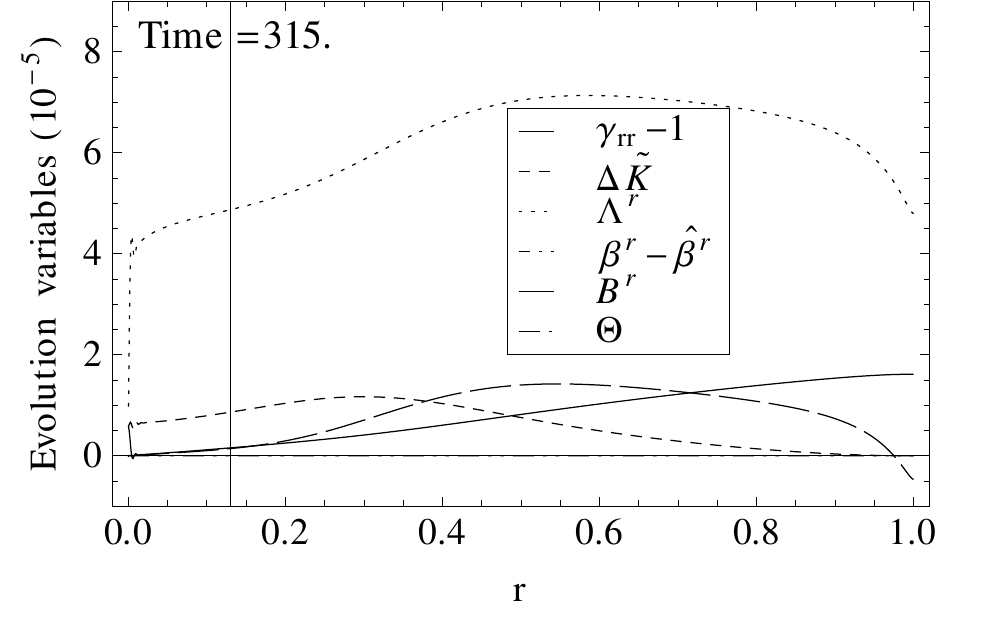}}&
\hspace{-4ex} \mbox{\includegraphics[width=0.55\linewidth]{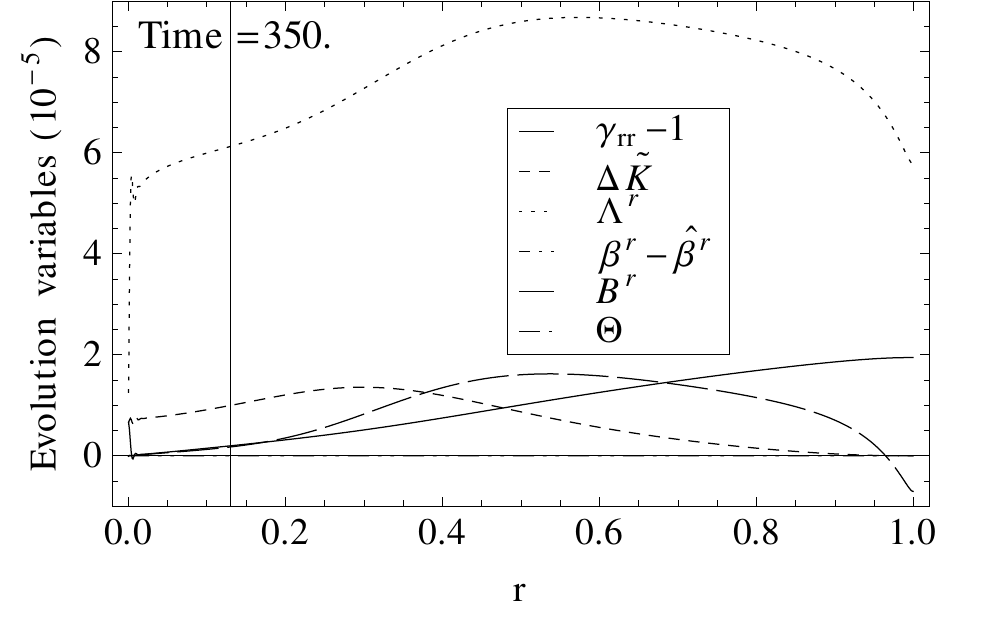}}
\end{tabular}\vspace{-2ex}
\caption{Variations of several variables at two late times corresponding to the \CZ{} ($C_{Z4c}=0$) case in \fref{fs:tailsconvc}. The variables are slowly drifting away from their stationary values.}
\label{fs:driftvars}
\end{figure}

Figure \ref{fs:Hgztail} shows two examples of the convergence of the Hamiltonian constraint. At this time the profile of \CZ{}'s $H$ is approximately fixed in time, but the value of GBSSN's $H$ is slowly increasing. However, the coincidence of the curves in the interior of the domain is good.
% Hamiltonian constraint for tails
\begin{figure}[htbp!!]
\center
\begin{tabular}{@{}c@{}@{}c@{}}
\mbox{\includegraphics[width=0.5\linewidth]{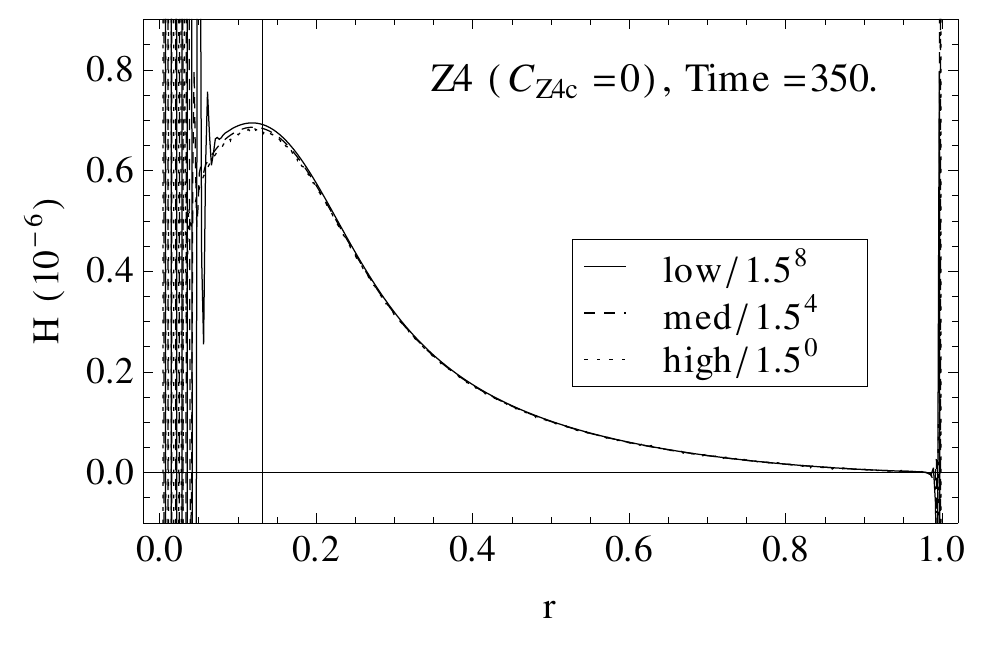}}&
\mbox{\includegraphics[width=0.5\linewidth]{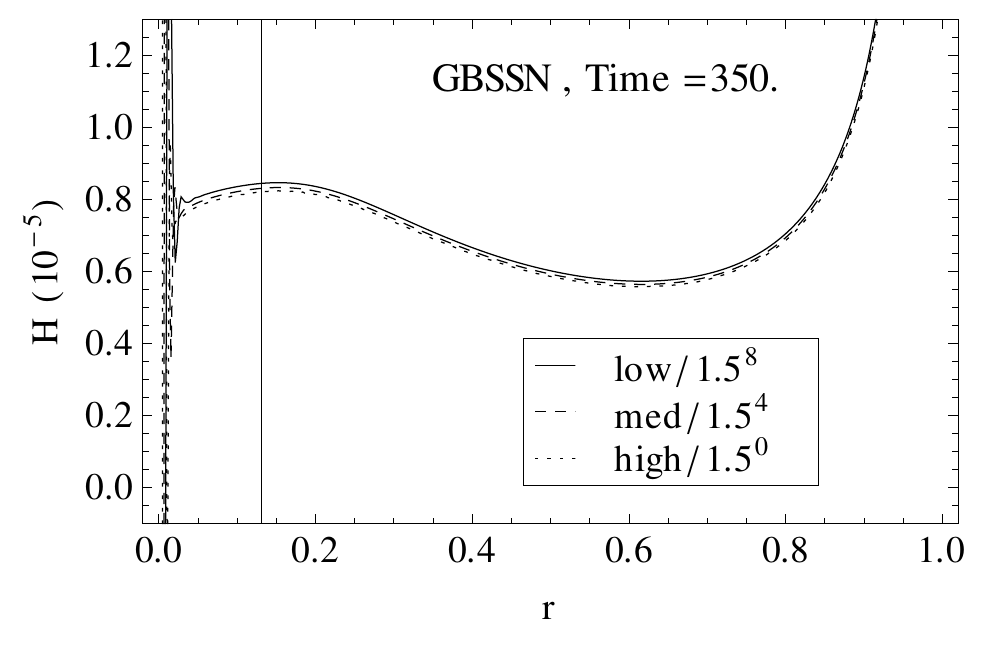}}
\end{tabular}\vspace{-3ex}
\caption{Convergence of the Hamiltonian constraint at $t=350$ for the \CZ{} ($C_{Z4c}=0$) and GBSSN cases with off-centered stencils in the advection terms.}
\label{fs:Hgztail}
\end{figure}

\begin{figure}[htbp!!]
\center
\mbox{\includegraphics[width=0.9\linewidth]{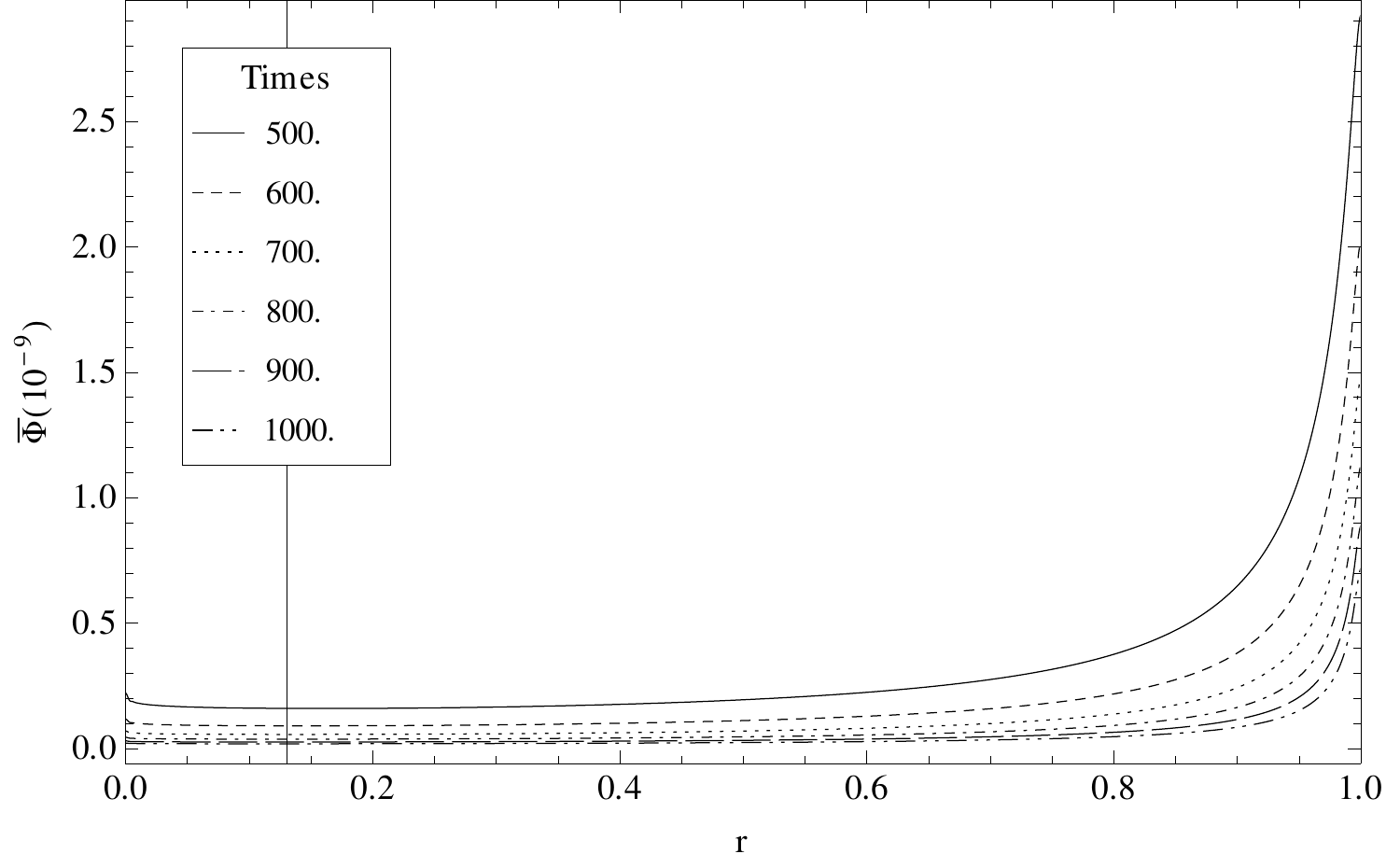}}\vspace{-2ex}
\caption{The scalar field $\bPhi$ at the given late times for a simulation with the \CZ{} ($C_{Z4c}=0$) case and 1200 gridpoints. The decay of the tail (rescaled by $10^{9}$) can be seen in the vicinity of $\scri^+$.}
\label{fs:philatetails}
\end{figure}

The profile of the scalar field at late times after an initial perturbation with $A_\Phi=10^{-4}$ is shown in \fref{fs:philatetails}. The change in the horizon location (indicated by the vertical line) for the interval of time indicated is smaller than 0.03\%. The rescaled scalar field $\bPhi$ does not develop any sharp features at $\scri^+$, but simply decays with the expected power-law, as was shown in figures \ref{fs:phitails} and \ref{fs:tailsr}.
% Profile of the scalar field at late times

\begin{figure}[htbp!!]
\center
\begin{tabular}{ m{0.5\linewidth}@{} @{}m{0.5\linewidth}@{} }
\hspace{-2ex} \mbox{\includegraphics[width=1.08\linewidth]{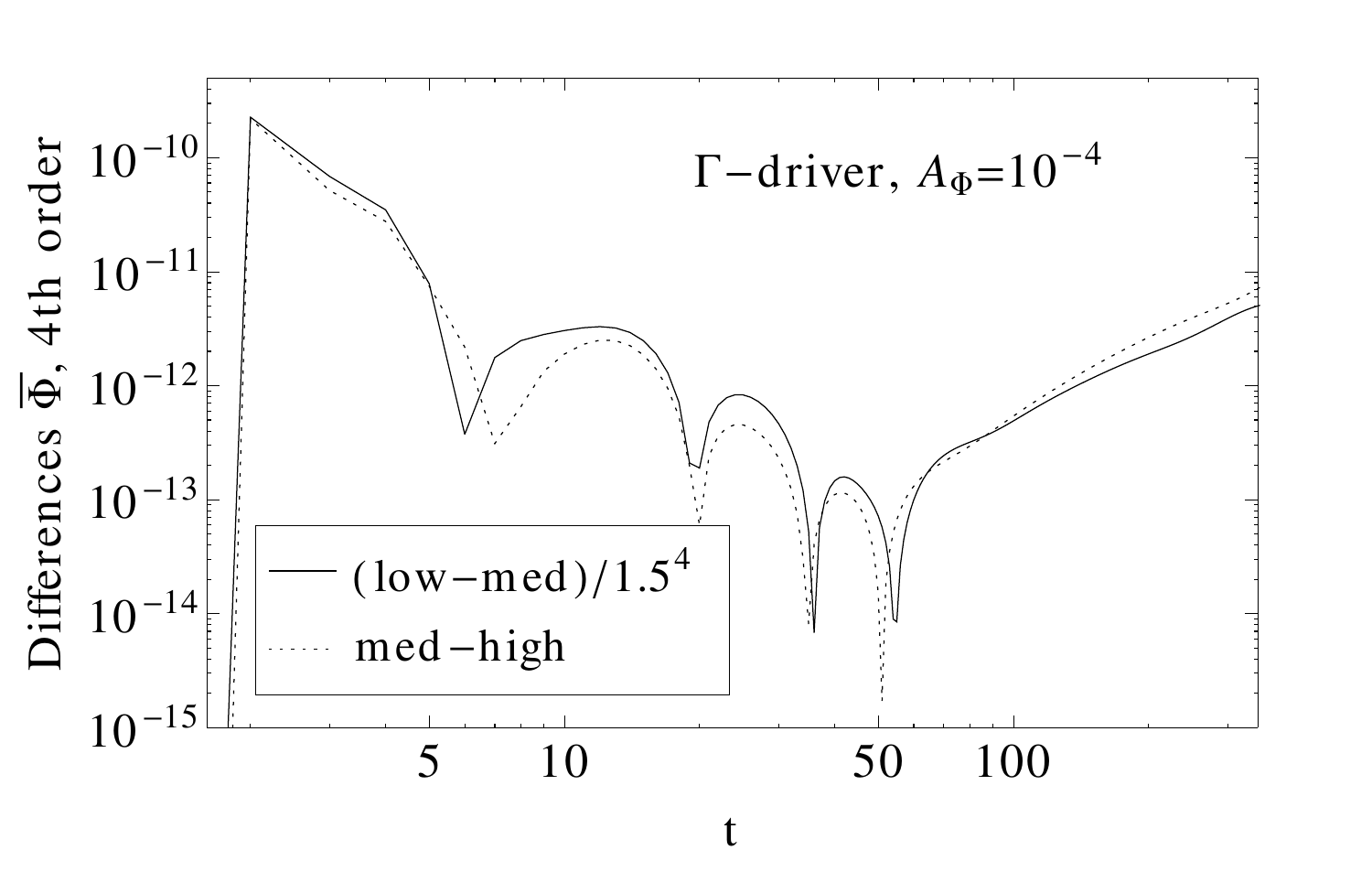}}&
\hspace{-1ex} \mbox{\includegraphics[width=1.08\linewidth]{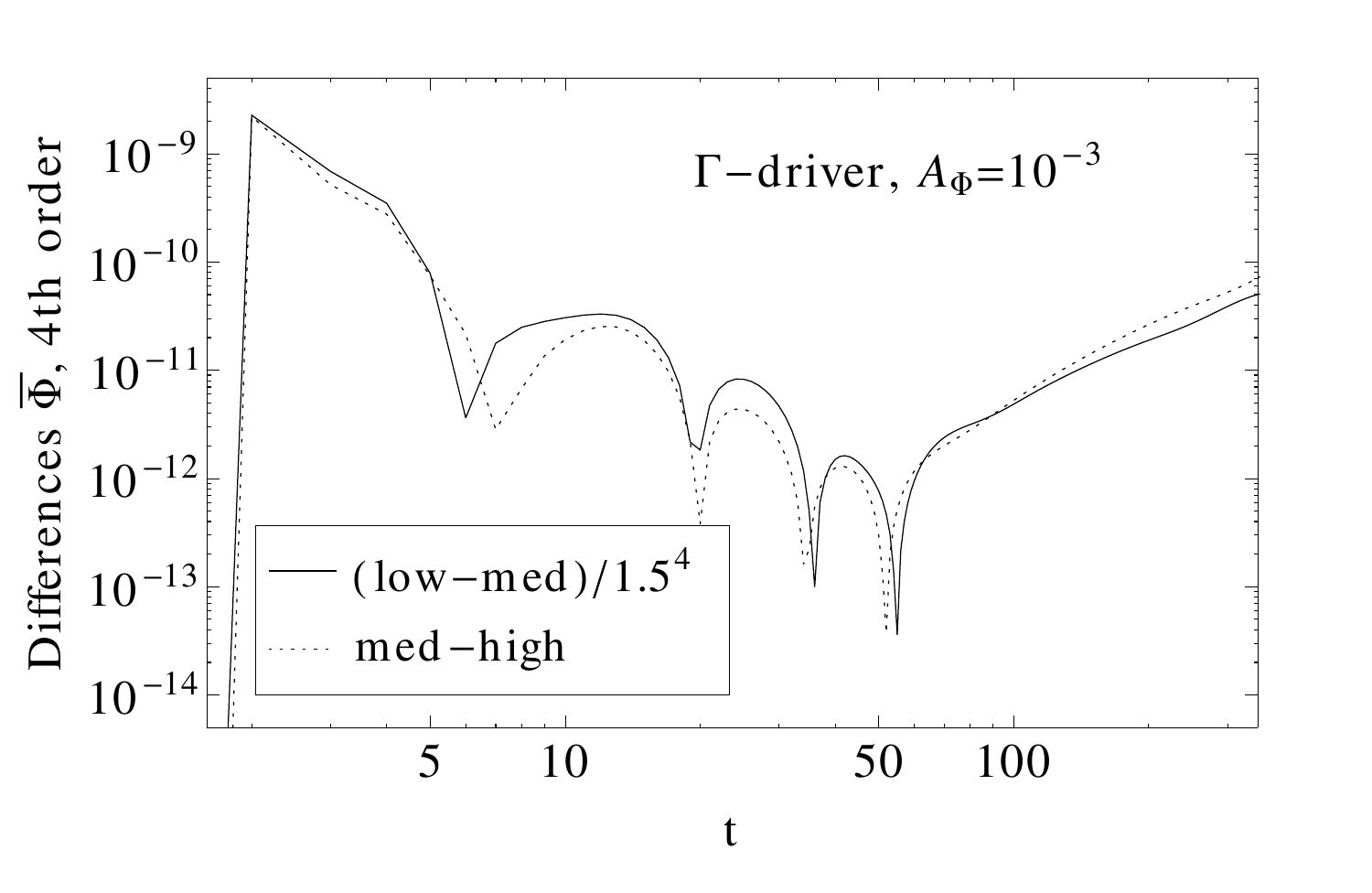}}\\
\hspace{-2ex} \mbox{\includegraphics[width=1.08\linewidth]{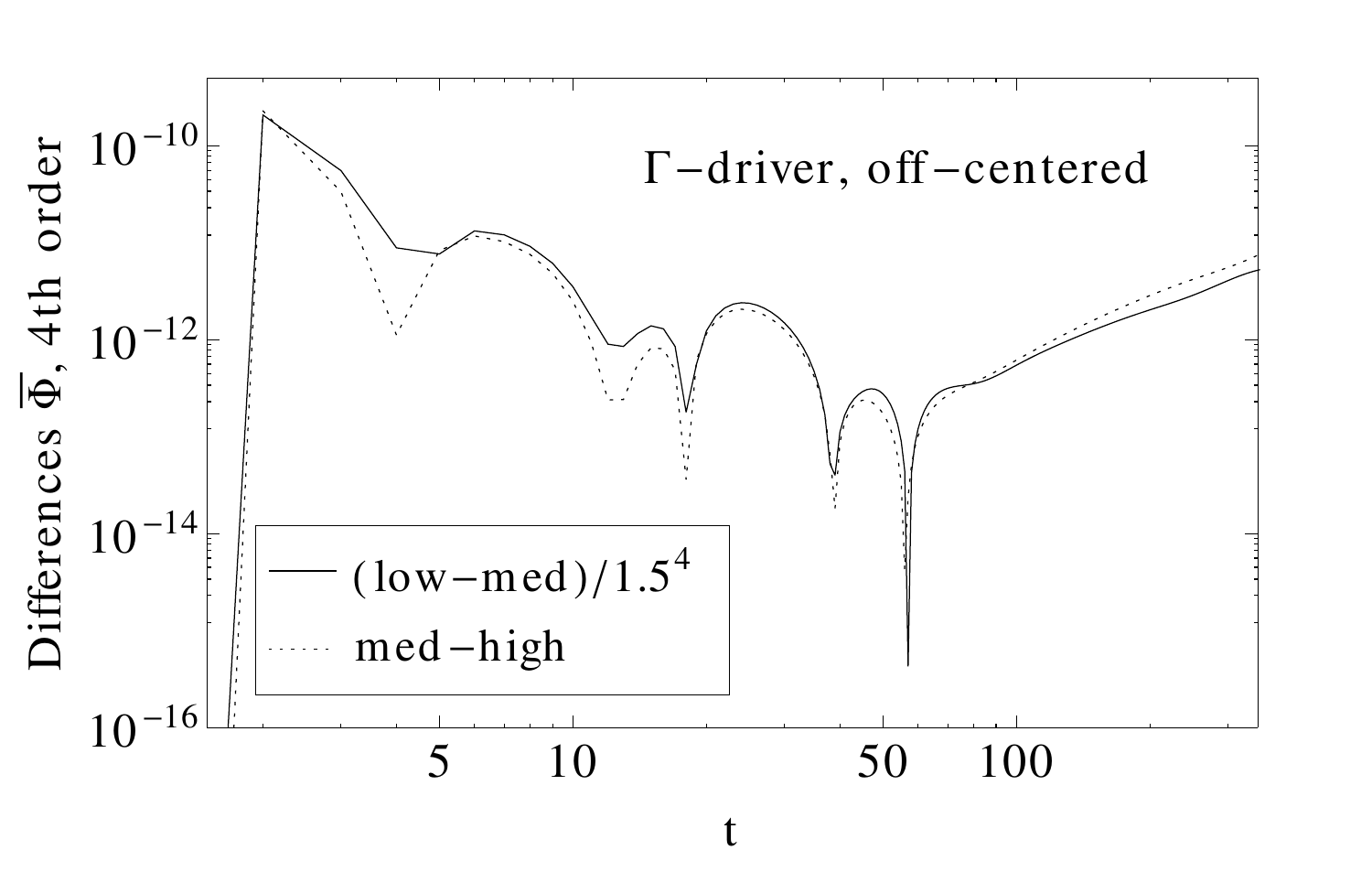}}&
\hspace{-1ex} \mbox{\includegraphics[width=1.08\linewidth]{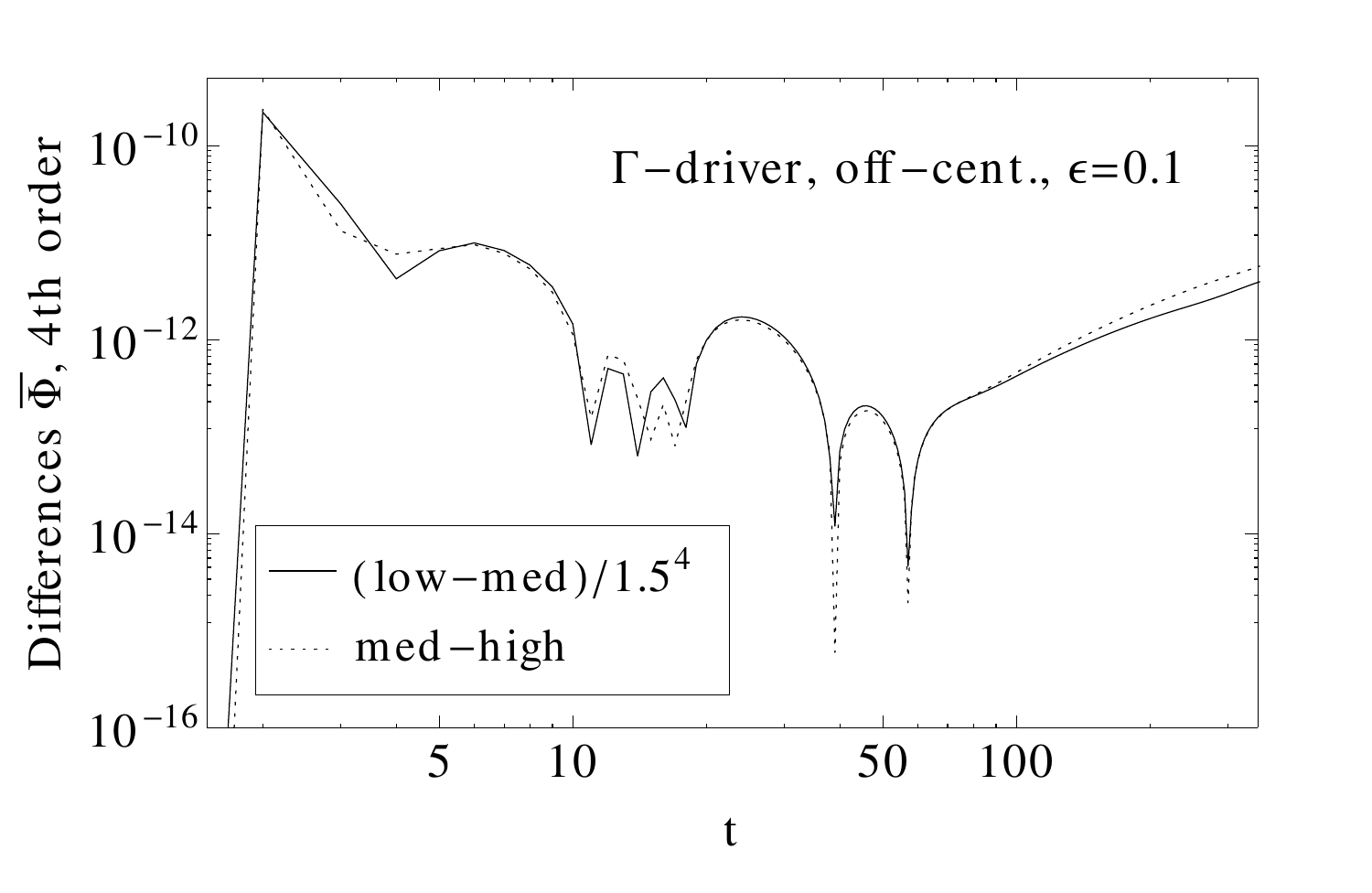}}\\
\hspace{-2ex} \mbox{\includegraphics[width=1.08\linewidth]{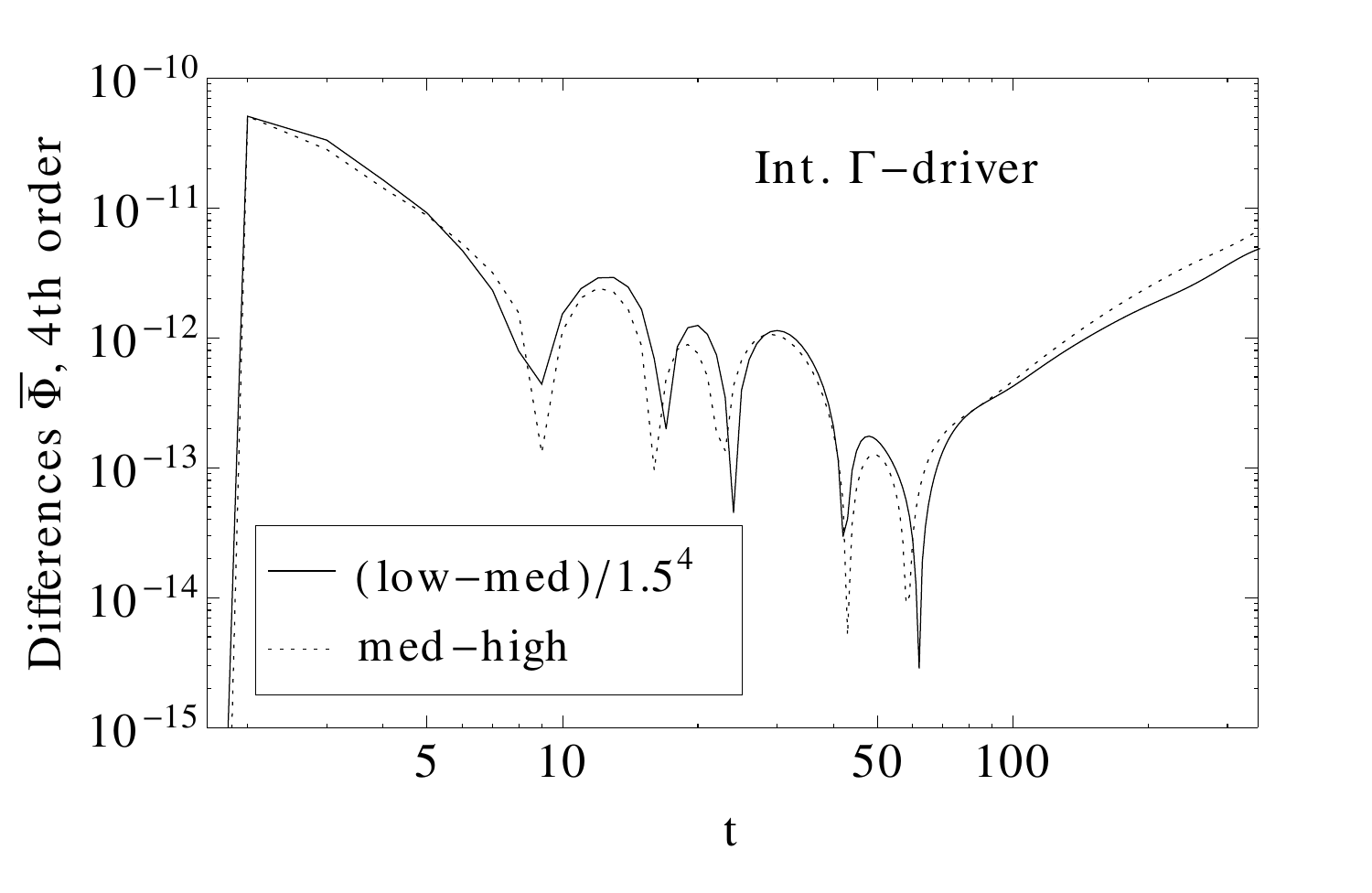}}&
\hspace{-2ex} \mbox{\includegraphics[width=1.08\linewidth]{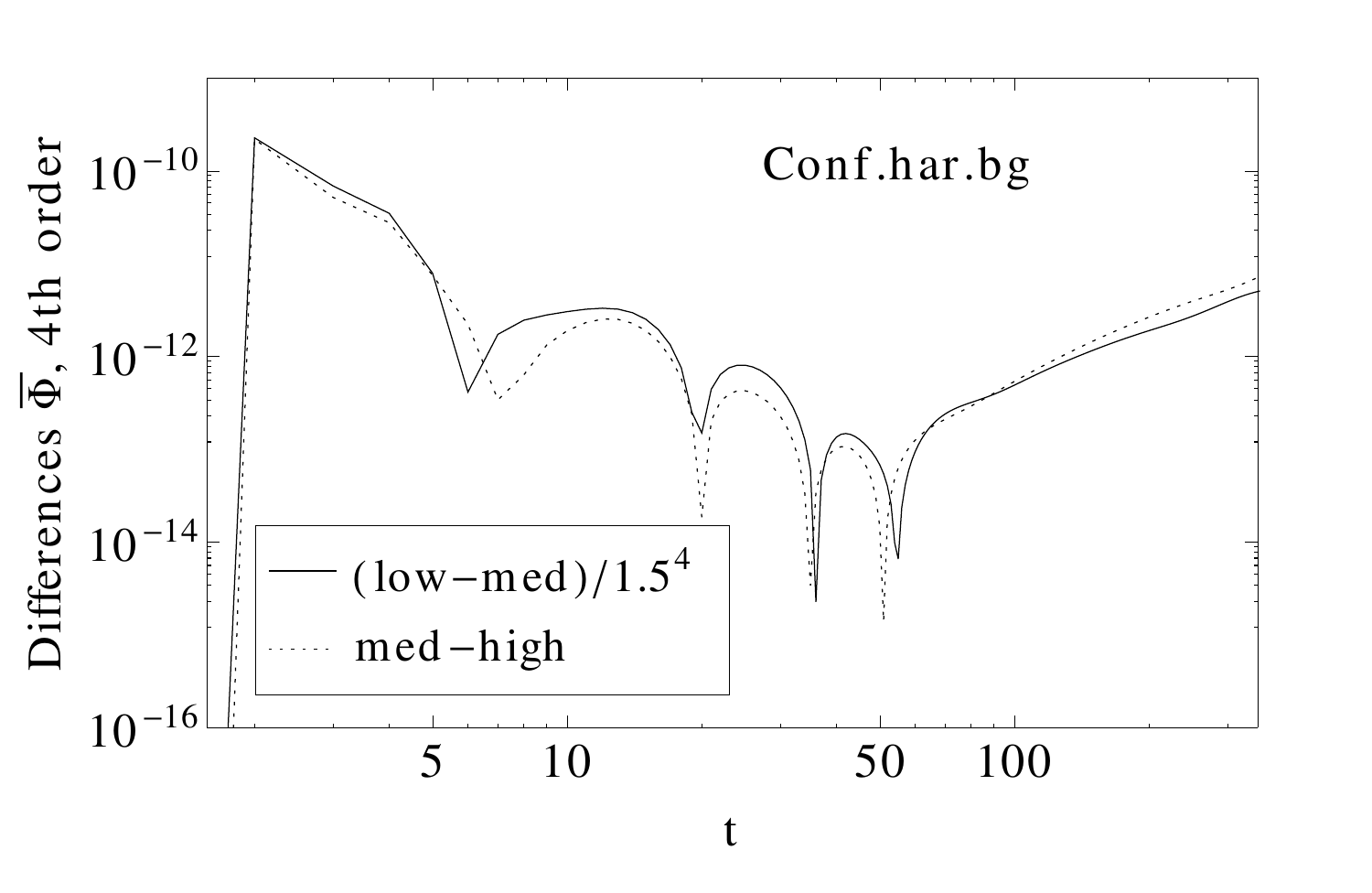}}
\end{tabular}
\vspace{-2ex}
\caption{The convergence of $\bPhi$ extrapolated at $\scri^+$ for an initial perturbation with $c=0.25$. Simulations performed with the \CZ{} ($C_{Z4c}=0$) equations.}
\label{fs:tailsconv}
\end{figure}

\begin{figure}[htbp!!]
\center
\begin{tabular}{ m{0.5\linewidth}@{} @{}m{0.5\linewidth}@{} }
\hspace{-2ex} \mbox{\includegraphics[width=1.08\linewidth]{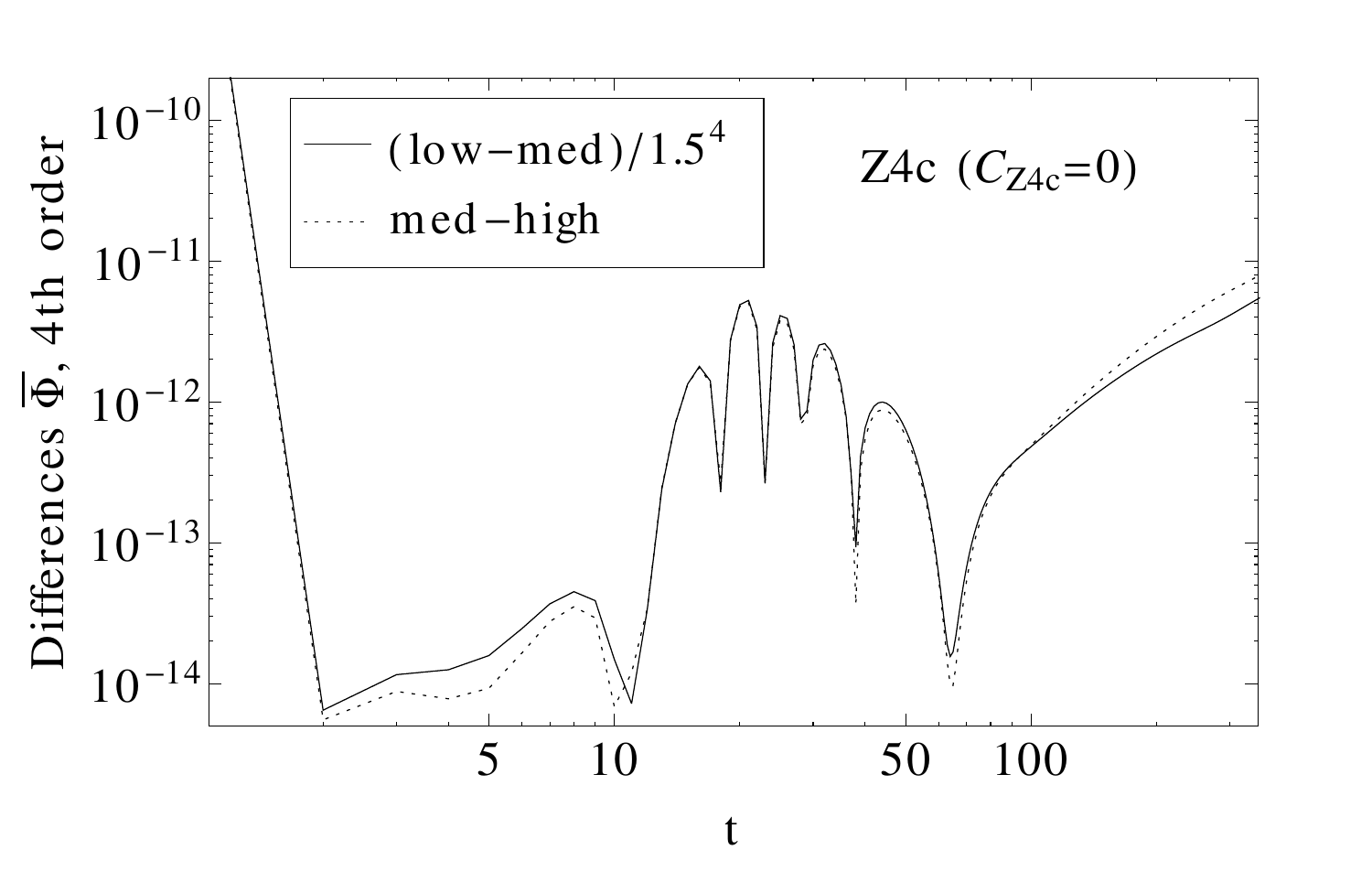}}&
\hspace{-1ex} \mbox{\includegraphics[width=1.08\linewidth]{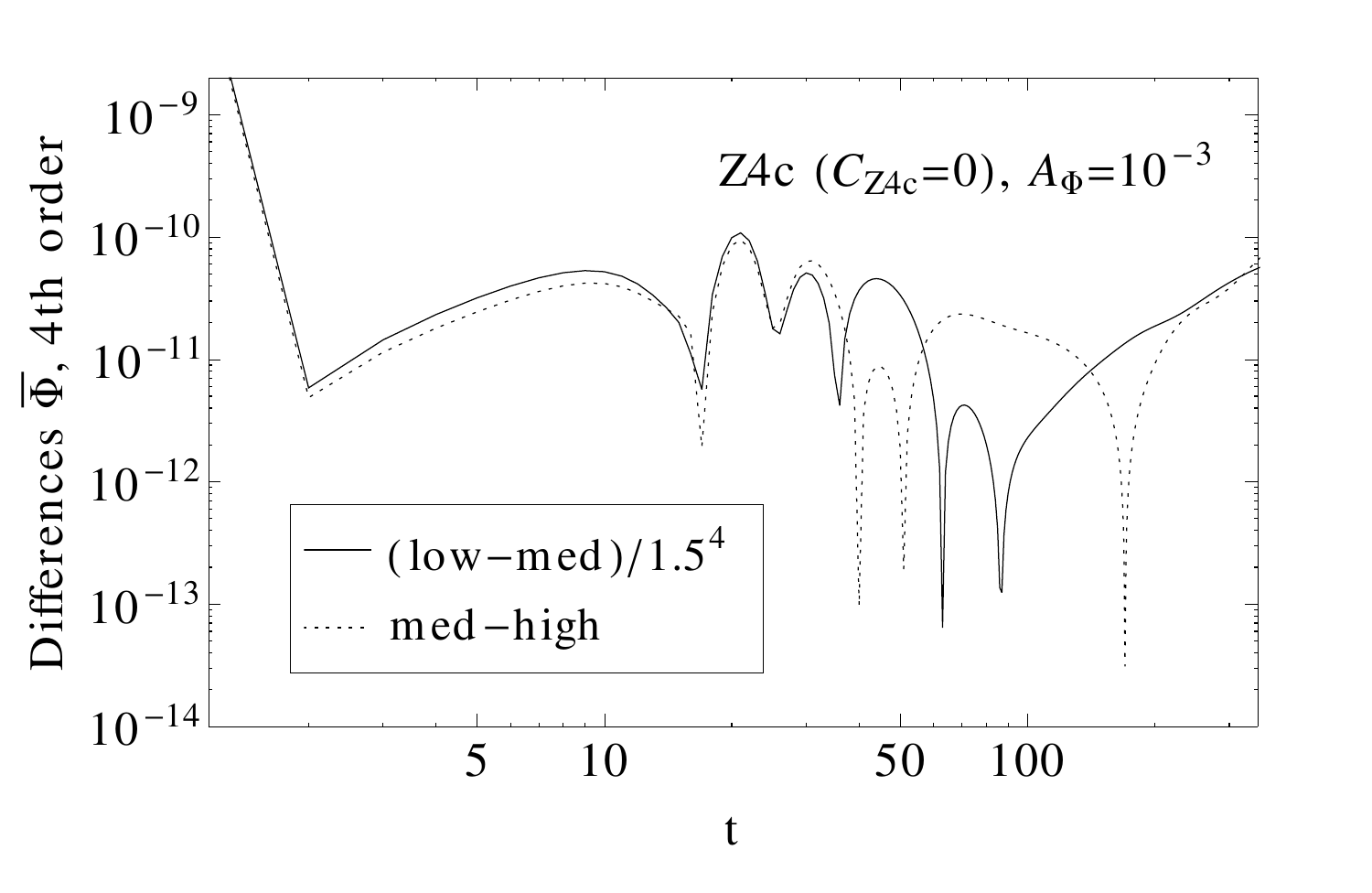}}\\
\vspace{-3ex} \hspace{-2ex} \mbox{\includegraphics[width=1.08\linewidth]{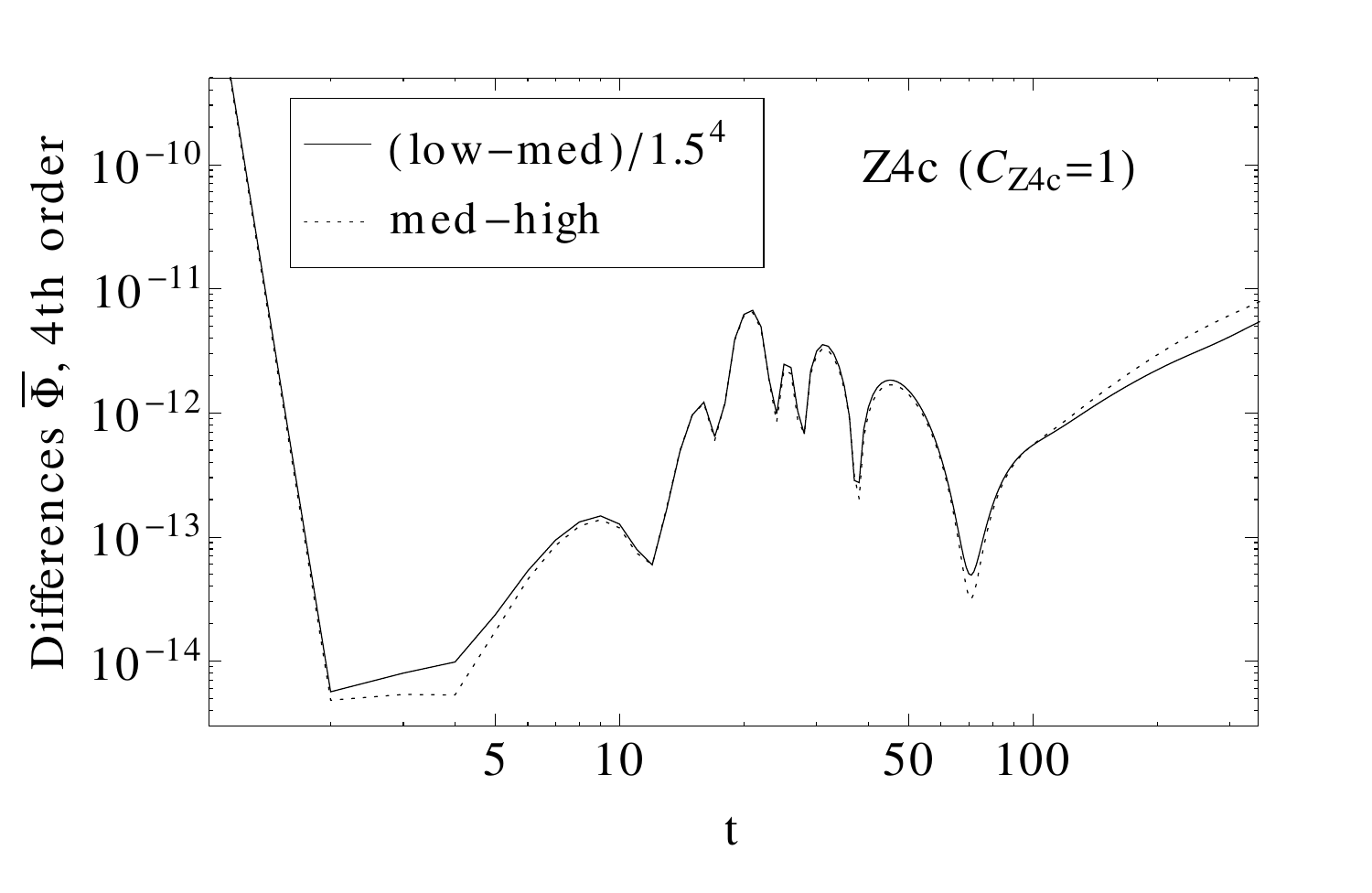}}&
\vspace{-3ex} \hspace{-1ex} \mbox{\includegraphics[width=1.08\linewidth]{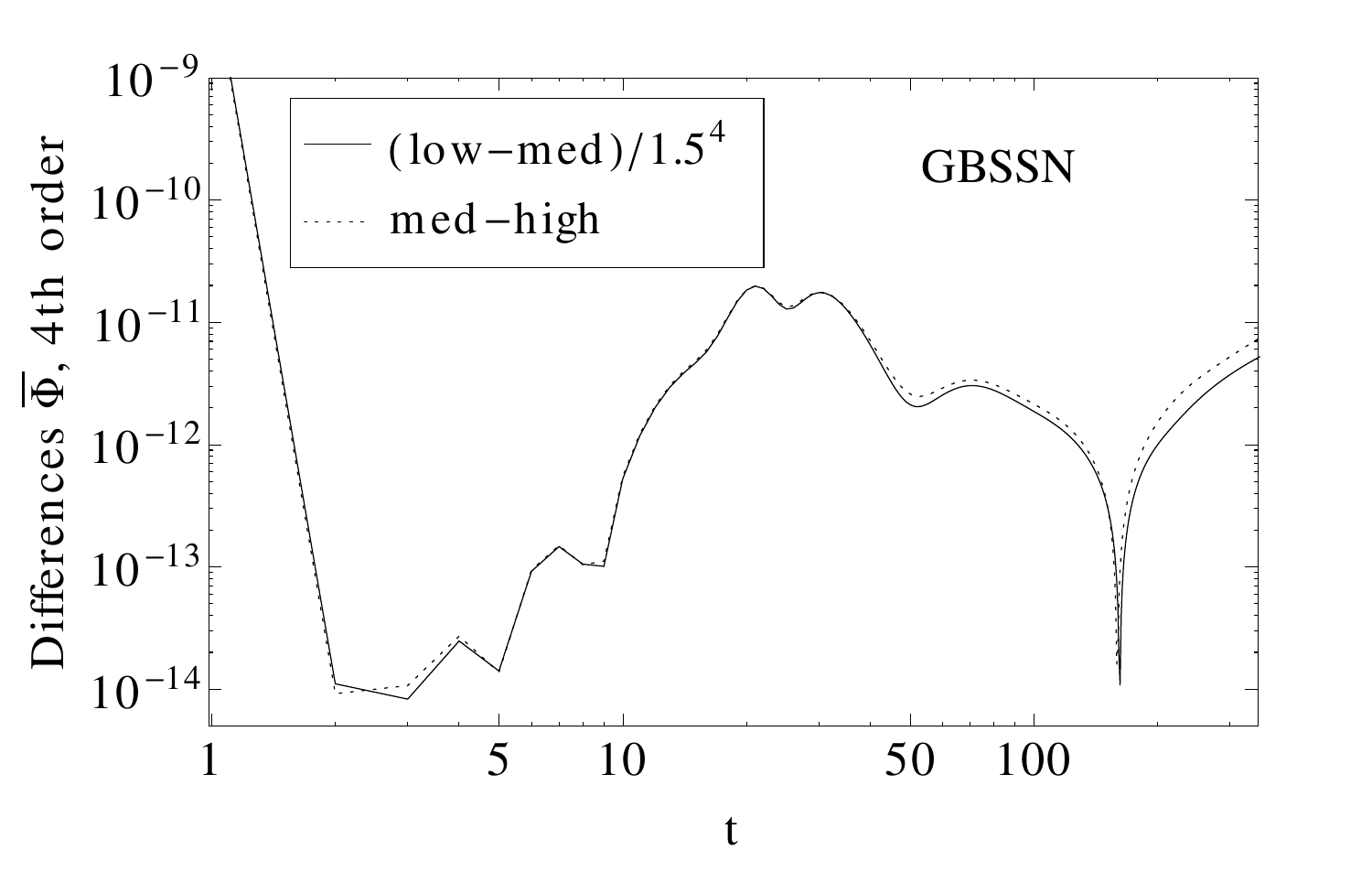}}\\
\vspace{-3ex} \hspace{-2ex} \mbox{\includegraphics[width=1.08\linewidth]{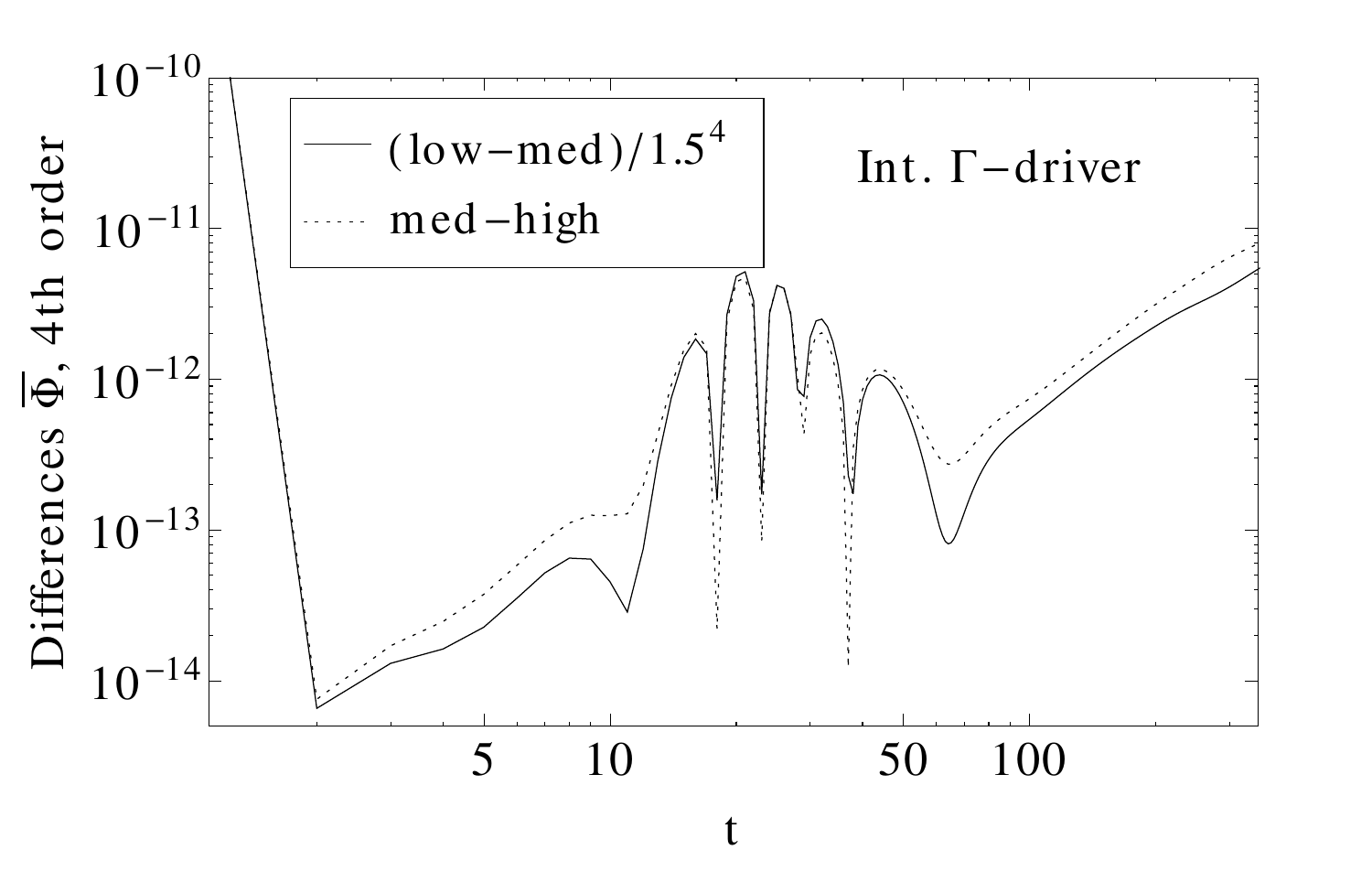}}&
\vspace{-3ex} \hspace{-2ex} \mbox{\includegraphics[width=1.08\linewidth]{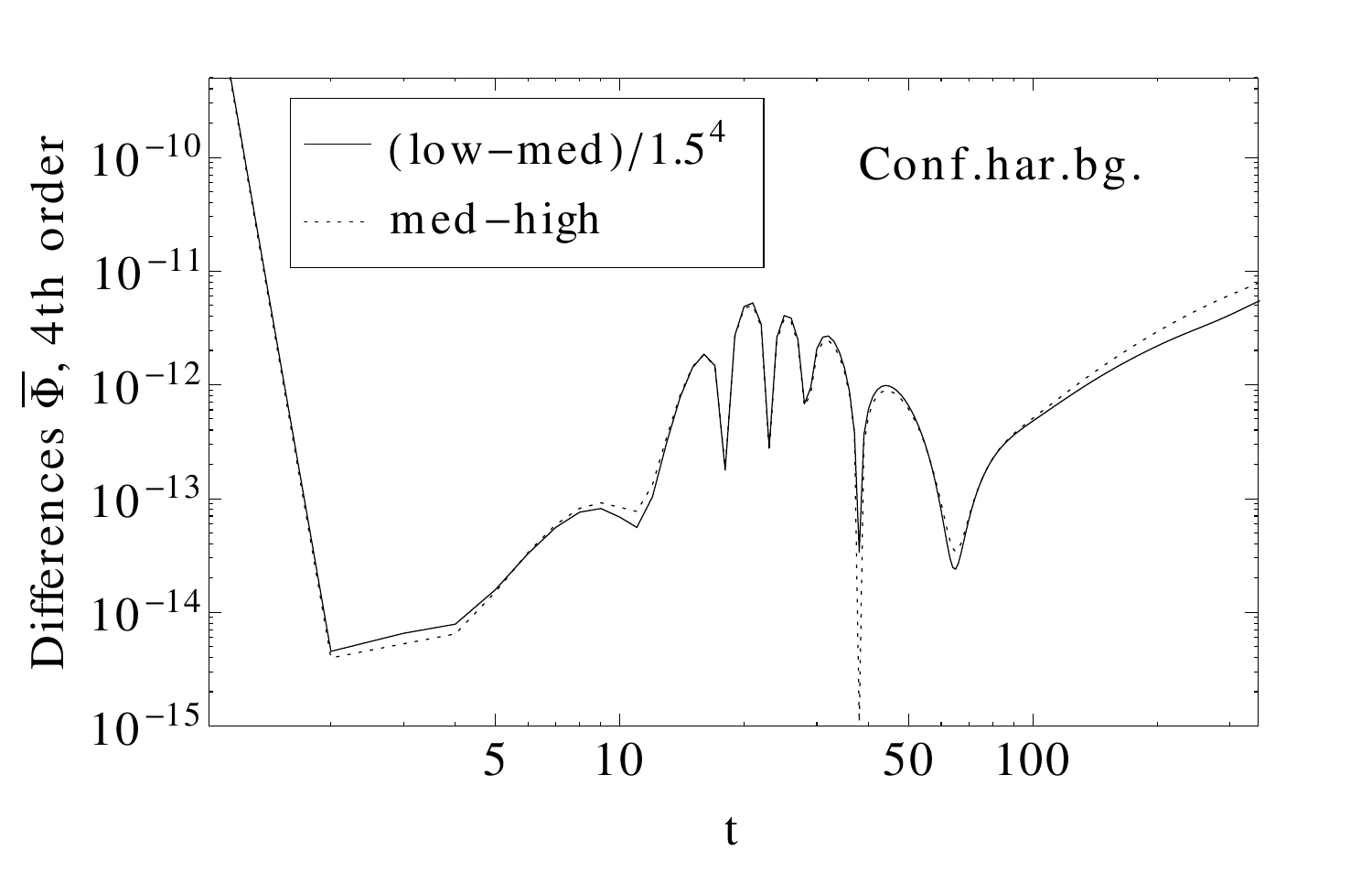}}\\
\vspace{-3ex} \hspace{-2ex} \mbox{\includegraphics[width=1.08\linewidth]{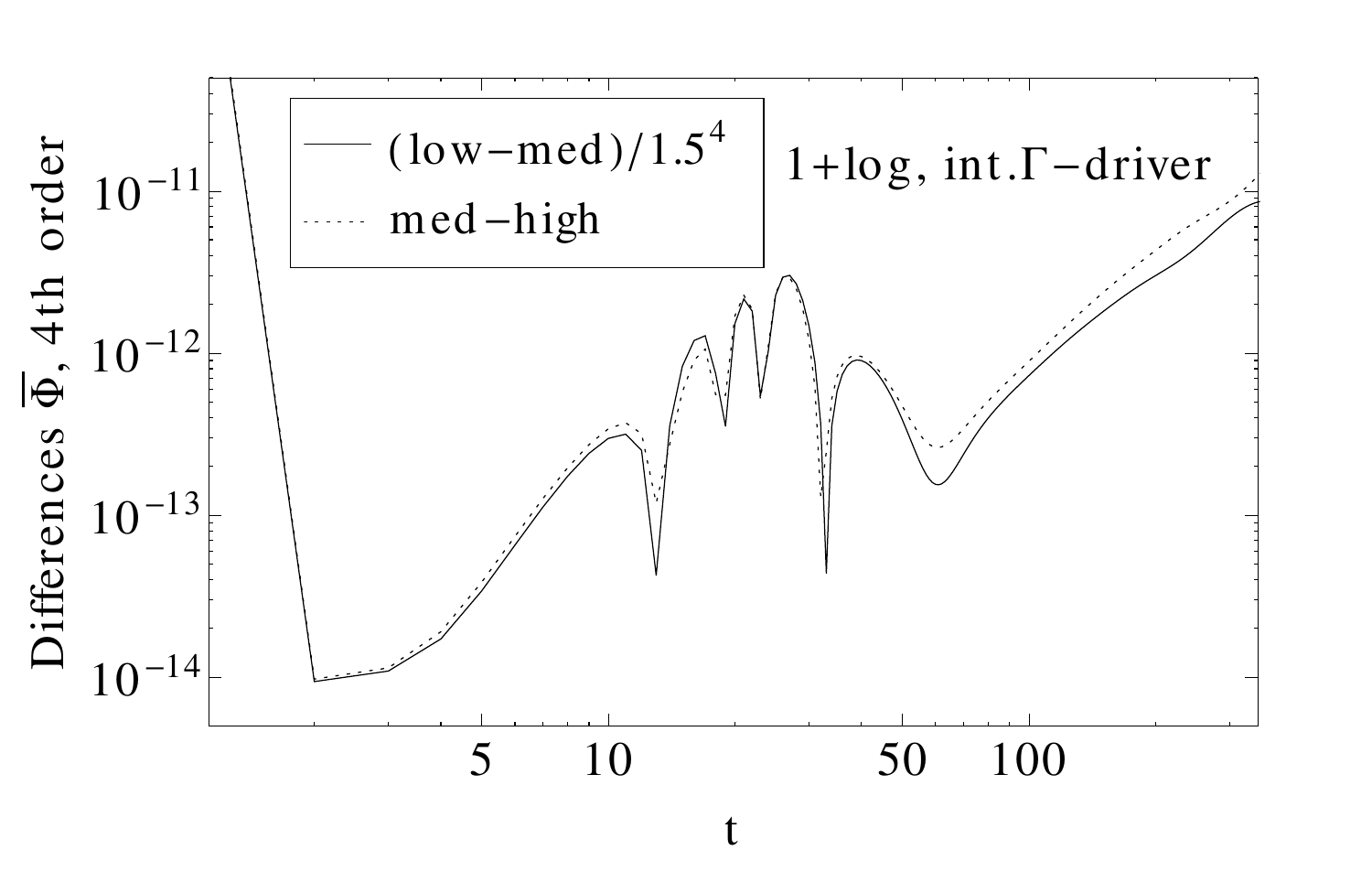}}&
\vspace{-3ex} \hspace{-2ex} \mbox{\includegraphics[width=1.08\linewidth]{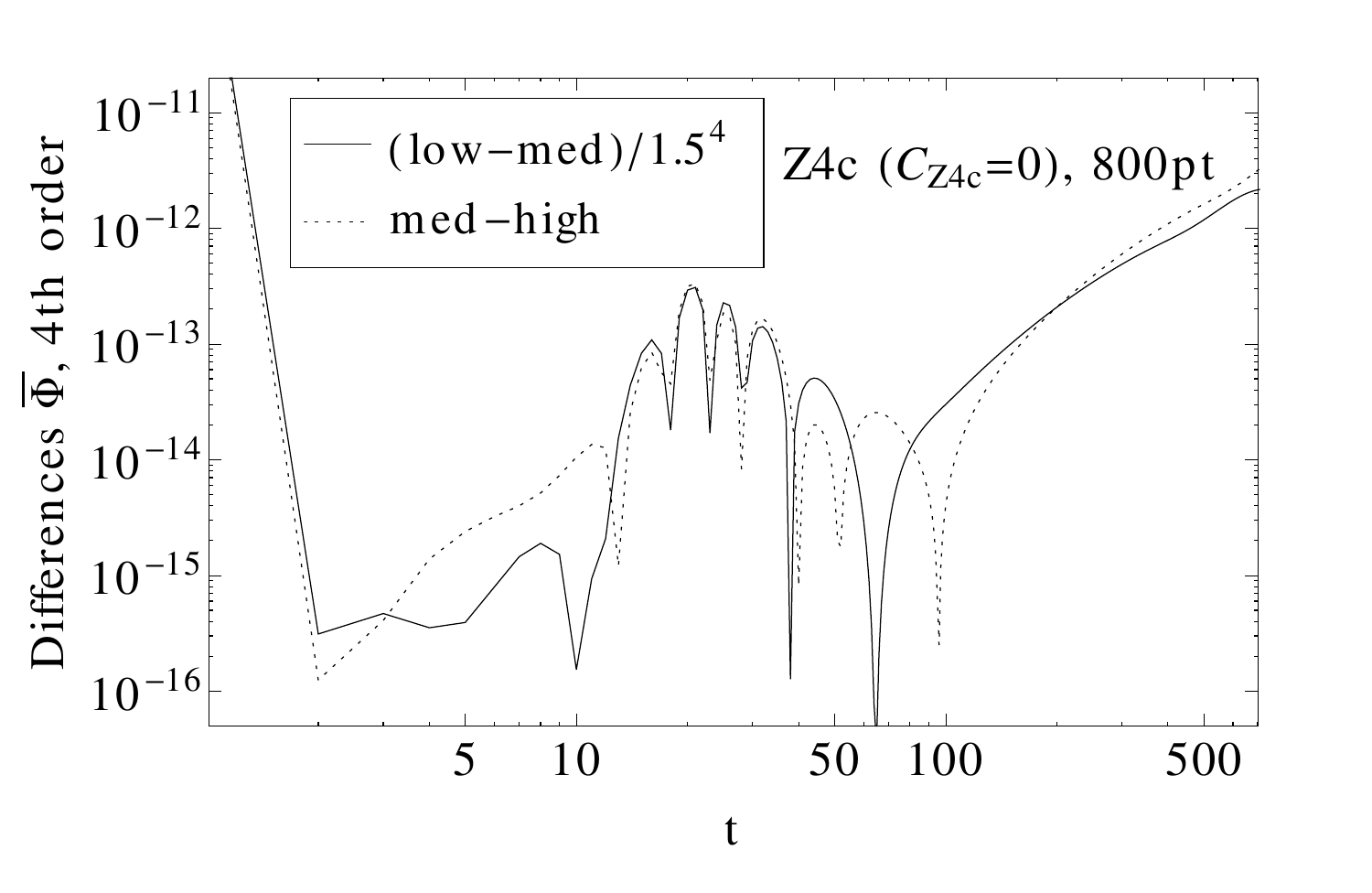}}
\end{tabular}
\vspace{-2ex}
\caption{The convergence of $\bPhi$ extrapolated at $\scri^+$ for an initial perturbation with $c=0.5$ and using off-centered stencils in the advection terms with $\epsilon=0.1$. The exception is the plot labeled with ``1+log, int.$\Gamma$-driver'', which does not use off-centered stencils and the dissipation is set as $\epsilon=0.5r$. The number of spatial gridpoints of the runs are (400,600,900), except for the bottom-right case that used (800,1200,1800).}
\label{fs:tailsconvc}
\end{figure}

The convergence of the rescaled scalar field extrapolated at $\scri^+$ is shown in figures \ref{fs:tailsconv} and \ref{fs:tailsconvc}. The simulations in \fref{fs:tailsconv} use mostly outgoing initial scalar field perturbations located at $r=0.25$, while the initial perturbations for the plots in \fref{fs:tailsconvc} are mostly ingoing and centered at $r=0.5$.
The common default choices and parameters for the simulations in both figures are: lowest resolution run with 400 points and $\Delta t=0.0005$; $\xi_\alpha=1$, $\xi_{\beta^r}=5$ (for Gamma-drivers), $A_{\bPhi}=10^{-4}$, $\sigma=0.1$ $\epsilon=0.25$, $\cL=4$, $\pphi=1$. The figures only show data up to $t=350$, because starting at $t\approx100$ the convergence order starts to decrease. %(for a lower resolution run with 400 points)

All the simulations in \fref{fs:tailsconv} have been performed with the \CZ{} ($C_{Z4c}=0$) equations with harmonic slicing (except when the conformal harmonic background gauge is used) and with the indicated shift conditions. The initial perturbation for the top-right plots in \fref{fs:tailsconv} and \fref{fs:tailsconvc} had an amplitude of $A_\Phi=10^{-3}$ (note the larger amplitude of the errors), as indicated. The convergence results of the $A_\Phi=10^{-3}$ case in \fref{fs:tailsconvc} is significantly worse than the rest. The middle row of \fref{fs:tailsconv} shows results of simulations where off-centered stencils were used in the derivatives of the advection terms (on the right the dissipation applied was less than the standard use in this work, namely $\epsilon=0.1$); the coincidence between the curves is better than in the other cases.

The nice convergence of the runs with off-centered stencils in the advection terms motivated to perform the simulations shown in \fref{fs:tailsconvc} with this choice, except the bottom-left plot that used centered stencils. Here the different formulations, as indicated in the corresponding plots, were tested and for this case all show quite good convergence (the errors in the GBSSN case have larger amplitude). The simulations of the first and second rows used the Gamma-driver condition, while the bottom ones used the indicated shift condition plus \CZ{} ($C_{Z4c}=0$). The bottom-right plot corresponds to a convergence test that used twice as many points (and time-steps) as the other ones. However, the convergence results are not as good (the coincidence between the curves is worse even if the errors are smaller) and the convergence order also starts to decrease after $t\approx100$. %This could possibly be a consequence of the variable drift, which does not depend on the resolution. - but the $A_{\Phi}=10^{-3}$ case is affected in the same way and not more ...

No case using the physical harmonic gauge condition with background source terms is displayed, because its stability and convergence problems are not yet solved and no comparable results could be obtained.

% Convergence of tails

% Convergence of tails - centered initial data

%\newpage

\subsection{Large scalar field perturbation on Schwarzschild trumpet}\label{ss:large}

%mention change in dissipation operator
Decreasing the amount of dissipation in the interior of the BH prevents the simulations with larger amplitudes of the initial perturbation to crash near the origin (inside of the BH): the largest value tested successfully (with the 1+log condition) is $A_\Phi=0.03$. The tuning of the dissipation was performed as indicated in \eref{ee:tunediss}, with $\epsilon_0=0$ and $\epsilon_1=0.5$.

\begin{figure}[htbp!!]
\center
\vspace{-2.5ex}
 \begin{tikzpicture}[scale=2.5]\draw (-0.7cm,0cm) node {};
		\draw (0cm, 0.1cm) node {\small Not recalc.}; \draw (0cm, -0.1cm) node {\small $\cL=0$}; \draw (0.5cm, 0cm) -- (0.8cm, 0cm);
		\draw (1.5cm, 0.1cm) node {\small Recalc.}; \draw (1.5cm, -0.1cm) node {\small $\cL=0$}; \draw [dashed] (2cm, 0cm) -- (2.3cm, 0cm);
		\draw (3cm, 0.1cm) node {\small Not recalc.}; \draw (3cm, -0.1cm) node {\small $\cL=4$};  \draw [dotted] (3.5cm, 0cm) -- (3.8cm, 0cm);
		\draw (4.5cm, 0.1cm) node {\small Recalc.}; \draw (4.5cm, -0.1cm) node {\small $\cL=4$}; \draw [dash pattern= on 4pt off 2pt on 1pt off 2pt] (5cm, 0cm) -- (5.3cm, 0cm);
	\end{tikzpicture}
\\
\begin{tabular}{ m{0.5\linewidth}@{} @{}m{0.5\linewidth}@{} }
\hspace{-1.0ex}\mbox{\includegraphics[width=1\linewidth]{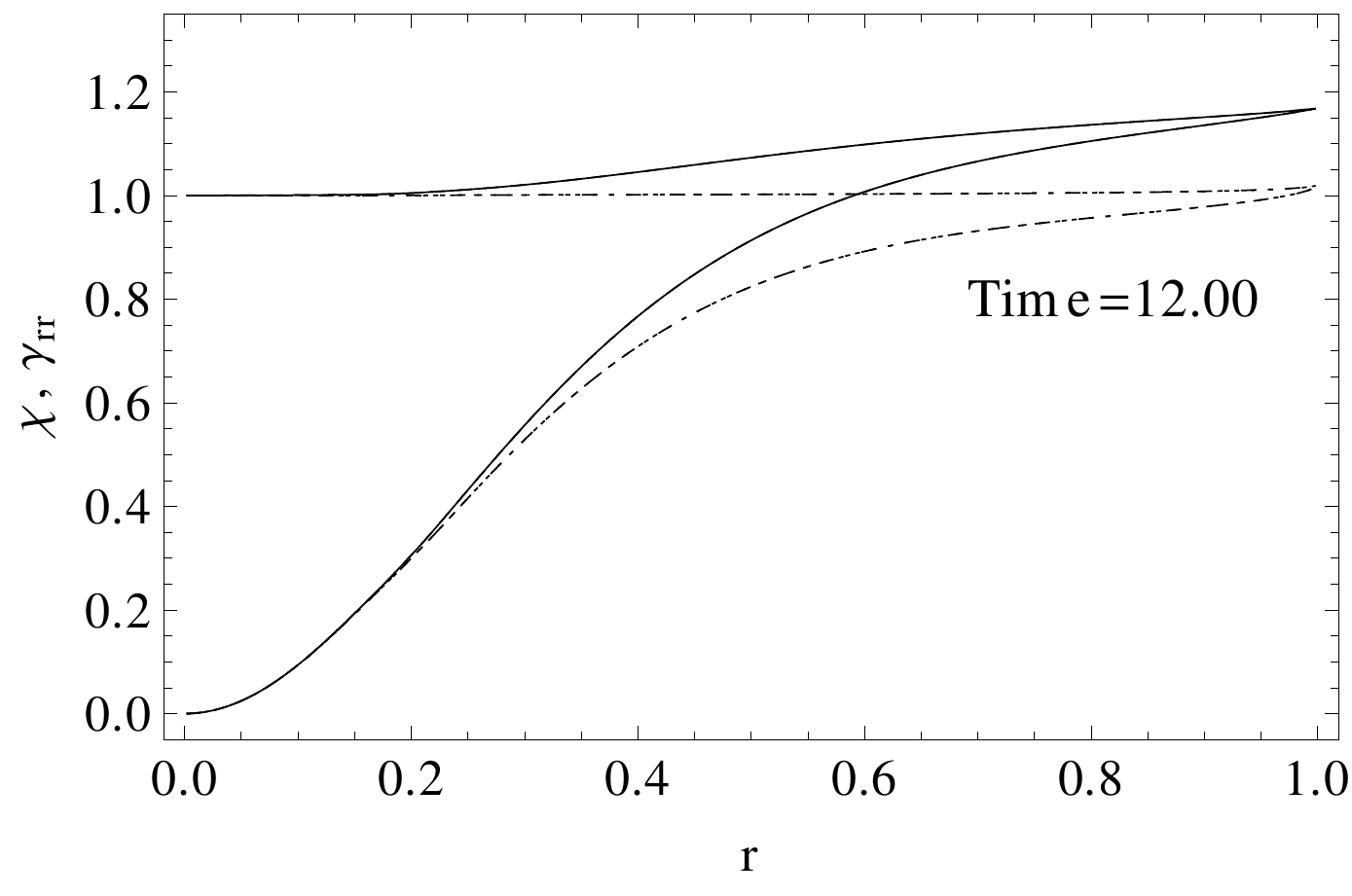}}&
\hspace{+0.8ex} \mbox{\includegraphics[width=1\linewidth]{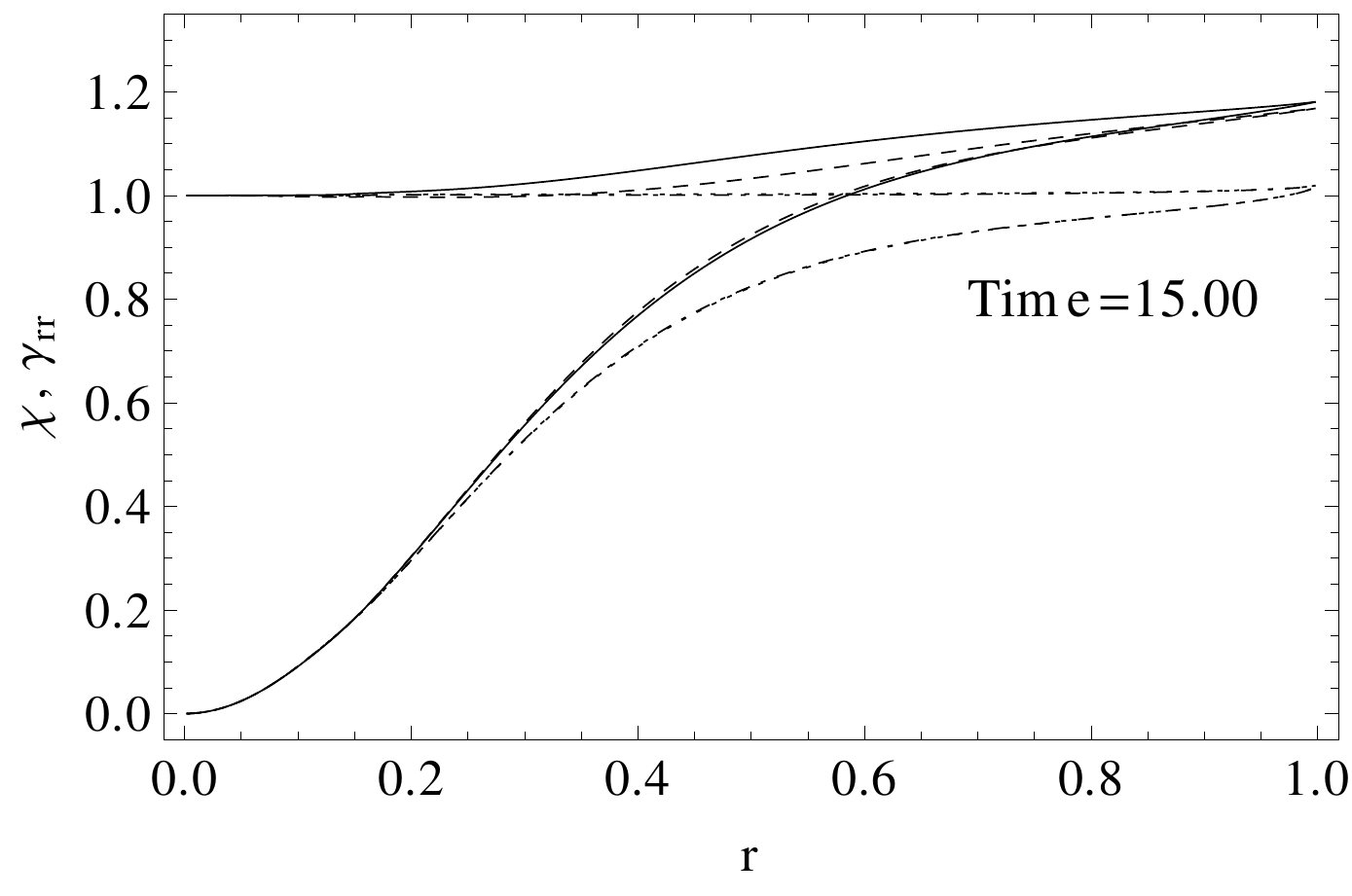}}\\
\vspace{-5.5ex} \hspace{-0.6ex}\mbox{\includegraphics[width=0.99\linewidth]{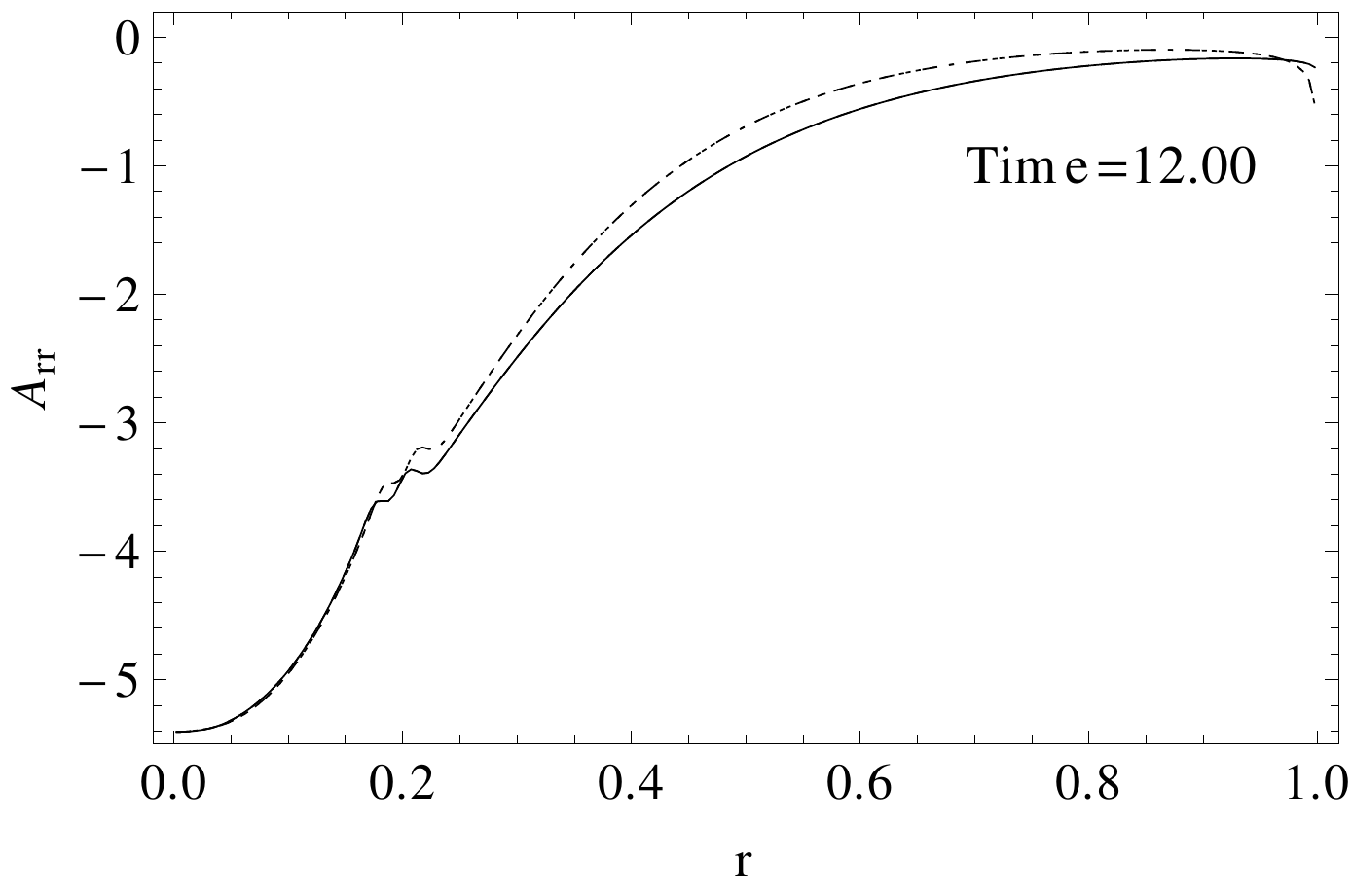}}&
\vspace{-5.5ex} \hspace{+1.2ex} \mbox{\includegraphics[width=0.99\linewidth]{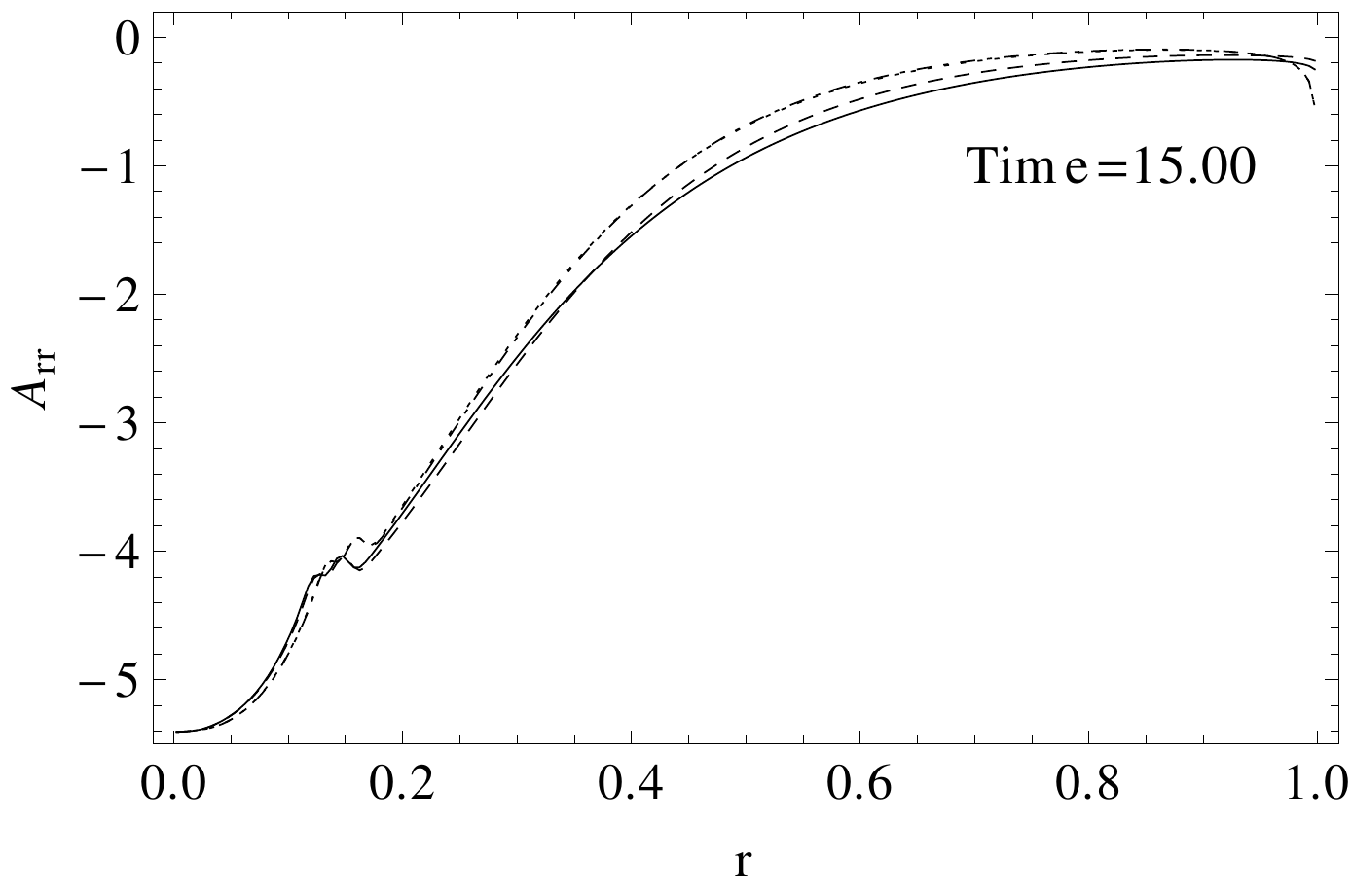}}\\
\vspace{-5.7ex} \hspace{-3.5ex}\mbox{\includegraphics[width=1.06\linewidth]{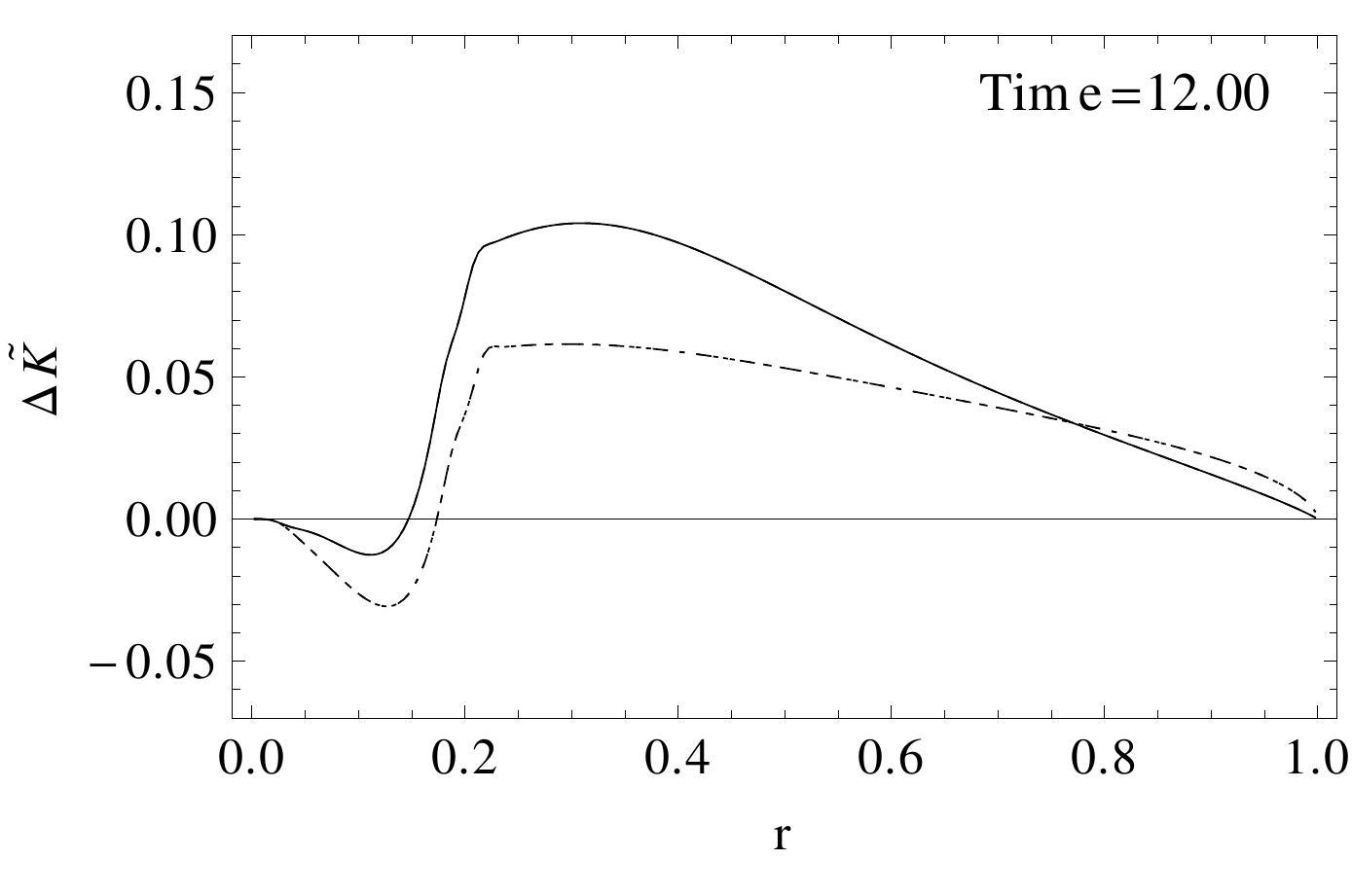}}&
\vspace{-5.7ex} \hspace{-1.8ex} \mbox{\includegraphics[width=1.06\linewidth]{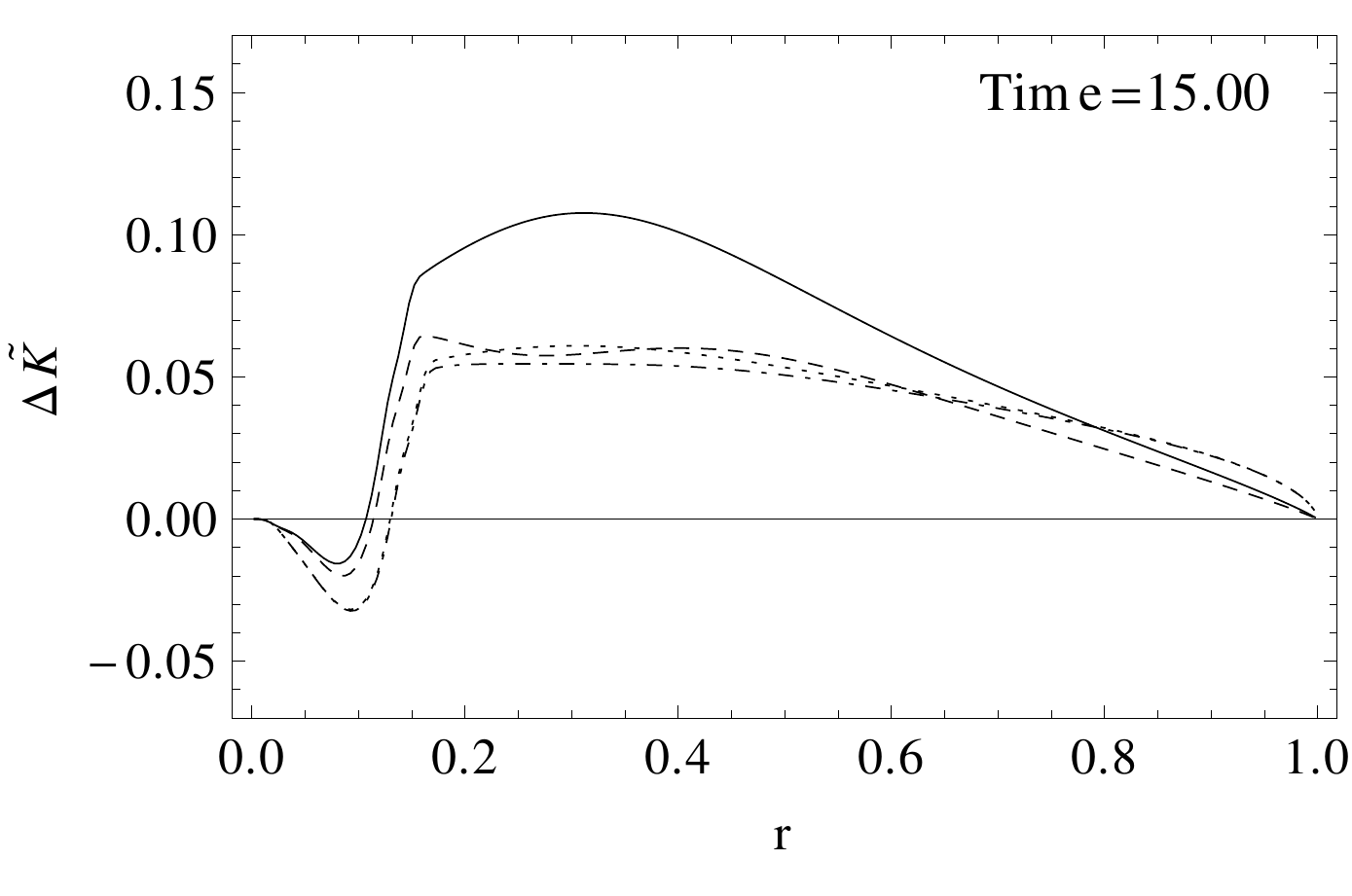}}\\
\vspace{-6.3ex} \hspace{-0.8ex}\mbox{\includegraphics[width=1\linewidth]{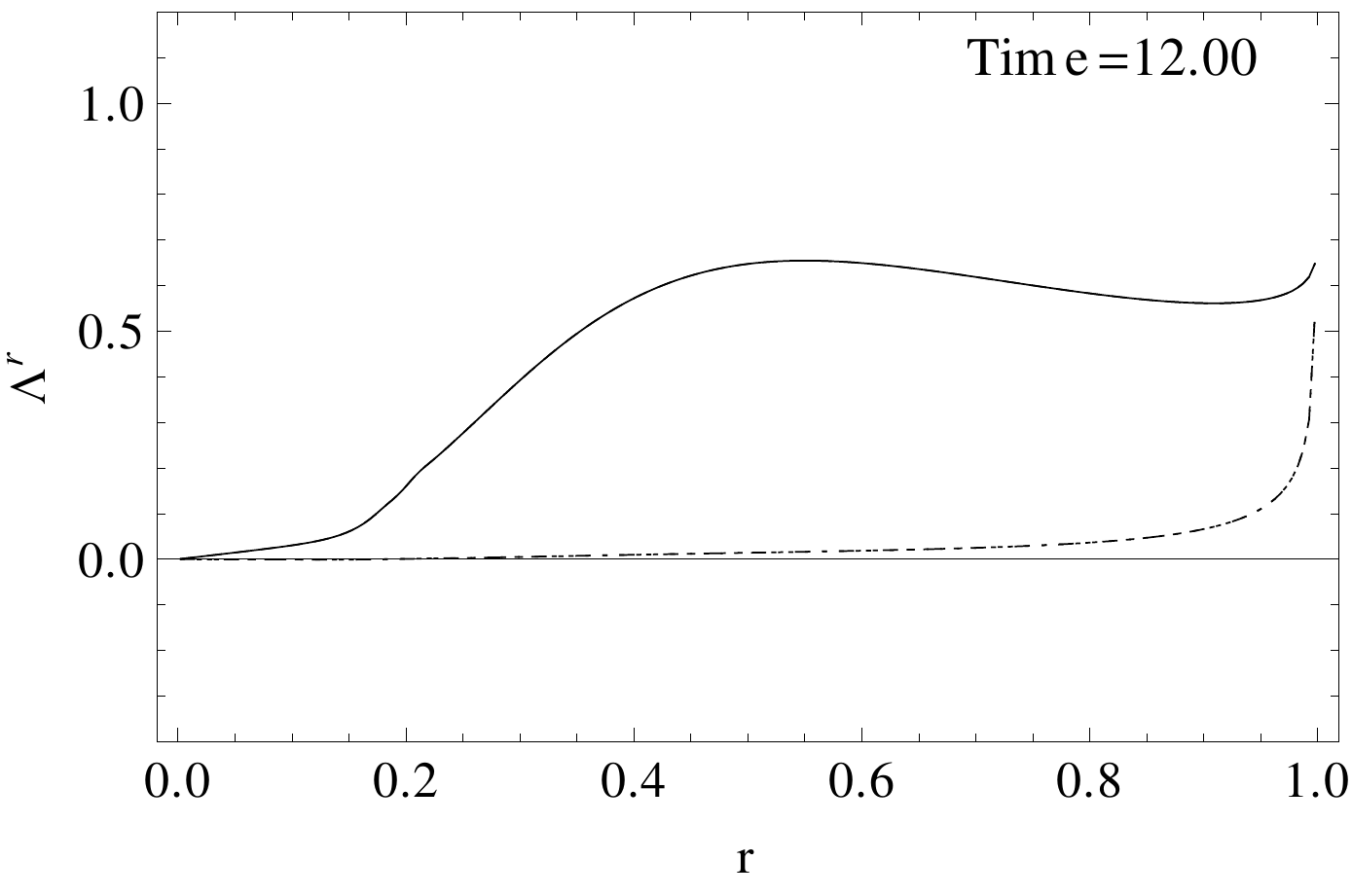}}&
\vspace{-6.3ex} \hspace{+1.0ex} \mbox{\includegraphics[width=1\linewidth]{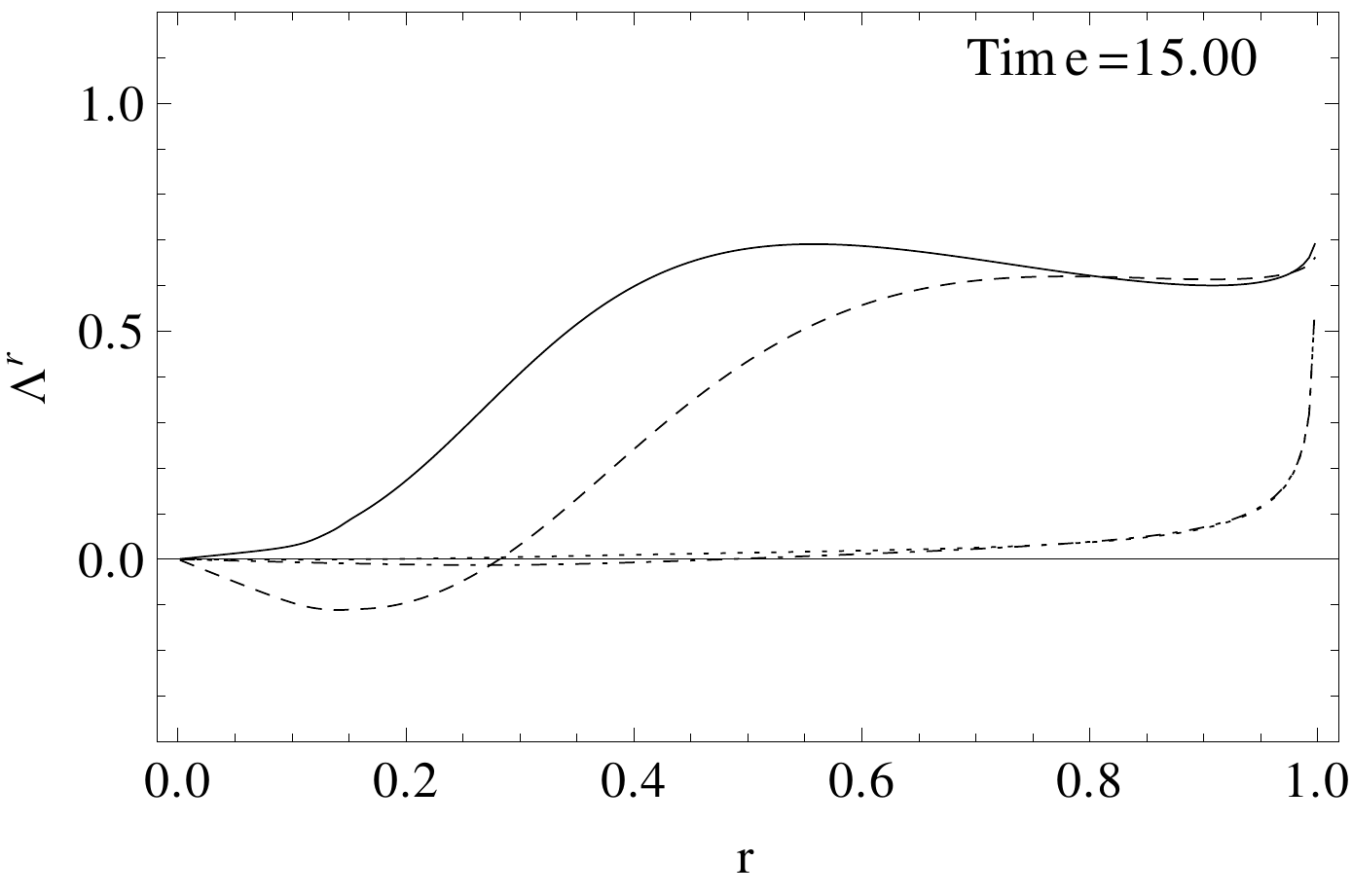}}\\
\vspace{-5.5ex} \hspace{-2.2ex}\mbox{\includegraphics[width=1.04\linewidth]{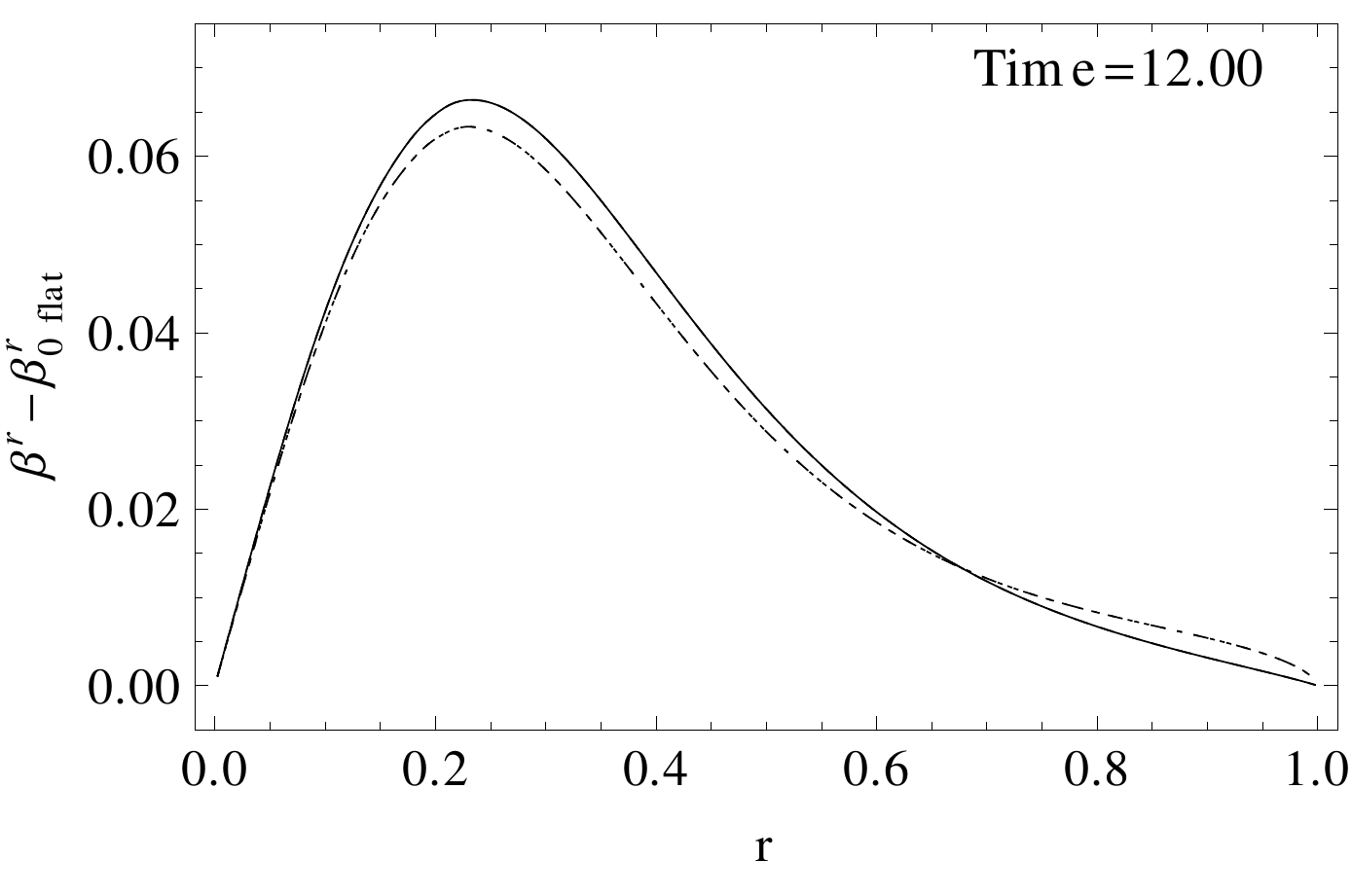}}&
\vspace{-5.5ex} \hspace{-0.6ex} \mbox{\includegraphics[width=1.04\linewidth]{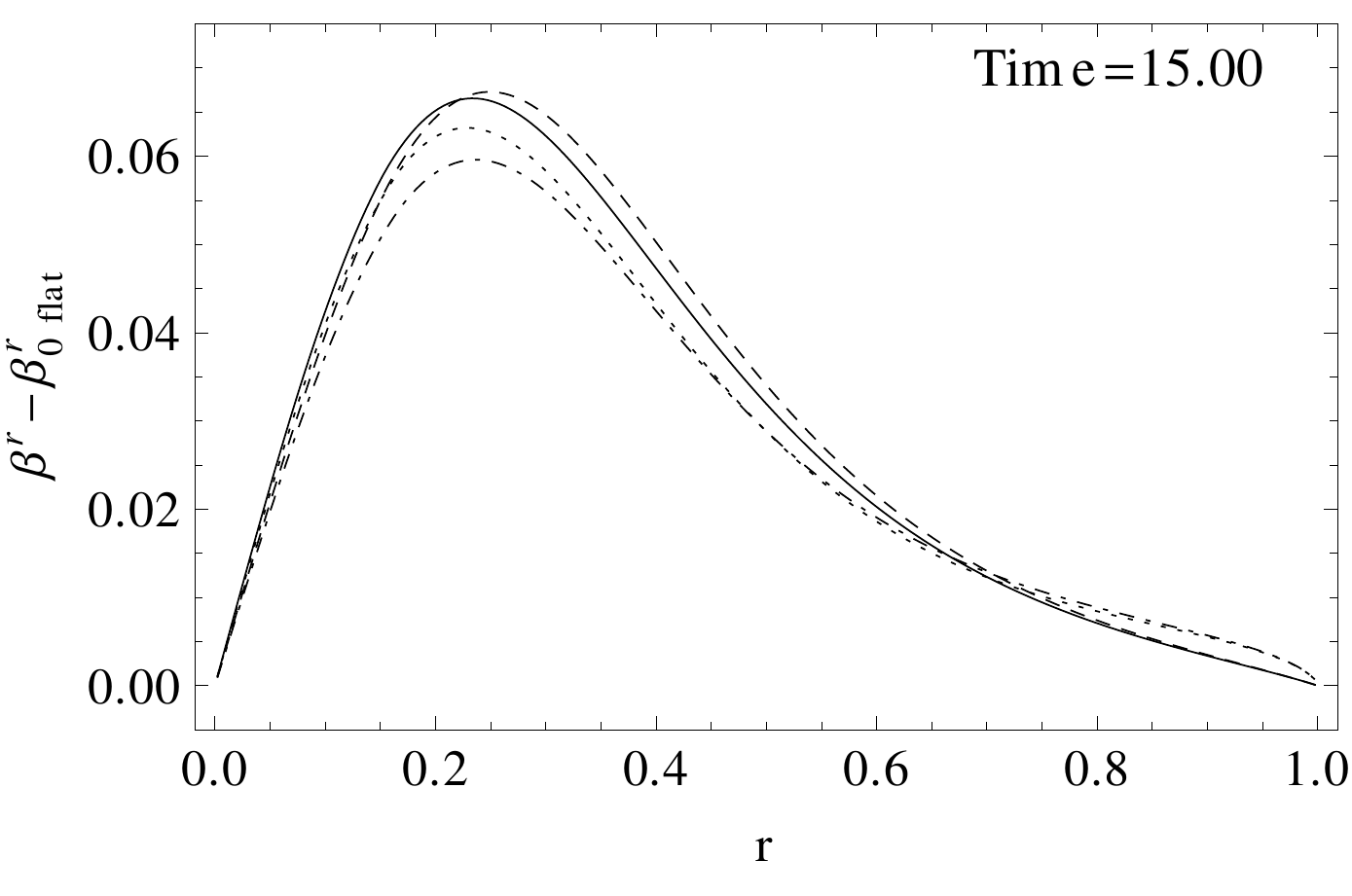}}
\end{tabular}
\vspace{-3ex}
\caption{Evolution of several variables under an initial perturbation with $c=0.5$ and $A_\Phi=0.03$ for the indicated value of $\cL$ and with or without recalculation of the gauge source functions.}
\label{fs:bhlarge1}
\end{figure}

\begin{figure}[htbp!!]
\center
\vspace{-2.5ex}
 \begin{tikzpicture}[scale=2.5]\draw (-0.7cm,0cm) node {};
		\draw (0cm, 0.1cm) node {\small Not recalc.}; \draw (0cm, -0.1cm) node {\small $\cL=0$}; \draw (0.5cm, 0cm) -- (0.8cm, 0cm);
		\draw (1.5cm, 0.1cm) node {\small Recalc.}; \draw (1.5cm, -0.1cm) node {\small $\cL=0$}; \draw [dashed] (2cm, 0cm) -- (2.3cm, 0cm);
		\draw (3cm, 0.1cm) node {\small Not recalc.}; \draw (3cm, -0.1cm) node {\small $\cL=4$};  \draw [dotted] (3.5cm, 0cm) -- (3.8cm, 0cm);
		\draw (4.5cm, 0.1cm) node {\small Recalc.}; \draw (4.5cm, -0.1cm) node {\small $\cL=4$}; \draw [dash pattern= on 4pt off 2pt on 1pt off 2pt] (5cm, 0cm) -- (5.3cm, 0cm);
	\end{tikzpicture}
\\
\begin{tabular}{ m{0.5\linewidth}@{} @{}m{0.5\linewidth}@{} }
\hspace{-1.0ex}\mbox{\includegraphics[width=1\linewidth]{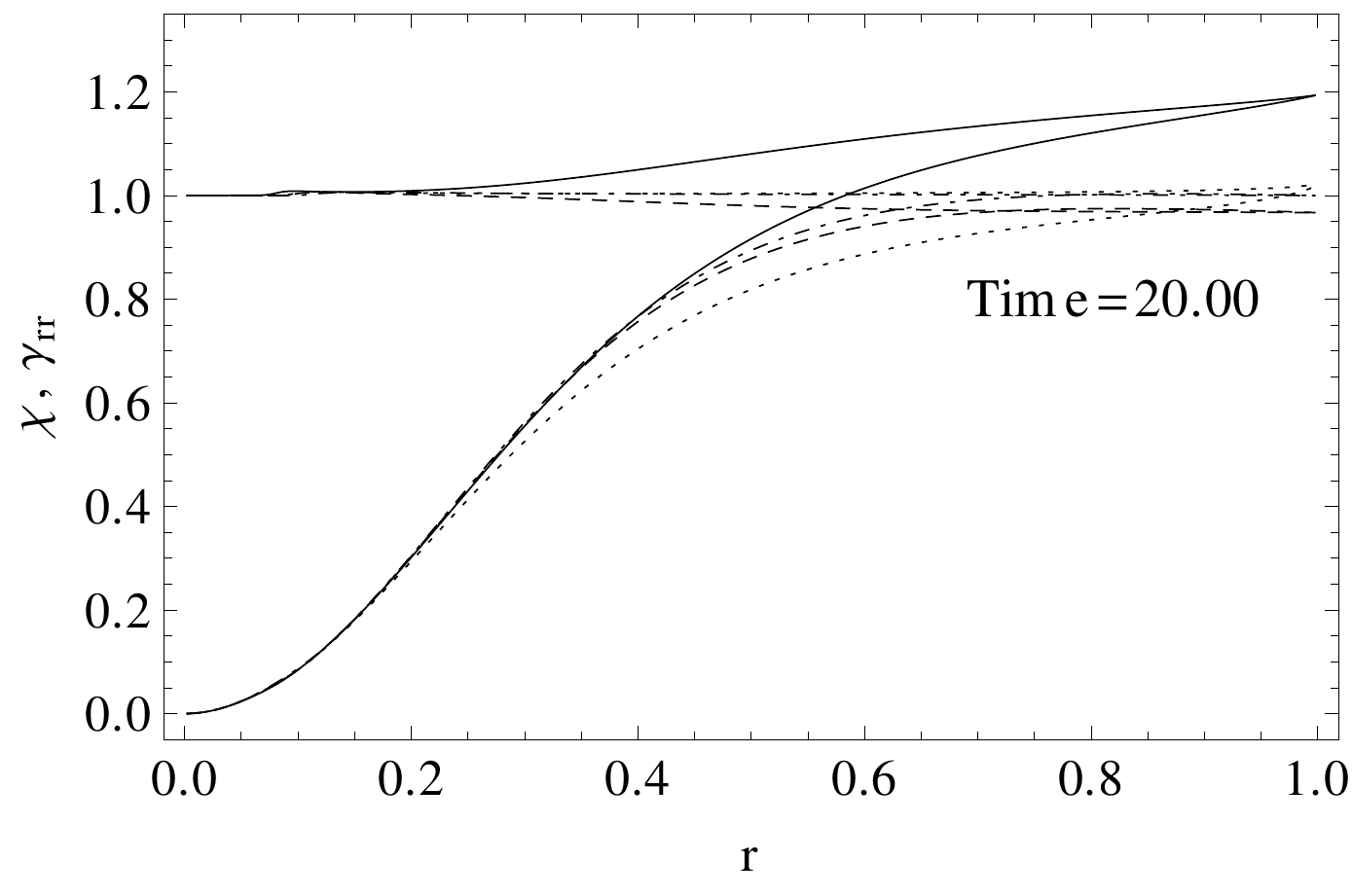}}&
\hspace{+0.8ex} \mbox{\includegraphics[width=1\linewidth]{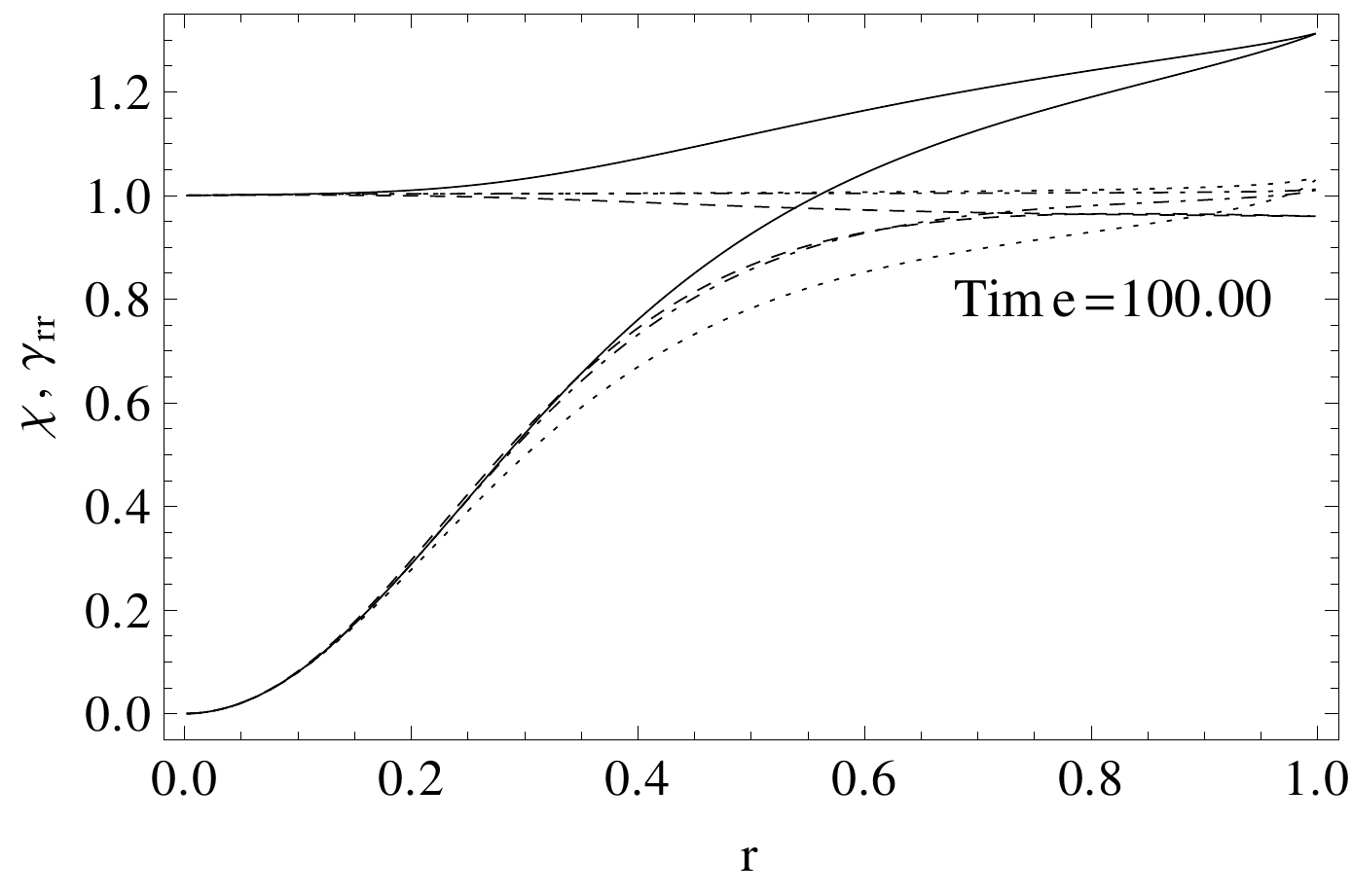}}\\
\vspace{-5.5ex} \hspace{-0.6ex}\mbox{\includegraphics[width=0.99\linewidth]{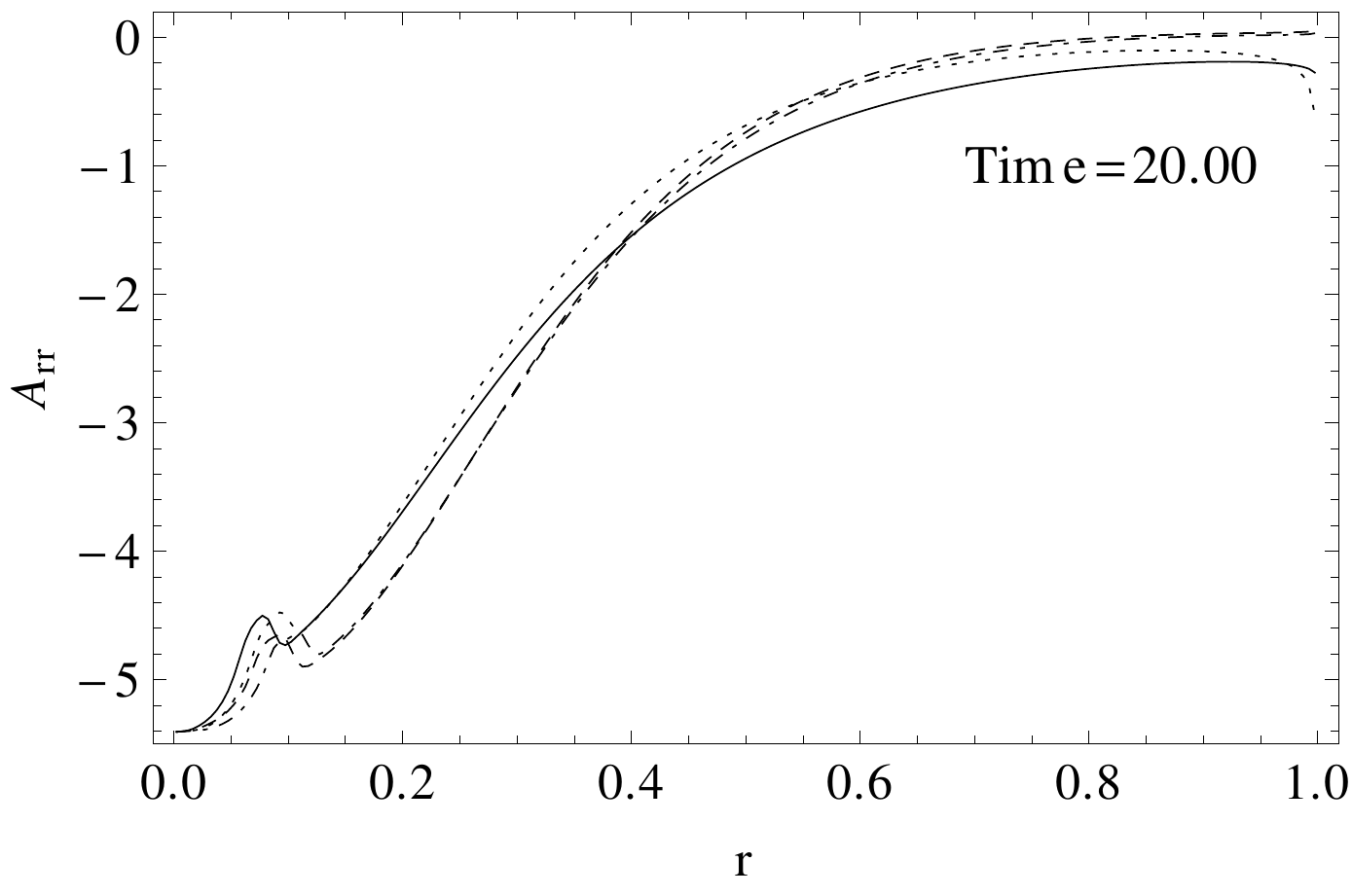}}&
\vspace{-5.5ex} \hspace{+1.2ex} \mbox{\includegraphics[width=0.99\linewidth]{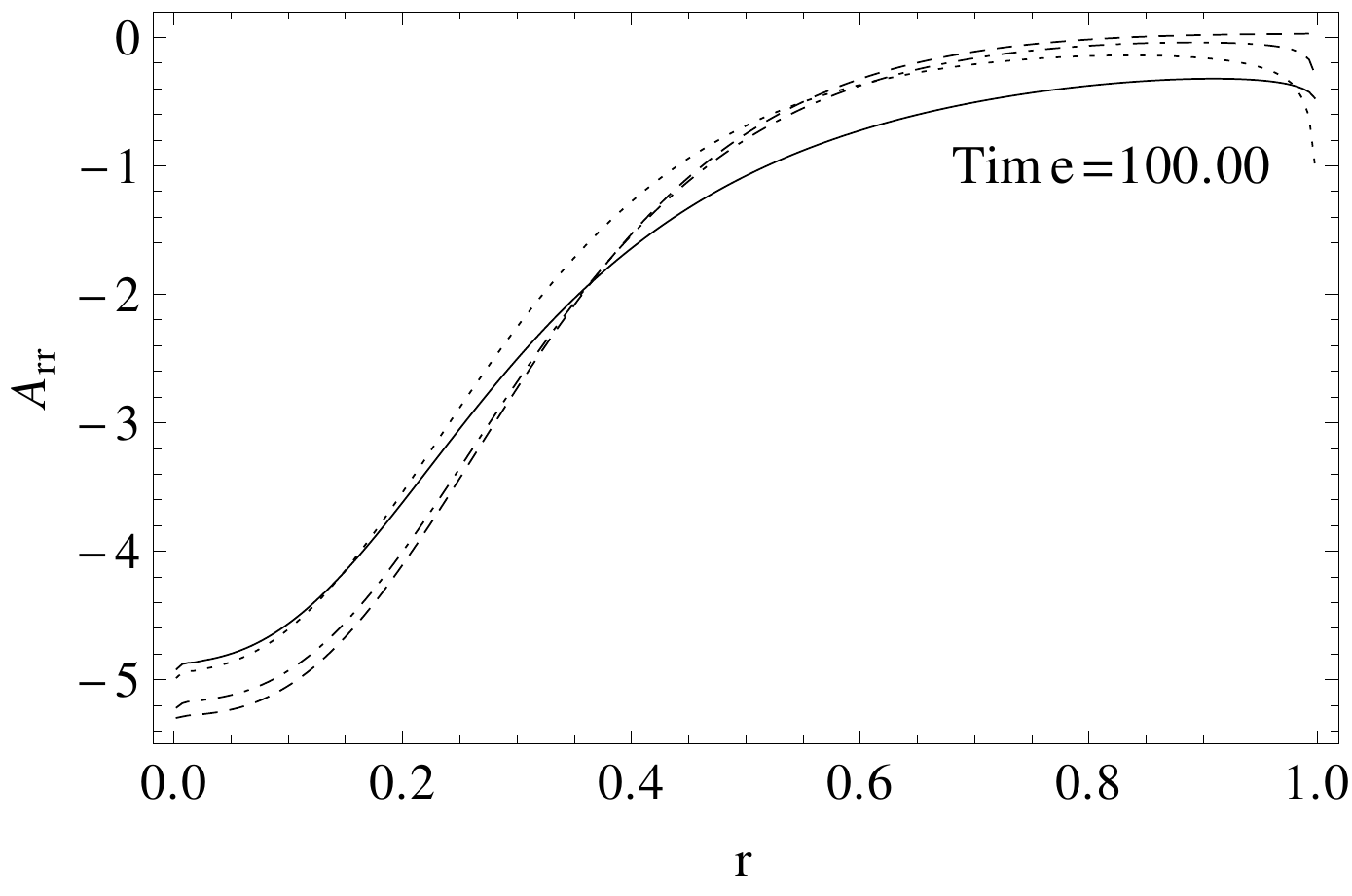}}\\
\vspace{-5.7ex} \hspace{-3.5ex}\mbox{\includegraphics[width=1.06\linewidth]{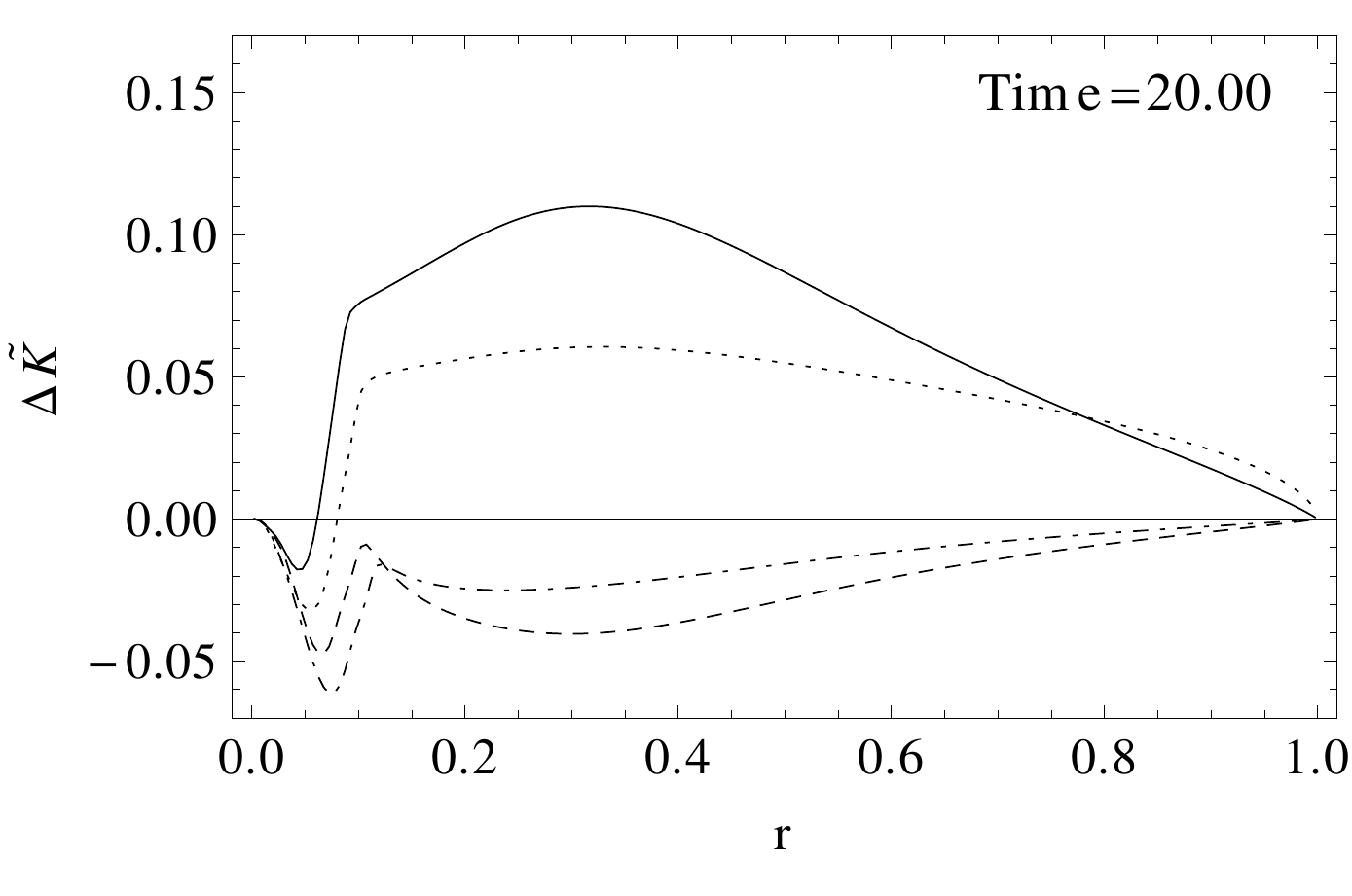}}&
\vspace{-5.7ex} \hspace{-1.8ex} \mbox{\includegraphics[width=1.06\linewidth]{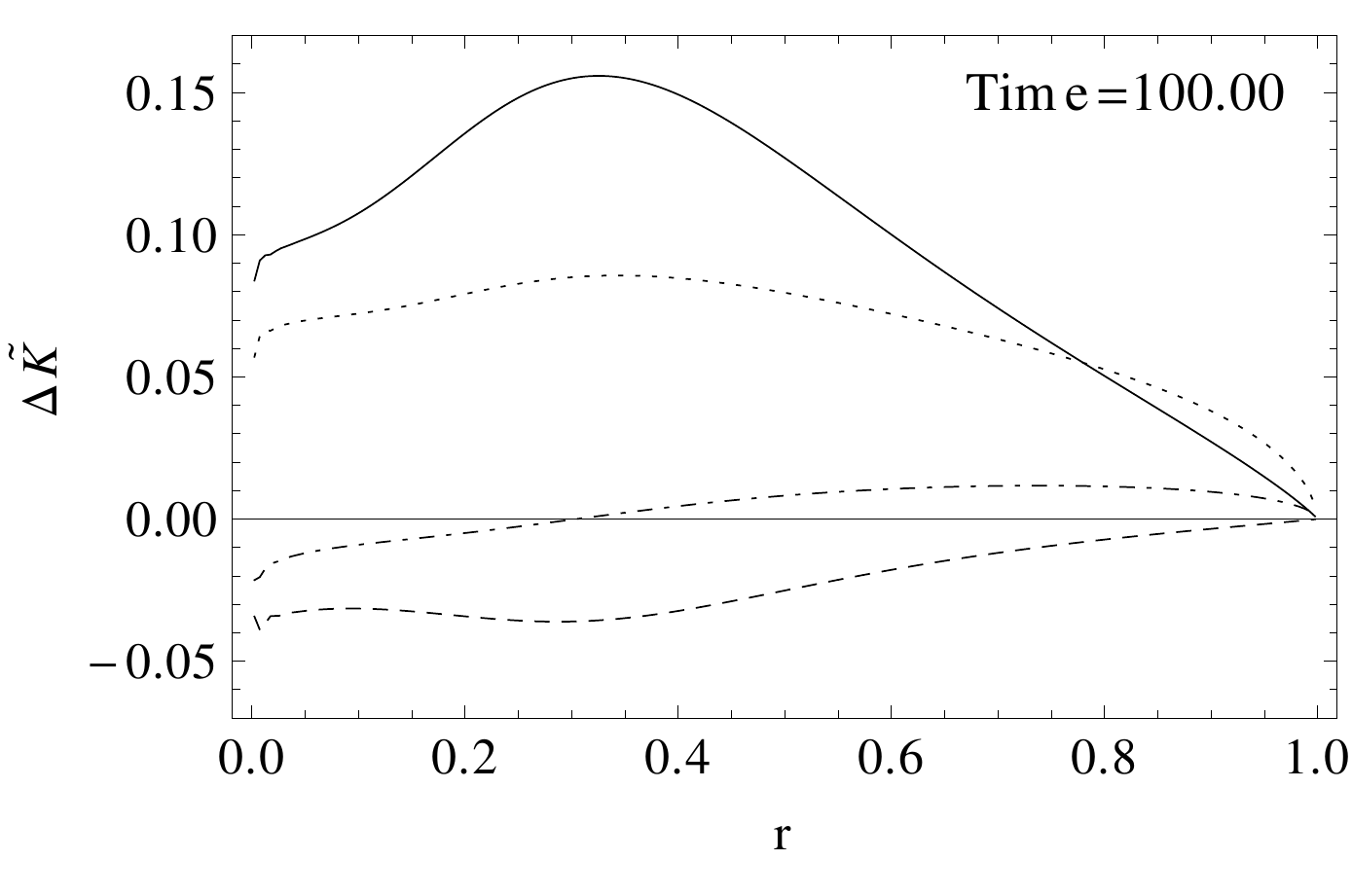}}\\
\vspace{-6.3ex} \hspace{-0.8ex}\mbox{\includegraphics[width=1\linewidth]{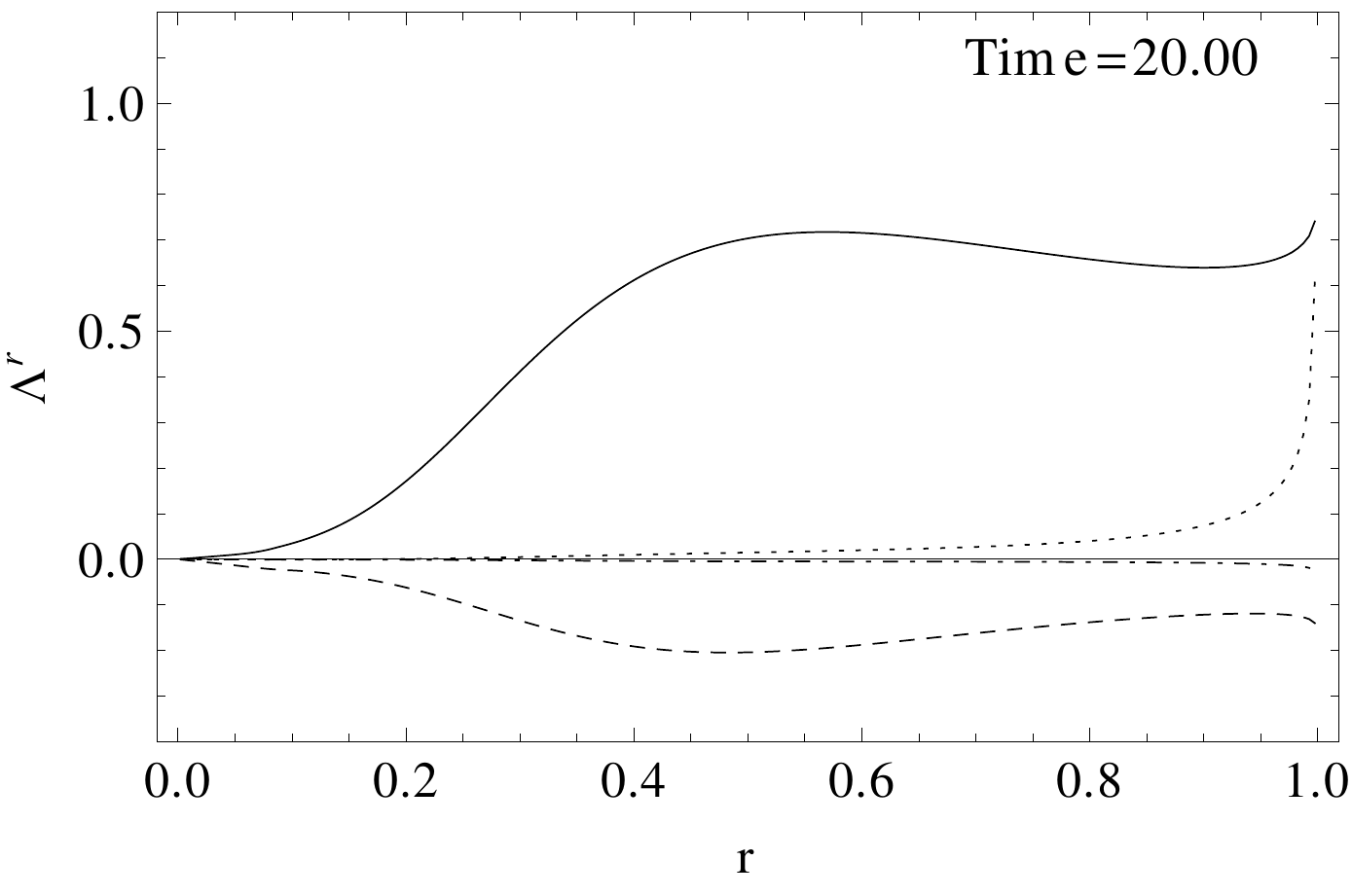}}&
\vspace{-6.3ex} \hspace{+1.0ex} \mbox{\includegraphics[width=1\linewidth]{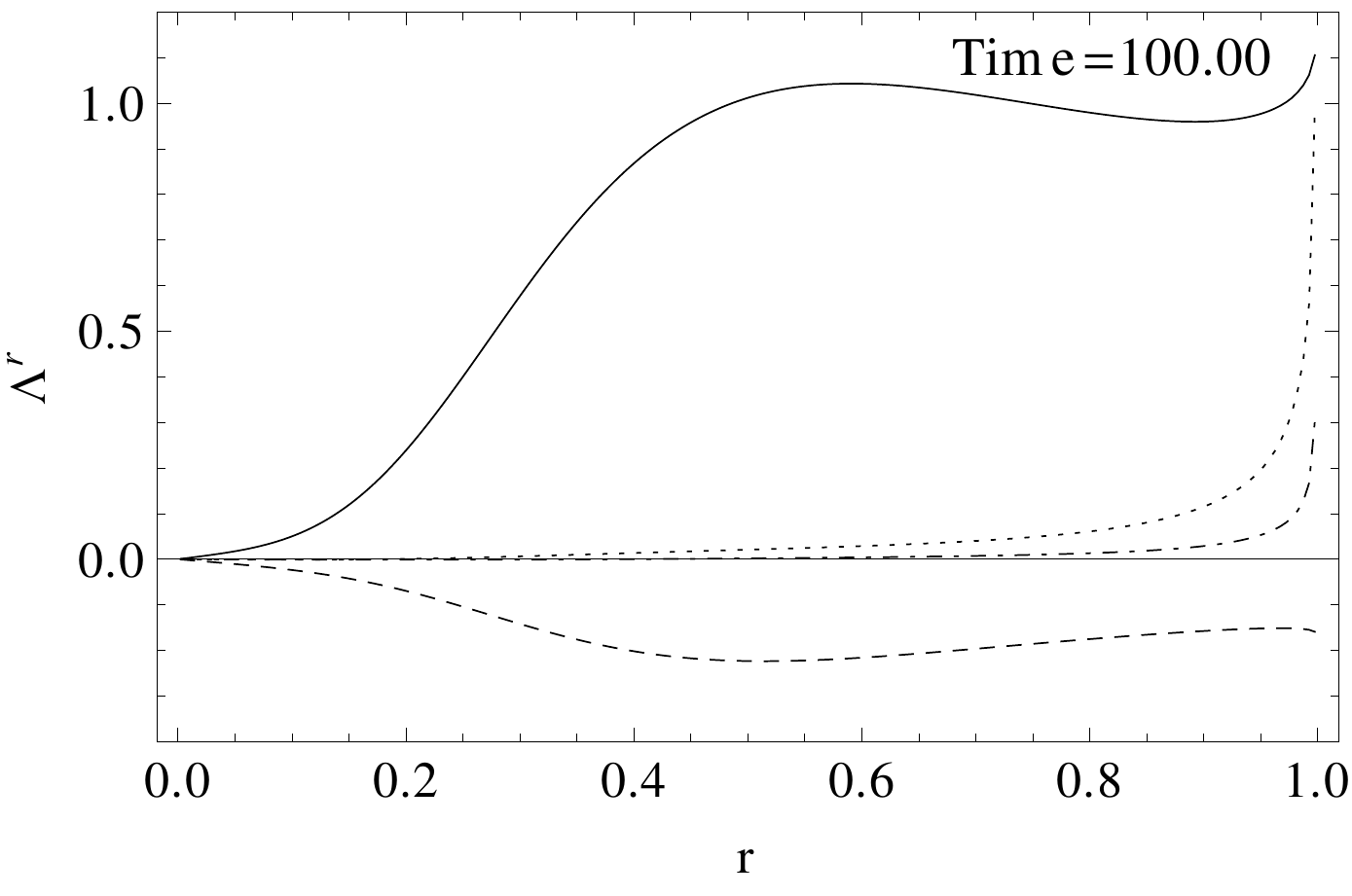}}\\
\vspace{-5.5ex} \hspace{-2.2ex}\mbox{\includegraphics[width=1.04\linewidth]{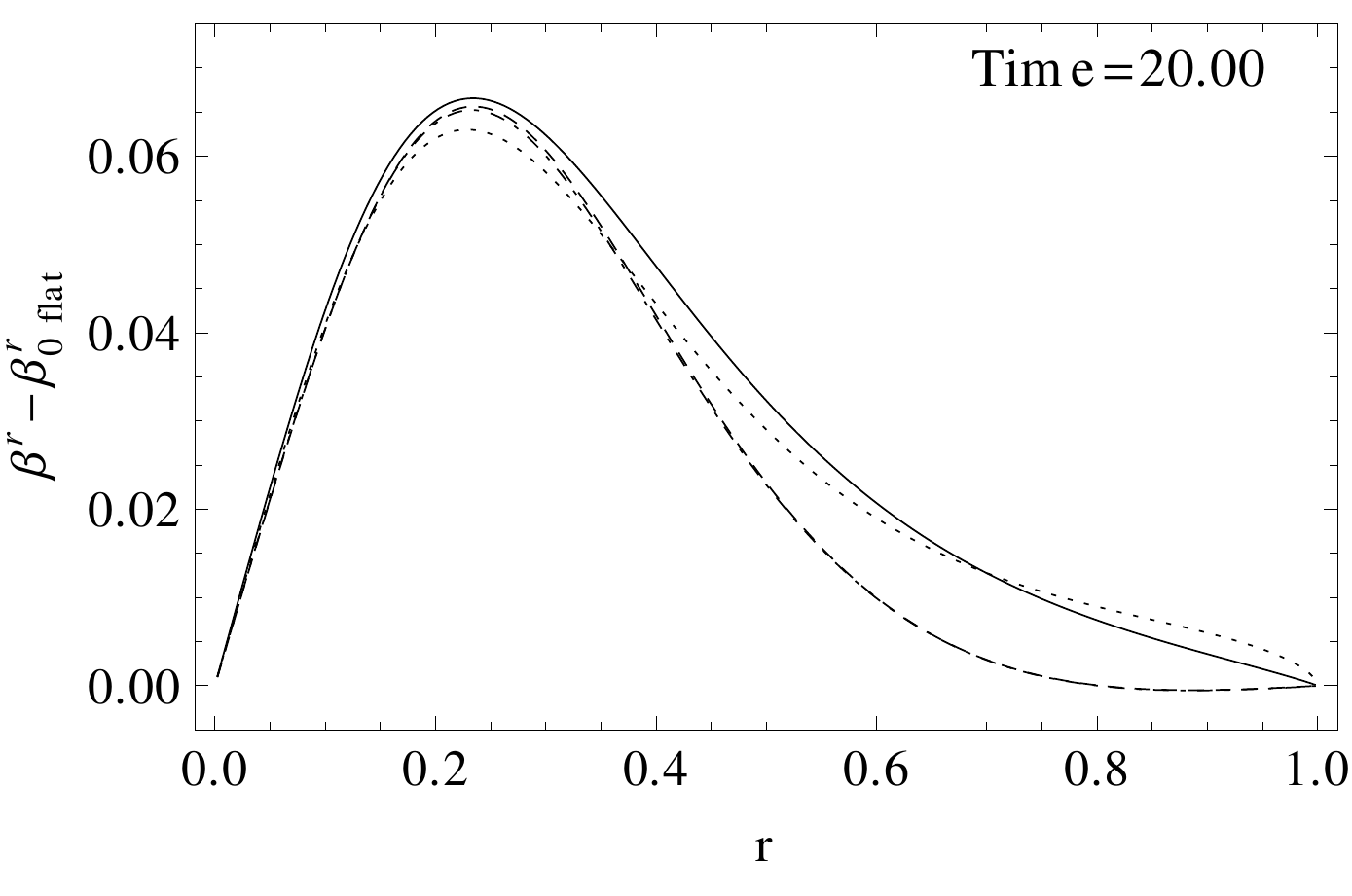}}&
\vspace{-5.5ex} \hspace{-0.6ex} \mbox{\includegraphics[width=1.04\linewidth]{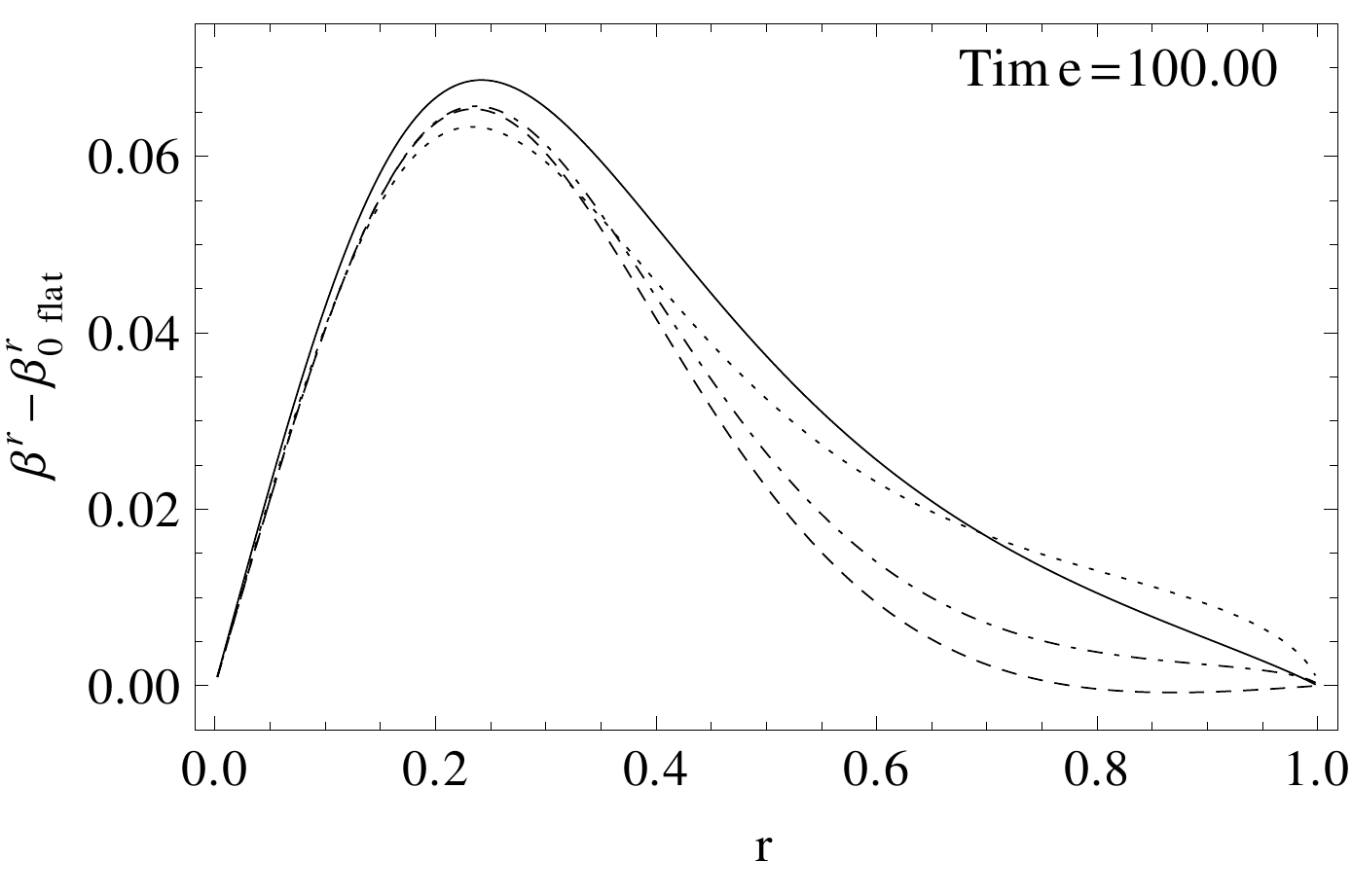}}
\end{tabular}
\vspace{-3ex}
\caption{Evolution of several variables: continuation. In the top row plots in the previous and in this figure, $\chi$ is the quantity that vanishes at the origin, while $\gamma_{rr}$ is unity there.}
\label{fs:bhlarge2}
\end{figure}

The behaviour of some of the variables in four different simulations with an initial perturbation $A_{\Phi}=0.03$, $\sigma=0.1$ and $c=0.5$ is presented in figures \ref{fs:bhlarge1} and \ref{fs:bhlarge2}. The GBSSN system was used with 1+log slicing condition ($\xi_{1+log}=2$) and integrated Gamma-driver shift condition ($\xi_{\beta^r}=5$, $\lambda=0.083$ and $\eta=0.1$). Other parameter choices common to the four simulations were $\kappa_1=1.5$, $\Kc=-1$, $M=1$, critical $\Cc=3.11$, 400 spatial gridpoints and $\Delta t=0.001$.

% Large scalar field perturbation

The differences among the simulations were that two of them used $\cL=0$ and the other two $\cL=4$ (the choice which was set to decrease the drift effect). In the evolution of one of the $\cL=0$ and one of the $\cL=4$ cases, the final mass of the BH (the initial $M$ + the energy brought in by the scalar field) was evaluated and set as the new total mass. Then the new critical value of $\Cc$ and the corresponding new compactification factor $\aconf$ were calculated and used to evaluate the source functions of the gauge conditions. These two cases are labeled with ``Recalc.'' (recalculated) in the figures, while the ones which continue using the initial source functions are denoted by ``Not recalc.'' (not recalculated).

Figures \ref{fs:bhlarge1} and \ref{fs:bhlarge2} show the the behaviour of several variables at four different times in five rows. The quantities $\chi$ and $\gamma_{rr}$ are shown in the same plots; $\gamma_{rr}$ corresponds to the curves that go to one at the origin, while $\chi$ vanishes there. Instead of plotting $\beta^r$, representing $\beta^r-\beta^r_{0\ flat}=\beta^r+\case{r}{3}$ is more convenient to see the changes in the profile.
The recalculation of the new total BH mass (and new critical value of $\Cc$) takes place at $t\approx14.1$ (this is when the scalar field pulse arrives at the BH's horizon). The final mass of the BH is $M\approx1.09$, almost a 10\% increase with respect to the initial $M=1$.
The location of the horizon is not shown in the plots, because it is different for each of the simulations. It is located in the range $r_{hor}\in[0.13,0.14]$.

At time $t=12$ (first column) the difference between the profiles is due only to the chosen value of $\cL$, because the recalculation has not taken place yet. The difference between both cases is especially large for $\Lambda^r$. Note that the relations $\atscrip{\chi}=\atscrip{\gamma_{rr}}$ and $\atscrip{\DPK}=0$ hold at all times, as required by the regularity conditions on the gauge variables. At $t=100$ the curves of $A_{rr}$ reach the origin at two different values: the profiles of the simulations with recalculated source functions are closer to the initial value at $r=0$ (displayed in \fref{fin:BiniArr}), while the others have a smaller absolute value there. Close to $\scri^+$ the recalculated case with $\cL=4$ seems to be the closest one to the expected $\atscrip{A_{rr}}=0$, but the other curves are drifting away towards more negative values. In $\DPK$ and $\Lambda^r$ the sign of the recalculated and not recalculated versions of $\cL=0$ is the opposite.

\begin{figure}[h!!]
\center\vspace{-4ex}
\begin{tabular}{ m{0.5\linewidth}@{} @{}m{0.5\linewidth}@{} }
\hspace{-3ex} \mbox{\includegraphics[width=1.1\linewidth]{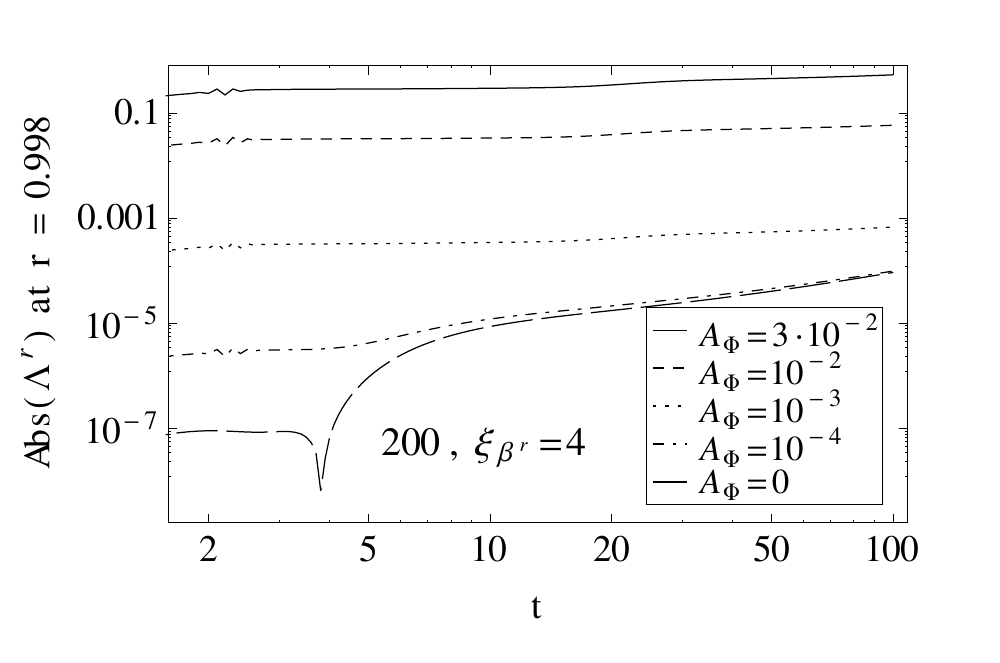}}&
\hspace{-2.5ex} \mbox{\includegraphics[width=1.1\linewidth]{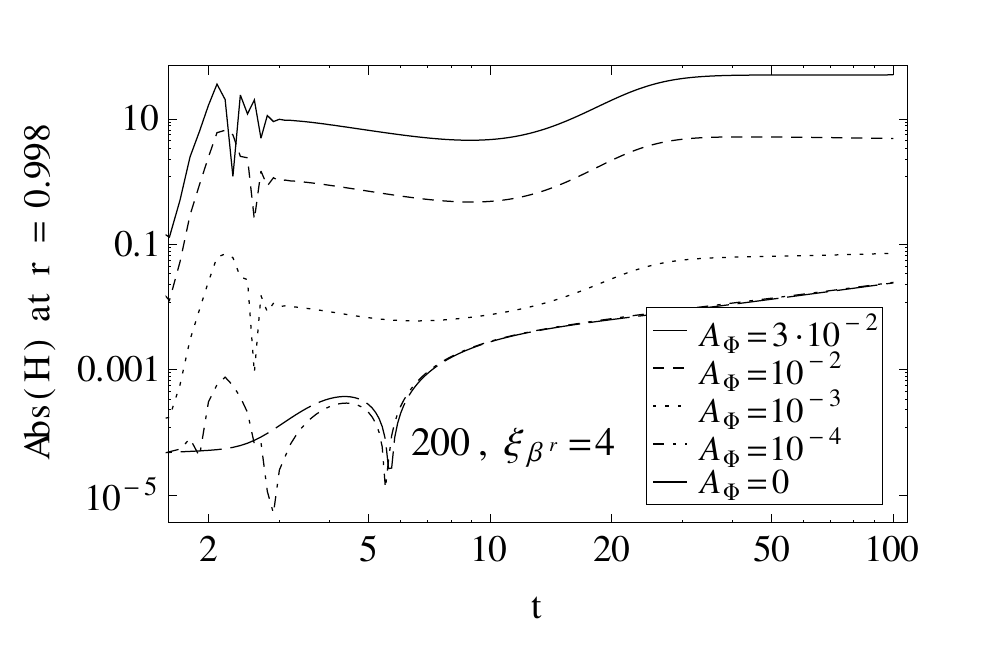}}\\
\vspace{-4ex}\hspace{-3ex} \mbox{\includegraphics[width=1.1\linewidth]{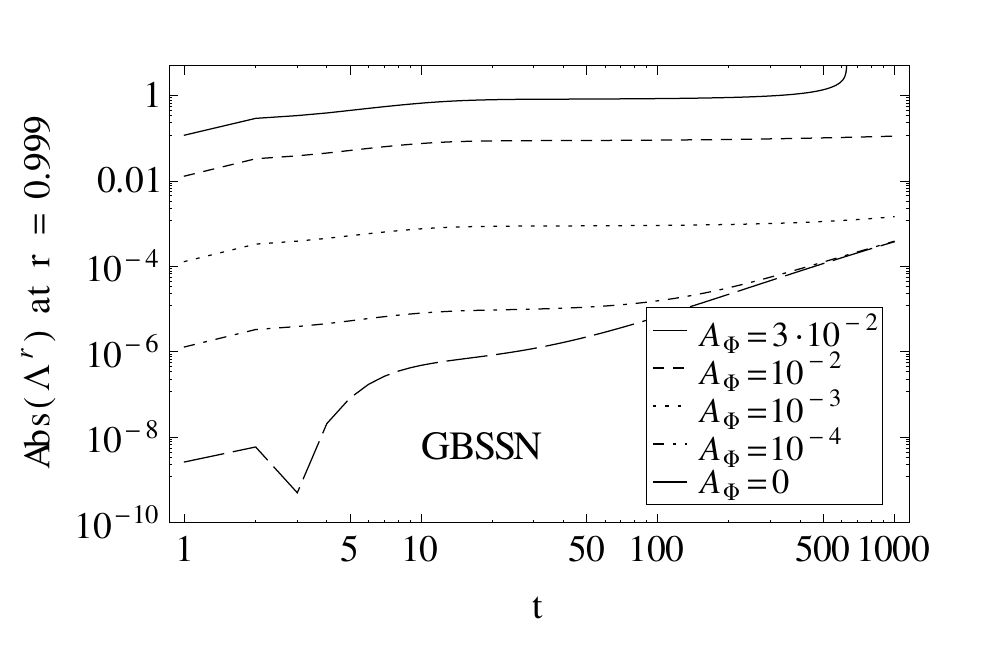}}&
\vspace{-4ex}\hspace{-2.5ex} \mbox{\includegraphics[width=1.1\linewidth]{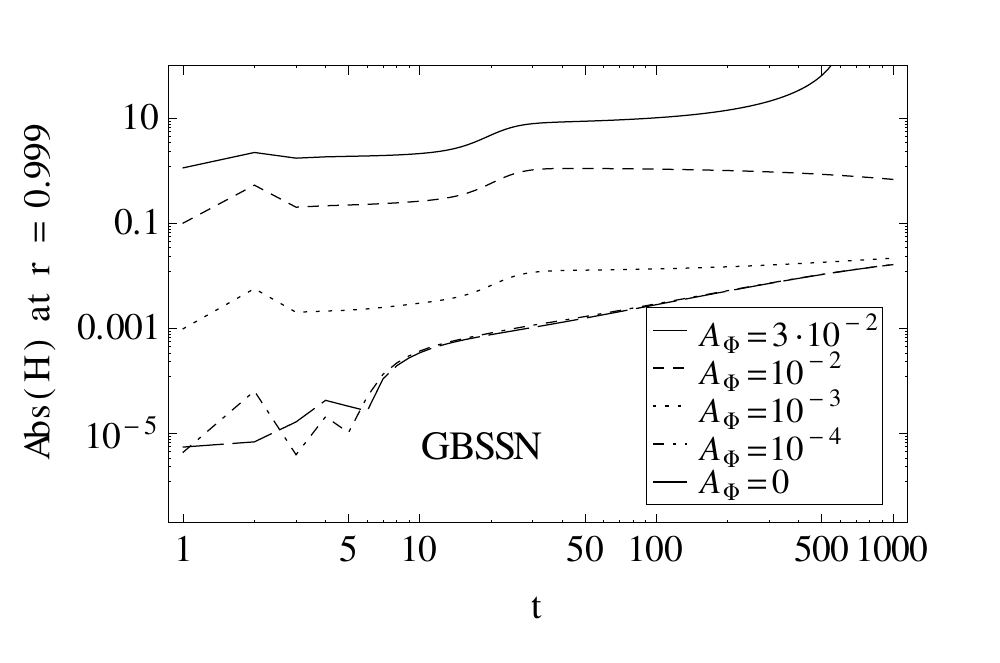}}\\
\vspace{-4ex}\hspace{-3ex} \mbox{\includegraphics[width=1.1\linewidth]{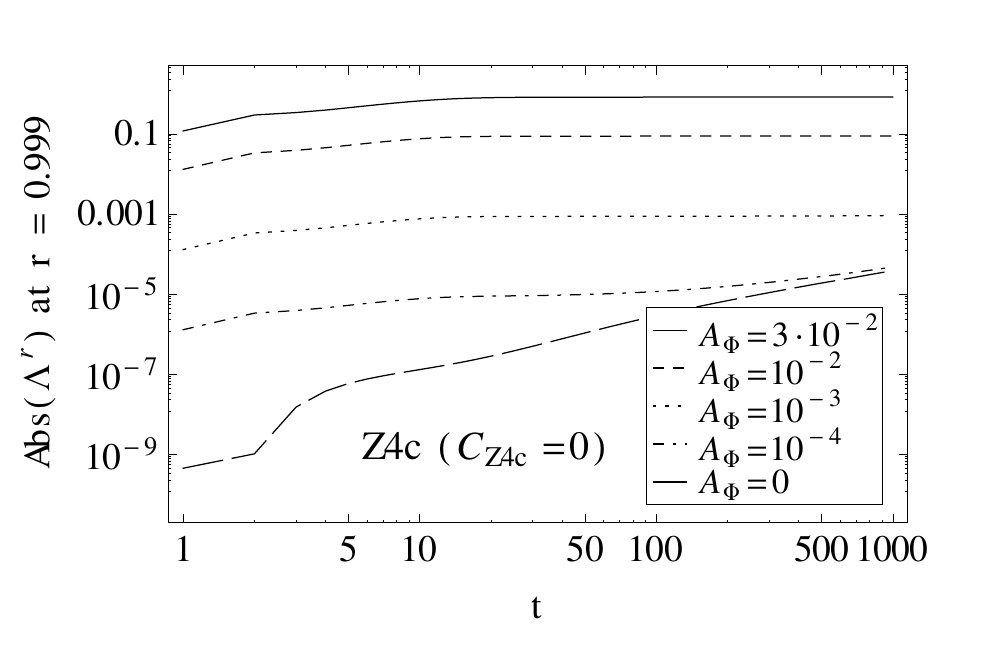}}&
\vspace{-4ex}\hspace{-2.5ex} \mbox{\includegraphics[width=1.1\linewidth]{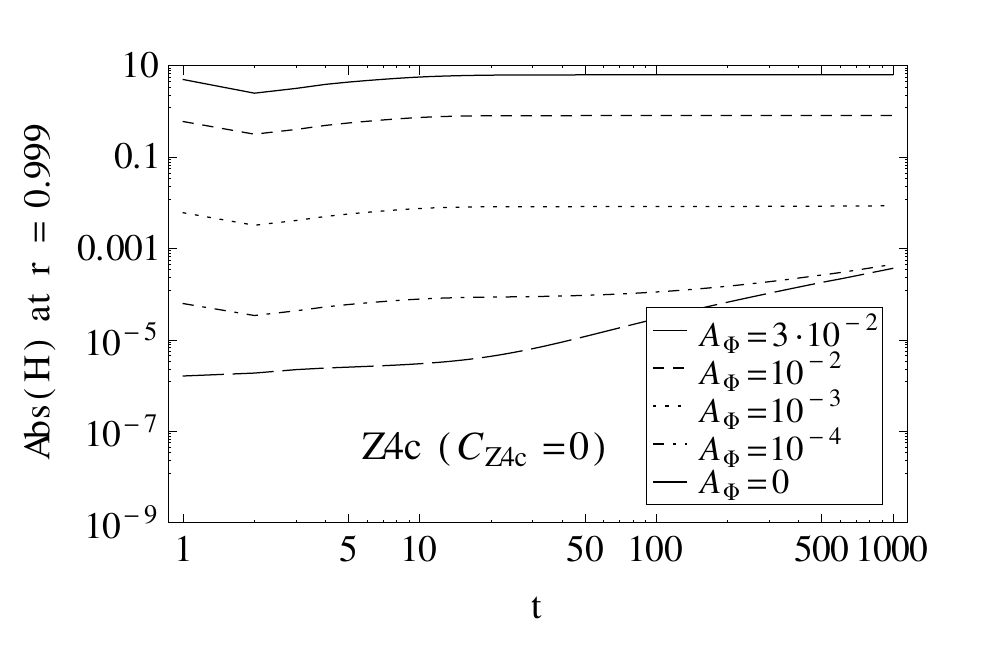}}
\end{tabular}
\vspace{-3.5ex}
\caption{Effect of the amplitude of the initial perturbation $A_\Phi$ on the $\Lambda^r$ variable and the Hamiltonian constraint at the closest gridpoint to $\scri^+$ over time. The first row shows data from a simulation with 200 gridpoints and $\cL=4$, while the other two rows correspond to simulations with GBSSN and \CZ{} ($C_{Z4c}=0$) with 400 points, $\cL=0$ and evolved up to $t=1000$.}
\label{fs:drift}
\end{figure}

In subsection \ref{se:onlySchw} the behaviour of unperturbed Schwarzschild trumpet initial data was studied. We now want to better understand the relation between the observed drift and the amplitude of the initial perturbation.
Figure \ref{fs:drift} shows the behaviour in time of $\Lambda^r$ and the Hamiltonian constraint $H$ at the closest gridpoint to $\scri^+$ ($r=0.998$ in the case of the 200 points simulations and $r=0.999$ for the 400 points one) for different values of the amplitude of the initial perturbation. The gauge conditions used are the 1+log and the integrated Gamma-driver with parameter choices $\xi_{1+log}=2$ and $\xi_{\beta^r}=5$.
%The equivalent plot for a different value of $r$ outside of the BH's horizon would be very similar.
The behaviour of the $A_\Phi=10^{-4}$ and $A_\Phi=0$ cases, especially in GBSSN's $H$, is very similar and, as time passes, the closer the curves come together.
The GBSSN simulation with the largest amplitude and with 400 points crashes at $t=633$; the value of $\Lambda^r$ at that time is about 6 times as large as its \CZ{} ($C_{Z4c}=0$) equivalent. The latter has actually arrived at a stationary value (it is attained at $t\approx100$), while the GBSSN evolution continues growing and the values of both quantities compared to the \CZ{} are larger.
The curve corresponding to the unperturbed initial state is the one with the fastest growth (in a similar way as the curves in \fref{fs:Schw} seem to grow without bound) and it looks like at a certain point in time it will overtake the effect of the curves that correspond to larger initial perturbations. If this was the case, a long term instability could develop, but no clear conclusion can be drawn from the data obtained so far.
Apparently, the larger the initial amplitude, the more the variables deviate from their initial states, but also the faster a stable stationary state seems to be reached.
% Effect of initial amplitude in drift

\begin{figure}[h!!]
\center\vspace{-1ex}
\begin{tabular}{ m{0.5\linewidth}@{} @{}m{0.5\linewidth}@{} }
\hspace{-3ex} \mbox{\includegraphics[width=1.05\linewidth]{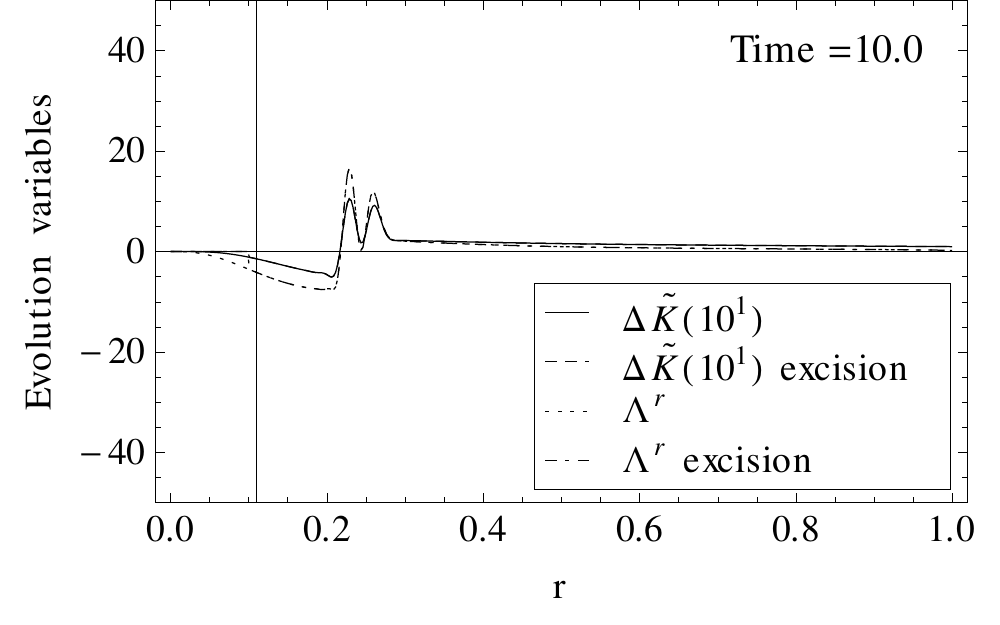}}&
\hspace{-2ex} \mbox{\includegraphics[width=1.05\linewidth]{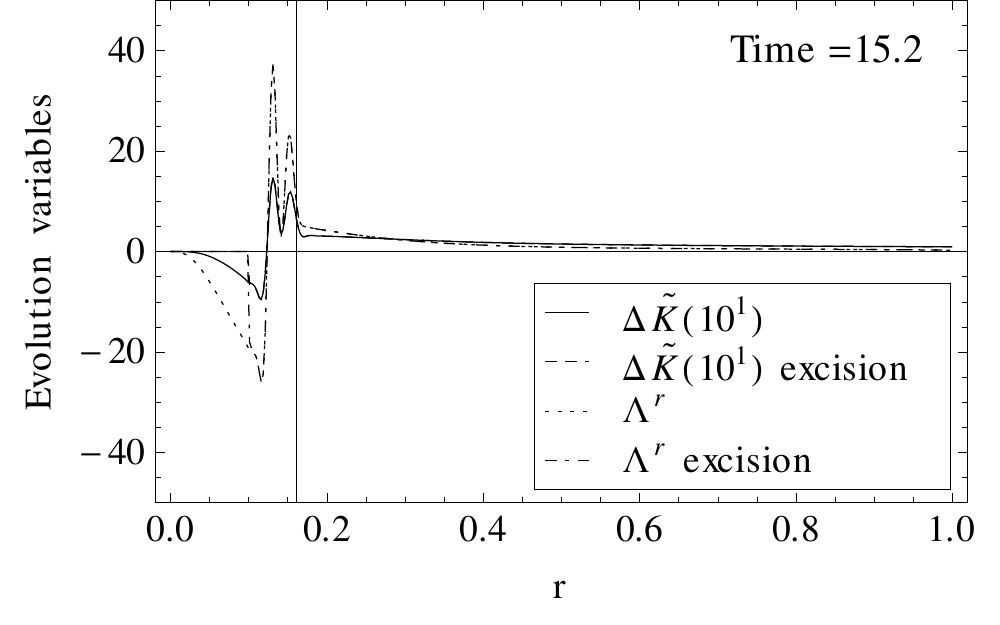}}\\
\hspace{-3ex} \mbox{\includegraphics[width=1.05\linewidth]{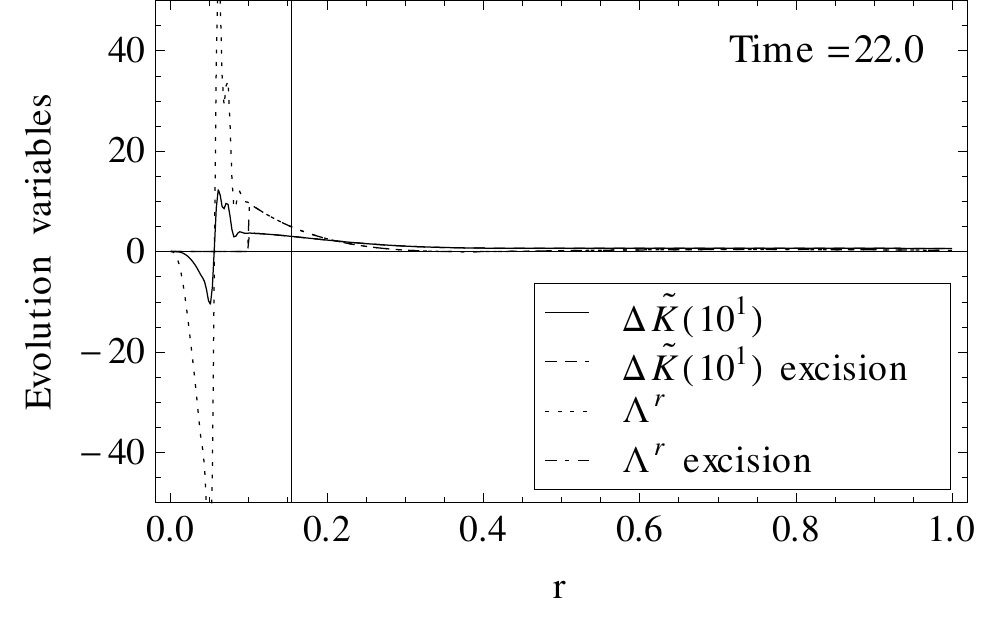}}&
\hspace{-2ex} \mbox{\includegraphics[width=1.05\linewidth]{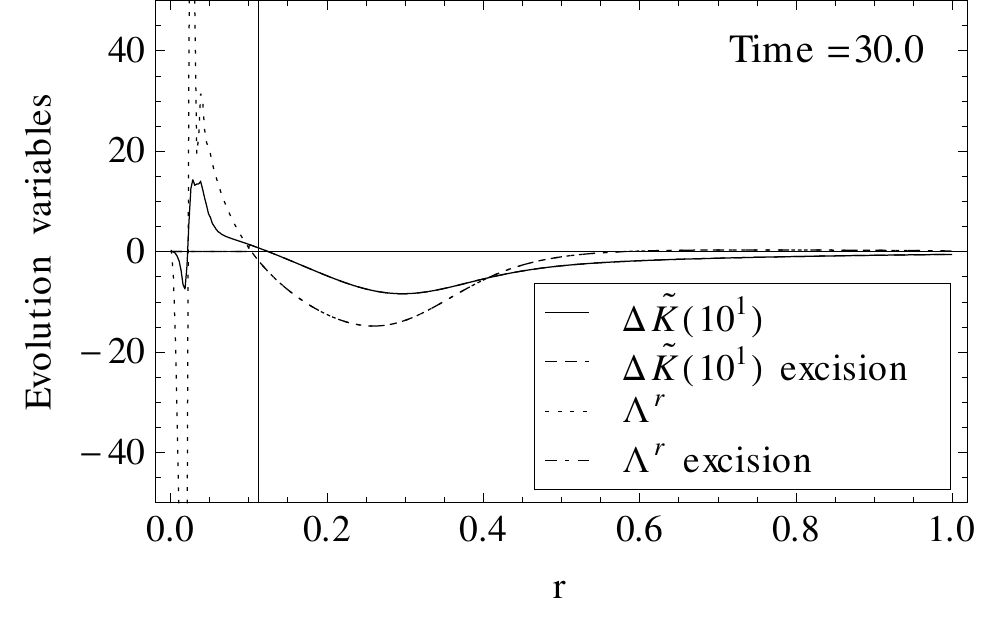}}
\end{tabular}
\vspace{-3ex}
\caption{Values of $\DPK$ and $\Lambda^r$ at the given times for a simulation with \CZ{} ($C_{Z4c}=0$) and harmonic gauge conditions with physical metric background source functions. The initial scalar field perturbation is $A_\Phi=0.03$. The vertical line denotes the position of the horizon and the excision boundary is located at $r=0.10$.}
\label{fs:excA03}
\end{figure}

Another comparison between excised and non-excised simulations, in this case for an initial perturbation of $A_\Phi=0.03$, is presented in \fref{fs:excA03}. The gauge conditions with physical source terms \eref{ee:improvedpbg} and the choice $\cL=0$ give the appropriate speeds at the horizon. The evolution equations are the \CZ{} ($C_{Z4c}=0$) ones.
The effect of the harmonic gauge condition chosen on the displayed quantities $\DPK$ and $\Lambda^r$ is quite different from the 1+log one (compare to the third and fourth rows in figures \ref{fs:bhlarge1} and \ref{fs:bhlarge2}): the 1+log profile is smoother, while the harmonic one presents sharper features.
The two simulations (with and without excision) crash at $t=32.7$ due to a bulk instability: an exaggerated growth in the interior of the domain of some of the variables makes the simulation crash. If a larger value of $\cL$ had been chosen, the simulation without excision might have run longer before the crash, but then the eigenspeeds at the horizon would not have been appropriate for a simulation with excision.

The conclusion from subsection \ref{se:onlySchw} that states that the evolution of the outer spacetime is not affected by the problems that arise from inside of the BH is also valid here: the agreement between the excised and non-excised data in \fref{fs:excA03} is good (the lines lie on top of each other) and the two simulations crash in the same way and at the same time. The origin of the instability thus has to have its origin in the continuum equations or in the geometry of the underlying hyperboloidal slice, more specifically the Schwarzschild CMC trumpet.
% Excision in A03

%From the previously obtained results that the end state of the collapse process in \fref{fs:collapse} is stable and different from the CMC trumpet initial data \fref{fin:Bini}, which have turned out not to be a stable stationary solution of the equations, it is possible that the observed drift is nothing but how the variables try to find their stationary final states.
%As the gauge source functions force $\alpha$ and $\beta^r$ to take values the CMC trumpet values that are possibly incompatible with the real stationary solution, this solution cannot be found and the simulation eventually crashes.
CMC trumpet initial data in an evolution with gauge source functions calculated from these initial data have turned out to be an unstable stationary solution of the system, whereas a simulation with flat spacetime source functions does find a stable stationary solution, as was exemplified by the collapse process in \fref{fs:collapse}. The observed drift seems to be the movement of the variables towards their stable stationary final states. The gauge source functions force $\alpha$ and $\beta^r$ to take the CMC trumpet values that are possibly incompatible with the real stationary solution, so that this solution cannot be found and the simulation eventually crashes.
%and this process seems to be more successful for source functions which are not calculated from the CMC trumpet data.
The experiments on which these hypotheses are based have only been performed successfully with the 1+log gauge condition, as the harmonic one was more likely to become unstable due its smaller characteristic speeds close to the BH, so that there is still much to test before a certain claim can be made.

\chapter{Discussion}\label{c:discussion}

\section{Summary}

% equations
This work represents a first step towards the numerical implementation of the hyperboloidal initial value problem of GR using conformal compactification methods in spherical symmetry. 
The Einstein equations for a regular conformally rescaled metric have been expressed in the commonly used BSSN and Z4 formulations. The divergent terms at $\scri^+$ that appear in the equations require that some relations between the variables, the regularity conditions, are satisfied so that the formally divergent terms attain a regular limit.  
The numerical experiments performed indicated that some changes at the continuum level had to be performed to the equations in order to obtain a well-behaved system. These consisted of choosing the trace of the physical extrinsic curvature, $\pK$, instead of the conformal one, $\cK$, as evolution variable, and of adding a constraint damping term to $\dot \Lambda^r$'s RHS. 
%Appropriately added constraint damping terms and useful variable transformations solved the instability problems and allowed to obtain long term evolutions. 
%derived the equations for the conformally rescaled metric in the BSSN and Z4 formalisms: divergent terms at $\scri^+$ ``turn into'' regularity conditions there. 
%changes at the continuum level (using physical $\pK$ and constraint damping term in $\dot \Lambda^r$) required to obtain a well-behaved system. 

% initial data
Spherically symmetric initial data for regular spacetimes and spacetimes including a BH have been derived. The values of the variables were taken from the line element expressed in terms of a compactified radial coordinate on a CMC hyperboloidal slice. 
In the case of BH initial data, the geometry of the slice is defined in terms of the mass $M$ and charge $Q$ of the BH, the mean curvature $\Kc$ and the integration constant $\Cc$. For the critical value of $\Cc$, a CMC trumpet geometry that connects $\scri^+$ with an infinitely long cylinder located at a certain value of the areal radius is obtained, and the derived initial data of all of the evolution variables is finite for all values of the compactified radial coordinate. 
The CMC trumpet slices of the Schwarzschild and RN spacetimes present a similar structure. %; the extreme RN case has the special property that the horizon and the location of the cylinder are mapped to the same point of the compactified radius. 
%initial data: , trumpet CMC slices in Schwarzschild and RN present similar basic structure - extreme case special; 

% gauge conds
The gauge conditions play a fundamental role in the simulations and require a very careful treatment. The main difference between standard Cauchy slices and hyperboloidal slices is that for the latter the mean extrinsic curvature does not vanish: this different background geometry introduces changes in the construction of the equation of motion for the lapse, where the sign of the extrinsic curvature has to be cancelled with an appropriate source term to avoid exponential growths. The remaining source terms in the slicing and shift conditions have to ensure that the arising regularity conditions are compatible with those of the Einstein equations and that the gauge variables attain stationary values at $\scri^+$ suitable for the hyperboloidal evolution. 
The harmonic and 1+log slicing conditions and the Gamma-driver shift conditions have been successfully adapted to the hyperboloidal evolution. In an effort to systematize the calculation of the source terms, harmonic gauge conditions for lapse and shift with source functions calculated from a conformal or physical background metric have been derived and tested, and a less artificial treatment of the gauge variables at $\scri^+$ has been obtained. % that did not so heavily rely on damping terms. 
%With the second condition it is possible to achieve the preferred conformal gauge. 
The harmonic and 1+log slicing conditions were defined in the conformal picture, but the preliminar results of the harmonic gauge with gauge source functions calculated from the physical metric, where the preferred conformal gauge is also satisfied, suggest to define them in the physical one.
%suggest that it may be a better idea to define them in the physical one. 
%: the conditions related to the treatment of future null infinity have been introduced into the setup and evolved gauge conditions for lapse and shift with convenient source functions have been implemented and tuned. There is however still much work to do regarding gauge conditions even in the simple setup of spherical symmetry. 
%gauge conditions. very delicate treatment, 

%results
%A stable code that can run forever with regular initial data - and up to a reasonable time with BH initial data - has been implemented with common techniques and has provided an important approach to evolutions on a compactified hyperboloidal foliation. It has allowed to study the behaviour of the quantities around $\scri^+$ (even if only a staggered grid has provided stable results so far) and to check their convergence. 
The numerical implementation of the conformally rescaled equations using common techniques, the adapted gauge conditions and the derived initial data required a detailed stabilizing procedure that involved a hierarchical analysis of the equations. This study shed some light on the continuum instabilities that affect the Einstein equations in the conformal hyperboloidal picture and helped develop some intuition about how to solve future problems.
The result is a stable spherically symmetric code that can run forever for some initial configurations and for a period of time long enough to allow the observation of the decay of the scalar field for others. It has provided an important approach to vacuum and scalar field evolutions on a compactified hyperboloidal foliation and allowed to study the global behaviour of the quantities from the origin to $\scri^+$. 

The evolution of gauge waves, even if they do not imply physical dynamics, have served as a useful and strong test on the robustness of the equation's setup, but probably the most relevant results obtained with regular and Schwarzschild initial data are the evolution of the Einstein equations coupled to a massless scalar field. 
The tested scenarios were the reflection of a wave pulse at the origin, the collapse of a larger pulse into a BH and the perturbation of Schwarzschild initial data by a scalar field perturbation. The power-law decay tails of the latter was studied in the vicinity of future null infinity and the expected exponent was obtained at $\scri^+$. 
The extraction of the scalar wave signal there was successful and its convergence order accurately corresponded to the expected one. 
The results obtained so far have used a staggered grid, so that the equations were never evaluated exactly on $\scri^+$. Although a non-staggered grid may be preferable for some configurations, it is not a necessary requirement at all, because the correct convergence order for the signal at $\scri^+$ has already been obtained using extrapolation on the closest gridpoints to $\scri^+$.
The convergence results of the signal at $\scri^+$ have given useful indications about the performance of the different formulations, gauge conditions and parameter choices. 

%Schwarzschild CMC trumpet was thought to be a stationary solution of the Einstein equations (and thus used to calculate the gauge source functions), but that does not seem the case after seeing the difference between the collapse end state (for tuned 1+log and integrated Gamma-driver) and the CMC trumpet initial data, and also to the drift detected in the BH evolutions that use CMC-trumpet-based initial data. 
%the hypothesis regarding the drift observed in Schwarzschild CMC trumpet initial data ... 
The conclusion regarding stationarity of the initial data is that the derived regular initial data are indeed stationary for the Einstein equations and the gauge conditions considered here, but the Schwarzschild CMC trumpet, which was used to calculate the gauge source functions and as initial data for the simulations, does not seem to be a stable stationary solution of the equations, at least when evolved with the harmonic or the 1+log slicing conditions adapted to the hyperboloidal slice. This is also indicated by the difference in the variables's profiles between the end state of the collapse process (obtained using the 1+log and integrated Gamma-driver gauge conditions) and the calculated CMC trumpet initial data. The drift detected in the BH evolutions that use CMC-trumpet-based initial data also seems to point in this direction. %\upda{Example of drift getting the variables to the stationary solution?}

Although the experiments performed in this work reduce to spherical symmetry, the treatment of the radial direction is mostly the same as in a more general setup, so that some conclusions obtained from this work are expected to carry over to more general numerical studies of the hyperboloidal initial value formulation. 

%\newpage

\section{Outlook}

\subsubsection{Gauge conditions}

%Among the problems to solve that are more directly related to the experiments presented in this thesis are the further study and improvement of the gauge conditions.
%problems that have been studied in a large variety of setups: BH simulations in spherical symmetry, in axial symmetry, full three-dimensional evolutions, etc. 
%1+log + pbg, Gamma-driver (integrated) + pbg
The treatment of the gauge conditions at $\scri^+$ is much better understood now, but other problems arise due to the presence of strong field initial data. The  next step is the tuning of the gauge conditions in the interior part of the integration domain according to common prescriptions and the improvement of some of the derived gauge conditions, especially those that satisfy the preferred conformal gauge. A possibility would be to test a matching of the 1+log slicing condition and the (integrated) Gamma-driver shift condition in the interior part with harmonic-like gauge conditions in the neighborhood of $\scri^+$. 
%The harmonic and 1+log slicing conditions were defined in the conformal picture, but the preliminar results of the harmonic gauge with gauge source functions calculated from the physical metric suggest that it may be a better idea to define them in the physical one. 
%
%find stationary solution to the Schwarzschild spacetime in the compactified hyperboloidal picture
Another problem to solve is to understand what stable stationary solutions for the Schwarzschild spacetime on a hyperboloidal slice exist and how they relate to the gauge source functions.

\subsubsection{Regularity conditions}

A brief description of the regularity conditions at $\scri^+$ required by the equations has been included in this work. 
They are especially important in a non-staggered implementation, because they have to be explicitly satisfied at $\scri^+$ so that the equations attain regular limits there. 
However, a more general study of of the regularity conditions in a [2+1]+1 decomposition will allow to understand much better the treatment required by $\scri^+$ and prepare the path for more complicated setups other than spherical symmetry. Work in this direction has already started. %at the level of the tensorial equations 

%important in the non-staggered case, which is required to impose boundary conditions

\subsubsection{Implementation into a three-dimensional code}

%The aim of this work is to contribute to the current efforts of modeling GW signals. As already explained in detail, the hyperboloidal initial value problem has considerable advantages over other approaches commonly used nowadays and a successful implementation of a hyperboloidal code would be very beneficial for the field. 

A three-dimensional implementation of the setup used in this work using the Einstein Toolkit framework \cite{einsteintoolkit} is one of the next steps planned. This will take advantage of all the knowledge about the problem obtained so far and will require a spherical boundary in the numerical setup. This can be implemented using the multipatch framework in the \texttt{Llama} Code \cite{llama}. Another possible option is to implement it using three-dimensional spherical polar coordinates \cite{Baumgarte:2012xy}, which allow for a clear separation of the radial direction that will hopefully simplify the regularizations at $\scri^+$. The coordinate singularities (not present in a Cartesian grid) can however pose difficulties. 
The development of more complicated hyperboloidal-based codes will require suitable initial data, which means that previous work \cite{Buchman:2009ew} on the hyperboloidal elliptic equations will be continued. %This means continuing with previous work on the hyperboloidal elliptic equations and developing new techniques. 

\subsubsection{Simulations in Anti-deSitter spacetimes} 

In the case of a negative cosmological constant $\Lambda$, $\scri^+$ is not a null surface but a timelike one and appropriate boundary conditions at $\scri^+$ have to be prescribed (this is one of the reasons for using a non-staggered grid). The required boundary conditions have to be reflecting-like and current numerical results in four-dimensional AdS of a constrained evolution of the Einstein equations coupled to a scalar field in spherical symmetry \cite{Bizon:2011gg} indicate that AdS is non-linearly unstable. 
Evolving AdS with the procedures presented in this thesis and appropriate boundary conditions in spherical symmetry could complement the stability results of AdS obtained so far and maybe even provide new results.

\appendix

\chapter{Construction of Penrose diagrams}\label{c:diagr}

Here I will show the explicit expressions used to create the Carter-Penrose diagrams presented in chapters \ref{c:introduction} and \ref{c:initial}. 

\section{Kruskal-Szekeres-like coordinates} % for the black hole spacetimes

Omitting angular dimensions, we consider a line element of the form
\begin{equation}\label{ed:}
d\tilde s^2 = -A(\tilde r)d\tilde t^2+\frac{1}{A(\tilde r)}d\tilde r^2 .
\end{equation}

First we eliminate the coordinate singularity at the horizon introducing a tortoise coordinate, in terms of which the line element takes the form
\begin{equation}
d\tilde s^2 = A(\tilde r)\left(-d\tilde t^2+d\rtort^2\right) , 
\end{equation}
and thus is related to the original radial coordinate $\tilde r$ as 
\begin{equation}
d\rtort = \frac{d\tilde r}{A(\tilde r)} . 
\end{equation}
The integrated expression of $\rtort$ depends explicitly on the form of $A(\tilde r)$ and includes an integration constant that will be set to convenience for each case. 

A transformation to the null coordinates $\tilde u$ and $\tilde v$ is performed
\begin{equation}\label{ed:uvtr}
\tilde u=\tilde t-\rtort , \qquad \tilde v=\tilde t+\rtort , 
\end{equation}
so that the line element reads 
\begin{equation}
d\tilde s^2 = -A(\tilde r)\,d\tilde u \,d\tilde v . 
\end{equation}
The quantity $A(\tilde r)$ is to be expressed in terms of $\tilde u$ and $\tilde v$. Its exact expression will depend on the specific form of $\rtort(\tilde r)$ and will be explicitly given in the following sections. 

It is convenient to perform a coordinate transformation on the null coordinates $\tilde u$ and $\tilde v$ that leaves the null cone structure invariant. This transformation will determine the precise form of the metric and can thus serve to simplify it. The transformations that will be considered here are of the form
\begin{equation}\label{ed:UVuv}
\tilde U=-e^{-\frac{\tilde u}{B}} , \qquad \tilde V=e^{\frac{\tilde v}{B}}  , 
\end{equation}
where $B$ is some given function of the parameters specific to each case. The choice $B=4M$ in the Schwarzschild spacetime gives the Kruskal-Szekeres coordinates. 

The following change will allow us to express the line element in a form similar to Minkowski spacetime: 
\begin{equation}\label{ed:TRUV}
\tilde T=\frac{1}{2}\left(\tilde V+\tilde U\right) , \qquad \tilde R=\frac{1}{2}\left(\tilde V-\tilde U\right) . 
\end{equation}
In terms of the original $\tilde t$ and $\tilde r$ coordinates, their expressions reduce to
\begin{equation}\label{ed:TR}
\tilde T=e^{\frac{\rtort}{B}}\sinh\left(\frac{\tilde t }{B}\right) , \qquad \tilde R=e^{\frac{\rtort}{B}}\cosh\left(\frac{\tilde t }{B}\right) , 
\end{equation}
with $\rtort$ expressed in terms of $\tilde r$. 

The final expressions of $\tilde T$ and $\tilde R$ are compactified in the same way as done with the Minkowski spacetime in subsection \ref{compMin}, i.e. using \eref{ei:uvtr} (with $\tilde T$ and $\tilde R$ instead of $\tilde t$ and $\tilde r$), \eref{ei:UVuv} and \eref{ei:TRUV}. To obtain the Carter-Penrose diagrams, $R$ is plotted in the horizontal axis and $T$ in the vertical one. % in my Mathematica notebook this transformation includes an extra division by 2, so all ranges in my notebook are half of those listed here. 

For some values of the radial coordinate - inside of the horizon $\tilde r<2M$ for the Schwarzschild case and among horizons $\tilde r_-<\tilde r<\tilde r_+$ for the non-extreme RN one - the radius $\tilde r$ becomes the timelike coordinate and the time $\tilde t$ turns into the spatial one, because $A(\tilde r)<0$ there. %In these ranges of $\tilde r$ the coordinate changes in \eref{ed:uvtr} and \eref{ed:TR} have to be substituted respectively with \begin{eqnarray}\label{ed:channeg}\tilde u=-\tilde t+\rtort , &\qquad& \tilde v=\tilde t+\rtort , \\\tilde T=\frac{1}{2}\left(\tilde V-\tilde U\right) , &\qquad& \tilde R=\frac{1}{2}\left(\tilde V+\tilde U\right) \end{eqnarray}
In these ranges of $\tilde r$, the sign of $\tilde u$ and $\tilde U$ in the coordinate transformations \eref{ed:uvtr}, \eref{ed:UVuv} and \eref{ed:TRUV} is the opposite one. However, the resulting $\tilde T$  and $\tilde R$ yield exactly the same expression as in \eref{ed:TR}. 
As $\tilde T$ is spacelike and $\tilde R$ is timelike, their expressions in this case have to be interchanged before applying the compactification procedure of subsection \ref{compMin}. 

\section{Schwarzschild spacetime} \label{cap:Schw}

The line element of the Schwarzschild spacetime is 
\begin{equation}\label{ed:schwle}
d\tilde s^2 = -\left(1-\frac{2M}{\tilde r}\right)d\tilde t^2+\left(1-\frac{2M}{\tilde r}\right)^{-1}d\tilde r^2 = \left(1-\frac{2M}{\tilde r}\right)\left(-d\tilde t^2+d\rtort^2\right) , 
\end{equation}
and the relation between the Schwarzschild radius $\tilde r$ and the corresponding tortoise coordinate $\rtort$ is
\begin{equation}\label{ed:rtortSchw}
\rtort = \tilde r + 2M\ln (\tilde r-2M) + C_{Schw} , 
\end{equation}
where $C_{Schw}$ is an integration constant that will be set to a convenient value. 
The substitution of $A(\tilde r) = 1-\frac{2M}{\tilde r}$ is performed as follows
\begin{eqnarray*}
\rtort = \frac{1}{2}\left(\tilde v-\tilde u\right) = \tilde r + 2M\ln (\tilde r-2M) +C_{Schw} , \\
\frac{1}{2M}\left[\frac{1}{2}\left(\tilde v-\tilde u\right)-\tilde r - C_{Schw} \right] = \ln \left(\tilde r-2M\right) , \\ 
e^{\frac{\tilde v-\tilde u}{4M}} e^{-\frac{\tilde r+C_{Schw}}{2M}}  = \tilde r-2M = \tilde r\left(1-\frac{2M}{\tilde r}\right), 
\end{eqnarray*}
so that 
\begin{equation}
d\tilde s^2 = -\left(1-\frac{2M}{\tilde r}\right)\,d\tilde u \,d\tilde v =  - \frac{1}{\tilde r} e^{-\frac{\tilde r}{2M}}e^{-\frac{C_{Schw}}{2M}} e^{-\frac{\tilde u}{4M}}d\tilde u\,e^{\frac{\tilde v}{4M}}d\tilde v . 
\end{equation}
Using the transformation \eref{ed:UVuv} with $B=4M$, the previous line element becomes
\begin{equation}
d\tilde s^2 =  - \frac{(4M)^2}{\tilde r} e^{-\frac{\tilde r}{2M}}e^{-\frac{C_{Schw}}{2M}} d\tilde U\,d\tilde V . 
\end{equation}
The final form of \eref{ed:TR} is obtained by calculating $e^{\frac{\rtort}{4M}}$ substituting \eref{ed:rtortSchw} and using $C_{Schw}=-2M\ln D_{Schw}$: 
\begin{equation}
e^{\frac{\rtort}{4M}} = e^{\frac{1}{4M}\left[\tilde r + 2M\ln (\tilde r-2M) -2M\ln D_{Schw}\right]} = e^{\frac{\tilde r}{4M}}\sqrt{\frac{r-2M}{D_{Schw}}} . 
\end{equation}
For $\tilde r>2M$, the appropriate choice is $D_{Schw}=2M$ and for $\tilde r<2M$, $D_{Schw}=-2M$ gives a real value. 

The final expressions to be compactified in the Schwarzschild case are the following. Note that the interchange of $\tilde T$ and $\tilde R$ inside of the BH's horizon has already been performed in \eref{ed:TRSchwin}: 
\begin{subequations}
\begin{itemize}
\item[$\tilde r>2M$] 
\begin{equation}\label{ed:TRSchwout}
\tilde T=\sqrt{\frac{\tilde r}{2M}-1}\ e^{\frac{\tilde r}{4M}}\sinh\left(\frac{\tilde t }{4M}\right) , \qquad 
\tilde R=\sqrt{\frac{\tilde r}{2M}-1}\ e^{\frac{\tilde r}{4M}}\cosh\left(\frac{\tilde t }{4M}\right) , 
\end{equation}
\item[$\tilde r<2M$]
\begin{equation}\label{ed:TRSchwin}
\tilde T=\sqrt{1-\frac{\tilde r}{2M}}\ e^{\frac{\tilde r}{4M}}\cosh\left(\frac{\tilde t }{4M}\right) , \qquad 
\tilde R=\sqrt{1-\frac{\tilde r}{2M}}\ e^{\frac{\tilde r}{4M}}\sinh\left(\frac{\tilde t }{4M}\right) . 
\end{equation}
\end{itemize}
\end{subequations}

The ranges of the compactified coordinates to include the outer spacetime (zone I in \fref{fi:Schw}) and the interior of the horizon (zone II) are $T\in[-\case{\pi}{2},\case{\pi}{2}]$ and $R\in[-\case{\pi}{2},\pi]$. The symmetric outer spacetime (zone IV) and the white hole (zone III) are obtained by compactifying \eref{ed:TRSchwout} and \eref{ed:TRSchwin} with a minus sign and plotting $T$ in the previous range and $R\in[-\pi,\case{\pi}{2}]$.

%\begin{eqnarray*} \rtort = \frac{1}{2}\left(\tilde v-\tilde u\right) = \tilde r + 2M\ln (r-2M) +C_{Schw} , \\ \frac{1}{2M}\left[\frac{1}{2}\left(\tilde v-\tilde u\right)-\tilde r  \right] = \ln \left(\frac{r-2M}{D_{Schw}}\right) , \\  e^{\frac{\tilde v-\tilde u}{4M}} e^{-\frac{\tilde r}{2M}}  = \frac{r-2M}{D_{Schw}} \end{eqnarray*}

\section{Reissner-Nordstr\"om spacetime}\label{cap:RN}
%follow \cite{Hawking:1973uf}

In the non-extreme RN case the line element is given by
\begin{equation}\label{ed:RNle}
d\tilde s^2 = -\left(1-\frac{2M}{\tilde r}+\frac{Q^2}{\tilde r^2}\right)d\tilde t^2+\left(1-\frac{2M}{\tilde r}+\frac{Q^2}{\tilde r^2}\right)^{-1}d\tilde r^2 = \left(1-\frac{2M}{\tilde r}+\frac{Q^2}{\tilde r^2}\right)\left(-d\tilde t^2+d\rtort^2\right), 
\end{equation}
and the integrated expression of the tortoise coordinate takes the form
\begin{equation}\label{ed:rtortRN}
\rtort = \tilde r + M\ln (\tilde r^2-2M\tilde r+Q^2) + \frac{2M^2-Q^2}{2\sqrt{M^2-Q^2}}\left[\ln\left(\tilde r-\tilde r_+\right)-\ln\left(\tilde r-\tilde r_-\right)\right] + C_{RN} , 
\end{equation}
where $\tilde r_\pm=M\pm\sqrt{M^2-Q^2}$, or equivalently as presented in \cite{Hawking:1973uf} %\upda{check constant is the same}
\begin{equation}\label{ed:rtortRN2}
\rtort = \tilde r + \frac{\tilde r_+^2}{\tilde r_+-\tilde r_-}\ln (\tilde r-\tilde r_+) - \frac{\tilde r_-^2}{\tilde r_+-\tilde r_-}\ln (\tilde r-\tilde r_-) + C_{RN} . 
\end{equation}

As there are two logarithmic terms in \eref{ed:rtortRN2}, expressing $A(\tilde r)=1-\case{2M}{\tilde r}+\case{Q^2}{\tilde r^2}$ can be done in two ways depending on the one which gets the positive sign from the following expression 
\begin{equation*}
\left[\frac{1}{2}\left(\tilde v-\tilde u\right)-\tilde r - C_{RN} \right](\tilde r_+-\tilde r_-) = \tilde r_+^2\ln \left(\tilde r-\tilde r_+\right) - \tilde r_-^2\ln \left(\tilde r-\tilde r_-\right). 
\end{equation*}
Taking the first logarithmic term positive and isolating gives a line element
\begin{equation}
d\tilde s^2 %= -\left(1-\frac{2M}{\tilde r}+\frac{Q^2}{\tilde r^2}\right)\,d\tilde u \,d\tilde v 
=  - \frac{1}{\tilde r^2} \frac{(\tilde r-\tilde r_-)^{\tilde r_-^2+1}}{(\tilde r-\tilde r_+)^{\tilde r_+^2-1}} e^{-\frac{(\tilde r_+-\tilde r_-)}{2}(\tilde r+C_{RN})} e^{-\frac{(\tilde r_+-\tilde r_-)}{2}\tilde u}d\tilde u\,e^{\frac{(\tilde r_+-\tilde r_-)}{2}\tilde v}d\tilde v . 
\end{equation}
Now the transformation \eref{ed:UVuv} with $B=\case{2}{r_+-r_-}$ can be performed and the line element becomes
\begin{equation}
d\tilde s^2=  - \left(\frac{2}{(\tilde r_+-\tilde r_-)\tilde r}\right)^2 \frac{(\tilde r-\tilde r_-)^{\tilde r_-^2+1}}{(\tilde r-\tilde r_+)^{\tilde r_+^2-1}} e^{-\frac{(\tilde r_+-\tilde r_-)}{2}(\tilde r+C_{RN})} d\tilde U\,d\tilde V . 
\end{equation}
The term $e^{\frac{(\tilde r_+-\tilde r_-)}{2}\rtort}$ that appears in the corresponding expressions to \eref{ed:TR} is expressed in terms of $\tilde r$, with the substitution $C_{RN}=-\case{\tilde r_+^2}{\tilde r_+-\tilde r_-}\ln D_{RN}+\case{\tilde r_-^2}{\tilde r_+-\tilde r_-}\ln E_{RN}$: 
\begin{equation}
e^{\frac{(\tilde r_+-\tilde r_-)}{2}\rtort}=e^{\frac{(\tilde r_+-\tilde r_-)}{2}\tilde r}\sqrt{\frac{\tilde r-\tilde r_+}{D_{RN}}}^{\,\tilde r_+^2}\sqrt{\frac{\tilde r-\tilde r_-}{E_{RN}}}^{\,-\tilde r_-^2} . 
\end{equation}
The appropriate choices for the constants $D_{RN}$ and $E_{RN}$ such that the expression will be real are: if $\tilde r>\tilde r_+\to D_{RN}=\tilde r_+$ and if $\tilde r<\tilde r_+\to D_{RN}=-\tilde r_+$; equivalently $\tilde r>\tilde r_-\to E_{RN}=\tilde r_-$ and if $\tilde r<\tilde r_-\to E_{RN}=-\tilde r_-$. 

Choosing the positive sign for the second logarithmic term will express the line element as 
\begin{equation}
d\tilde s^2 =  - \frac{1}{\tilde r^2} \frac{(\tilde r-\tilde r_+)^{\tilde r_+^2+1}}{(\tilde r-\tilde r_-)^{\tilde r_-^2-1}} e^{+\frac{(\tilde r_+-\tilde r_-)}{2}(\tilde r+C_{RN})} e^{\frac{(\tilde r_+-\tilde r_-)}{2}\tilde u}d\tilde u\,e^{-\frac{(\tilde r_+-\tilde r_-)}{2}\tilde v}d\tilde v . 
\end{equation}
This suggests to, instead of using \eref{ed:UVuv} as it is, interchange the roles of $\tilde U$ and $\tilde V$, so using
\begin{equation}
\tilde U=e^{\frac{(\tilde r_+-\tilde r_-)}{2}\tilde u} \qquad \textrm{and} \qquad \tilde V=-e^{-\frac{(\tilde r_+-\tilde r_-)}{2}\tilde v}  , 
\end{equation}
now gives according to \eref{ed:TRUV}
\begin{equation}
\tilde T=e^{-\frac{(\tilde r_+-\tilde r_-)}{2}\rtort}\sinh\left(\frac{(\tilde r_+-\tilde r_-)}{2}\tilde t\right) , \qquad \tilde R= - e^{-\frac{(\tilde r_+-\tilde r_-)}{2}\rtort}\cosh\left(\frac{(\tilde r_+-\tilde r_-)}{2}\tilde t\right) , 
\end{equation}
with 
\begin{equation}
e^{-\frac{(\tilde r_+-\tilde r_-)}{2}\rtort}=e^{-\frac{(\tilde r_+-\tilde r_-)}{2}\tilde r}\sqrt{\frac{\tilde r-\tilde r_+}{D_{RN}}}^{\,-\tilde r_+^2}\sqrt{\frac{\tilde r-\tilde r_-}{E_{RN}}}^{\,\tilde r_-^2} 
\end{equation}
and the same choices for $D_{RN}$ and $E_{RN}$ described before. 

The final expressions used in the construction of the RN Penrose diagrams are given by the following cases. The interchange between spatial and timelike coordinates in the region between the horizons has already been performed. The choice of the first positive logarithmic term has been made for the first two sets of expressions, while the other choice was made for the remaining two sets: 
\begin{subequations}
\begin{itemize}
\item[$\tilde r>\tilde r_+$] 
\begin{eqnarray}%\label{ed:TRRNout}
\tilde T&=&\sqrt{\frac{\tilde r}{\tilde r_+}-1}^{\,\tilde r_+^2}\sqrt{\frac{\tilde r}{\tilde r_-}-1}^{\,-\tilde r_-^2}e^{\frac{(\tilde r_+-\tilde r_-)}{2}\tilde r}\sinh\left(\frac{(\tilde r_+-\tilde r_-)}{2}\tilde t\right) , \\
\tilde R&=&\sqrt{\frac{\tilde r}{\tilde r_+}-1}^{\,\tilde r_+^2}\sqrt{\frac{\tilde r}{\tilde r_-}-1}^{\,-\tilde r_-^2}e^{\frac{(\tilde r_+-\tilde r_-)}{2}\tilde r}\cosh\left(\frac{(\tilde r_+-\tilde r_-)}{2}\tilde t\right) , 
\end{eqnarray}
\item[$\tilde r_-<\tilde r<\tilde r_+$] 
\begin{eqnarray}%\label{ed:TRRNout}
\tilde T&=&\sqrt{1-\frac{\tilde r}{\tilde r_+}}^{\,\tilde r_+^2}\sqrt{\frac{\tilde r}{\tilde r_-}-1}^{\,-\tilde r_-^2}e^{\frac{(\tilde r_+-\tilde r_-)}{2}\tilde r}\cosh\left(\frac{(\tilde r_+-\tilde r_-)}{2}\tilde t\right) , \\
\tilde R&=&\sqrt{1-\frac{\tilde r}{\tilde r_+}}^{\,\tilde r_+^2}\sqrt{\frac{\tilde r}{\tilde r_-}-1}^{\,-\tilde r_-^2}e^{\frac{(\tilde r_+-\tilde r_-)}{2}\tilde r}\sinh\left(\frac{(\tilde r_+-\tilde r_-)}{2}\tilde t\right) , 
\end{eqnarray}
or
\begin{eqnarray}%\label{ed:TRRNout}
\tilde T&=&-\sqrt{\frac{\tilde r}{\tilde r_-}-1}^{\,\tilde r_-^2}\sqrt{1-\frac{\tilde r}{\tilde r_+}}^{\,-\tilde r_+^2}e^{-\frac{(\tilde r_+-\tilde r_-)}{2}\tilde r}\cosh\left(\frac{(\tilde r_+-\tilde r_-)}{2}\tilde t\right) , \label{ed:TRNmi} \\
\tilde R&=&\sqrt{\frac{\tilde r}{\tilde r_-}-1}^{\,\tilde r_-^2}\sqrt{1-\frac{\tilde r}{\tilde r_+}}^{\,-\tilde r_+^2}e^{-\frac{(\tilde r_+-\tilde r_-)}{2}\tilde r}\sinh\left(\frac{(\tilde r_+-\tilde r_-)}{2}\tilde t\right) ,  \label{ed:RRNmi}
\end{eqnarray}
\item[$\tilde r<\tilde r_-$]
\begin{eqnarray}%\label{ed:TRRNout}
\tilde T&=&\sqrt{1-\frac{\tilde r}{\tilde r_-}}^{\,\tilde r_-^2}\sqrt{1-\frac{\tilde r}{\tilde r_+}}^{\,-\tilde r_+^2}e^{-\frac{(\tilde r_+-\tilde r_-)}{2}\tilde r}\sinh\left(\frac{(\tilde r_+-\tilde r_-)}{2}\tilde t\right) , \label{ed:TRNi} \\
\tilde R&=&-\sqrt{1-\frac{\tilde r}{\tilde r_-}}^{\,\tilde r_-^2}\sqrt{1-\frac{\tilde r}{\tilde r_+}}^{\,-\tilde r_+^2}e^{-\frac{(\tilde r_+-\tilde r_-)}{2}\tilde r}\cosh\left(\frac{(\tilde r_+-\tilde r_-)}{2}\tilde t\right) .  \label{ed:RRNi}
\end{eqnarray}
\end{itemize}
\end{subequations}

When plotting the compactified coordinates of the Penrose diagram, in order to obtain the correct location of the zones, a $\pi$ term has to be added to the compactified $T$ calculated from \eref{ed:TRNmi} and \eref{ed:TRNi} (and also respectively mixed with \eref{ed:RRNmi} and \eref{ed:RRNi}), so that the implicitly plotted quantities are $(R,T+\pi)$. The range of $R$ in the Penrose diagrams is the same as in the Schwarzschild case, but for a $T$ covering the complete spacetime like in \fref{fin:RNCcompl} we have $T\in[-\case{3\pi}{2},\case{3\pi}{2}]$.

\section{Extreme Reissner-Nordstr\"om spacetime}\label{cap:eRN}

In the case $Q=M$, the RN line element reduces to
\begin{equation}\label{ed:RNle}
d\tilde s^2 = -\left(1-\frac{M}{\tilde r}\right)^2d\tilde t^2+\left(1-\frac{M}{\tilde r}\right)^{-2}d\tilde r^2 =\left(1-\frac{M}{\tilde r}\right)^2\left(-d\tilde t^2+d\rtort^2\right) , 
\end{equation}
and the corresponding tortoise coordinate is given by
\begin{equation}\label{ed:rtorteRN}
\rtort = \tilde r + M\ln (\tilde r-M)^2 -\frac{M^2}{\tilde r-M} + C_{eRN} . 
\end{equation}
The calculation to express $A(\tilde r) = \left(1-\case{M}{\tilde r}\right)^2$ is now
\begin{eqnarray*}
%\rtort = \frac{1}{2}\left(\tilde v-\tilde u\right) = \tilde r + M\ln (\tilde r-M)^2 -\frac{M^2}{\tilde r-M} + C_{eRN} , \\
\frac{1}{M}\left[\frac{1}{2}\left(\tilde v-\tilde u\right)-\tilde r +  \frac{M^2}{\tilde r-M} - C_{eRN} \right] = \ln \left(\tilde r-M\right)^2 , 
%\\ e^{\frac{\tilde v-\tilde u}{2M}} e^{-\frac{\tilde r- \frac{M^2}{\tilde r-M}+C_{eRN}}{2M}}  = \left(\tilde r-M\right)^2 = \tilde r^2\left(1-\frac{M}{\tilde r}\right)^2, 
\end{eqnarray*}
and using the transformation \eref{ed:UVuv} with $B=2M$, the line element becomes
\begin{equation}
d\tilde s^2 =  - \frac{(2M)^2}{\tilde r^2} e^{-\frac{\tilde r-\frac{M^2}{\tilde r-M}}{M}}e^{-\frac{C_{eRN}}{M}} d\tilde U\,d\tilde V . 
\end{equation}

To obtain the final form of \eref{ed:TR} we substitute \eref{ed:rtorteRN} and $C_{eRN}=-M\ln D_{eRN}-\case{M}{2}$ (the last term is included to make the singularity $\tilde r=0$ coincide with the vertical line at $R=0$) to calculate 
\begin{equation}
e^{\frac{\rtort}{2M}} = e^{\frac{1}{2M}\left(\tilde r-\frac{M^2}{\tilde r-M}-M\right)}\frac{r-M}{D_{eRN}} , 
\end{equation}
with the choices $\tilde r>M\to D_{eRN}=M$ and $\tilde r<M\to D_{eRN}=-M$. 
The expressions used for creating the extreme RN diagrams are: 
\begin{subequations}
\begin{itemize}
\item[$\tilde r>M$] 
\begin{eqnarray}\label{ed:TReRNout}
\tilde T&=& \left(\frac{\tilde r}{M}-1\right) e^{\frac{1}{2M}\left(\tilde r-\frac{M^2}{\tilde r-M}-M\right)}\sinh\left(\frac{\tilde t }{2M}\right) , \\
\tilde R&=& \left(\frac{\tilde r}{M}-1\right) e^{\frac{1}{2M}\left(\tilde r-\frac{M^2}{\tilde r-M}-M\right)}\cosh\left(\frac{\tilde t }{2M}\right) , 
\end{eqnarray}
\item[$\tilde r<M$]
\begin{eqnarray}\label{ed:TReRNin}
\tilde T&=& \left(1-\frac{\tilde r}{M}\right) e^{\frac{1}{2M}\left(\tilde r-\frac{M^2}{\tilde r-M}-M\right)}\sinh\left(\frac{\tilde t }{2M}\right) , \\
\tilde R&=& \left(1-\frac{\tilde r}{M}\right) e^{\frac{1}{2M}\left(\tilde r-\frac{M^2}{\tilde r-M}-M\right)}\cosh\left(\frac{\tilde t }{2M}\right) . 
\end{eqnarray}
\end{itemize}
\end{subequations}
Outside of the horizon the compactified quantities to plot are simply $T$ versus $R$, but in the interior $\tilde r<M$,  $R-\case{\pi}{2}$ is plotted on the horizontal axis and $\case{\pi}{2}-T$ on the vertical one. The range used in the diagrams in \fref{fin:eRNK} in $T\in[-\case{\pi}{2},\pi]$ and $R\in[0,\pi]$. 

%\upda{mention add pi when plotting diagram and -1/2 in constant for appropriate location of the singularity}

% if non-critical $\Cc$, the height function changes infinitely at $\tilde r = M$. 

\section{Numerical calculation of the height function for black hole spacetimes}
%\subsection{Critical $\Cc$}%\subsection{Non-critical $\Cc$}

The height function, whose derivative is given in its general form by \eref{ein:heightp}, can be integrated numerically (once the values of the parameters $M$, $Q$, $\Kc$ and $\Cc$ have been set) and used to display hyperboloidal foliations in the Penrose diagrams in chapter \ref{c:initial}. 
This has been done using the {\tt NIntegrate} function of {\tt Mathematica}.

The only critical part of the calculation is the treatment of the divergences of the height function: it becomes infinite at the horizon(s) of the BHs ($\tilde r_\pm=M\pm\sqrt{M^2-Q^2}$, the roots of $A(\tilde r)$) and at the roots of the expression in the square root ($R_0$ for the critical $\Cc$ value and $R_1$ and $R_2$ for a value smaller than the critical one, except in the extreme RN case, where no $R_1$ and $R_2$ exist). 
As the integration is only performed for visualization purposes, I have not tried to find a more sophisticated way of integrating that avoids the divergences, but have simply set the integration limits such that they come quite close to the divergence from both sides. How close the limits have to be is given by how continuous the slices look at the horizons and $R_i$ in the diagrams. 
As an example, let us consider diagram b) in \fref{fin:RNvarC}: the height function is integrated in the parts $(0,\tilde r_-)$, $(\tilde r_-,R_1)$, $(R_2,r_+)$, $(\tilde r_+,\tilde r_\textrm{large})$, where $\tilde r_\textrm{large}\approx200$ and the parenthesis indicate that the given value of the radial coordinate is not reached. The height function is not integrated in the region $(R_1,R_2)$ because it is imaginary there. The distance between the closest $\tilde r$ and the divergent points is of the order of $10^{-5}-10^{-8}$ in most cases. 

The numerical points $(\tilde r_i,h_i)$ are interpolated into a function $h(\tilde r)$ and introduced into the expressions of $R(\tilde t, \tilde r)$ and $T(\tilde t, \tilde r)$ \eref{ei:TRUV} using the substitution 
\begin{equation}
\tilde t= t+h(\tilde r) . 
\end{equation}
The result is a hyperboloidal foliation that depends on the coordinates $\tilde r$ and $t$. 
The hyperboloidal slices in the Penrose diagrams are obtained by plotting implicitly $T$ in the vertical axis and $R$ in the horizontal one, in terms of $\tilde r$ (in the parts where the height function has been integrated) and for several fixed values of $t$.

\cleardoublepage
\phantomsection
\addcontentsline{toc}{chapter}{Bibliography}
\bibliographystyle{tocplain}
%%\bibliographystyle{../../master/thesis/tocunsrt}
%%\bibliography{references}
\bibliography{../articles/hypcomp}

% Define empty bibhead if not already defined
\providecommand{\bibhead}[1]{}
% Define tocrefpdfbookmark if not already defined
\expandafter\ifx\csname pdfbookmark\endcsname\relax%
  \providecommand{\tocrefpdfbookmark}{}
\else\providecommand{\tocrefpdfbookmark}{%
   \hypertarget{tocreferences}{}%
   \pdfbookmark[1]{References}{tocreferences}}%
\fi

\tocrefpdfbookmark
\begin{thebibliography}{100}

\bibitem{einsteintoolkit}\bibhead{einsteintoolkit}
{Einstein Toolkit}.
\newblock {\tt http://einsteintoolkit.org/}.

\bibitem{llama}\bibhead{llama}
{Llama Code}.
\newblock {\tt http://llamacode.bitbucket.org/}.

\bibitem{spec}\bibhead{spec}
{Spectral Einstein Code ({\tt SpEC})}.
\newblock {\tt https://www.black-holes.org/SpEC.html}.

\bibitem{TheLIGOScientific:2014jea}\bibhead{TheLIGOScientific:2014jea}
J.~Aasi et~al.: {Advanced LIGO}.
\newblock {\em Class.Quant.Grav.}, 32:074001, 2015.
\newblock [\epfmtdoi{10.1088/0264-9381/32/7/074001}, \epfmt{arxiv}{1411.4547}].

\bibitem{Abbott:2007kv}\bibhead{Abbott:2007kv}
B.P. Abbott et~al.: {LIGO: The Laser interferometer gravitational-wave
  observatory}.
\newblock {\em Rept.Prog.Phys.}, 72:076901, 2009.
\newblock [\epfmtdoi{10.1088/0034-4885/72/7/076901}, \epfmt{arxiv}{0711.3041}].

\bibitem{Acernese:2008zzf}\bibhead{Acernese:2008zzf}
F.~Acernese, M.~Alshourbagy, P.~Amico, F.~Antonucci, S.~Aoudia, et~al.: {Virgo
  status}.
\newblock {\em Class.Quant.Grav.}, 25:184001, 2008.
\newblock [\epfmtdoi{10.1088/0264-9381/25/18/184001}].

\bibitem{TheVirgo:2014hva}\bibhead{TheVirgo:2014hva}
F.~Acernese et~al.: {Advanced Virgo: a second-generation interferometric
  gravitational wave detector}.
\newblock {\em Class.Quant.Grav.}, 32(2):024001, 2015.
\newblock [\epfmtdoi{10.1088/0264-9381/32/2/024001}, \epfmt{arxiv}{1408.3978}].

\bibitem{Alcubierre}\bibhead{Alcubierre}
Miguel Alcubierre: {\em {Introduction to 3+1 Numerical Relativity}}.
\newblock Oxford University Press, 2008.
\newblock [\epfmtdoi{10.1093/acprof:oso/9780199205677.001.0001}].

\bibitem{Alcubierre:2002kk}\bibhead{Alcubierre:2002kk}
Miguel Alcubierre, Bernd Bruegmann, Peter Diener, Michael Koppitz, Denis
  Pollney, et~al.: {Gauge conditions for long term numerical black hole
  evolutions without excision}.
\newblock {\em Phys.Rev.}, D67:084023, 2003.
\newblock [\epfmtdoi{10.1103/PhysRevD.67.084023},
  \epfmt{arxiv}{gr-qc/0206072}].

\bibitem{Alcubierre:2001vm}\bibhead{Alcubierre:2001vm}
Miguel Alcubierre, Bernd Bruegmann, Denis Pollney, Edward Seidel, and Ryoji
  Takahashi: {Black hole excision for dynamic black holes}.
\newblock {\em Phys.Rev.}, D64:061501, 2001.
\newblock [\epfmtdoi{10.1103/PhysRevD.64.061501},
  \epfmt{arxiv}{gr-qc/0104020}].

\bibitem{Alcubierre:2005gh}\bibhead{Alcubierre:2005gh}
Miguel Alcubierre, Alejandro Corichi, Jose~A. Gonzalez, Dario Nunez, Bernd
  Reimann, et~al.: {Generalized harmonic spatial coordinates and hyperbolic
  shift conditions}.
\newblock {\em Phys.Rev.}, D72:124018, 2005.
\newblock [\epfmtdoi{10.1103/PhysRevD.72.124018},
  \epfmt{arxiv}{gr-qc/0507007}].

\bibitem{Alcubierre:2010is}\bibhead{Alcubierre:2010is}
Miguel Alcubierre and Martha~D. Mendez: {Formulations of the 3+1 evolution
  equations in curvilinear coordinates}.
\newblock {\em Gen.Rel.Grav.}, 43:2769--2806, 2011.
\newblock [\epfmtdoi{10.1007/s10714-011-1202-x}, \epfmt{arxiv}{1010.4013}].

\bibitem{Alic:2011gg}\bibhead{Alic:2011gg}
Daniela Alic, Carles Bona-Casas, Carles Bona, Luciano Rezzolla, and Carlos
  Palenzuela: {Conformal and covariant formulation of the Z4 system with
  constraint-violation damping}.
\newblock {\em Phys.Rev.}, D85:064040, 2012.
\newblock [\epfmtdoi{10.1103/PhysRevD.85.064040}, \epfmt{arxiv}{1106.2254}].

\bibitem{Andersson:2002gn}\bibhead{Andersson:2002gn}
Lars Andersson: {Construction of hyperboloidal initial data}.
\newblock {\em Lect.Notes Phys.}, 604:183--194, 2002.
\newblock [\epfmt{arxiv}{gr-qc/0205083}].

\bibitem{Andersson:1992yk}\bibhead{Andersson:1992yk}
Lars Andersson, Piotr Chrusciel, and Helmut Friedrich: {On the Regularity of
  solutions to the Yamabe equation and the existence of smooth hyperboloidal
  initial data for Einsteins field equations}.
\newblock {\em Commun.Math.Phys.}, 149:587--612, 1992.
\newblock [\epfmtdoi{10.1007/BF02096944}].

\bibitem{Anninos:1995am}\bibhead{Anninos:1995am}
Peter Anninos, Karen Camarda, Joan Masso, Edward Seidel, Wai-Mo Suen, et~al.:
  {Three-dimensional numerical relativity: The Evolution of black holes}.
\newblock {\em Phys.Rev.}, D52:2059--2082, 1995.
\newblock [\epfmtdoi{10.1103/PhysRevD.52.2059}, \epfmt{arxiv}{gr-qc/9503025}].

\bibitem{Anninos:1994dj}\bibhead{Anninos:1994dj}
Peter Anninos, Greg Daues, Joan Masso, Edward Seidel, and Wai-Mo Suen: {Horizon
  boundary condition for black hole space-times}.
\newblock {\em Phys.Rev.}, D51:5562--5578, 1995.
\newblock [\epfmtdoi{10.1103/PhysRevD.51.5562}, \epfmt{arxiv}{gr-qc/9412069}].

\bibitem{Arbona:1999ym}\bibhead{Arbona:1999ym}
A.~Arbona, C.~Bona, J.~Mass\'o, and J.~Stela: {Robust evolution system for
  numerical relativity}.
\newblock {\em Phys.Rev.}, D60:104014, 1999.
\newblock [\epfmtdoi{10.1103/PhysRevD.60.104014},
  \epfmt{arxiv}{gr-qc/9902053}].

\bibitem{Arnowitt:1962hi}\bibhead{Arnowitt:1962hi}
Richard~L. Arnowitt, Stanley Deser, and Charles~W. Misner: {The Dynamics of
  general relativity}.
\newblock {\em Gen.Rel.Grav.}, 40:1997--2027, 2008.
\newblock [\epfmtdoi{10.1007/s10714-008-0661-1}, \epfmt{arxiv}{gr-qc/0405109}].

\bibitem{Ascher:1995:IEM}\bibhead{Ascher:1995:IEM}
Uri~M. Ascher, Steven~J. Ruuth, and Brian T.~R. Wetton: Implicit-explicit
  methods for time-dependent partial differential equations.
\newblock 32(3):797--823, June 1995.

\bibitem{Babiuc:2007vr}\bibhead{Babiuc:2007vr}
M.C. Babiuc, S.~Husa, D.~Alic, I.~Hinder, C.~Lechner, et~al.: {Implementation
  of standard testbeds for numerical relativity}.
\newblock {\em Class.Quant.Grav.}, 25:125012, 2008.
\newblock [\epfmtdoi{10.1088/0264-9381/25/12/125012},
  \epfmt{arxiv}{0709.3559}].

\bibitem{Baker:2006yw}\bibhead{Baker:2006yw}
John~G. Baker, Joan Centrella, Dae-Il Choi, Michael Koppitz, and James van
  Meter: {Binary black hole merger dynamics and waveforms}.
\newblock {\em Phys.Rev.}, D73:104002, 2006.
\newblock [\epfmtdoi{10.1103/PhysRevD.73.104002},
  \epfmt{arxiv}{gr-qc/0602026}].

\bibitem{Baker:2005vv}\bibhead{Baker:2005vv}
John~G. Baker, Joan Centrella, Dae-Il Choi, Michael Koppitz, and James van
  Meter: {Gravitational wave extraction from an inspiraling configuration of
  merging black holes}.
\newblock {\em Phys.Rev.Lett.}, 96:111102, 2006.
\newblock [\epfmtdoi{10.1103/PhysRevLett.96.111102},
  \epfmt{arxiv}{gr-qc/0511103}].

\bibitem{Bardeen:2011ip}\bibhead{Bardeen:2011ip}
James~M. Bardeen, Olivier Sarbach, and Luisa~T. Buchman: {Tetrad formalism for
  numerical relativity on conformally compactified constant mean curvature
  hypersurfaces}.
\newblock {\em Phys.Rev.}, D83:104045, 2011.
\newblock [\epfmtdoi{10.1103/PhysRevD.83.104045}, \epfmt{arxiv}{1101.5479}].

\bibitem{Baumgarte:2012xy}\bibhead{Baumgarte:2012xy}
Thomas~W. Baumgarte, Pedro~J. Montero, Isabel Cordero-Carrion, and Ewald
  Muller: {Numerical Relativity in Spherical Polar Coordinates: Evolution
  Calculations with the BSSN Formulation}.
\newblock {\em Phys.Rev.}, D87(4):044026, 2013.
\newblock [\epfmtdoi{10.1103/PhysRevD.87.044026}, \epfmt{arxiv}{1211.6632}].

\bibitem{Baumgarte:2007ht}\bibhead{Baumgarte:2007ht}
Thomas~W. Baumgarte and Stephen~G. Naculich: {Analytical representation of a
  black hole puncture solution}.
\newblock {\em Phys.Rev.}, D75:067502, 2007.
\newblock [\epfmtdoi{10.1103/PhysRevD.75.067502},
  \epfmt{arxiv}{gr-qc/0701037}].

\bibitem{Baumgarte:1998te}\bibhead{Baumgarte:1998te}
Thomas~W. Baumgarte and Stuart~L. Shapiro: {On the numerical integration of
  Einstein's field equations}.
\newblock {\em Phys.Rev.}, D59:024007, 1999.
\newblock [\epfmtdoi{10.1103/PhysRevD.59.024007},
  \epfmt{arxiv}{gr-qc/9810065}].

\bibitem{Bernuzzi:2009ex}\bibhead{Bernuzzi:2009ex}
Sebastiano Bernuzzi and David Hilditch: {Constraint violation in free evolution
  schemes: Comparing BSSNOK with a conformal decomposition of Z4}.
\newblock {\em Phys.Rev.}, D81:084003, 2010.
\newblock [\epfmtdoi{10.1103/PhysRevD.81.084003}, \epfmt{arxiv}{0912.2920}].

\bibitem{Beyer:2004sv}\bibhead{Beyer:2004sv}
Horst~R. Beyer and Olivier Sarbach: {On the well posedness of the
  Baumgarte-Shapiro-Shibata-Nakamura formulation of Einstein's field
  equations}.
\newblock {\em Phys.Rev.}, D70:104004, 2004.
\newblock [\epfmtdoi{10.1103/PhysRevD.70.104004},
  \epfmt{arxiv}{gr-qc/0406003}].

\bibitem{CBO9780511524639A010}\bibhead{CBO9780511524639A010}
N.~T. Bishop: Some aspects of the characteristic initial value problem in
  numerical relativity.
\newblock In Ray d'Inverno, editor, {\em Approaches to Numerical Relativity},
  pp. 20--33. Cambridge University Press, 1992.
\newblock Cambridge Books Online.

\bibitem{Bishop:1997ik}\bibhead{Bishop:1997ik}
Nigel~T. Bishop, Roberto Gomez, Luis Lehner, Manoj Maharaj, and Jeffrey
  Winicour: {High powered gravitational news}.
\newblock {\em Phys.Rev.}, D56:6298--6309, 1997.
\newblock [\epfmtdoi{10.1103/PhysRevD.56.6298}, \epfmt{arxiv}{gr-qc/9708065}].

\bibitem{Bishop:1998ukk}\bibhead{Bishop:1998ukk}
Nigel~T. Bishop, Roberto Gomez, Luis Lehner, Bela Szilagyi, Jeffrey Winicour,
  et~al.: {Cauchy characteristic matching}.
\newblock In Bala Iyer and Biplab Bhawal, editors, {\em {On the Black Hole
  Trail}}, pp. 383--408. Kluver, 1998.
\newblock [\epfmt{arxiv}{gr-qc/9801070}].

\bibitem{Bishop:1996gt}\bibhead{Bishop:1996gt}
Nigel~T. Bishop, Roberto Gomez, Luis Lehner, and Jeffrey Winicour: {Cauchy
  characteristic extraction in numerical relativity}.
\newblock {\em Phys.Rev.}, D54:6153--6165, 1996.
\newblock [\epfmtdoi{10.1103/PhysRevD.54.6153}].

\bibitem{Bizon:2011gg}\bibhead{Bizon:2011gg}
Piotr Bizoń and Andrzej Rostworowski: {On weakly turbulent instability of
  anti-de Sitter space}.
\newblock {\em Phys.Rev.Lett.}, 107:031102, 2011.
\newblock [\epfmtdoi{10.1103/PhysRevLett.107.031102},
  \epfmt{arxiv}{1104.3702}].

\bibitem{bona-2003-67}\bibhead{bona-2003-67}
C.~Bona, T.~Ledvinka, C.~Palenzuela, and M.~Zacek: {General-covariant evolution
  formalism for Numerical Relativity}.
\newblock {\em {Phys. Rev.}}, D67:104005, 2003.
\newblock [\epfmtdoi{10.1103/PhysRevD.67.104005}].

\bibitem{Bona:2003qn}\bibhead{Bona:2003qn}
C.~Bona, T.~Ledvinka, C.~Palenzuela, and M.~Zacek: {A symmetry-breaking
  mechanism for the Z4 general-covariant evolution system}.
\newblock {\em Phys. Rev.}, D69:064036, 2004.
\newblock [\epfmtdoi{10.1103/PhysRevD.69.064036},
  \epfmt{arxiv}{gr-qc/0307067}].

\bibitem{Bona:1992zz}\bibhead{Bona:1992zz}
Carles Bona and Joan Mass\'o: {Hyperbolic evolution system for numerical
  relativity}.
\newblock {\em Phys.Rev.Lett.}, 68:1097--1099, 1992.
\newblock [\epfmtdoi{10.1103/PhysRevLett.68.1097}].

\bibitem{Bona:1994dr}\bibhead{Bona:1994dr}
Carles Bona, Joan Mass\'o, Edward Seidel, and Joan Stela: {A New formalism for
  numerical relativity}.
\newblock {\em Phys.Rev.Lett.}, 75:600--603, 1995.
\newblock [\epfmtdoi{10.1103/PhysRevLett.75.600},
  \epfmt{arxiv}{gr-qc/9412071}].

\bibitem{PhysRevD.40.1022}\bibhead{PhysRevD.40.1022}
{Bona, Carles and Mass\'o, Joan}: {Einstein's evolution equations as a system
  of balance laws}.
\newblock {\em Phys. Rev. D}, 40:1022--1026, Aug 1989.
\newblock [\epfmtdoi{10.1103/PhysRevD.40.1022}].

\bibitem{bonmas}\bibhead{bonmas}
{Bona, Carles and Mass\'o Joan}: {Numerical relativity: evolving spacetime}.
\newblock {\em International Journal of Modern Physics C}, 04(04):883--907,
  1993.
\newblock [\epfmtdoi{10.1142/S0129183193000690}].

\bibitem{Bonazzola1999433}\bibhead{Bonazzola1999433}
S.~Bonazzola, E.~Gourgoulhon, and J.-A. Marck: Spectral methods in general
  relativistic astrophysics.
\newblock {\em Journal of Computational and Applied Mathematics}, 109(1-2):433
  -- 473, 1999.
\newblock [\epfmtdoi{10.1016/S0377-0427(99)00167-3}].

\bibitem{Bondi:1962px}\bibhead{Bondi:1962px}
H.~Bondi, M.G.J. van~der Burg, and A.W.K. Metzner: {Gravitational waves in
  general relativity. VII. Waves from axisymmetric isolated systems}.
\newblock {\em Proc.Roy.Soc.Lond.}, A269:21--52, 1962.
\newblock [\epfmtdoi{10.1098/rspa.1962.0161}].

\bibitem{Bonnor66}\bibhead{Bonnor66}
W.~B. Bonnor and M.~A. Rotenberg: {\em Proc.\ R.\ Soc. London}, A289:247, 1966.

\bibitem{Bowen:1980yu}\bibhead{Bowen:1980yu}
Jeffrey~M. Bowen and James~W. York, Jr.: {Time asymmetric initial data for
  black holes and black hole collisions}.
\newblock {\em Phys.Rev.}, D21:2047--2056, 1980.
\newblock [\epfmtdoi{10.1103/PhysRevD.21.2047}].

\bibitem{Boyle:2009vi}\bibhead{Boyle:2009vi}
Michael Boyle and Abdul~H. Mroue: {Extrapolating gravitational-wave data from
  numerical simulations}.
\newblock {\em Phys.Rev.}, D80:124045, 2009.
\newblock [\epfmtdoi{10.1103/PhysRevD.80.124045}, \epfmt{arxiv}{0905.3177}].

\bibitem{Brandt:1997tf}\bibhead{Brandt:1997tf}
Steven Brandt and Bernd Bruegmann: {A Simple construction of initial data for
  multiple black holes}.
\newblock {\em Phys.Rev.Lett.}, 78:3606--3609, 1997.
\newblock [\epfmtdoi{10.1103/PhysRevLett.78.3606},
  \epfmt{arxiv}{gr-qc/9703066}].

\bibitem{1963PhRv..131..471B}\bibhead{1963PhRv..131..471B}
D.~R. {Brill} and R.~W. {Lindquist}: {Interaction Energy in Geometrostatics}.
\newblock {\em Physical Review}, 131:471--476, July 1963.
\newblock [\epfmtdoi{10.1103/PhysRev.131.471}].

\bibitem{Brown:2005aq}\bibhead{Brown:2005aq}
J.~David Brown: {Conformal invariance and the conformal-traceless decomposition
  of the gravitational field}.
\newblock {\em Phys. Rev.}, D71:104011, 2005.
\newblock [\epfmtdoi{10.1103/PhysRevD.71.104011},
  \epfmt{arxiv}{gr-qc/0501092}].

\bibitem{Brown:2007nt}\bibhead{Brown:2007nt}
J.~David Brown: {BSSN in Spherical Symmetry}.
\newblock {\em Class. Quant. Grav.}, 25:205004, 2008.
\newblock [\epfmtdoi{10.1088/0264-9381/25/20/205004},
  \epfmt{arxiv}{0705.3845}].

\bibitem{Brown:2009dd}\bibhead{Brown:2009dd}
J.~David Brown: {Covariant formulations of BSSN and the standard gauge}.
\newblock {\em Phys. Rev.}, D79:104029, 2009.
\newblock [\epfmtdoi{10.1103/PhysRevD.79.104029}, \epfmt{arxiv}{0902.3652}].

\bibitem{Bruegmann:1997uc}\bibhead{Bruegmann:1997uc}
Bernd Bruegmann: {Binary black hole mergers in 3-d numerical relativity}.
\newblock {\em Int.J.Mod.Phys.}, D8:85, 1999.
\newblock [\epfmtdoi{10.1142/S0218271899000080}, \epfmt{arxiv}{gr-qc/9708035}].

\bibitem{Bruegmann:2006at}\bibhead{Bruegmann:2006at}
Bernd Bruegmann, Jose~A. Gonzalez, Mark Hannam, Sascha Husa, Ulrich Sperhake,
  et~al.: {Calibration of Moving Puncture Simulations}.
\newblock {\em Phys.Rev.}, D77:024027, 2008.
\newblock [\epfmtdoi{10.1103/PhysRevD.77.024027},
  \epfmt{arxiv}{gr-qc/0610128}].

\bibitem{Buchman:2009ew}\bibhead{Buchman:2009ew}
Luisa~T. Buchman, Harald~P. Pfeiffer, and James~M. Bardeen: {Black hole initial
  data on hyperboloidal slices}.
\newblock {\em Phys.Rev.}, D80:084024, 2009.
\newblock [\epfmtdoi{10.1103/PhysRevD.80.084024}, \epfmt{arxiv}{0907.3163}].

\bibitem{Calabrese:2005fp}\bibhead{Calabrese:2005fp}
Gioel Calabrese and Carsten Gundlach: {Discrete boundary treatment for the
  shifted wave equation}.
\newblock {\em Class.Quant.Grav.}, 23:S343--S368, 2006.
\newblock [\epfmtdoi{10.1088/0264-9381/23/16/S04},
  \epfmt{arxiv}{gr-qc/0509119}].

\bibitem{Calabrese:2005ft}\bibhead{Calabrese:2005ft}
Gioel Calabrese, Ian Hinder, and Sascha Husa: {Numerical stability for finite
  difference approximations of Einstein's equations}.
\newblock {\em J.Comput.Phys.}, 218:607--634, 2006.
\newblock [\epfmtdoi{10.1016/j.jcp.2006.02.027}, \epfmt{arxiv}{gr-qc/0503056}].

\bibitem{Campanelli:2005dd}\bibhead{Campanelli:2005dd}
Manuela Campanelli, C.O. Lousto, P.~Marronetti, and Y.~Zlochower: {Accurate
  evolutions of orbiting black-hole binaries without excision}.
\newblock {\em Phys.Rev.Lett.}, 96:111101, 2006.
\newblock [\epfmtdoi{10.1103/PhysRevLett.96.111101},
  \epfmt{arxiv}{gr-qc/0511048}].

\bibitem{Campanelli:2006gf}\bibhead{Campanelli:2006gf}
Manuela Campanelli, C.O. Lousto, and Y.~Zlochower: {The Last orbit of binary
  black holes}.
\newblock {\em Phys.Rev.}, D73:061501, 2006.
\newblock [\epfmtdoi{10.1103/PhysRevD.73.061501},
  \epfmt{arxiv}{gr-qc/0601091}].

\bibitem{Chirvasa:2008xx}\bibhead{Chirvasa:2008xx}
M.~Chirvasa and S.~Husa: {Finite Difference Methods for Second Order in Space,
  First Order in Time Hyperbolic Systems and the Linear Shifted Wave Equation
  as a Model Problem in Numerical Relativity}.
\newblock {\em {Journal of Computational Physics}}, 229(7):2675 -- 2696, 2010.
\newblock [\epfmtdoi{10.1016/j.jcp.2009.12.016}, \epfmt{arxiv}{0812.3752}].

\bibitem{Choptuik:1992jv}\bibhead{Choptuik:1992jv}
Matthew~W. Choptuik: {Universality and scaling in gravitational collapse of a
  massless scalar field}.
\newblock {\em Phys.Rev.Lett.}, 70:9--12, 1993.
\newblock [\epfmtdoi{10.1103/PhysRevLett.70.9}].

\bibitem{Cook:2000vr}\bibhead{Cook:2000vr}
Gregory~B. Cook: {Initial data for numerical relativity}.
\newblock {\em Living Rev.Rel.}, 3:5, 2000.
\newblock [\epfmt{arxiv}{gr-qc/0007085}].

\bibitem{Cook:1989fb}\bibhead{Cook:1989fb}
Gregory~B. Cook and James~W. York, Jr.: {Apparent Horizons for Boosted or
  Spinning Black Holes}.
\newblock {\em Phys.Rev.}, D41:1077, 1990.
\newblock [\epfmtdoi{10.1103/PhysRevD.41.1077}].

\bibitem{CorderoCarrion:2012ic}\bibhead{CorderoCarrion:2012ic}
Isabel Cordero-Carrion and Pablo Cerda-Duran: {Partially implicit Runge-Kutta
  methods for wave-like equations in spherical-type coordinates}.
\newblock 2012.
\newblock [\epfmt{arxiv}{1211.5930}].

\bibitem{Courant1928Uber}\bibhead{Courant1928Uber}
R.~Courant, K.~Friedrichs, and H.~Lewy: {\"{U}ber die partiellen
  Differenzengleichungen der mathematischen Physik}.
\newblock {\em Mathematische Annalen}, 100(1):32--74, December 1928.
\newblock [\epfmtdoi{10.1007/BF01448839}].

\bibitem{Dain:2002ee}\bibhead{Dain:2002ee}
Sergio Dain, Carlos~O. Lousto, and Ryoji Takahashi: {New conformally flat
  initial data for spinning black holes}.
\newblock {\em Phys.Rev.}, D65:104038, 2002.
\newblock [\epfmtdoi{10.1103/PhysRevD.65.104038},
  \epfmt{arxiv}{gr-qc/0201062}].

\bibitem{1918SPAW.......154E}\bibhead{1918SPAW.......154E}
Albert {Einstein}: {{\"U}ber Gravitationswellen}.
\newblock {\em Sitzungsberichte der K{\"o}niglich Preu{\ss}ischen Akademie der
  Wissenschaften (Berlin)}, pp. 154--167, 1918.

\bibitem{PhysRevD.7.2814}\bibhead{PhysRevD.7.2814}
Frank Estabrook, Hugo Wahlquist, Steven Christensen, Bryce DeWitt, Larry Smarr,
  and Elaine Tsiang: Maximally slicing a black hole.
\newblock {\em Phys. Rev. D}, 7:2814--2817, May 1973.
\newblock [\epfmtdoi{10.1103/PhysRevD.7.2814}].

\bibitem{Field:2010mn}\bibhead{Field:2010mn}
Scott~E. Field, Jan~S. Hesthaven, Stephen~R. Lau, and Abdul~H. Mroue:
  {Discontinuous Galerkin method for the spherically reduced BSSN system with
  second-order operators}.
\newblock {\em Phys.Rev.}, D82:104051, 2010.
\newblock [\epfmtdoi{10.1103/PhysRevD.82.104051}, \epfmt{arxiv}{1008.1820}].

\bibitem{Fornberg:1998:CWF}\bibhead{Fornberg:1998:CWF}
Bengt Fornberg: Calculation of weights in finite difference formulas.
\newblock 40(3):685--691, 1998.
\newblock [\epfmtdoi{http://dx.doi.org/10.1137/S0036144596322507}].

\bibitem{Frauendiener:1997zc}\bibhead{Frauendiener:1997zc}
J\"org Frauendiener: {Numerical treatment of the hyperboloidal initial value
  problem for the vacuum Einstein equations. 1. The Conformal field equations}.
\newblock {\em Phys.Rev.}, D58:064002, 1998.
\newblock [\epfmtdoi{10.1103/PhysRevD.58.064002},
  \epfmt{arxiv}{gr-qc/9712050}].

\bibitem{Frauendiener:1997ze}\bibhead{Frauendiener:1997ze}
J\"org Frauendiener: {Numerical treatment of the hyperboloidal initial value
  problem for the vacuum Einstein equations. 2. The Evolution equations}.
\newblock {\em Phys.Rev.}, D58:064003, 1998.
\newblock [\epfmtdoi{10.1103/PhysRevD.58.064003},
  \epfmt{arxiv}{gr-qc/9712052}].

\bibitem{Frauendiener:1998yi}\bibhead{Frauendiener:1998yi}
J\"org Frauendiener: {Numerical treatment of the hyperboloidal initial value
  problem for the vacuum Einstein equations. 3. On the determination of
  radiation}.
\newblock {\em Class.Quant.Grav.}, 17:373--387, 2000.
\newblock [\epfmtdoi{10.1088/0264-9381/17/2/308},
  \epfmt{arxiv}{gr-qc/9808072}].

\bibitem{Frauendiener:2002iw}\bibhead{Frauendiener:2002iw}
J\"org Frauendiener: {Some aspects of the numerical treatment of the conformal
  field equations}.
\newblock {\em Lect.Notes Phys.}, 604:261--282, 2002.
\newblock [\epfmt{arxiv}{gr-qc/0207093}].

\bibitem{lrr-2004-1}\bibhead{lrr-2004-1}
J\"org Frauendiener: Conformal infinity.
\newblock {\em Living Reviews in Relativity}, 7(1), 2004.
\newblock [\epfmtdoi{10.12942/lrr-2004-1}].

\bibitem{Frauendiener1999475}\bibhead{Frauendiener1999475}
Jörg Frauendiener: {Calculating initial data for the conformal Einstein
  equations by pseudo-spectral methods}.
\newblock {\em Journal of Computational and Applied Mathematics},
  109(1–2):475 -- 491, 1999.
\newblock [\epfmtdoi{http://dx.doi.org/10.1016/S0377-0427(99)00168-5},
  \epfmt{arxiv}{gr-qc/9806103}].

\bibitem{Fredrich:aaa}\bibhead{Fredrich:aaa}
Helmut Friedrich: {On the Regular and the Asymptotic Characteristic Initial
  Value Problem for Einstein{\textquoteright}s Vacuum Field Equations}.
\newblock {\em {Proc.Roy.Soc.Lond.}}, A375(1761):169--184, 1981.
\newblock [\epfmtdoi{10.1098/rspa.1981.0045}].

\bibitem{Friedrich:1981wx}\bibhead{Friedrich:1981wx}
Helmut Friedrich: {The asymptotic characteristic initial value problem for
  Einstein's vacuum field equations as an initial value problem for a
  first-order quasilinear symmetric hyperbolic system}.
\newblock {\em Proc.Roy.Soc.Lond.}, A378:401--421, 1981.
\newblock [\epfmtdoi{10.1098/rspa.1981.0159}].

\bibitem{friedrich1983}\bibhead{friedrich1983}
Helmut Friedrich: Cauchy problems for the conformal vacuum field equations in
  general relativity.
\newblock {\em Comm. Math. Phys.}, 91(4):445--472, 1983.

\bibitem{Friedrich:2003fq}\bibhead{Friedrich:2003fq}
Helmut Friedrich: {Radiative gravitational fields and asymptotically static or
  stationary initial data}.
\newblock {\em Lect. Notes Phys.}, 604, 2002.
\newblock [\epfmt{arxiv}{gr-qc/0304003}].

\bibitem{1983RSPSA.385..345F}\bibhead{1983RSPSA.385..345F}
Helmut {Friedrich} and John~M. {Stewart}: {Characteristic Initial Data and
  Wavefront Singularities in General Relativity}.
\newblock {\em Royal Society of London Proceedings Series A}, 385:345--371,
  February 1983.
\newblock [\epfmtdoi{10.1098/rspa.1983.0018}].

\bibitem{Gentle:2000aq}\bibhead{Gentle:2000aq}
Adrian~P. Gentle, Daniel~E. Holz, Arkady Kheyfets, Pablo Laguna, Warner~A.
  Miller, et~al.: {Constant crunch coordinates for black hole simulations}.
\newblock {\em Phys.Rev.}, D63:064024, 2001.
\newblock [\epfmtdoi{10.1103/PhysRevD.63.064024},
  \epfmt{arxiv}{gr-qc/0005113}].

\bibitem{Gundlach:2004jp}\bibhead{Gundlach:2004jp}
Carsten Gundlach and Jose~M. Martin-Garcia: {Symmetric hyperbolicity and
  consistent boundary conditions for second order Einstein equations}.
\newblock {\em Phys.Rev.}, D70:044032, 2004.
\newblock [\epfmtdoi{10.1103/PhysRevD.70.044032},
  \epfmt{arxiv}{gr-qc/0403019}].

\bibitem{Gundlach:2006tw}\bibhead{Gundlach:2006tw}
Carsten Gundlach and Jose~M. Martin-Garcia: {Well-posedness of formulations of
  the Einstein equations with dynamical lapse and shift conditions}.
\newblock {\em Phys.Rev.}, D74:024016, 2006.
\newblock [\epfmtdoi{10.1103/PhysRevD.74.024016},
  \epfmt{arxiv}{gr-qc/0604035}].

\bibitem{Gundlach:2005eh}\bibhead{Gundlach:2005eh}
Carsten Gundlach, Jose~M. Martin-Garcia, Gioel Calabrese, and Ian Hinder:
  {Constraint damping in the Z4 formulation and harmonic gauge}.
\newblock {\em Class.Quant.Grav.}, 22:3767--3774, 2005.
\newblock [\epfmtdoi{10.1088/0264-9381/22/17/025},
  \epfmt{arxiv}{gr-qc/0504114}].

\bibitem{PhysRevD.49.883}\bibhead{PhysRevD.49.883}
Carsten Gundlach, Richard~H. Price, and Jorge Pullin: {Late-time behavior of
  stellar collapse and explosions. I. Linearized perturbations}.
\newblock {\em Phys. Rev. D}, 49:883--889, Jan 1994.
\newblock [\epfmtdoi{10.1103/PhysRevD.49.883}].

\bibitem{gustafsson1995time}\bibhead{gustafsson1995time}
B.~Gustafsson, H.O. Kreiss, and J.~Oliger: {\em Time dependent problems and
  difference methods}.
\newblock Pure and applied mathematics. Wiley, 1995.

\bibitem{1964AnPhy..29..304H}\bibhead{1964AnPhy..29..304H}
S.~G. {Hahn} and R.~W. {Lindquist}: {The two-body problem in geometrodynamics}.
\newblock {\em Annals of Physics}, 29:304--331, September 1964.
\newblock [\epfmtdoi{10.1016/0003-4916(64)90223-4}].

\bibitem{Hannam:2006xw}\bibhead{Hannam:2006xw}
Mark Hannam, Sascha Husa, Bernd Bruegmann, Jose~A. Gonzalez, Ulrich Sperhake,
  et~al.: {Where do moving punctures go?}
\newblock {\em J.Phys.Conf.Ser.}, 66:012047, 2007.
\newblock [\epfmtdoi{10.1088/1742-6596/66/1/012047},
  \epfmt{arxiv}{gr-qc/0612097}].

\bibitem{Hannam:2009ib}\bibhead{Hannam:2009ib}
Mark Hannam, Sascha Husa, and Niall~O Murchadha: {Bowen-York trumpet data and
  black-hole simulations}.
\newblock {\em Phys.Rev.}, D80:124007, 2009.
\newblock [\epfmtdoi{10.1103/PhysRevD.80.124007}, \epfmt{arxiv}{0908.1063}].

\bibitem{Hannam:2006vv}\bibhead{Hannam:2006vv}
Mark Hannam, Sascha Husa, Denis Pollney, Bernd Bruegmann, and Niall
  O'Murchadha: {Geometry and regularity of moving punctures}.
\newblock {\em Phys.Rev.Lett.}, 99:241102, 2007.
\newblock [\epfmtdoi{10.1103/PhysRevLett.99.241102},
  \epfmt{arxiv}{gr-qc/0606099}].

\bibitem{Harms:2014dqa}\bibhead{Harms:2014dqa}
Enno Harms, Sebastiano Bernuzzi, Alessandro Nagar, and An~Zenginoglu: {A new
  gravitational wave generation algorithm for particle perturbations of the
  Kerr spacetime}.
\newblock {\em Class.Quant.Grav.}, 31(24):245004, 2014.
\newblock [\epfmtdoi{10.1088/0264-9381/31/24/245004},
  \epfmt{arxiv}{1406.5983}].

\bibitem{Hawking:1968qt}\bibhead{Hawking:1968qt}
Stephen Hawking: {Gravitational radiation in an expanding universe}.
\newblock {\em J.Math.Phys.}, 9:598--604, 1968.
\newblock [\epfmtdoi{10.1063/1.1664615}].

\bibitem{Hawking:1973uf}\bibhead{Hawking:1973uf}
S.W. Hawking and G.F.R. Ellis: {\em {The Large scale structure of space-time}}.
\newblock {Cambridge Monographs on Mathematical Physics}. {Cambridge University
  Press}, 1973.

\bibitem{Hubner:1999th}\bibhead{Hubner:1999th}
Peter H\"ubner: {A Scheme to numerically evolve data for the conformal Einstein
  equation}.
\newblock {\em Class.Quant.Grav.}, 16:2823--2843, 1999.
\newblock [\epfmtdoi{10.1088/0264-9381/16/9/302},
  \epfmt{arxiv}{gr-qc/9903088}].

\bibitem{Hubner:1998hn}\bibhead{Hubner:1998hn}
Peter H\"ubner: {How to avoid artificial boundaries in the numerical
  calculation of black hole spacetimes}.
\newblock {\em Class.Quant.Grav.}, 16:2145, 1999.
\newblock [\epfmtdoi{10.1088/0264-9381/16/7/301},
  \epfmt{arxiv}{gr-qc/9804065}].

\bibitem{Hubner:2000pb}\bibhead{Hubner:2000pb}
Peter H\"ubner: {From now to timelike infinity on a finite grid}.
\newblock {\em Class.Quant.Grav.}, 18:1871--1884, 2001.
\newblock [\epfmtdoi{10.1088/0264-9381/18/10/305},
  \epfmt{arxiv}{gr-qc/0010069}].

\bibitem{Hubner:2000zn}\bibhead{Hubner:2000zn}
Peter H\"ubner: {Numerical calculation of conformally smooth hyperboloidal
  data}.
\newblock {\em Class.Quant.Grav.}, 18:1421--1440, 2001.
\newblock [\epfmtdoi{10.1088/0264-9381/18/8/302},
  \epfmt{arxiv}{gr-qc/0010052}].

\bibitem{Hubner:1994pd}\bibhead{Hubner:1994pd}
{H\"ubner, Peter}: {A Method for calculating the structure of (singular)
  space-times in the large}.
\newblock {\em Phys.Rev.}, D53:701--721, 1996.
\newblock [\epfmtdoi{10.1103/PhysRevD.53.701}, \epfmt{arxiv}{gr-qc/9409029}].

\bibitem{Hulse:1974eb}\bibhead{Hulse:1974eb}
R.A. Hulse and J.H. Taylor: {Discovery of a pulsar in a binary system}.
\newblock {\em Astrophys.J.}, 195:L51--L53, 1975.
\newblock [\epfmtdoi{10.1086/181708}].

\bibitem{Husa:2002kk}\bibhead{Husa:2002kk}
Sascha Husa: {Problems and successes in the numerical approach to the conformal
  field equations}.
\newblock {\em Lect.Notes Phys.}, 604:239--260, 2002.
\newblock [\epfmt{arxiv}{gr-qc/0204043}].

\bibitem{Husa:2002zc}\bibhead{Husa:2002zc}
Sascha Husa: {Numerical relativity with the conformal field equations}.
\newblock {\em Lect.Notes Phys.}, 617:159--192, 2003.
\newblock [\epfmt{arxiv}{gr-qc/0204057}].

\bibitem{Husa:2007zz}\bibhead{Husa:2007zz}
Sascha Husa: {Numerical modeling of black holes as sources of gravitational
  waves in a nutshell}.
\newblock {\em Eur. Phys. J. ST}, 152:183--207, 2007.
\newblock [\epfmtdoi{10.1140/epjst/e2007-00381-6}, \epfmt{arxiv}{0812.4395}].

\bibitem{Husa:2007hp}\bibhead{Husa:2007hp}
Sascha Husa, Jose~A. Gonzalez, Mark Hannam, Bernd Bruegmann, and Ulrich
  Sperhake: {Reducing phase error in long numerical binary black hole
  evolutions with sixth order finite differencing}.
\newblock {\em Class.Quant.Grav.}, 25:105006, 2008.
\newblock [\epfmtdoi{10.1088/0264-9381/25/10/105006},
  \epfmt{arxiv}{0706.0740}].

\bibitem{Husa:2005ns}\bibhead{Husa:2005ns}
Sascha Husa, Carsten Schneemann, Tilman Vogel, and An{\i}l Zengino\u{g}lu:
  {Hyperboloidal data and evolution}.
\newblock {\em AIP Conf.Proc.}, 841:306--313, 2006.
\newblock [\epfmtdoi{10.1063/1.2218186}, \epfmt{arxiv}{gr-qc/0512033}].

\bibitem{Iriondo:1995ar}\bibhead{Iriondo:1995ar}
Mirta Iriondo, Edward Malec, and Niall~O. Murchadha: {The Constant mean
  curvature slices of asymptotically flat spherical space-times}.
\newblock {\em Phys.Rev.}, D54:4792--4798, 1996.
\newblock [\epfmtdoi{10.1103/PhysRevD.54.4792}, \epfmt{arxiv}{gr-qc/9503030}].

\bibitem{Kennefick:1997kb}\bibhead{Kennefick:1997kb}
Daniel Kennefick: {Controversies in the history of the radiation reaction
  problem in general relativity}.
\newblock 1997.
\newblock [\epfmt{arxiv}{gr-qc/9704002}].

\bibitem{Kidder:2001tz}\bibhead{Kidder:2001tz}
Lawrence~E. Kidder, Mark~A. Scheel, and Saul~A. Teukolsky: {Extending the
  lifetime of 3-D black hole computations with a new hyperbolic system of
  evolution equations}.
\newblock {\em Phys.Rev.}, D64:064017, 2001.
\newblock [\epfmtdoi{10.1103/PhysRevD.64.064017},
  \epfmt{arxiv}{gr-qc/0105031}].

\bibitem{kreiss1973methods}\bibhead{kreiss1973methods}
H.O. Kreiss and J.~Oliger: {\em Methods for the approximate solution of time
  dependent problems}.
\newblock GARP publications series No. 10. International Council of Scientific
  Unions, World Meteorological Organization, 1973.

\bibitem{CPA:CPA3160090206}\bibhead{CPA:CPA3160090206}
P.~D. Lax and R.~D. Richtmyer: Survey of the stability of linear finite
  difference equations.
\newblock {\em Communications on Pure and Applied Mathematics}, 9(2):267--293,
  1956.
\newblock [\epfmtdoi{10.1002/cpa.3160090206}].

\bibitem{Lichnerowicz:d}\bibhead{Lichnerowicz:d}
A.~Lichnerowicz: {L’integration des \'equations de la gravitation relativiste
  et le probl\'eme des n corps}.
\newblock {\em J. Math. Pures Appl.}, 23:37–63, 1944.

\bibitem{Lovelace:2010ne}\bibhead{Lovelace:2010ne}
Geoffrey Lovelace, Mark.A. Scheel, and Bela Szilagyi: {Simulating merging
  binary black holes with nearly extremal spins}.
\newblock {\em Phys.Rev.}, D83:024010, 2011.
\newblock [\epfmtdoi{10.1103/PhysRevD.83.024010}, \epfmt{arxiv}{1010.2777}].

\bibitem{Malec:2003dq}\bibhead{Malec:2003dq}
Edward Malec and Niall~O Murchadha: {Constant mean curvature slices in the
  extended Schwarzschild solution and collapse of the lapse. Part I}.
\newblock {\em Phys.Rev.}, D68:124019, 2003.
\newblock [\epfmtdoi{10.1103/PhysRevD.68.124019},
  \epfmt{arxiv}{gr-qc/0307046}].

\bibitem{Marsa:1996fa}\bibhead{Marsa:1996fa}
R.L. Marsa and M.W. Choptuik: {Black hole scalar field interactions in
  spherical symmetry}.
\newblock {\em Phys.Rev.}, D54:4929--4943, 1996.
\newblock [\epfmtdoi{10.1103/PhysRevD.54.4929}, \epfmt{arxiv}{gr-qc/9607034}].

\bibitem{xAct}\bibhead{xAct}
José~M. Martín-García: {{\em xAct:} Efficient tensor computer algebra for
  {\it Mathematica}}.
\newblock {\tt http://www.xact.es/}.

\bibitem{Misner:1964je}\bibhead{Misner:1964je}
Charles~W. Misner and David~H. Sharp: {Relativistic equations for adiabatic,
  spherically symmetric gravitational collapse}.
\newblock {\em Phys.Rev.}, 136:B571--B576, 1964.
\newblock [\epfmtdoi{10.1103/PhysRev.136.B571}].

\bibitem{talkmoncrief}\bibhead{talkmoncrief}
{Moncrief, Vincent}: {talk given at the workshop on Mathematical Issues in
  Numerical Relativity held at ITP Santa Barbara from January 10 – 28}, 2000.
\newblock Online version available at
  http://online.itp.ucsb.edu/online/numrel00/moncrief/.

\bibitem{Montero:2012yr}\bibhead{Montero:2012yr}
Pedro~J. Montero and Isabel Cordero-Carrion: {BSSN equations in spherical
  coordinates without regularization: vacuum and non-vacuum spherically
  symmetric spacetimes}.
\newblock {\em Phys.Rev.}, D85:124037, 2012.
\newblock [\epfmtdoi{10.1103/PhysRevD.85.124037}, \epfmt{arxiv}{1204.5377}].

\bibitem{Nagy:2004td}\bibhead{Nagy:2004td}
Gabriel Nagy, Omar~E. Ortiz, and Oscar~A. Reula: {Strongly hyperbolic second
  order Einstein's evolution equations}.
\newblock {\em Phys.Rev.}, D70:044012, 2004.
\newblock [\epfmtdoi{10.1103/PhysRevD.70.044012},
  \epfmt{arxiv}{gr-qc/0402123}].

\bibitem{NOK}\bibhead{NOK}
T.~Nakamura, K.~Oohara, and Y.~Kojima: {General relativistic collapse to black
  holes and gravitational waves from black holes}.
\newblock {\em Prog. Theor. Phys. Suppl.}, 90:1--218, 1987.

\bibitem{newman1962approach}\bibhead{newman1962approach}
Ezra Newman and Roger Penrose: An approach to gravitational radiation by a
  method of spin coefficients.
\newblock {\em Journal of Mathematical Physics}, 3(3):566--578, 1962.

\bibitem{Ohme:2009gn}\bibhead{Ohme:2009gn}
Frank Ohme, Mark Hannam, Sascha Husa, and Niall~O Murchadha: {Stationary
  hyperboloidal slicings with evolved gauge conditions}.
\newblock {\em Class.Quant.Grav.}, 26:175014, 2009.
\newblock [\epfmtdoi{10.1088/0264-9381/26/17/175014},
  \epfmt{arxiv}{0905.0450}].

\bibitem{O'Murchadha:1974nc}\bibhead{O'Murchadha:1974nc}
Niall O'Murchadha and James~W. York: {Initial - value problem of general
  relativity. 1. General formulation and physical interpretation}.
\newblock {\em Phys.Rev.}, D10:428--436, 1974.
\newblock [\epfmtdoi{10.1103/PhysRevD.10.428}].

\bibitem{O'Murchadha:1974nd}\bibhead{O'Murchadha:1974nd}
Niall O'Murchadha and James~W. York: {Initial-value problem of general
  relativity. 2. Stability of solutions of the initial-value equations}.
\newblock {\em Phys.Rev.}, D10:437--446, 1974.
\newblock [\epfmtdoi{10.1103/PhysRevD.10.437}].

\bibitem{Ossokine:2013zga}\bibhead{Ossokine:2013zga}
Serguei Ossokine, Lawrence~E. Kidder, and Harald~P. Pfeiffer:
  {Precession-tracking coordinates for simulations of compact-object-binaries}.
\newblock {\em Phys.Rev.}, D88:084031, 2013.
\newblock [\epfmtdoi{10.1103/PhysRevD.88.084031}, \epfmt{arxiv}{1304.3067}].

\bibitem{PhysRevLett.10.66}\bibhead{PhysRevLett.10.66}
Roger Penrose: Asymptotic properties of fields and space-times.
\newblock {\em Phys. Rev. Lett.}, 10:66--68, Jan 1963.
\newblock [\epfmtdoi{10.1103/PhysRevLett.10.66}].

\bibitem{Penrose:1965am}\bibhead{Penrose:1965am}
Roger Penrose: {Zero rest mass fields including gravitation: Asymptotic
  behavior}.
\newblock {\em Proc.Roy.Soc.Lond.}, A284:159, 1965.
\newblock [\epfmtdoi{10.1098/rspa.1965.0058}].

\bibitem{9780511564048}\bibhead{9780511564048}
Roger Penrose and Wolfgang Rindler: {\em Spinors and Space-Time}, volume 1, 2.
\newblock Cambridge University Press, 1984, 1986.
\newblock Cambridge Books Online.

\bibitem{Pollney:2009ut}\bibhead{Pollney:2009ut}
Denis Pollney, Christian Reisswig, Nils Dorband, Erik Schnetter, and Peter
  Diener: {The Asymptotic Falloff of Local Waveform Measurements in Numerical
  Relativity}.
\newblock {\em Phys.Rev.}, D80:121502, 2009.
\newblock [\epfmtdoi{10.1103/PhysRevD.80.121502}, \epfmt{arxiv}{0910.3656}].

\bibitem{Pollney:2009yz}\bibhead{Pollney:2009yz}
Denis Pollney, Christian Reisswig, Erik Schnetter, Nils Dorband, and Peter
  Diener: {High accuracy binary black hole simulations with an extended wave
  zone}.
\newblock {\em Phys.Rev.}, D83:044045, 2011.
\newblock [\epfmtdoi{10.1103/PhysRevD.83.044045}, \epfmt{arxiv}{0910.3803}].

\bibitem{PhysRevD.5.2419}\bibhead{PhysRevD.5.2419}
Richard~H. Price: Nonspherical perturbations of relativistic gravitational
  collapse. i. scalar and gravitational perturbations.
\newblock {\em Phys. Rev. D}, 5:2419--2438, May 1972.
\newblock [\epfmtdoi{10.1103/PhysRevD.5.2419}].

\bibitem{Reisswig:2009us}\bibhead{Reisswig:2009us}
C.~Reisswig, N.T. Bishop, D.~Pollney, and B.~Szilagyi: {Unambiguous
  determination of gravitational waveforms from binary black hole mergers}.
\newblock {\em Phys.Rev.Lett.}, 103:221101, 2009.
\newblock [\epfmtdoi{10.1103/PhysRevLett.103.221101},
  \epfmt{arxiv}{0907.2637}].

\bibitem{Rinne:2009qx}\bibhead{Rinne:2009qx}
Oliver Rinne: {An Axisymmetric evolution code for the Einstein equations on
  hyperboloidal slices}.
\newblock {\em Class.Quant.Grav.}, 27:035014, 2010.
\newblock [\epfmtdoi{10.1088/0264-9381/27/3/035014}, \epfmt{arxiv}{0910.0139}].

\bibitem{Rinne:2013qc}\bibhead{Rinne:2013qc}
Oliver Rinne and Vincent Moncrief: {Hyperboloidal Einstein-matter evolution and
  tails for scalar and Yang-Mills fields}.
\newblock {\em Class.Quant.Grav.}, 30:095009, 2013.
\newblock [\epfmtdoi{10.1088/0264-9381/30/9/095009}, \epfmt{arxiv}{1301.6174}].

\bibitem{Sachs:1961zz}\bibhead{Sachs:1961zz}
R.K. Sachs: {Gravitational waves in general relativity. VI. The outgoing
  radiation condition}.
\newblock {\em Proc.Roy.Soc.Lond.}, A264:309--338, 1961.
\newblock [\epfmtdoi{10.1098/rspa.1961.0202}].

\bibitem{Sachs:1962wk}\bibhead{Sachs:1962wk}
R.K. Sachs: {Gravitational waves in general relativity. VIII. Waves in
  asymptotically flat space-times}.
\newblock {\em Proc.Roy.Soc.Lond.}, A270:103--126, 1962.
\newblock [\epfmtdoi{10.1098/rspa.1962.0206}].

\bibitem{Sarbach:2002bt}\bibhead{Sarbach:2002bt}
Olivier Sarbach, Gioel Calabrese, Jorge Pullin, and Manuel Tiglio:
  {Hyperbolicity of the BSSN system of Einstein evolution equations}.
\newblock {\em Phys.Rev.}, D66:064002, 2002.
\newblock [\epfmtdoi{10.1103/PhysRevD.66.064002},
  \epfmt{arxiv}{gr-qc/0205064}].

\bibitem{Scheel:1997kb}\bibhead{Scheel:1997kb}
Mark~A. Scheel, Thomas~W. Baumgarte, Gregory~B. Cook, Stuart~L. Shapiro, and
  Saul~A. Teukolsky: {Numerical evolution of black holes with a hyperbolic
  formulation of general relativity}.
\newblock {\em Phys.Rev.}, D56:6320--6335, 1997.
\newblock [\epfmtdoi{10.1103/PhysRevD.56.6320}, \epfmt{arxiv}{gr-qc/9708067}].

\bibitem{Scheel:2014ina}\bibhead{Scheel:2014ina}
Mark~A. Scheel, Matthew Giesler, Daniel~A. Hemberger, Geoffrey Lovelace, Kevin
  Kuper, et~al.: {Improved methods for simulating nearly extremal binary black
  holes}.
\newblock {\em Class.Quant.Grav.}, 32(10):105009, 2015.
\newblock [\epfmtdoi{10.1088/0264-9381/32/10/105009},
  \epfmt{arxiv}{1412.1803}].

\bibitem{Schinkel:2013zm}\bibhead{Schinkel:2013zm}
David Schinkel, Marcus Ansorg, and Rodrigo Panosso~Macedo: {Initial data for
  perturbed Kerr black holes on hyperboloidal slices}.
\newblock {\em Class.Quant.Grav.}, 31:165001, 2014.
\newblock [\epfmtdoi{10.1088/0264-9381/31/16/165001},
  \epfmt{arxiv}{1301.6984}].

\bibitem{Schneemann}\bibhead{Schneemann}
Carsten Schneemann: {Numerische Berechnung von hyperboloidalen Anfangsdaten
  f\"ur die Einstein-Gleichungen}.
\newblock Master's thesis, 2006.

\bibitem{Seidel:1992vd}\bibhead{Seidel:1992vd}
Edward Seidel and Wai-Mo Suen: {Towards a singularity proof scheme in numerical
  relativity}.
\newblock {\em Phys.Rev.Lett.}, 69:1845--1848, 1992.
\newblock [\epfmtdoi{10.1103/PhysRevLett.69.1845},
  \epfmt{arxiv}{gr-qc/9210016}].

\bibitem{PhysRevD.52.5428}\bibhead{PhysRevD.52.5428}
Masaru Shibata and Takashi Nakamura: Evolution of three-dimensional
  gravitational waves: Harmonic slicing case.
\newblock {\em Phys. Rev. D}, 52:5428--5444, Nov 1995.
\newblock [\epfmtdoi{10.1103/PhysRevD.52.5428}].

\bibitem{Smarr:1976qy}\bibhead{Smarr:1976qy}
L.~Smarr, A.~Cadez, Bryce~S. DeWitt, and K.~Eppley: {Collision of Two Black
  Holes: Theoretical Framework}.
\newblock {\em Phys.Rev.}, D14:2443--2452, 1976.
\newblock [\epfmtdoi{10.1103/PhysRevD.14.2443}].

\bibitem{Smarr:1977uf}\bibhead{Smarr:1977uf}
Larry Smarr and James~W. York, Jr.: {Kinematical conditions in the construction
  of space-time}.
\newblock {\em Phys.Rev.}, D17:2529--2551, 1978.
\newblock [\epfmtdoi{10.1103/PhysRevD.17.2529}].

\bibitem{Smarr:1978dia}\bibhead{Smarr:1978dia}
Larry Smarr and James~W. York, Jr.: {Radiation gauge in general relativity}.
\newblock {\em Phys.Rev.}, D17(8):1945--1956, 1978.
\newblock [\epfmtdoi{10.1103/PhysRevD.17.1945}].

\bibitem{Sorkin:2009bc}\bibhead{Sorkin:2009bc}
Evgeny Sorkin and Matthew~W. Choptuik: {Generalized harmonic formulation in
  spherical symmetry}.
\newblock {\em Gen. Rel. Grav.}, 42:1239--1286, 2010.
\newblock [\epfmtdoi{10.1007/s10714-009-0905-8}, \epfmt{arxiv}{0908.2500}].

\bibitem{stewart1997advanced}\bibhead{stewart1997advanced}
John Stewart: {\em Advanced General Relativity}.
\newblock Mir, 1997.

\bibitem{stefried}\bibhead{stefried}
John~M. Stewart and Helmut Friedrich: {Numerical Relativity. I. The
  Characteristic Initial Value Problem}.
\newblock {\em Proceedings of the Royal Society of London A: Mathematical,
  Physical and Engineering Sciences}, 384(1787):427--454, 1982.
\newblock [\epfmtdoi{10.1098/rspa.1982.0166}].

\bibitem{Tamburino:1966zz}\bibhead{Tamburino:1966zz}
Louis~A. Tamburino and Jeffrey~H. Winicour: {Gravitational Fields in Finite and
  Conformal Bondi Frames}.
\newblock {\em Phys.Rev.}, 150:1039--1053, 1966.
\newblock [\epfmtdoi{10.1103/PhysRev.150.1039}].

\bibitem{Taylor:2013zia}\bibhead{Taylor:2013zia}
Nicholas~W. Taylor, Michael Boyle, Christian Reisswig, Mark~A. Scheel, Tony
  Chu, et~al.: {Comparing Gravitational Waveform Extrapolation to
  Cauchy-Characteristic Extraction in Binary Black Hole Simulations}.
\newblock {\em Phys.Rev.}, D88(12):124010, 2013.
\newblock [\epfmtdoi{10.1103/PhysRevD.88.124010}, \epfmt{arxiv}{1309.3605}].

\bibitem{Tuite:2013hza}\bibhead{Tuite:2013hza}
Patrick Tuite and Niall~Ó Murchadha: {Constant Mean Curvature Slices of the
  Reissner-Nordstr\"{o}m Spacetime}.
\newblock 2013.
\newblock [\epfmt{arxiv}{1307.4657}].

\bibitem{ValienteKroon:2004gj}\bibhead{ValienteKroon:2004gj}
Juan~Antonio Valiente~Kroon: {Asymptotic expansions of the Cotton-York tensor
  on slices of stationary space-times}.
\newblock {\em Class.Quant.Grav.}, 21:3237--3250, 2004.
\newblock [\epfmtdoi{10.1088/0264-9381/21/13/009},
  \epfmt{arxiv}{gr-qc/0402033}].

\bibitem{vanMeter:2006vi}\bibhead{vanMeter:2006vi}
James~R. van Meter, John~G. Baker, Michael Koppitz, and Dae-Il Choi: {How to
  move a black hole without excision: Gauge conditions for the numerical
  evolution of a moving puncture}.
\newblock {\em Phys.Rev.}, D73:124011, 2006.
\newblock [\epfmtdoi{10.1103/PhysRevD.73.124011},
  \epfmt{arxiv}{gr-qc/0605030}].

\bibitem{Vano-Vinuales:2014ada}\bibhead{Vano-Vinuales:2014ada}
Alex Va{\~n}{\'o}-Vi{\~n}uales and Sascha Husa: {Unconstrained hyperboloidal
  evolution of black holes in spherical symmetry with GBSSN and Z4c}.
\newblock {\em J.Phys.Conf.Ser.}, 600(1):012061, 2015.
\newblock [\epfmtdoi{10.1088/1742-6596/600/1/012061},
  \epfmt{arxiv}{1412.4801}].

\bibitem{Vano-Vinuales:2014koa}\bibhead{Vano-Vinuales:2014koa}
Alex Va{\~n}{\'o}-Vi{\~n}uales, Sascha Husa, and David Hilditch: {Spherical
  symmetry as a test case for unconstrained hyperboloidal evolution}.
\newblock {\em Class. Quant. Grav.}, 32(17):175010, 2015.
\newblock [\epfmtdoi{10.1088/0264-9381/32/17/175010},
  \epfmt{arxiv}{1412.3827}].

\bibitem{Wald}\bibhead{Wald}
Robert~M. Wald: {\em {General Relativity}}.
\newblock The University of Chicago Press, 1984.

\bibitem{Weyhausen:2011cg}\bibhead{Weyhausen:2011cg}
Andreas Weyhausen, Sebastiano Bernuzzi, and David Hilditch: {Constraint damping
  for the Z4c formulation of general relativity}.
\newblock {\em Phys.Rev.}, D85:024038, 2012.
\newblock [\epfmtdoi{10.1103/PhysRevD.85.024038}, \epfmt{arxiv}{1107.5539}].

\bibitem{Winicour:2005ge}\bibhead{Winicour:2005ge}
Jeffrey Winicour: {Characteristic evolution and matching}.
\newblock {\em Living Rev.Rel.}, 1:5, 1998.
\newblock [\epfmtdoi{10.12942/lrr-2012-2}, \epfmt{arxiv}{gr-qc/0102085}].

\bibitem{yamabe1960}\bibhead{yamabe1960}
Hidehiko Yamabe: {On a deformation of Riemannian structures on compact
  manifolds}.
\newblock {\em Osaka Math. J.}, 12(1):21--37, 1960.

\bibitem{York:1971hw}\bibhead{York:1971hw}
James~W. York, Jr.: {Gravitational degrees of freedom and the initial-value
  problem}.
\newblock {\em Phys.Rev.Lett.}, 26:1656--1658, 1971.
\newblock [\epfmtdoi{10.1103/PhysRevLett.26.1656}].

\bibitem{York:1972sj}\bibhead{York:1972sj}
James~W. York, Jr.: {Role of conformal three geometry in the dynamics of
  gravitation}.
\newblock {\em Phys.Rev.Lett.}, 28:1082--1085, 1972.
\newblock [\epfmtdoi{10.1103/PhysRevLett.28.1082}].

\bibitem{York:1973ia}\bibhead{York:1973ia}
James~W. York, Jr.: {Conformatlly invariant orthogonal decomposition of
  symmetric tensors on Riemannian manifolds and the initial value problem of
  general relativity}.
\newblock {\em J.Math.Phys.}, 14:456--464, 1973.
\newblock [\epfmtdoi{10.1063/1.1666338}].

\bibitem{York:1974psa}\bibhead{York:1974psa}
James~W. York, Jr.: {Covariant decompositions of symmetric tensors in the
  theory of gravitation}.
\newblock {\em Ann.Inst.Henri Poincare}, A21:319--332, 1974.

\bibitem{york}\bibhead{york}
James~W. York, Jr.: Kinematics and dynamics of general relativity.
\newblock In L.L. Smarr, editor, {\em Sources of Gravitational Radiation}, pp.
  83--126, Cambridge; New York, 1979. Cambridge University Press.

\bibitem{1979sgrr.work...83Y}\bibhead{1979sgrr.work...83Y}
James~W. {York}, Jr.: {Kinematics and dynamics of general relativity}.
\newblock In L.~L. {Smarr}, editor, {\em Sources of Gravitational Radiation},
  pp. 83--126, 1979.

\bibitem{York:1998hy}\bibhead{York:1998hy}
James~W. York, Jr.: {Conformal 'thin sandwich' data for the initial-value
  problem}.
\newblock {\em Phys.Rev.Lett.}, 82:1350--1353, 1999.
\newblock [\epfmtdoi{10.1103/PhysRevLett.82.1350},
  \epfmt{arxiv}{gr-qc/9810051}].

\bibitem{1982sag..conf..147Y}\bibhead{1982sag..conf..147Y}
James~W. {York}, Jr. and T.~{Piran}: {The Initial Value Problem and Beyond}.
\newblock In R.~A. {Matzner} and L.~C. {Shepley}, editors, {\em Spacetime and
  Geometry}, p. 147, 1982.

\bibitem{Zenginoglu:2011zz}\bibhead{Zenginoglu:2011zz}
Anil Zenginoglu and Gaurav Khanna: {Null infinity waveforms from
  extreme-mass-ratio inspirals in Kerr spacetime}.
\newblock {\em Phys.Rev.}, X1:021017, 2011.
\newblock [\epfmtdoi{10.1103/PhysRevX.1.021017}, \epfmt{arxiv}{1108.1816}].

\bibitem{Zenginoglu:2007it}\bibhead{Zenginoglu:2007it}
An{\i}l Zengino\u{g}lu: {\em {A conformal approach to numerical calculations of
  asymptotically flat spacetimes}}.
\newblock PhD thesis, 2007.
\newblock [\epfmt{arxiv}{0711.0873}].

\bibitem{Zenginoglu:2008wc}\bibhead{Zenginoglu:2008wc}
An{\i}l Zengino\u{g}lu: {A Hyperboloidal study of tail decay rates for scalar
  and Yang-Mills fields}.
\newblock {\em Class.Quant.Grav.}, 25:175013, 2008.
\newblock [\epfmtdoi{10.1088/0264-9381/25/17/175013},
  \epfmt{arxiv}{0803.2018}].

\bibitem{Zenginoglu:2008pw}\bibhead{Zenginoglu:2008pw}
An{\i}l Zengino\u{g}lu: {Hyperboloidal evolution with the Einstein equations}.
\newblock {\em Class.Quant.Grav.}, 25:195025, 2008.
\newblock [\epfmtdoi{10.1088/0264-9381/25/19/195025},
  \epfmt{arxiv}{0808.0810}].

\bibitem{Zenginoglu:2007jw}\bibhead{Zenginoglu:2007jw}
An{\i}l Zengino\u{g}lu: {Hyperboloidal foliations and scri-fixing}.
\newblock {\em Class.Quant.Grav.}, 25:145002, 2008.
\newblock [\epfmtdoi{10.1088/0264-9381/25/14/145002},
  \epfmt{arxiv}{0712.4333}].

\bibitem{Zenginoglu:2008uc}\bibhead{Zenginoglu:2008uc}
An{\i}l Zengino\u{g}lu, Dario Nunez, and Sascha Husa: {Gravitational
  perturbations of Schwarzschild spacetime at null infinity and the
  hyperboloidal initial value problem}.
\newblock {\em Class.Quant.Grav.}, 26:035009, 2009.
\newblock [\epfmtdoi{10.1088/0264-9381/26/3/035009}, \epfmt{arxiv}{0810.1929}].

\bibitem{Zlochower:2003yh}\bibhead{Zlochower:2003yh}
Yosef Zlochower, Roberto Gomez, Sascha Husa, Luis Lehner, and Jeffrey Winicour:
  {Mode coupling in the nonlinear response of black holes}.
\newblock {\em Phys.Rev.}, D68:084014, 2003.
\newblock [\epfmtdoi{10.1103/PhysRevD.68.084014},
  \epfmt{arxiv}{gr-qc/0306098}].

\end{thebibliography}
%%http://springerlink.com/content/r8j337151325/#section=183880&page=5&locus=75
%%http://grwiki.physics.ncsu.edu/wiki/Main_Page

\end{document}